%% file: lhc.tex
\documentclass[twoside]{article}

\pdfoutput=1

\usepackage{times}
\usepackage{amsmath,amssymb,mathrsfs}
\usepackage{graphicx}
\usepackage{feynmp}
\usepackage{makeidx}
\usepackage{url}
\usepackage{hyperref}
\makeatletter
\@ifundefined{pdfoutput}{}{\DeclareGraphicsRule{*}{mps}{*}{}}
\makeatother

\numberwithin{equation}{section}
\makeindex
\raggedright

\input{declare}

\begin{document}

\title{Lectures on LHC Physics}

\author{Tilman Plehn \\[10mm]
        Institut f\"ur Theoretische Physik \\ Universit\"at Heidelberg \\[15mm]
        {\small arXiv:0910.4182, always preliminary version, this one from February 17, 2014}
        \footnote{Lecture Notes
          in Physics, Vol 844, Springer 2012; 
          second edition planned Spring 2014;
          updated
          under \url{www.thphys.uni-heidelberg.de/\~plehn}}}

\date{}

\maketitle
\thispagestyle{empty}

\begin{abstract}
  With the discovery of the Higgs boson the LHC experiments have
  closed the most important gap in our understanding of fundamental
  interactions. We now know that the interactions between elementary
  particles can be described by quantum field theory, more
  specifically by a renormalizable gauge theory. This theory is valid
  to arbitrarily high energy scales and do not require an ultraviolet
  completion. In these notes I cover three aspects to help understand
  LHC results in the Higgs sector and in searches for physics beyond
  the Standard Model: many facets of Higgs physics, QCD as it is
  relevant for LHC measurements, and standard phenomenological
  background knowledge.  The lectures should put young graduate
  students into a position to really follow advanced writeups and first
  research papers. In that sense they can serve as a starting point
  for a research project in LHC physics. With this new, significantly
  expanded version I am confident that also some more senior
  colleagues will find them useful and interesting.
\end{abstract}

\newpage

\null\vspace{\stretch{1}}
\begin{flushleft}
for Thomas Binoth
\end{flushleft}
\vspace{\stretch{2}}\null
\thispagestyle{empty}

\newpage

\tableofcontents 

\begin{fmffile}{feynman}

\newpage

\section*{From the author}

These notes are based on lectures at Heidelberg University between the
Summer 2009 and in the Winter 2013/2014, written up in coffee shops
around the world. Obviously, in the Fall of 2012 they were heavily
adapted to the new and exciting experimental realities. It felt great
to rewrite the Higgs chapter from a careful description of possible
experimental signals into a description of an actual experimental
research program. I promise I will do it again once the LHC discovers
physics beyond the Standard Model.\bigskip

To those familiar with the German system it will be obvious that the
target audience of the lecture are students who know field theory and
are starting to work on their master thesis; carefully studying these
notes should put you into a position to start actual research in LHC
physics. The way I prefer to learn physics is by calculating things on
a piece of paper or on the blackboard. This is why the notes look the
way they look.  Because this is not a text book there is less text in
the notes than actual talk during the lecture. So when reading these
notes, take a break here and there, get a coffee and think about the
physics behind the calculation you just followed.\bigskip

The text is divided into three main parts:
\begin{itemize}
\item[--] In the first part I focus on Higgs physics and collider
  searches. To understand what we are looking for I start with the
  most minimalistic and not renormalizable models describing massive
  gauge bosons. I then slowly advance to the usual fundamental Higgs
  scalar we are really searching for.  At the end of this part what
  everybody should understand is the usual set of ATLAS or CMS graphs
  shown in Figure~\ref{fig:higgs_sig}, where many colored lines
  represent different search channels and their discovery
  potential. Many QCD issues affecting Higgs searches I will skip in
  the Higgs part and postpone to the...

\item[--] ...QCD part. Here, I am taking at least one step back and
  study the theory which describes Higgs production and everything
  else at the LHC. Two core discussions shape this part: first, I
  derive the DGLAP equation by constructing the splitting
  kernels. This leads us to the parton shower and to the physical
  interpretation of resumming different logarithms in the QCD
  perturbation series. Second, there are two modern approaches
  combining parton shower and matrix element descriptions of jet
  radiation, which I introduce at least on the level of simplified
  models. Throughout the QCD discussion I avoid the more historically
  interesting deep inelastic scattering process and instead rely on
  the Drell--Yan process or its inverted $R$ ratio process for
  motivation and illustration. Because the first two parts of the
  lecture notes are really advanced quantum field theory,
  something is missing: there are...

\item[--] ...many aspects of LHC physics we need to take into account
  once we look at experimental LHC results. Some of them, like old
  fashioned jets and fat jets, helicity amplitudes, or missing
  transverse energy I cover in the third part. This part will expand
  in the online version over the coming years while I will keep these
  lecture notes up to date with actual discussions of LHC data.
\end{itemize}

At the end there follows a brief sketch of how to compute a cross
section from a Lagrangian and without using Feynman rules. This is of
course not what we do, but the brief writeup has proven useful many
times.

What is almost entirely missing is an introduction to searches for new
physics completing the Standard Model of particle physics beyond the
weak scale. Covering this topic appropriately would at least double
the length of these notes. For the structure of such models and their
signatures I instead refer to our review article~\cite{bsm_review}
and in particular to its second chapter.\bigskip

Last, but not least, the literature listed at the end of each part
is not meant to cite original or relevant research papers. Instead, I
collected a set of review papers or advanced lecture notes
supplementing these lecture notes in different directions. 
Going through some of these mostly introductory papers will
be instructive and fun once the basics have been covered by these
lecture notes.\bigskip \bigskip \bigskip \bigskip

{\sl I am still confident that these notes are far
  from bug free. So if you read them and you
  did not email me at least a few typos and complaints about poor
  explanations, you did not read them carefully enough.}

\newpage

\input{chapters}

\newpage

\begin{center}
{\bf Acknowledgments}
\end{center}

The list of people I would like to thank is long and still growing:
starting with Peter Zerwas, Wim Beenakker, Roland H\"opker, Michael
Spira and Michael Kr\"amer I would like to thank all the people who
taught me theoretical physics and phenomenology over many years. This
later on included Tao Han, Dieter Zeppenfeld, Uli Baur, and Thomas
Binoth. The latter two hopefully got to watch the Higgs discovery from
their clouds up there. Since moving to Heidelberg it has been great
fun to benefit from the physics knowledge in our happy institute on
Philosophenweg. To all of these people I would like to say: I am very
sorry, but what you are holding in your hands is the best I could
do.\bigskip

Of all the great experimentalists who have taught me LHC physics I
would like to single out Dirk Zerwas and thank him continuous insight
into experimental physics, from our first semester at Heidelberg to
the last day of writing these notes. Another experimentalist, Kyle
Cranmer, taught me enough statistics to avoid major disasters, which
I am very grateful for.  On the other Heidelberg hill I would like to
thank Manfred Lindner for his many comments on my black board notes
when his class started right after mine. For anything to do with Higgs
couplings and statistics I would like to thank Michael Rauch for many
years of fun collaboration and for helping me with the text. As a
long--term collaborator I am missing Dave Rainwater who should have
stayed in LHC physics and who would now be the leading Higgs expert in
the US.\bigskip

The QCD part of this lecture is based on my 2008 TASI lectures, and I
would like to thank Tom DeGrand and Ben Allanach for their comments on
the TASI notes and Sally Dawson for her encouragement to put these
notes on the web. For this longer version I am indebted to Steffen
Schumann for helping me out on many QCD questions over the past years,
to Jan Pawlowski for uncountable cups of coffee on field theory and
QCD, to Fabio Maltoni and Johan Alwall for teaching me jet merging, to
Michelangelo Mangano for many detailed comments, and to Peter
Schichtel for helping me sort out many QCD topics. Alert readers like
David Lopez--Val, Sebastian Bock, Florian G\"ortz, Michael Spannowsky,
Martin Weber, Anja Butter, Malte Buschmann, and Manuel Scinta followed
my appeal to give me lists of typos, and Manuela Wirschke carefully
read the notes removing language mistakes --- thank you very much to
everyone who helped make this writeup more correct and more
readable.\bigskip

Most importantly, I would like to thank all the people who have
convinced me that theoretical physics even including QCD is fun --- at
least most of the time. Not surprisingly this includes many US
colleagues from our TASI year '97.

\end{fmffile}
 
\newpage

\printindex

\end{document}

%% file: declare.tex
\newcommand{\sop}{\mathcal{S}}
\newcommand{\hop}{\mathcal{H}}
\newcommand{\bra}[1]{\langle #1|}
\newcommand{\ket}[1]{|#1\rangle}
\newcommand{\braket}[2]{\langle #1|#2\rangle}
\newcommand{\vk}{\vec{k}}
\newcommand{\vp}{\vec{p}}

\newcommand{\vx}{\vec{x}}

\newcommand{\qlb}{\overline{Q}_L}
\newcommand{\qrb}{\overline{Q}_R}
\newcommand{\ql}{Q_L}
\newcommand{\qr}{Q_R}
\newcommand{\elb}{\overline{L}_L}
\newcommand{\erb}{\overline{L}_R}
\newcommand{\el}{L_L}
\newcommand{\er}{L_R}
\newcommand{\psib}{\overline{\psi}}
\newcommand{\Psib}{\overline{\Psi}}

\newcommand\one{\leavevmode\hbox{\small1\normalsize\kern-.33em1}}
\newcommand{\p}{\partial}
\newcommand{\w}{\vec{w}}
\newcommand{\sdag}{\Sigma^{\dag}}
\newcommand{\s}{\Sigma}
\newcommand{\msbar}{\overline{\text{MS}}}
\newcommand{\qns}{f_q^\text{NS}}
\newcommand{\met}{\slashchar{E}_T}
\newcommand{\mst}{m_{\tilde {t}}}

\newcommand{\pro}{\mathbb{P}}
\newcommand{\prol}{\mathbb{P}_L}
\newcommand{\pror}{\mathbb{P}_R}
\newcommand{\prolr}{\mathbb{P}_{L,R}}
\newcommand{\prorl}{\mathbb{P}_{R,L}}

\newcommand{\lag}{\mathscr{L}}
\newcommand{\lumi}{\mathcal{L}}
\newcommand{\ope}{\mathcal{O}}
\newcommand{\matx}{\left |\mathcal{M} \right|^2}
\newcommand{\mat}{\mathcal{M}}
\newcommand{\qqquad}{\qquad \qquad}
\newcommand{\qqqquad}{\qquad \qquad \qquad}

\newcommand{\lqcd}{\Lambda_\text{QCD}}
\newcommand{\ltc}{\Lambda_\text{T}}
\newcommand{\letc}{\Lambda_\text{ETC}}
\newcommand{\ftc}{f_\text{T}}
\newcommand{\tetc}{T_\text{ETC}}
\newcommand{\really}{\stackrel{!}{=}}

\newcommand{\mev}{\text{MeV}}
\newcommand{\gev}{\text{GeV}}
\newcommand{\tev}{\text{TeV}}

\newcommand{\br}{\text{BR}}
\newcommand{\sign}{\text{sign}}
\newcommand{\ifb}{\text{fb}^{-1}}

\def\slashchar#1{\setbox0=\hbox{$#1$}           
   \dimen0=\wd0                                 
   \setbox1=\hbox{/} \dimen1=\wd1               
   \ifdim\dimen0>\dimen1                        
      \rlap{\hbox to \dimen0{\hfil/\hfil}}      
      #1                                        
   \else                                        
      \rlap{\hbox to \dimen1{\hfil$#1$\hfil}}   
      /                                         
   \fi}
\newcommand{\dslash}{\slashchar{\partial}}
\newcommand{\Dslash}{\slashchar{D}}

\def\eg{{\sl e.g.} \,}
\def\ie{{\sl i.e.} \,}
\DeclareMathOperator{\tr}{Tr}

%% file: chapters.tex
\section{Higgs physics}
\label{sec:higgs}

Understanding the nature of electroweak symmetry breaking --- or
slightly more specifically deciphering the Higgs mechanism --- is the
main goal of the ATLAS and CMS experiments at the LHC. Observing some
kind of Higgs boson and studying its properties involves many
experimental and theoretical issues focused around understanding
hadron collider data and QCD predictions to unprecedented
precision. The latter will be the main topic of the second half of
this lecture.

On the other hand, before we discuss the details of Higgs signatures,
backgrounds, and related QCD aspects we should start with a discussion
of electroweak symmetry breaking. Higgs physics at the LHC means much
more than just finding a light fundamental Higgs boson as predicted by
the Standard Model of particle physics.  As a matter of fact, the
discovery of a light Higgs boson was announced on July 4th, 2012, and
we will briefly discuss it in
Section~\ref{sec:higgs_discovery}.\bigskip

In our theory derivation in Section~\ref{sec:higgs_ewsb} we prefer to
follow an effective theory approach. This means we do not start by
writing down the Higgs potential and deriving the measured gauge boson
and fermion masses. Instead, we step by step include gauge boson and
fermion masses in our gauge theories, see what this means for the
field content, and show how we can embed this mechanism in a
renormalizable fundamental gauge theory. Only this last step will lead
us to the Standard Model and the Higgs potential. In
Section~\ref{sec:higgs_sm} we will return to the usual path and
discuss the properties of the renormalizable Standard Model including
high energy scales. This includes new physics effects in terms of
higher--dimensional operators in Section~\ref{sec:higgs_pot}, an
extended supersymmetric Higgs sector in Section~\ref{sec:higgs_2hdm},
and general effects of new particles in the Higgs potential in
Section~\ref{sec:higgs_coleman}.

In Section~\ref{sec:higgs_decays} we will start discussing Higgs
physics at colliders, leading us to the Higgs discovery papers
presented in Section~\ref{sec:higgs_discovery}. Higgs production in
gluon fusion, weak boson fusion, and in association with a gauge boson
will be in the focus of
Sections~\ref{sec:higgs_gf} to~\ref{sec:higgs_wh}, with a special focus
on QCD issues linked to jet radiation in
Section~\ref{sec:higgs_cjv}. The LHC experiments have shown that they
can not only discover a Higgs resonance, but also study many Higgs
properties, some of which we discuss in
Section~\ref{sec:higgs_beyond}.

In our approach to the Higgs mechanism it is clear that a fundamental
Higgs particle is not the only way to break the electroweak
symmetry. We therefore discuss alternative embeddings in strongly
interacting physics in Section~\ref{sec:higgs_alternatives}. This part
will also include a very brief introduction to the hierarchy
problem. Finally, in Section~\ref{sec:higgs_inflation} we will touch
on a slightly more speculative link between Higgs physics and
inflation.

\subsection{Electroweak symmetry breaking}
\label{sec:higgs_ewsb}

As a starting point, let us briefly remind ourselves what the Higgs
boson really is all about and sketch the Standard Model Lagrangian
with  mass terms for gauge bosons and
fermions. As a matter of fact, in a first step in
Section~\ref{sec:higgs_qed} we will try to make a photon
massive without introducing a physical Higgs field. Even for the
$SU(2)$ gauge theory of the electroweak Standard Model we might get
away without a fundamental Higgs boson, as we will show in
Section~\ref{sec:higgs_doublets}. Then, we will worry about quantum
fluctuations of the relevant degrees of freedom, which leads us to the
usual picture of the Higgs potential, the Higgs boson, and the
symmetries related to its implementation. This approach is not only
interesting because it allows us to understand the history of the
Higgs boson and identify the key steps in this triumph of quantum
field theory, it also corresponds to a modern effective field theory
picture of this sector. Such an effective field theory approach will
make it easy to ask the relevant questions about the experimental
results, guiding us towards the experimental confirmation of the Higgs
mechanism as part of the Standard Model of elementary particles.

\subsubsection{What masses?}
\label{sec:masses}

The relevance of the experimental Higgs discovery cannot be
over--stated. The fact that local gauge theories describe the
fundamental interactions of particles has been rewarded with a whole
list of experimental and theoretical Nobel prizes. What appears a
straightforward construction to us now has seen many challenges by
alternative approaches, some justified and some not all that
justified. The greatest feature of such a gauge theory lead to the
Nobel prize given to Gerald 't Hooft and Martinus Veltman: the absence
of a cutoff scale. Mathematically we call this validity to arbitrarily
high energy scales renormalizability, physically it means that the
Standard Model describing the interactions of quarks and leptons is
truly fundamental. There is the usual argument about the Planck scale
as an unavoidable cutoff and the apparent non--renormalizability of
gravity, but the final word on that is still missing.

What is important to notice is that the Nobel prize for `t Hooft and
Veltman would have had to be exchanged for a Fields medal without the
experimental discovery of the Higgs boson. Massive local gauge
theories are not fundamental without the Higgs mechanism, \ie without
spontaneous symmetry breaking in a relativistic framework with doublet
fields and hence predicting the existence of a narrow and relatively
light Higgs boson. The Higgs discovery is literally the keystone to a
fundamental theory of elementary particles --- without it the whole
construction would collapse.\bigskip


When people say that the Higgs mechanism is responsible for the masses
of (all) particles they actually mean very specific masses. These are
not the proton or neutron masses which are responsible for the masses
of people, furniture, or the moon. In a model describing the
fundamental interactions, the mass of the weak gauge bosons as the
exchange particles of the weak nuclear force has structural
impact. Let us briefly remind ourselves of the long history of
understanding these masses.

The first person to raise the question about the massive structure of
the weak gauge boson, without knowing about it, was Enrico Fermi. In
1934 he wrote a paper on {\sl Versuch einer Theorie der
  $\beta$-Strahlen}, proposing an interaction responsible for the
neutron decay into a proton, an electron, and an antineutrino. He
proposed an interaction Lagrangian which we nowadays write as 
\begin{alignat}{5}
\lag \supset 
G_F \; \left( \overline{\psi}_1 * \psi_2 \right) \; 
       \left( \overline{\psi}_3 * \psi_4 \right) \; .
\label{eq:intro_fermi}
\end{alignat}
The four external fermions, in our case quarks and leptons, are
described by spinors $\psi$. The star denotes the appropriate scalar,
vector, or tensor structure of the interaction current, which we will
leave open at this stage. On the one hand we know that spinors have
mass dimension 3/2, on the other hand the Lagrangian density has to
integrate to the action and therefore has to have mass dimension
four. This means that in the low energy limit the Fermi coupling constant has to have mass
dimension $G_F \sim 1/\Lambda^2$ with an appropriate mass scale
$\Lambda$. We now know that in the proper
theory, where the Fermi interaction should include an exchange
particle, this scale $\Lambda$ should be linked to the mass of this
exchange particle, the $W$ boson.

The key to understanding such a dimensionful coupling in terms of
massive exchange particles was published by Hideki Yukawa in 1935
under the title {\sl On the interaction of elementary particles}. He
links the mass of exchange particles to the potential they
generate after Fourier transform,
\begin{alignat}{5}
V(r) &= - \frac{e}{r} && \text{massless particle exchange} \notag \\ 
V(r) &= - g^2 \; \frac{e^{-mr}}{r} \qqquad &&\text{massive particle exchange with $m$.}
\label{eq:intro_yuk}
\end{alignat}
Yukawa did not actually talk about the weak nuclear force at the quark
level. His model was based on fundamental protons and neutrons, and
his exchange particles were pions. But his argument applies perfectly
to Fermi's theory at the quark level. Using Eq.\eqref{eq:intro_yuk} we
can link the mass of the exchange particle, in units of $c=1$ and
$\hbar=1$, to the reach of the interaction. For radii above $1/m$ the
massive Yukawa potential is suppressed exponentially. For the weak
nuclear force this is the structure we need, because the force which
links quarks for example to protons and neutrons is incredibly hard to
observable at larger distances. 

What is still missing in our argument is the link between a coupling
strength with mass dimension and massive exchange particles. Since the
1920s many physicists had been using a theory with a quantized
electromagnetic field to compute processes involving charged
particles, like electrons, and photons. The proper, renormalizable
quantum field theory of charged particles and the photon as the
exchange particle was proposed by Sin-Itiro Tomonaga in 1942. Julian
Schwinger independently developed the same theory, for example in his
papers {\sl Quantum electrodynamics I A covariant formulation} and
{\sl On quantum electrodynamics and the magnetic moment of the
  electron}, both published in 1948. The
development of quantum electrodynamics as the theory of massless
photon exchange was from the beginning driven by experimental
observations. For example the calculation of the Lamb shift was a key
argument to convince physicists that such a theory was not only
beautiful, but also useful or `correct' by experimental standards. The
extension of QED to a non--abelian $SU(2)$ gauge group was proposed by
Sheldon Glashow, Julian Schwinger's student, in 1961, but without any
hint of the Higgs mechanism.

Combining these three aspects gives us a clear picture of what people in
the 1950s knew as the basis to solve the puzzle of the weak nuclear
force: interactions between fundamental particles are very
successfully described by massless photon exchange; interactions with
a finite geometric range correspond to a massive exchange particle;
and in the low energy limit such a massive particle exchange can
reproduce Fermi's theory.\bigskip

The main question we will discuss in these lecture notes 
is how to construct a QED--like quantum theory of massive
exchange particles, the $W^\pm$ and $Z^0$ bosons. Usually, the first
reason we give to why the photon is massless is the local $U(1)$ gauge
invariance which essentially defines the QED Lagrangian. However, we
will see in Section~\ref{sec:higgs_qed} that making the
photon massive requires much more fundamental changes to the
theory. This problem, linked to work by Yoichiro Nambu and
specifically to Goldstone's theorem, was what Peter Higgs and his
contemporaries solved for Lorentz-invariant gauge theories in
1964. The idea of spontaneous symmetry breaking was well established
in solid state physics, going back to the work by Landau and Ginzburg,
by Bardeen, Cooper, Schrieffer, and by Anderson on
super-conductivity. However, these systems did not have a Higgs state.
Historically, Walter Gilbert triggered Peter Higgs' first paper in
1964 by making the wrong statement that spontaneous symmetry breaking
would fail for Lorentz-invariant theories.  This fundamental
mistake did not keep him from receiving a Nobel Prize 1980, but for
chemistry. Statements of this kind were very popular at the
time, based on more and more rigorous proofs of Goldstone's
theorem. Needless to say, the Higgs discovery is a good indication
that all of them are wrong for local gauge theories.

The actual Higgs particle only features in Peter Higgs' second paper
in 1964. The very clear prediction of
the new particle in this paper is supposedly due to Yoichiro Nambu,
the journal referee for the paper. The same mechanism of spontaneous
symmetry breaking in high energy physics was, independently of and
even slightly before Peter Higgs' papers, proposed by Francois Englert
and Robert Brout. It is probably fair to assume that Robert Brout, had
he not passed away in 2011, would have been the third Nobel Laureate
in Physics, 2013. Still in 1964 the group of Gerald Guralnik, Carl
Hagen, and Thomas Kibble published a more detailed and rigorous field
theoretical study of the Higgs mechanism. In 1966 Peter Higgs wrote a
third paper, in which he worked out many details of the Higgs mechanism
and the Higgs boson, including scattering rates and decay
widths. Still without linking the Higgs mechanism to the weak force,
this can be considered the first phenomenological study of the Higgs
boson.\bigskip

Combining the Higgs mechanism with QED and applying this combination
to the weak interaction is the birth of the Standard Model of
elementary particles. Steven
Weinberg proposed {\sl A model of leptons} in 1967, for the first time
including fermion masses generated by the Higgs mechanism. Together
with Abdus Salam's paper on {\sl Weak and Electromagnetic
  Interactions} from 1968 the Standard Model, as we know it today, was
now complete. However, most physics aspects which we link to the Higgs
boson nowadays, did not feature at the time. One reason is that the
actual proof of renormalizability by Gerald 't Hooft and Martinus
Veltman still had to be given in 1972, so asking questions about the
high energy behavior of the Standard Model was lacking the formal
basis. Because in these lecture notes we are to some degree following
the historic or effective field theory approach the discussion of the
renormalizable field theory and its ultraviolet behavior will have to
wait until Section~\ref{sec:higgs_unitarity}.\bigskip

If we want to follow this original logic of the Higgs mechanism and
the prediction of the new particle we now know that to do. First, we
need to understand what really keeps the photon from acquiring even a
tiny mass. This will allow us to construct a gauge theory Lagrangian
for massive weak bosons. Finally, we will see how the same mechanism
will allow us to include massive fermions in the electroweak theory.

\subsubsection{Massive photon}
\label{sec:higgs_qed}

Even though this is not the physical problem we are interested in, we
start by breaking electrodynamics and giving a mass to the (massless)
photon of our usual locally $U(1)_Q$-symmetric Lagrangian.  To its
kinetic $F \cdot F$ term we would like to add a photon mass term $m^2
A^2/2$, which we know is forbidden by the gauge symmetry.  We will see
that adding such a photon mass term to the Lagrangian requires a bit of
work, but it not a very hard problem.  The key idea it to also add
an innocent looking real (uncharged) scalar field without a mass and
without a coupling to the photon, but with a \underline{scalar--photon
  mixing}\index{scalar--photon mixing} term and a non--trivial gauge
transformation.  The result is called the Boulware-Gilbert model
\begin{alignat}{5}
\lag &= - \frac{1}{4} F_{\mu\nu} F^{\mu\nu}
        + \frac{1}{2} e^2 f^2 A_\mu^2 
        + \frac{1}{2} \left( \p_\mu \phi \right)^2 
        - e f A_\mu \p^\mu \phi \notag \\
     &= - \frac{1}{4} F_{\mu\nu} F^{\mu\nu}
        + \frac{1}{2} e^2 f^2 \left( A_\mu - \frac{1}{ef} \p_\mu \phi 
                        \right)^2 \; , 
\label{eq:massive_photon}
\end{alignat}
where $f$ is a common mass scale for the photon mass and the mixing.
It ensures that all terms in the Lagrangian have mass dimension four ---
remembering that bosonic fields like $A_\mu$ and $\phi$ have mass
dimension one.  The additional factor $e$ will become the usual
electric charge, but at this stage it is a dimensionless number
without any specific relevance in this interaction-less Lagrangian.
Because all terms in Eq.\eqref{eq:massive_photon} have mass dimension
four and there are not inverse powers of mass our theory should be renormalizable.

We can define a simultaneous gauge transformation of both fields in
the Lagrangian
\begin{alignat}{5}
 A_\mu \longrightarrow A_\mu + \frac{1}{ef} \p_\mu \chi
 \qqqquad
 \phi  \longrightarrow \phi + \chi \; ,
\label{eq:qed_gauge}
\end{alignat}
under which the Lagrangian is indeed invariant: the kinetic term for
the photon we leave untouched, so it will be gauge invariant just as it
was before. The simultaneous gauge invariance is then defined to keep
the second term in Eq.\eqref{eq:massive_photon} invariant.  If we now
re-define the photon field as $B_\mu = A_\mu - \p_\mu \phi/(ef)$ we
need to compare the new and the old kinetic terms
\begin{alignat}{5}
F_{\mu \nu} \Big|_B 
&= \p_\mu B_\nu - \p_\nu B_\mu 
 = \p_\mu \left( A_\nu - \frac{1}{ef} \p_\nu \phi \right)
  -\p_\nu \left( A_\mu - \frac{1}{ef} \p_\mu \phi \right) \notag \\
&= \p_\mu A_\nu - \p_\nu A_\mu 
 = F_{\mu \nu} \Big|_A \; ,
\end{alignat}
and then rewrite the Lagrangian of Eq.\eqref{eq:massive_photon} as
\begin{alignat}{5}
\boxed{
\lag = - \frac{1}{4} F_{\mu\nu} F^{\mu\nu}
       + \frac{1}{2} e^2 f^2 B_\mu^2 
     = - \frac{1}{4} F_{\mu\nu} F^{\mu\nu}
       + \frac{1}{2} m_B^2 B_\mu^2 
} \; .
\end{alignat}
This Lagrangian effectively describes a \underline{massive photon
  field}\index{massive photon} $B_\mu$, which has absorbed the real
scalar $\phi$ as its additional longitudinal component. This is
because a massless gauge boson $A_\mu$ has only two on--shell degrees
of freedom, a left handed and a right handed polarization,
while the massive $B_\mu$ has an additional longitudinal polarization
degree of freedom. Without any fundamental Higgs boson appearing, the
massive photon has `eaten' the real scalar field $\phi$. Of course,
the new field $B_\mu$ is not simply a photon with a mass term, because
this is still forbidden by gauge invariance. Our way out is to split
the massive photon field into the transverse degrees of freedom
$A_\mu$ and the longitudinal mode $\phi$ with their different gauge
transformations given by Eq.\eqref{eq:qed_gauge}.\bigskip

What kind of properties does this field $\phi$ need to have, so that
we can use it to provide a photon mass? From the combined gauge
transformation in Eq.\eqref{eq:qed_gauge} we immediately see that any
additional purely scalar term in the Lagrangian, like a scalar
potential $V(\phi)$, needs to be symmetric under the linear
  shift $\phi \to \phi + \chi$, to not spoil gauge invariance.  This
means that we cannot write down polynomial terms $\phi^n$, like a mass
or a self coupling of $\phi$. An interaction term $\phi A A$ would not
be possible, either. Only \underline{derivative interactions}\index{derivative interaction}
proportional to $\p \phi$ which are attached to other (conserved) currents are allowed. In
that case we can absorb the shift by $\chi$ into a total derivative in
the Lagrangian.\bigskip

This example illustrates a few vital properties of
\underline{Nambu--Goldstone bosons}\index{Goldstone boson} (NGB). Such
massless physical states appear in many areas of physics and are
described by \underline{Goldstone's theorem}\index{Goldstone's theorem}. It applies to global continuous symmetries of the
Lagrangian which are violated by a non--symmetric vacuum state, a
mechanism called spontaneous symmetry breaking\index{spontaneous symmetry breaking}. 
Based on Lorentz invariance and states with a
positively definite norm we can then prove:

If a \underline{global symmetry} group is spontaneously broken into a group of
lower rank, its broken generators correspond to physical Goldstone
modes. These scalar fields transform non--linearly under the larger and
linearly under the smaller group. This way they are massless and
cannot form a potential, because the non--linear transformation only
allows derivative terms in the Lagrangian.

One common modification of this situation is an explicit breaking of
the smaller symmetry group.  In that case the Nambu-Goldstone bosons
become pseudo--Goldstones and acquire a mass of the size of this
hard-breaking term.\bigskip

Before Peter Higgs and his colleagues proposed their mechanism of
electroweak symmetry breaking they were caught between two major no-go
theorems. First, they needed an additional degree of freedom to make
massless gauge bosons massive. Secondly, the spontaneous breaking of a
gauge symmetry supposedly predicted massless scalar states which were
clearly ruled out experimentally. These two problems solve each other
once we properly treat the special case that the spontaneously
broken symmetry is a \underline{local gauge symmetry}. It turns out
that the Goldstone theorem does not apply, because a local gauge theory cannot
be Lorentz invariant and only have positively defined states
simultaneously.  Instead of becoming massless scalars the Goldstone
modes are then `eaten' by the additional degrees of freedom of the
massive gauge bosons. This defines the incredibly elegant Higgs
mechanism.  The gauge boson mass is given by the vacuum expectation
value breaking the larger symmetry. A massive additional scalar degree
of freedom, the \underline{Higgs boson}, appears if there are more
Goldstone modes than degrees of freedom for the massive gauge
bosons\index{Higgs boson}.

\subsubsection{Standard Model doublets}
\label{sec:higgs_doublets}

One of the complications of the Standard Model is its $SU(2)$ doublet
structure.  In the last section we have chosen not to introduce a
charged $SU(2)$ doublet, which is why there are no degrees of freedom
left after the photon gets its mass. This means that our toy model is
not going to be well suited to provide the three degrees of freedom
needed to make $SU(2)$ gauge bosons massive. What it illustrates is
only how by introducing a neutral scalar particle without an
interaction but with a mixing term we make gauge bosons heavy, in
spite of gauge invariance.\bigskip

Fermion fields have mass dimension 3/2, so we know how mass and
interaction terms in the renormalizable dimension-4 Lagrangian have to
look. For example, the interaction of fermions with gauge bosons is
most easily written in terms of \underline{covariant
  derivatives}\index{covariant derivative}. The terms
\begin{alignat}{5}
 \lag_{D4} = \qlb i\Dslash \ql
           +\qrb i\Dslash \qr
           +\elb i\Dslash \el
           +\erb i\Dslash \er
           -\frac{1}{4}F_{\mu\nu}F^{\mu\nu}...
\label{eq:lag4}
\end{alignat}
describe electromagnetic interactions introducing a covariant
derivative $D_\mu =\p_\mu +ieqA_\mu $ with the photon field also
appearing in the field strength tensor $F_{\mu\nu}=\p_\mu A_\nu-\p_\nu
A_\mu $. The same form works for the weak $SU(2)$ interactions, except
that the weak interaction\index{weak interaction} knows about the
chirality of the fermion fields, so we have to distinguish $\Dslash_L$
and $\Dslash_R$. The covariant derivatives we write in terms of the
$SU(2)$ basis matrices or \underline{Pauli matrices}\index{Pauli matrices} $\tau_{1,2,3}$ or
$\tau_{+,-,3}$, with $\tau_\pm = (\tau_1 \pm i \tau_2)/2$.
\begin{alignat}{9}
 &D_{L\mu}
 &=& \p_\mu + ig' \left( q-\frac{\tau_3}{2} \right) B_\mu
      + ig \sum_{a=1,2,3} \, W^a_\mu \frac{\tau_a }{2} \notag \\
&&=&\p_\mu + ieqA_\mu 
                 + ig_Z\left( -qs_w^2+\frac{\tau_3}{2}\right) Z_\mu
                 + i\frac{g}{2}\left( \tau_1 W^1_\mu +\tau_2 W^2_\mu \right) \notag \\
&&\equiv&\p_\mu + ieqA_\mu 
                 + ig_Z\left( -qs_w^2+\frac{\tau_3}{2}\right) Z_\mu
                 + i\frac{g}{\sqrt{2}}\left( \tau_+W^+_\mu +\tau_-W^-_\mu \right) \notag \\
 &D_{R\mu}&=&D_{L\mu}\bigg|_{\tau = 0}\notag \\
 &&& \tau_+=\begin{pmatrix}
0 & 1  \\
0 & 0 \end{pmatrix} \qquad  \tau_-=\begin{pmatrix}
0 & 0  \\
1 & 0 \end{pmatrix} \notag \\
 &&& \tau_1=\begin{pmatrix}
0 & 1  \\
1 & 0 \end{pmatrix} \qquad  \tau_2=\begin{pmatrix}
0 & -i  \\
i & 0 \end{pmatrix} \qquad  \tau_3=\begin{pmatrix}
1 & 0  \\
0 & -1 \end{pmatrix} \; ,
\label{eq:cov_der}
\end{alignat}
The explicit sum in the first line we will omit in the rest of this
lecture. All indices appearing twice include an implicit sum. The
fields $B_\mu$ and $W^a_\mu$ are the new gauge bosons.  In the second
line we re-write the covariant derivative in the
photon $A_\mu$ and the $Z$ boson mass eigenstates. 
What is not obvious from this argument
is that we can actually write the ratio $g'/g$ in terms of a rotation
angle, which implicitly assumes that we can rotate the $B$ and $W^3$
fields into the physical mass-eigenstate photon and $Z$ fields
\begin{alignat}{5}
 \begin{pmatrix}
         A_\mu \\ Z_\mu
        \end{pmatrix} 
=\begin{pmatrix}
         c_w & s_w \\ -s_w & c_w
        \end{pmatrix} 
 \begin{pmatrix}
         B_\mu \\ W^3_\mu
        \end{pmatrix} \; .
\label{eq:higgs_weakrot}
\end{alignat}
The details of this rotation do not matter for the Higgs sector.  The
normalization of the charged gauge fields we will fix later. At this
level the two weak couplings $g$ and $g_Z$ do not necessarily
coincide, but we will get back to this issue in
Section~\ref{sec:higgs_custodial}.\bigskip

Before we generalize the Boulware-Gilbert model to the weak gauge
symmetry of the Standard Model it is instructive to review the form of
the mass term for massive gauge bosons following from
Eq.\eqref{eq:cov_der}. In particular, there will appear a relative
factor two between the two bases of the Pauli matrices, \ie in terms
of $W^{1,2}$ and $W^\pm$, which often causes confusion.  For later use
we also need a sum rule for the $SU(2)$ generators or 
\underline{Pauli matrices}\index{Pauli matrices} $\tau_{1,2,3}$ as 
written out in Eq.\eqref{eq:cov_der}. They satisfy the relation
$\tau_a \tau_b = \delta_{ab} + i \epsilon_{abc} \tau_c$ or the
commutator relation $[\tau_a, \tau_b] = 2 i \epsilon_{abc} \tau_c$.
Summing over indices we see that
\begin{alignat}{5}
\sum_{a,b} \tau^a \tau^b = 
\sum_{a,b} \left( \delta^{ab} + i \epsilon^{abc} \tau_c \right) = 
\sum \delta^{ab} + i \sum_{a \ne b} \epsilon^{abc} \tau_c = 
\sum \delta^{ab} + i \sum_{a < b} \left( \epsilon^{abc} + \epsilon^{bac} \right)  \tau_c = 
\sum \delta^{ab} \; .
\label{eq:paulimat}
\end{alignat}
The basis of three Pauli matrices we can write in terms of
$\tau_{1,2,3}$ as well as in terms of $\tau_{+,-,3}$. The latter 
correspond to two charged and one neutral
vector bosons. While the usual basis is written in terms of complex
numbers, the second set of generators reflects the fact that for
$SU(2)$ as for any $SU(N)$ group we can find a set of real generators
of the adjoint representation. When we switch between the two bases we
only have to make sure we get the standard normalization of all fields
as shown in Eq.\eqref{eq:cov_der},
\begin{alignat}{8}
 \sqrt{2}\left( \tau_+W^+_\mu +\tau_-W^-_\mu \right)
    &=\sqrt{2}\begin{pmatrix} 0 & W_\mu^+             \\
                                     0 & 0 
                   \end{pmatrix}
    +\sqrt{2} \begin{pmatrix} 0 & 0                    \\
                                     W_\mu^- & 0 
              \end{pmatrix} \notag \\
  &\really  \tau_1W^1_\mu +\tau_2W^2_\mu
    =\begin{pmatrix} 0 & W^1_\mu        \\
                             W^1_\mu & 0 
     \end{pmatrix}
    +\begin{pmatrix} 0 & -iW_\mu^2     \\
                              iW_\mu^2 & 0 
     \end{pmatrix} \notag \\
 \qqquad \Leftrightarrow \qqquad
  & W^+_\mu =\frac{1}{\sqrt{2}}\left(W^1_\mu -iW^2_\mu \right)
  \qqqquad 
  W^-_\mu =\frac{1}{\sqrt{2}}\left(W^1_\mu +iW^2_\mu \right) \; .
\label{eq:higgs_gaugebosons}
\end{alignat}
To track these factors of two in the definitions of the weak gauge
field we have a close look at the dimension-2 mass term for charged
and neutral gauge bosons
\begin{alignat}{5}
 \lag_{D2} = -\frac{m_W^2}{2} \left( W^{1,\mu}W^1_\mu + W^{2,\mu}W^2_\mu
                           \right)
            -\frac{m^2_Z}{2} Z^\mu Z_\mu
= - m^2_W W^{+,\mu}W^-_\mu - \frac{m^2_Z}{2} Z^\mu Z_\mu \; .
\label{eq:wz_lag}
\end{alignat}
The relative factor two in front of the $W$ mass appears because the
$Z$ field is neutral and the $W$ field is charged. This difference
also appear for neutral and charged scalars discussed in field theory.
In our conventions it corresponds to the factors $1/\sqrt{2}$ in the
$SU(2)$ generators $\tau_\pm$.

Of course, in the complete Standard Model Lagrangian there are many
additional terms involving the massive gauge bosons, \eg kinetic terms
of all kinds, but they do not affect our discussion of $U(1)_Y$ and
$SU(2)_L$ gauge invariance.\bigskip

Guessing the form of the fermion masses the one thing we have to ensure
is that we combine the left handed and right handed doublets ($\ql,
\el$) and singlets ($\qr, \er$) properly:
\begin{alignat}{5}
 \lag_{D3} = -\qlb m_Q\qr-\elb m_L\er+...
\label{eq:lag3}
\end{alignat}
This form strictly speaking requires a doublet structure of the Higgs--Goldstone
fields, which we will briefly comment on later. For now we ignore this
notational complication.  Following our labeling scheme by mass
dimension fermion masses will be included as $\lag_{D3}$.
\underline{Dirac mass}\index{mass!fermion mass, Dirac mass} terms
simply link $SU(2)$ doublet fields for leptons and quarks with right
handed singlets and give mass to all fermions in the Standard
Model. This helicity structure of mass terms we can easily derive by
introducing left handed and right handed projectors\index{chiral projectors}
\begin{alignat}{5}
\psi_L = \frac{\one - \gamma_5}{2} \, \psi \equiv \prol \psi
\qqqquad 
\psi_R = \frac{\one + \gamma_5}{2} \, \psi \equiv \pror \psi \; ,
\label{eq:def_prolr}
\end{alignat}
where $\psi$ is a generic Dirac spinor and $\prolr$ are projectors in
this $4\times 4$ Dirac space. At this stage we do not need the
explicit form of the gamma matrices which we will introduce in 
Eq.\eqref{eq:dirac_matrices}. The mass term for a Dirac fermion
reads
\begin{alignat}{5}
\psib \, \one \psi 
    & = \psib \, \left( \prol + \pror \right) \psi \notag \\
    & = \psib \, \left( \prol^2 + \pror^2 \right) \psi \notag \\
    & = \psi^\dag \gamma_0 \, \left( \prol^2 + \pror^2 \right) \psi 
      \qqquad
      &\text{with}& \qquad \psib = \psi^\dag \gamma^0 \notag \\
    & = \psi^\dag \left( \pror \gamma^0 \prol + \prol \gamma^0 \pror \right) \psi 
      \qqquad
      &\text{with}& \qquad \{ \gamma_5, \gamma_\mu \} = 0 \notag \\
    & = (\pror \psi)^\dag \gamma^0 (\prol \psi)
       +(\prol \psi)^\dag \gamma^0 (\pror \psi)  
      \qqquad
      &\text{with}& \qquad \gamma_5^\dag = \gamma_5, \prolr^\dag = \prolr \notag \\
    & = ( \overline{\pror \psi} ) \one (\prol \psi)
       +( \overline{\prol \psi} ) \one (\pror \psi) \notag \\
    & = \psib_R \, \one \psi_L 
       +\psib_L \, \one \psi_R \; .
\label{eq:higgs_chiral_mass}
\end{alignat}
The kinetic term stays diagonal
\begin{alignat}{5}
\psib \, \dslash \psi 
    & = \psib \, \dslash \left( \prol^2 + \pror^2 \right) \psi \notag \\
    & = \psib \, \left( \pror \dslash \prol + \prol \dslash \pror \right) \psi \notag \\
    & = ( \overline{\prol \psi} ) \dslash (\prol \psi)
       +( \overline{\pror \psi} ) \dslash (\pror \psi) \notag \\
    & = \psib_L \, \dslash \psi_L 
       +\psib_R \, \dslash \psi_R \; .
\label{eq:higgs_chiral_kin}
\end{alignat}
The link between the chiral projectors and coupling structures we 
discuss in Section~\ref{sec:field_theory}.

In general, these mass terms can be matrices in generation space,
which implies that we might have to rotate the fermion fields from an
interaction basis into the mass basis, where these mass matrices are
diagonal. Flavor physics dealing with such $3\times 3$ mass matrices
is its own field of physics with its own reviews and lecture notes, so
we will omit this complication here.  For our discussion of
electroweak symmetry breaking it is sufficient to study one fermion
generation at a time.\bigskip

The well known problem with the mass terms in Eq.\eqref{eq:lag3} is
that they are not gauge invariant.  To understand this issue of
\underline{fermion masses}\index{mass!fermion mass, Dirac mass} we
check the local weak $SU(2)_L$ transformation
\begin{alignat}{5}
U(x) = \exp\left(i\alpha^a(x)\frac{\tau_a}{2}\right)
     \equiv e^{i (\alpha \cdot \tau)/2} \; ,
\end{alignat}
which only transforms the left handed fermion fields and leaves the
right handed fields untouched
\begin{alignat}{4}
&  \el  \stackrel{U}{\to} U\el  \qquad\qquad && \ql  \stackrel{U}{\to} U\ql\notag \\
&  \er  \stackrel{U}{\to} \er   \qquad\qquad && \qr  \stackrel{U}{\to} \qr \; .
\label{eq:su2_trafo}
\end{alignat}
It is obvious that there is no way we can make left--right mixing
fermion mass terms as shown in Eq.\eqref{eq:lag3} invariant under this
left handed $SU(2)_L$ gauge transformation, where one of the fermion
field picks up a factor $U$ and the other is unchanged,
\begin{alignat}{5}
 \qlb m_Q\qr 
 \stackrel{U}{\to}
\,\qlb U^{-1} m_Q\qr\,\neq\,\qlb m_Q\qr \; .
\end{alignat}
In analogy to the massive photon case, to write a gauge--invariant
Lagrangian for massive fermions we have to add something else to our
minimal Standard Model Lagrangian. Note that this addition does not
have to be a fundamental scalar Higgs field, dependent on how picky we
are with the properties of our new Lagrangian beyond the issue of
gauge invariance.\bigskip

To see what we need to add let us also look at the local $U(1)$
transformations involved. We start with a slightly complicated-looking
way of writing the abelian \underline{hypercharge}\index{hypercharge} $U(1)_Y$ and
\underline{electric charge}\index{electric charge}  $U(1)_Q$ transformations, making it more
obvious how they mix with the neutral component of $SU(2)_L$ to give the
electric charge.

Let us start with the neutral component of the $SU(2)_L$ transformation
$V = \exp\left( i \beta \tau_3/2 \right)$.  Acting on a field with an
$SU(2)_L$ charge this it not a usual $U(1)$ transformation. What we can
do is combine it with another, appropriately chosen
transformation. This hypercharge transformation is
proportional to the unit matrix and hence commutes with all other
matrices
\begin{alignat}{4}
  \exp(i \beta q) \; V^\dag
&=\exp( i\beta q) \; 
  \exp\left( - \frac{i}{2} \beta\tau_3\right) 
\qqquad &&\text{with} \quad  V
= U(x) \Big|_{\tau_3}
= \exp \left( \frac{i}{2} \beta \tau_3 \right)
\notag \\
&=\exp\left( i\beta \frac{y \one+\tau_3}{2}\right) \;
  \exp\left( - \frac{i}{2} \beta\tau_3\right) 
\qqquad &&\text{with} \quad 
\boxed{q\equiv\frac{y\one+\tau_3}{2}} \qquad
 y_Q=\frac{1}{3}\quad y_L=-1 \notag \\
&=\exp\left( i\frac{\beta}{2}y\one\right) \; .
\end{alignat}
The relation between the charge $q$, the hypercharge $y$, and the
isospin $\tau_3$ is called the Gell-Mann--Nishijima formula. The
indices $Q$ and $L$ denote quark and lepton doublets. Acting on a left
handed field the factor $\tau_3$ above is replaced by its eigenvalue
$\pm 1$ for up--type and down--type fermions.  The $U(1)_Y$ charges or
quantum numbers $y$ are the quark and lepton hypercharges of the
Standard Model. As required by the above argument, properly combined
with the isospin they give the correct electric charges
$q_{Q,L}$. Since $\tau_3$ and the unit matrix commute with each other
the combined exponentials have no additional factor a la
Baker--Campbell--Hausdorff $e^A e^B = e^{A+B} e^{[A,B]/2}$. In analogy
to Eq.\eqref{eq:su2_trafo} left handed and right handed quark and
lepton fields transform under this $U(1)_Y$ symmetry as
\begin{alignat}{4}
&\el\to\, \exp\left( i \beta q_L \right) V^\dag \el
        = \exp\left( i \frac{\beta}{2} y_L\one\right) \el  \qqquad &&
 \ql\to   \exp\left( i \beta q_Q \right) V^\dag \ql
        = \exp\left( i \frac{\beta}{2} y_Q\one\right) \ql  \notag \\ 
&\er\to\, \exp\left( i \beta q_L \right) \er &&
 \qr\to   \exp\left( i \beta q_Q \right) \qr \; .
\label{eq:u1_trafo}
\end{alignat}
Under a combined $SU(2)_L$ and $U(1)_Y$ transformation the left handed
fermions see the hypercharge, while the right handed fermions only see
the electric charge. Just as for the $SU(2)_L$ transformation $U$ we do
not have to compute anything to see that such different
transformations of the left handed and right handed fermion fields do
not allow for a Dirac mass term.

\subsubsection{Sigma model}
\label{sec:higgs_sigma}

One way of solving this problem  is to introduce an additional field $\s(x)$. This field
will in some way play the role of the real scalar field we used for
the photon mass generation. Its physical properties will become clear
piece by piece from the way it appears in the Lagrangian and from the
required gauge invariance. The equation of motion for the $\s$ field
will also have to follow from the way we introduce it in the
Lagrangian.

Following the last section, we first introduce $\s$ into the
\underline{fermion mass}\index{mass!fermion mass, Dirac mass} term.
This will tell us what it takes to make this mass term gauge invariant
under the weak $SU(2)_L$ transformation defined in Eq.\eqref{eq:su2_trafo}
\begin{alignat}{5}
 \qlb\s m_Q\qr
 \stackrel{U}{\to}\,\qlb U^{-1}\s^{(U)} m_Q\qr\,
 \really \,\qlb\s m_Q\qr
\qquad \Leftrightarrow \qquad \s \to \s^{(U)} = U\s \; .
\label{eq:u_trafo}
\end{alignat}
If the result should be a dimension-4 Lagrangian the mass dimension of
$\s$ has to be $m^0=1$.  The same we can do for the $U(1)_Y$
transformation $V$ as described in Eq.\eqref{eq:u1_trafo}
\begin{alignat}{4}
   \qlb \s m_Q\qr \stackrel{V}{\to} \, & 
   \qlb \exp\left(-i \frac{\beta}{2} y \one \right) \s^{(V)} m_Q
        \exp\left( i\beta q\right) \qr                  
\notag \\
&= \qlb \s^{(V)} \exp\left(-i \frac{\beta}{2} y \one \right) \;
        \exp\left(i \beta q \right) m_Q \qr 
        \qqquad && \text{$\exp\left (i \frac{\beta}{2} y \one \right)$ always commuting} \notag \\
&= \qlb \s^{(V)} V  m_Q \qr \notag \\
&\really \qlb \s \, m_Q \qr 
  \qquad \Leftrightarrow \qquad 
  \s \to \s^{(V)} = \s V^\dag \; .
\label{eq:sigma_trafo}
\end{alignat}
Combining this with Eq.\eqref{eq:u_trafo} gives us the 
\underline{transformation property} we need
\begin{alignat}{5}
\boxed{\s \to U\s V^\dag} \; .
\label{eq:higgs_sigmatrafo}
\end{alignat}
For any $\s$ with this property the $\lag_{D3}$ part of the Lagrangian
has the required $U(1)_Y \times SU(2)_L$ symmetry, independent of what
this $\s$ field really means. From the way it transforms we see that $\s$
is a $2\times 2$ matrix with mass dimension zero. In other words,
including a $\s$ field in the fermion mass terms gives a $U(1)_Y$ and
$SU(2)_L$-invariant Lagrangian, without saying anything about possible
representations of $\s$ in terms of physical fields
\begin{alignat}{5}
\boxed{
 \lag_{D3} = -\qlb\s m_Q\qr-\elb \s m_L\er+ \text{h.c.}+...
}
\label{eq:higgs_d3}
\end{alignat}
Fixing the appropriate transformations of the $\s$ field allows us 
to include fermion masses without any further complication.\bigskip

In a second step, we deal with the \underline{gauge boson masses}\index{mass!gauge boson mass}.
We start with the left handed covariant derivative
already used in Eq.\eqref{eq:cov_der}

\begin{alignat}{5}
  D_{L\mu} 
   = \p_\mu + ig' \left( q-\frac{\tau_3}{2} \right) B_\mu
      + igW^a_\mu \frac{\tau_a }{2}
   = \p_\mu +ig' \frac{y}{2} B_\mu +igW^a_\mu \frac{\tau_a }{2} \; .
\end{alignat}
Instead of deriving the gauge transformation of $\Sigma$ let us start
with a well--chosen ansatz and work backwards step by step, to check
that we indeed arrive at the correct masses. First, we consistently
require that the covariant derivative acting on the $\s$ field in the
gauge-symmetric Lagrangian reads
\begin{alignat}{5}
 D_\mu \s  =  \p_\mu \s
               + ig' \s B_\mu \frac{y}{2} \Big|_{q=0}
               + i g W^a_\mu \frac{\tau_a}{2}\s 
           =  \p_\mu \s
               - ig' \s B_\mu \frac{\tau_3}{2}
               + i g W^a_\mu \frac{\tau_a}{2}\s \; ,
\label{eq:cov_sigma}
\end{alignat}
If we introduce the abbreviations $V_\mu \equiv \s(D_\mu \s)^\dag $
and $T \equiv \s\tau_3\s^\dag $ we claim we can write the gauge boson
mass terms as
\begin{alignat}{5}
\boxed{
 \lag_{D2} = \frac{v^2}{4} \; \tr[V_\mu V^\mu ]
          + \Delta \rho \frac{v^2}{8} \; \tr[TV_\mu ] \; \tr[TV^\mu ]
} \; .
\label{eq:betaprime1}
\end{alignat}
The trace acts on the $2 \times 2$ $SU(2)$ matrices. The parameter
$\Delta \rho$ is conventional and will be the focus of
Section~\ref{sec:higgs_custodial}. We will show below that this form
is gauge invariant and gives the correct gauge boson masses.\bigskip

Another structural question is what additional terms of mass dimension
four we can write down using the dimensionless field $\s$ and which
are gauge invariant. Our first attempt of a building block
\begin{alignat}{5}
 \sdag \s \stackrel{U,V}{\to} \, (U\s V^\dag )^\dag (U\s V^\dag )
          =       V \sdag U^\dag U \s V^\dag  
          =       V \sdag \s V^\dag \neq\sdag\s
\end{alignat}
is forbidden by $SU(2)_L$ gauge invariance according to
Eq.\eqref{eq:sigma_trafo}. On the
other hand, a circularly symmetric trace $\tr(\sdag\s) \to \tr(V\sdag
\s V^\dag) =\tr(\sdag\s)$ changes this into a gauge invariant
combination, which allows for the additional \underline{potential
  terms}\index{Higgs potential}, meaning terms with no derivatives
\begin{alignat}{5}
\boxed{
 \lag_\s = - \frac{\mu^2 v^2}{4} \tr(\sdag \s)
           - \frac{\lambda v^4}{16} \left( \tr(\sdag\s) \right)^2
           + \cdots
} \; ,
\label{eq:sigma_pot}
\end{alignat}
with properly chosen prefactors $\mu ,v,\lambda$. This fourth term
finalizes our construction of the relevant weak Lagrangian
\begin{alignat}{5}
\lag = \lag_{D2} + \lag_{D3} +\lag_{D4} +\lag_\s \; ,
\end{alignat}
organized by mass dimension.

As rule of thumb we will later notice that once we express the
potential of Eq.\eqref{eq:sigma_pot} in terms of the usual Higgs
doublet $|\phi|^2$, the prefactors will just be $\mu$ and
$\lambda$. The parameter $\mu$ and the factor $v$ appearing with every
power of $\s$ have mass dimension one, while $\lambda$ has mass
dimension zero. Higher--dimensional terms in a dimension-4 Lagrangian
are possible as long as we limit ourselves to powers of $\tr (\sdag
\s)$. However, they lead to higher powers in $v$ which we will see
makes them higher--dimensional operators in our complete quantum
theory.\bigskip

To check that Eq.\eqref{eq:betaprime1} gives the correct masses in the
Standard Model we start with $\tr(\sdag\s)$ and assume it acquires a
finite (expectation) value after we properly deal with $\s$. The
definitely simplest way to achieve this is to assume
\begin{alignat}{5}
\boxed{ \s(x)= \one } \; .
\end{alignat}
This choice is called \underline{unitary gauge}\index{unitary gauge}.
It looks like a dirty trick to first introduce $\s(x)= \one$ and then
use this field for a gauge invariant implementation of gauge boson
masses. Clearly, a constant does not exhibit the correct
transformation property under the $U$ and $V$ symmetries, but we can
always work in a specific gauge and only later check the physical
predictions for gauge invariance. The way the sigma field breaks
our gauge symmetry we can schematically see from
\begin{alignat}{5}
 \s  \to  U \s V^\dag  =  U \one V^\dag  = 
 U V^\dag \really \one \; ,
\label{eq:sigma_break}
\end{alignat}
which requires $U=V$ to be the remaining $U(1)$ gauge symmetry after
including the $\s$ field.  Certainly, $\s = \one$ gives the correct
fermion masses in $\lag_{D3}$ and makes the potential $\lag_\s$ an
irrelevant constant. What we need to check is $\lag_{D2}$ which is
supposed to reproduce the correct gauge boson masses. Using the
covariant derivative from Eq.\eqref{eq:cov_sigma} acting on a constant
field we can compute the auxiliary field $V_\mu$ in unitary gauge
\begin{alignat}{5}
 V_\mu = \s (D_\mu \s)^\dag
       & = \one (D_\mu \s)^\dag \notag \\
       & = - igW^a_\mu \frac{\tau_a }{2}
           + ig'B_\mu \frac{\tau_3}{2}           \notag \\
       & = - igW^+_\mu \frac{\tau_+ }{\sqrt{2}}
           - igW^-_\mu \frac{\tau_-}{\sqrt{2}}
           - igW^3_\mu \frac{\tau_3}{2}
           + ig'B_\mu \frac{\tau_3}{2}           \notag \\
       & = - i\frac{g}{\sqrt{2}} \left(  W^+_\mu \tau_+
                                        +W^-_\mu \tau_- \right)
           - ig_Z Z_\mu \frac{\tau_3}{2}  \, ,
\label{eq:v_unitary}
\end{alignat}
with $Z_\mu =c_wW^3_\mu -s_wB_\mu$ and the two coupling constants
$g_Z=g/c_w$ and $g'=g s_w/c_w$ as defined in
Eq.\eqref{eq:higgs_weakrot}. This gives us the first of the two terms
in the gauge boson mass Lagrangian $\lag_{D2}$
\begin{alignat}{5}
  \tr[ V_\mu V^\mu ] & = -2 \, \frac{g^2}{2} W^+_\mu {W^-}^\mu \tr(\tau_+\tau_-)
                         - \frac{g^2_Z}{4} Z_\mu Z^\mu \tr(\tau_3^2) \notag \\
                     & = - g^2 W^+_\mu {W^-}^\mu - \frac{g^2_Z}{2}Z_\mu Z^\mu  \; ,
\end{alignat}
using $\tau_\pm^2 = 0$, $\tr (\tau_3 \tau_\pm) = 0$, $\tr (\tau_\pm
\tau_\mp) = 1$, and $\tr (\tau_3^2) = \tr \one = 2$. The second mass term in
$\lag_{D2}$ proportional to $\Delta \rho$ is equally simple in unitary gauge
\begin{alignat}{3}
&&  T            &= \s\tau_3\sdag = \tau_3        \notag \\
&  \Rightarrow & \quad
  \tr(TV_\mu ) &= \tr\left( -ig_Z Z_\mu \frac{\tau_3^2}{2} \right)
                = -i g_Z Z_\mu                    \notag \\
&  \Rightarrow & \quad
  \tr(TV_\mu) \; \tr(TV^\mu) &= -g^2_Z Z_\mu Z^\mu \; .
\end{alignat}
Inserting both terms into Eq.\eqref{eq:betaprime1} yields the complete
gauge boson mass term
\begin{alignat}{3}
\lag_{D2} &= \frac{v^2}{4} \left( - g^2 W_\mu^+ W^{-\mu}
                                - \frac{g^2_Z}{2}Z_\mu Z^\mu \right)
          +\Delta \rho \frac{v^2}{8} \left( - g^2_Z Z_\mu Z^\mu \right)   \notag \\
       &= -\frac{v^2 g^2}{4} W_\mu ^+W^{-\mu}
          -\frac{v^2 g^2_Z}{8} Z_\mu Z^\mu 
          -\Delta \rho \frac{v^2 g^2_Z}{8} Z_\mu Z^\mu \notag \\
       &= -\frac{v^2 g^2}{4} W_\mu ^+W^{-\mu}
          -\frac{v^2 g^2_Z}{8} \left( 1+\Delta \rho\right) Z_\mu Z^\mu \; .
\label{eq:lag_d2final}
\end{alignat}
Identifying the masses with the form given in Eq.\eqref{eq:wz_lag} and
assuming universality of \underline{neutral and charged current}\index{weak interaction!neutral current}\index{weak interaction!charged current} interactions
($\Delta \rho=0$) we find
\begin{alignat}{5}
 \boxed{m_W = \frac{g v}{2}} \qqqquad 
 \boxed{m_Z=\sqrt{1+\Delta \rho} \; \frac{g_Z v}{2} 
        \stackrel{\Delta \rho=0}{=} \frac{g_Z v}{2}
        = \frac{g v}{2 c_w} } \; .
\label{eq:betaprime2}
\end{alignat}
The role of a possible additional and unwanted $Z$-mass contribution
$\Delta \rho$ we will discuss in Section~\ref{sec:higgs_custodial} on
custodial symmetry. Given that we know the heavy gauge boson masses
($m_W \sim 80~\gev$) and the weak coupling ($g \sim 0.7$) from
experiment, these relations experimentally tell us $v \sim
246~\gev$.\bigskip

Let us just recapitulate what we did until now --- using this $\s$
field with its specific transformation properties and its finite
constant value $\s = \one$ in unitary gauge we have made the fermions
and electroweak gauge boson massive. Choosing this constant finite
field value for $\s$ is not the only and not the minimal assumption
needed to make the gauge bosons heavy, but it leads to the most
compact Lagrangian. From the photon mass example, however, we know
that there must be more to this mechanism. We should for example be
able to see the additional degrees of freedom of the longitudinal
gauge boson modes if we step away from unitary gauge.\bigskip

If a finite expectation value of the terms in the potential $\lag_\s$ should be 
linked to electroweak symmetry breaking and the gauge boson masses we
can guess that the 
\underline{minimal assumption} leading to finite gauge boson masses 
is
$\langle \tr(\sdag(x)\s(x))\rangle\neq 0$ in the vacuum. Every
parameterization of $\s$ with this property will lead to the same
massive gauge bosons, so they are all physically equivalent --- as
they should be given that they are only different gauge choices. In
the canonical normalization we write
\begin{alignat}{5}
\boxed{\frac{1}{2}\; \langle \tr(\sdag(x)\s(x))\rangle=1} 
 \qquad \forall x \; ,
\label{eq:higgs_tracesigma}
\end{alignat}
which instead of our previous $\s(x)= \one$ we can also fulfill
through
\begin{alignat}{5}
  \sdag(x)\s(x) = \one \qquad \forall x \; .
\end{alignat}
This means that $\s(x)$ is a unitary matrix which like any $2\times 2$
unitary matrix can be expressed in terms of the Pauli matrices.
This solution still forbids fluctuations which in the original
condition Eq.\eqref{eq:higgs_tracesigma} on the expectation value only
vanish on average. However, in contrast to $\s(x) = \one$ it allows a
non--trivial $x$ dependence. A unitary matrix $\s$ with the appropriate
normalization can for example be written as a simple linear combination
of the basis elements, \ie in the \underline{linear representation}\index{Goldstone boson!linear representation} 
\begin{alignat}{5}
\s(x) = \frac{1}{\sqrt{1+\dfrac{w_aw_a}{v^2}}}
        \left( \one - \dfrac{i}{v} \vec{w}(x) \right) 
 \qquad \text{with}\quad  \vec{w}(x)=w_a(x)\tau_a \; ,
\label{eq:higgs_goldstone_lin}
\end{alignat}
where $\vec{w}(x)$ has mass dimension one which is absorbed by the
mass scale $v$. These fields are a set of scalar
\underline{Nambu-Goldstone modes}\index{Goldstone boson}. From the
photon mass example for Goldstone's theorem we know that they will
become the missing degrees of freedom for the three now massive gauge
bosons $W^\pm$ and $Z$. The normalization scale $v$ fixes the relevant
energy scale of our Lagrangian.\bigskip

Another way of parameterizing the unitary field $\s$ in terms of the
Pauli matrices\index{Pauli matrices} is
\begin{alignat}{5}
\boxed{
 \s(x) = \exp\left( - \frac{i}{v} \vec{w}(x)\right) }
 \qquad \text{with}\quad  \vec{w}(x)=w_a(x)\tau_a \; ,
\label{eq:higgs_goldstone_exp}
\end{alignat}
Because the relation between $\s$ and $\vec{w}$ is not
linear, this is referred to as a \underline{non--linear representation}\index{Goldstone boson!non--linear representation} 
of the $\s$ field. Using the commutation properties of the Pauli
matrices we can expand $\s$ as
\begin{alignat}{3}
 \s &= \one - \frac{i}{v}\w
         + \frac{1}{2} \frac{(-1)}{v^2} w_a \tau_a w_b \tau_b
         + \frac{1}{6} \frac{i}{v^3} w_a \tau_a w_b\tau_b w_c \tau_c 
         + \ope(w^4) \notag \\
    &= \one - \frac{i}{v}\w
         - \frac{1}{2v^2} w_a w_a \one
         + \frac{i}{6v^3} w_a w_a \w + \ope(w^4) 
    \qqquad \text{using Eq.\eqref{eq:paulimat}} \notag \\
    &= \left( 1 - \frac{1}{2v^2} w_a w_a + \ope(w^4) \right) \one
       - \frac{i}{v} \left( 1 - \frac{1}{6v^2} w_a w_a + \ope(w^4) \right) \w \; .
\label{eq:feynman_goldstone}
\end{alignat}
From this expression we can for example read off Feynman rules for the
longitudinal gauge fields $\vec{w}$, which we will use later.  The
different ways of writing the $\s$ field in terms of the Pauli
matrices of course cannot have any impact on the physics.\bigskip

Before we move on and introduce a physical Higgs boson we briefly
discuss different gauge choices and the appearance of Goldstone
modes. If we break the full electroweak gauge symmetry $SU(2)_L \times
U(1)_Y \rightarrow U(1)_Q$ we expect three Goldstone bosons which
become part of the weak gauge bosons and promote those from massless
gauge bosons (with two degrees of freedom each) to massive gauge
bosons (with three degrees of freedom each). This is the point of view
of the unitary gauge, in which we never see Goldstone
modes\index{unitary gauge}.

In the general renormalizable $R_\xi$ gauge we can actually see
these Goldstone modes appear separately in the \underline{gauge boson
  propagators}\index{propagator!gauge boson}
\begin{alignat}{5}
 \Delta^{\mu\nu}_{VV}(q)
&=
 \frac{-i}{q^2-m_V^2+i\epsilon}
 \left[g^{\mu\nu}+(\xi-1)\frac{q^\mu q^\nu}{q^2-\xi m_V^2}\right]
 \notag \\
&=
\left\{%
\begin{array}{ll}
 \dfrac{-i}{q^2-m_V^2+i\epsilon}
 \left[g^{\mu\nu}-\dfrac{q^\mu q^\nu}{m_V^2}\right]
     \qqquad & \text{unitary gauge} \; \xi \to \infty \\[4mm]
 \dfrac{-i}{q^2-m_V^2+i\epsilon} \; g^{\mu\nu}
     & \text{Feynman gauge} \; \xi =1 \\[2mm]
 \dfrac{-i}{q^2-m_V^2+i\epsilon}
 \left[g^{\mu\nu}-\dfrac{q^\mu q^\nu}{q^2}\right]
     & \text{Landau gauge} \; \xi =0 \; .\\
\end{array}
\right.
\end{alignat}
\index{unitary gauge}Obviously, these gauge choices are physically equivalent.  However,
something has to compensate, for example, for the fact that in Feynman
gauge the whole Goldstone term vanishes and the polarization sum looks
like a massless gauge boson, while in unitary gauge we can see the
effect of these modes directly.  The key is the Goldstone propagator\index{propagator!Goldstone boson},
with its additional propagating scalar degrees of freedom
\begin{alignat}{5}
 \Delta_{VV}(q^2)=\frac{-i}{q^2-\xi m_V^2+i\epsilon} \; ,
\label{eq:higgs_goldstoneprop}
\end{alignat}
for both heavy gauge bosons $(V=Z,W^+)$. The Goldstone mass $\sqrt{\xi}
m_V$ depends on the gauge: in unitary gauge the infinitely heavy
Goldstones do not propagate ($\Delta_{VV} (q^2) \to 0$), while in
Feynman gauge and in Landau gauge we have to include them as
particles.  From the form of the Goldstone propagators we can guess
that they will indeed cancel the second term of the gauge boson
propagators.

These different gauges have different Feynman rules and Green's
functions, even a different particle content, so for a given problem
one or the other might be the most efficient to use in computations or
proofs. For example, the proof of renormalizability was first
formulated in unitary gauge. Loop calculations might be more efficient
in Feynman gauge, because of the simplified propagator structure,
while many QCD processes benefit from an explicit projection on the
physical external gluons. Tree level helicity amplitudes are usually
computed in unitary gauge, etc...

\subsubsection{Higgs boson}
\label{sec:higgs_higgsboson}

At this stage we have defined a perfectly fine electroweak theory with
massive gauge bosons. All we need is a finite vacuum expectation value
for $\s$, which means this field \underline{spontaneously
  breaks}\index{spontaneous symmetry breaking} the electroweak
symmetry not by explicit terms in the Lagrangian but via the vacuum.
The origin of this finite vacuum expectation
value is not specified.  This aspect that the Higgs mechanism does not actually specify
where the vacuum expectation value $v$ comes from is emphasized by Peter Higgs 
in his original paper.  If we are interested in physics at
or above the electroweak energy scale $E \sim v$ some kind of
ultraviolet completion of this $\s$ model should tell us what the $\s$
field's properties as a quantum object are.\bigskip

If we consider our $\s$ model itself the fundamental theory and
promote the $\s$ field to a quantum field like all other Standard
Model fields, we need to allow for \underline{quantum fluctuations}\index{Higgs field!quantum fluctuations} of
$\tr(\sdag\s)$ around the vacuum value $\tr(\sdag\s)=2$. Omitting the 
Goldstone modes we can
parameterize these new degrees of freedom as a real scalar field
\begin{alignat}{5}
\boxed{
\s\,\to\,\left(1+\frac{H}{v}\right)\s
} \; ,
\label{eq:higgs_define}
\end{alignat}
as long as this physical field $H$ has a vanishing vacuum expectation
value and therefore 
\begin{alignat}{5}
 \frac{1}{2} \; \langle \tr(\sdag\s) \rangle
=\left< \left(1+\frac{H}{v}\right)^2 \right> = 1
\qqquad \Leftrightarrow \qqquad 
\left< H \right> = 0 \; .
\end{alignat}
This real Higgs field is the fourth direction in the basis choice for
the unitary matrix $\s$ for example shown in
Eq.\eqref{eq:higgs_goldstone_lin}, where only $w_a$ are originally
promoted to quantum fields. 

The factor in front of the fluctuation term $H/v$ is not fixed until
we properly define the physical Higgs field and make sure that
its kinetic term does not come with an unexpected prefactor.
On the other hand, if we assume that the neutral Goldstone
mode $w_3$ has the correct normalization, the Higgs field should be
added to $\s$ such that it matches this Goldstone, as we will see
later in this section and then in more detail in
Section~\ref{sec:higgs_sm}.

The non--dynamical limit of this Higgs ansatz is indeed our sigma model
in unitary gauge $\sdag \s = \one$, equivalent to $H =0$.
Interpreting the fluctuations around the non--trivial vacuum as a
\underline{physical Higgs field}\index{Higgs field} is the usual Higgs
mechanism.\bigskip

For this new Higgs field the Lagrangian $\lag_\s$ defines a potential
following the original form of Eq.\eqref{eq:sigma_pot}
\begin{alignat}{5}
\lag_\s = - \frac{\mu^2v^2}{2} \left( 1 + \frac{H}{v} \right)^2
          - \frac{\lambda v^4}{4} \left( 1 + \frac{H}{v} \right)^4 + ...
\label{eq:sigma_pot_h}
\end{alignat}
The dots stand for higher--dimensional terms which might or might not
be there. We will have a look at them in
Section~\ref{sec:higgs_pot}. Some of them are not forbidden by any
symmetry, but they are not realized at tree level in the Standard
Model either.  The minimum of this potential occurs at $H=0$, but this
potential is not actually needed to give mass to gauge bosons and
fermions.  Therefore, we postpone a detailed study of the Higgs
potential to Section~\ref{sec:higgs_pot}.\bigskip

Let us recall one last time how we got to the Higgs mechanism from a static
gauge invariant theory, the $\s$ model.  From an effective field
theory point of view we can introduce the Goldstone modes and
with them gauge boson masses without introducing a fundamental Higgs
scalar. All we need is the finite vacuum expectation value for $\s$ to
spontaneously break electroweak symmetry.  For this symmetry breaking
we do not care about quantum fluctuations of the $\s$ field, which means we do
not distinguish between the invariant $\tr (\sdag \s)$ and its
expectation value.  Any properties of the $\s$ field as a quantum
field are left to the ultraviolet completion, which has to decide for
example if $\s$ is a fundamental or composite field.  This way, the
Higgs field could just be one step in a ladder built out of effective
theories.  Such a non--fundamental Higgs field is the basis for
so-called strongly interacting light Higgs models where the Higgs
field is a light composite field with a different canonical
normalization as compared to a fundamental scalar.\bigskip

Counting degrees of freedom we should be able to write $\s$ as a
complex doublet with four degrees of freedom, three of which are eaten
Goldstones and one is the fundamental Higgs scalar. On the pure
Goldstone side we can choose for example between the linear
representation of Eq.\eqref{eq:higgs_goldstone_lin} and the non--linear
representation of Eq.\eqref{eq:higgs_goldstone_exp}. If we extend the
linear representation and for now ignore the normalization we find
\begin{alignat}{5}
 \s  =  \left( 1 + \frac{H}{v} \right) \one
         - \frac{i}{v}\w
     =     \frac{1}{v} 
           \begin{pmatrix}
                 v+H-i w_3 & -w_2-i w_1  \\
                  w_2-i w_1 & v+H+iw_3 
                 \end{pmatrix}
     =     \frac{\sqrt{2}}{v}\, ( \tilde{\phi} \phi ) \; .
\label{eq:higgs_vshift}
\end{alignat}
The last step is just another way to write the $2 \times 2$ matrix as
a bi-doublet in terms of the two $SU(2)_L$ doublets containing the
physical Higgs field and the Goldstone modes for the massive vector
bosons $W$ and $Z$,
\begin{alignat}{5}
       \phi = 
             \frac{1}{\sqrt{2}}
             \begin{pmatrix}
                    -w_2 - i w_1 \\ v+H+iw_3 
             \end{pmatrix} 
\qqquad
\tilde{\phi}= - i \tau_2 \; \phi^* =
             \frac{1}{\sqrt{2}}
             \begin{pmatrix}
                     v+H-iw_3 \\ w_2 - i w_1
             \end{pmatrix} \; .
\label{eq:def_phi}
\end{alignat}
This description of the Higgs field as part of the $SU(2)_L$ doublet
in the \underline{linear representation}\index{Goldstone boson!linear
  representation} has a profound effect on the form of the Lagrangian:
we can only include the Higgs field in $SU(2)_L$-invariant ways, for
example using the combination $\phi^\dag \phi$.  In the presence of a
new mass scale $\Lambda$ the structure and mass dimension of the
doublet field $\phi$ will help us organize the most general
electroweak and Higgs Lagrangians, for example allowing for additional
terms $\phi^\dag \phi/\Lambda^2$.  In contrast, the \underline{non--linear
representation} of Eq.\eqref{eq:higgs_goldstone_exp} cannot be cast
into such a $SU(2)$ invariant form, and the Higgs field appears as a
singlet under the weak gauge symmetry. In an extended Lagrangian we
can simply add a general power series in $H/v$ or $H/\Lambda$ to any
gauge operator.\bigskip

The vacuum expectation value $v$ appearing in the $\phi$ and
$\tilde{\phi}$ doublets corresponds to $\langle \s \rangle = \one$.
In this form the normalization of the two real scalars $w_3$ and $H$
is indeed the same, so their kinetic terms will be of the same form.
The over--all factors $1/\sqrt{2}$ in the definition of the doublets
are purely conventional and sometimes lead to confusion when some
people define $v = 246~\gev$ while others prefer $v = 174~\gev$.  The
latter choice is a little less common but has the numerological
advantage of $v \sim m_t$.  For the fermion sector Eq.\eqref{eq:lag3}
this bi-doublet structure is important, because it means that we give
mass to up--type fermions and down--type fermions not with the same
field $\phi$, but with $\phi$ and $\tilde{\phi}$.\bigskip

Following Eq.\eqref{eq:higgs_vshift} we can for example derive the
couplings of the physical Higgs boson to the massive $W$ and $Z$ gauge
bosons from Eq.\eqref{eq:lag_d2final} with custodial symmetry,
\begin{alignat}{3}
\lag_{D2} 
&=   -\frac{(v+H)^2 g^2}{4} W_\mu ^+W^{-\mu}
     -\frac{(v+H)^2 g^2_Z}{8} \left( 1 + \Delta \rho \right) Z_\mu Z^\mu 
\notag \\
&\supset
  -\frac{2vH g^2}{4} W_\mu ^+W^{-\mu}
  -\frac{2vH g^2_Z}{8} \left( 1 + \Delta \rho \right) Z_\mu Z^\mu 
\notag \\
&=
  -g m_W \, H W_\mu ^+W^{-\mu}
  -\frac{g_Z m_Z}{2} \left( 1 + \Delta \rho \right) H Z_\mu Z^\mu 
\; .
\label{eq:lag_wwh}
\end{alignat}
The same we can do for each fermion, where Eq.\eqref{eq:lag3} in the
diagonal limit and with the appropriate normalization of the Yukawa
coupling $y_f = \sqrt{2} m_f/v$ becomes
\begin{alignat}{5}
\lag_{D3} 
&\to - y_f \frac{(v+H)}{\sqrt{2}} \; \psib_f \psi_f
&\supset - \frac{y_f}{\sqrt{2}} \; H \psib_f \psi_f \; .
\label{eq:lag_ffh}
\end{alignat}
The couplings of the scalar Higgs boson are completely determined by
the masses of the particles it is coupling to. This includes the
unwanted correction $\Delta \rho$ to the
$Z$ mass.

Apart from problems arising when we ask for higher precision and
quantum corrections, the effective sigma model clearly breaks down at
large enough energies which can excite the fluctuations of the sigma
field and for example produce a Higgs boson. This is the
\underline{job of the LHC}, which is designed and built to take us
into an energy range where we can observe the structure of electroweak
symmetry breaking beyond the effective theory and the Goldstone
modes. The observation of a light and narrow Higgs resonance roughly
compatible with its Standard Model definition in
Eq.\eqref{eq:higgs_define} is only a first step into this
direction.\bigskip

\subsubsection{Custodial symmetry}
\label{sec:higgs_custodial}

Analyzing the appearance of $\Delta \rho$ in Eq.\eqref{eq:betaprime1}
and Eq.\eqref{eq:betaprime2} we will see that not only higher
energies, but also higher precision leads to a breakdown of the
effective sigma model. At some point we start seeing how the relative
size of the $W$ and $Z$ masses are affected by quantum fluctuations of
the sigma field, \ie the three Goldstone modes and the Higgs boson
itself. Diagrammatically, we can compute these quantum effects by
evaluating Higgs contributions to the one-loop form of the $W$ and $Z$
propagators.

From the construction in Section~\ref{sec:higgs_sigma} we know that
electroweak symmetry breaking by a sigma field or Higgs doublet links
the couplings of neutral and charged currents firmly to the masses of
the $W$ and $Z$ bosons. On the other hand, the general renormalizable
Lagrangian for the gauge boson masses in Eq.\eqref{eq:betaprime1}
involves two terms, both symmetric under $SU(2)_L \times U(1)_Y$ and
hence allowed in the electroweak Standard Model.  The mass values
coming from $\tr[V_\mu V^\mu ]$ give $m_W$ and $m_Z$ proportional to
$g \equiv g_W$ and $g_Z.$ The second term involving $(\tr[T V_\mu])^2$
only contributes to $m_Z$.  

The relative size of the two gauge boson
masses can be expressed in terms of the weak mixing angle $\theta_w$,
together with the assumption that $G_F$ or $g$ universally govern
charged current ($W^\pm$) and neutral-current ($W^3$) interactions. At
tree level this experimentally very well tested relation corresponds
to $\Delta \rho=0$ or
\begin{alignat}{5}
 \frac{m_W^2}{m_Z^2} = \frac{g^2}{g_Z^2} = \cos^2 \theta_w \equiv c_w^2 \; .
\label{eq:custo0}
\end{alignat}
In general, we can introduce a free parameter $\rho$\index{electroweak precision measurements!$\rho$ parameter} which breaks this relation
\begin{alignat}{5}
 \boxed{g_Z^2 \to g_Z^2 \; \rho} \qqqquad
 m_Z \to m_Z \; \sqrt{\rho}  = m_Z \; \sqrt{1 + \Delta \rho}\; ,
\end{alignat}
which from measurements is very strongly constrained to be unity.  It
is defined to correspond to our theoretically known allowed deviation
$\Delta \rho$.  In $\lag_{D2}$ the $Z$-mass term precisely predicts
this deviation.  To bring our Lagrangian into agreement with
measurements we better find a reason to constrain $\Delta \rho$ to
zero, and the $SU(2)_L \times U(1)_Y$ gauge symmetry unfortunately
does not do the job.\bigskip

Looking ahead, we will find that in the Standard Model $\rho = 1$ is
actually violated at the one-loop level.  This means we are looking
for an \underline{approximate symmetry} of the Standard Model. What we
can hope for is that this symmetry is at least a good symmetry in the
$SU(2)_L$ gauge sector and slightly broken elsewhere.  One possibility
along those lines is to replace the $SU(2)_L \times U(1)_Y$
symmetry\index{weak interaction} with a larger \underline{$SU(2)_L
  \times SU(2)_R$} symmetry. At this stage this extended symmetry does
not have to be a local gauge symmetry, a global version of $SU(2)_L$
combined with a global $SU(2)_R$ is sufficient. This global symmetry
would have to act like
\begin{alignat}{2}
 & \s\to \, U\s V^\dag \qqquad U\in SU(2)_L \qqquad V\in SU(2)_R \notag \\
 & \tr(\sdag\s)\to \, 
   \tr \left( V\sdag U^\dag U \s V^\dag \right) = 
   \tr( \s^\dag \s) \; .
\label{eq:custo1}
\end{alignat}
In this setup, the three components of $W^\mu$ form a triplet under
$SU(2)_L$ and a singlet under $SU(2)_R$.  If we cannot extract
$\tau_3$ as a special generator of $SU(2)_L$ and combine it with the
$U(1)_Y$ hypercharge the $W$ and $Z$ masses have to be identical,
corresponding to $c_w =1$ at tree level.

In the gauge boson and fermion mass terms computed in unitary gauge
the $\s$ field becomes identical to its vacuum expectation value
$\one$\index{unitary gauge}. The combined
global $SU(2)_L$ transformations act on the symmetry breaking vacuum expectation value the same way as shown in
Eq.\eqref{eq:higgs_sigmatrafo},
\begin{alignat}{5}
 \langle \s \rangle \to 
 \langle U \s V^\dag \rangle = 
 \langle U \one V^\dag \rangle = 
 U V^\dag \really \one \; .
\label{eq:custo2}
\end{alignat}
The last step, \ie the symmetry requirement for the Lagrangian can
only be satisfied if we require $U=V$. In other words, the vacuum
expectation value for $\s$ breaks $SU(2)_L \times SU(2)_R$ to the
\underline{diagonal subgroup} $SU(2)_{L+R}$.  The technical term is
precisely defined this way --- the two $SU(2)$ symmetries reduce to
one remaining symmetry which can be written as $U=V$.  Depending on if
we look at the global symmetry structure in the unbroken or broken
phase the \underline{custodial symmetry} group either refers to $SU(2)_R$ or
$SU(2)_{L+R}$.\bigskip

Even beyond tree level the global $SU(2)_L \times SU(2)_R$ symmetry
structure can protect the relation $\rho = 1$ between the gauge boson masses
shown in Eq.\eqref{eq:custo0}.
From Eq.\eqref{eq:custo1} we immediately see that it allows all terms
in the Higgs potential $\lag_\s$, but it changes the picture not only
for gauge boson but also for fermion masses. If fermions reside in
$SU(2)_L$ as well as $SU(2)_R$ doublets we cannot implement any
difference between up--type and down--type fermions in the Lagrangian.
The custodial symmetry is only intact in the limit for example of
identical third generation fermion masses $m_b = m_t$.

The measured masses $m_t \gg m_b$ change the protected tree level
value $\rho =1$: self energy loops in the $W$ propagator involve a
mixture of the bottom and top quark, while the $Z$ propagator includes
pure bottom and top loops.  Skipping the loop calculation we quote
their different contributions to the gauge boson masses as
\index{electroweak precision measurements!$\rho$ parameter} 
\begin{alignat}{5}
 \Delta\rho & \supset \frac{3 G_F}{8\sqrt{2}\pi^2}
                    \left(   m_t^2+m_b^2
                           - 2 \frac{m_t^2m_b^2}{m_t^2-m_b^2} 
                               \log\frac{m_t^2}{m_b^2}
                    \right) \notag \\
             & = \frac{3 G_F}{8\sqrt{2}\pi^2}
                    \left(   2 m_b^2 + m_b^2 \delta
                           - 2 m_b^2 \frac{1+\delta}{\delta}
                               \log \left( 1+ \delta \right)
                    \right) \qqquad \text{defining} \quad m_t^2=m_b^2 (1+\delta) \notag \\
            & =     \frac{3 G_F}{8\sqrt{2}\pi^2}
                    \left(   2 m_b^2 + m_b^2 \delta
                           - 2 m_b^2 \left( \frac{1}{\delta} + 1 \right)
                               \left( \delta
                                     -\frac{\delta^2}{2}
                                     +\frac{\delta^3}{3}
                                     +\ope(\delta^4) \right)
                    \right) \notag \\
            & =    \frac{3 G_F}{8\sqrt{2}\pi^2} m_b^2 \;
                    \left(  2 + \delta 
                          - 2 - 2 \delta 
                          + \delta + \delta^2
                          - \frac{2}{3} \delta^2
                          + \ope(\delta^3)
                    \right) \notag \\
            & =    \frac{3 G_F}{8\sqrt{2}\pi^2} m_b^2 \; 
                    \left( \frac{1}{3} \delta^2
                          + \ope(\delta^3) 
                    \right) \notag \\
            & =    \frac{G_F m_W^2}{8\sqrt{2}\pi^2} \; 
                    \left( \frac{\left( m_t^2 - m_b^2 \right)^2}{m_W^2 m_b^2} 
                          + \cdots \right) \; .
\label{eq:rho_top}
\end{alignat}
In the Taylor series above the assumption of $\delta$ being small is
of course not realistic, but the result is nevertheless instructive:
the shift vanishes very rapidly towards the chirally symmetric limit
$m_t \sim m_b$.  The sign of the contribution of a chiral fermion
doublet to $\Delta \rho$ is always positive.  In terms of realistic
Standard Model mass ratios it scales like
\begin{alignat}{5}
 \Delta\rho  & \supset \frac{3 G_F}{8\sqrt{2}\pi^2} m_t^2 
                    \left(   1
                           - 2 \frac{m_b^2}{m_t^2} 
                               \log\frac{m_t^2}{m_b^2}
                    \right)
= \frac{3 G_Fm_W^2}{8\sqrt{2}\pi^2} \; \frac{m_t^2}{m_W^2} \; 
                    \left(   1 + \ope \left( \frac{m_b^2}{m_t^2} \right) \right) \; ,
\end{alignat}
remembering that the Fermi coupling constant has a mass dimension
fixed by $G_F \propto 1/m_W^2$.\bigskip

We have already argued that hypercharge or electric charge break
custodial symmetry. From the form of the covariant derivative $D_\mu
\s$ including a single $\tau_3$ we can guess that the $SU(2)_R$
symmetry will not allow $B$ field interactions which are proportional
to $s_w \sim \sqrt{1/4}$.  A second contribution to the $\rho$
parameter therefore arises from Higgs loops in the presence of $g' \neq
0$
\begin{alignat}{5}
\boxed{
 \Delta\rho\supset - \frac{11 G_Fm_Z^2s_w^2}{24\sqrt{2}\pi^2}
                \; \log\frac{m_H^2}{m_Z^2}.
}
\label{eq:rho_higgs}
\end{alignat}
The loop diagrams responsible for the contribution are simply virtual
Higgs exchanges in the $W$ and $Z$ self energies, which not only have
a different factorizing couplings, but also different $W$ and $Z$
masses inside the loop. These masses inside loop diagrams appear as
logarithms. The sign of this contribution implies that larger Higgs
masses give increasingly negative contributions to the $\rho$ parameter.
\bigskip

There is another parameterization of the same effect, namely the $T$
parameter. It is part of an effective theory parameterization of
deviations from the tree level relations between gauge boson masses,
mixing angles, and neutral and charged current couplings. 
If we allow for deviations from the Standard Model gauge sector
induced by vacuum polarization corrections $\Pi(p^2)$ and their
momentum derivatives $\Pi'(p^2)$ we can write down additional Lagrangian
terms
\begin{alignat}{5}
\Delta \lag =
&- \frac{\Pi'_{\gamma \gamma}}{4} \hat{F}_{\mu \nu} \hat{F}^{\mu \nu} \; 
- \frac{\Pi'_{WW}}{2} \hat{W}_{\mu \nu} \hat{W}^{\mu \nu} \; 
- \frac{\Pi'_{ZZ}}{4} \hat{Z}_{\mu \nu} \hat{Z}^{\mu \nu} \notag \\
&- \frac{\Pi'_{\gamma Z}}{4} \hat{F}_{\mu \nu} \hat{Z}^{\mu \nu} \; 
- \Pi_{WW} \, \hat{m}_W^2 \hat{W}^+_\mu \hat{W}^-{}^\mu 
- \frac{\Pi_{ZZ}}{2} \, \hat{m}_Z^2 \hat{Z}_\mu \hat{Z}^\mu \; .
\label{eq:higgs_stu_lag1}
\end{alignat}
The field strengths $\hat{F}_{\mu \nu},\hat{W}_{\mu \nu},\hat{Z}_{\mu
  \nu}$ are based the fields
$\hat{A}_\mu,\hat{W}_\mu,\hat{Z}_\mu$. The hats are necessary, because
the kinetic terms and hence the fields do not (yet) have the canonical
normalization. To compute the proper field normalization we assume
that all $\Pi'$ are small, so we can express the hatted gauge fields
in terms of the properly normalized fields as
\begin{equation}
\hat{A}_\mu = \left( 1 - \frac{\Pi'_{\gamma \gamma}}{2} \right) A_\mu 
            + \Pi'_{\gamma Z} Z_\mu 
\qqquad 
\hat{W}_\mu = \left( 1 - \frac{\Pi'_{WW}}{2} \right) W_\mu 
\qqquad 
\hat{Z}_\mu = \left( 1 - \frac{\Pi'_{ZZ}}{2} \right) Z_\mu \; .
\label{eq:higgs_stu_fields}
\end{equation}
To check this ansatz, we can 
for example extract all terms proportional to the photon--$Z$
mixing $\Pi'_{\gamma Z}$ arising from Eq.\eqref{eq:higgs_stu_fields}
and find
\begin{alignat}{5}
- \frac{1}{4} \hat{F}_{\mu \nu} \hat{F}^{\mu \nu} \Big|_{\gamma Z}
&=
 - \frac{1}{4}
   \left( \p_\mu \hat{A}_\nu - \p_\nu \hat{A}_\mu \right) \,
   \left( \p_\mu \hat{A}_\nu - \p_\nu \hat{A}_\mu \right) \Big|_{\gamma Z}
   \notag \\
&=
 - \frac{1}{4}
   \left( \p_\mu (A+\Pi'_{\gamma Z} Z)_\nu 
        - \p_\nu (A+\Pi'_{\gamma Z} Z)_\mu \right) \,
   \left( \p_\mu (A+\Pi'_{\gamma Z} Z)_\nu 
        - \p_\nu (A+\Pi'_{\gamma Z} Z)_\mu \right) \Big|_{\gamma Z}
   \notag \\
&=
 - \frac{\Pi'_{\gamma Z}}{4}
   \left( \p_\mu A_\nu - \p_\nu A_\mu \right) \,
   \left( \p_\mu Z_\nu - \p_\nu Z_\mu \right)
 - \frac{\Pi'_{\gamma Z}}{4}
   \left( \p_\mu Z_\nu - \p_\nu Z_\mu \right) \,
   \left( \p_\mu A_\nu - \p_\nu A_\mu \right)
 + \ope(\Pi'{}^2) \notag \\
&=
 - \frac{\Pi'_{\gamma Z}}{2}
   \left( \p_\mu Z_\nu - \p_\nu Z_\mu \right) \,
   \left( \p_\mu A_\nu - \p_\nu A_\mu \right) 
 + \ope(\Pi'{}^2)
   \notag \\
&= 
 - \frac{\Pi'_{\gamma Z}}{2} Z_{\mu \nu} F^{\mu \nu}  
   + \ope(\Pi'{}^2)
 = 
 - \frac{\Pi'_{\gamma Z}}{2} \hat{Z}_{\mu \nu} \hat{F}^{\mu \nu}  
   + \ope(\Pi'{}^2) \; .
\end{alignat}
The field shift in Eq.\eqref{eq:higgs_stu_fields} indeed absorbs the
explicit new contribution in Eq.\eqref{eq:higgs_stu_lag1}. Assuming
that this works for all gauge field combinations the Lagrangian
including the loop--induced $\Delta \lag$ gets the canonical form
\begin{alignat}{5}
 \lag \supset 
&- \frac{1}{4} F_{\mu \nu} F^{\mu \nu} \; 
- \frac{1}{2} W_{\mu \nu} W^{\mu \nu} \; 
- \frac{1}{4} Z_{\mu \nu} Z^{\mu \nu} \notag \\
&- ( 1 + \Pi_{WW} -\Pi'_{WW} ) \, \hat{m}_W^2 W^+_\mu W^-{}^\mu 
- \frac{1}{2} (1+\Pi_{ZZ}+\Pi'_{ZZ}) \, \hat{m}_Z^2 Z_\mu Z^\mu \; .
\label{eq:higgs_stu_lag2}
\end{alignat}
The physical $Z$ mass now has to be $m_Z^2 = (1+\Pi_{ZZ}+\Pi'_{ZZ}) \,
\hat{m}_Z^2$. Just as in the usual Lagrangian we can link the two
gauge boson masses through the (hatted) weak mixing angle $\hat{m}_W =
\hat{c}_w \hat{m}_Z$. In terms of this mixing angle we can compute the
muon decay constant, the result of which we quote as
\begin{alignat}{5}
\frac{\hat{s}_w^2}{s_w^2} &=
1 + \frac{c_w^2}{c_w^2 - s_w^2}
         \; \left( \Pi'_{\gamma \gamma} - \Pi'_{ZZ} - \Pi_{WW} + \Pi_{ZZ} 
            \right) \notag \\
\text{or} \qquad 
\frac{\hat{c}_w^2}{c_w^2} &=
1 - \frac{s_w^2}{c_w^2 - s_w^2}
         \; \left( \Pi'_{\gamma \gamma} - \Pi'_{ZZ} - \Pi_{WW} + \Pi_{ZZ} 
            \right) \; .
\end{alignat}
With this result for example the complete $W$--mass term in
Eq.\eqref{eq:higgs_stu_lag2} reads
\begin{alignat}{5}
\lag 
&\supset
 - ( 1+\Pi_{WW}-\Pi'_{WW} ) \; \hat{c}_w^2 \; \hat{m}_Z^2 \; 
 W^+_\mu W^-{}^\mu \notag \\
&=
 - ( 1+\Pi_{WW}-\Pi'_{WW} ) 
    \left[ 1 - \frac{s_w^2}{c_w^2 - s_w^2}
         \left( \Pi'_{\gamma \gamma} - \Pi'_{ZZ} - \Pi_{WW} + \Pi_{ZZ} \right) \right] 
    c_w^2 
    ( 1-\Pi_{ZZ}+\Pi'_{ZZ} ) \; m_Z^2 \; W^+_\mu W^-{}^\mu \notag \\
& = -\left[ 1-\Pi'_{WW}+\Pi'_{ZZ}+\Pi_{WW}-\Pi_{ZZ}  
          -\frac{s_w^2}{c_w^2-s_w^2} 
           \left( \Pi'_{\gamma \gamma}-\Pi'_{ZZ}-\Pi_{WW}+\Pi_{ZZ} \right)
    \right] \; m_Z^2 \; W^+_\mu W^-{}^\mu \notag \\
& = -\left[ 1 
          - \frac{\alpha S}{2(c_w^2-s_w^2)}
          + \frac{c_w^2 \alpha T}{c_w^2-s_w^2}
          + \frac{\alpha U}{4 s_w^2}
    \right] \; m_Z^2 \; W^+_\mu W^-{}^\mu \; .
\end{alignat}
In the last step we have defined three typical combinations of the
different correction factors as
\begin{alignat}{5}
S &=
\frac{4 s_w^2 c_w^2}{\alpha} \; \left( -\Pi'_{\gamma \gamma} + \Pi'_{ZZ} 
                        -\Pi'_{\gamma Z} \frac{c_w^2-s_w^2}{c_w s_w} \right) 
\notag \\
T &=
\frac{1}{\alpha} \left( \Pi_{WW} - \Pi_{ZZ} \right)
\notag \\
U &=
\frac{4 s_w^4}{\alpha} \; \left( \Pi'_{\gamma \gamma} - \frac{\Pi'_{WW}}{s_w^2} 
                    + \Pi'_{ZZ} \frac{c_w^2}{s_w^2} 
                    - 2 \Pi'_{\gamma Z} \frac{2 c_w}{s_w}  \right) \; .
\end{alignat}
Two of these so-called \underline{Peskin--Takeuchi parameters} can be
understood fairly easily: the $S$-parameter corresponds to a shift of
the $Z$ mass. This is not completely obvious because it seems to also
involve photon terms. We have to remember that the weak mixing angle
is defined such that the photon is massless, while all mass terms are
absorbed into the $Z$ boson. The $T$ parameter obviously compares
contributions to the $W$ and $Z$ masses. The third parameter $U$ is
less important for most models.\bigskip

To get an idea how additional fermions contribute to $S$
and $T$ we quote the contributions from the heavy fermion doublet:
\begin{alignat}{5}
\Delta S =
\frac{N_c}{6 \pi} \; \left( 1 - 2 Y \log \frac{m_t^2}{m_b^2} \right) 
\qquad \qquad 
\Delta T &=
\frac{N_c}{4 \pi s_w^2 c_w^2 m_Z^2} \;
\left( m_t^2 + m_b^2
      -\frac{2 m_t^2 m_b^2}{m_t^2-m_b^2} 
       \log \frac{m_t^2}{m_b^2} 
\right) \; ,
\label{eq:s_t_sm}
\end{alignat}
with $Y=1/6$ for quarks and $Y=-1/2$ for leptons. 
While the parameter $S$ has nothing to do with our custodial symmetry,
$\rho$ and $T \sim \Delta \rho/\alpha$ are closely linked.  Their main
difference is the reference point, where $\rho =1$ refers to its
tree level value and $T=0$ is often chosen for some kind of light
Higgs mass and including the Standard Model top-bottom corrections.
A similar third set of parameters going back to Altarelli and Barbieri
consists of $\epsilon_{1,2,3}$, where the leading effect on the
custodial symmetry can be translated via $\epsilon_1 = \alpha
T$.\bigskip

Typical experimental constraints form an ellipse in the $S$ vs $T$
plane along the diagonal. They are usually quoted as $\Delta T$ with
respect to a reference Higgs mass. Compared to a 125~GeV Standard
Model Higgs boson the measured values range around $T \sim 0.1$ and $S
\sim 0.05$. Additional contributions $\Delta T \sim 0.1$ tend to be
within the experimental errors, much larger contributions are in
contradiction with experiment.

\begin{figure}[t]
\begin{center}
\includegraphics[width=0.3\hsize]{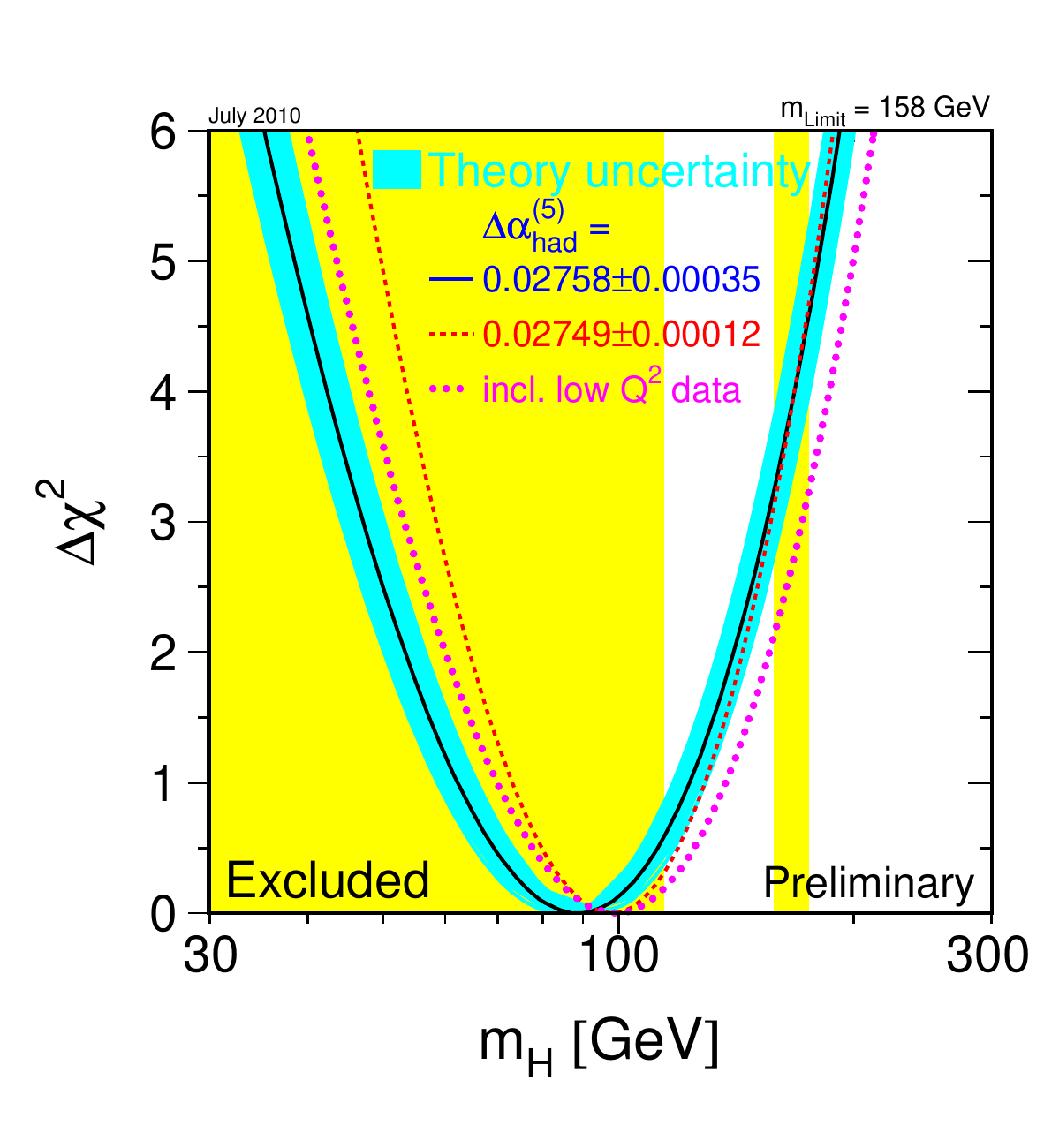}
\end{center}
\caption{Allowed range of Higgs masses in the Standard Model after
  taking into account electroweak precision data, most notably the
  $\rho$ parameter contribution from the Higgs itself,
  Eq.\eqref{eq:rho_higgs}. Figure from the LEP electroweak working
  group, with updates available under
  \url{http://lepewwg.web.cern.ch/LEPEWWG}.}
\label{fig:higgs_blueband}
\end{figure}

There are two reasons to discuss these loop contributions breaking the
custodial symmetry in the Standard Model. First, $\Delta \rho$ is
experimentally very strongly constrained by \underline{electroweak
  precision measurements}\index{electroweak precision measurements!Higgs mass}, which means that alternative models for
electroweak symmetry breaking usually include the same kind of
approximate custodial symmetry by construction.  As a matter of fact,
this constraint is responsible for the death of technicolor
models, which describe the Higgs boson as a bound state under a new
QCD-like interaction and which we will discuss in Section~\ref{sec:higgs_technicolor}.

Even more importantly, in the Standard Model we can measure the
symmetry violations from the heavy quarks and from the Higgs sector
shown in Eqs.\eqref{eq:rho_top} and~\eqref{eq:rho_higgs} in electroweak
precision measurements. Even though the Higgs contributions depend on
the Higgs mass only logarithmically, we can then derive an
\underline{upper bound on the Higgs mass}\index{Higgs
  mass!experimental upper bound} of the order of $\ope(200~\gev)$, as
shown in Figure~\ref{fig:higgs_blueband}.  This strongly suggests that
if we are faced (mostly) with the Standard Model at the weak scale the
Tevatron and at the LHC will be looking for a fairly light Higgs boson
--- or something that very much looks like a light fundamental Higgs
boson. This is the reason why in the absence of other hints for new
physics at the LHC the discovery of a light Standard--Model--like Higgs
boson is not unexpected. Any significant deviation of a Higgs boson
from the Standard Model prediction would have to be compensated by
additional yet unobserved particles in the relevant self energy
diagrams.

Turning this argument around, we should firmly keep in mind that the
$\rho$ parameter only points to a light fundamental Higgs boson if we
assume the Standard Model Higgs mechanism. For any other model it
might point to something similar to a light Higgs field, but does not
have to be fundamental. Including additional fields in a model can
even turn around this prediction and prefer a heavy Higgs
state. By now, studying electroweak precision data given the
measured Higgs mass is one of the most sensitive consistency tests of 
the Standard Model.\bigskip

The symmetry breaking pattern we describe in this section is a nice
example to check the predictions from the Goldstone theorem quoted in
Section~\ref{sec:higgs_qed}, so our last question is: how do physical
modes which we introduce as $\s(x)=\exp(-i\w /v)$ transform under the
different broken and unbroken global $SU(2)$ symmetries which make up
the custodial symmetry and can we construct a model of electroweak
symmetry breaking around the custodial symmetry?  This brings us back
to the example of the photon mass, where we first saw Goldstone's
theorem\index{Goldstone's theorem} at work.

Under the usual $SU(2)_L$ symmetry we know the transformation reads
$\s \to U \s$ with $U=\exp(i \alpha \cdot \tau/2)$. The
transformation properties of the Goldstone modes
$\vec{w}$\index{Goldstone boson} follow from the infinitesimal
transformations
\begin{alignat}{5}
 \one - i \frac{w \cdot \tau}{v} 
       & \to \left( \one + i \frac{\alpha \cdot \tau}{2} \right)
             \left( \one - i \frac{w \cdot \tau}{v} \right) \notag \\
       & =  \one + \frac{i}{v} \left( 
              - w \cdot \tau + \frac{v}{2} \alpha \cdot \tau \right) + \cdots \notag \\
       & \really  \one - i \frac{w' \cdot \tau}{v} 
\qqquad \text{implying} \quad
\boxed{w \to w' = w - \frac{v}{2} \alpha} \; ,
\label{eq:nonlinear}
\end{alignat}
so $U$ is a \underline{non--linear transformation},
since $w'_a$ is not
proportional to $w_a$.  The same independent structure we find for the
$SU(2)_R$ transformation. This model of electroweak symmetry breaking
we call a non--linear sigma model. In our discussion of the Goldstone
theorem we already quoted its most important feature: when we
construct a Lagrangian this non--linear symmetry transformation forbids
mass terms, gauge interactions, Yukawa couplings, and quadratic
potential terms for these modes in $\s$. As discussed in
Section~\ref{sec:higgs_qed} only derivative terms like the kinetic
term and derivative couplings are allowed under the $SU(2)_L$ and
$SU(2)_R$ symmetries.\bigskip

Similarly, we can evaluate the transformation of these physical modes
under the remaining diagonal symmetry group $SU(2)_{L+R}$ with $\s \to
U \s U^\dag$ and instead find
\begin{alignat}{5}
 \one - i \frac{w \cdot \tau}{v} 
       & \to \left( \one + i \frac{\alpha \cdot \tau}{2} \right)
             \left( \one - i \frac{w \cdot \tau}{v} \right) 
             \left( \one - i \frac{\alpha \cdot \tau}{2} \right) \notag \\
       & =   \left( \one + i \frac{\alpha \cdot \tau}{2} \right)
             \left( \left[ \left( \one - i \frac{w \cdot \tau}{v} \right) 
                          ,\left( \one - i \frac{\alpha \cdot \tau}{2} \right) \right] 
                   +\left( \one - i \frac{\alpha \cdot \tau}{2} \right)
                    \left( \one - i \frac{w \cdot \tau}{v} \right) 
             \right)  \notag \\
       & =   \left( \one + i \frac{\alpha \cdot \tau}{2} \right)
             \; \left[  -i \frac{w \cdot \tau}{v} 
                      , -i \frac{\alpha \cdot \tau}{2} \right] 
             +  \left( \one - i \frac{w \cdot \tau}{v} \right) + \cdots \notag \\
       & =   \left( \one + i \frac{\alpha \cdot \tau}{2} \right)
             \; \frac{1}{2v} \; 
                   2i \tau \, (\alpha \times w) 
             +  \left( \one - i \frac{w \cdot \tau}{v} \right) + \cdots \notag \\
       & = \one - i \frac{w \cdot \tau}{v}   
           + i \frac{\tau \, (\alpha \times w)}{v} + \cdots \notag \\
& 
\qqquad \text{implying} \quad
 \boxed{w_i \to w_i' = w_i -\varepsilon_{ijk} \alpha_j w_k} \; ,
\end{alignat}
which is a \underline{linear transformation}.  In the fourth line we
use the commutator
\begin{alignat}{2}
 [\tau_a,\tau_b] = 2i \varepsilon_{abc} \tau_c \quad 
 & \Rightarrow \quad (\alpha \cdot \tau)(w \cdot \tau)
                    = \alpha \cdot w 
                      + i \tau \, (\alpha \times w)
 \qqquad \text{using Eq.\eqref{eq:paulimat}}
                    \notag \\
 & \Rightarrow \quad \left[ ( \alpha \cdot \tau),
                            ( w \cdot \tau)
                     \right]
                    = 2i \tau \, (\alpha  \times w) \; .
\end{alignat}
In other words, when we transform the physical modes corresponding to
the broken generators in $\s$ by the larger symmetry $SU(2)_L \times
SU(2)_R$ we find a non--linear transformation, while the approximate
symmetry $SU(2)_{L+R}$ leads to a linear transformation. This is
precisely what Goldstone's theorem predicts for the spontaneous
breaking of a global electroweak symmetry.

\subsection{The Standard Model}
\label{sec:higgs_sm}

Before we discuss all the ways we can look for a Higgs Boson and go
through the Higgs discovery in the summer of 2012 we briefly review
the Higgs mechanism in the Standard Model. This will link the somewhat
non--standard effective theory approach we used until now to the
standard textbook arguments.  In the last sections we have seen that
there does not really need to be such a fundamental scalar, but
electroweak precision data tells us whatever it is the Higgs should
look very similar to a light fundamental scalar, unless we see some
seriously new states and effects around the weak scale.

To make it a little more interesting and since we are already in the
mood of not taking the Standard Model Higgs sector too literally, in
Section~\ref{sec:higgs_pot} we include higher--dimensional operators on
top of the usual \underline{renormalizable}\index{renormalizable operators} dimension-4 operators in the Higgs potential.  Such
operators generally occur in effective theories based on ultraviolet
completions of our Standard Model, but their effects are often small
or assumed to be small.

Once we want to analyze the behavior of the Higgs sector over a wide
range of energy scales, like we will do in
Sections~\ref{sec:higgs_unitarity} and~\ref{sec:higgs_rge}, we need to
take the Standard Model seriously and in turn find constraints on the
structure and the validity of our Standard Model with a fundamental
Higgs boson.

\subsubsection{Higgs potential to dimension six}
\label{sec:higgs_pot}

In the \underline{renormalizable Standard Model} all terms in the
Lagrangian are defined to be of mass dimension four, like
$m_{f}\bar{\psi}\psi$ or $\bar{\psi}\p_\mu \psi$ or
$F_{\mu\nu}F^{\mu\nu}$. This mass dimension is easy to read off if we
remember that for example scalar fields or vector-boson fields
contribute mass dimension one while fermion spinors carry mass
dimension 3/2. The same renormalizability assumption we usually make
for the Higgs potential, even though from the previous discussion it
is clear that higher--dimensional terms --- stemming from higher powers
of $\tr(\sdag \s)$ --- can and should exist.

Starting from the Higgs doublets introduced in Eq.\eqref{eq:def_phi}
and for now ignoring the Goldstone modes the simplified Higgs--only doublet
\begin{alignat}{5}
\qquad \phi =\frac{1}{\sqrt{2}} 
             \begin{pmatrix}
                    -w_2 - iw_1 \\ v+H+i w_3 
                   \end{pmatrix}
       \sim  \frac{1}{\sqrt{2}}
             \begin{pmatrix}
                    0 \\ v+H
                   \end{pmatrix}
\label{eq:higgs_phi_simp}
\end{alignat}
leaves us with only two renormalizable potential terms in
Eq.\eqref{eq:sigma_pot_h}, now written in terms of the Higgs doublet
and with $\mu^2$ and $\lambda$ as prefactors
\begin{alignat}{5}
-\lag_\s = V_\text{SM}=\mu^2|\phi|^2+\lambda|\phi|^4+\text{const} \; .
\end{alignat}
To emphasize that renormalizability is a strong and not necessarily
very justified theoretical assumption in LHC Higgs physics, we allow
for more operators in the Higgs potential.  Following the discussion
in Section~\ref{sec:higgs_higgsboson} we will use the linear
representation in terms of the doublet $\phi$ to organize the extended
Lagrangian by mass dimensions.  If we expand the possible mass
dimensions and the operator basis, there are exactly \underline{two
  gauge--invariant operators}\index{Higgs potential!dimension-6 operators} of dimension six we can write down in terms of the Higgs
doublet $|\phi|^2$, \ie before electroweak symmetry breaking
\begin{alignat}{5}
\ope_1=\frac{1}{2}\p_\mu (\phi^{\dagger}\phi) \;
                           \p^\mu (\phi^{\dagger}\phi) 
\qqqquad
\ope_2 =-\frac{1}{3}(\phi^{\dagger}\phi)^3  \; .
\end{alignat}
There exists one more possible operator $(D_\mu
\phi)^\dagger \phi \; \phi^\dagger (D^\mu \phi)$, but it
violates custodial symmetry,
so we ignore it in our
analysis. The prefactors in the Lagrangian are conventional, because to
construct a Lagrangian we have to multiply these operators with
general coefficients of mass dimension minus two, parameterized in terms of an
unknown mass scale $\Lambda$
\begin{alignat}{5}
\lag_{D6} = 
 \sum_{i=1}^{2} \dfrac{f_i}{\Lambda^2} \ope_i \; .
\end{alignat}
As long as the typical energy scale $E$ in the numerator
in our matrix element is small $(E \ll \Lambda)$, the corrections from
the additional operators are small as well.\bigskip

Before we compute the Higgs potential including $\ope_2$ we look at
the effects of the dimension-6 operator $\ope_1$.  It contributes to
the kinetic term of the Higgs field in the Lagrangian, before or after
symmetry breaking
\begin{alignat}{5}
 \ope_1 
&= 
 \frac{1}{2}\p_\mu (\phi^{\dagger}\phi) \;
            \p^\mu (\phi^{\dagger}\phi)  
 \notag \\
&=
 \frac{1}{2}\p_\mu \left(\frac{(\hat{H}+v)^2}{2}\right)
 \p^\mu \left(\frac{(\hat{H}+v)^2}{2}\right)       
 \notag \\
&=
 \frac{1}{2} \; (\hat{H}+v)^2 \;
            \p_\mu \hat{H} \;
            \p^\mu \hat{H} \; .
\label{eq:higgs_o1}
\end{alignat}
We use the symbol $\hat{H}$ for the Higgs field as part of $\phi$.
From the similar case of the gauge fields in Eq.\eqref{eq:higgs_stu_lag1}
we can guess that there will be a
difference between $\hat{H}$ and the physical Higgs field $H$ at the
end of the day.  The contribution from $\ope_1$ leaves us with a
combined kinetic term
\begin{alignat}{5}
 \lag_\text{kin} = \frac{1}{2}\p_\mu \hat{H}
                                         \p^\mu \hat{H}
 \left(1+\frac{f_1 v^2}{\Lambda^2} \right) 
 \really 
 \frac{1}{2} \p_\mu H \; \p^\mu H
 \qquad  \Leftrightarrow \qquad 
 \boxed{H=\sqrt{1+\frac{f_1 v^2}{\Lambda^2}} \hat{H}
} \; .
\label{eq:higgs_kinetic}
\end{alignat}
This is a simple rescaling to define the \underline{canonical kinetic
  term} in the Lagrangian, corresponding to a finite wave function
renormalization which ensures that the residuum of the Higgs
propagator is one. This kind of condition is well known from the LSZ
equation and the proper definition of outgoing states. It means we
have to eventually replace $\hat{H}$ with $H$ in the entire Higgs
sector. In most cases such a wave function renormalization will not 
lead to observable physics effects, because it can be absorbed for example
in coupling renormalization. However, in this case the setup of the electroweak
sector does not give us enough freedom to absorb this scaling factor, 
so it will appear in the observable couplings similar to a form 
factor in strongly interacting models.\bigskip

Taking into account the additional dimension-6 operator $\ope_2$ we
can write the Higgs potential as
\begin{alignat}{5}
\boxed{
V=\mu^2|\phi|^2+\lambda|\phi|^4+\frac{f_2}{3\Lambda^2}|\phi|^6 
} \; .
\label{eq:higgs_pot}
\end{alignat}
The positive sign in the last term of the potential $V$ 
ensures that for
$f_2>0$ the potential is bounded from below for large field values
$\phi$. The non--trivial minimum at $\phi \ne 0$ is given by
\begin{alignat}{5}
 \frac{\p V}{\p \, |\phi|^2} = 
 \mu^2
 +2 \lambda|\phi|^2
 +\frac{3 f_2}{3\Lambda^2}|\phi|^4 \really 0
\qquad \Leftrightarrow \qquad 
 |\phi|^4
 +\frac{2\lambda\Lambda^2}{f_2 }|\phi|^2
 +\frac{\mu^2\Lambda^2}{f_2} \really 0  \; ,
\end{alignat}
defining the minimum position $|\phi|^2 = v^2/2$. The two solutions of
the quadratic equation for $v^2/2$ are
\begin{alignat}{5}
\frac{v^2}{2} 
&
 = -\frac{\lambda\Lambda^2}{f_2 } \pm 
 \left[\left(\frac{\lambda\Lambda^2}{f_2}\right)^2
 -\frac{\mu^2\Lambda^2}{f_2}\right]^{\frac{1}{2}} \;
 =\;\frac{\lambda\Lambda^2}{f_2 }
 \left[-1 \;\pm\; \sqrt{1-\frac{\mu^2 f_2}{\Lambda^2 \lambda^2}}\right]
 \notag \\
&
 =
 \frac{\lambda\Lambda^2}{f_2} \left[-1\;\pm\;\left(1-\frac{f_2 \mu^2}
 {2\lambda^2\Lambda^2}-\frac{f_2^2\mu^4}{8\lambda^4\Lambda^4}
 +\ope \left( \frac{1}{\Lambda^6} \right) \right)\right]    
 \notag \\
&
 =
\left\{%
\begin{array}{ll}
 -\dfrac{\mu^2}{2\lambda}
 -\dfrac{f_2 \mu^4}{8\lambda^3 \Lambda^2}
 +\ope \left( \dfrac{1}{\Lambda^4} \right)
 =-\dfrac{\mu^2}{2\lambda}
   \left(1+\dfrac{f_2\mu^2}{4\lambda^2\Lambda^2}
  +\ope \left( \dfrac{1}{\Lambda^4} \right) \right)
 \equiv \dfrac{v_0^2}{2}
   \left(1+ \dfrac{f_2 v_0^2}{4\lambda\Lambda^2}
 +\ope \left( \dfrac{1}{\Lambda^4} \right) \right)
\\
 -\dfrac{2\lambda\Lambda^2}{f_2^2} + \ope(\Lambda^0)
\\
\end{array}
\right.
\label{eq:higgs_vev}
\end{alignat}
The first solution we have expanded around the Standard Model minimum,
$v_0^2=-\mu^2/\lambda$.  The second, high--scale solution is not the
vacuum relevant for our Standard Model. Note that from the $W,Z$
masses we know that $v=246$~GeV so $v$ is really our first observable
in the Higgs sector, sensitive to the higher--dimensional
operators.\bigskip

To compute the Higgs mass as the second observable we could study the second
derivative of the potential in the different directions, but we can
also simply collect all quadratic terms contributing to the Lagrangian
by hand. The regular dimension-4 contributions in terms of the shifted
Higgs field $\hat{H}$ are\index{Higgs potential}
\begin{alignat}{5}
V_\text{SM} = \mu^2 \frac{(\hat{H}+v)^2}{2} 
           + \lambda \frac{(\hat{H}+v)^4}{4}
           = \frac{\mu^2}{2} \left( \hat{H}^2 \cdots \right)
           + \frac{\lambda}{4} \left( \cdots 6\hat{H}^2 v^2 \cdots \right) \; .
\label{eq:higgsmass_sm}
\end{alignat}
Only the terms in the parentheses contribute to the Higgs mass in
terms of $\mu$, $v$ and $\lambda$.  Including the additional
potential operator in terms of $\hat{H}$ gives
\begin{alignat}{5}
 \ope_2 
&=
 -\frac{1}{3}(\phi^{\dagger}\phi)^3 \notag \\
&=
 -\frac{1}{3} \frac{(\hat{H}+v)^6}{8} \notag \\
&=
 -\frac{1}{24} \left( \hat{H}^6+6\hat{H}^5v+15\hat{H}^4v^2
 +20\hat{H}^3v^3+15\hat{H}^2v^4+6\hat{H}v^5+v^6 \right) \; .
\end{alignat}
Combining both gives us the complete quadratic mass term to dimension
six
\begin{alignat}{5}
 \lag_\text{mass} &=
 -\frac{\mu^2}{2} \hat{H}^2-\frac{3}{2}\lambda v^2 \hat{H}^2
 -\frac{f_2 }{\Lambda^2}\frac{15}{24}v^4\hat{H}^2 \notag \\
&= 
 -\frac{1}{2} \left(\mu^2+3\lambda v^2 + 
 \frac{5}{4}\frac{f_2 v^{4}}{\Lambda^2}\right)\hat{H}^2
 \notag \\
&= 
 -\frac{1}{2} 
  \left(-\lambda v^2 \left(1+\frac{f_2 v^2}{4\lambda\Lambda^2}\right)
        +3\lambda v^2+ \frac{5}{4}\frac{f_2 v^4}{\Lambda^2}\right)\hat{H}^2
 \qqquad && \text{replacing $\mu^2$ using Eq.\eqref{eq:higgs_vev} twice}
 \notag \\
&=
 -\frac{1}{2} 
  \left(2\lambda v^2-\frac{f_2 v^4}{4\Lambda^2} 
       + \frac{5}{4}\frac{f_2 v^4}{\Lambda^2}\right)
 \left(1+\frac{f_1 v^2}{\Lambda^2}\right)^{-1}H^2
 \qqquad && \text{replacing $\hat{H}$ using Eq.\eqref{eq:higgs_kinetic}}
 \notag \\
&=
 -\frac{1}{2} 
  \left(2\lambda v^2+\frac{f_2 v^4}{\Lambda^2} \right)
  \left(1-\frac{f_1 v^2}{\Lambda^2}+\ope \left( \frac{1}{\Lambda^4} \right) \right)
  H^2
 \notag \\
&=
 -\lambda v^2
  \left(1 + \frac{f_2 v^2}{2 \lambda \Lambda^2} \right)
  \left(1-\frac{f_1 v^2}{\Lambda^2}+\ope \left( \frac{1}{\Lambda^4} \right) \right)
  H^2
 \notag \\
&=
 -\lambda v^2 \left(1-\frac{f_1 v^2}{\Lambda^2}+
 \frac{f_2 v^2}{2 \Lambda^2\lambda}+
 \ope \left( \frac{1}{\Lambda^4} \right) \right)H^2
\really - \frac{m_H^2}{2} H^2 \notag \\
\Leftrightarrow \qquad
m_H^2 &= 
  2 \lambda v^2 \left(1-\frac{f_1 v^2}{\Lambda^2}+
 \frac{f_2 v^2}{2 \Lambda^2\lambda} \right) \; .
\end{alignat}
Including dimension-6 operators the relation between the vacuum
expectation value, the Higgs mass and the factor in front of the
$|\phi|^4$ term in the potential changes. Once we measure the
Higgs mass at the LHC, we can compute the trilinear and quadrilinear
\underline{Higgs self couplings}\index{Higgs coupling!self coupling}
by collecting the right powers of $H$ in the Higgs potential, in
complete analogy to the Higgs mass above. We find
\begin{alignat}{5}
 \lag_\text{self}=
&-
 \frac{m_H^2}{2v}\left[
 \left(1-\frac{f_1 v^2}{2\Lambda^2}
 +\frac{2f_2 v^4}{3 \Lambda^2 {m_H^2}}\right)H^3
 -\frac{2f_1 v^2}{\Lambda^2 {m_H^2}}
 H \, \p_\mu H \, \p^\mu H\right]
 \notag \\
&-
 \frac{m_H^2}{8v^2}\left[\left(1-\frac{{f_1}v^2}{\Lambda^2}
 +\frac{4f_2 v^4}{{\Lambda^2}m_H^2}\right)H^4
 -\frac{4f_1 v^2}{\Lambda^2 m_H^2} H^2 \, \p_\mu  \, H\p^\mu H\right] \; .
 \end{alignat}
This gives the Feynman rules
\begin{alignat}{5}
\parbox{25mm}{
\begin{fmfgraph*}(50,50)
 \fmfset{arrow_len}{2mm}
 \fmfleft{in}
 \fmf{dashes,width=0.5}{in,v1}
 \fmf{dashes,width=0.5}{out1,v1}
 \fmf{dashes,width=0.5}{out2,v1}
 \fmfright{out1,out2}
\end{fmfgraph*}
}
 -i\frac{3m_H^2}{v}\left(1-\frac{f_1v^2}{2\Lambda^2}+\frac{2f_2 v^4}
 {3\Lambda^2m_H^2}+\frac{2f_1 v^2}{3\Lambda^2m_H^2}
 \sum_{j<k}^3 (p_j p_k) \right)
\end{alignat}
and 
\begin{alignat}{5}
\parbox{25mm}{
\begin{fmfgraph*}(50,50)
 \fmfset{arrow_len}{2mm}
 \fmfleft{in1,in2}
 \fmf{dashes,width=0.5}{in1,v1}
 \fmf{dashes,width=0.5}{in2,v1}
 \fmf{dashes,width=0.5}{out1,v1}
 \fmf{dashes,width=0.5}{out2,v1}
 \fmfright{out1,out2}
\end{fmfgraph*}
}
 -i\frac{3m_H^2}{v^2}\left(1-\frac{f_1v^2}{\Lambda^2}+\frac{4f_2 v^4}
 {\Lambda^2m_H^2}+\frac{2f_1 v^2}{3\Lambda^2m_H^2}
  \sum_{j<k}^4 (p_j p_k) \right)
\label{eq:higgs_selfcoup}
\end{alignat}
\bigskip

From this discussion we see that in the Higgs sector the Higgs self
couplings as well as the Higgs mass can be computed from the Higgs
potential and depend on the operators we take into account. As
mentioned before, in the Standard Model we use only the dimension-4
operators which appear in the renormalizable Lagrangian and which give
us the Higgs mass and self couplings
\begin{alignat}{5}
\boxed{m_H^2=2\lambda v^2=-2\mu^2}
\qqquad \text{and} \qqquad 
  \lag_\text{self}=
   - \frac{m_H^2}{2v} \, H^3
   - \frac{m_H^2}{8v^2} \, H^4 \; ,
\label{eq:higgs_d4}
\end{alignat}
with $v = v_0 = 246$~GeV. Given the measured Higgs mass the Higgs self
coupling comes out as $\lambda \simeq 1/8$. When the Higgs sector becomes more
complicated, not the existence but the form of such relations between
masses and couplings will change. With this information we could now
start computing Higgs observables at the LHC, but let us first see
what else we can say about the Higgs potential from a theoretical
point of view.\bigskip

The Higgs self couplings computed in Eq.\eqref{eq:higgs_selfcoup} are
structurally different from their dimension-4 counter parts in that
their higher dimensional modifications are momentum dependent.  They
are proportional to $f_1$, which means they arise from the operator
$\ope_1$ shown in Eq.\eqref{eq:higgs_o1}. For large momenta
such terms will cause problems, because the momentum in the numerator
can exceed the suppression $1/\Lambda$.  In our derivation part of
this operator is absorbed into a \underline{wave function
  renormalization}, ensuring the appropriate kinetic term of the Higgs
scalar dictated by the definition of outgoing states in an interacting
field theory. The question is if we can define a wave function
renormalization which also removes the momentum dependent terms for
the Higgs self couplings. If $\hat{H}$ is the original Higgs field as
part of $\phi$ we can relate it to the physical Higgs field $H$ using
the generally parameterization
\begin{alignat}{5}
H = 
\left( 1+\frac{a_0 v^2}{\Lambda^2} \right) \hat{H} 
        + \frac{a_1v}{\Lambda^2} \hat{H}^2 
        + \frac{a_2}{\Lambda^2} \hat{H}^3 \; .
\label{eq:higgs_wfnew1}
\end{alignat}
The powers of $v$ and $\Lambda$ are chosen such that all additional
terms are related to a dimension-6 operator ($1/\Lambda^2$) but the
$a_j$ have no mass dimension. Unlike for our first attempt we now
include powers of the Higgs field in the wave function
renormalization. Even higher terms in $\hat{H}$ would be allowed,
but it will turn out that we do not need them. The canonically
normalized kinetic term for the real scalar field $H$ is
\begin{alignat}{5}
 \lag_\text{kin} &= 
 \frac{1}{2} \p_\mu H \; \p^\mu H
\notag \\
&= \frac{1}{2} \;
 \p_\mu 
 \left[ \left( 1+\frac{a_0 v^2}{\Lambda^2} \right) \hat{H}  
      + \frac{a_1v}{\Lambda^2} \hat{H}^2 
      + \frac{a_2}{\Lambda^2} \hat{H}^3 \right] 
 \p^\mu 
 \left[ \left( 1+\frac{a_0 v^2}{\Lambda^2} \right) \hat{H} 
      + \frac{a_1v}{\Lambda^2} \hat{H}^2 
      + \frac{a_2}{\Lambda^2} \hat{H}^3 \right]
\notag \\
&= 
 \left[ 1 + \frac{a_0 v^2}{\Lambda^2} 
      + \frac{2a_1v}{\Lambda^2} \hat{H} 
      + \frac{3a_2}{\Lambda^2} \hat{H}^2 
 \right]^2 \frac{\p_\mu \hat{H} \p^\mu \hat{H}}{2}
\notag \\
 &= 
 \left[ 1 + \frac{2 a_0 v^2}{\Lambda^2} + \frac{a_0^2 v^4}{\Lambda^4} 
      + \left( 1+\frac{a_0 v^2}{\Lambda^2} \right) \frac{4a_1v}{\Lambda^2} \hat{H} 
      + \left( \frac{6a_2}{\Lambda^2} + \frac{2 a_0 v^2}{\Lambda^2} \frac{3a_2}{\Lambda^2} 
             + \frac{4a_1^2v^2}{\Lambda^4} \right) \hat{H}^2 
      + \ope(\hat{H}^3)
 \right] \frac{\p_\mu \hat{H} \p^\mu \hat{H}}{2}
\notag \\
 &= 
 \left[ 1 + \frac{2a_0 v^2}{\Lambda^2} 
          + \frac{4a_1v}{\Lambda^2} \hat{H} 
          + \frac{6a_2}{\Lambda^2} \hat{H}^2 
          + \ope(\hat{H}^3)
          + \ope \left( \frac{1}{\Lambda^4} \right)
  \right] \frac{\p_\mu \hat{H} \p^\mu \hat{H}}{2} \; .
\label{eq:higgs_kinetic1}
\end{alignat}
Terms of higher mass dimension or including higher powers of
$\hat{H}$ will only appear once we go to dimension-8 operators.
This general form based on Eq.\eqref{eq:higgs_wfnew1} we should use
to remove all contributions from the dimension-6 operator
$\ope_1$ to the kinetic term of the Higgs field $\hat{H}$
\begin{alignat}{5}
\lag_\text{kin} 
&= 
\frac{\p_\mu \hat{H} \p^\mu \hat{H}}{2} + \frac{f_1}{\Lambda^2} \; \ope_1
\notag \\
&=
\left[ 1 + 
       \frac{f_1}{\Lambda^2} \; \left( v^2 + 2 v \hat{H} + \hat{H}^2 \right) 
\right] \;
\frac{\p_\mu \hat{H}  \p^\mu \hat{H}}{2} \; .
\label{eq:higgs_kinetic2}
\end{alignat}
Comparing Eq.\eqref{eq:higgs_kinetic1} and
Eq.\eqref{eq:higgs_kinetic2} we can identify the general pre-factors
$a_j$ with the specific $f_1$ from our dimension-6 ansatz. This gives
us $a_0 = f_1/2$, $a_1 = f_1/2$, and $a_2 = f_1/6$.  The wave function
renormalization Eq.\eqref{eq:higgs_wfnew1} then reads
\begin{alignat}{5}
\boxed{
H = \left( 1 + \frac{f_1 v^2}{2 \Lambda^2} \right) \hat{H}
  + \frac{f_1v}{2\Lambda^2} \hat{H}^2
  + \frac{f_1}{6\Lambda^2} \hat{H}^3
  + \ope(\hat{H}^4)
  + \ope \left( \frac{1}{\Lambda^4} \right)
} \; .
\label{eq:higgs_wfnew2}
\end{alignat}
In this alternative, \underline{generalized canonical normalization}
of the Higgs field we avoid any momentum dependent contributions to
the Higgs self couplings. The prize we pay is that we have to apply the shift defined in
Eq.\eqref{eq:higgs_wfnew2} throughout the
entire Standard Model Lagrangian. This means that the higher dimensional
operator $\ope_1$ leads to multiple Higgs couplings to any pair
of massive gauge bosons or massive fermions. This observation is at
the heart of the so-called strongly interacting light Higgs (SILH).

Because the wave function renormalization is not a physical
observable; the two approaches are physically equivalent and 
predict the same physical observables to a precision
$1/\Lambda^2$. Beyond dimension-6 operators they will become
different. However, to properly define the external Higgs states in
the interacting theory we need to ensure that the wave function and
its commutators are properly defined. Looking at
Eq.\eqref{eq:higgs_wfnew2} we see that either the field $H$ or the
field $\hat{H}$ will induce a wave function normalization and hence
a field commutator dependent on the field value.\bigskip

We will see later that multiple Higgs couplings to other
states are a serious challenge to the LHC, as are the momentum
dependent terms in the Higgs self couplings. This means that for the
interpretation of LHC data these dimension-6 interactions do not pose
a serious problem.

\subsubsection{Mexican hat}
\label{sec:higgs_mexicanhat}

To understand the well known picture of a Mexican hat which usually
illustrates the Higgs mechanism we have to include Goldstone modes
again. The Higgs doublets including all degrees of freedom are defined
in Eq.\eqref{eq:def_phi}. For our illustration it is sufficient to
extend Eq.\eqref{eq:higgs_phi_simp} by including all
\underline{neutral degrees of freedom} in the Higgs doublet
\begin{alignat}{5}
       \phi =
             \frac{1}{\sqrt{2}}
             \begin{pmatrix} -w_2 - i w_1 \\ v+H+iw_3 \end{pmatrix} 
\sim
             \frac{1}{\sqrt{2}}
             \begin{pmatrix} 0 \\ v+H+iw_3 \end{pmatrix} \; .
\label{eq:def_phi_neut}
\end{alignat}
In this approximation we can again compute the potential defined in
Eq.\eqref{eq:higgs_pot}, but omitting the dimension-6 terms
\begin{alignat}{5}
V &=\mu^2 (\phi^\dag \phi)+\lambda (\phi^\dag \phi)^2
\notag \\
  &= \frac{\mu^2}{2} \; \left( (v+H)^2 + w_3^2 \right)
   + \frac{\lambda}{4} \; \left( (v+H)^2 + w_3^2 \right)^2 \; .
\label{eq:higgs_pot_neut}
\end{alignat}
In the \underline{unbroken phase}, \ie in the absence of a vacuum
expectation value or for $v=0$ the potential reads
\begin{alignat}{5}
V &= \frac{\mu^2}{2} \; \left( H^2 + w_3^2 \right)
   + \frac{\lambda}{4} \; \left( H^2 + w_3^2 \right)^2 \; .
\end{alignat}
This form depends only on a radius in the two-dimensional plane formed
by the Higgs and Goldstone field values $|\phi| =
\sqrt{H^2+w_3^2}/\sqrt{2}$. Its first derivative with respect to
$|\phi|$ is $V' = \mu^2 |\phi| + 3\lambda |\phi|^3/4$, so there exists
only a minimum at $|\phi|=0$ or equivalently at $H = w_3 = 0$, where
the potential becomes $V=0$. Towards larger field values the potential
increases first proportional to $|\phi|^2$ and finally proportional to
$|\phi|^4$, always rotationally symmetric in the $H$-$w_3$
plane.\bigskip

\begin{figure}[t]
\begin{center}
\includegraphics[width=0.60\hsize]{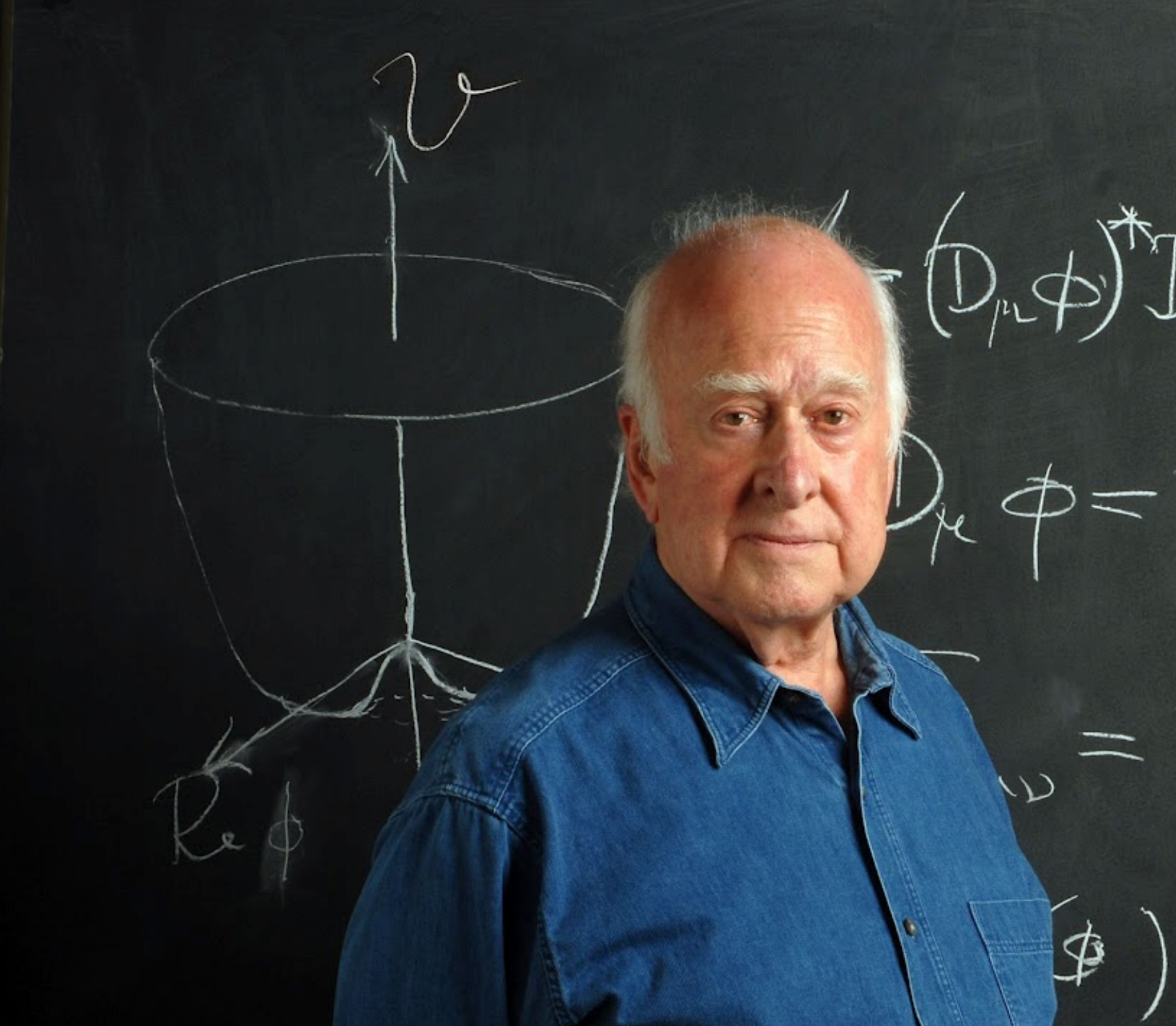}
\end{center}
\caption{Mexican hat form of the Higgs potential, figure from the 
University of Edinburgh website.}
\label{fig:higgs_mexican}
\end{figure}

In the \underline{broken phase} with $v = 246$~GeV the form of the
potential changes. We can follow exactly the derivation following
Eq.\eqref{eq:higgs_pot} where the minimum condition in the complex
plane is $|\phi| = v/\sqrt{2}$. This minimum is rotationally symmetric;
on the real $H$ axis it requires $H = \pm v$ while on the imaginary
$w_3$ axis it appears at $w_3 = \pm iv$. For large field values
$|\phi| \gg v$ the potential rapidly increases proportional to
$|\phi|^4$. This form of the potential is shown in
Figure~\ref{fig:higgs_mexican}.

To determine the masses of the particles corresponding to the Higgs
and Goldstone fields we choose one point on the degenerate vacuum
circle, \eg $\phi = v$. There we can compute the \underline{mass
  matrix} from the second derivatives of the potential at the minimum
\begin{alignat}{5}
\boxed{
\left( \mat_H^2 \right)_{jk}
= \frac{\p^2 V}{\p^2 \left\{ H w_3 \right\} } \Bigg|_\text{minimum}
} \; .
\label{eq:higgs_massmatrix}
\end{alignat}
For real fields, where the mass term is proportional to $\lag \supset
-V \supset - m^2 H^2/2$ there is no factor $1/2$ in this relation.
Inserting the full form of the neutral potential given in
Eq.\eqref{eq:higgs_pot_neut} we first find
\begin{alignat}{5}
\frac{\p V}{\p H} 
&= \mu^2 (v+H) + \frac{\lambda}{2} \left( (v+H)^2 + w_3^2 \right) 2 (v+H) 
\notag \\
&= \mu^2 (v+H) + \lambda (v+H) \left( (v+H)^2 + w_3^2 \right)
\notag \\
\frac{\p^2 V}{\p H^2} \Bigg|_\text{minimum}
&= \mu^2 + \lambda \left( (v+H)^2 + w_3^2 \right)
         + 2\lambda (v+H)^2 \Bigg|_\text{minimum}
\notag \\
&= \mu^2 + 3 \lambda v^2 = 2 \lambda v^2 \; .
\end{alignat}
In the last step we use the relation $\mu^2 = - \lambda v^2$ at the
minimum. This second derivative is identical to the known Higgs mass,
but to be save we still compute the complete mass matrix and determine
the mass eigenvalues. The second diagonal entry in the neutral
Higgs--Goldstone mass matrix is
\begin{alignat}{5}
\frac{\p V}{\p w_3} 
&= \mu^2 w_3 + \frac{\lambda}{2} \left( (v+H)^2 + w_3^2 \right) 2 w_3
\notag \\
&= \mu^2 w_3 + \lambda w_3 \left( (v+H)^2 + w_3^2 \right)
\notag \\
\frac{\p^2 V}{\p w_3^2} \Bigg|_\text{minimum}
&= \mu^2 + \lambda \left( (v+H)^2 + w_3^2 \right)
         + 2\lambda w_3^2 \Bigg|_\text{minimum}
\notag \\
&= \mu^2 + \lambda v^2 = 0 \; .
\end{alignat}
The off-diagonal entry of the symmetric mass matrix is 
\begin{alignat}{5}
\frac{\p^2 V}{\p H \p w_3} \Bigg|_\text{minimum}
&= \frac{\p}{\p w_3} \, 
   \left[ \mu^2 (v+H) + \lambda (v+H) \left( (v+H)^2 + w_3^2 \right) \right]
   \Bigg|_\text{minimum}
&= 2 \lambda (v+H) w_3 \Bigg|_\text{minimum} = 0 
\end{alignat}
Putting all this together we find that the Higgs and Goldstone basis
is identical with the mass eigenstates of the symmetric mass matrix
\begin{alignat}{5}
\boxed{
\mat_H^2
= \begin{pmatrix}
         2 \lambda v^2 & 0 \\ 0 & 0 
  \end{pmatrix} 
}
\end{alignat}
If we fix our vacuum to the positive real axis $\phi = v$ the Higgs
mode living on the real axis is massive while the Goldstone mode
living orthogonally in the direction of the \underline{flat potential
  valley} is massless. The fact that we had to fix the vacuum to the
real axis is a direct consequence of our linear gauge choice in
Eq.\eqref{eq:higgs_vshift}. In the unitary form of
Eq.\eqref{eq:higgs_goldstone_exp} the Goldstone mode is aligned with
the potential valley over the entire Higgs--Goldstone field
plane.\bigskip

This result confirms Goldstone's theorem, at least in the first
step. If we consider the Goldstone modes simply additional scalars
they are massless. Historically, this was the big problem with
spontaneous symmetry breaking, because such massless scalars with weak
charge would have been observed. From
Eq.\eqref{eq:higgs_goldstoneprop} we know that after breaking a local
gauge symmetry the Goldstones become part of the massive gauge fields,
and their mass is not a physical parameter.

\subsubsection{Unitarity}
\label{sec:higgs_unitarity}

If we want to compute transition amplitudes at very high energies the
Goldstone modes become very useful. In the $V$ rest frame we can write
the three polarization vectors of a massive gauge boson as
\begin{alignat}{5}
 \epsilon^\mu_{T,1}=
 \left( 
 \begin{array}{c}
 0  \\[-1mm]
 1  \\[-1mm]
 0  \\[-1mm]
 0 \end{array} \right)  
 \qquad
 \epsilon^\mu_{T,2}=
 \left( 
 \begin{array}{c}
 0  \\[-1mm]
 0  \\[-1mm]
 1  \\[-1mm]
 0 \end{array} \right) 
 \qquad
 \epsilon^\mu_{L}=
 \left( 
 \begin{array}{c}
 0  \\[-1mm]
 0  \\[-1mm]
 0  \\[-1mm]
 1 \end{array} \right) \; .
\end{alignat}
If we boost $V$ into the $z$ direction, giving it a four-momentum
$p^\mu=(E,0,0,|\vec{p}|)$, the polarization vectors become
\begin{alignat}{5}
\epsilon^\mu_{T,1}=
\left( 
\begin{array}{c}
0  \\[-1mm]
1  \\[-1mm]
0  \\[-1mm]
0 \end{array} \right) 
 \qquad
 \epsilon^\mu_{T,2}=
\left( 
\begin{array}{c}
0  \\[-1mm]
0  \\[-1mm]
1  \\[-1mm]
0 \end{array} \right)  
 \qquad
\epsilon^\mu_{L}=
\frac{1}{m_{V}}
\left( 
\begin{array}{c}
|\vec{p}|  \\[-1mm]
0          \\[-1mm]
0          \\[-1mm]
E \end{array} \right)
\stackrel{E \gg m_V}{\longrightarrow}
\frac{1}{m_{V}}
\left(
\begin{array}{c}
|\vec{p}|  \\[-1mm]
0          \\[-1mm]
0          \\[-1mm]
|\vec{p}| 
\end{array} \right) 
\equiv \frac{1}{m_V} \, p^\mu \; .
\end{alignat}
Very relativistic gauge bosons are dominated by their longitudinal
polarization $|\vec{\epsilon}_L| \sim E/m_V \gg 1$. This longitudinal
degree of freedom is precisely the Goldstone boson, so 
at high energies we can approximate the complicated vector bosons
$Z,W^\pm$ as scalar Goldstone bosons $w_0,w_{\pm}$. The problem which
gauge--dependent mass value $\xi m_V$ to assign to the Goldstone fields
does not occur, because in the high energy limit we automatically assume $m_V \to
0$. This simplification comes in handy for example when we talk
about unitarity as a constraint on the Higgs sector.  This relation
between Goldstones and gauge bosons at very high energies is called
the \underline{equivalence theorem}\index{equivalence theorem}.\bigskip

Based on the equivalence theorem we can compute the amplitude for
$W^+ W^- \rightarrow W^+ W^-$ scattering at very high energies ($E \gg
m_W$) in terms of scalar Goldstones bosons\index{Goldstone boson!Feynman rules}.  Three diagrams contribute to this processes:
a four-point vertex, the $s$-channel Higgs exchange and the
$t$-channel Higgs exchange:
\begin{equation*}
\parbox{25mm}{
\begin{fmfgraph*}(50,50)
 \fmfset{arrow_len}{2mm}
 \fmfleft{in1,in2}
 \fmf{dashes,width=0.5}{in1,v1}
 \fmf{dashes,width=0.5}{in2,v1}
 \fmf{dashes,width=0.5}{out1,v1}
 \fmf{dashes,width=0.5}{out2,v1}
 \fmfright{out1,out2}
\end{fmfgraph*}
} \quad + \quad
\parbox{25mm}{
\begin{fmfgraph*}(60,50)
 \fmfset{arrow_len}{2mm}
 \fmfleft{in1,in2}
 \fmf{dashes,width=0.5}{in1,v1}
 \fmf{dashes,width=0.5}{in2,v1}
 \fmf{dashes,width=0.5}{v1,v2}
 \fmf{dashes,width=0.5}{out1,v2}
 \fmf{dashes,width=0.5}{out2,v2}
 \fmfright{out1,out2}
\end{fmfgraph*}
} \quad + \quad
\parbox{25mm}{
\begin{fmfgraph*}(50,50)
 \fmfset{arrow_len}{2mm}
 \fmfleft{in1,in2}
 \fmf{dashes,width=0.5}{in1,v1}
 \fmf{dashes,width=0.5}{in2,v2}
 \fmf{dashes,width=0.5}{v1,v2}
 \fmf{dashes,width=0.5}{out1,v1}
 \fmf{dashes,width=0.5}{out2,v2}
 \fmfright{out1,out2}
\end{fmfgraph*}
}
\end{equation*}
To confirm these Feynman diagrams and to compute the corresponding
amplitude we need some basic Feynman rules, for example the Goldstone
couplings to the Higgs boson and the four-Goldstone couplings. We
start with the Higgs doublet, again including the Goldstone modes in
analogy to Section~\ref{sec:higgs_mexicanhat}
\begin{alignat}{5}
 \phi=\; \frac{1}{\sqrt{2}}
         \begin{pmatrix}
           -w_2-iw_1  \\ v+H+iw_3 
         \end{pmatrix}
  \qquad \Rightarrow \qqquad
 \phi^{\dagger}\phi &=
    \; \frac{1}{2} \left( w_1^2+w_2^2+w_3^2+(v+H)^2 \right) \\
 \left( \phi^{\dagger}\phi \right)^2 &=
 \frac{1}{4} \left(\sum_i w_i^2\right)^2
 + \frac{1}{2} \left( v+H \right)^2 \sum_i w_i^2
 + \frac{1}{4} (v+H)^4
 \notag \\
 &= \frac{1}{4} \left( \sum_i w_i^2\right)^2
 + \left( v\,H + \frac{v^2}{2} + \frac{H^2}{2} \right) \, \sum_i w_i^2+ \ope(w^0) \notag \; .
\end{alignat}
In the last step we neglect all terms without the Goldstone
fields. Note that there are no three-Goldstone vertices, only triple
dimension-four couplings including the Higgs and a coupling factor
$v$.  Only keeping the relevant terms contributing to the
four-Goldstone and Higgs--Goldstone--Goldstone couplings at dimension
four the potential becomes
\begin{alignat}{5}
V
&=
 \mu^2|\phi|^2 + \lambda |\phi|^4
\supset \lambda |\phi|^4
 = \frac{m_H^2}{2 v^2} \; |\phi|^4
 \notag \\ 
&=
 \frac{m_H^2}{2v^2} \left[
    \frac{1}{4} \left( \sum_i w_i ^2\right)^2
  +  vH \sum_i w_i^2
  + \ope(w^0) 
                    \right]
 \notag \\
&=
 \frac{m_H^2}{8v^2} \left( \sum_i w_i^2\right)^2
 +\frac{m_H^2}{2v}\,H \sum_i w_i^2
 + \ope(w^0) \; .
\label{eq:pot_goldstone}
\end{alignat}
Focussing on the scattering of charged Goldstones $w_\pm w_\pm \to
w_\pm w_\pm$ we use the corresponding fields $w_\pm=(w_1 \pm iw_2
)/\sqrt{2}$ following Eq.\eqref{eq:higgs_gaugebosons}.  They appear in
the above expression as $w_1^2+w_2^2 = 2 w_+w_-$, so we find the terms
\begin{alignat}{5}
V &\supset
 \frac{m_H^2}{2v^2}
 w_+w_-w_+w_-+
 \frac{m_H^2}{v} H w_+w_- \; ,
\end{alignat}
which fix the two Feynman rules we need. Linking the Lagrangian to the
Feynman rule for the quartic coupling involves one complication: for
each positively charged Goldstone in the vertex there are two ways we
can identify them with the Lagrangian fields. In addition, there are
also two choices to identify the two negatively charged Goldstones,
which implies an additional combinatorial factor four in the Feynman
rule. Including a common factor $(-i)$ the two Feynman rules then
become $-2im_H^2/v^2$ and $-im_H^2/v$.

The potential in Eq.\eqref{eq:pot_goldstone} has an interesting feature
which has recently lead to some discussions on the computation of
Higgs decays to two photons.  The question is if the one-loop
$H\gamma\gamma$ amplitude mediated by a closed $W$ boson loop should
vanish in the limit $m_W \to 0$. This is indeed the case for a closed
fermion loop contributing to the same process through the Yukawa
coupling. The $W$ loop, in contrast, consists of transverse and
longitudinal $W$ modes. The latter we can describe in terms of
Goldstone modes which couple to the external Higgs field following
Eq.\eqref{eq:pot_goldstone}. Because $m_W$ never appears in this
potential there is no reason why the Goldstone modes should decouple,
and indeed they do not.\bigskip

The amplitude for the Goldstone scattering process is given in terms
of the Mandelstam variables $s$ and $t$ which describe the momentum
flow $p^2$ through the two Higgs propagators and which we will
properly introduce in Section~\ref{sec:qcd_dy_r}
\begin{alignat}{5}
 A&=i  \;    \frac{-2 i m_H^2}{v^2}
          + \left(\frac{-i m_H^2}{v}\right)^2 \frac{i}{s-m_H^2}
          + \left(\frac{-i m_H^2}{v}\right)^2 \frac{i}{t-m_H^2} \notag \\
  &=\frac{m_H^2}{v^2}
    \left[  2
          + \frac{m_H^2}{s-m_H^2}
          + \frac{m_H^2}{t-m_H^2}
    \right] \; .
\label{eq:gold_scat}
\end{alignat}
The factor $i$ which ensures that the amplitude is real appears
between the transition rate computed from the Feynman rules and the
actual transition amplitude, as shown in
Eq.\eqref{eq:qft_nofeyn1a}.\bigskip

For this process we want to test the unitarity of the $S$ matrix,
which we write in terms of a transition amplitude $S = \one + i A$.
The $S$ matrix should be unitary to conserve probability
\begin{alignat}{5}
 \one \really S^\dag S 
 = ( \one - i A^\dag ) ( \one + i A )
 = \one + i (A - A^\dag) + A^\dag A
\qquad \Leftrightarrow \qquad 
 A^\dag A = - i (A - A^\dag)  \; .
\end{alignat}
If we sandwich $(A - A^\dag)$ between \underline{identical
  asymptotically free fields}, which means that we are looking at
forward scattering with a scattering angle $\theta \to 0$, we find in
the high energy limit or for massless external particles
\begin{alignat}{5}
  -i \langle j | A - {A^*}^T | j \rangle
= -i \langle j | A - A^* | j \rangle
= 2 \, \text{Im} A(\theta=0)
\qquad \Rightarrow \qquad 
\boxed{
\sigma \equiv \frac{1}{2s} \langle j |A^\dag A| j \rangle 
= \frac{1}{s} \; \text{Im} A(\theta=0)
} \; .
\end{alignat}
Assuming that our Lagrangian is hermitian this imaginary part
corresponds only to absorptive terms in the scattering amplitude.
This is the usual formulation of the \underline{optical
  theorem}\index{optical theorem} reflecting unitarity in terms of the
transition amplitude $A$.\bigskip

To include the dependence on the scattering angle $\theta$ we
decompose the transition amplitude into partial waves
\begin{alignat}{5}
 A = 16\pi \sum_{l=0}^{\infty}(2l+1) \, P_l(\cos \theta) \, a_{l}
\qqquad \text{with} \qqquad
 \int_{-1}^1dx\,P_l(x)P_{l'}(x)=\frac{2}{2l+1}\delta_{ll'} \; ,
\end{alignat}
ordered by the orbital angular momentum $l$.  $P_{l}$ are the Legendre
polynomials of the scattering angle $\theta$, which obey an
orthogonality condition.  The scattering cross section including all
prefactors and the phase space integration is then given by
 \begin{alignat}{5}
 \sigma
&=
 \int d\Omega \, \frac{|A|^2}{64\pi^2 s} \notag \\
&= \frac{(16\pi)^2}{64\pi^2 s} \, 2 \pi 
 \int_{-1}^1 \, d\cos{\theta} \sum_{l}
 \sum_{l'}(2l+1)(2l'+1) \; a_l a_{l'}^* \;
 P_l(\cos \theta)P_{l'}(\cos \theta)
 \notag \\
&=
 \frac{8\pi}{s} \sum_{l}2(2l+1) \; |a_l|^2
 =\frac{16\pi}{s} \sum_{l}(2l+1) \; |a_l|^2 \; .
\end{alignat}
The relation between the integral over the scattering angle $\theta$ and
the Mandelstam variable $t$ we will discuss in more detail in
Section~\ref{sec:qcd_dy_r}.  Applied to each term in the partial
wave expansion the optical theorem requires
\begin{alignat}{5}
 \frac{16\pi}{s}(2l+1) \; |a_l|^2 
&= 
 \frac{1}{s} \; \text{Im} A(\theta=0) \Bigg|_l \notag \\
&=
 \frac{1}{s} \; 16\pi(2l+1) \; \text{Im} \, a_l
 \qqquad \Leftrightarrow \qqquad
 |a_l|^2 \really \text{Im} \, a_l \; ,
 \end{alignat}
using $P_l(\cos{\theta}=1)=1$.  This condition we can rewrite as
\begin{alignat}{5}
(\text{Re}\; a_{l})^2+\left(\text{Im} \; a_{l}-\frac{1}{2}\right)^2
              =\frac{1}{4}
 \qqquad \Rightarrow \qqquad 
 \boxed{ \left| \mbox{Re}\:a_{l} \right| \:<\: \frac{1}{2}}  \; ,
\end{alignat}
once we recognize that the condition on $\mbox{Im}\:a_l$ and on
$\mbox{Re}\:a_l$ is a circle around $a_l=(0,1/2)$ with radius
$1/2$.\bigskip

It is important to remember that in the above argument we have
formulated the constraint for each term in the sum over the Legendre
polynomials. Mathematically, this is well justified, but of course
there might be physics effects which lead to a systematic cancellation
between different terms. This is why the constraint we compute is
referred to as \underline{perturbative unitary}\index{perturbative unitarity}. For Goldstone scattering we compute the supposedly
leading first term in the partial wave expansion from the amplitude
\begin{alignat}{5}
 a_0 = \frac{1}{16\pi s}\int_{-s}^0 dt\,|A|\,
&=
 \, \frac{1}{16\pi s}
 \int_{-s}^0 dt \, 
 \frac{m_H^2}{v^2} \left[ 2+\frac{m_H^2}{s-m_H^2}
                            +\frac{m_H^2}{t-m_H^2}
                            \right]
 \notag \\
&=
 \, \frac{m_H^2}{16\pi v^2} \left[ 2+\frac{m_H^2}{s-m_H^2}
                                    -\frac{m_H^2}{s}
                                     \log{\left(1+\frac{s}{m_H^2}\right)}
                            \right] \notag \\
&=\, \frac{m_H^2}{16 \pi v^2} \left[ 2+
                              \ope\left(\frac{m_H^2}{s}\right)
                                   \right] \; .
\end{alignat}
In the high energy limit $s \gg m_H^2$ this translates into an upper
limit on the Higgs mass which in Eq.\eqref{eq:gold_scat} enters as
the Goldstone coupling in the numerator
\begin{alignat}{5}
 \frac{m_H^2}{8\pi v^2}\,<\,\frac{1}{2} \qquad \Leftrightarrow
 \quad \boxed{m_H^2<4\pi v^2 = (870~\mbox{GeV})^2} \; .
\label{eq:unitarity_mass}
\end{alignat}
This is the maximum value of $m_H$ consistent with perturbative
unitarity for $WW\rightarrow WW$ scattering. Replacing the Higgs mass
by the self coupling we can formulate the same constraint as $\lambda
< 2 \pi$. The leading term in our analysis of perturbative unitarity
is simply the size of the four-Goldstone coupling, the two Higgs
diagrams are sub-leading in $m_H^2/s$.  This means that perturbative
unitarity seriously probes the limitations of perturbation theory, so
we should include higher order effects as well as higher dimensional
operators to get a reliable numerical prediction in the range of $m_H
\lesssim 1$~TeV. 

Of course, if we limit $s$ to a finite value this bound changes, and
we can compute a maximum scale $s_\text{max}$ which leaves
$WW\rightarrow WW$ perturbatively unitary for fixed $m_H$: for $m_H
\lesssim v$ this typically becomes $\sqrt{s_\text{max}}\sim
1.2$~TeV. This number is one of the motivations to build the LHC as a
high energy collider with a partonic center--of--mass energy in the
few-TeV range. If something had gone wrong with the Standard--Model--like
Higgs sector we could have expected to see something else curing unitarity
around the TeV scale. A Higgs boson too heavy to be produced at the
LHC would essentially not been able to function as a Higgs
boson.\bigskip

In many discussions of unitarity and the Higgs sector we explain the
role of the Higgs boson in the unitarization of $WW$ scattering as a
cancellation of the leading divergences through virtual Higgs
exchange.  Clearly, this is not what we see in our
argument. Nevertheless, both answers are correct, because the
separation of a gauge--invariant transition amplitude it gauge
dependent. Our Higgs--Goldstone gauge assumes the existence of a Higgs
boson when we include the coupling strength $m_H$ in the Feynman rules
for the Goldstones. With that assumption we are no longer allowed to
test the assumption that the Higgs not be there. All we can do is
decouple the Higgs by making it heavier, which gives us the limit
shown in Eq.\eqref{eq:unitarity_mass}. In the unitary gauge, where the
gauge bosons are massive and the Goldstones are eaten, the Higgs is an
additional state which we can remove at the expense of ruining
renormalizability. At tree level this gauge gives us a cancellation of
the leading divergences from gauge boson exchange with the help of
the Higgs diagrams. Because the transition amplitude is gauge
invariant the limit on the Higgs mass will be identical.\bigskip

One last but very important comment we need to make: this unitarity
argument only works if the $WWH$ coupling is exactly what it should
be.  While perturbative unitarity only gives us a fairly rough upper
limit on $m_H$, it also uniquely fixes $g_{WWH}$ to its Standard Model
value. Any sizeable deviation from this value again means new physics
appearing at the latest around the mass scales of
Eq.\eqref{eq:unitarity_mass}.

Looking at processes like $WW\rightarrow f\bar{f}$ or $WW\rightarrow
WWH$ or $WW\rightarrow HHH$ we can fix \underline{all Higgs
  couplings}\index{Higgs coupling} in the Standard Model, including
$g_{Hff}$, $g_{HHH}$, $g_{HHHH}$, using exactly the same argument.
The most important result of the unitarity test is probably not the
upper bound on the Higgs mass, but the underlying assumption that the
unitarity test only works in the presence of one Higgs boson if all
Higgs couplings look exactly as predicted by the Standard Model.

\subsubsection{Renormalization group analysis}
\label{sec:higgs_rge}

The unitarity condition derived above is the first of a series of
theoretical constraints which we can derive as self consistency
conditions on a Higgs boson turning the Standard Model with its
particle masses into a renormalizable theory.  
We can derive two additional
theoretical constraints from the renormalization group
equation of the Higgs potential, specifically from the
\underline{renormalization scale
  dependence}\index{scales!renormalization scale} of the self coupling
$\lambda(Q^2)$.  Such a scale dependence arises automatically when we
encounter ultraviolet divergences and absorb the $1/\epsilon$ poles
into a minimal counter term. We will discuss this running of couplings
in more detail in Section~\ref{sec:qcd_counter_terms} focussing on the
running QCD coupling $\alpha_s$. In the case of a running quartic
Higgs coupling $\lambda$ the one-loop $s$, $t$ and $u$-channel
diagrams only depending on $\lambda$ itself are
\begin{equation*}
\parbox{25mm}{
\begin{fmfgraph*}(50,50)
 \fmfset{arrow_len}{2mm}
 \fmfleft{in1,in2}
 \fmf{dashes,width=0.5}{in1,v1}
 \fmf{dashes,width=0.5}{in2,v1}
 \fmf{dashes,width=0.5}{out1,v1}
 \fmf{dashes,width=0.5}{out2,v1}
 \fmfright{out1,out2}
\end{fmfgraph*}
} \quad + \quad
\parbox{20mm}{
\begin{fmfgraph*}(30,50)
 \fmfset{arrow_len}{2mm}
 \fmfleft{in1,in2}
 \fmf{dashes,width=0.5}{in1,v1}
 \fmf{dashes,width=0.5}{in2,v2}
 \fmf{dashes,width=0.5}{out1,v1}
 \fmf{dashes,width=0.5}{out2,v2}
 \fmf{dashes,width=0.5,left,tension=.5}{v1,v2,v1}
 \fmfright{out1,out2}
\end{fmfgraph*}
} \quad + \quad 
\parbox{25mm}{
\begin{fmfgraph*}(60,40)
 \fmfset{arrow_len}{2mm}
 \fmftop{in1,in2}
 \fmf{dashes,width=0.5}{in1,v1}
 \fmf{dashes,width=0.5}{in2,v2}
 \fmf{dashes,width=0.5}{out1,v1}
 \fmf{dashes,width=0.5}{out2,v2}
 \fmf{dashes,width=0.5,left,tension=.5}{v1,v2,v1}
 \fmfbottom{out1,out2}
\end{fmfgraph*}
} \quad + \quad 
\parbox{25mm}{
\begin{fmfgraph*}(50,50)
 \fmfset{arrow_len}{2mm}
 \fmfleft{in1,in2}
 \fmf{dashes,width=0.5}{in1,v2}
 \fmf{dashes,width=0.5}{in2,v1}
 \fmf{dashes,width=0.5}{out1,v3}
 \fmf{dashes,width=0.5}{out2,v1}
 \fmf{dashes,width=0.5,tension=.5}{v2,v1,v3}
 \fmf{dashes,width=0.5,left,tension=.05}{v2,v3,v2}
 \fmfright{out1,out2}
\end{fmfgraph*}
}
\end{equation*}
Skipping the calculation we quote the complete renormalization group
equation including diagrams with the Higgs boson, the top quark and
the weak gauge bosons inside the loops
\begin{alignat}{5}
\boxed{
 \frac{d\,\lambda}{d\,\log{Q^2}}=\frac{1}{16\pi^2}
 \left[12\lambda^2+6\lambda y_t^2 - 3y_t^4
 -\frac{3}{2}\lambda \left( 3g_2^2+g_1^2 \right)
 +\frac{3}{16} \left( 2g_2^4+(g_2^2+g_1^2)^2 \right) \right]
} \; ,
\label{eq:lambda_rge}
\end{alignat}
\index{renormalization group equation!Higgs self coupling}with
$y_t=\sqrt{2}m_t/v$. This formula will be the basis of the
discussion in this section.\bigskip

The first regime we study is where the Higgs self coupling $\lambda$
becomes strong. Fixed order perturbation theory as we use it in the
unitarity argument runs into problems in this regime and the
renormalization group equation is the appropriate tool to describe
it. If we reside in a somewhat strongly interacting regime the leading term in Eq.\eqref{eq:lambda_rge} reads
\begin{alignat}{5}
 \frac{d\,\lambda}{d\,\log{Q^2}}
= \frac{1}{2 Q} \, \frac{d\,\lambda}{d\,Q} 
=\frac{1}{16\pi^2} \, 12\lambda^2+\ope(\lambda)
=\frac{3}{4\pi^2}{\lambda^2}+\ope(\lambda) \; .
\end{alignat}

\begin{figure}[t]
\begin{center}
\includegraphics[width=0.35\hsize]{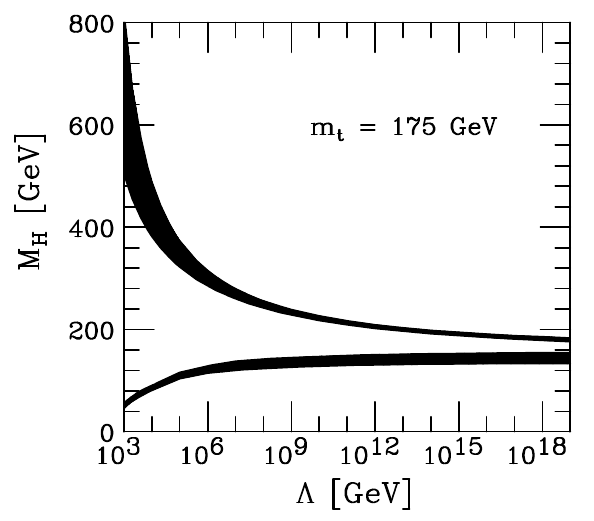}
\end{center}
\caption{Triviality or Landau pole (upper) and stability bounds
  (lower) for the Standard Model Higgs boson in the $m_H - Q$
  plane. Similar arguments first appeared in Ref.~\cite{Cabibbo:1979ay},
  the actual scale dependence can be seen in 
  Refs.~\cite{christof,Lindner:1985uk}.}
\label{fig:higgs_scale}
\end{figure}

Because of the positive sign on the right hand side the quartic
coupling will become stronger and eventually diverge for
large scales $Q^2$. Obviously, this divergence should not happen in a
physical model and will give us a constraint on the maximum value of
$\lambda$ allowed. The approximate renormalization group equation we
can solve by replacing $\lambda=g^{-1}$
\begin{alignat}{5}
& 
 \frac{d\,\lambda}{d\log{Q^2}}=\frac{d}{d\log{Q^2}}
 \frac{1}{g}=-\frac{1}{g^2}\frac{d\,g}{d\log{Q^2}}
 \really \frac{3}{4\pi^2}\frac{1}{g^2}
 \notag \\
\Leftrightarrow&
 \qquad 
\frac{d\,g}{d\log{Q^2}}=-\frac{3}{4\pi^2}
 \qquad \Leftrightarrow \qquad 
g(Q^2)=-\frac{3}{4\pi^2} \log{Q^2}+C \; .
\end{alignat}
The boundary condition $\lambda(Q^2=v^2)=\lambda_0$ fixes the
integration constant $C$
\begin{alignat}{5}
 g_0 = \frac{1}{\lambda_0} 
     = -\frac{3}{4\pi^2}\log{v^2}+C
 \quad\Leftrightarrow\quad 
 C = g_0 + \frac{3}{4\pi^2}\log{v^2} \notag \\
 \Rightarrow\quad 
 g(Q^2) = -\frac{3}{4\pi^2}\log{Q^2}
          + g_0 + \frac{3}{4\pi^2}\log{v^2}
 =-\frac{3}{4\pi^2}\log{\frac{Q^2}{v^2}}+g_0
\notag \\
 \Leftrightarrow\quad 
\boxed{\lambda(Q^2)=\left[ - \frac{3}{4\pi^2}\log{\frac{Q^2}{v^2}}
                           + \frac{1}{\lambda_0} 
                    \right]^{-1}
 =\lambda_0 \left[ 1 - \frac{3}{4\pi^2}
                       \lambda_0\log{\frac{Q^2}{v^2}}\right]^{-1}} \; .
\label{eq:higgs_solvelambda}
\end{alignat}
We start from scales $Q \sim v$ where the expression in brackets is
close to one. Moving towards larger scales the denominator becomes
smaller until $\lambda$ hits a pole at the critical value
$Q_\text{pole}$
\begin{alignat}{5}
 1-\frac{3}{4\pi^2}\lambda_{0}\log{\frac{Q_\text{pole}^2}{v^2}} \really 0
\qquad &\Leftrightarrow& \qquad
 \frac{3}{4\pi^2}\lambda_{0}\log{\frac{Q_\text{pole}^2}{v^2}} &= 1
 \notag \\
       &\Leftrightarrow&
 \log{\frac{Q_\text{pole}^2}{v^2}} &= \frac{4\pi^2}{3\lambda_0}
 \notag \\
        &\Leftrightarrow&
 Q_\text{pole} &= v \; \exp{\frac{2\pi^2}{3\lambda_0}}
    =v \; \exp{\frac{4\pi^2 v^2}{3m_H^2}}
\label{eq:landau_pole}
\end{alignat}
Such a pole is called a \underline{Landau pole}\index{Landau pole} and
gives us a maximum scale beyond which we cannot rely on our
perturbative theory to work.  In the upper line of
Figure~\ref{fig:higgs_scale} we show $Q_\text{pole}$ versus the Higgs
mass, approximately computed in Eq.\eqref{eq:landau_pole}. As a
function of the Higgs mass $Q_\text{pole}$ gives the maximum scale
were our theory is valid, so we have to reside below and to
the left of the upper line in Figure~\ref{fig:higgs_scale}. Turning
the argument around, for given $Q_\text{pole}$ we can read off the
maximum allowed Higgs mass which in the limit of large cutoff values
around the Planck scale $10^{19}$~GeV becomes $m_H \lesssim 180$~GeV,
in good agreement with the observed Higgs mass around 125~GeV.

This limit is often referred to as the \underline{triviality
  bound}\index{Higgs mass!triviality bound}, which at first glance is
precisely not what this theory is --- trivial or non--interacting.  The
name originates from the fact that if we want our Higgs potential to
be perturbative at all scales, the coupling $\lambda$ can only be zero
everywhere. Any finite coupling will hit a Landau pole at some
scale. Such a theory with zero interaction is called trivial.\bigskip

After looking at the ultraviolet regime we can go back to the full
renormalization group equation of Eq.\eqref{eq:lambda_rge} and ask a
completely different question: if the Higgs coupling $\lambda$ runs as
a function of the scale, how long will $\lambda>0$ ensure that our
Higgs potential is bounded from below?

This bound is called the \underline{stability bound}\index{Higgs mass!stability bound}. On the right
hand side of Eq.\eqref{eq:lambda_rge} there are two terms with a
negative sign which in principle drive $\lambda$ through zero. One of
them vanishes for small $\lambda\sim 0$, so we can neglect it under
the assumption that we only study very weakly interacting Higgs sectors. In the
small-$\lambda$ regime we therefore encounter two finite competing
terms
\begin{alignat}{5}
 \frac{d\,\lambda}{d\log{Q^2}} &=
 \frac{1}{16\pi^2} \left[-3\frac{4m_t^4}{v^4}
                         +\frac{3}{16} \left( 2g_2^4+ \left(g_2^2+g_1^2
                                                      \right)^2
                                       \right) + \ope(\lambda)
                   \right]
 \notag \\
 \Leftrightarrow\quad
 \lambda(Q^2) &\simeq
  \lambda(v^2)
 +\frac{1}{16\pi^2}\left[-\frac{12m_t^4}{v^4}
                         +\frac{3}{16} \left( 2g_2^4+\left(g_2^2+g_1^2
                                                     \right)^2
                                       \right)
                   \right] \; \log \frac{Q^2}{v^2} \; .
\label{eq:stability1}
\end{alignat}
The usual boundary condition at $\lambda(v^2)=m_H^2/(2 v^2)$ is the
starting point from which the top Yukawa coupling drives $\lambda$
through zero. This second critical scale
$\lambda(Q_\text{stable}^2)=0$ also depends on the Higgs mass
$m_H$. The second (smaller) contribution from the weak gauge coupling
ameliorates this behavior. The condition for a zero Higgs self
coupling is
\begin{alignat}{5}
 \lambda(v^2)=
 \frac{m_H^2}{2v^2} &\really
 -\frac{1}{16\pi^2}\left[ - \frac{12m_t^4}{v^4}
                          + \frac{3}{16} \left( 2g_2^4+\left(g_2^2+g_1^2
                                                       \right)^2
                                         \right)
                   \right] \; \log \frac{Q_\text{stable}^2}{v^2}
 \notag \\
 \Leftrightarrow\quad
 \frac{m_H^2}{v^2}\,&= \,
 \frac{1}{8\pi^2}\left[ \frac{12m_t^4}{v^4}
                       -\frac{3}{16} \left( 2g_2^4+\left(g_2^2+g_1^2
                                                       \right)^2
                                         \right)
                   \right] \; \log \frac{Q_\text{stable}^2}{v^2} \notag \\
 \Leftrightarrow\quad
 m_H &= \left\{
 \begin{array}{cc}
  70~\mbox{GeV}   \quad \text{for} \quad Q_\text{stable}=10^3~\mbox{GeV} \\
  130~\mbox{GeV}  \quad \text{for} \quad Q_\text{stable}=10^{16}~\mbox{GeV}
 \end{array}\right. \; .
\end{alignat}
From Eq.\eqref{eq:stability1} we see that only for energy scales below
$Q_\text{stable}(m_H)$ the Higgs potential is bounded from below and
our vacuum stable. For a given maximum validity scale
$Q_\text{stable}$ this stability bound translates into a minimum Higgs
mass balancing the negative slope in Eq.\eqref{eq:stability1} for
which our theory is then well defined.  In
Figure~\ref{fig:higgs_scale} we show $Q_\text{stable}$ as the lower
curve, above which our consistent theory has to reside.

Our discussion of the triviality bound and of the stability of our
vacuum has a weak spot: it follows from Eq.\eqref{eq:lambda_rge} and
assumes that only the renormalizable couplings enter the behavior of
the Higgs vacuum at large energy scales. On the other hand, if we
start from low energies we should at some point reach energy scales
where higher--dimensional operators enter the picture. Even in the
Standard Model such operators get induced by loops, and
non--perturbative studies indicate that they stabilize the Higgs potential
and prevent the sign change in $\lambda$.  If that should be true it
would mean that our perturbative approximation only considering the
leading renormalizable operators does not allow us to extrapolate to
energy scales beyond $10^{10}$~GeV or more and that vacuum stability
is simply not an issue.\bigskip

Summarizing what we know about the Higgs mass in the Standard Model we
already have indirect experimental as well as theoretical constraints on
this otherwise free parameter in the Higgs sector.

Strictly in the Standard Model, electroweak precision data points to
the mass range $m_H\lesssim 200$~GeV. This means at the LHC we were
either looking for a light Higgs boson or we should have expected a
drastic modifications of our Standard Model, altering this
picture significantly. If a discovery of a Higgs boson around $m_H = 125$~GeV
means good or bad news is in the eye of the beholder. Certainly, at
this mass we do not have to expect huge deviations from the Standard
Model motivated by the Higgs sector.

From the renormalization group we have two pieces of information on
the Higgs mass, again in the renormalizable Standard Model: the Landau
pole or triviality bound gives an upper limit on $m_H$ as a function
of the cutoff scale.  Vacuum stability gives a lower bound on $m_H$ as
a function of the cutoff scale.  Running both cutoff scales towards
the Planck mass $Q_\text{pole}, Q_\text{stable} \rightarrow
10^{19}$~GeV, we see in Figure~\ref{fig:higgs_scale} that only Higgs
mass values around $m_H=130\cdots180$~GeV are allowed for a truly
fundamental and stable Standard Model. Above this parameter range the Higgs
sector interacts too strongly and soon develops a Landau pole, while below
this Higgs mass window the Higgs sector is too weakly interacting to give
a stable vacuum.
The exact numbers including
renormalization group running at two loops gives the stability
condition in two forms~\cite{Buttazzo:2013uya}
\begin{alignat}{5}
m_H &> 129.6~\gev 
     + 2 \left( m_t - 173.35~\gev \right)
     - \frac{\alpha_s(m_Z) - 0.1184}{0.0014}~\gev 
     \pm 0.3~\gev  
\notag \\
m_t &< \left( 171.36 \pm 0.46 \right)~\gev \; ,
\end{alignat}
where the error bars have to be taken with a grain of salt, if we
follow the strict rules about combining experimental and theory
uncertainties described in Section~\ref{sec:sim_errors}. 
If vacuum stability should really be a problem, the
observed value of the Higgs mass around 125~GeV is at the edge of the
vacuum stability bound, again leaving everything open.

\subsubsection{Top--Higgs renormalization group}
\label{sec:higgs_fixedpoint}

The two critical scales in the running of the Higgs self coupling are
not the only interesting feature of the renormalization group equation
Eq.\eqref{eq:lambda_rge}.  If we limit ourselves to only the top and
Higgs sector it reads
\begin{alignat}{5}
 \frac{d\,\lambda}{d\,\log{Q^2}}=\frac{1}{16\pi^2}
 \left( 12\lambda^2+6\lambda y_t^2 - 3y_t^4 \right)
 \; .
\label{eq:lambda_rge2}
\end{alignat}
The definition of a fixed point $\lambda_*$ is
that the function $\lambda(Q^2)$ has to stick to this value
$\lambda_*$ once it reaches it. An attractive fixed point is a value
$\lambda_*$ which is automatically approached when the argument $Q^2$
reaches the corresponding infrared or ultraviolet regime.
If we assume that the Higgs self coupling is
closely related to the Higgs mass, $m_H = \sqrt{2 \lambda} v$, a
fixed point really tells us something about the 
observable Higgs mass.

Including all couplings in Eq.\eqref{eq:lambda_rge} we see that there
is no obvious fixed point of $\lambda$ for either large (UV) or small (IR)
scales $Q$.  The solution of the RGE for $\lambda$ alone we compute in
Eq.\eqref{eq:higgs_solvelambda}.  In the infrared the scalar four
point couplings as well as its derivative vanish,
\begin{alignat}{5}
\lim_{\log Q^2 \to -\infty} \lambda(Q^2) = \lambda_* = 0 
\qqqquad
\lim_{\log Q^2 \to -\infty} \frac{d\lambda}{d \log Q^2} 
= \lim_{\log Q^2 \to -\infty} \frac{3 \lambda^2}{4 \pi^2}
= 0 \; .
\end{alignat}
This means that in the infrared the scalar self coupling alone would
approach zero. Such a vanishing fixed point $\lambda_* = 0$ is called
a \underline{Gaussian fixed point}. Obviously, higher powers of
$\lambda$ in the RGE will not change this infrared pattern.  The
triviality bound is the first example of an attractive IR fixed point
in renormalization group running.\bigskip

The question is if we can find something interesting when we go
beyond the pure Higgs system. The largest electroweak coupling
is the top Yukawa, already included in Eq.\eqref{eq:lambda_rge2}.  In
complete analogy we can compute the Higgs loop corrections to the
running of the top Yukawa coupling
\begin{alignat}{5}
 \frac{d\,y_t^2}{d\,\log{Q^2}}=
  \frac{9}{32\pi^2} \; y_t^4
 \; .
\label{eq:yuk_rge}
\end{alignat}
Again, the top Yukawa is closely related to the top mass, $m_t = y_t
v/\sqrt{2}$. The top Yukawa also has an attractive Gaussian IR fixed
point at $y_{t,*} = 0$, but this is not what we are after. Instead, we
define the \underline{ratio of the two couplings} as
%
\begin{alignat}{5}
R &= \frac{\lambda}{y_t^2}
\end{alignat}
and compute the running of that ratio,
\begin{alignat}{5}
\frac{dR}{d\log Q^2} &=
\frac{d\lambda}{d\log Q^2} \; \frac{1}{y_t^2} 
+ \lambda \frac{(-1)}{y_t^4} \; \frac{d y_t^2}{d\log Q^2}
\notag \\
&=
\frac{1}{16\pi^2 y_t^2}
\left( 12\lambda^2+6\lambda y_t^2 - 3y_t^4 \right)
- \frac{9 \lambda}{32\pi^2} 
\notag \\
&= 
\frac{1}{16\pi^2} \;
\left( 12 \lambda R + \frac{3}{2} \lambda - 3 y_t^2 
\right) 
\notag \\
&= 
\frac{\lambda}{16\pi^2} \;
\left( 12 R + \frac{3}{2} - 3 \frac{1}{R}
\right) 
\notag \\
&= 
\frac{3\lambda}{32\pi^2R} \;
\left( 8R^2 + R - 2
\right) \really 0 
\qqquad \Leftrightarrow \qqquad
\boxed{ R_* = \frac{\sqrt{65} -1}{16} \simeq 0.44 } \; .
\label{eq:higgs_cwfixedpoint}
\end{alignat}
This is not a fixed point in any of the two couplings involved, but a
\underline{fixed point in the ratio} of the two. It is broken by the
gauge couplings, most notably by the $\alpha_s$ correction to the
running top Yukawa or top mass. With the non--Gaussian IR fixed point for the
coupling ratio $R$ as well as the Gaussian fixed points for the individual couplings the question is
how they are approached. It turns out that the system first approaches
the fixed-point region for $R$ and on that line approaches the double
zero-coupling limit.  In the far infrared this predicts a ratio of the
top mass to the Higgs mass of
\begin{alignat}{5}
\frac{\lambda}{y_t^2} 
= \frac{m_H^2}{2v^2} \; \frac{v^2}{2m_t^2} \Bigg|_\text{IR}
= \frac{m_H^2}{4 m_t^2} \Bigg|_\text{IR}
= 0.44  
\qqquad \Leftrightarrow \qqquad
\frac{m_H}{m_t} \Bigg|_\text{IR}
= 1.33 
\label{eq:higgs_tophiggsratio}
\end{alignat}
At first sight this is not in good agreement with the Standard Model
value. On the other hand, the top and Higgs masses we usually quote
are not running masses in the far infrared. If the analysis leading to
Eq.\eqref{eq:higgs_tophiggsratio} is done properly, including
gravitational effects in the ultraviolet, it predicts a pole mass of
$m_t=172$~GeV and a Higgs mass $m_H = 126$~GeV. Puzzling.\bigskip

While the Wetterich fixed point in Eq.\eqref{eq:higgs_cwfixedpoint} is the most
obvious to discuss in these lecture notes we should also mention that
there exists a prototypical fixed point of this kind: the
\underline{Pendleton-Ross fixed point} relates the strong coupling and
the top mass $\alpha_s/y_t^2$ in the infrared. It is strictly speaking
only valid for non--perturbatively large strong coupling, making it
hard to predict a value for the top mass.  Together with a more
detailed analysis of the actual running the link to the strong
couplings predicts a top pole mass in the 100-200~GeV range. What does
the obvious quantitative applicability of these fixed points really
mean?  They suggest that our Standard Model is rooted at high scales
and our weak--scale infrared parameters are simply fixed by a
renormalization group analysis. Unfortunately, infrared fixed points
imply that during the renormalization group evolution we forget all of
the high--scale physics. Someone up at high scales wants to know that
he/she is in charge, but does not want to reveal any additional
information. If we do not find any signature of new physics at the LHC
we will have to study such predictions and extract the underlying
high--scale structures from the small effect around general fixed point
features.

\subsubsection{Two Higgs doublets and supersymmetry}
\label{sec:higgs_2hdm}

In Section~\ref{sec:higgs_higgsboson} we indicate how in the Standard
Model the $SU(2)_L$ doublet structure of the fermions really involves
the Higgs field $H$ and its conjugate $H^\dag$ to give mass to up--type
and down--type fermions. One motivation to use two Higgs doublets
instead of one is to avoid the conjugate Higgs field and instead give
mass up--type and down--type fermions with one $\s^{(u,d)}$ field
each. Such a setup is particularly popular because it appears in
supersymmetric theories\index{supersymmetry}.\bigskip

There are (at least) two reasons why supersymmetric models require
additional Higgs doublets: first, supersymmetry invariance does not
allow us to include superfields and their conjugates in the same
Lagrangian. The role which $H^\dag$ plays in the Standard Model has to
be taken on by a second Higgs doublet.  Second, the moment we
postulate fermionic partners to the weakly charged scalar Higgs bosons
we will generate a chiral anomaly at one loop. This means that quantum
corrections for example to the effective couplings of two gluons to a
pseudoscalar Higgs or Goldstone boson violate the symmetries of the
Standard Model. This anomaly should not appear in the limit of
massless fermions inside the loop, but it does once we include one
supersymmetric Higgsino state. The easiest way to cancel this anomaly
is through a second Higgsino with opposite hypercharge.

Two-Higgs--doublet models of type~II
are not the only way to extend the Standard Model Higgs sector by
another doublet. The crucial boundary condition is that the second
Higgs doublet should not induce flavor changing neutral currents which
have never been observed and are forbidden in the Standard Model. The
simplest flavor--compatible approach is to generate all fermion masses with one Higgs
doublet and add the second doublet only in the gauge sector. This
ansatz is forbidden in supersymmetry and usually referred to as
type~I. As mentioned above, type~II models separate the up--type and
down--type masses in the quark and lepton sector and link them to one
of the doublets each. Im both cases, type~I and type~II we can
disconnect the lepton sector from the quark sector and for example
flip the assignment of the two doublet in the lepton sector. Models of
type~III use a different way to avoid large flavor changing neutral
currents. In an appropriate basis only one of the Higgs doublets
develops a vacuum expectation value. It does not induce any flavor
changing mass terms. The other doublet couples to two fermions of
masses $m_{1,2}$ proportional to $\sqrt{m_1 m_2}$. This way, flavor
changing neutral currents are sufficiently suppressed. In the gauge
boson sector is it important that none of these three models allow for
a $Z^0 W^+ H^-$ coupling, so they obey the custodial symmetry
discussed in Section~\ref{sec:higgs_custodial} at tree level.  In our
discussion of fermion masses and couplings we will limit ourselves to
a model of type~II, though. A supersymmetrized Standard Model
Lagrangian together with a type~II two-Higgs--doublet model we refer to
as the minimal supersymmetric Standard Model (MSSM)\index{MSSM}. \bigskip

Using \underline{two sigma fields}\index{two Higgs doublet model} to
generate the gauge boson masses is a straightforward generalization of
Eq.\eqref{eq:betaprime1},
\begin{alignat}{5}
 \lag_{D2} =  \frac{v_u^2}{2} \; \tr \left[V_\mu^{(u)} V^{(u) \mu}\right]
            + \frac{v_d^2}{2} \; \tr \left[V_\mu^{(d)} V^{(d) \mu}\right] \; .
\end{alignat}
As for one Higgs doublet we define $V_\mu^{(u,d)}=\Sigma^{(u,d)}
(D_\mu \Sigma^{(u,d)})^\dagger$ for two $\s$ fields.  In unitary gauge
we can compute the corresponding gauge boson masses following
Eq.\eqref{eq:v_unitary}. The squares of the individual vacuum
expectation values add to the observed value of $v = 246$~GeV. This
structure can be generalized to any number of Higgs doublets. For two
Higgs doublets it allows us to use the known value of $v$ and a new
mixing angle $\theta$ as a parameterization:
\begin{alignat}{5}
v_u^2 + v_d^2 = v^2 
\qquad \Leftrightarrow \qquad 
v_u=v\sin \beta \quad \text{and} \quad v_d=v\cos \beta \; .
\end{alignat}
For our type II setup the fermion mass terms in Eq.\eqref{eq:higgs_d3}
include the two Higgs doublets separately
\begin{alignat}{5}
 \lag_{D3} = -\qlb m_{Qu} \s_u \frac{\one+\tau_3}{2} \qr
              -\qlb m_{Qd} \s_d \frac{\one-\tau_3}{2} \qr +... \; ,
\end{alignat}
with the isospin projectors $(\one \pm \tau_3)/2$.\bigskip  

To study the physical Higgs bosons we express each of the two sigma
fields in the usual representation
\begin{alignat}{5}
 \s^{(u,d)}= \one +\frac{H^{(u,d)}}{v_{u,d}} 
         -\frac{i \vec{w}^{(u,d)}}{v_{u,d}} 
 \qqquad \vec{w}^{(u,d)} = w^{(u,d)}_a \tau_a \; ,
\end{alignat}
which means that the longitudinal vector bosons are 
\begin{alignat}{5}
\vec{w}=  \cos \beta \; \vec{w}^{(u)}
         +\sin \beta \; \vec{w}^{(d)} \; .
\end{alignat}
Following Eq.\eqref{eq:def_phi} we can parameterize each of the Higgs
doublets in terms of their physical Goldstone and Higgs modes. 
We first recapitulate the available degrees of
freedom. Following the structure Eq.\eqref{eq:def_phi} we
parameterize the two Higgs doublets, now in terms of $H$ and
consistently omitting the prefactor $1/\sqrt{2}$.
\begin{alignat}{5}
&\begin{pmatrix}H^+_u \\ H^0_u\end{pmatrix}
= 
\begin{pmatrix}\text{Re} H^+_u + i \, \text{Im} H^+_u\\
                  v_u+\text{Re} H^0_u + i \, \text{Im} H^0_u
\end{pmatrix}
\qquad \qquad \qquad
\begin{pmatrix} H^0_d\\H^-_d\end{pmatrix}
=
\begin{pmatrix} v_d+ \text{Re} H^0_d + i \, \text{Im} H^0_d\\
                        \text{Re} H^-_d + i \, \text{Im} H^-_d 
\end{pmatrix}
\end{alignat}
As required by electroweak symmetry breaking we have three Goldstone
modes, a linear combination of $\text{Im} H_u^0$ and $\text{Im} H_d^0$
gives the longitudinal $Z$ while a linear combination of $H_u^+$ and
$H_d^-$ gives the longitudinal polarization of $W^\pm$.  The remaining
five degrees of freedom form physical scalars, one charged Higgs boson
$H^\pm$, two neutral CP-even Higgs bosons $H_u^0,H_d^0$ mixing into
the mass eigenstates $h^0$ and $H^0$, and a pseudo-scalar Higgs boson
$A^0$ from the remaining imaginary part.\bigskip

In addition to introducing two Higgs doublets the supersymmetric
Standard Model fixes the quartic Higgs coupling $\lambda$.  From
Eq.\eqref{eq:higgs_d4} we know that the quartic coupling fixes the
Higgs mass to $m_H^2=2\lambda v^2$, which means that supersymmetry
fixes the Higgs boson mass(es).  In broken supersymmetry we have to
consider three different sources of scalar self interactions in the
Lagrangian:
\begin{enumerate}
\item \underline{$F$ terms} from the SUSY-conserving scalar potential
  $W \supset \mu\cdot H_u H_d$ include four-scalar interactions
  proportional to Yukawa couplings,
\begin{alignat}{5}
\lag_W &= -\frac{|\mu|^2}{2}
\left( |H^+_u|^2 + |H^-_d|^2 + |H^0_u|^2 + |H^0_d|^2 \right) \; .
\label{eq:higgs_fterm}
\end{alignat}
Note that there is the usual relative sign between the definition of
the scalar potential and the Lagrangian.

\item in the Higgs sector the gauge--coupling mediated SUSY-conserving
  \underline{$D$ terms} involve abelian $U(1)_Y$ terms $D=g H^\dag H$ as
  well as non--abelian $SU(2)_L$ terms $D^\alpha=g' H^\dag \tau^\alpha H$
  with the Pauli matrices as $SU(2)_L$ generators,
\begin{alignat}{5}
\lag_D =&
-\frac{g^2}{16} \left[
  \left( |H^+_u|^2+|H^0_u|^2-|H^-_d|^2-|H^0_d|^2
  \right)^2
  + 4 \; | H^+_u H^0_d+H^0_u H^-_d|^2 \right] \notag\\
&-
\frac{g'^2}{16} 
  \left( |H^+_u|^2+|H^0_u|^2-|H^-_d|^2-|H^0_d|^2
  \right)^2 \; .
\label{eq:higgs_dterm}
\end{alignat}
The sign of the $D$ terms in the Lagrangian is indeed predicted to be
negative.

\item last, but not least scalar masses and self couplings appear as
\underline{soft SUSY breaking} parameters
\begin{alignat}{5}
\lag_\text{soft} =
 -\frac{m^2_{H_u}}{2} \left( |H^+_u|^2+|H^0_u|^2 \right)
 -\frac{m^2_{H_d}}{2} \left( |H^-_d|^2+|H^0_d|^2 \right)
 -\frac{b}{2} \left( H^+_u H^-_d-H^0_u H^0_d + \text{h.c.} \right)
\end{alignat}
\end{enumerate}

All these terms we can collect into the Higgs potential for a two
Higgs doublet model
\begin{alignat}{5}
V 
=& \frac{|\mu|^2+m^2_{H_u}}{2} \left( |H^+_u|^2+|H^0_u|^2 \right)
+  \frac{|\mu|^2+m^2_{H_d}}{2} \left( |H^0_d|^2+|H^-_d|^2 \right) 
\notag \\
&+  \frac{b}{2} \left( H^+_u H^-_d - H^0_u H^0_d + \text{h.c.} \right) 
\notag\\
&+ \frac{g^2+g'^2}{16}
   \left( |H^+_u|^2+|H^0_u|^2-|H^-_d|^2-|H^0_d|^2 \right)^2
   +\frac{g^2}{4} |H^+_u H^0_d+H^0_u H^-_d|^2
\label{eq:higgs_susypot1}
\end{alignat}
This full form we would like to simplify a little before focussing on
the neutral states.  Because now we have two Higgs doublets to play
with we can first rotate them simultaneously without changing the
potential $V$. We choose $H^+_u=0$ at the minimum of $V$, \ie at the
point given by
\begin{alignat}{5}
\frac{\p V}{\p H^+_u} 
=& |H^+_u|\left( |\mu|^2+m^2_{H_u} \right) 
 + \frac{b}{2} H^-_d 
\notag \\ 
&+ \frac{g^2+g'^2}{4}
   |H^+_u| \left( |H^+_u|^2+|H^0_u|^2-|H^-_d|^2-|H^0_d|^2 \right) 
 + \frac{g^2}{2} H^0_d \left( H^+_u H^0_d+H^0_u H^-_d \right) 
\notag \\
\stackrel{H^+_u = 0}{\longrightarrow}&
   \frac{b}{2} H^-_d 
 + \frac{g^2}{2} H^0_d H^0_u H^-_d \really 0 \; .
\end{alignat}
This minimization condition can be fulfilled either as $H^-_d=0$ or as
$b + g^2 H^0_d H^0_u = 0$. Choosing a field dependent
value of the SUSY breaking parameter $b$ is hard to justify --- our
minimum condition should be a condition on the fields and not on the
Lagrangian parameters.
The condition $H^-_d=0$ simplifies the functional form of the
potential at the minimum to
\begin{alignat}{5}
V \Big|_\text{minimum}
&= \frac{|\mu|^2+m^2_{H_u}}{2} |H^0_u|^2
  +\frac{|\mu|^2+m^2_{H_d}}{2} |H^0_d|^2
 -b |H^0_u||H^0_d|
  +\frac{g^2+g'^2}{16} \left( |H^0_u|^2-|H^0_d|^2 \right)^2 \; .
\label{eq:higgs_susypot2}
\end{alignat}
At the minimum we absorb the phase of $b$ into a rotation of
$H^0_d H^0_u$, so the entire $b$ term then becomes real.\bigskip

In this re-rotation we have simply removed the charged Higgs and
Goldstone states from the potential. Because there are no charged
vacuum expectation values this should not affect the rest of the
neutral spectrum.  We will use the \underline{simplified
  supersymmetric Higgs potential} for our study of the neutral Higgs
states.  Looking for the minimum of the neutral part of the Higgs
potential will allow us to relate the two vacuum expectation values
$v_{u,d}$ to the parameters in the potential. The minimum conditions
are
\begin{alignat}{5}
0\stackrel{!}{=} 
\frac{\p  V}{\p |H^0_u|}\Bigg|_{H^0_j=v_j}
=& \left( |\mu|^2+m^2_{H_u} \right) |H^0_u|
 - b |H^0_d|
 + \frac{g^2+g'^2}{4} 
   |H^0_u| \left( |H^0_u|^2-|H^0_d|^2\right)  \; \Bigg|_{H^0_i=v_i}
\notag\\
0\stackrel{!}{=} 
\frac{\p  V}{\p |H^0_d|}\Bigg|_{H^0_j=v_j}
=& \left( |\mu|^2+m^2_{H_d} \right) |H^0_d|
 - b |H^0_u|
 - \frac{g^2+g'^2}{4} 
   |H^0_d| \left( |H^0_u|^2-|H^0_d|^2\right)  \; \Bigg|_{H^0_i=v_i}
\notag\\
\Leftrightarrow \qquad
0 =& |\mu|^2+m^2_{H_u}
    - b \frac{v_d}{v_u}+ \frac{g^2+g'^2}{4} \; \left (v^2_u-v^2_d \right ) 
\notag\\
0 =& |\mu|^2+m^2_{H_d}
    - b \frac{v_u}{v_d}- \frac{g^2+g'^2}{4} \; \left (v^2_u-v^2_d \right ) 
\label{eq:pot_stable}
\end{alignat}
From the Standard Model Higgs sector with custodial symmetry,
Eq.\eqref{eq:betaprime2}, we know how to replace the gauge couplings
squared by the gauge boson masses
\begin{alignat}{5}
m_Z^2 = \frac{g^2+g'^2}{2}\; \left( v^2_u+v^2_d \right)
\qquad \qquad \qquad 
m^2_W = \frac{g^2}{2} \; \left( v^2_u+v^2_d \right) \; .
\end{alignat}
The minimum conditions then read 
\begin{alignat}{5}
|\mu|^2+m^2_{H_u}=b~\cot \beta+\frac{m_Z^2}{2}\cos (2\beta)
\qquad \qquad 
|\mu|^2+m^2_{H_d}=b~\tan \beta-\frac{m_Z^2}{2}\cos (2\beta) \; .
\label{eq:higgs_stable}
\end{alignat}
These relations can be used to express $b$ and $|\mu|$ in terms of the
gauge boson masses and the angle $\beta$. This suggests that the
extended Higgs sector will be governed by two independent mass
scales, $m_Z \sim v_{u,d}$ and $\sqrt{b}$.  For now, we will still
keep $b$ and $|\mu|$ to shorten our expressions.\bigskip

The masses of all physical modes as fluctuations around the vacuum
state are given by the quadratic approximation to the potential around
the vacuum. Because the \underline{interaction eigenstates $H_{u,d}$}
do not have to be \underline{mass eigenstates} for their real or
imaginary parts the matrix of second derivatives defines a scalar mass
matrix just like in Eq.\eqref{eq:higgs_massmatrix}
\begin{alignat}{5}
\boxed{
\left( \mat^2 \right)_{jk}
= \frac{\p^2 V}{\p H^0_j \p H^0_k} \Bigg|_\text{minimum}
}
\end{alignat}
We will compute the masses of all three scalar Higgs states, beginning
with the compute pseudoscalar mass $m_A$. If this state is a
superposition of $\text{Im} H^0_u$ and $\text{Im} H^0_d$ the relevant
terms in Eq.\eqref{eq:higgs_susypot2} are
\begin{alignat}{5}
V &\supset
 \frac{|\mu|^2+m^2_{H_u}}{2} (\text{Im} H^0_u)^2
+\frac{|\mu|^2+m^2_{H_d}}{2} (\text{Im} H^0_d)^2
+b \; \text{Im} H^0_u \; \text{Im} H^0_d \\
&+
 \frac{g^2+g'^2}{16}
 \left[ (\text{Re} H^0_u)^2 + (\text{Im} H^0_u)^2
       -(\text{Re} H^0_d)^2 - (\text{Im} H^0_d)^2
 \right]^2 \notag\\
\frac{\p  V}{\p  (\text{Im} H^0_u)} &\supset 
   \left( |\mu|^2+m^2_{H_u} \right) \text{Im}H^0_u
 + b \; \text{Im} H_d^0 
 +\frac{g^2+g'^2}{4} \text{Im} H^0_u \; 
 \left[ (\text{Re} H^0_u)^2 + (\text{Im} H^0_u)^2
       -(\text{Re} H^0_d)^2 - (\text{Im} H^0_d)^2 \right]
\notag\\
\frac{\p^2 V}{\p (\text{Im} H^0_u)^2} &=
  \left( |\mu|^2+m^2_{H_u} \right)
 +\frac{g^2+g'^2}{4} 
 \left[ (\text{Re} H^0_u)^2 + (\text{Im} H^0_u)^2
       -(\text{Re} H^0_d)^2 - (\text{Im} H^0_d)^2 \right]
 + \frac{g^2+g'^2}{2} (\text{Im} H^0_u)^2 \notag \; .
\label{eq:higgs_pota}
\end{alignat}
Evaluating this derivative at the minimum of the potential and with
both scalar fields replaced by their vacuum expectation values gives us the masses
\begin{alignat}{5}
m_{\text{Im} H_u}^2 
&= \left( |\mu|^2+m^2_{H_u} \right)
  +\frac{g^2+g'^2}{4} \left( v^2_u-v^2_d \right) 
\notag\\
&= b \; \frac{v_d}{v_u} = b \cot \beta 
\qqquad \text{and} \qquad
m_{\text{Im} H_d}^2 = b \tan \beta \; ,
\end{alignat}
where we use the minimum condition Eq.\eqref{eq:higgs_stable} and the
symmetry under the exchange $H_u \leftrightarrow H_d$.  The parameter
$b$ indeed has mass dimension two.  For the mixed second derivative we
find
\begin{alignat}{5}
\frac{\p^2 V}{\p (\text{Im} H^0_u)\p (\text{Im} H^0_d)} \Bigg|_\text{minimum} = b \; .
\end{alignat}
Without any assumptions the mass matrix for the two CP-odd Higgs and
Goldstone mode is symmetric and has the form
\begin{alignat}{5}
\mat^2_A
= b \; 
\begin{pmatrix}
      \cot \beta&1\\1&\ \tan \beta
      \end{pmatrix}
\qquad \text{with the eigenvalues }
\quad \left\{\begin{array}{ll}
  m_G^2 = 0 & \\
  m_A^2 = \dfrac{2b}{\sin (2\beta)} 
\end{array}\right. \; .
\label{eq:higgs_massmat_odd}
\end{alignat}
The massive state $A^0$ is a \underline{massive pseudoscalar Higgs},
while the Goldstone is massless, as expected, and will be absorbed by
the massive $Z$ boson.  The mixing angle between these two
Goldstone/Higgs modes is given by $\beta$,
\begin{alignat}{5}
\dfrac{2 (\mat_A^2)_{12}}{m_A^2 - m_G^2}
= \dfrac{2b}{\dfrac{2b}{\sin (2\beta)}-0} 
= \sin (2\beta) \; .
\end{alignat}
Without going into any details we can assume that the Yukawa couplings
of the heavy pseudoscalar $A^0$ will depend on the mixing angle $\tan
\beta$. It turns out that its coupling to bottom quarks is enhanced by
$\tan\beta$ while the coupling to the top is reduced by the same
factor.\bigskip

Exactly the same calculation as in Eq.\eqref{eq:higgs_pota} we can
follow for the two CP-even scalar Higgs bosons, starting with
$\text{Re} H_u^0$. The relevant quadratic terms in the potential now
are
\begin{alignat}{5}
V &\supset 
 \frac{|\mu|^2 + m^2_{H_u}}{2} (\text{Re} H^0_u)^2
 + \frac{g^2+g'^2}{16}
   \left[ (\text{Re} H^0_u)^2 + (\text{Im} H^0_u)^2
         -(\text{Re} H^0_d)^2 - (\text{Im} H^0_d)^2 \right]^2 
 \\
\frac{\p  V}{\p  (\text{Re} H^0_u)} &\supset
   \left( |\mu|^2 + m^2_{H_u} \right) \text{Re} H^0_u
 + \frac{g^2+g'^2}{4} \; \text{Re} H^0_u \;
   \left[ (\text{Re} H^0_u)^2 + (\text{Im} H^0_u)^2
         -(\text{Re} H^0_d)^2 - (\text{Im} H^0_d)^2 \right]
\notag\\
\frac{\p^2V}{\p (\text{Re} H^0_u)^2} &=
  \left( |\mu|^2 + m^2_{H_u} \right)
+ \frac{g^2+g'^2}{4} \;
  \left[ (\text{Re} H^0_u)^2 + (\text{Im} H^0_u)^2
         -(\text{Re} H^0_d)^2 - (\text{Im} H^0_d)^2 \right]
+ \frac{g^2+g'^2}{2} \; (\text{Re} H^0_u)^2 \; .\notag 
\end{alignat}
The mass follows once we evaluate this second derivative at the
minimum, which means with both real parts of the scalar Higgs fields replaced
by vacuum expectation values:
\begin{alignat}{5}
m_{\text{Re} H_u}^2 
&= \left( |\mu|^2 + m^2_{H_u} \right)
 + \frac{g^2+g'^2}{4} \left( v^2_u-v^2_d+2v^2_u \right) \notag \\
&= |\mu|^2 + m^2_{H_u} 
 + \frac{g^2+g'^2}{4} \left(3v^2_u - v^2_d\right)\notag\\
&= b \cot\beta
 -\frac{g^2+g'^2}{4} \left( v^2_u-v^2_d \right)
 +\frac{g^2+g'^2}{4} \left( 3v^2_u-v^2_d \right) 
 \qquad \qquad \text{using Eq.\eqref{eq:higgs_stable}} \notag\\
&= b \cot\beta
 +\frac{g^2+g'^2}{2} \, v^2_u \notag\\
&= b\cot \beta + m_Z^2 \sin^2\beta  \notag \\
m_{\text{Re} H_d}^2 
&= b\tan \beta + m_Z^2 \cos^2\beta \; . \notag 
\end{alignat}
As mentioned above, $b$ has the dimension mass squared.  Going back to
Eq.\eqref{eq:higgs_susypot2} we see that the mixed derivative
includes two terms
\begin{alignat}{5}
\frac{\p^2V}{\p (\text{Re} H^0_u)\p (\text{Re} H^0_d)} \Bigg|_\text{minimum} =& 
- b 
 + \frac{g^2+g'^2}{16} \; 4 \; \text{Re} H^0_u \; (-2) \; \text{Re} H^0_d \notag\\
=& - b - \frac{g^2+g'^2}{2} v^2 \sin \beta \cos \beta \notag\\
=& - b - \frac{m_Z^2}{2} \sin (2\beta) \; .
\end{alignat}
Collecting all double derivatives with respect to the real part of the
scalar fields we arrive at the mass matrix for the two CP-even Higgs
bosons $\text{Re} H^0_u$ and $\text{Re} H^0_d$. The Lagrangian
parameter $b$ we can replace by the physical Higgs and gauge boson
masses and the mixing angle $\beta$,
\begin{alignat}{5}
\mat^2_{h,H} & =
\begin{pmatrix}
   b\cot\beta+m_Z^2\sin^2\beta&-b- \dfrac{m_Z^2}{2} \sin (2\beta)\\
  -b-\dfrac{m_Z^2}{2} \sin (2\beta) &b \tan \beta+m_Z^2 \cos^2\beta
\end{pmatrix} \notag \\
& =
\begin{pmatrix}
   \dfrac{m_A^2}{2} \sin (2\beta) \cot\beta+m_Z^2\sin^2\beta&
   - \dfrac{m_A^2+m_Z^2}{2} \sin (2\beta) \\
   - \dfrac{m_A^2+m_Z^2}{2} \sin (2\beta) & 
   \dfrac{m_A^2}{2} \sin (2\beta) \tan \beta+m_Z^2 \cos^2\beta
\end{pmatrix} \notag \\
& =
\begin{pmatrix}
   m_A^2 \cos^2 \beta + m_Z^2\sin^2\beta&
   - \dfrac{m_A^2+m_Z^2}{2} \sin (2\beta) \\
   - \dfrac{m_A^2+m_Z^2}{2} \sin (2\beta) & 
   m_A^2 \sin^2\beta +m_Z^2 \cos^2\beta
\end{pmatrix}\; .
\label{eq:higgs_massmat_even}
\end{alignat}
The mass values for the \underline{mass eigenstates $h^0,H^0$},
ordered by mass $m_h < m_H$, are
\begin{alignat}{5}
2 m^2_{h,H} 
&=
m_A^2+m_Z^2 \mp 
\sqrt{\left( m_A^2+m_Z^2 \right)^2 -4m_A^2 m_Z^2 \cos^2(2\beta)}
\notag \\
&\simeq  
m_A^2 \mp
\sqrt{m^4_A+ 2m_A^2 m_Z^2 \left( 1 - 2 \cos^2(2\beta) \right)}
\qquad \qquad \qquad \text{for} \; m_A\gg m_Z \notag\\
&\simeq 
m_A^2 \mp 
m_A^2 \sqrt{1-\frac{4m_Z^2}{m_A^2}\cos (4\beta)} \; .
\end{alignat}
In the limit of a heavy pseudoscalar the
supersymmetric Higgs sector with its fixed quartic couplings predicts
one light and one heavy scalar mass eigenstate,
\begin{alignat}{5}
\boxed{
m^2_{h,H} 
=
\frac{m_A^2}{2} \mp 
\frac{m_A^2}{2} \left[ 1-\frac{2m_Z^2}{m_A^2}\cos (4\beta) \right]
=  
\left\{\begin{array}{ll}
m_Z^2\cos (4\beta) \\[2mm]
m_A^2               
\end{array}\right. } \; .
\end{alignat}
The mass of the lighter of these two states depends on the parameter
$\beta$, but it is bounded from above to $m_h < m_Z$. As we will see
in the following section this upper bound is modified by loop
corrections, but it is fair to say that supersymmetry
\underline{predicts one light Higgs}. If that is sufficient to claim
that the discovery of a 125~GeV Higgs boson is the first discovery
predicted by supersymmetry is a little controversial, though.\bigskip

The mixing angle between the two CP-even scalar states is in general
independent of the pseudoscalar mixing angle $\beta$. We denote it as
$\alpha$, and it can be computed from the mixing matrix shown in
Eq.\eqref{eq:higgs_massmat_even}. The couplings of the light and heavy
scalar Higgs to up--type and down--type quarks are modified both in
terms of $\alpha$ and in terms of $\beta$, where $\alpha$ appears in
the numerator through Higgs mixing and $\beta$ appears in the
denominator of the Yukawas $m_q/v$, replacing $v$ by $v_u$ and
$v_d$. The correction factors for the light Higgs boson $h^0$ are
$\cos \alpha/\sin \beta$ for up--type quarks and $-\sin \alpha/\cos
\beta$ for down--type quarks. The same factors for the heavy Higgs
$H^0$ are $\sin \alpha/\sin \beta$ when coupling to up--type quarks and
$\cos \alpha/\cos \beta$ when coupling to down--type quarks.\bigskip

To keep the equations simple we ignore the \underline{charged
  Higgs} entirely, even though its existence would be the most
striking sign of an extended Higgs sector with (at least) one
additional doublet. At tree level a full
analysis of the Higgs potential in Eq.\eqref{eq:higgs_susypot1} gives
us a massless Goldstone and a massive charged Higgs scalar with
\begin{alignat}{5}
m_{H^\pm} = \sqrt{m_W^2 + m_A^2} \; .
\end{alignat}
Its Yukawa coupling include up--type and down--type contributions,
dependent on the chiralities of the fermions. However, after adding
all chiralities the coupling factor typically becomes $(m_d^2 \tan^2
\beta + m_u^2 \cot^2 \beta)$, very similar to the pseudoscalar
case.\bigskip

\begin{figure}[t]
\begin{center}
\includegraphics[width=0.35\hsize]{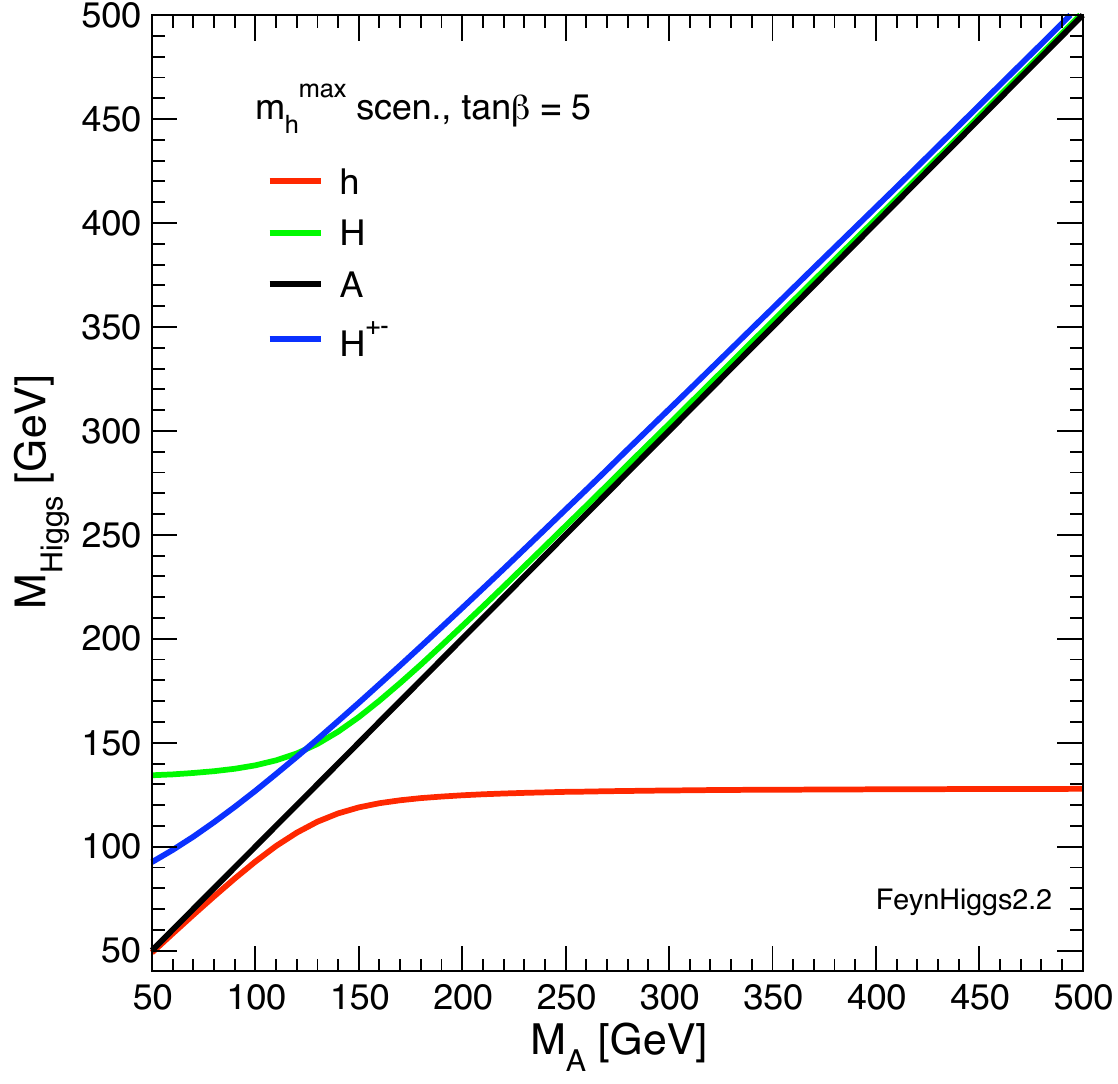}
\end{center}
\caption{Masses of all supersymmetric Higgs states as a function of
  the pseudoscalar Higgs mass, computed with FeynHiggs. Figures of
  this kind can be found for example in Ref.~\cite{Hahn:2009zz}.}
\label{fig:higgs_mssm}
\end{figure}

In the large-$m_A$ limit $m_A^2 \sim b \gg m_Z^2$ the Higgs mass
matrix as shown the first line of Eq.\eqref{eq:higgs_massmat_even}
is simplified and essentially aligns with its pseudoscalar counter part
Eq.\eqref{eq:higgs_massmat_odd},
\begin{alignat}{5}
\mat^2_{h,H}
\simeq b \, 
\begin{pmatrix}
   \cot\beta & -1 \\
  -1 & \tan \beta
\end{pmatrix}
\qquad \qquad 
\Rightarrow \quad
\alpha = \pi - \beta 
\end{alignat}
This means $\cos \alpha = \sin \beta$ and $\sin \alpha = - \cos
\beta$. The correction factors for the $h^0$ Yukawa couplings
become unity while the couplings of the heavy Higgs $H^0$ are $\tan
\beta$ enhanced for down--type quarks and $\tan \beta$ suppressed for
up--type quarks. From a phenomenological point of view the light
supersymmetric Higgs scalar behaves just like a Standard Model Higgs
boson while the heavy scalar and pseudoscalar Higgs bosons are hardly
distinguishable. Both of them and the charged Higgs have large masses
of order $m_A$. In Figure~\ref{fig:higgs_mssm} we show all masses
of the physics Higgs bosons, now including radiative corrections which
we will discuss in Section~\ref{sec:higgs_coleman}. For large
pseudoscalar masses we clearly see the decoupling of all three heavy
states from the one light Higgs boson.

As a matter of fact, this \underline{decoupling regime} where the
light supersymmetric Higgs boson is indistinguishable from a Standard
Model Higgs of the same mass is exact and includes all couplings and
properties. Small deviations from the Standard Model couplings,
suppressed by a finite mass ratio $m_h/m_A$, we need to look for in
the Higgs coupling analysis discussed in
Section~\ref{sec:higgs_couplings}.

\subsubsection{Coleman--Weinberg potential}
\label{sec:higgs_coleman}

In Section~\ref{sec:higgs_pot} we discuss the form of the Higgs
potential, as defined by all allowed renormalizable operators in the
Lagrangian. We make it a little more interesting
by including dimension-6 operators which are not renormalizable, but
we stick to a power series in $\phi^\dag \phi$. A different kind of
contribution to the Higgs potential\index{Higgs potential} can arise from loops of states
which couple to the Higgs boson. 

We start by limiting ourselves to dimension-4 operators and replace
the tree level potential Eq.\eqref{eq:higgs_pot} by an
\underline{effective potential}, including a tree level and a loop
contribution.  The question is if we can induce spontaneous symmetry
breaking with a non--trivial vacuum position of the Higgs field through
a loop--induced potential.\bigskip

Our toy model is a $\phi^4$ theory of a single real massive scalar
field, a little simpler than the complex Higgs--Goldstone field in the
Standard Model
\begin{alignat}{5}
\lag 
= \frac{1}{2} (\partial_\mu \phi )^2 
- \frac{m_0^2}{2} \phi^2
- \frac{\lambda_0}{4} \phi^4 
\label{eq:higgs_phi4}
\end{alignat}
Using some basic field theory we can elegantly describe this
alternative source of spontaneous symmetry breaking. We first review
the \underline{generating functional} for a free real scalar field
theory (following Mark Srednicki's conventions changed to our metric)
\begin{alignat}{5}
Z_0(J) 
&= \int \mathcal{D} \phi \; e^{i S_0(\phi) + i \int d^4x \, J \phi}
\notag \\
&= \int \mathcal{D} \phi \; e^{i \int d^4x \, \left( \lag_0 + J \phi \right)}
\notag \\
&= e^{\frac{i}{2} \int d^4x_1 d^4x_2 \, J(x_1) \Delta(x_1-x_2) J(x_2)} 
\notag \\ 
\text{with} \qquad 
\Delta(x_1-x_2) &= - \int \frac{d^4k}{(2 \pi)^4} \; 
                          \frac{e^{-ik(x_1-x_2)}}{k^2-m_0^2} 
\notag \\
\tilde{\Delta}(k^2) &= - \frac{1}{k^2-m_0^2} \; .                          
\end{alignat}
In this form we see how we can compute propagators or other time
ordered products of field operators using \underline{functional
  derivatives} on the generating functional
\begin{alignat}{5}
i \Delta(x_1 - x_2) \equiv 
\langle 0 | T \phi(x_1) \phi(x_2) | 0 \rangle
&= \frac{1}{i} \frac{\delta}{\delta J(x_1)} \; 
   \frac{1}{i} \frac{\delta}{\delta J(x_2)} \; 
  Z_0(J) \Bigg|_{J=0} \; .
\label{eq:higgs_genfunc}
\end{alignat}
The vacuum expectation value of the free field itself is zero, as is
the expectation value for any odd number of scalar fields. This is
because there will always be one factor $J$ left after the functional
derivative which then gets set to zero.\bigskip

Once we switch on an interaction $\lambda_0$ this does not have to be
true any longer.  Moving from $Z$ to $iW = \log Z$ means we omit the
unconnected interaction diagrams.  In analogy to the free theory we
define an \underline{effective action} $\Gamma$ in terms of exact
propagators and exact vertices as
\begin{alignat}{5}
Z_\Gamma(J) 
= \int \mathcal{D} \phi \; e^{i \Gamma(\phi) +  i \int d^4x \, J \phi}
\equiv e^{i W_\Gamma(J)} \; .
\label{eq:higgs_effaction}
\end{alignat}
This effective action defines a \underline{stationary field configuration}
$\phi_J$ through
\begin{alignat}{5}
\frac{\delta}{\delta \phi(x)} \; 
\left( \Gamma(\phi) + \int d^4 x' J(x') \phi(x')
\right) = 0 
\qquad \Leftrightarrow \qquad
\frac{\delta \Gamma(\phi)}{\delta \phi(x)} \Bigg|_{\phi_J}
= - J(x) \; .
\label{eq:higgs_saddle}
\end{alignat}
Such a stationary point of the exponential allows us to expand the
effective action and the corresponding generating functional defined
in Eq.\eqref{eq:higgs_effaction} in terms of the field fluctuations.

At this point it would help if we could make physics sense out of the
field configuration $\phi_J(x)$. We only quote that to
leading terms in $\hbar$ (\ie at tree level) we can express the
interacting generating functional for connected diagrams in terms of
the effective action at this stationary point as a Legendre
transform,
\begin{alignat}{5}
W(J) = \Gamma(\phi_J) + \int d^4x \, J(x) \phi_J(x) \; .
\label{eq:higgs_leadinghbar}
\end{alignat}
A proper derivation of this formula can be found in Chapter~21 of Mark
Srednicki's field theory book.  Using this relation we can speculate
about a non--trivial expectation value of an interacting scalar field
in the presence of a finite source $J$. In analogy to
Eq.\eqref{eq:higgs_genfunc} we need to compute
\begin{alignat}{5}
\langle 0 | \phi(x) | 0 \rangle_J 
&= \frac{\delta}{\delta J(x)} \; W(J) \notag \\
&= \frac{\delta \Gamma(\phi_J)}{\delta J(x)} \; 
   + \phi_J(x) 
   + \int d^4x' \, J(x') \frac{\delta \phi_J(x')}{\delta J(x)} 
\qqquad &&\text{using Eq.\eqref{eq:higgs_leadinghbar}}
\notag \\
&= \int d^4x' \, \frac{\delta \Gamma(\phi_J)}{\delta \phi_J(x')} \; 
                 \frac{\delta \phi_J(x')}{\delta J(x)} \; 
   + \phi_J(x) 
   + \int d^4x' \, J(x') \frac{\delta \phi_J(x')}{\delta J(x)} \notag \\
&= \int d^4x' \, 
   \left( \frac{\delta \Gamma(\phi_J)}{\delta \phi_J(x')} + J(x') 
   \right) \; \frac{\delta \phi_J(x')}{\delta J(x)} \; 
   + \phi_J(x) 
\qqquad &&\text{using Eq.\eqref{eq:higgs_saddle}}
\notag \\
& = \phi_J(x) \; .
\end{alignat}
On the way we apply the definition of the stationary point in
Eq.\eqref{eq:higgs_saddle}. The \underline{expectation value} we are
looking for is nothing but the stationary point of the effective
action $\Gamma$.  In the limit $J=0$ this value $\phi_J(x)$ becomes a
vacuum expectation value.\bigskip

Motivated by the general expectation that a classical solution will
change much more slowly than the quantum field we assume that $\phi_J$
is constant,
\begin{alignat}{5}
\boxed{\phi(x) = \phi_J + \eta(x)}
\qquad \text{and} \quad 
e^{i W_\Gamma(J)}
= \int \mathcal{D} \eta \; e^{i \Gamma(\phi_J + \eta) +  i \int d^4x \, J \; (\phi_J+\eta)} \; .
\end{alignat}
The path integral over $\phi_J$ is trivial and only changes the
irrelevant normalization of the generating functional. The expanded
exponential around the saddle point with its vanishing first
derivative reads
\begin{alignat}{5}
\Gamma(\phi) + \int d^4x \; J(x) \phi(x)
&= \Gamma(\phi_J) + \int d^4x \; J(x) \phi_J(x)
 + \frac{1}{2} \int d^4x_1 d^4x_2 \; \eta(x_1)  
   \left( \frac{\delta^2 \Gamma(\phi)}{\delta \phi(x_1) \delta \phi(x_2)} \right)_{\phi = \phi_J}
   \eta(x_2) \notag \\
&\equiv \Gamma(\phi_J) + \int d^4x \; J(x) \phi_J(x)
 + \frac{1}{2} \int d^4x_1 d^4x_2 \; \eta(x_1) \Gamma^{(2)}(\phi_J) \eta(x_2) \; .
\label{eq:higgs_defgam2}
\end{alignat}
This means the linear term vanishes by definition around the
stationary point while the source term does not contribute beyond the
linear term. The last step is the definition of
$\Gamma^{(2)}(\phi_J)$.  Exponentiating this action we can make use of
the definition of the functional determinant for real scalar fields,
\begin{alignat}{5}
Z(\eta) 
= \int \mathcal{D} \eta 
  \; e^{-i \int d^nx_1 \, d^nx_2 \, \eta(x_1) M \eta(x_2)} 
\equiv \frac{2 (2 \pi)^n}{\det M} \; .
\label{eq:higgs_funcdet}
\end{alignat}
Inserting this formula into the definition of the generating
functional for connected Green functions $W_\Gamma$ gives us
immediately
\begin{alignat}{5}
i W_\Gamma(J)
&= \log \left[ 
   e^{i \Gamma(\phi_J) + i \int d^4x J(x)\phi_J(x)}
   \int \mathcal{D} \eta \; 
        e^{\frac{i}{2} \int d^4x_1 d^4x_2 \; \eta(x_1) \Gamma^{(2)}(\phi_J) \eta(x_2) }
   \right]
\notag \\
&= i \Gamma(\phi_J) + i \int d^4x \; J(x)\phi_J(x)
  + \log \left[ \frac{2 (2\pi)^n}{\det \left( -\Gamma^{(2)}(\phi_J) \right)} \right]^{1/2}
\qqquad \text{using Eq.\eqref{eq:higgs_funcdet}}
\notag \\
&= i \Gamma(\phi_J) + i \int d^4x \; J(x)\phi_J(x)
  + \frac{1}{2} \log \det \left( 2^{n+1} \pi^n \right) 
  - \frac{1}{2} \log \det \left( -\Gamma^{(2)}(\phi_J) \right)
\notag \\
&= i \left[ \Gamma(\phi_J) + \int d^4x \; J(x)\phi_J(x) 
  + \frac{i}{2} \tr \log \left( -\Gamma^{(2)}(\phi_J) \right) 
  + \text{const} \right] \; .
\end{alignat}
In the last step we exploit the general operator identity commuting
the logarithm and the trace.  Finite terms in a potential we can
ignore.  Comparing this result to Eq.\eqref{eq:higgs_leadinghbar} we
see that the exact generating functional $W_\Gamma$ includes an
additional loop--induced term,
\begin{alignat}{5}
W_\Gamma(J) = W(J) + \frac{i}{2} \tr \log \left( -\Gamma^{(2)}(\phi_J) \right) \; .
\end{alignat}
In other words, the Legendre transform of the full effective
connected generating functional $W_\Gamma$ includes an
\underline{additional $\tr \log$ contribution}. We need to translate this loop--induced
contribution to the effective action into
something we can evaluate for our model. The underlying concept is the
effective potential.  If $\phi_J$ does not propagate, its effective
action only includes potential terms and no kinetic term. In other
words, we can define an effective potential which the propagating
field $\eta(x)$ feels as
\begin{alignat}{5}
\boxed{
V_\text{eff} = V_0 + V_\text{loop} = - \frac{1}{L^4} \, 
\left[ \Gamma(\phi_J)
+\frac{i}{2} \tr \log \left( -\Gamma^{(2)}(\phi_J) \right)\right]
} \; .
\label{eq:higgs_veff_def}
\end{alignat}
The relative factor $L^4$ is the phase space volume which
distinguishes the action from the Lagrangian. It will drop out once we
compute $V_\text{loop}$. In this definition of the effective potential
we naively assume that both terms are finite and well defined. It will
turn out that this is not the case, so we should add to the definition
in Eq.\eqref{eq:higgs_veff_def} something like `finite terms of' or
`renormalized'.\bigskip

Until now our argument has been very abstract, so let us see if
computing the effective potential for our real scalar field
Eq.\eqref{eq:higgs_phi4} clarifies things.  Following the definition
in Eq.\eqref{eq:higgs_defgam2} we find
\begin{alignat}{5}
- \Gamma^{(2)}(\phi_J) 
&=
- 
\frac{\delta^2}{\delta \eta(x_1) \delta \eta(x_2)}
\int d^4x \;
\left[ 
- \frac{1}{2} \eta \partial_\mu^2 \eta 
- \frac{m_0^2}{2} (\phi_J+\eta)^2
- \frac{\lambda_0}{4} (\phi_J+\eta)^4 
\right] \Bigg|_{\eta=0}
\notag \\
&=
- \int d^4x \;
\frac{\delta^2}{\delta \eta(x_1) \delta \eta(x_2)}
\left[ 
- \frac{1}{2} \eta \partial_\mu^2 \eta
- \frac{m_0^2}{2} \left( \eta^2 + \cdots \right) 
- \frac{\lambda_0}{4} \left( \eta^4 + 4 \eta^3 \phi_J + 6 \eta^2 \phi_J^2 + \cdots \right) 
\right] \Bigg|_{\eta=0}
\notag \\
&= 
\partial_1^2 + m_0^2 + 3 \lambda_0 \phi_J^2  \; .
\end{alignat}
The $\tr \log$ combination we know how to compute once we assume we
know the eigenvalues of the d'Alembert operator $\p^2$. Because it
will turn out that in four space--time dimensions we need to remove
ultraviolet divergences through renormalization we compute it in
$n=4-2\epsilon$ dimensions. The formula for this $n$-dimensional
\underline{scalar loop integral} is standard in the literature:
\begin{alignat}{5}
\tr \log \left( \partial^2 + C \right)
&= \sum_p \log \left( -p^2 + C \right) 
\notag \\
&= L^n \int \frac{d^np}{(2\pi)^n} \log \left( -p^2 + C \right)
\notag \\
&= -i L^n \; 
   \Gamma \left( - \dfrac{n}{2} \right) \dfrac{C^n}{(4\pi)^{n/2}} \; .
\label{eq:higgs_onepointint}
\end{alignat}
The loop--induced contribution to the effective potential, now
including the renormalization scale to protect the mass dimension, is
then
\begin{alignat}{5}
V_\text{loop} 
&= - \frac{i}{2L^4} \tr \log 
   \left( \partial^2 + m_0^2 + 3 \lambda_0 \phi_J^2 \right)
\notag \\
&= - \mu_R^{4-n} L^{4-n} \frac{1}{2 (4\pi)^{n/2}} 
   \Gamma \left( - \dfrac{n}{2} \right) 
   \left( m_0^2 + 3 \lambda_0 \phi_J^2 \right)^{n/2}
\notag \\
&= - \mu_R^{2 \epsilon} \frac{1}{2 (4\pi)^{2-\epsilon}} 
   \Gamma \left( -2 + \epsilon \right) 
   \left( m_0^2 + 3 \lambda_0 \phi_J^2 \right)^{2-\epsilon}
\notag \\
&= - \mu_R^{2 \epsilon} \frac{1}{2 (4\pi)^{2-\epsilon}} 
   \frac{\Gamma \left( \epsilon \right)}{(-2+\epsilon)(-1+\epsilon)} 
   \left( m_0^2 + 3 \lambda_0 \phi_J^2 \right)^{2-\epsilon}
\notag \\
&= - \frac{1}{2 (4\pi)^2}\;
   \frac{1}{2 - 3\epsilon} 
   \left( \frac{1}{\epsilon} - \gamma_E + \log(4\pi) \right) 
   \left( m_0^2 + 3 \lambda_0 \phi_J^2 \right)^2
   \left( 1 - \epsilon \log \frac{m_0^2 + 3 \lambda_0 \phi_J^2}{\mu_R^2}
            + \ope(\epsilon^2)
   \right)
\notag \\
&= - \frac{1}{64 \pi^2} 
   \left( \frac{1}{\epsilon} - \gamma_E + \log(4\pi) + \frac{3}{2} \right) 
   \left( m_0^2 + 3 \lambda_0 \phi_J^2 \right)^2
   + \frac{1}{64 \pi^2} 
   \left( m_0^2 + 3 \lambda_0 \phi_J^2 \right)^2
   \log \frac{m_0^2 + 3 \lambda_0 \phi_J^2}{\mu_R^2} \; .
\label{eq:higgs_vloop}
\end{alignat}
In the second to last line we use the simple trick
\begin{alignat}{5}
C^\epsilon 
= e^{\log C^\epsilon}
= e^{\epsilon \log C}
= 1 + \epsilon \log C + \ope(\epsilon^2) \; .
\end{alignat}
The expression for $V_\text{loop}$ is divergent in the limit $\epsilon \to 0$, so we need
to renormalize it. In the $\msbar$ scheme this simply means
subtracting the pole $1/\epsilon - \gamma_E + \log(4\pi)$, so the
renormalized effective potential or \underline{Coleman--Weinberg
  potential} becomes
\begin{alignat}{5}
\boxed{
V_\text{eff} = 
V_0 + V_\text{loop}^\text{(ren)}
= V_0 + \frac{1}{64\pi^2} 
   \left( m^2 + 3 \lambda \phi_J^2 \right)^2
   \left( \log \frac{m^2 + 3 \lambda \phi_J^2}{\mu_R^2} 
        - \frac{3}{2}
   \right) 
} \; .
\label{eq:higgs_veff1}
\end{alignat}
The bare mass and coupling appearing in Eq.\eqref{eq:higgs_vloop}
implicitly turn into their renormalized counter parts in the $\msbar$
scheme.  Combining the tree level and the loop--induced potentials we
see how this additional contribution affects our real scalar $\phi^4$
theory defined by Eq.\eqref{eq:higgs_phi4} in its massless limit
\begin{alignat}{5}
V_\text{eff} 
&= 
\frac{\lambda}{4} \phi_J^4
   + \frac{9 \lambda^2}{64\pi^2} \phi_J^4 
   \left( \log \frac{3 \lambda \phi_J^2}{\mu_R^2} 
        - \frac{3}{2}
   \right) 
\notag \\
&= 
\frac{\lambda}{4} \phi_J^4 \; 
\left[ 
1 + \frac{9 \lambda}{16 \pi^2} \;
   \left( \log \frac{3 \lambda \phi_J^2}{\mu_R^2} 
        - \frac{3}{2}
   \right)  
\right] \; .
\label{eq:higgs_veff2}
\end{alignat}
In the limit where the logarithm including a physical mass scale
$\mu_R \equiv M$ becomes large enough to overcome the small coupling
$\lambda$ we can compute where the expression in brackets and hence
the whole effective potential passes through zero.
Close to this point the potential also develops a
\underline{non--trivial minimum}, \ie a minimum at finite field values,
\begin{alignat}{5}
\frac{d}{d\phi_J^2} \; V_\text{eff}(\phi_J)
&= 
\frac{\lambda}{2} \phi_J^2 \; 
\left[ 
1 + \frac{9 \lambda}{16 \pi^2} \;
   \left( \log \frac{3 \lambda \phi_J^2}{M^2} 
        - \frac{3}{2}
   \right)  
\right]
+ 
\frac{\lambda}{4} \phi_J^4 \; 
   \frac{9 \lambda}{16 \pi^2} \;
   \frac{1}{\phi_J^2}
\notag \\
&\simeq
\frac{\lambda}{2} \phi_J^2 \; 
\left[ 
1 + \frac{9 \lambda}{16 \pi^2} \;
    \log \frac{3 \lambda \phi_J^2}{M^2} 
+ \frac{9 \lambda}{32 \pi^2} 
\right] 
\qquad && \text{with} \; - \log \frac{\phi_J^2}{M^2} \gg 1
\notag \\
&\simeq 
\frac{\lambda}{2} \phi_J^2 \; 
\left[ 
1 
+ \frac{9 \lambda}{16 \pi^2} \;
  \log \frac{3 \lambda \phi_J^2}{M^2} 
\right] 
&& \text{with} \; \frac{\lambda}{4\pi^2} \ll 1
\notag \\
&\equiv 0 
\qqquad \Leftrightarrow \qquad
\phi_{J,\text{min}}^2 = \frac{M^2}{3 \lambda} e^{-16\pi^2/(9\lambda)} \; .
\label{eq:higgs_cwminimum}
\end{alignat}
The finite term $-3/2$ in
Eq.\eqref{eq:higgs_veff2} is numerically sub-leading and hence often
omitted.  Moreover, compared with Eq.\eqref{eq:higgs_veff2} the
leading contribution only applies the derivative to the over--all
factor $\phi_J^4$, not to the argument inside the logarithm. This
minimum is exclusively driven by the loop contribution to the scalar
potential. This means that the loop--induced Coleman--Weinberg potential
can break electroweak symmetry, when applied to the Higgs field in the
Standard Model. However, the position of this minimum we should take
with a grain of salt, because logarithms of the kind $\log \phi_J/M$
will appear in many places of the higher order corrections.  The
mechanism of generating a physical mass scale through a strong
interaction combined with a renormalization group analysis or
renormalization scale is called \underline{dimensional
  transmutation}\index{dimensional transmutation}.\bigskip

In the Standard Model we can compute the size of the Higgs self
coupling, $\lambda = m_H^2/(2 v^2)=0.13$. Forgetting the fact that our
toy model is a real scalar theory we can then compute the
corresponding field values at the minimum or vacuum expectation
value. In the loop--induced minimum it comes out very small,
\begin{alignat}{5}
\frac{\phi_{J,\text{min}}^2}{M^2} 
= 2.6 \times e^{-135}
\simeq 10^{-60} 
 \; .
\end{alignat}
To explain the gauge boson masses and the Higgs boson in the Standard
Model the Coleman--Weinberg potential is not well suited. However, in
the \underline{supersymmetric Higgs sector} discussed in
Section~\ref{sec:higgs_2hdm} it is very useful to compute the mass of
the lightest supersymmetric Higgs boson beyond tree level. We left
this scenario with the prediction $m_h < m_Z$, which is clearly ruled
out by the measured value of $m_H = 125$~GeV. The question is if loop
corrections to the supersymmetric Higgs potential can increase this
Higgs mass bound, such that it agrees with the measurement.\bigskip

The toy model we will study is the light supersymmetric Higgs boson
combined with a second heavy scalar, the scalar partner of the top
quark.  \underline{Top squarks} are the supersymmetric partners of the
chiral left handed and right handed top quarks. They can mix, which
means that we have to define a set of mass eigenstates
$\tilde{t}_{1,2}$. From Section~\ref{sec:higgs_2hdm} we know that
there are two independent kinds of four-scalar couplings between the
stops and the Higgs bosons: $F$-term Yukawa interactions proportional
to the top Yukawa $y_t$ and $D$-term gauge interactions. If we limit
ourselves to the large top Yukawa corrections and neglect stop mixing
we only have to consider one scalar state $\tilde{t}$. To simplify
things further we also assume that this scalar be real, neglecting the
imaginary part of the electrically and weakly charged supersymmetric
top partner.

The Lagrangian we study is the purely real stop--Higgs system with a
stop--stop--Higgs--Higgs coupling $y_t$, extended from the Higgs system
given in Eq.\eqref{eq:higgs_d4} to
\begin{alignat}{5}
\lag = 
  \frac{1}{2} \left( \partial_\mu H \right)^2
+ \frac{1}{2} \left( \partial_\mu \tilde{t} \right)^2
- \frac{\mst^2}{2} \tilde{t}^2
- \frac{y_t^2}{4} \, \tilde{t}^2 H^2 
+ \text{Higgs terms} \; .
\label{eq:higgs_stop}
\end{alignat}
The last term is the renormalizable four-point interaction between the
two scalar fields. Its leading coupling strength is the supersymmetric
Yukawa coupling, \ie the $F$ term scalar interaction introduced in
Eq.\eqref{eq:higgs_fterm} with $y_t=\sqrt{2}m_t/v$.\bigskip

The form of the Higgs potential at tree level is unchanged compared to
the Standard Model because the stop $\tilde{t}$ does not have a finite
vacuum expectation value. To see what the Coleman--Weinberg effective
potential Eq.\eqref{eq:higgs_veff1} tells us about this case we need
to briefly \underline{recapitulate its derivation}. The basis of our
derivation is an expansion of the Legendre transformed effective
action Eq.\eqref{eq:higgs_defgam2} around a stationary point. This
stationary point is the classical or tree level solution,
Eq.\eqref{eq:higgs_leadinghbar}, so we can assume that the
Coleman--Weinberg potential comes from a loop diagram. Because the only
coupling in our scalar $\phi^4$ theory is the scalar self coupling
$\lambda$ the relevant diagram must be the scalar one-point diagram.
The trace we compute in Eq.\eqref{eq:higgs_onepointint} is linked to a
loop integral with one scalar propagator, confirming this
interpretation. Finally, the $\msbar$ renormalization of the loop mass
$m$ and the coupling $\lambda$ in Eq.\eqref{eq:higgs_veff1} is exactly
what we would expect from such a calculation. In the final expression
we see that the mass renormalization as well as the coupling
renormalization can trigger symmetry breaking.  In the example shown
in Eq.\eqref{eq:higgs_cwminimum} we limit ourselves to the coupling
$\lambda$ alone, to illustrate the loop--induced effect in addition to
the tree level $\phi^4$ term.\bigskip

Looking at the Lagrangian Eq.\eqref{eq:higgs_stop} we are instead
interested in the effect which an additional massive scalar has on the
Higgs potential. This means we have to consider the general
Coleman--Weinberg form in Eq.\eqref{eq:higgs_veff1} in the limit
$\lambda = 0$. The mass which appears in the loop--induced potential is
the mass which appears in the relevant one-point loop integral. Now,
this integral is a closed stop loop coupling to the Higgs propagator
through the four-point coupling $y_t^2$. More generally, the
loop--induced or Coleman--Weinberg potential derived in
Eq.\eqref{eq:higgs_veff1} induced only by a massive loop contributing
to the Higgs propagator is
\begin{alignat}{5}
\boxed{ 
V_\text{eff} = \frac{1}{64 \pi^2} \; 
\sum \; (-1)^S \; n_\text{dof} \;
m^4(\phi_J) \left( 
\log \frac{m^2(\phi_J)}{\mu_R^2} - \frac{3}{2}
\right)
} \; .
\label{eq:higgs_cw}
\end{alignat}
Spin effects in the closed loop are taken care of by $(-1)^S = +1$ for
bosons and $(-1)^S = -1$ for fermions. The number of degrees of
freedom is $n_\text{dof} =1$ for a real scalar, 2 for a complex
scalar, and 4 for a fermion. The mass $m$ is the $\msbar$ mass of the
particle running inside the loop.  One widely used approximation is
the generalization of Eq.\eqref{eq:higgs_veffsm} to the Standard Model
including the Higgs mode, the Goldstone modes, and the large top
Yukawa $m_t = y_t v \sqrt{2} = y_t \phi_J$. In that case the Higgs
potential in Eq.\eqref{eq:higgs_pot} includes a negative mass term
$m^2 \to - 2\mu^2$ and a unit prefactor instead of $\lambda/4$,
\begin{alignat}{5}
V_\text{eff} &= V_0 
\notag \\
&+ \frac{1}{64\pi^2} \left( -2\mu^2 + 12 \lambda
\phi_J^2 \right)^2 \left( \log \frac{-\mu^2 + 12 \lambda
  \phi_J^2}{\mu_R^2} - \frac{3}{2} \right) + \frac{3}{64\pi^2} \left(
-2\mu^2 + 4\lambda \phi_J^2 \right)^2 \left( \log \frac{-2\mu^2 + 4\lambda
  \phi_J^2}{\mu_R^2} - \frac{3}{2} \right) 
\notag \\ 
&-
\frac{N_c}{16\pi^2} \, ( y_t \phi_J )^4 \left(
\log \frac{y_t^2 \phi_J^2}{\mu_R^2} - \frac{3}{2} \right) \; .
\label{eq:higgs_veffsm}
\end{alignat}
The MSSM\index{MSSM} differs from our toy model in two ways. First, the stop is
not a single neutral scalar, but a set of two charged scalars. If both
of them couple proportional to $y_t^2$ to the Higgs boson and we omit
the sub-leading finite term $-3/2$ we find
\begin{alignat}{5}
V_\text{loop}
= 
\frac{2 N_c}{64 \pi^2} \; 
\left( y_t \phi_J \right)^4 
\left( \log \frac{m_{\tilde{t}_1}^2}{\mu_R^2} + \log \frac{m_{\tilde{t}_2}^2}{\mu_R^2} 
      \right)
= 
\frac{N_c}{32 \pi^2} \; 
\left( y_t \phi_J \right)^4 
\log \frac{m_{\tilde{t}_1}^2 m_{\tilde{t}_2}^2}{\mu_R^2} \; .
\end{alignat}
The prefactor reflects the complex stop field with its two degrees of
freedom. Second, for this model to be complete we need to also take
into account the top quark contribution from
Eq.\eqref{eq:higgs_veffsm}.  By definition this includes both
chiralities with $n_\text{dof} = 4$, so altogether we find
\begin{alignat}{5}
V_\text{loop}^\text{(MSSM)} 
&= \frac{N_c}{32 \pi^2} \; (y_t \phi_J)^4 \;
\left( 
 \log \frac{m_{\tilde{t}_1}^2}{\mu_R^2}
+\log \frac{m_{\tilde{t}_2}^2}{\mu_R^2}
-2\log \frac{m_t^2}{\mu_R^2}
\right) \notag \\
&= \frac{N_c}{32 \pi^2} \; (y_t \phi_J)^4 \;
 \log \frac{m_{\tilde{t}_1}^2 m_{\tilde{t}_2}^2}{m_t^4} \; .
\label{eq:higgs_higgsnlo}
\end{alignat}
This is the leading loop correction to the \underline{lightest Higgs
  mass} in the MSSM, lifting the allowed mass range at tree level,
$m_h^2 < m_Z^2$, to include the measured value of 125~GeV. Looking in
more detail, in Eq.\eqref{eq:higgs_higgsnlo} we assume the stop
mass matrix to be diagonal. If we allow for a non--diagonal stop mass
matrix the value increases even further and we find power corrections
to $m_h$ proportional to the off-diagonal entries in the stop mass
matrix. All these corrections allow the observed Higgs mass around
125~GeV to be consistent with the MSSM prediction --- remembering that
the light Higgs mass is actually a prediction from the quartic gauge
couplings in the MSSM. However, the observed Higgs mass suggests that
the additional Higgs bosons is heavy and that the mixing between the
stop interaction eigenstates is strong.

\subsection{Higgs decays and signatures}
\label{sec:higgs_decays}

\begin{figure}[t]
\begin{center}
\includegraphics[width=0.5\hsize]{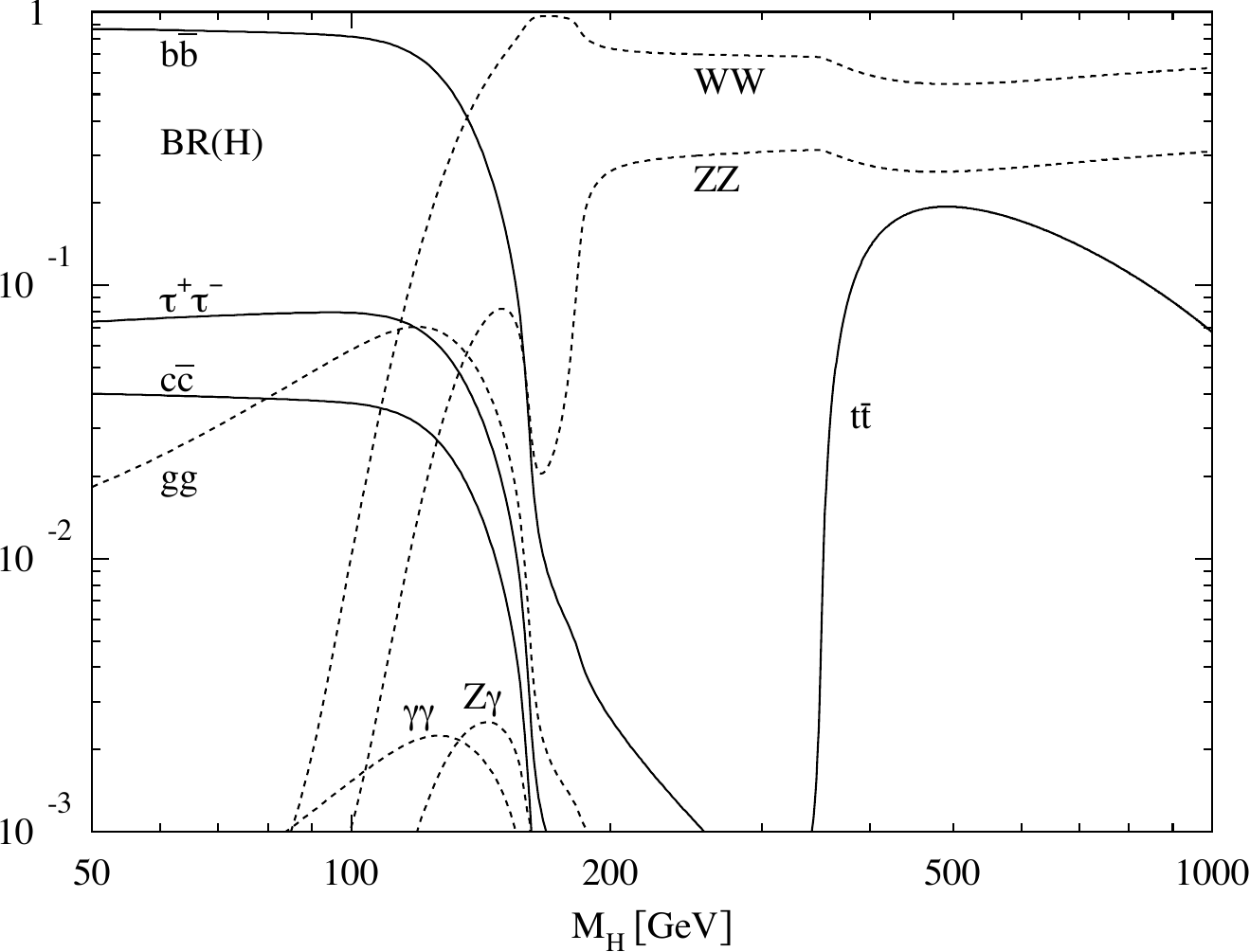}
\end{center}
\caption{Branching ratios of the Standard-Model Higgs
  boson\index{Higgs boson!branching ratios} as a function of its mass,
  computed with HDECAY. Off--shell effects in the decays to $WW$ and
  $ZZ$ are taken into account. Figure found for example in
  Refs.\cite{Spira:1997dg,Djouadi:2005gi}.}
\label{fig:higgs_decays}
\end{figure}

Signatures for new particles at colliders consist of a production
process and a decay pattern. Both, the production and the decay
can contribute to unique kinematic features which we can use to
extract signal from background events. The actual new particle is then
described by a Breit--Wigner propagator\index{particle width!Breit--Wigner propagator} for unstable particles which we
will discuss in detail in Section~\ref{sec:qcd_bw}. Since the Higgs
boson is a scalar there are no correlations between production and
decay process, which simplifies the calculation and simulation of
Higgs signatures. For backgrounds this factorization might of 
course not hold.\bigskip

Unlike the production processes the Higgs decay pattern is as simple
as it can be. At tree level all decay rates are determined by the
Higgs coupling to Standard Model particles, which are \underline{fixed
  by unitarity}\index{perturbative unitarity}.  The rule for different
Higgs decays is simple; because by definition the Higgs field couples
to all particles (including itself) proportional to their masses it
will preferably decay to the heaviest states allowed by phase
space. This goes back to the condition $\langle \s \rangle = \one$
translated into the appearance of the combination $(v+H)$ in
the Higgs field $\phi$ and in the Lagrangian.

This behavior we see in Figure~\ref{fig:higgs_decays}. Starting at low
masses this first applies to decays to $\tau\tau$ and $b\bar{b}$. The
relative size of their branching ratios around $10 \%:90 \%$ is given
by their Yukawa couplings in the appropriate renormalization scheme
($y_b/y_\tau \sim 1.4$), times an additional color factor $N_c=3$ for
the bottom quarks. Once the off--shell decays to $WW$ are allowed, they very
soon dominate. The dominant decays to bottom pairs and $W$ pairs
become equal for Higgs masses around 130~GeV. This is why we can
consider ourselves lucky with an observed Higgs mass around 125~GeV:
nature has chosen exactly the Higgs mass which allows us to observe
the largest number of different Higgs decays and this way extensively
study the Higgs sector, as discussed in
Section~\ref{sec:higgs_couplings}.

Because of the small mass difference between the $W$ and $Z$ bosons
the decay to $ZZ$ is not as dominant, compared to the $WW$ decay which
has two degrees of freedom ($W^+W^-$ and $W^-W^+$) in the final
state. In particular in the region where the $W$ decays first becomes
on--shell we see a drop of in the still off--shell $Z$ decays. For large
Higgs masses the ratio of $H \to WW$ and $H \to ZZ$ decays is fixed by
the relative factor of two, corresponding to the number of degrees of
freedom forming the final state. Above the top threshold the
$t\bar{t}$ decay becomes sizeable, but never really dominates.\bigskip

We can roughly estimate the Higgs width from its decay channels: in
general, we know that particles decaying through the weak interaction
have a width--to--mass ratio of $\Gamma/m \sim 1/100$. The main Higgs
decay is to bottom quarks, mediated by a small bottom Yukawa coupling,
$m_b/m_W \lesssim 1/30$. First, this means that in the Standard Model
we expect $\Gamma_H/m_H \sim 10^{-5}$, consistent with the exact
prediction $\Gamma_H \sim 4$~MeV. Second, loop--induced couplings can
compete with such a small tree level decay width.  In particular the
loop--induced decay to two photons plays an important role in LHC
phenomenology. It proceeds via a top and a $W$ triangle which enter
with opposite signs in the amplitude and hence interfere
destructively. The larger $W$ contribution fixes the sign of the
loop--induced coupling.

The structure of the $\gamma \gamma H$ coupling is similar to the
production via the loop--induced $ggH$ coupling which we will discuss
in Section~\ref{sec:higgs_gf} and then generalize to the photon case in
Section~\ref{sec:higgs_gamma}.  The reason for considering this decay
channel are the LHC detectors. To extract a Higgs signal from the
backgrounds we usually try to measure the four-momenta of the Higgs
decay products and reconstruct their invariant mass. The signal should
then peak around $m_H$ while the backgrounds we expect to be more or
less flat.  The LHC detectors are designed to
measure the photon momentum and energy particularly well. The
resolution in $m_{\gamma \gamma}$ will at least be a factor of 10
better than for any other decay channel, except for muons. Moreover,
photons do not decay, so we can use all photon events in the Higgs
search, while for example hadronically decaying $W/Z \to 2$~jets are
not particularly useful at the LHC. These enhancement factors make the
Higgs decay to two photons a promising signature, in spite of its
small branching ratio around $2 \cdot 10^{-3}$. More details of the
different decay channels we will give in
Section~\ref{sec:higgs_gf_lhc}.\bigskip

Because an observed Higgs sector can deviate from the minimal Standard
Model assumptions in many ways the LHC or other future colliders will
study the different Higgs decays and, as a function of $m_H$, answer
the questions
\begin{itemize}
\item[--] are gauge--boson couplings proportional to $m_{W,Z}$?  
\item[--] are these couplings dimension-3 operators?
\item[--] are fermion Yukawa couplings proportional to $m_f$?  
\item[--] is there a Higgs self coupling, \ie a remnant of the Higgs
  potential?
\item[--] do $\lambda_{HHH}$ and $\lambda_{HHHH}$ show signs of
  higher--dimensional operators?
\item[--] are there any other unexpected effects, like a Higgs decay
  to invisible particles?
\end{itemize}
But before we study the Higgs we need to discover it...

\subsection{Higgs discovery}
\label{sec:higgs_discovery}

Of course we cannot discover particles which do not get produced, and
for such a discovery we need to understand the production
mechanism. On the other hand, once we know the decay signatures of a
Higgs boson we should be able to at least roughly understand what the
LHC has been looking for. In that sense there is no need to further
delay a brief account of the Higgs discovery, as announced on the 4th
of July, 2012.\bigskip

Without knowing any theoretical particle physics we first need to
discuss the main feature or problem of hadron collider physics: there
is no such thing as a signal without a background. More precisely,
there is no kinematic configuration which is unique to signal events
and cannot appear as an unlucky combination of uninteresting Standard
Model or QCD processes and detector effects. This implies that any LHC
measurement will always be a \underline{statistics exercise} based on
some kind of event counting combined with a probability estimate for
the signal nature of a given event.\bigskip

Because signals are new things we have not seen before, they are rare
compared to backgrounds. Digging out signal events from a large number
of background events is the art of LHC physics. To achieve this we
need to understand all backgrounds with an incredible precision, at
least those background events which populate the signal region of
phase space. Such a background description will always be a
combination of experimental and theoretical knowledge.  The high
energy community has agreed that we call a $5\sigma$ excess over the
known backgrounds a signal discovery
\begin{alignat}{5}
\frac{S}{\sqrt{B}} &= \#\{ \sigma \} > 5 \qquad \qquad \qquad 
&&\text{(Gaussian limit)} \notag \\
P_\text{fluct} &< 5.8 \times 10^{-7} 
&&\text{(fluctuation probability)} \; .
\label{eq:higgs_significance}
\end{alignat}
\index{Gaussian distribution}More details on this probability measure
we will give later in this section. This statistical definition of a
`discovery' goes back to Enrico Fermi, who asked for $3\sigma$.
The number of researchers and analyses in high energy
physics has exploded since those days, so nowadays we do not trust anybody
who wants to sell you a $3\sigma$ evidence as a discovery.
Everyone who has been around for a few years has seen a great number
of those go away. People usually have reasons to advertize such
effects, like a need for help by the community, a promotion, or a wish
to get the Stockholm call, but all they are really saying is that
their errors do not allow them to make a conclusive statement. On the
other hand, in the Gaussian limit the statistical significance
improves with the integrated luminosity as $\sqrt{\lumi}$. So all we need to do is take more data
and wait for a $3\sigma$ anomaly to hit $5\sigma$, which is what
ATLAS and CMS did between the Moriond conference in the Spring of 2012
and the ICHEP conference in the Summer of 2012.\bigskip

In this section we will go through the results presented by ATLAS (and
CMS) after the ICHEP conference 2012. During the press conference
following the scientific presentations on the 4th of July 2012 the
ATLAS and CMS spokes-people and the CERN general director announced the
\underline{discovery of a new particle}, consistent with the Standard
Model Higgs boson. To keep it simple, we will limit ourselves to the
ATLAS discovery paper~\cite{atlas_discovery} --- the corresponding CMS
publication~\cite{cms_discovery} is very similar.\bigskip

To understand the numbers quoted in the Higgs discovery paper we need
some basic statistical concepts.  This leads us to the general
question on how to statistically test hypotheses for example
predicting an event rate $B_\text{theo}$ (predicted background) or
$(S+B)_\text{theo}$ (predicted signal plus background), where the 
corresponding measured number of events is
$N$.  The actual ATLAS and CMS analyses are much more
complicated than our argument in terms of event numbers, but our
illustration captures most relevant points.  For simplicity we assume
the usual situation where $(S+B)_\text{theo} > B_\text{theo}$.  If we
would like to know how badly our background--only
prediction is ruled out we need to know if
there is any chance a fluctuation around $B_\text{theo}$ would be
consistent with a measured value $N$. Note that the index `theo' does
not mean that these predictions are entirely based on the
underlying theory. If we largely understand a data set for example in
terms of the Standard Model without a Higgs, we can use measurements
in regions where we do not expect to see a Higgs effect to improve or
even replace the theoretical predictions.

First, an experimental outcome $N < B_\text{theo}$ means that the
background prediction is closer than the signal plus background
prediction. This means we are done with that signal hypothesis. It
gets a little harder when we observe $B_\text{theo} \lesssim N <
(S+B)_\text{theo}$. In this situation we need to define a measure
which allows us to for example rule out a signal prediction because
the measured event rates are close to the background prediction. In
this \underline{ruling-out mode} we ask the following question: `Given
that the background prediction and the measurement largely agree, how
sure are we that there is no small signal there?'. To answer this
question we compute the statistical distribution of event counts $N'$
around the predicted background value $B_\text{theo}$. In the
\underline{Gaussian limit} this is symmetric curve centered around
$B_\text{theo}$ with a standard deviation $\sigma_B$,
\begin{equation} 
f(N';B_\text{theo},\sigma_B) = \frac{1}{\sqrt{2 \pi}
  \sigma_B} \;
e^{-(N'-B_\text{theo})^2/(2\sigma_B^2)} \; .
\label{eq:higgs_gaussian}
\end{equation}
For small event numbers we need to replace this Gaussian with an
asymmetric Poisson distribution, which has the advantage that by
definition it does not cover negative event numbers. The entire
Gaussian integral is normalized to unity, and $68.3\%$ or $95.4\%$ of
it falls within one or two standard deviations $\sigma_B$ around
$B_\text{theo}$.  This number of standard deviations is a little
misleading because symmetric cuts around the central value
$B_\text{theo}$ is not what we are interested in.  A measure of how
well $(S+B)_\text{theo}$ is ruled out by exactly observing $N =
B_\text{theo}$ events is the normalized distance from the observed
background $S_\text{theo}/\sigma_B$.  To quantify which kinds of small
signals would be consistent with the observation of essentially the
background rate $N = B_\text{theo}$ we make a choice: if
$(S+B)_\text{theo}$ does not fall into the right 5\% tail of the
background observation it is fine, if it falls into this tail it is
ruled out. Given $N = B_\text{theo}$ this defines a critical number of
expected signal events. Any model predicting more signal events is
ruled out at the \underline{95\% confidence level} (CL). These 95\% of
the probability distribution are not defined symmetrically, but
through integrating from $N' > -\infty$ in
Eq.\eqref{eq:higgs_gaussian}, so the 95\% confidence level corresponds
to something like $S_\text{theo}/\sigma_B < 1.5$. In practice, this
makes it relatively easy to translate limits from one signal
interpretation to another: all we need to do is compute the number of
expected signal events in a given analysis, without any error analysis
or other complications. Whenever it comes out above the published
critical value the model is ruled out.

A variation of the ruling-out mode is when the observed
number of events lies above or below the background prediction $N \ne
B_\text{theo}$.  In this case we still apply the 95\% confidence level
condition following Eq.\eqref{eq:higgs_gaussian}, but replace the
central value $B_\text{theo}$ by the number of observed events
$N$. This gives us two critical values for $S_\text{theo}$, the
expected exclusion limit computed around $B_\text{theo}$ and the
observed exclusion limit around $N$. If the observed background
fluctuates below the prediction $N < B_\text{theo}$ we rule our more
models than expected, when it fluctuates above $N > B_\text{theo}$ the
exclusion starts to fail. This is the moment when we statistically
switch from ruling-out mode to \underline{discovery mode}.\bigskip

The problem of a discovery is the fundamental insight that it is not
possible to prove any scientific hypothesis correct. All we can do is
prove all alternatives wrong. In other words, we discover a
signal by solidly ruling out the background--only hypothesis. Following
the above argument we now observe $B_\text{theo} < N \sim
(S+B)_\text{theo}$. The question becomes: `How likely is it that the
background alone would have fluctuated to the observed number of
events?'. To answer this question we need to again compute the
statistical probability around the background hypothesis,
Eq.\eqref{eq:higgs_gaussian}. The difference to the argument above is
that in the discovery mode this distribution is entirely hypothetical.
A $5\sigma$ discovery of a given signal is claimed if at maximum a
fraction of $5.8 \times 10^{-7}$ expected events around
the predicted value $B_\text{theo}$ lie above the measured value $N \sim
(S+B)_\text{theo}$,
\begin{alignat}{5}
p_0 \equiv \int_N^\infty dN' \; f(N';B_\text{theo},\sigma_B)
    < 5.8 \times 10^{-7} \; .
\label{eq:higgs_pvalue}
\end{alignat}
The function $f$ can be close to a Gaussian but does not have to
be. For example for small event numbers it should also have a Poisson
shape, to avoid negative event numbers contributing to the integral.
One interesting aspects in this argument is worth noting: backgrounds
at the LHC are usually extracted from data with the help of standard
theory tools. An obvious advantage of Eq.\eqref{eq:higgs_pvalue} is
that we can immediately generalize it to more than one dimension, with
a complicated function $f$ indicating the correlated event numbers for
several search channels.

Finally, in Eq.\eqref{eq:higgs_pvalue} the signal does not feature at
all. Theorist hardly participate in the actual discovery of a new
particle once they have suggested what to look for and delivered an
understanding of the background in terms of simulations. On the other
hand, experimentalists really only discover for example the `Higgs
boson' because it shows up in the search for a Higgs boson without 
any obviously weird features. To claim
the discovery of a Higgs boson we need the $5\sigma$ deviation from
the background expectations and a solid agreement of the observed
features with the signal predictions. For the Higgs boson this for
example means that the observed rates can be mapped on Higgs couplings
which agree with the list in Section~\ref{sec:higgs_decays}. Such an
analysis we present in Section~\ref{sec:higgs_couplings}.\bigskip

\begin{figure}[t]
\begin{center}
  \includegraphics[width=0.35\hsize]{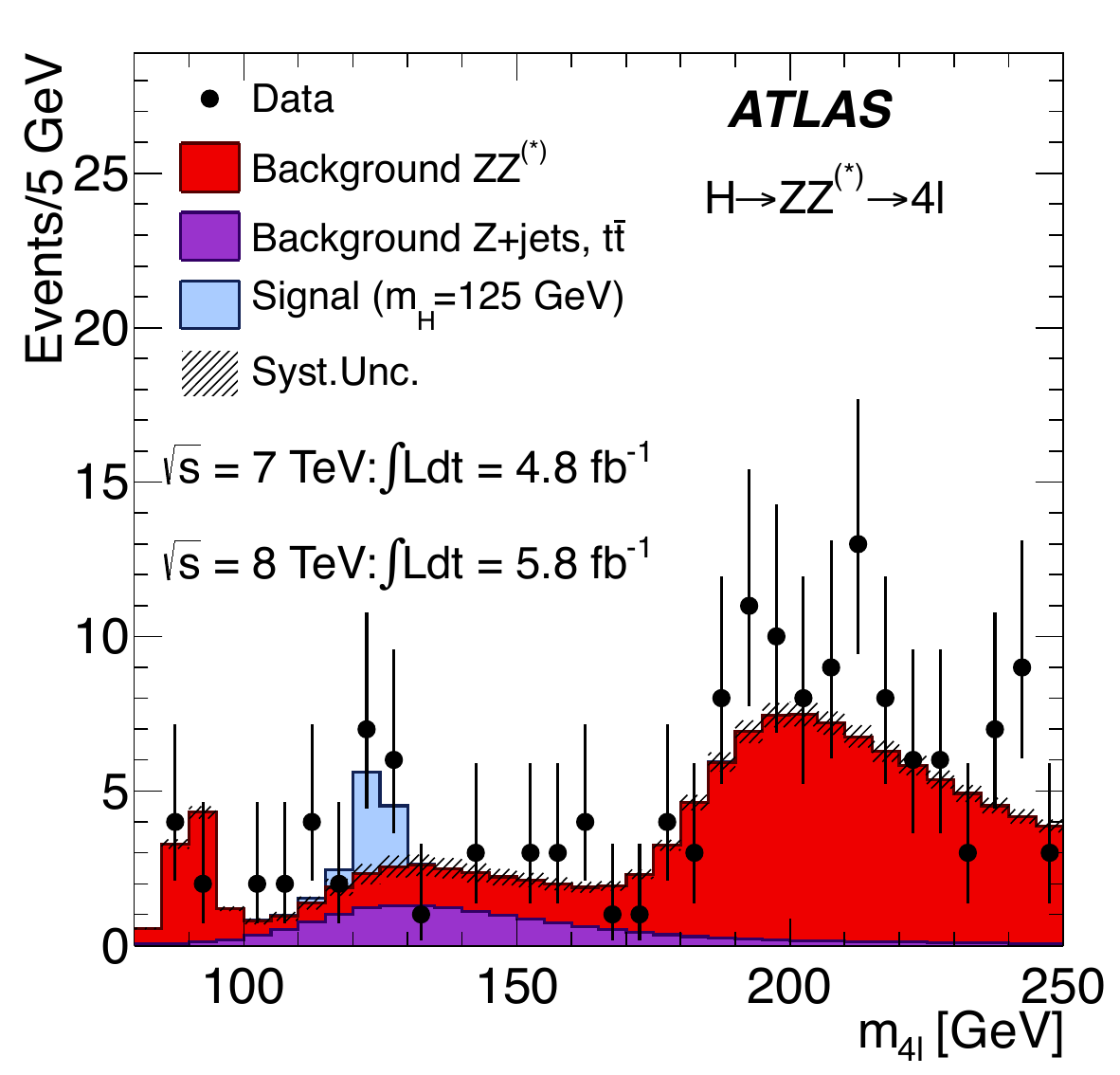}
  \hspace*{0.15\textwidth}
  \raisebox{0mm}{\includegraphics[width=0.35\hsize]{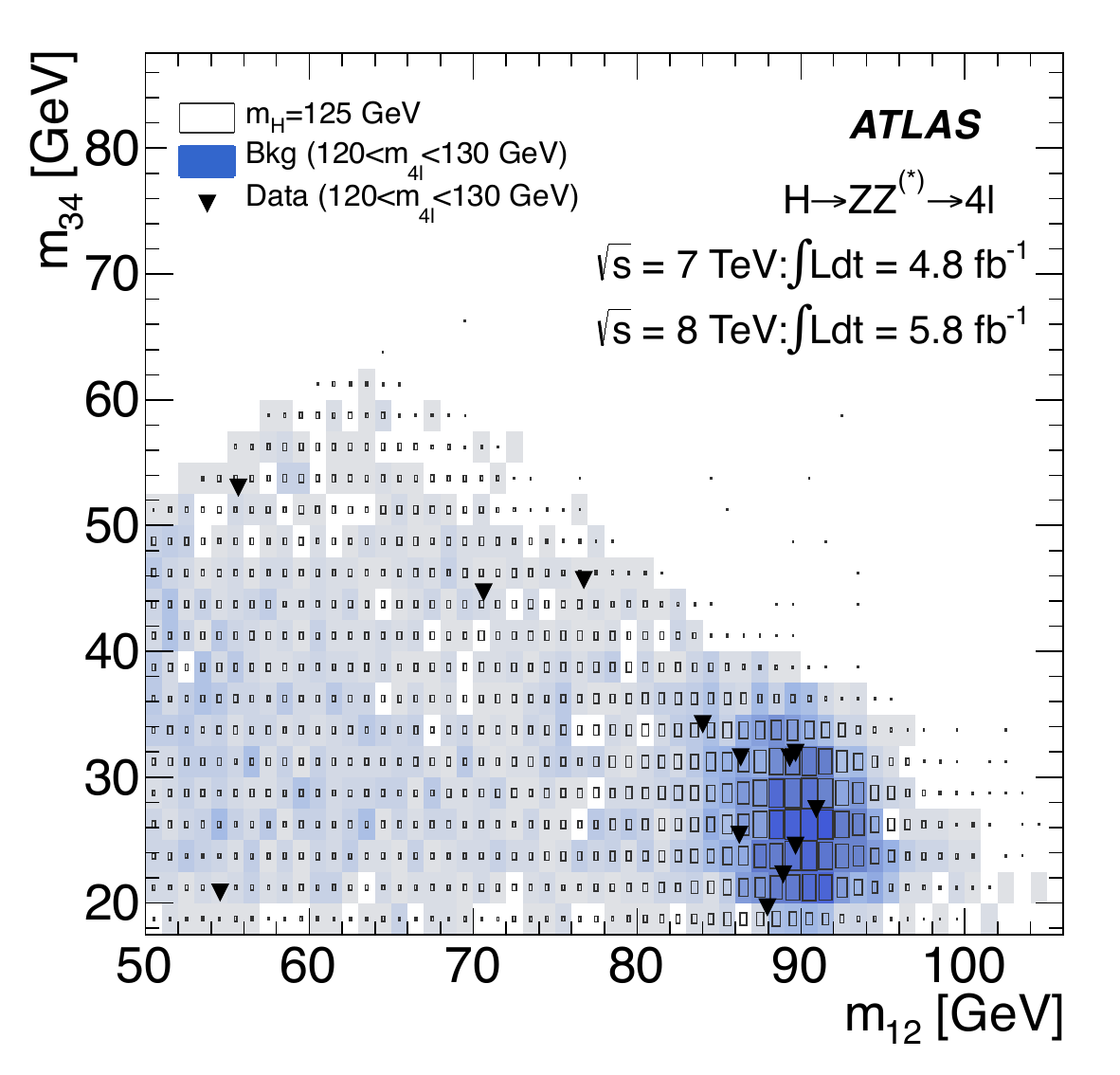}}
\end{center}
\caption{Left: Higgs mass peak in the $m_{4 \ell}$ spectrum, as shown
  in the ATLAS Higgs discovery paper~\cite{atlas_discovery}. Right:
  the correlation of the reconstructed $Z$ and $Z^*$ masses in the
  same analysis.}
\label{fig:higgs_zz}
\end{figure}

Going back to the \underline{ATLAS discovery paper} --- the most
important information is included in the abstract: ATLAS has
discovered something in their search for the Standard Model Higgs
boson. The analysis uses the data sets collected in 2011 and 2012,
with their respective proton--proton energies of 7~TeV and 8~TeV. The
channels which contribute to the statistical discovery are Higgs
decays $H \to ZZ \to 4\ell$, $H \to \gamma \gamma$, and $H \to W^+
W^-\to 2\ell \, 2\nu$.  In addition, ATLAS includes the results from
the $\tau^+ \tau^-$ and $b\bar{b}$ decays, but their impact is
negligible. What ATLAS observes is a peak with an invariant $\gamma
\gamma$ or $4 \ell$ mass of 126~GeV and a combined significance of
$5.9\sigma$.\bigskip

What follows after the short introduction are a section on the ATLAS
detector (rather useless for us), and a section on the simulation of
the signal and background event samples (not all that relevant for the
outcome). Next comes the first of the three discovery channels,
\underline{$H \to ZZ \to 4\ell$}, where the four leptons include all
possible combinations of two opposite-sign electrons and two
opposite-sign muons. The idea of this analysis is to reconstruct the
invariant mass of the four leptons $m_{4\ell}$ and observe a signal
peak at the Higgs mass value over a relatively flat and well
understood background.  We discuss more details on this analysis in
Section~\ref{sec:higgs_gf_lhc}. In the left panel of
Figure~\ref{fig:higgs_zz} we see that $m_{4\ell}$ distribution. The
clearly visible low peak around $m_{4\ell} \sim m_Z \sim 91$~GeV
arises from an on--shell $Z$ decay into two leptons plus a radiated
photon, which in turn splits into two leptons. Above the threshold for
the continuum production process $pp \to ZZ$ the background cross
section increases again. In between, we cannot even speak of a flat
background distribution, with typically one or two events per mass
bin. Nevertheless, the \underline{signal peak} is clearly visible, and
we can proceed to compute the signal significance, making sure that we
use Poisson statistics instead of Gaussian statistics. The result is
quoted in Table~7 of the ATLAS paper --- the $ZZ$ decay channel
contributes $3.6\sigma$ to the Higgs discovery, with a central Higgs
mass value around 125~GeV. This number of sigmas really is the $p_0$
value defined in Eq.\eqref{eq:higgs_pvalue} translated into a Gaussian
equivalent number of standard deviations $N/\sigma_B$.  In the right
panel of Figure~\ref{fig:higgs_zz} we show an important consistency
check of the $ZZ$ sample supposedly coming from a Higgs decay.  By
definition, all signal events have a combined value around $m_{4 \ell}
= 125$~GeV. In addition, we know that the four leptons come from,
possibly off--shell, $Z$ bosons.  In Section~\ref{sec:qcd_bw} we will
discuss the functional form of $m_{\ell \ell}$ around the $Z$-mass
pole. Quantum mechanics requires the unstable $Z$ boson to decay
exponentially, which corresponds to a Breit--Wigner shape\index{particle width!Breit--Wigner propagator} around the
resonance. This shape drops rapidly in the vicinity of the pole, but
further out develops linear tails. For our case $M_H < 2 m_Z$ 
it is most likely that one of the two $Z$ bosons is on its
mass shell while the other one decays into two leptons around $m_{\ell
  \ell} \sim m_H - m_Z =35$~GeV. This is precisely what we see in
Figure~\ref{fig:higgs_zz}. Just as a side remark: in CMS this
distribution was originally very different from what we would expect from quantum
mechanics.\bigskip

\begin{figure}[t]
\begin{center}
  \includegraphics[width=0.43\hsize]{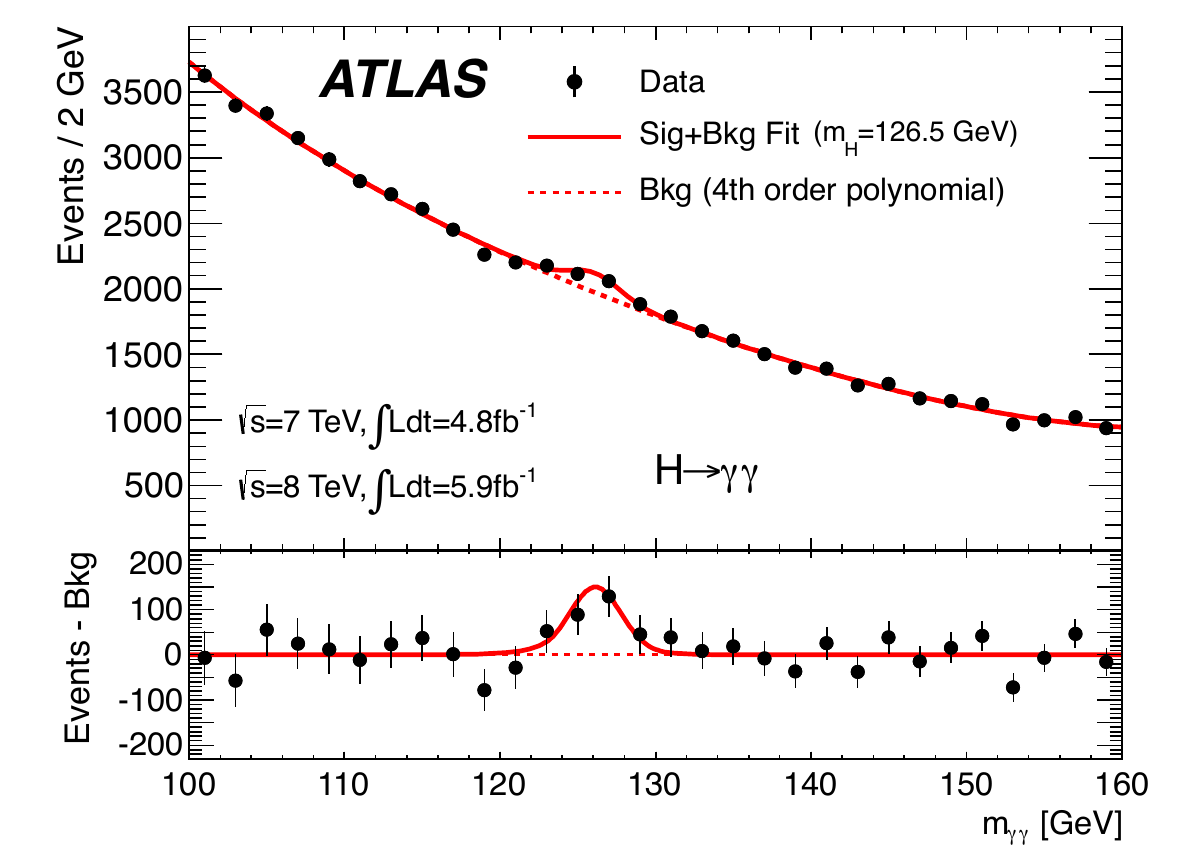}
  \hspace*{0.1\textwidth}
  \raisebox{0mm}{\includegraphics[width=0.43\hsize]{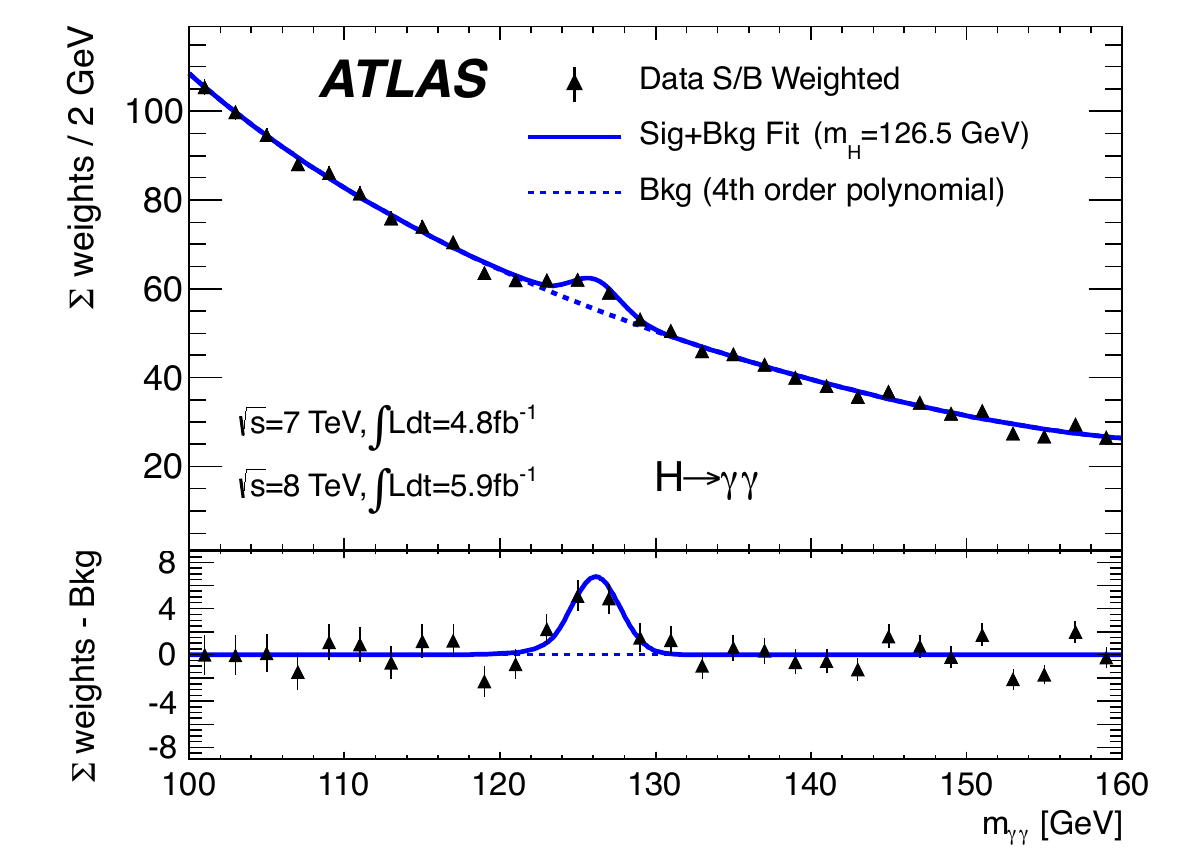}}
\end{center}
\caption{Left: Higgs mass peak in the $m_{\gamma \gamma}$ spectrum.
  Right: the same distribution where all events are re-weighted with
  the value of $S/B$ for the corresponding sub-selection. Both figures
  from the ATLAS discovery paper~\cite{atlas_discovery}.}
\label{fig:higgs_photons}
\end{figure}

The second analysis presented in the ATLAS paper is the search for
rare \underline{$H \to \gamma \gamma$} decays. The basic strategy is
to reconstruct the invariant mass of two photons $m_{\gamma \gamma}$
and check if it is flat, as expected from the background, or peaked at
the Higgs mass. In the left panel of Figure~\ref{fig:higgs_photons} we
show this distribution. The functional shape of the background is
flat, so without any derivation from first principles we can
approximate the curve outside the peak region by a polynomial. This
fit to a flat background we can subtract from the measured data
points, to make the peak more accessible to the eye. For the peak
shown in the left panel of Figure~\ref{fig:higgs_photons} we could now
compute a signal significance, \ie the probability that the flat
background alone fluctuates into the observed data points. 

However, this is not how the analysis is done. One piece of
information we should include is that we are not equally sure
that an experimentally observed photon
really is a photon everywhere in the detector. Some of the events entering the distribution shown
in the left panel of Figure~\ref{fig:higgs_photons} are more valuable
than others. Therefore, ATLAS ranks the photon phase space or the
relevant detector regions by their reliability of correctly
identifying two photons and measuring their invariant mass. In each of
these ten regions, listed in the Table~4 of their paper, they look at
the $m_{\gamma \gamma}$ distribution and determine the individual peak
significance. If the detector performance were the same in all ten
regions the combination of these ten significances would be the same as
the significance computed from the left panel of
Figure~\ref{fig:higgs_photons}. Because the individual phase space and
detector regimes have different signal and background efficiencies
some events shown in the left panel of Figure~\ref{fig:higgs_photons}
are more equal than others, \ie they contribute with a larger weight
to the combined significance. This is what ATLAS illustrates in the
right panel of Figure~\ref{fig:higgs_photons}: here, all events are
weighted with the signal--to--background ratio $S/B$ for the respective
sub-analysis. It is an interesting development that such a
\underline{purely illustrational figure} without any strict scientific
value enters a Higgs discovery paper. Obviously, the ATLAS and CMS
collaborations feel that their actual results are not beautiful enough
even for the scientific community, so they also deliver the public
relations version.  What is scientifically sound is the measured
signal significance of a mass peak centered at 126.5~GeV, quoted as
$4.5\sigma$ in their Table~7, but it cannot be computed from either of
the two curves shown in Figure~\ref{fig:higgs_photons}.\bigskip

The two $\gamma \gamma$ and $ZZ$ analyses are the main ATLAS results
shown in the Higgs discovery talk on July 4th. In the discovery paper
ATLAS adds a third channel, namely \underline{leptonic $H \to WW$}
decays. Obviously, we cannot reconstruct the $2 \ell 2\nu$ mass, which
means we need to rely on the transverse momentum balance in the
detector to approximately reconstruct the Higgs mass. The details of
this transverse mass measurement you can find in
Section~\ref{sec:sim_met}. Similar to the photon case, the analysis is
then split into different regimes, now defined by the number of jets
recoiling against the Higgs. The motivation for this observable is
first to reject the top pair background with its additional two $b$
jets and second to use the sensitivity of signal and background
kinematics to the transverse momentum of the Higgs. However, in
Section~\ref{sec:qcd_solve_dglap} we will see that perturbative QCD
does not allow us to separately study collinear jets, \ie jets with
transverse momentum below the Higgs mass, beyond leading order in
$\alpha_s$. Such an observable violates collinear factorization, induces 
possibly large logarithms, and this way 
spoils the application of precision QCD predictions. After fitting the
transverse mass distribution instead of simply cutting out the signal
region the $WW$ channel contributes $2.8\sigma$ to the final
significance, but without a good Higgs mass determination.\bigskip

\begin{figure}[t]
\begin{center}
  \includegraphics[width=0.45\hsize]{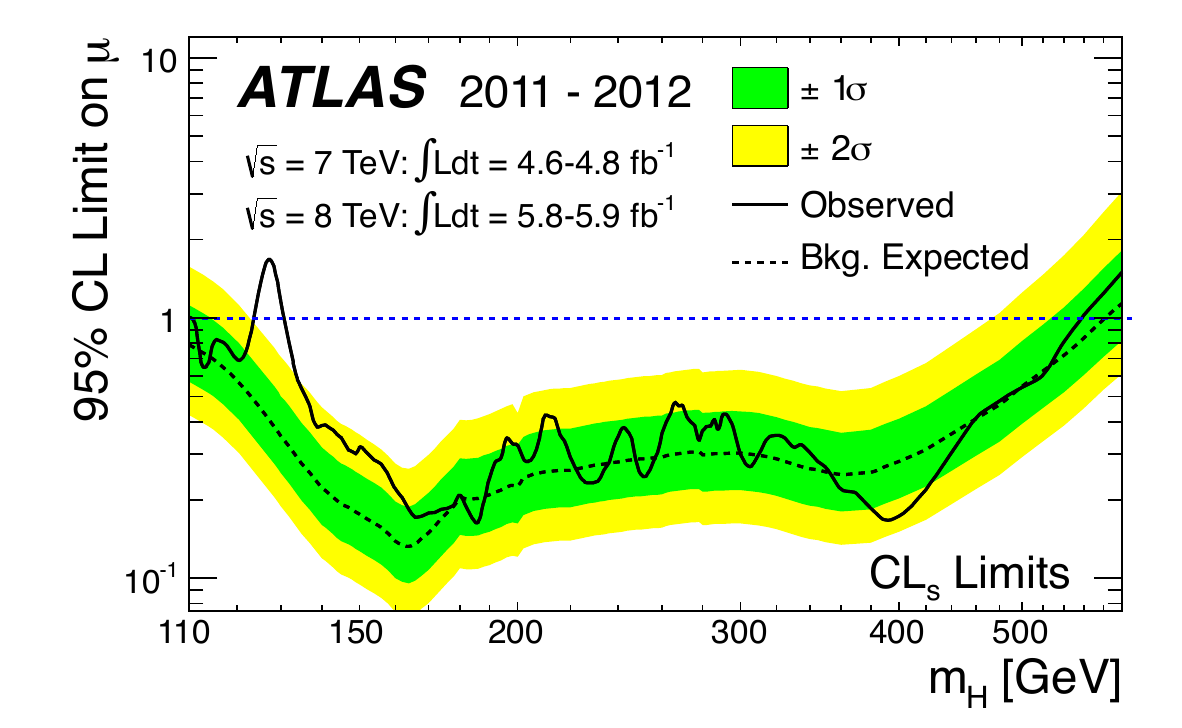}
  \hspace*{0.05\textwidth}
  \raisebox{0mm}{\includegraphics[width=0.42\hsize]{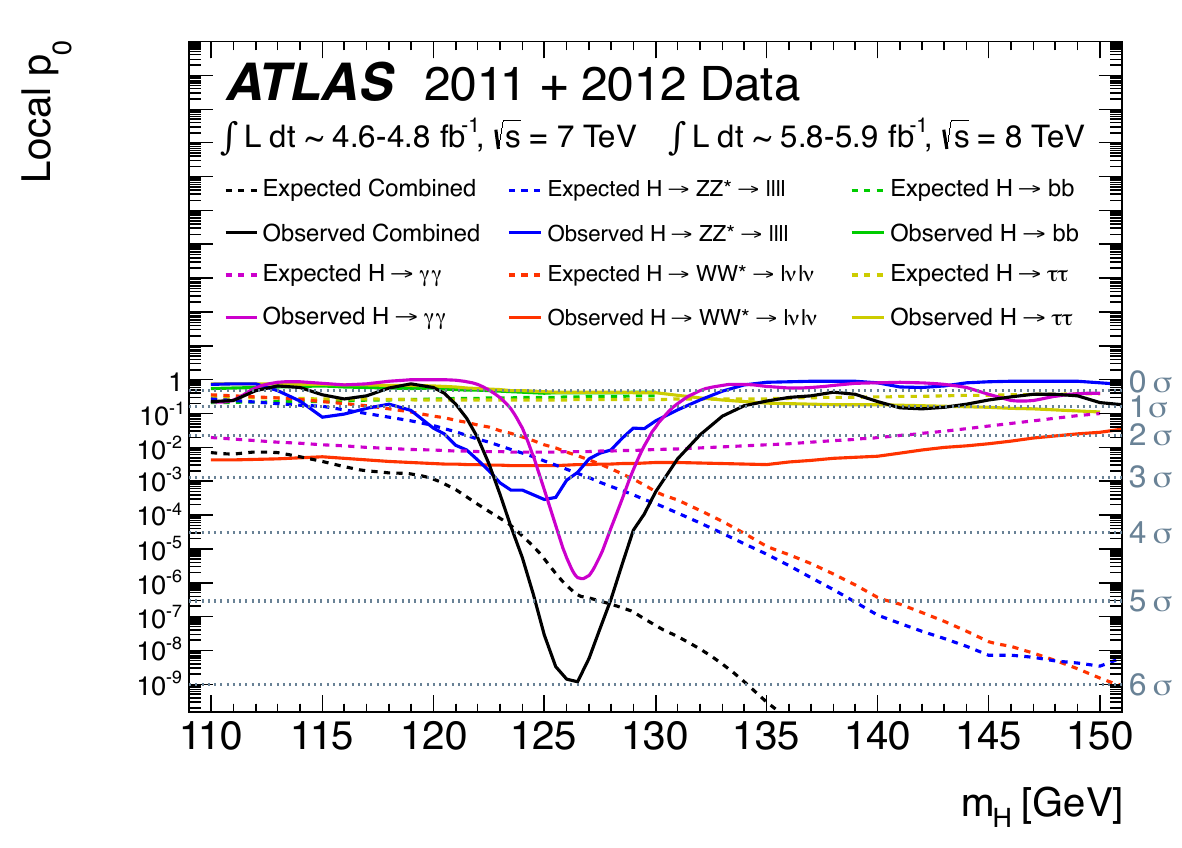}}
\end{center}
\caption{Left: exclusion limits on a hypothetical signal strength as a
  function of the Higgs mass~\cite{atlas_discovery}.  Right: signal
  significance as computed from the probability of a background
  fluctuation for the different channels. This figure is from the
  supplementary documentation to Ref.~\cite{atlas_discovery}.}
\label{fig:higgs_discovery}
\end{figure}

The next two sections in the ATLAS discovery paper discuss details of
the statistical analysis and the correlation of systematic
uncertainties. After that, ATLAS combines the three analyses and
interprets the result in terms of a \underline{Standard Model Higgs
  boson}. First, in the \underline{ruling--out mode} described above
ATLAS gets rid of models with a Standard Model Higgs boson in the mass
ranges $111-122$~GeV and $131-559$~GeV. This kind of exclusions from
LHC (and Tevatron) Higgs analyses are shown in the colors of a great
soccer nation.  If we know the expected signal and background numbers
and the detector performance, we can compute the number of signal
events which we expect to exclude with a given amount of data if there
were only background and no Higgs events. In the left panel of
Figure~\ref{fig:higgs_discovery} the dashed line shows the expected
exclusion limit as a function of the assumed Higgs mass and in terms
of the signal strength normalized to the Standard Model Higgs rate
$\mu = (\sigma \times \br)/(\sigma \times \br)_\text{SM}$. With the
quoted amount of data we would expect to exclude the entire mass range
from 110~GeV to 580~GeV, provided this Higgs has Standard Model
production and decay rates. For a hypothetical Higgs boson with only
half the number of expected events we only expect to exclude Higgs
masses from 120 to 460~GeV. Because this expected exclusion limit is a
statistical computation it has error bars which are shown in green
($1\sigma$) and yellow ($2\sigma$). Replacing our expected event rates
with data we find the solid curve in the left panel of
Figure~\ref{fig:higgs_discovery}. Around a hypothetical Higgs mass of
125~GeV the two significantly deviate, so we need to switch to
\underline{discovery mode}.\bigskip

In the right panel of Figure~\ref{fig:higgs_discovery} we show the
$p_0$ value computed by ATLAS as a function of the hypothetical Higgs
mass. This Higgs mass is not actually needed to predict the background
rates, but it enters because we optimize the signal searches for
assumed Higgs masses. For example, we attempt to rule out the $\gamma
\gamma$ background by determining its shape from a wide
$m_{\gamma\gamma}$ range and test it for deviations with a resolution
of few GeV. We see that the $H \to \gamma\gamma$ search as well as the $H \to
ZZ$ search point towards Higgs masses around 125~GeV. For the $H \to
WW$ analysis the assumed Higgs mass is less relevant, so the $p_0$
value shows a broad excess. Combining all channels gives us the solid
black line in the right panel of Figure~\ref{fig:higgs_discovery},
with a minimum $p_0$ value around $10^{-9}$ or $5.9\sigma$.

Again, we can compute the signal significance which we would have
predicted for this situation, shown as the dashed black line. The prediction
reaches only $5\sigma$, which means that assuming we are observing a
Standard Model Higgs boson ATLAS' signal significance is slightly
enhanced by upwards fluctuations in the event numbers.\bigskip

One final technical term in the ATLAS discovery paper is not yet
clear: this number of $5.9\sigma$ is described as the `local
significance'. The quoted local $p_0$ value is the probability of the
background fluctuating into the observed signal configuration with a
Higgs mass around 125~GeV. As mentioned above, ruling out the
background should naively not depend on signal properties like the
Higgs mass, but it does.  Let us assume that we search for a Higgs
boson in $m_{\gamma \gamma}$ bins of $\pm 2$~GeV and in the mass range
of 110~GeV to 150~GeV. If all ten analyses in the different mass
windows have identical $p_0$ values of $p_{0,j} = 10^{-9}$ and if they
are statistically independent, we can approximately compute the probability
of the background faking a signal in at least one of them as
\begin{alignat}{5}
p_0^\text{(global)} 
= \sum_{j=1}^{10} \; p_{0,j} 
= 10 \times 10^{-9} 
= 10^{-8} \; .
\end{alignat}
This means that for a global 5-sigma discovery with $p_0^\text{global}
< 5.8 \times 10^{-7}$ we need to require a significantly smaller
values for the combination of the $p_0^\text{local}$ for a given Higgs
mass.  In the above approximation the reduction is simply an effect of
independent Poisson processes, its proper treatment is significantly more
complicated. It is called \underline{look-elsewhere effect}, where some people correctly point out that the more
appropriate name would be look-everywhere effect.  Obviously, if we
combine the $\gamma \gamma$ and $ZZ$ analyses with the flat $p_0$
value of the $WW$ search the result is not as easy
anymore. The global value
$p_0^\text{(global)}$ which ATLAS quotes corresponds to
$5.1\sigma$ for an initial Higgs mass range of 110~GeV to 600~GeV. The
remaining discussion of the Higgs excess in the ATLAS paper we
postpone to Section~\ref{sec:higgs_couplings}.\bigskip

Essentially the same details we can find in the CMS discovery
paper~\cite{cms_discovery}. The observed local significance in the
three main channels is $5.1\sigma$, with an expected $5.2\sigma$. The
additional decay channels $H \to \tau\tau$ and $H \to b\bar{b}$ do not
contribute, either by bad luck or expectedly. The main psychological
difference between ATLAS and CMS seems to be that by the ordering of
the references ATLAS starts with the well established Standard Model,
of which the Higgs mechanism is a generic part which one would not
mind discovering --- while CMS starts with the references to the
prediction of the Higgs boson.

\subsection{Higgs production in gluon fusion}
\label{sec:higgs_gf}

\begin{figure}[t]
\begin{center}
  \includegraphics[width=0.48\hsize]{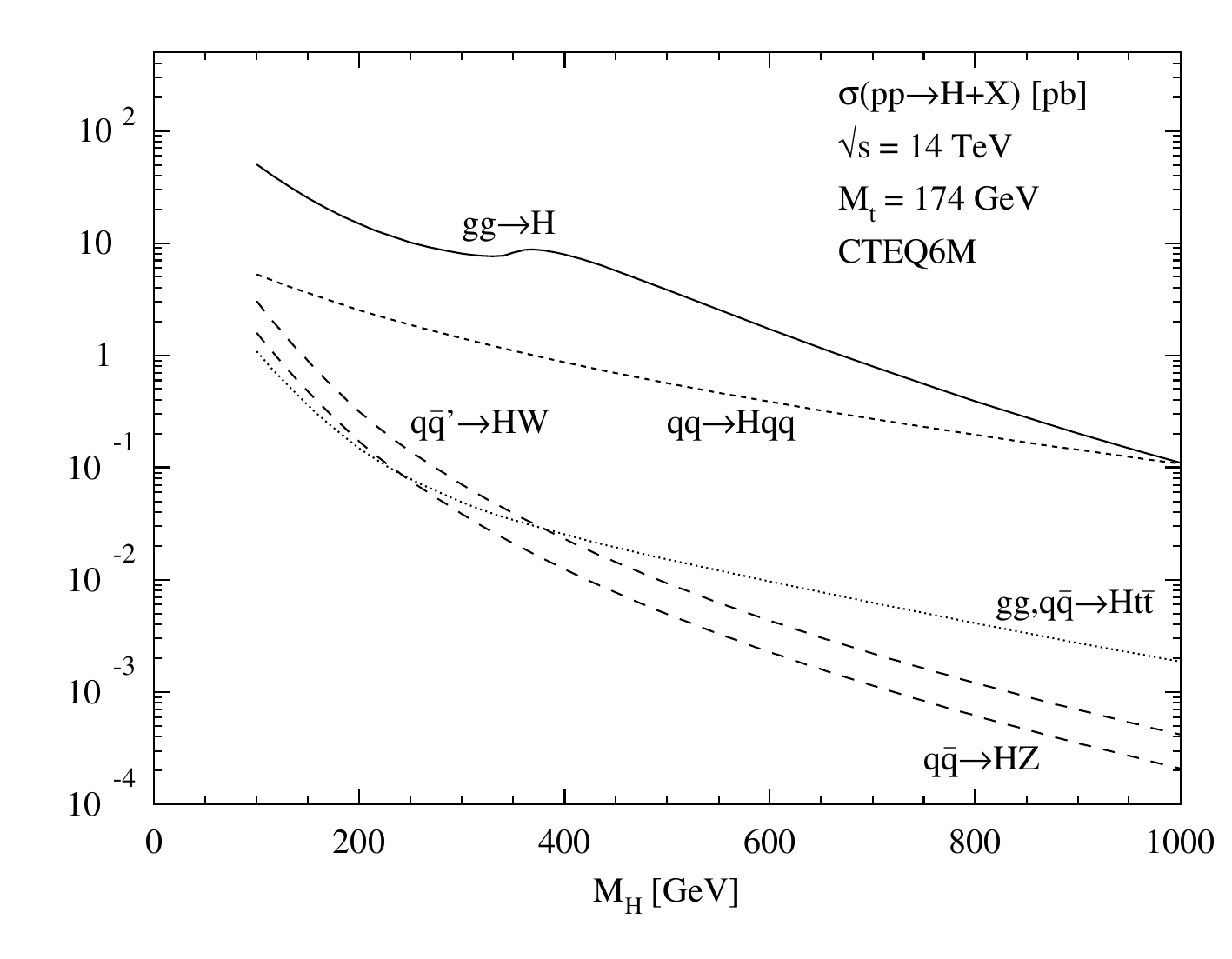}
  \hspace*{0.1\textwidth}
  \raisebox{2.5mm}{\includegraphics[width=0.375\hsize]{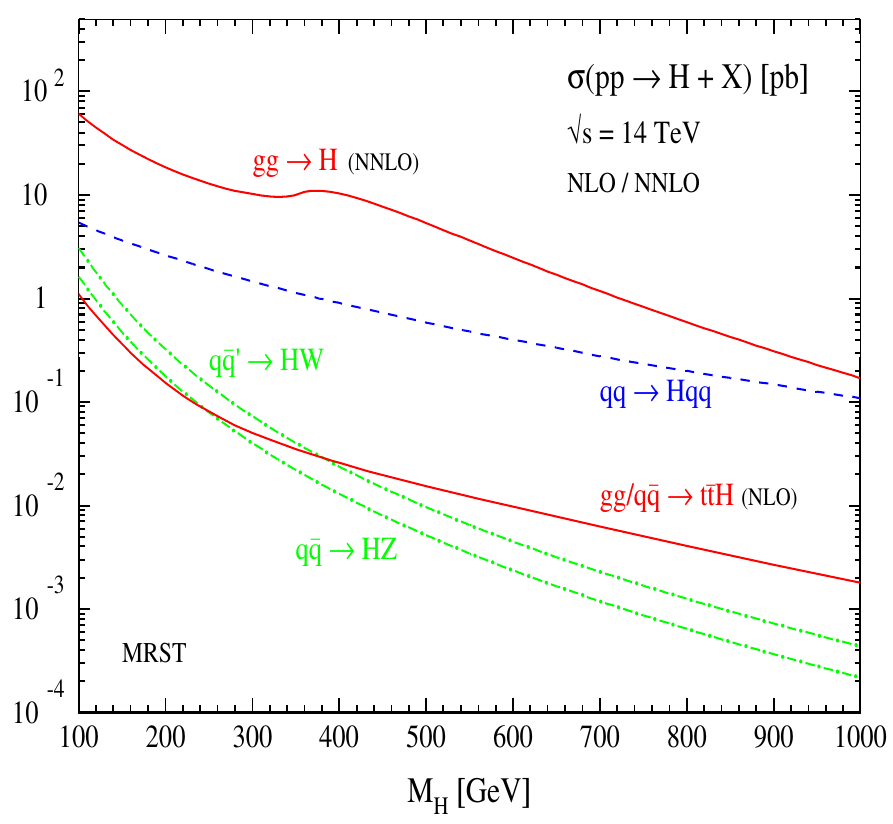}}
\end{center}
\caption{Left: production cross section for a Standard-Model Higgs
  boson at the LHC\index{Higgs boson!LHC cross sections}, as a
  function of the Higgs mass. Figure from
  Ref.~\cite{Spira:1997dg}. Right: updated version including higher
  order corrections.}
\label{fig:higgs_rates}
\end{figure}

After discussing the main aspects of the Higgs discovery we now go
back to some theoretical physics background.
Looking for the Higgs boson at hadron colliders starts with bad news:
at tree level the Higgs hardly couples to light-flavor quarks and has
no coupling to gluons. This is because the Higgs boson couples to all
Standard Model particles proportional to their mass --- this is the
same operator they get their mass from. Because the $SU(3)_C$ symmetry
of QCD is not broken, there is no coupling to gluons at all. 

On the other hand, the protons at the LHC contain a lot of gluons,
again something we will talk about in more detail in
Section~\ref{sec:qcd}, so the question is if we can find and use a
loop--induced coupling of two gluons to the Higgs. In spite of the
expected suppression of the corresponding cross section by a one-loop
factor $(g^2/(16 \pi^2))^2$ we would hope to arrive at an observable
production cross section $pp \to H$.  Numerically, it will turn out
that the production of Higgs bosons in gluon fusion is actually the
dominant process at the LHC, as shown in Figure~\ref{fig:higgs_rates}.

\subsubsection{Effective gluon--Higgs coupling}
\label{sec:higgs_gluon}

If an effective $ggH$ coupling should be mediated by a closed Standard
Model \underline{particle loop}\index{Higgs coupling!loop--induced} the top is the perfect candidate: on
the one hand it has a strong coupling to gluons, and on the other hand
it has the largest of all Standard Model couplings to the Higgs boson,
$m_t/v \sim 0.7$. The corresponding Feynman diagram is

\begin{center}
\begin{fmfgraph*}(90,60)
 \fmfset{arrow_len}{2mm}
 \fmfleft{in1,in2}
 \fmf{gluon,width=0.5}{in1,v1}
 \fmf{gluon,width=0.5}{in2,v2}
 \fmf{fermion,width=0.5,lab.side=right,label=$q$,tension=0.5}{v1,v3}
 \fmf{fermion,width=0.5,tension=0.5}{v3,v2}
 \fmf{fermion,width=0.5,tension=0.3}{v2,v1}
 \fmf{dashes,width=0.5,lab.side=right,label=$p$}{v3,out1}
 \fmfright{out1}
 \fmflabel{$k_2$}{in2}
 \fmflabel{$k_1$}{in1}
\end{fmfgraph*}
\end{center}
We construct this effective coupling in three steps, starting with the
Dirac trace occurring the top loop. All momenta are defined as
incoming with $k_1^2=k_2^2=0$ and $p^2=m_H^2$. The Dirac indices of
the two gluons are $\mu$, $\nu$ and the loop momentum is $q$, so in
the first step we need to compute
\begin{alignat}{5}
T^{\mu\nu}=
\tr \left[ (\slashchar{q}+m_t) \, \gamma^\mu \,
          (\slashchar{q}+\slashchar{k}_1 +m_t) \, \gamma^\nu \,
          (\slashchar{q}+\slashchar{k}_1+\slashchar{k}_2+m_t) 
    \right] \; .
\label{eq:higgs_tensor}
\end{alignat}
The calculational problem is the tensor structure of this
trace. Because of gauge invariance we can neglect terms proportional
to $k_1^\mu$ and $k_2^\nu$; they would not survive the
multiplication with the transverse gluon polarization
$(k\cdot\epsilon=0)$. In a so-called axial gauge\index{axial gauge} we could also get rid
of the remaining terms proportional to $k_1^\nu$ and
$k_2^\mu$.\bigskip

However, there is a better way to compute this trace. We know that
there is no tree level Higgs coupling of the Higgs to two gluons,
which would correspond to the fields $H A_\mu A^\mu$ with mass
dimension three in the Lagrangian. So we need to find another operator
mediating such a coupling, keeping in mind that it is loop induced and
can therefore include a mass suppression by powers of the top
mass. The Higgs can also couple to the \underline{field
  strength}\index{QCD field strength} in the invariant form $H G_{\mu
  \nu} G^{\mu \nu}$ with $G_{\mu \nu} \equiv \p_\mu A_\nu - \p_\nu
A_\mu + \ope(A^2)$. This operator has mass dimension five and arises from the
dimension-6 gauge--invariant object $\phi^\dag \phi \, G_{\mu \nu}
G^{\mu \nu}$ after breaking $SU(2)_L$.

The factor in front of this term is the effective coupling we are
going to compute in this section. Before that it pays to briefly look
at the operator itself. Switching from position space and its momentum
operator to momentum space $\p \to i k$ shows that the gauge
invariant operator linking exactly two gluon fields to a Higgs field
has to be proportional the tensor
\begin{alignat}{5}
G^{\mu\nu}G_{\mu\nu} \; \stackrel{\text{F.T.}}{\longrightarrow} & \;
i \left( {k_1}_\mu {A_1}_\nu  - {k_1}_\nu {A_1}_\mu \right)
 i \left( {k_2}_\mu {A_2}_\nu  - {k_2}_\nu {A_2}_\mu \right)
  + \ope(A^3)  \notag\\
=& \;
- 2 \left[ (k_1 k_2) (A_1 A_2)
        -(k_1 A_2)(k_2 A_1) \right]
  + \ope(A^3)  \notag\\
=& \;
- 2 (k_1 k_2) \, A_{1 \mu} A_{2 \nu}
  \left[ g^{\mu \nu} - \frac{k_1^\nu k_2^\mu}{k_1 k_2} \right]
  + \ope(A^3)  \notag\\
=& \;
- \sqrt{2} m_H^2 \, A_{1 \mu} A_{2 \nu}
  \; P_T^{\mu \nu} 
  + \ope(A^3) \; ,
\label{eq:higgs_op}
\end{alignat}
where $P_T^{\mu \nu}$ is the transverse
tensor\index{transverse tensor}
\begin{alignat}{5}
P_T^{\mu \nu} &= \frac{1}{\sqrt{2}} \; 
               \left[ g^{\mu \nu} - \frac{k_1^\nu k_2^\mu}{(k_1 k_2)}
               \right] \notag \\ 
P_T^{\mu \nu} P_{T \mu \nu} &= 1 
\qquad \text{and} \quad
P_T^{\mu \nu} k_{1 \mu} = 0 = P_T^{\mu \nu} k_{2 \nu} 
\qqquad ( k_1^2 = 0 = k_2^2 ) \; . 
\label{eq:higgs_transverse}
\end{alignat}
Based on this known tensor structure of $T^{\mu \nu}$ we can extract
the scalar \underline{form factor}\index{Higgs coupling!form factor}
$F$ which corresponds to the Dirac trace of
Eq.\eqref{eq:higgs_tensor} 
\begin{alignat}{5}
T^{\mu \nu} \sim F \; P_T^{\mu \nu}
 \qquad \Leftrightarrow \qquad
P_{T \mu \nu} T^{\mu \nu} \sim P_{T \mu \nu} P_T^{\mu \nu} \; F = F \; .
\end{alignat}
The exact definition of the full form factor $F$ in the Higgs--gluon
coupling will obviously include all prefactors and the loop integral.
This way we project out the relevant gluon tensor structure or the
relevant degrees of freedom of the two gluons contributing to this
effective coupling. Terms involving a larger number of gluon fields
are related to this $ggH$ coupling by non--abelian $SU(3)$ gauge
invariance.\bigskip

Our projection requires that we first compute $P_{T \mu \nu} T^{\mu
  \nu}$ based on Eq.\eqref{eq:higgs_tensor}. One thing to mention at
this stage is that nobody in the world really computes Dirac traces by
hand anymore. There are powerful programs, like FORM\index{FORM},
which do this job for us. Using it we find the form factor
\begin{alignat}{5}
P_{T \mu \nu} T^{\mu \nu} = \frac{4 m_t}{\sqrt{2}} \; 
\left(  - m_H^2 + 3 m_t^2 
        - \frac{8}{m_H^2} (k_1 q) (k_2 q) 
        - 2 (k_1 q) 
        + q^2
\right) \; .
\label{eq:numerator}
\end{alignat}
\bigskip

Inside this trace there appears the loop momentum $q$, which in our
second step we have to consider as part of the loop integration. The
effective $ggH$ vertex includes the loop integral with the tensor
structure from Eq.\eqref{eq:higgs_tensor} in the numerator,
\begin{alignat}{5}
&\int\frac{d^4q}{16 \pi^4}
\frac{P_{T \mu \nu} T^{\mu \nu}}{[q^2-m_t^2][(q+k_1)^2-m_t^2][(q+k_1+k_2)^2-m_t^2]} 
\notag \\
&=\frac{4 m_t}{\sqrt{2}} \; 
\int\frac{d^4q}{16 \pi^4} \;
\frac{q^2 - 2 (k_1 q)         
     - 8/m_H^2 (k_1 q) (k_2 q) 
     - m_H^2 + 3 m_t^2}{[q^2-m_t^2][(q+k_1)^2-m_t^2][(q+k_1+k_2)^2-m_t^2]} 
\; .
\label{eq:tensorint}
\end{alignat}
The non--trivial $q^\mu$ dependence of the numerator observed in
Eq.\eqref{eq:tensorint} we can take care of using a few tricks. For
example in the first term we use the relation $q^2/(q^2-m_t^2) = 1 +
m_t^2/(q^2-m_t^2)$ and then shift $q$, knowing that the final result
for this integral will be finite. This is non--trivial piece of
information, because most loop calculations lead to ultraviolet
divergences, which need to be removed by first regularizing the
integral and then renormalizing the parameters. The reason why we do
not see any divergences in this process is that for a renormalization
we would need something to renormalize, \ie a leading order process
which receives quantum corrections. However, we only compute this
one-loop amplitude because there is no tree level vertex. There is nothing
to renormalize, which means there are no ultraviolet divergences.

While these tricks help a little, we still do not know how to remove
$(k_1 q) (k_2 q)$ in the third term of
Eq.\eqref{eq:tensorint}. The method of choice is a
\underline{Passarino-Veltman reduction} which turns tensor integrals
into scalar integrals, where a scalar integral does not have powers of
the loop momentum in the numerator. For example, the scalar
three-point function is given by
\begin{alignat}{5}
C(k_1^2,k_2^2,m_H^2;m_t,m_t,m_t) \equiv  
\int\frac{d^4q}{i\pi^2}
\frac{1}{[q^2-m_t^2][(q+k_1)^2-m_t^2][(q+k_1+k_2)^2-m_t^2]} \; .
\label{eq:higgs_threepoint}
\end{alignat}
The integral measure is $d^4q/(i \pi^2)$, or $d^nq/(i
\pi^{n/2})$ for an arbitrary number of space--time dimensions. This
removes any over--all factors $2$ and $\pi$ from the expression for the
scalar integrals, as we will see below.  Applied to the tensor
integral in Eq.\eqref{eq:tensorint} the reduction algorithm gives us
\begin{alignat}{5}
\int\frac{d^4q}{i\pi^2}
    \frac{P_{T \mu \nu} T^{\mu \nu}}{[...][...][...]} 
= \frac{4 m_t}{\sqrt{2}}
  \left[ 2 + \left(  4 m_t^2 - m_H^2 \right) \, C(0,0,m_H^2;m_t,m_t,m_t)
  \right] \; .
\label{eq:formfac1}
\end{alignat}
The first term not proportional to any scalar integral has a curious
origin. It comes from a combination of $\ope(\epsilon)$ terms from the
Dirac trace in $n=4-2\epsilon$ dimensions and a two-point function
which using the integration measure $1/i \pi^{n/2}$ always includes an
ultraviolet divergence $1/\epsilon$. Note that these terms appear in
the calculation in spite of the fact that the final result for the
effective gluon--Higgs coupling is finite.\bigskip

Scalar integrals we can for example calculate using the Feynman
parameterization
\begin{alignat}{5}
\frac{1}{A_1 A_2 \cdots A_n} =
   \int_0^1 \, dx_1 \cdots dx_n \, 
   \delta \left( \sum x_i -1 \right)
   \frac{(n-1)!}{(x_1 A_1 + x_2 A_2 + \cdots + x_n A_n)^n} \; ,
\end{alignat}
but we usually obtain shorter analytical expressions using the
\underline{Cuskosky cutting rule}\index{Cuskosky cut rules} which links the imaginary part of a
diagram or a scalar integral to the sum of all cut Feynman
graphs.\bigskip

The cut rule is directly related to the unitarity of the $S$ matrix
and the optical theorem discussed in
Section~\ref{sec:higgs_unitarity}.  Limiting ourselves to scalar
amplitudes/integrals, the cut rule tells us that the sum of all cut
one-loop or squared scalar diagrams has to vanish, including the two
external cuts which correspond to the untouched amplitude $A$ and its
complex conjugate $A^*$. This gives us a very useful expression for
the imaginary part of the amplitude
\begin{alignat}{5}
- i \left( A - A^* \right) = 2 \, \text{Im} A
\really 16 \pi^2 \sum_\text{cut graphs} A \; .
\end{alignat}
The factor $16 \pi^2$ arises from the generic integral measure $1/(16
\pi^4)$ which we replace by $d^4 q/(i \pi^2)$ such that the scalar
integrals have a typical prefactor of one. Cutting diagrams means
replacing all internal propagators by $1/(q^2 - m^2) \to 2 \pi \,
\theta(q_0) \, \delta(q^2-m^2)$\index{propagator!cutting}. Of the four
dimensions of the loop integral $d^4q$ the two on--shell conditions
cancel two, leaving us with an simple angular integral. This angular
integral does not include any kinematic information affecting the pole
or cut structure of the scalar diagram.

From this imaginary part we compute the complete amplitude or scalar
integral.  If we know the pole or cut structure of the amplitude after
cutting it, we can make use of the \underline{Cauchy
  integral}\index{Cauchy integral} for a complex analytic function
$A(z)$
\begin{alignat}{5}
A(z) &= \frac{1}{2 \pi i} \oint_\text{counter--clockwise} d z' \; \frac{A(z')}{z' - z}  \; ,
\label{eq:higgs_cauchy}
\end{alignat}
and compute the unknown real part of $A(q^2)$. As an example, let us
consider a scalar integral which like a full propagator has a cut on
the real axis above $q^2 = m_1^2 + m_2^2$. This cut should not lie
inside the integration contour of the Cauchy integral
Eq.\eqref{eq:higgs_cauchy}, so we deform the circle to instead follow
the cut right above and below the real axis. If no other poles occur
in the integral we find
\begin{alignat}{5}
\text{Re} A(q^2) &= \frac{1}{2 \pi i} \oint d{q'}^2 \; \frac{i \; \text{Im} A({q'}^2)}{{q'}^2 - q^2} 
 \notag \\
       &= 
          \frac{1}{2 \pi} \int_\infty^{(m_1^2+m_2^2)} d{q'}^2 \;
                           \frac{\text{Im} A({q'}^2-i \epsilon)}{{q'}^2 - q^2} 
        + \frac{1}{2 \pi} \int_{(m_1^2+m_2^2)}^\infty d{q'}^2 \;
                           \frac{\text{Im} A({q'}^2+i \epsilon)}{{q'}^2 - q^2} 
 \notag \\
       &= \frac{1}{2\pi} \int_{(m_1^2+m_2^2)}^\infty d{q'}^2 \;
                            \frac{\text{Im} A({q'}^2+i\epsilon) - \text{Im} A({q'}^2-i\epsilon)}{{q'}^2 - q^2} 
 \notag \\
       &\equiv \frac{1}{2\pi} \int_{(m_1^2+m_2^2)}^\infty d{q'}^2 \;
                            \frac{\text{Im}_+ A({q'}^2) }{{q'}^2 - q^2} \; .
\end{alignat}
This step assumes a sufficiently fast convergence on the integration
contour for large momenta. This method of
computing for example scalar integrals is known to produce the most
compact results.\bigskip

The expression for the finite scalar three point function appearing in our effective
coupling Eq.\eqref{eq:formfac1} has the form
\begin{alignat}{5}
C(0,0,m_H^2;m_t,m_t,m_t) 
=& \frac{1}{m_H^2} 
   \int_0^1\frac{dx}{x} \; \log \frac{m_H^2x(1-x)-m_t^2}{(-m_t^2)}
    \notag\\
=& \frac{1}{m_H^2} 
   \int_0^1\frac{dx}{x} \; \log\left(1 - x (1-x) \frac{m_H^2}{m_t^2}\right)
    \notag\\
=& \frac{1}{2m_H^2} \; \log^2 \left( -\frac{1+\sqrt{1-4 m_t^2/m_H^2}}
                                  {1-\sqrt{1-4 m_t^2/m_H^2}}
                          \right)
 \qqquad \frac{4m_t^2}{m_H^2} \equiv \tau <1 \; .
\end{alignat}
For general top and Higgs masses it reads
\begin{alignat}{5}
C(0,0,m_H^2;m_t,m_t,m_t) =
- \frac{2 f(\tau)}{m_H^2} 
\qquad \text{with} \qquad
f(\tau)= \left\{\begin{array}{lr}
                 \left( \arcsin \sqrt{\dfrac{1}{\tau}}\right)^2
                &\tau>1 \\
                 -\dfrac{1}{4} \left( \log \dfrac{1+\sqrt{1-\tau}}
                                                 {1-\sqrt{1-\tau}}
                 - i\pi\right)^2 \quad
                &\tau<1
                 \end{array} \right. \; ,
\label{eq:threepoint}
\end{alignat}
including imaginary or absorptive terms for $\tau<1$. The
dimensionless variable $\tau$ is the appropriate parameter to describe
the behavior of this scalar integral. For example the low energy limit
of the scalar integral, \ie the limit in which the top loop becomes
heavy and cannot be resolved by the external energy of the order of
the Higgs mass, will be given by $\tau \gtrsim 1$ which means $m_H < 2
m_t$.  In contrast to what many people who use such effective vertices
assume, the expression in Eq.\eqref{eq:threepoint} is valid
for \underline{arbitrary Higgs and top masses}, not just in the
heavy top limit.

Expressing our Dirac trace and loop integral in terms of this
function $f(\tau)$ we find for our effective coupling in
Eq.\eqref{eq:formfac1}
\begin{alignat}{5}
\int\frac{d^4q}{i\pi^2}
    \frac{P_{T \mu \nu} T^{\mu \nu}}{[...][...][...]} 
&= \frac{4 m_t}{\sqrt{2}}
  \left( 2 - \left(  4 m_t^2 - m_H^2 \right) \frac{2 f(\tau)}{m_H^2}
  \right) \notag \\
&= \frac{4 m_t}{\sqrt{2}}
  \left( 2 - 2 \left( \tau - 1 \right) f(\tau)
  \right) \notag \\
&= \frac{8 m_t}{\sqrt{2}}
  \left( 1 + ( 1- \tau ) f(\tau) \right) \; .
\label{eq:formfac2}
\end{alignat}
\bigskip

Using this result we can as the third and last step of our calculation
collect all factors from the propagators and couplings in our Feynman
diagram and compute the \underline{effective $ggH$ coupling} 
now including all pre-factors,
\begin{alignat}{5}
F 
=& - \; i^3 \; (-ig_s)^2 \; \frac{im_t}{v} \; \tr(T^aT^b) \;
 \frac{i\pi^2}{16\pi^4} \; 
 \int\frac{d^4q}{i\pi^2}
    \frac{P_{T \mu \nu} T^{\mu \nu}}{[...][...][...]} 
 \notag \\
=& - \; i^3 \; (-ig_s)^2 \; \frac{im_t}{v} \; \tr(T^aT^b) \;
 \frac{i\pi^2}{16\pi^4} \; 
 \frac{8 m_t}{\sqrt{2}} \left( 1 + ( 1- \tau ) f(\tau) \right)
 \notag \\
=& \; \frac{g_s^2 m_t}{v} \; \frac{\delta^{ab}}{2} \;
 \frac{i}{16 \pi^2} \;
 \frac{8 m_t}{\sqrt{2}} \left( 1 + ( 1- \tau ) f(\tau) \right)
 \notag \\
=& \; \frac{g_s^2}{v} \; \frac{\delta^{ab}}{2} \;
 \frac{i}{16 \pi^2} \;
 \frac{8}{\sqrt{2}} \frac{m_H^2 \tau}{4} \left( 1 + ( 1- \tau ) f(\tau) \right)
 \notag \\
=& \; i g_s^2 \; \delta^{ab} \;
 \frac{1}{16 \sqrt{2} \pi^2} \;
 \frac{m_H^2}{v} \; \tau \; 
 \left( 1 + ( 1- \tau ) f(\tau) \right)
 \notag \\
=& \; i \alpha_s \; \delta^{ab} \;
 \frac{1}{4 \sqrt{2} \pi} \;
 \frac{m_H^2}{v} \; \tau \; 
 \left( 1 + ( 1- \tau ) f(\tau) \right) \; .
\label{eq:effcoup_factors}
\end{alignat}
The numerical factors originate from the closed fermion loop,
the three top propagators, the two top-gluon couplings, the top Yukawa
coupling, the color trace, the unmatched loop integration measure,
and finally the result computed in Eq.\eqref{eq:formfac2}.\bigskip

Based on Eq.\eqref{eq:higgs_op} we can write in momentum space as well
as in position space
\begin{alignat}{5}
F \;  P_T^{\mu \nu} A_{1 \mu} A_{2 \nu} 
&= F \; \frac{- G^{\mu\nu}G_{\mu\nu}}{\sqrt{2} m_H^2} \; .
\end{alignat}
In this form
we can include the form factor $F$ in an effective Lagrangian
and finally define the Feynman rule we are interested in
\begin{alignat}{5}
\boxed{
\lag_{ggH} \supset \; \frac{1}{v} \;
                          g_{ggH} \; H \, G^{\mu\nu}G_{\mu\nu} 
} 
\qqquad \text{with} \qquad
\frac{1}{v} \, g_{ggH} 
 = -i \; \frac{\alpha_s}{8 \pi} \; 
   \frac{1}{v} \; \tau \left[1+(1-\tau)f(\tau)\right] \; ,
\label{eq:higgs_eff1}
\end{alignat}
after dropping $\delta^{ab}$. It is important to notice that the
necessary factor in front of the dimension-5 operator is $1/v$ and not
$1/m_t$. This is a particular feature of this coupling, which does not
decouple for heavy top quarks because we have included the top Yukawa
coupling in the numerator. Without this Yukawa coupling, the heavy top
limit $\tau \to \infty$ of the expression would be zero, as we will
see in a minute. Unlike one might expect from a general effective
theory point of view, the higher dimensional operator inducing the
Higgs--gluon coupling is not suppressed by a large energy scale. This
means that for example a fourth generation of heavy fermions will
contribute to the effective Higgs--gluon coupling as strongly as the
top quark, with no additional suppression by the heavy new masses. The
breaking of the usual decoupling by a large Yukawa coupling makes it
easy to experimentally rule out such an additional generation of
fermions, based on the Higgs production rate.\bigskip

Of course, just like we have three-gluon and four-gluon couplings in
QCD we can compute the $gggH$ and the $ggggH$ couplings from the $ggH$
coupling simply using gauge invariance defining the terms we omit in
Eq.\eqref{eq:higgs_op}. This set of $n$-gluon couplings to the Higgs
boson is again not an approximate result in the top mass. Gauge
invariance completely fixes the $n$-gluon coupling to the Higgs via
one exact dimension-5 operator in the Lagrangian. These additional
gluon field arise from the commutator of two gluon field in the field
strength tensor, so they only exist in non--abelian QCD and cannot be
generalized to the photon-photon-Higgs coupling. 

\subsubsection{Low--energy theorem}
\label{sec:higgs_low}

The general expression for $g_{ggH}$ is not particularly handy, but for
light Higgs bosons we can write it in a more compact form. We start
with a Taylor series for $f(\tau)$ in the \underline{heavy-top limit}
$\tau\gg 1$
\begin{alignat}{5}
f(\tau) = \left[ \arcsin \frac{1}{\tau^{1/2}} \right]^2
        = \left[ \frac{1}{\tau^{1/2}} 
                +\frac{1}{6 \tau^{3/2}}
                +\ope \left( \frac{1}{\tau^{5/2}} \right) 
          \right]^2
        = \frac{1}{\tau}
         +\frac{1}{3\tau^2}
         +\ope \left( \frac{1}{\tau^3} \right) 
\stackrel{\tau \to \infty}{\longrightarrow} 0 \; ,
\label{eq:higgs_f_limit}
\end{alignat}
and combine it with all other $\tau$-dependent terms 
from Eq.\eqref{eq:higgs_eff1}
\begin{alignat}{5}
\tau \, \left[1+(1-\tau)f(\tau) \right] 
=& \tau \, \left[1+(1-\tau) \left( \frac{1}{\tau}
                                  +\frac{1}{3\tau^2}
                                  +\ope \left( \frac{1}{\tau^3} \right) \right)
           \right] \notag\\
=& \tau \, \left[ 1 + \frac{1}{\tau} 
                 -1 - \frac{1}{3\tau} + \ope \left( \frac{1}{\tau^2} \right) \right] \notag\\
=& \tau \, \left[ \frac{2}{3\tau} +\ope \left( \frac{1}{\tau^2} \right) \right] \notag\\
=& \frac{2}{3} + \ope \left( \frac{1}{\tau} \right) \; ,
\qqquad \text{implying} \qquad
\boxed{ g_{ggH} = -i \; \frac{\alpha_s}{12 \pi} } \; .
\end{alignat}
In this low energy or heavy top limit we have \underline{decoupled the
  top quark} from the set of propagating Standard Model particles. The
$ggH$ coupling does not depend on $m_t$ anymore and gives a finite
result. Computing this finite result in Eq.\eqref{eq:effcoup_factors}
we had to include the top Yukawa coupling from the numerator.  We
emphasize again that while this low energy approximation is very
compact to analytically write down the effective $ggH$ coupling, it is
not necessary to numerically compute processes involving the effective
$ggH$ coupling. \bigskip

In this low energy limit we can easily add more Higgs bosons to the
loop.  Attaching an external Higgs leg to the gluon self energy
diagram simply means replacing one of the two top propagators with two
top propagators and adding a Yukawa coupling
\begin{alignat}{5}
\frac{i}{\slashchar{q}-m_t} \; \to \; 
\frac{i}{\slashchar{q}-m_t} \; \frac{-i\sqrt{2}m_t}{v} \; \frac{i}{\slashchar{q}-m_t} \; 
\end{alignat}
We can compare this replacement to a differentiation with respect to $m_t$
\begin{alignat}{5}
\frac{\p}{\p m_t} \; \frac{1}{\slashchar{q}-m_t}
&= \frac{\p}{\p m_t} \; \frac{\slashchar{q}+m_t}{q^2-m_t^2} \notag \\
&= \frac{(q^2-m_t^2) - (\slashchar{q}+m_t)(-2m_t)}{(q^2-m_t^2)^2} \notag \\
&= \frac{q^2 + 2 m_t \slashchar{q} + m_t^2}{(q^2-m_t^2)^2} 
 = \frac{\slashchar{q}^2 + 2 m_t \slashchar{q} + m_t^2}{(q^2-m_t^2)^2} 
 = \frac{(\slashchar{q} + m_t)(\slashchar{q} + m_t)}{(q^2-m_t^2)^2} \notag \\
\Rightarrow \qquad 
\frac{i}{\slashchar{q}-m_t} \; &\to \; 
\frac{-i \sqrt{2} m_t}{v} \;
\frac{1}{\slashchar{q}-m_t} \; \frac{1}{\slashchar{q}-m_t} 
= 
\frac{-i \sqrt{2} m_t}{v} \; \frac{\p}{\p m_t} \; \frac{1}{\slashchar{q}-m_t} \; .
\label{eq:higgs_lowenergy}
\end{alignat}
This means that we can replace one propagator by two propagators using
a derivative with respect to the heavy propagator mass.  The correct
treatment including the gamma matrices in $\slashchar{q} = \gamma_\mu
q^\mu$ involves carefully adding unit matrices in this slightly
schematic derivation.  However, our shorthand notation gives us an
idea how we can in the limit of a heavy top derive the $ggH^{n+1}$
couplings from the $ggH^n$ coupling
\begin{alignat}{5}
g_{ggH^{n+1}} = m_t^{n+1} \; 
              \frac{\p}{\p m_t} \left( \frac{1}{m_t^n}  g_{ggH^n} \right) \; .
\end{alignat}
This relation holds for the scattering amplitude before squaring and
assuming that the tensor structure is given by the same transverse
tensor in Eq.\eqref{eq:higgs_transverse}.  We can for example use this
relation to link the two effective couplings with one or two external
Higgs legs, \ie the triangle form factor $g_{ggH} = -i \alpha_s/(12 \pi)$ and the box form factor
$g_{ggHH} = + i \alpha_s/(12 \pi)$. Corrections to this relation
appear at the order $1/m_t$. 

The question arises if we can even link
this form factor to the gluon self energy without any Higgs
coupling. The relation in Eq.\eqref{eq:higgs_lowenergy} suggests that
there should not be any problem as long as we keep the momenta of the
incoming and outgoing gluon different. In Section~\ref{sec:qcd_run_coup}
we will see that the gluon self energy loops have a transverse tensor
structure, so there is no reason not to use the top loop in the gluon
self energy to start the series of $ggH^n$ couplings in the heavy top
limit. Note that this does not include the entire gluon self energy
diagram, but only the top loop contributing. The so-called beta
function does not appear in this effective coupling. If we want to
combine the effects of more than just one particle in the gluon self
energy we need to integrate out one state after the other. The
appropriate effective theory framework then becomes the
Coleman--Weinberg potential discussed in
Section~\ref{sec:higgs_coleman}.\bigskip

To obtain the correct mass dimension each external Higgs field appears
as $H/v$.  Using $\log (1+x) = - \sum_{n=1} (-x)^n/n$ we can
eventually resum this series of effective couplings in the form
\begin{alignat}{5}
\boxed{
\lag_{ggH} =
G^{\mu\nu}G_{\mu\nu} \; \frac{\alpha_s}{\pi} \;
\left( \frac{H}{12v}-\frac{H^2}{24v^2} + \ldots \right)
=\frac{\alpha_s}{12\pi} \; G^{\mu\nu}G_{\mu\nu} \;
 \log \left(1+\frac{H}{v} \right)
} \; .
\label{eq:higgs_eff2}
\end{alignat}

In Section~\ref{sec:higgs_unitarity} we note that there should be a
relative factor between the Lagrangian and the Feynman rule accounting
for more than one way to identify the external legs of the Feynman
rule with the fields in the Lagrangian. For $n$ neutral Higgs fields
the effective coupling has to include an additional factor of $1/n$
which is precisely the denominator of the logarithm's Taylor
series.\bigskip

Such a closed form of the Lagrangian is very convenient for simple
calculations and gives surprisingly exact results for the $gg \to H$
production rate at the LHC, as long as the Higgs mass does not exceed
roughly twice the top mass. However, for example for $gg \to H+$jets
production its results only hold in the limit that {\sl all} jet
momenta are much smaller than $m_t$. It also becomes problematic for
example in the pair production process $gg \to HH$ close to threshold, where the
momenta of slow--moving Higgs bosons lead to an additional scale in the
process. We will come back to this process later.

\subsubsection{Effective photon-Higgs coupling}
\label{sec:higgs_gamma}

The Higgs coupling to two photons can be computed exactly the same way
as the effective coupling to gluons. Specifically, we can compute a
form factor in analogy to Eq.\eqref{eq:effcoup_factors} and
Eq.\eqref{eq:higgs_eff1}. The only difference is that there exist two
particles in the Standard Model which have a sizeable Higgs coupling
(or mass) and electric charge: heavy quarks and the $W$ boson. Both
contribute to the partial width of a Higgs decaying to two
photons.\bigskip

To allow for some more generality we give the results for the Higgs
decay width for a general set of fermions, gauge bosons, and scalars
in the loop:
\begin{alignat}{5}
\Gamma (H \to \gamma \gamma) &= 
\frac{G_F \alpha^2 m_H^3}{256 \sqrt{2} \pi^3}
\left|
  \sum_f \; N_c Q_f^2 \; g_{Hff} A_{1/2}
 + g_{HWW} A_1
 + \sum_s N_c Q_s^2 \; \frac{g_{Hss}}{m_s^2} A_0 
\right|^2 \; .
\label{eq:higgs_gamma}
\end{alignat}
The color factor for fermions without a color charge should be
replaced by unity. In the literature, the form factors $A$ are often
defined with an additional factor 2, so the prefactor will only include
$1/128$. The factor $G_F$ describes the coupling of the loop particles
to the Higgs boson, while each factor $\alpha = e^2/(4 \pi)$ arises
because of the QED coupling of the loop particles to a photon. In the
Standard Model the two leading contributions are
\begin{alignat}{5}
Q_t^2 g_{Htt} = \left(\frac{2}{3} \right)^2 \times 1 \; 
\qqqquad
g_{HWW} = 1
\qqqquad 
g_{Hss} = 0 
\end{alignat}
This means that the Standard Model Higgs couplings proportional to the
particle masses are absorbed into the form factors $A$. From the
discussion in Section~\ref{sec:higgs_low} we know that the top quark
form factor does not decouple for heavy quarks because of the Yukawa
coupling in the numerator. The same is true for the $W$ boson where
the entire $W$ mass dependence is absorbed into $G_F$.\bigskip

The form factors $A$ for scalars, fermions, and gauge bosons we can
only quote here. They all include $f(\tau)$ as defined in
Eq.\eqref{eq:threepoint}, which is the scalar three-point function and
can be computed without any further approximations.  As before, we
define $\tau = 4 m^2/m_H^2$ given the mass $m$ of the particle
running in the loop.  From Eq.\eqref{eq:higgs_f_limit} we know that in
the limit of heavy loop particles $\tau \to \infty$ the scalar
integral scales like $f(\tau) \sim 1/\tau + 1/(3 \tau^2)\cdots$, which
allows us to compute some basic properties of the different loops
contributing to the effective Higgs--photon coupling.
\begin{alignat}{5}
&A_0    &&= - \frac{\tau}{2} \left[ 1 - \tau f(\tau) \right]  
          &&\stackrel{\tau \to \infty}{\longrightarrow} - \frac{\tau}{2} + \frac{\tau}{2} + \frac{1}{6} = \frac{1}{6} \notag \\ 
&A_{1/2} &&= \tau \left[ 1 + (1 -\tau) \; f(\tau) \right] 
          &&\longrightarrow \; \tau + 1 - \tau - \frac{1}{3} = \frac{2}{3} \notag \\
&A_1    &&= - \frac{1}{2} \left[ 2 + 3 \tau + 3 (2 \tau - \tau^2 ) \; f(\tau) \right] 
         &&\longrightarrow \; -1 - \frac{3}{2} \tau - 3 + \frac{3}{2} \tau + \frac{1}{2} = - \frac{7}{2} \; .
\end{alignat}
We see that unless a sign appear in the prefactors --- and from
Eq.\eqref{eq:higgs_gamma} we know that it does not --- the top quark
and $W$ boson loops interfere destructively. This is not an effect of
the spin alone, because the scalar form factor has the same sign as
the fermion expression. In addition, we observe that for equal charges
the gauge boson will likely dominate over the fermion, which is
correct for the top quark vs $W$ boson in the Standard Model.

\subsubsection{Signatures}
\label{sec:higgs_gf_lhc}

Different Higgs production and decay processes are not only important
for the Higgs discovery, they also allow us to test many properties
of the recently discovered new particle. Many aspects of such 
measurements go beyond our knowledge of hadron collider physics and
QCD. For example
to discuss the Higgs production in gluon fusion we would
normally need to know how to deal with gluons inside the incoming
protons, how to parameterize the phase space of the Higgs decay
products, and how to kinematically distinguish interesting events
from the rest. All of this we will piece by piece introduce in
Section~\ref{sec:qcd_dy}. On the other hand, we can try to 
understand the LHC capabilities in Higgs physics already at this point.
In the following
sections on Higgs production at the LHC we will therefore limit ourselves to
some very basic phenomenological features and postpone any discussion
on how to compute these features.\bigskip

The first quantity we can compute and analyze at colliders is the
total number of events expected from a certain production process in a
given time interval. For example for our current Higgs studies the
event numbers in different Higgs production and decay channels are the
crucial input.  Such a number of events is the product of the
proton--proton LHC luminosity\index{luminosity} measured in inverse
femtobarns, the total production cross section measured in femtobarns,
and the detection efficiency measured in per-cent. In other words, a
predicted event rate it is split into a collider--specific number
describing the initial state, a process--specific number describing the
physical process, and a detector--specific efficiency for each particle
in the final state.

The latter is the easiest number to deal with: over the sensitive
region of the detector, the \underline{fiducial volume}, the detection efficiency
is a set of numbers depending on the nature of the detected particle
and its energy. This number is very good for muons, somewhere between
90\% and 100\%, and less than 1/3 for tau leptons. Other particles typically
range somewhere in between.

\begin{figure}[t]
\begin{center}
\includegraphics[width=0.45\hsize]{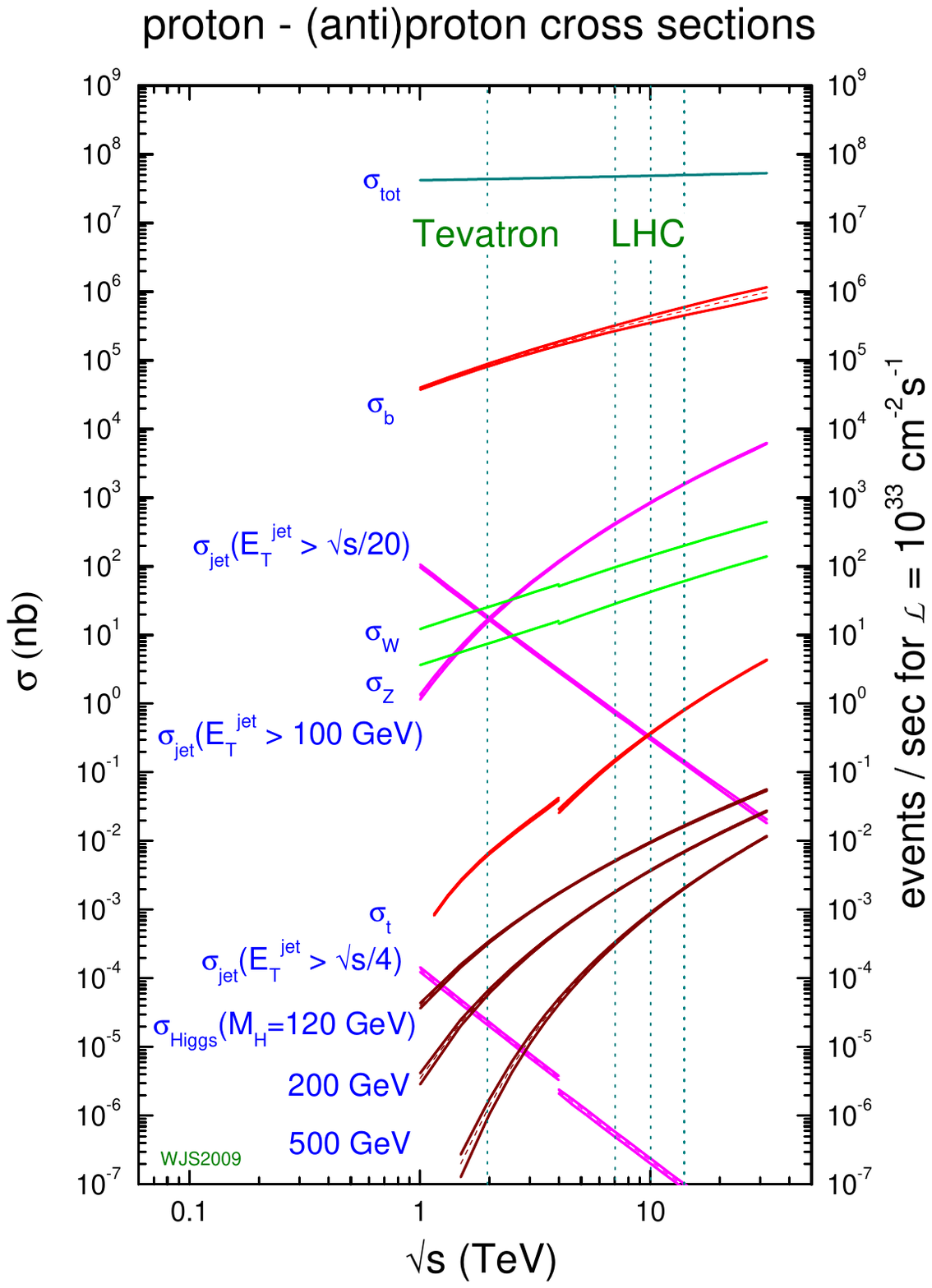}
\end{center}
\caption{Production rates for signal and background processes at
  hadron colliders\index{cross section!Tevatron and LHC processes}. The discontinuity is due to the Tevatron being a
  proton--antiproton collider while the LHC is a proton--proton
  collider. The two colliders correspond to the $x$--axis values of
  2~TeV and something between 7~TeV and 14~TeV. Figure from
  Ref.~\cite{Campbell:2006wx}.}
\label{fig:higgs_lhcall}
\end{figure}

For theorists luminosity is simply a conversion number between cross
sections which we compute for a living and event numbers. People who
build colliders use units involving seconds and square meters, but for
us inverse femtobarns work better. Typical numbers are: a year
of LHC running at design luminosity could deliver up to 10 inverse
femtobarns per year in the first few years and three to ten times that
later. 
The key numbers and their orders of magnitude for typical
signals are
\begin{alignat}{5}
N_\text{events} = \sigma_\text{tot} \cdot \lag \qquad \quad
\lag = 10 \cdots 300 \, \text{fb}^{-1} \qquad \quad
\sigma_\text{tot} = 1 \cdots 10^4 \, \text{fb} \; .
\end{alignat}
Different cross sections for Tevatron and LHC processes are shown in
Figure~\ref{fig:higgs_lhcall}.\bigskip

Finally, talking about cross sections and how to compute them we need
to remember that at the LHC there exist two kinds of processes. The
first involves all particles which we know and love, like
old-fashioned electrons or slightly more modern $W$ and $Z$ bosons or
most recently top quarks. All of these processes we call
\underline{backgrounds}. They are described by QCD, which means QCD is
the theory of evil. Top quarks have an interesting history,
because when I was a graduate student they still belonged to the
second class of processes, the \underline{signals}. These either
involve particles we have not seen before or particles we want to know
something about.  By definition, signals are very rare compared to
backgrounds.  As an example, Figure~\ref{fig:higgs_lhcall} shows that
at the LHC the production cross section for a pair of bottom quarks is
larger than $10^5$~nb or $10^{11}$~fb, the typical production rate for
$W$ or $Z$ bosons ranges around 200~nb or $2 \times 10^8$~fb, the rate
for a pair of 500~GeV supersymmetric\index{supersymmetry} gluinos would have been
$4 \times 10^4$~fb, and the Higgs rate can be as big as $2 \times
10^5$~fb.  This really rare Higgs signal was extracted by ATLAS and
CMS with a $5\sigma$ significance in the Summer of 2012.
If we see such a new particles someone gets a call from Stockholm,
while for the rest of the community the corresponding processes
instantly turn into backgrounds.\bigskip

One last aspect we have to at least mention is the
\underline{trigger}\index{trigger}. Because of the sheer mass of data
at the LHC, we will not be able to write every LHC event on tape. As a
matter of fact, we could not even write every top pair event on
tape. Instead, we have to decide very fast if an event has the
potential of being interesting in the light of the physics questions
we are asking at the LHC. Only these events we keep. Before 
a mis-understanding occurs: while experimentalists are reluctant to change
triggers these are not carved in stone, so as a functioning high
energy physics community we will not miss great new physics just
because we forgot to include it in the trigger menu. For now we can
safely assume that above an energy threshold we will keep all events
with leptons or photons, plus as much as we can events with missing
energy, like neutrinos in the Standard Model and dark matter particles
in new physics models\index{dark matter, WIMP miracle} and jets with
high energy coming from resonance decays. This trigger menu reflects
the general attitude that the LHC is not built to study QCD, and that
very soft final states for example from bottom decays are best studied
by the LHCb experiment instead of ATLAS and CMS. \bigskip

With this minimal background of collider phenomenology we can look at
Higgs production in gluon fusion, combined with different Higgs
decays. This is the production channels which dominated the Higgs
discovery discussed in Section~\ref{sec:higgs_discovery}, based on LHC
runs with a center--of--mass energy of 7~TeV (2011) and 8~TeV (2012). The
total 14~TeV Higgs production cross section through the loop--induced
$ggH$ coupling we show in Figure~\ref{fig:higgs_rates}. For a
reasonably light Higgs boson the cross section ranges around at least
30~pb, which for relevant luminosities starting around $30~\ifb$ means
$10^6$ events. The question is: after multiplying with the relevant
branching ratio and detection efficiencies, which of the decays can be
distinguished from the Standard Model background statistically? Since
gluon fusion really only produces the Higgs boson the details of the
production process do not help much with the background
suppression. The results of experimental simulations, for example by
the ATLAS collaboration, are shown in Figure~\ref{fig:higgs_sig}. The
complete list of possible Higgs decays, ordered by decreasing
branching ratio according to Figure~\ref{fig:higgs_decays}, is:

\begin{figure}[t]
\begin{center}
  \includegraphics[width=0.4\hsize]{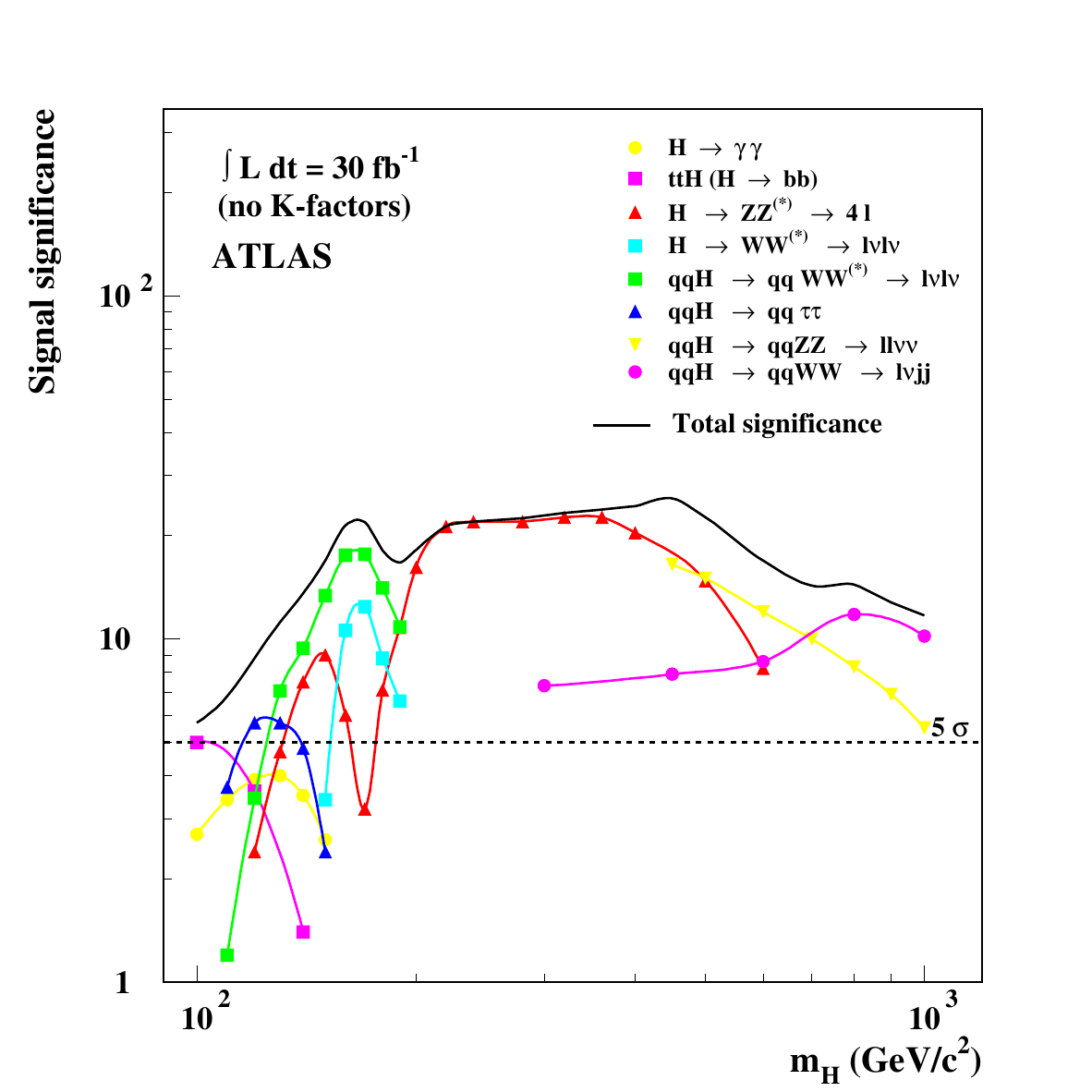}
  \hspace*{0.15\hsize}
  \raisebox{-1.8mm}{\includegraphics[width=0.405\hsize]{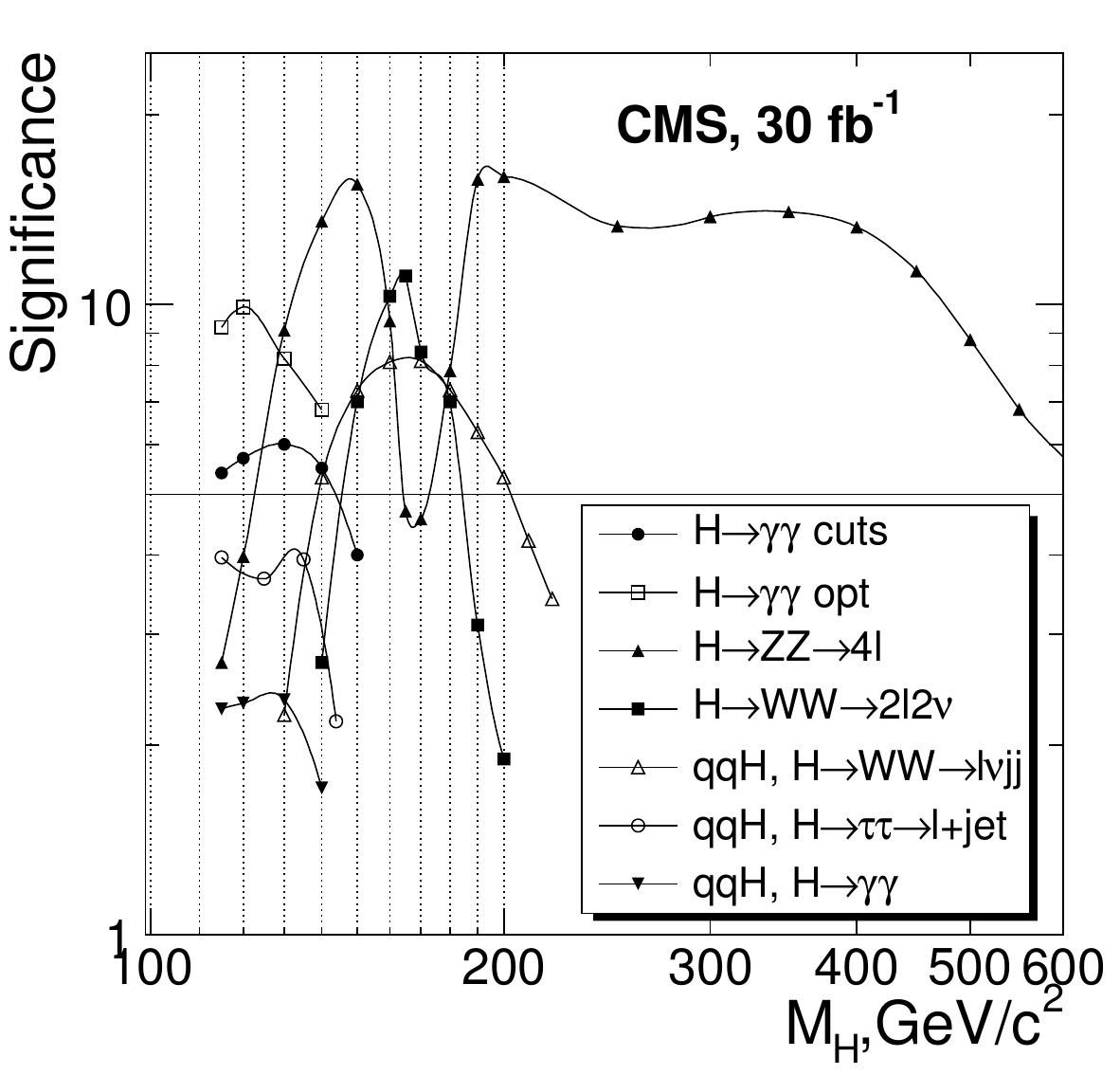}}
\end{center}
\caption{Simulated statistical significance for different Higgs
  production and decay channels for an integrated luminosity of
  $30~\text{fb}^{-1}$ (left: ATLAS~\cite{atlas_higgs}; right:
  CMS~\cite{cms_tdr}). Five standard deviations over the backgrounds
  are required for discovery. Given the measured Higgs mass around
  125~GeV these figures are of mostly historic interest, except that
  they illustrate how lucky we are that the Higgs mass lies right in
  between the fermion and bosonic decays.}
\label{fig:higgs_sig}
\end{figure}
%
\begin{itemize}
\item[--] $gg\rightarrow H\rightarrow b\bar{b}$ is hopeless, because of
  the sheer size of the QCD continuum background $gg\rightarrow b\bar{b}$,
  which according to Figure~\ref{fig:higgs_lhcall} exceeds the signal
  by roughly eight orders of magnitude. In gluon fusion there is
  little to cut on except for the invariant mass of the $b\bar{b}$
  pair with an $\ope(10\%)$ mass resolution. Such a cut will not
  reduce the background by more than two or three orders of magnitude,
  so the signal--to--background ratio will be tiny. Pile--up, \ie
  different scattering events between the proton bunches might in
  addition produce unwanted structures in the $m_{bb}$ distribution
  for the pure QCD background. The final blow might be that this
  channel will, as it stands, not be triggered on.
\item[--] $gg\rightarrow H\rightarrow \tau^+\tau^-$ is problematic.  If
  taus decay leptonically we can identify them in the detector, but
  there will appear one or two neutrinos in their decay. This means
  that we cannot reconstruct the tau momentum.  We will discuss this
  decay and an approximate mass reconstruction in detail in
  Section~\ref{sec:higgs_approx_mass}. This approximate reconstruction
  only works when the neutrinos lead to a measurable two-dimensional
  missing energy vector. If the Higgs decays at rest its decay
  production will be mostly back to back, so its low velocity makes
  the reconstruction of $m_{\tau\tau}\sim m_H$ hard.  It is widely
  assumed that Higgs production in gluon fusion is too close to
  threshold to see many decays to tau leptons, but in combination with
  other production channels and including hard jet recoil this channel should have
  some power.
\item[--] $gg\rightarrow H\rightarrow \gamma\gamma$ is, in spite of the
  small rate, the main Higgs discovery channel. Because
  $m_{\gamma\gamma}$ can be reconstructed to $\ope(1\%)$ this
  observable has incredibly precise side bins to the left and the to
  right of the Higgs peak. This is what for example the
  electromagnetic calorimeter of CMS has been designed for. The main
  problem is backgrounds for example from pions mistaken for photons,
  while theory input will play no role in this analysis. A slight
  problem is that in gluon fusion there is again little to cut on
  except for $m_{\gamma \gamma}$. The only additional observables
  which can reduce the physical two-photon background exploit a
  slightly boosted Higgs kinematics, either through the opening angles
  between the photons or the transverse momentum of the two photon
  system. In spite of the small branching ratios to photons the
  $\gamma \gamma$ channels is therefore limited by physical,
  irreducible backgrounds and our understanding of their $m_{\gamma
    \gamma}$ distribution.

  The peak in the invariant mass of the photons is great to measure
  the \underline{Higgs mass}: once we see a Gaussian peak we can
  determine its central value with a precision of
  $\Gamma_\text{detector}/\sqrt{S}$ (in a signal dominated sample with
  $S$ signal events), which translates into the per-mille level. The only issue in
  this measurement are systematic effects from the photon energy
  calibration. Unlike for electrons and photons there is no $Z$ peak
  structure in $m_{\gamma \gamma}$, so we would have to calibrate the
  photon energy from a strongly dropping distribution like $d \sigma/d
  E_\gamma$. An alternative approach is to use the calibration of
  another electromagnetic particle, like the electron, and convert the
  electron energy scale into a photon energy scale using Monte Carlo
  detector simulations or to look for $\ell^+ \ell^- \gamma$ decays
  of the $Z$.
\item[--] $gg\rightarrow H \rightarrow W^+W^-$ has a large rate, but
  once one of the $W$ bosons decays leptonically the Higgs mass is
  hard to reconstruct. All we can do is reconstruct a transverse mass
  variable, which we discuss in Section~\ref{sec:sim_met}. On the
  other hand, the backgrounds are electroweak and therefore small. A
  very dangerous background is top pair production which gives us two
  $W$ bosons and two relatively hard bottom quarks with typical
  transverse momenta $p_{T,b} \gtrsim 40$~GeV. We can strongly reduce
  this background by vetoing jets in additional to the two
  leptonically decaying $W$ bosons. As we will learn in
  Section~\ref{sec:qcd} such a jet veto is a problem once it covers
  collinear jet radiation from the incoming hadrons. In that case it
  breaks collinear factorization, a principle underlying any precision
  computation of the Higgs production rate at the LHC.

  The $H \to WW$ analysis strongly relies on angular correlations ---
  if the two gauge bosons come from a spin-zero resonance they have to
  have opposite polarization; because the $W$ coupling to fermions is
  purely left handed this implies that the two leptons prefer to move
  into the same direction as opposed to back--to--back. This effect can
  be exploited either by asking for a small opening angle of the two
  leptons or asking for a small invariant mass of the two
  leptons. Note that once we apply this cut we have determined the spin
  structure of the extracted signal to be a scalar.

  In the original ATLAS and CMS analyses the $WW$ decay looked not
  very useful for Higgs masses below 150~GeV, \ie for far off--shell
  Higgs decays.  Because there is not much more to cut on the expected
  significance dropped sharply with the decreasing branching
  ratio. However, in the 7~TeV and 8~TeV run this effect was countered
  by lowering the minimum transverse momentum requirements for the
  leptons, so the Higgs mass range covered by the $WW$ analysis now
  extends to the observed mass of 125~GeV.
\item[--] $gg\rightarrow H\rightarrow ZZ$ works great for $ZZ$ to four
  leptons, in particular muons, because of the fully reconstructed
  $m_{4 \ell}\sim m_{ZZ}\sim m_H$. Of all Higgs channels it requires
  the least understanding of the LHC detectors. Therefore it is
  referred to as \underline{`golden channel'}. Experimentally at least
  the four-muon channel is relatively easy. The electron decays can
  serve as a useful cross check.

  Its limitation are the leptonic $Z$ branching ratio and the 
  sharp drop in the off--shell Higgs branching ratio towards smaller
  Higgs masses. Once we include the leptonic $Z$ decays the over--all
  $H \to 4\ell$ branching ratio for a 125~GeV Higgs is tiny. The good
  news is that unlike in the two-photon channel there are essentially
  no irreducible backgrounds. The continuum production $q\bar{q} \to
  ZZ$ is an electroweak ($2\to 2$) process and as rare as the Higgs
  production process. The loop--induced $gg \to ZZ$ process is
  kinematically very similar to the signal, but even more rare. One
  useful cut based on the Breit--Wigner propagator shape is the
  distribution of the two invariant masses of the lepton pairs
  $m_{\ell \ell}$. For their Higgs discovery ATLAS and CMS asked for
  one pair of leptons with $m_{12} \sim m_Z$, all four leptons with
  $m_{4\ell} = 125$~GeV, and the second pair of leptons off--shell,
  $m_{34} \ll m_Z$.

\item[--] $gg\rightarrow H\rightarrow Z\gamma$ has recently been
  advertized as theoretically interesting once we link it to the
  observed loop--induced $H \to \gamma \gamma$ and tree level $H \to
  ZZ$ decays.  It behaves a little like $\gamma\gamma$, but with a
  smaller rate and a further reduced branching ratio of $Z\rightarrow
  \ell^+\ell^-$.  Instead of combining the advantages of $H \to ZZ$
  and $H \to \gamma \gamma$ this channel combines more of the
  disadvantages, so it is not likely to be measured soon. Of course,
  as for any channel seeing it will give us more information on the
  Higgs boson, so we should not give up. In addition, for some
  theoretical ideas it might be useful to determine an upper limit on
  the $H \to Z\gamma$ branching ratio.
\item[--] $gg\rightarrow H \rightarrow \mu^+ \mu^-$ might be the only hope we will
  ever have to measure a second-generation Yukawa coupling at the
  LHC. Because of its clear signature and its huge $q\bar{q} \to
  Z,\gamma \to \mu^+ \mu^-$ background this analysis resembles the
  photons channel, but with a much more rare signal. Eventually, other
  production processes might help with the Higgs decays to muons,
  similar to the $H \to \tau^+ \tau^-$ case.
\item[--] $gg\rightarrow H \rightarrow$~invisible is not predicted in
  Standard Model; it is obviously hopeless if only the Higgs is
  produced, because we would be trying to extract a signal of missing
  energy and nothing else. `Absolutely nothing' in addition to some QCD remnant is not a good
  signature for the trigger.
\end{itemize}

From the list of above channels we understand that the Higgs discovery
is dominated by the `golden' $H \to 4\ell$ and the `silver' $H \to
\gamma \gamma$ channels. The off--shell and hardly reconstructable $H
\to WW$ channel adds only little in terms of a distinctive signal. If
we want to learn more about the Higgs boson, we need additional
production mechanisms opening more decay signatures. Moreover, at this
point it is still not clear why ATLAS and CMS in their Higgs discovery
papers separate a Higgs--plus--two--jets signal for example in the photon
decay channel.

\subsection{Higgs production in weak boson fusion}
\label{sec:higgs_wbf}

Going back to Figure~\ref{fig:higgs_rates} we see that while gluon
fusion gives the largest Higgs production rate at the LHC, there are
other promising channels to study.  In the Standard Model the Higgs
has sizeable couplings only to the $W$ and $Z$ bosons and to the top
quark, so instead of via the top Yukawa coupling we can produce Higgs
bosons via their gauge boson couplings. This induces two channels, the
larger of which is weak boson fusion $qq\rightarrow qqH$: two incoming
quarks each radiate a $W$ or $Z$ boson which merge and form a Higgs.
Because the LHC is a $pp$ collider and because the proton mostly
contains the valence quarks $(uud)$ and low-$x$ gluons it is important
that this process can proceed as $ud\rightarrow du H$, where the $u$
radiates a $W^+$ and the $d$ radiates a $W^-$. The Feynman diagram for
this process is
\begin{center}
\begin{fmfgraph*}(90,60)
 \fmfset{arrow_len}{2mm}
 \fmfleft{in1,in2}
 \fmf{fermion,width=0.5,tension=1.2}{in1,v1}
 \fmf{fermion,width=0.5,tension=1.2}{in2,v2}
 \fmf{fermion,width=0.5,tension=.8}{v1,out1}
 \fmf{fermion,width=0.5,tension=.8}{v2,out3}
 \fmf{photon,width=0.5}{v1,v3}
 \fmf{photon,width=0.5}{v2,v3}
 \fmf{dashes,width=0.5,tension=.8}{v3,out2}
 \fmfright{out1,out2,out3}
\end{fmfgraph*}
\end{center}
If the Higgs were a $Z$ boson, it could also bremsstrahlung off the
incoming or outgoing quarks, but for Higgs production at colliders we
safely assume that the first two generation fermions are massless. That is
at least unless we discuss a muon collider as a specific way to
produce Higgs bosons. 

In a way, weak boson fusion looks like double
deep inelastic scattering, one from each of the protons. This is one
of the key observations which in Section~\ref{sec:higgs_cjv} we will
use for background suppression via the central jet veto. The double
deep inelastic scattering approximation is also a good way to compute
corrections to the weak boson fusion production rate, at least provided we
neglect kinematic distributions. Just a final comment: the LHC experiments
refer to weak boson fusion as vector boson fusion (VBF). However, vector 
boson fusion includes incoming gluons, which have very different kinematic 
properties, so in the following we strictly mean weak boson fusion 
mediated by massive $W$ and $Z$ exchange.

\subsubsection{Production kinematics}
\label{sec:higgs_kinematics}

In the Feynman diagrams for weak boson fusion Higgs production we
encounter intermediate massive gauge boson propagators. They induce a
particular shape of the kinematic distributions of the final--state
jet.  First, we need to quote the exact calculation showing that in the
matrix element squared we will usually find one power of $p_T$ in the
numerator.  With this information we can look for the maximum in the
$p_{T,j} = p_{T,W}$ spectrum as a function of the momentum in the beam
direction, $p_3$, and the absolute value of the two-dimensional
transverse momentum $p_T$
\begin{alignat}{5}
0 \; \really \; & 
\frac{\p}{\p p_T} \; \frac{p_T}{E^2-p_T^2-p_L^2-m_W^2}  \notag\\
=&
\frac{1}{E^2-p_T^2-p_L^2-m_W^2} 
  + p_T \; \frac{(-1)}{(E^2-p_T^2-p_L^2-m_W^2)^2} \; (-2p_T)\notag\\
=&
\frac{E^2+p_T^2-p_L^2-m_W^2}{(E^2-p_T^2-p_L^2-m_W^2)^2} \notag\\
\sim&
\frac{C m_W^2 +p_T^2-m_W^2}{(E^2-p_T^2-p_L^2-m_W^2)^2} 
\qqquad \text{with} \quad E^2 \sim p_L^2 \gg m_W^2 
\quad \text{but} \quad E^2 - p_L^2 = C \; m_W^2 \notag\\
=&
\frac{p_T^2-(1-C)m_W^2}{(E^2-p_T^2-p_L^2-m_W^2)^2} \notag\\
\Leftrightarrow \qquad & \boxed{ p_T^2 = (1-C) m_W^2 }
\end{alignat}
at the maximum and for some number $C < 1$.  This admittedly hand-waving argument shows
that in weak boson fusion Higgs production the transverse momenta of
the outgoing jets peak at values below the $W$ mass. In reality, the
peak occurs around $p_T \sim 30$~GeV. This transverse momentum
scale we need to compare to the longitudinal momentum given by the
energy scale of valence quarks at the LHC, \ie several hundreds of
GeV.

These two forward jets are referred to as \underline{tagging
  jets}. They offer a very efficient cut against QCD backgrounds:
because of their back--to--back geometry and their very large
longitudinal momentum, their invariant mass $m_{jj}$ will for a 14~TeV
collider energy easily exceed a TeV. For any kind of QCD background
this will not be the case. Compared to Higgs production in gluon
fusion the tagging jets\index{tagging jet} are an example how features
of the production process which have little or nothing to do with the
actual Higgs kinematics can help reduce backgrounds --- the largest
production rate does not automatically yield the best signatures.
The only problem with the weak boson fusion channel is that
its distinctive $m_{jj}$ distribution requires a large collider
energy, so running the LHC at 7~TeV and 8~TeV for a Higgs discovery
was very bad news for this channel.

Moving on to the Higgs kinematics, in contrast to the jets the Higgs
and its decay products are expected to hit the detector centrally. 
We are looking for two forward jets and for example two $\tau$ leptons
or two $W$ bosons in the central detector. Last but not least, the
Higgs is produced with finite transverse momentum which is largely determined
by the acceptance cuts on the forward jets and their
typical transverse momentum scale ${p_T}_H \sim m_W$.\bigskip

Compared to Higgs production in gluon fusion we buy this
distinctive signature and its efficient extraction from the background
at the expense of the rate. Let us start with the partonic cross
sections: the one-loop amplitude for $gg\rightarrow H$ is suppressed
by $\alpha_s y_t/(4\pi) \sim (1/10) \, (2/3) \, (1/12) = 1/180$.  For
the production cross section this means a factor of $(1/180)^2 \sim
1/40000$. The cross section for weak boson fusion is proportional to
$g^6$, but with two additional jets in the final state. Including the
additional phase space for two jets this roughly translates into
$g^6/(16 \pi)^2 \sim (2/3)^6 \, 1/(16\pi)^2 = (64/729) \, (1/2500) \,
\sim 1/25000$.  These two numbers governing the main LHC production
cross sections roughly balance each other.

The difference in rate which we see in Figure~\ref{fig:higgs_rates} instead
arises from the quark and gluon luminosities. In weak boson fusion the
two forward jets always combine to a large partonic center--of--mass
energy $x_1 x_2 s > (p_{j,1}+p_{j,2})^2 = 2(p_{j,1} p_{j,2})$, with
the two parton momentum fractions $x_{1,2}$ and the hadronic center of
mass energy $\sqrt{s} = 14$~TeV.  Producing a single Higgs in gluon
fusion probes the large gluon parton density at typical parton
momentum fractions $x \sim m_H/\sqrt{s} \sim 10^{-3}$. This means that
each of the two production processes with their specific incoming
partons probes its most favorable parton momentum fraction: low-$x$
for gluon fusion and high-$x$ for valence quark scattering. Looking at
typical LHC energies, the gluon parton density grows very steeply for
$x\lesssim 10^{-2}$. This means that gluon fusion wins: for a 125~GeV
Higgs the gluon fusion rate of $\sim 50$~pb clearly exceeds the weak
boson fusion rate of $\sim 4.2$~pb. On the other hand, these numbers
mean little when we battle an 800~pb $t\bar{t}$ background relying on
kinematic cuts either on forward jets or on Higgs decay
products.\bigskip

In Figure~\ref{fig:higgs_sig} we see that for large Higgs mass the weak
boson fusion rate approaches the gluon fusion rate. The two
reasons for this behavior we mentioned already in this section: first
of all, for larger $x$ values the rate for $gg\rightarrow H$ decreases
steeply with the gluon density, while in weak boson fusion the already
huge partonic center of mass energy due to the tagging jets\index{tagging jet} ensures
that an increase in $m_H$ makes no difference anymore.  Even more
importantly, there appear \underline{large logarithms} because the
low-$p_T$ enhancement of the quark--$W$ splitting.  If we neglect $m_W$
in the weak boson fusion process the $p_{T,j}$ distributions will
diverge for small $p_{T,j}$ like $1/p_{T,j}$, as we will see in Section~\ref{sec:qcd_dglap} After integrating over
$p_{T,j}$ this yields a $\log p_{T,j}^\text{max}/p_{T,j}^\text{min}$
dependence of the total rate.  With the $W$ mass cutoff and a typical
hard scale given by $m_H$ this logarithm becomes
\begin{alignat}{5}
\sigma_\text{WBF} \propto 
\left( \log \frac{p_{T,j}^\text{max}}{p_{T,j}^\text{min}}\right)^2 \sim 
\left( \log \frac{m_H}{m_W} \right)^2 \; .
\label{eq:coll_wbf}
\end{alignat}
For $m_H = \ope(\tev)$ this logarithm gives us an enhancement
by factors of up to 10, which makes weak boson fusion the dominant
Higgs production process.

Motivated by such logarithms, we will talk about partons inside the
proton and their probability distributions for given momenta in
Section~\ref{sec:qcd_dglap}. In the \underline{effective $W$
  approximation}\index{effective W approximation} we can resum the logarithms appearing in
Eq.\eqref{eq:coll_wbf} or compute such a probability for $W$ bosons inside
the proton. This number is a function of the partonic momentum
fraction $x$ and can be evaluated as a function of the transverse
momentum $p_T$.  Because the incoming quark inside the proton has
negligible transverse momentum, the transverse momenta of the $W$
boson and the forward jet are identical. These transverse momentum
distributions in $p_{T,W} = p_{T,j}$ look different for transverse and
longitudinal gauge bosons
\begin{alignat}{5}
P_T(x,p_T) &\sim \frac{g_V^2 + g_A^2}{4 \pi^2} \;
                 \frac{1+(1-x)^2}{x} \;
                 \frac{p_T^3}{(p_T^2 + (1-x) \, m_W^2)^2} 
          &\quad \to \quad&
                 \frac{g_V^2 + g_A^2}{4 \pi^2} \;
                 \frac{1+(1-x)^2}{x} \;
                 \frac{1}{p_T} 
                 \notag \\
P_L(x,p_T) &\sim \frac{g_V^2 + g_A^2}{4 \pi^2} \;
                 \frac{1-x}{x} \;
                 \frac{2 m_W^2 p_T}{(p_T^2 + (1-x) \, m_W^2)^2} 
          &\quad \to \quad&
                 \frac{g_V^2 + g_A^2}{4 \pi^2} \;
                 \frac{1-x}{x} \;
                 \frac{2 m_W^2}{p_T^3} \; .
\label{eq:higgs_ptj}
\end{alignat}
The couplings $g_{A,V}$ describe the gauge coupling of the $W$ bosons
to the incoming quarks. Looking at large transverse momenta $p_T\gg
m_W$ the radiation of
longitudinal $W$ bosons falls off sharper than the radiation of
transverse $W$ bosons.  This different behavior of transverse and
longitudinal $W$ bosons is interesting, because it allows us to gain
information on the centrally produced particle and which modes it
couples to just from the transverse momentum spectrum of the forward
jets and without looking at the actual central particle.

However, numerically the effective $W$ approximation does not work well
for a 125~GeV Higgs at the LHC. The simple reason is that
the Higgs mass is of the order of the $W$ mass, as are the transverse
momenta of the $W$ and the final--state jets, and none of them are very
small. Neglecting for example the transverse momentum of the $W$
bosons or the final--state jets will not give us useful predictions for
the kinematic distributions, neither for the tagging jets nor for the
Higgs.  For the SSC, the competing design to the LHC in Texas which
unfortunately was never built, this might have been a different story,
but at the LHC we should not describe $W$ bosons (or for that matter top
quarks) as essentially massless partons inside the proton.

\subsubsection{Jet ratios and central jet veto}
\label{sec:higgs_cjv}

From the Feynman diagram for weak boson fusion we see that the diagram
describing a gluon exchange between the two quark lines multiplied
with the Born diagram is proportional to the color factor $\tr T^a \tr
T^b \delta^{ab}=0$.  The only way to avoid this suppression is the
interference of two identical final--state quarks, for example in $ZZ$
fusion. First, this does not involve only valence quarks and
second, this assumes a phase space configuration where one of the two
supposedly forward jets turns around and goes backwards, so
the interfering diagrams contribute in the same phase space
region. This means that virtual gluon exchange in weak boson fusion
is practically absent.

In Section~\ref{sec:qcd} we will see that virtual gluon exchange and
real gluon emission are very closely related. Radiating a gluon off
any of the quarks in the weak boson fusion process will lead to a
double infrared divergence, one because the gluon can be radiated at
small angles and one because the gluon can be radiated with vanishing
energy. The divergence at small angles is removed by redefining the
quark parton densities in the proton. The soft, non--collinear divergence has to
cancel between real gluon emission and virtual gluon
exchange. However, if virtual gluon exchange does not appear,
non--collinear soft gluon radiation cannot appear either. This means that
additional QCD jet activity as part of the weak boson fusion process
is limited to collinear radiation, \ie radiation along the beam line
or at least in the same direction as the far forward tagging
jets\index{tagging jet}. Gluon radiation into the central detector is
suppressed by the color structure of the weak boson fusion
process.\bigskip

While it is not immediately clear how to quantify such a statement it
is a very useful feature, for example looking at the top pair
backgrounds. The $W W b\bar{b}$ final state as a background to $qqH, H
\to WW$ searches includes two bottom jets which can mimic the signal's
tagging jets. At the end, it turns out that it is much more likely
that we will produce another jet through QCD jet radiation, \ie $pp
\to t\bar{t}$+jet, so only one of the two bottom jets from the top
decays needs to be forward. In any case, the way to isolate the Higgs
signal is to look at additional central jets.

As described above, for the signal additional jet activity is limited
to small-angle radiation off the initial--state and final--state quarks.
For a background like top pairs this is not the case, which means we can
reduce all kinds of background by vetoing jets in the central region
above $p_{T,j}\gtrsim 30$~GeV. This strategy is referred to as \underline{central
jet veto} or \underline{mini-jet veto}. Note that
it has nothing to do with rapidity gaps at HERA or pomeron exchange,
it is a QCD feature completely accounted for by standard perturbative
QCD.\bigskip

From QCD we then need to compute the probability of not observing
additional central jets for different signal and background
processes. Postponing the discussion of QCD parton splitting to
Section~\ref{sec:qcd_splitting} we already know that for small
transverse momenta the $p_{T,j}$ spectra for massless states will
diverge, as shown in Eq.\eqref{eq:coll_wbf}.  Looking at some kind of
$n$-particle final state and an additional jet radiation we can
implicitly define a reference point $p_T^\text{crit}$ at which the
divergent rate for one jet radiation $\sigma_{n+1}$ starts to exceed
the original rate $\sigma_n$, whatever the relevant process might be
\begin{alignat}{5}
\sigma_{n+1}(p_T^\text{crit}) \; = \;
\int_{p_T^\text{crit}}^\infty d p_{T,j} \;
\frac{d \sigma_{n+1}}{d p_{T,j}}
\really \sigma_n \; .
\end{alignat}
This condition defines a point in $p_T$ below which our perturbation
theory in $\alpha_s$, \ie in counting the number of external partons,
breaks down.  For weak boson fusion Higgs production we find
$p_T^\text{crit} \sim 10$~GeV, while for QCD processes like $t\bar{t}$
production it becomes $p_T^\text{crit}=40$~GeV.  In other words, jets
down to $p_T=$10~GeV are perturbatively well defined for Higgs
signatures, while for the QCD backgrounds jets below 40~GeV are much
more frequent than they should be looking at the perturbative series
in $\alpha_s$. This fixes the $p_T$ range where a central jet veto will
be helpful to reject backgrounds
\begin{alignat}{5}
p_{T,j}> 30~\text{GeV} 
\qqquad \text{and} \qqquad 
\eta_j^\text{(tag 1)} < \eta_j < \eta_j^\text{(tag 2)} \; .
\end{alignat}
The second condition reminds us of the fact that only central jets
will be rare in weak boson fusion. The smaller the $p_T$ threshold
the more efficient the central jet veto becomes, but at some point
experimental problems as well as non--perturbative QCD effects will
force us to stay above 20 or 30 or even 40~GeV.\bigskip

\begin{figure}[t]
\begin{center}
\includegraphics[width=0.40\hsize]{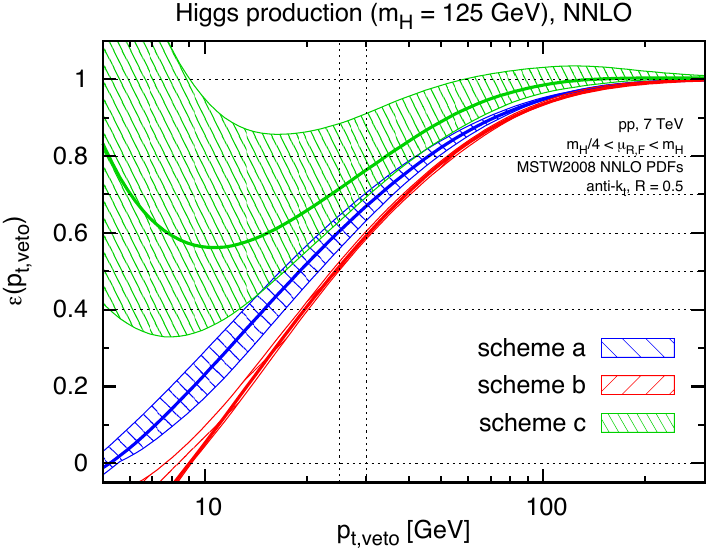}
\end{center}
\caption{Different predictions for the jet veto survival probability
  $P_\text{pass}$ as a function of the maximum allowed $p_{T,j}$. The
  example process chosen is Higgs production in gluon fusion. The
  shaded regions indicate the independent variation of the
  factorization and renormalization scales within $[m_H/4,m_H]$
  requiring $\mu_R/\mu_F$ to lie within $[0.5,2]$. The figure and the
  corresponding physics argument are taken from Ref.~\cite{Banfi:2012yh}.}
\label{fig:higgs_pveto}
\end{figure}

If we assign a probability pattern to the radiation of jets from the
core process we can compute the \underline{survival probability}
$P_\text{pass}$ of such a jet veto. For many years we have been told
that higher orders in the perturbative QCD series for the Higgs
production cross section is the key to understanding LHC rates. For
multi--jet observables like a jet veto this is not necessarily
true. As an example we assume NNLO or two-loop precision for the
Higgs production rate $\sigma = \sigma_0 + \alpha_s \sigma_1 +
\alpha_s^2 \sigma_2$ where we omit the over--all factor $\alpha_s^2$ in
$\sigma_0$.  Consequently, we define the cross section passing the jet
veto $\sigma^\text{(pass)} = P_\text{pass} \; \sigma = \sum_j
\alpha_s^j \sigma_j^\text{(pass)}$. Because the leading order
prediction only includes a Higgs in the final state we know that
$\sigma_0^\text{(pass)} = \sigma_0$.  Solving this definition for the
veto survival probability we can compute
\begin{alignat}{5}
P_\text{pass}^\text{(a)} = \frac{\sigma^\text{(pass)}}{\sigma}
\; = \;
\frac{\sigma_0 + \alpha_s \sigma_1^\text{(pass)}+\alpha_s^2 \sigma_2^\text{(pass)}}
     {\sigma_0 + \alpha_s \sigma_1 + \alpha_s^2 \sigma_2} \; ,
\label{eq:higgs_cjva}
\end{alignat}
motivated by including the maximum number of terms (NNLO) in the numerator
and denominator. The result as a function of the maximum allowed
$p_{T,j}$ is shown as `scheme a' in Figure~\ref{fig:higgs_pveto}. The
shaded region is an estimate of the theoretical uncertainty of this
prediction.

Alternatively, we can argue that the proper perturbative observable is
the fraction of vetoed events $(1 - P_\text{pass})$. Indeed, for small
values of $\alpha_s$ the jet radiation probability vanishes and with
it $(1 - P_\text{pass}) \sim \alpha_s \to 0$. This vetoed event
fraction we can compute as $\sigma_j - \sigma_j^\text{(pass)}$ for
$j \ge 0$.  However, we need to keep in mind that in the presence of an
additional jet the NNLO prediction for the inclusive Higgs production
rate reduces to NLO accuracy, so we include the two leading
terms in the numerator and denominator,
\begin{alignat}{5}
1 - P_\text{pass}^\text{(b)}
&= \frac{\alpha_s (\sigma_1 - \sigma_1^\text{(pass)})
        +\alpha_s^2 (\sigma_2 - \sigma_2^\text{(pass)})}
        {\sigma_0 + \alpha_s \sigma_1}
\notag \\
P_\text{pass}^\text{(b)}
&= 1 - \frac{\alpha_s (\sigma_1 - \sigma_1^\text{(pass)})
            +\alpha_s^2 (\sigma_2 - \sigma_2^\text{(pass)})}
            {\sigma_0 + \alpha_s \sigma_1}
 = \frac{\sigma_0 + \alpha_s \sigma_1^\text{(pass)}
        +\alpha_s^2 \sigma^\text{(pass)}_2 - \alpha_s^2 \sigma_2}
        {\sigma_0 + \alpha_s \sigma_1} \; .
\label{eq:higgs_cjvb}
\end{alignat}
Obviously, in Eq.\eqref{eq:higgs_cjvb} we can move the term
$-\alpha_s^2 \sigma_2$ into the denominator and arrive at
Eq.\eqref{eq:higgs_cjva} within the uncertainty defined by the unknown
$\alpha_s^3$ terms.  This defines `scheme b' in
Figure~\ref{fig:higgs_pveto}.

Finally, we can consistently Taylor expand the definition of
$P_\text{pass}$ as the ratio given in Eq.\eqref{eq:higgs_cjva}. The
two leading derivatives of a ratio read
\begin{alignat}{5}
\left( \frac{f}{g} \right)' 
&= \frac{f'g - fg'}{g^2}
\stackrel{f=g}{=} \frac{f' - g'}{g}
\notag \\
\left( \frac{f}{g} \right)'' 
&= \left( \frac{f'g}{g^2} - \frac{fg'}{g^2} \right)'
 = \frac{(f'g)' g^2 - f'g 2g g'}{g^4}
  -\frac{(fg')' g^2 - fg' 2g g'}{g^4}
\notag \\
&= \frac{(f'g)' - 2 f' g'}{g^2}
  -\frac{(fg')' g - 2 fg' g'}{g^3}
= \frac{f''g - f' g'}{g^2}
  -\frac{f g''g - fg' g'}{g^3} 
\stackrel{f=g}{=} 
   \frac{f'' - g''}{g} - \frac{g'(f'- g')}{g^2}
\end{alignat}
In the last steps we assume $f=g$ at the point where we evaluate the 
Taylor expansion.
Applied to the perturbative QCD series for $(1 - P_\text{pass})$ around
the zero-coupling limit this gives us
\begin{alignat}{5}
1 - P_\text{pass}^\text{(c)} 
&= 1- \frac{\sigma_0 + \alpha_s \sigma_1^\text{(pass)}+\alpha_s^2 \sigma_2^\text{(pass)} + \cdots}
     {\sigma_0 + \alpha_s \sigma_1 + \alpha_s^2 \sigma_2 + \cdots} 
\notag \\
P_\text{pass}^\text{(c)} &= 1
+ \alpha_s \;
  \frac{\sigma_1^\text{(pass)} - \sigma_1}{\sigma_0}
+ \alpha_s^2 \; \frac{\sigma_2^\text{(pass)} -  \sigma_2}{\sigma_0}
- \alpha_s^2 \;
 \frac{\sigma_1 (\sigma_1^\text{(pass)}- \sigma_1)}{\sigma_0^2} \; ,
\label{eq:higgs_cjvc}
\end{alignat}
defining `scheme c' in Figure~\ref{fig:higgs_pveto}. The numerical
results indicate that the three schemes are inconsistent within their
theoretical uncertainties, and that the most consistent Taylor
expansion around perfect veto survival probabilities is doing
particularly poorly. Towards small $p_{T,j}$ veto ranges the fixed
order perturbative approach clearly fails. The way to improve the
theoretical prediction is a re-organization of the perturbation theory
for small jet transverse momenta. We introduce this approach with its
leading term, the parton shower, in
Section~\ref{sec:qcd_shower}. For now we conclude that our theoretical
approach has to go beyond a fixed number of (hard) jets and include
the production of any number of jets in some kind of modified
perturbative series.\bigskip

One ansatz for the distribution of any number of radiated jets is
motivated by soft photon emission off a hard electron. In
Section~\ref{sec:qcd_multiple} we derive the \underline{Poisson distribution} in the numbers of jet which follows in the soft limit.
If we for now assume a Poisson distribution, the probability of
observing exactly $n$ jets given an expected $\bar{n}$ jets is\index{Poisson scaling}
\begin{alignat}{5}
f(n;\bar{n}) = \frac{\bar{n}^ne^{-\bar{n}}}{n!}
\qqquad \Rightarrow \qqquad
\boxed{
P_\text{pass} \equiv f(0;\bar{n}) = e^{-\bar{n}}
} \; .
\label{eq:mjv}
\end{alignat}
Note that this probability links rates for exactly $n$ jets, no at
least $n$ jets, \ie it described the exclusive number of jets.  The
Poisson distribution is normalized to unity, once we sum over all
possible jet multiplicities $n$.  It defines the so-called
exponentiation model. We consistently fix the expectation value in
terms of the inclusive cross sections producing at least zero or at
least one jet,
\begin{alignat}{5}
\langle n \rangle \equiv \bar{n} 
= \frac{\sigma_1(p_T^\text{min})}{\sigma_0} \; .
\label{eq:higgs_ratio}
\end{alignat}
This ensures that the inclusive jet ratio $\sigma_1/\sigma_0$ is
reproduced by the ratio of the corresponding Poisson distributions.
Including
this expectation value $\bar{n}$ into Eq.\eqref{eq:mjv} returns a veto
survival probability of $\exp(-\sigma_1/\sigma_0)$. This comes out
roughly as 88\% for the weak boson fusion signal and as 24\% for the
$t\bar{t}$ background. For the signal--to--background ratio this implies
a three-fold increase.\bigskip

An alternative model starts from a constant probability of radiating a
jet, which in terms of the inclusive cross sections $\sigma_n$, \ie
the production rate for the radiation of at least $n$ jets, reads
\begin{alignat}{5}
\frac{\sigma_{n+1}(p_T^\text{min})}
     {\sigma_n(p_T^\text{min})}  
= R^\text{(incl)}_{(n+1)/n} (p_T^\text{min}) \; .
\end{alignat}
We derive this pattern in Section~\ref{sec:qcd_jets}.  The expected
number of jets is then given by
\begin{alignat}{5}
\langle n \rangle 
&= \dfrac{1}{\sigma_0} \; \sum_{j=1} j (\sigma_j - \sigma_{j+1})
 = \dfrac{1}{\sigma_0} \; \left( \sum_{j=1} j \sigma_j 
                               - \sum_{j=2} (j-1) \sigma_j \right)
 = \dfrac{1}{\sigma_0} \; \sum_{j=1} \sigma_j  \notag \\
&= \dfrac{\sigma_1}{\sigma_0} \; \sum_{j=0} ( R^\text{(incl)}_{(n+1)/n})^j  
 = \dfrac{R^\text{(incl)}_{(n+1)/n}}{1- R^\text{(incl)}_{(n+1)/n}} \; ,
\label{eq:staircase_incl}
\end{alignat}
if $R^\text{(incl)}_{(n+1)/n}$ is a constant.  Assuming the series
converges this turns into a requirement on $p_T^\text{min}$. Radiating
jets with such a constant probability has been observed at many
experiments, including most recently the LHC, and is in the context of
$W+$jets referred to as \underline{staircase scaling}\index{staircase scaling}.
We will derive both, the Poisson scaling and the staircase scaling from QCD
in Section~\ref{sec:qcd_radiation}. Even without saying anything on how to
calculate exclusive processes with a fixed number of jets we can
derive a particular property of the constant probability of staircase
scaling: the ratios of the $(n+1)$-jet rate to the $n$-jet rate for
inclusive and exclusive jet rates are identical. We can see this by
computing the inclusive $R^\text{(incl)}_{(n+1)/n}$ in terms of
exclusive jet rates
\begin{alignat}{5}
R^\text{(incl)}_{(n+1)/n}
&= \dfrac{\sigma_{n+1}}{\sigma_n} 
 = \dfrac{\sum_{j=n+1}^\infty \sigma_j^{(\text{excl})}}
         {\sigma_n^{(\text{excl})} + \sum_{j=n+1}^\infty \sigma_j^{(\text{excl})}} 
   \notag \\
&= \dfrac{\sigma_{n+1}^{(\text{excl})} \sum_{j=0}^\infty R_{(n+1)/n}^j}{\sigma_n^{(\text{excl})} + \sigma_{n+1}^{(\text{excl})} \sum_{j=0}^\infty R_{(n+1)/n}^j} 
 \qqquad \qqquad \text{with} \quad 
 R_{(n+1)/n} = \frac{\sigma_{n+1}^{(\text{excl})}}{\sigma_n^{(\text{excl})}}\notag \\
&= \dfrac{\dfrac{R_{(n+1)/n} \sigma_n^{(\text{excl})}}{1-R_{(n+1)/n}}}{\sigma_n^{(\text{excl})} + \dfrac{R_{(n+1)/n} \sigma_n^{(\text{excl})}}{1-R_{(n+1)/n}}} 
 = \dfrac{R_{(n+1)/n}}{1-R_{(n+1)/n} + R_{(n+1)/n}} \notag \\
&= R_{(n+1)/n} \; .
\label{eq:staircase_excl}
\end{alignat}
\bigskip

To show that the exponentiation model and staircase scaling are not
the only assumptions we can make to compute jet rates we show yet
another, but similar ansatz which tries to account for an increasing
number of legs to radiate jets off. Based on
\begin{alignat}{5}
\frac{\sigma_{j+1}(p_T^\text{min})}
     {\sigma_j(p_T^\text{min})}
= \frac{j+1}{j} \; R^\text{(incl)}_{(n+1)/n}(p_T^\text{min}) \; ,
\end{alignat}
the expectation for the number of jets radiated gives, again following Eq.\eqref{eq:staircase_incl}
\begin{alignat}{5}
\langle n \rangle 
&= \dfrac{1}{\sigma_0} \; \sum_{j=1} j \sigma_j  
 = \dfrac{1}{\sigma_0} \; \sigma_0 \sum_{j=1} j (R^\text{(incl)}_{(n+1)/n})^j  \notag \\
&= R^\text{(incl)}_{(n+1)/n} \; \sum_{j=1} j (R^\text{(incl)}_{(n+1)/n})^{j-1}  
 = \dfrac{R^\text{(incl)}_{(n+1)/n}}{(1- R^\text{(incl)}_{(n+1)/n})^2} \; .
\end{alignat}
All of these models are more or less well motivated statistical
approximations.  The do not incorporate experimental effects or the
non--perturbative underlying event, \ie additional energy dependent but
process independent jet activity in the detectors from many not
entirely understood sources. For many reasons none of them is
guaranteed to give us a final and universal number. However, by the time we get to
Section~\ref{sec:qcd_ckkw} we will at least be able to more accurately describe the
central jet veto in QCD.\bigskip

For the Poisson distribution and the staircase distribution we can
summarize the main properties of the $n$-jet rates in terms of the
upper incomplete gamma function $\Gamma(n,\bar{n})$:

\begin{center}
\begin{tabular}{l|cc}
\hline
& staircase scaling & Poisson scaling \\
\hline
$\sigma_n^\text{(excl)}$ & 
       $\sigma^\text{(excl)}_0 \; e^{-bn}$ & $\sigma_0 \; \dfrac{e^{-\bar{n}} \bar{n}^n}{n!}$ \\[5mm] 
$R_{(n+1)/n} = \dfrac{\sigma_{n+1}^\text{(excl)}}{\sigma_n^\text{(excl)}}$ & $e^{-b}$ & $\dfrac{\bar{n}}{n+1}$ \\[2mm]
$R^\text{(incl)}_{(n+1)/n} = \dfrac{\sigma_{n+1}}{\sigma_n}$ & 
       $e^{-b}$ & $\left( \dfrac{(n+1) \, e^{-\bar{n}} \, \bar{n}^{-(n+1)}}
                                    {\Gamma(n+1) - n \Gamma(n,\bar{n})}
                  + 1 \right)^{-1}$ \\[5mm]
$\langle n \rangle$ & $\dfrac{1}{2} \dfrac{1}{\cosh b -1}$ & $\bar{n}$ \\[2mm]
$P_\text{pass}$ &
       $1-e^{-b}$ & $e^{-\bar{n}}$ \\[1mm] \hline
\end{tabular}
\end{center}

\subsubsection{Decay kinematics and signatures}
\label{sec:higgs_approx_mass}

For most of the Higgs decays discussed in
Section~\ref{sec:higgs_decays} it does not matter how the Higgs is
produced, as long as we only care about the signal events.  Many
aspects discussed in Section~\ref{sec:higgs_gf_lhc} for Higgs
production in gluon fusion can be directly applied to weak boson
fusion. Serious differences appear only when we also include
backgrounds and the kinematic cuts required to separate signal and
background events.\bigskip

The only fundamental difference appears in the reconstruction of Higgs
decay into $\tau$ pairs.  The sizeable transverse momentum of the
Higgs in the weak boson fusion process allows us to reconstruct the
invariant mass of a $\tau\tau$ system in the \underline{collinear
  approximation}\index{Higgs boson!collinear decay}: if we assume that
a $\tau$ with momentum $\vec{p}$ decays into a lepton with the
momentum $x \vec{p}$ and a neutrino, both moving into the same
direction as the tau, we can write the two-dimensional transverse
momentum of the two taus from the Higgs decay in two ways
\begin{alignat}{5}
\vec{p_1}+\vec{p_2}
\equiv \frac{\vec{k_1}}{x_1} +\frac{\vec{k_2}}{x_2}
\really \vec{k_1}+\vec{k_2}+\vec{\slashchar{k}} \; .
\label{eq:met_taus}
\end{alignat}
The missing transverse momentum $\vec{\slashchar{k}}$ is the measured
vector sum of the two neutrino momenta.  This equation is useful
because we can measure the missing energy vector at the LHC in the
transverse plane, \ie as two components, which means
Eq.\eqref{eq:met_taus} is really two equations for the two unknowns
$x_1$ and $x_2$. Skipping the calculation of solving these two
equations for $x_1 x_2$ we quote the result for the invariant
$\tau\tau$ mass 
\begin{alignat}{5}
\boxed{m_{\tau\tau}^2 = 2 \, (p_1 p_2) = 2 \frac{(k_1 k_2)}{x_1 x_2}} \; .
\end{alignat}
For the signal this corresponds to the Higgs mass.  From the
formula above it is obvious that this approximation does not only
require a sizeable $p_i \gg m_\tau$, but also that back--to--back taus
will not work --- the two vectors contributing to
$\vec{\slashchar{k}}$ then largely cancel and the computation
fails. This is what happens for the inclusive production channel $gg\rightarrow
H\rightarrow \tau\tau$, where the Higgs boson is essentially produced
at rest.\bigskip

Again, we can make a list of \underline{signatures} which work more or
less well in connection to weak boson fusion production. These
channels are also included in the summary plot by ATLAS, shown in
Figure~\ref{fig:higgs_sig}.
\begin{itemize}
\item[--] $qq \to qqH, H\rightarrow b\bar{b}$ is problematic because
  of large QCD backgrounds and because of the trigger in ATLAS. The
  signal--to--background ratio is not quite as bad as in the gluon
  fusion case, but still not encouraging. The most worrisome
  background is overlapping events, one producing the two tagging jets
  and the other one two bottom jets.  This overlapping scattering
  gives a non--trivial structure to the background events, so a brute
  force side-bin analysis will not work.
\item[--] $qq \to qqH, H\rightarrow\tau^+\tau^-$ had the potential to be a
  discovery channel for a light Higgs boson with $m_H\lesssim
  130$~GeV, at least for an LHC energy of 14~TeV. The 2012 run at
  8~TeV gives us the opportunity to at least see a small Higgs
  signal. In early analyses we can limit ourselves to only one jet
  recoiling against the Higgs, to increase the
  sensitivity. Eventually, the approximate mass reconstruction might
  be as good as $\sim 5$~GeV, because we can measure the peak position
  of a Gaussian distribution with a precision of
  $\Gamma_\text{detector}/\sqrt{S}$. This channel is particularly
  useful in scenarios beyond the Standard Model, like its minimal
  supersymmetric extension. It guaranteed the discovery of one Higgs
  boson over the entire supersymmetric\index{supersymmetry} parameter
  space without a dedicated SUSY search. 

  Like for almost all weak boson fusion analyses there are two
  irreducible backgrounds: $Z+n$~jets production at order $G_F
  \alpha_s^n$ and the same final state at order $G_F^3
  \alpha_s^{n-2}$. The latter has a smaller rate, but because it
  includes weak boson fusion $Z$ production as one subprocess it is
  much more similar to Higgs production for example from a kinematical
  or a QCD point of view.
\item[--] $qq \to qqH, H\rightarrow\gamma\gamma$ should be almost comparable
  with $gg\rightarrow H \rightarrow \gamma\gamma$ with its smaller
  rate but improved background suppression. For 14~TeV the two-jet
  topology which is already part of the discovery analysis presented in
  Section~\ref{sec:higgs_discovery} will become more and more
  important.  As a matter of fact, the weak boson fusion channel is
  usually included in $H \to \gamma\gamma$ analyses, and for
  neural net analyses zooming in on large Higgs transverse momenta it
  will soon dominate the inclusive analysis.
\item[--] $qq \to qqH, H\rightarrow W^+W^-$ contributes to the discovery
  channel for $m_H\gtrsim 125$~GeV, ideally at 14~TeV collider energy. 
  In comparison to $gg \rightarrow H
  \rightarrow W^+W^-$ it works significantly better for off--shell $W$
  decays, \ie for Higgs masses below 150~GeV. There, the multitude of
  background rejection cuts and a resulting better
  signal--to--background ratio win over the large rate in gluon
  fusion. Apart from the two tagging jets the analysis proceeds the
  same way as the gluon fusion analysis. The key background to get rid
  of are top pairs.
\item[--] $qq \to qqH, H\rightarrow ZZ$ is likely to work in spite of
  the smaller rate compared to gluon fusion. It might even be possible
  with one hadronic $Z$ decay, but there are not many detailed studies
  available. On the other hand, already the gluon--fusion $H \to ZZ$
  search, which is one of the backbones of the Higgs discovery, has
  essentially no backgrounds. This takes away the biggest advantage of
  weak boson fusion as a Higgs production channel --- the improved
  background reduction.
\item[--] $qq \to qqH, H\rightarrow Z\gamma$ is difficult due to a too
  small event rate and no apparent experimental advantages compared to
  the gluon--fusion Higgs production.
\item[--] $qq \to qqH, H\rightarrow \mu^+\mu^-$ sounds very hard, but
  it might be possible to observe at high luminosities. For gluon
  fusion the Drell--Yan background $Z \to \mu^+ \mu^-$ is very hard to
  battle only using the reconstructed mass of the muon pair. The two
  tagging jets and the central jet veto very efficiently remove the
  leading $Z+$jets backgrounds and leave us with mostly the $G_F^3$
  process. However, because there is no single highly efficient
  background rejection cut this analysis will require modern analysis
  techniques.
\item[--] $qq \to qqH, H\rightarrow$~invisible is the only discovery
  channel for an invisible Higgs which really works at the LHC. It
  relies on the pure tagging-jet signature, which means it is
  seriously hard and requires a very good understanding of the
  detector and of QCD effects. The irreducible $Z \to \nu \bar{\nu}$
  background is not negligible and has to be controlled essentially
  based on its QCD properties. Jet kinematics as well as jet counting
  are the key elements of this analysis.
\end{itemize}
\bigskip

Just a side remark for younger LHC physicists: weak boson fusion was
essentially unknown as a production mode for light Higgses until
around 1998 and started off as a very successful PhD project. This
meant that for example the Higgs chapter in the ATLAS TDR had to be
re-written. While it sometimes might appear that way, there is no such
thing as a completely understood field of LHC physics. Every aspect of
LHC physics continuously moves around and your impact only depends on
how good your ideas and their technical realizations are.

\subsection{Associated Higgs production}
\label{sec:higgs_wh}

In Figure~\ref{fig:higgs_rates} there appears a third class of
processes at smaller rates: associated Higgs production with heavy
Standard Model particles, like $W$ or $Z$ bosons or $t\bar{t}$ pairs.
Until the summer of 2008 the Higgs community at the LHC was convinced
that (a) we would not be able to make use of $WH$ and $ZH$ production
at the LHC and (b) we would not be able to see most of the light Higgs
bosons produced at the LHC, because $H \to b\bar{b}$ is no
promising signature in gluon fusion or weak boson fusion production.

One key to argument (a) are the two different initial states for
signal and background: at the Tevatron the processes $q\bar{q} \to
~ZH$ and $q'\bar{q} \to WH$ arise from valence quarks. At the LHC with
its proton--proton beam this is not possible, so the signal rate will
suffer when we go from Tevatron to LHC. The QCD background at leading
order is $q \bar{q} \to Zg^* \to Z b \bar{b}$ production, with an
off--shell gluon splitting into a low mass bottom quark pair. At
next--to--leading order, we also have to consider the $t$-channel
process $q \bar{q} \to Z \bar{b} b g$ and its flipped counter part $q
g \to Z \bar{b} b q$. This background becomes more dangerous for
larger initial--state gluon densities. Moving from
Tevatron to LHC the Higgs signal will decrease while the background
increases --- not a very promising starting point.\bigskip

With Ref.~\cite{Butterworth:2008iy} the whole picture changed. We will
discuss this search strategy in detail in Section~\ref{sec:sim_fatjet}
in the context of jets and jet algorithms at the LHC. It turns out
that searches for \underline{boosted Higgs bosons}\index{jet!Higgs tagger} are not only promising in the $VH, H \to b\bar{b}$ channel,
but might also resurrect the $t\bar{t}H, H \to b\bar{b}$
channel. These new channels are not yet included in the ATLAS list of
processes shown in Figure~\ref{fig:higgs_sig} because the simulations
are still at an early stage. But we can expect them to play a role in
LHC searches for a light Higgs boson.

This is another example of what is possible and what not in LHC
phenomenology: it all really depends only on how creative you are;
that even applies to a field like Standard Model Higgs searches, which
is supposedly studied to death.

\subsection{Beyond Higgs discovery}
\label{sec:higgs_beyond}

The prime goal of the LHC was to discover a new light scalar particle
which we could then experimentally confirm to be a Higgs boson, either
as predicted by the Standard Model or with modifications due to new
physics. This has worked great. The discovery of a new particle which
was predicted in 1964 purely on the grounds of quantum field theory
gives us great confidence in field theories as a description of
elementary particles. For the description of this new state we will
therefore consistently rely on the Lagrangian as the most basic object
of perturbative field theory.\bigskip

The Standard Model Lagrangian makes many predictions concerning the
properties of a Higgs boson; as a matter of fact, all its properties
except for its mass are fixed in the minimal one-doublet Higgs sector
of the Standard Model. The question is, can we test at least some of
these predictions?

In this section we will briefly touch on a few interesting questions
relevant to the Higgs Lagrangian.  This is where we have seen the most
progress in LHC Higgs physics over recent years: not only will we be
able to see a light Higgs boson simultaneously in different production
and decay channels, as discussed in Sections~\ref{sec:higgs_gf}
and~\ref{sec:higgs_wbf}, we can also study many of its properties. In
a way this section ties in with the effective theory picture we use to
introduce the Higgs mechanism: the obvious requirements to include
massive gauge bosons in an effective electroweak gauge theory leads us
towards the Standard Model Higgs boson only step by step, and at any
of these steps we could have stopped and postulated some alternative
ultraviolet completion of our theory.

\subsubsection{Coupling measurement}
\label{sec:higgs_couplings}

In Section~\ref{sec:higgs_discovery} we present the ATLAS Higgs
discovery paper in some detail. While it is clear that the observed
$ZZ \to 4\ell$, $\gamma \gamma$, and $WW
\to 2\ell \, 2\nu$ signals point towards the discovery of a Higgs
boson, the nature of the observed excess is not at all clear. For the
statistical analysis leading to the discovery the interpretation plays
no role. Only some very preliminary information on the observed
resonance can be deduced from the fact that it appears in analyses
which are designed to look for the Higgs boson.\bigskip

As a first step in analyzing the Higgs Lagrangian we can for example
assume the operator structure of the Standard Model and ask the
question how well each of the associated couplings agrees with the
Standard Model prediction.  If we for a moment forget about the
ultraviolet completion of our electroweak theory we observe a scalar
particle with a mass around 125~GeV which couples to $W$ and $Z$
bosons, photons, and probably gluons. To describe the Higgs discovery
in terms of a Lagrangian we need at least these four terms.  Because
the Higgs mechanism breaks the weak gauge symmetry the individual
operators in terms of the Higgs field do not have to be gauge
invariant. Moreover, the coupling measurement in the Standard Model
Lagrangian mixed renormalizable couplings to massive gauge bosons and
fermions with loop--induced couplings to massless gauge bosons. This
mix of renormalizable and dimension-6 operators cannot be expected to
descent from a proper effective field theory. If we are interested in
the ultraviolet structure of this free couplings model we can for
example consider the observed Higgs particle the lightest state in an
extended Higgs sector. This way its couplings are allowed to deviate
from the Standard Model predictions while renormalizability and
unitarity are ensured once we include the additional, heavier Higgs
states.\bigskip

Alternatively, we can define an effective field theory based on all
possible Higgs operators to a given mass dimension. In the case of a
linear representation this Lagrangian will be based on the Higgs
doublet $\phi$ and by construction $SU(2)_L$ gauge invariant.  Some of
the dimension-6 potential operators forming this effective theory we
study in Section~\ref{sec:higgs_pot}. Couplings to $W$ and $Z$ bosons
start at mass dimension four, but there exist of course
higher--dimensional operators linking the same particles; the same is
true for Yukawa couplings. Higgs couplings to gluons and photons, as
derived in Section~\ref{sec:higgs_gluon} and
Section~\ref{sec:higgs_gamma}, start at dimension six and could be
supplemented by operators with even higher mass dimension. In an even
more general approach we do not even assume that the Higgs boson has
spin zero. Instead, we can define effective theories also for spin-one
and spin-two Higgs impostors.

Each operator on this extensive list we can equip with a free coupling
factor, and this set of couplings we can fit to the LHC
measurements. Note that we really mean `measurements' and not `event
rates', because different operators lead to different kinematic
behavior and hence significantly different efficiencies in the LHC
analyses. As an example we can quote the angle between the two leptons
in the $H \to WW$ analysis or the structure of the tagging jets, which
work best for a spin-zero Higgs. Of course, adding all kinds of
kinematic distributions will add a huge amount of additional
information to the Lagrangian determination, but it is obvious that at
least at this stage such an measurement is unrealistic.\bigskip

Because of this complication we return to the original question,
comparing the LHC results to the \underline{Standard--Model--like
  Higgs Lagrangian}. The different operators essentially fix all Higgs
quantum numbers, the production and decay kinematics, and the
experimental efficiencies.  In our first attempt we will assume a
CP-even scalar Higgs boson, where dimension-4 terms in the Lagrangian
in general dominate over higher--dimensional terms. Effective couplings
to gluons and photons are included at dimension-6 because tree level
dimension-4 couplings do not exist. Deviations from this assumptions
we discuss in Section~\ref{sec:higgs_cp}.  Higgs potential terms
including the triple and quartic self-couplings we can ignore for now,
because LHC analyses will not be sensitive to them for a while. Again,
we will discuss possible measurements of the Higgs self-coupling in
Section~\ref{sec:higgs_self}.  Our basic Lagrangian with Standard
model couplings is a combination of Eq.\eqref{eq:lag_wwh},
Eq.\eqref{eq:lag_ffh}, and Eq.\eqref{eq:higgs_eff1}:
\begin{alignat}{5}
\lag 
\supset& \, \frac{1}{2} \left( \p_\mu H \right)^2 
- g m_W \, H W_\mu ^+W^{-\mu} 
- \frac{g m_Z}{2 c_w} H Z_\mu Z^\mu 
\notag \\
&+ g_{\gamma \gamma H} \frac{H}{v} A^{\mu \nu} A_{\mu \nu} 
+ g_{ggH} \frac{H}{v} G^{\mu \nu} G_{\mu \nu} 
- \sum_\text{fermions} \frac{y_f}{\sqrt{2}} \; H \psib_f \psi_f 
\notag \\
\equiv& \, \frac{1}{2} \left( \p_\mu H \right)^2 
- g_W \, H W_\mu ^+W^{-\mu} 
- g_Z \, H Z_\mu Z^\mu 
+ g_\gamma \, H A^{\mu \nu} A_{\mu \nu} 
 + g_g \, H G^{\mu \nu} G_{\mu \nu} 
 - \sum_\text{fermions} g_f \, H \psib_f \psi_f \; .
\label{eq:higgs_lag_coup}
\end{alignat}
In the second line we have defined each Higgs coupling to a particle
$x$ as $g_x$, which is commonly done in Higgs couplings analyses.
Given this ansatz we can measure the \underline{Higgs couplings} or
ratios of them, for example in relation to their Standard Model values
\begin{alignat}{5}
\boxed{ 
g_x
= \left( 1 + \Delta_x \right) \; g_x^\text{SM} 
}
\qqquad \text{and} \qqquad
\frac{g_x}{g_y}
= \left( 1 + \Delta_{x/y} \right) \; \left( \frac{g_x}{g_y} \right)^\text{SM} \; .
\label{eq:higgs_defcoup1}
\end{alignat}
For the effective couplings we need to separate the parametric
dependence on the Standard Model couplings which enter the one-loop
expression, so we find for example
\begin{alignat}{5}
g_\gamma = 
\left( 1 + \Delta_\gamma^\text{SM} + \Delta_\gamma \right) \; g_\gamma^\text{SM} \; .
\label{eq:higgs_defcoup2}
\end{alignat}
The contribution $\Delta_\gamma^\text{SM}$ is the deviation of the
effective couplings based on a possible shift in the values of the top
and $W$ couplings consistent with their tree level counterparts. This
term is crucial once we extract all Higgs couplings from data
consistently. Our ansatz in terms of the $\Delta_x$ is motivated by
the fact that at first sight the observed Higgs signal seems to be in
rough agreement with a Standard Model Higgs boson, so the measured
values of $\Delta_x$ should be small.\bigskip

\begin{figure}[t]
\begin{center}
\includegraphics[width=0.40\hsize]{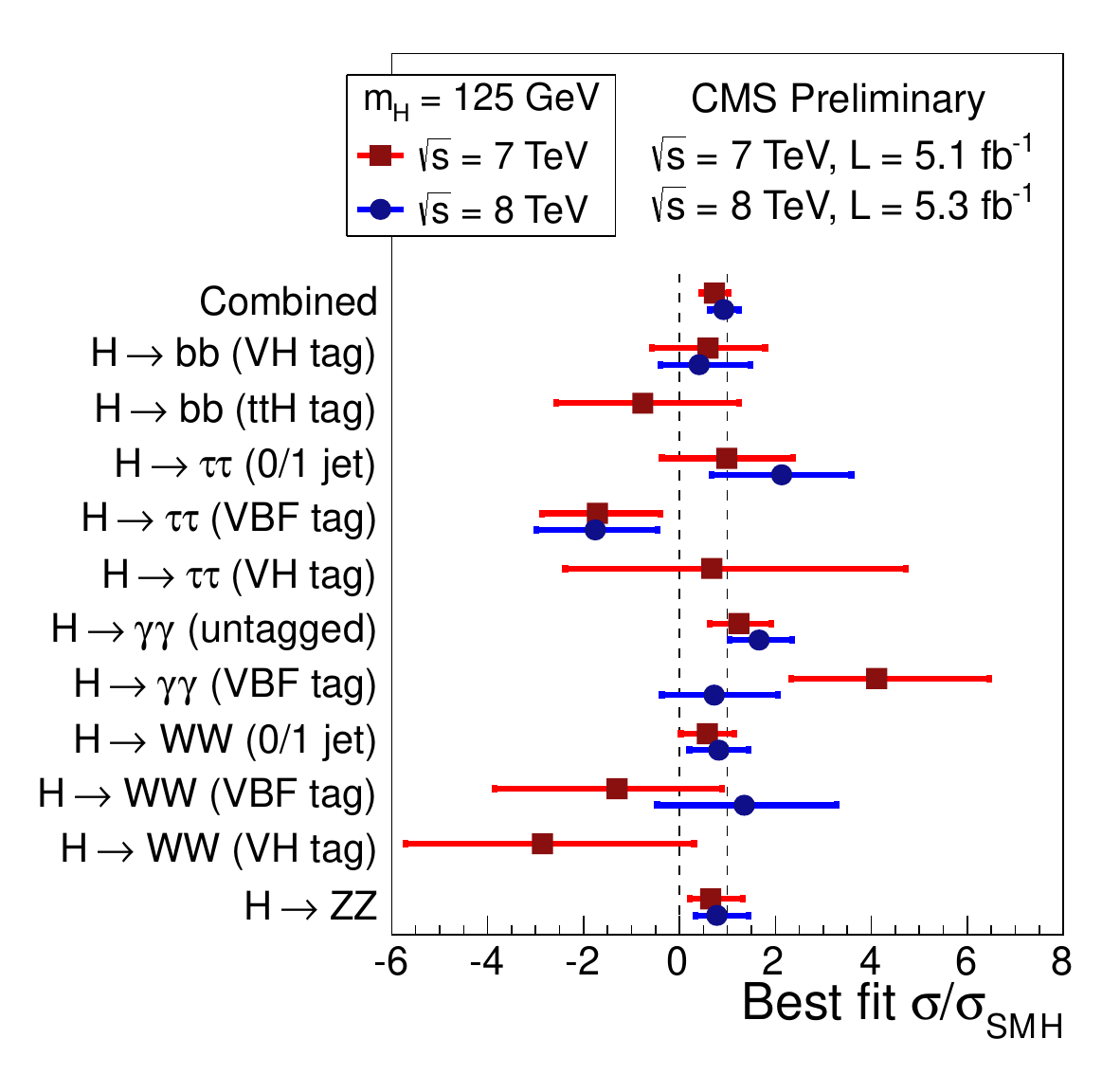}
\end{center}
\caption{Event rates for the different Higgs signatures relative to
  the Standard Model expectations as measured by CMS. Figure from the
  supplementary material to Ref.~\cite{cms_discovery}.}
\label{fig:higgs_coup0}
\end{figure}

A serious complication in the Higgs coupling extraction from event
numbers arises through the \underline{total Higgs width}. At the LHC, we will
mainly measure event rates in the different Higgs channels as shown in
Figure~\ref{fig:higgs_coup0}. Even though it enters all event rates we
will not be able to measure the width of a light Higgs boson as long
as it does not exceed $\ope(1~\gev)$. For small deviations from the
Standard Model couplings this means that we need to construct the
total Higgs width from other observables. The functional dependence of
the event count in a production channel $p$ and a decay channel $d$ is
\begin{alignat}{5}
N_\text{events} 
= \epsilon \times \sigma_p \times \br_d
\; \sim \; \frac{g_p^2 g_d^2}{\Gamma_\text{tot}(\{g_j\})} \; .
\label{eq:higgs_coup_nev}
\end{alignat}
The combined efficiencies and the fiducial detector volume we denote
as $\epsilon$. The couplings entering the production and decay
channels we denote as $g_{p,d}$. The total width, defined as the sum
of all partial widths, depends on all relevant Higgs couplings. This
functional behavior means that any LHC event number will depend on the
Higgs couplings $g_j$ in a highly correlated way, highly sensitive to
what we assume for the unobservable Higgs width. An interesting
question is if we can scale all Higgs couplings simultaneously without
affecting the observables. This means
\begin{alignat}{5}
N_\text{events} 
= \lim_{g \to 0} \; \frac{g^4}{\Gamma_\text{tot}}
= \lim_{g \to 0} \; \dfrac{g^4}{g^2 \dfrac{\Gamma_\text{obs}}{g^2} + \Gamma_\text{unobs}}
= \lim_{g \to 0} \; \dfrac{g^4}{g^2 \dfrac{\Gamma_\text{obs}}{g^2} + \Gamma_\text{unobs}}
= 0 \; .
\end{alignat}
The total width we have generally split into observable and
unobservable channels, where the observable channels scale like
$g_d^2$. This means that event numbers are sensitive to more than just
the ratio of Higgs couplings. Nevertheless, we need to make some kind
of assumption about the Higgs width.

The above argument suggests a theoretically sound assumption we can
make on the total Higgs width, once we observe a number of Higgs decay
channels and measure the underlying Higgs coupling
\begin{alignat}{5}
\Gamma_\text{tot} > \sum_\text{observed} \Gamma_j \; .
\label{eq:higgs_width1}
\end{alignat}
Each partial Higgs width is independently computed and corresponds to
a positive transition amplitude squared. If we assume the Higgs
Lagrangian Eq.\eqref{eq:higgs_lag_coup} we can compute each partial
width in terms of its unknown coupling.  There are no interference
effects between Higgs decay channels with different final--state
particles, so the total width is strictly bounded from below by the
sum of observed channels.

A slightly more tricky assumption gives us an upper limit on at least
one Higgs partial width. From our calculation in
Section~\ref{sec:higgs_unitarity} we know that the Standard Model
Higgs boson unitarizes the $WW \to WW$ scattering rate. If we
overshoot for example in the $s$-channel Higgs exchange we would need
an additional particle which compensates for this effect. However, the
amplitude of an such additional particle would be proportional to its
Higgs coupling squared, which means it is not clear where the required
minus sign would come from. Taken with a grain of salt this tells us
\begin{alignat}{5}
g_W < g_W^\text{SM} 
\qqquad \text{or} \qqquad
\Gamma_{H \to WW} < \Gamma_{H \to WW}^\text{SM} \; .
\end{alignat}
Given that correlations between different Higgs couplings will become
a problem in the coupling extraction such a constraint can be very
useful.\bigskip

In the analysis which we present in this section we do not assume an
upper limit to any Higgs partial width. Instead, we promote the
constraint in Eq.\eqref{eq:higgs_width1} to an exact relation:
\begin{alignat}{9}
\boxed{
\Gamma_\text{tot} = \sum_\text{observed} \; \Gamma_x(g_x) 
+ \text{2nd generation} < 2\,\gev
} \; .
\label{eq:higgs_width2}
\end{alignat}
Because at the LHC we will not observe any Higgs couplings to a second
generation fermion any time soon we correct for the charm quark
contribution using $g_c = m_c/m_t \times g_t^\text{SM} (1 +
\Delta_t)$. This avoids systematic offsets in the results. The total
upper limit on the Higgs width corresponds to very large individual
couplings and is an estimate of visible effects in the $H \to \gamma
\gamma$ analyses.\bigskip

\begin{table}[t]
\begin{center}
\begin{tabular}{l|cccc}
experiment & $H$ inclusive & $H$+2~jets & $H$+lepton(s) & $H$+top(s)
\\
theory & $gg \to H$ & $qq \to qqH$ & $q\bar{q} \to VH$ & $gg/q\bar{q} \to t\bar{t}H$ 
\\ \hline 
$H \to ZZ$ & 2011/2012 & $\ge$2015 & --- & ---
\\
$H \to \gamma \gamma$ & 2011/2012 & 2011/2012 & --- & ?
\\
$H \to WW$ & 2011/2012 & 2011/2012 & --- & ?
\\
$H \to \tau \tau$ & ? & 2012 & --- & ?
\\ 
$H \to \mu \mu$ & --- & ? & --- & ---
\\
$H \to b\bar{b}$ & --- & ? & $\ge$2015 & $\ge$2015
\\\hline
\end{tabular}
\end{center}
\caption{Higgs signatures with significant impact on the Higgs
  coupling determination in 2011/2012 and beyond. Question marks
  indicates that beyond 2015 these channels might contribute, but no
  reliable ATLAS or CMS analysis is currently available. In the top
  line we indicate the experimental signature while in the second line
  we show the leading production mode.}
\label{tab:higgs_coup}
\end{table}

In Table~\ref{tab:higgs_coup} we list the channels which we can rely
on in the extraction of the Higgs couplings. The details of the 2011
results in these channels we discuss with the Higgs discovery paper in
Section~\ref{sec:higgs_discovery}. Note that in
Table~\ref{tab:higgs_coup} we list the different Higgs production
channels which are theoretically well defined but not experimentally
observable. For example, the $H \to \gamma\gamma$ analysis is separated
into the jet--inclusive and the two-jet analysis, where the inclusive
rate is dominated by the gluon--fusion production process while the
two-jet rate is dominated by weak boson fusion. Once we can determine
the efficiencies for each of the production processes contributing to
the different analyses we can rotate the Higgs rates from the
experimental basis to the theoretical basis.  In the 2011/2012 data set
we then have six observable rates from which we want to extract as
many Higgs couplings as possible. Before we discuss the physics result
of such a coupling extraction let us introduce some of the
techniques.\bigskip

The naive approach to a parameter extraction is a \underline{$\chi^2$ minimization}, 
experimentally known as `running MINUIT'\index{Minuit}. The
variable $\chi^2$ measures the quality of a fit or the quality of
the theoretical assumptions compared to the measurements for the best
possible set of theoretical model parameters.  Given an
$n_\text{meas}$-dimensional vector of measurements
$\vec{x}_\text{meas}$ and the corresponding model predictions
$\vec{x}_\text{mod}(\vec{m})$, which in turn depend on an
$n_\text{mod}$-dimensional vector of model parameters, we define
\begin{alignat}{5}
\chi^2 (\vec{m}) &= 
\sum_{j=1}^{n_\text{meas}} 
\frac{\left| \vec{x}_\text{meas} - \vec{x}_\text{mod}(\vec{m}) \right|_j^2}{\sigma_j^2} \; ,
\label{eq:higgs_chi2simple}
\end{alignat}
where $\sigma_j^2$ is the variance of the channel $\vec{x}_j$. The
best fit point in model parameter space is the vector $\vec{m}$ for
which $\chi^2(\vec{m})$ assumes a minimum.  In the
Gaussian limit, we can not only compute the minimum value of $\chi^2$
but compare the entire $\chi^2$ distribution for variable model
parameters to our expectation.  If the theoretical predictions
$\vec{x}_\text{mod}$ depend on $n_\text{mod}$ model parameters we can
define the normalized or reduced $\chi^2$ distribution
\begin{alignat}{5}
\chi^2_\text{red}(\vec{m}) 
= \frac{\chi^2(\vec{m})}{n_\text{meas} - n_\text{mod} + 1}
\equiv \frac{\chi^2(\vec{m})}{\nu} \; .
\end{alignat}
For this definition of the number of degrees of freedom
$\nu$ we assume $n_\text{meas} > n_\text{mod}+1$, which means that
the system is over--constrained and a measure for the quality of the
fit makes sense. If we find $\min \chi_\text{red} \sim 1$ the error
estimate of $\sigma_j$ entering Eq.\eqref{eq:higgs_chi2simple} is
reasonable. The problem with the $\chi^2$ test is that it requires us
to know not only the variance $\sigma^2_j$, but also the form of the
actual $\vec{x}_\text{meas}$ distributions to apply any test on
the final value of $\chi^2$. The natural distribution to assume is a
Gaussian, defined in Eq.\eqref{eq:higgs_gaussian} with the covariance
$\sigma^2$. Technically, a common assumption in
the determination of the best fit is that the functional form of
$\chi^2$ be quadratic. In that case we can compute the confidence
level of a distribution, as defined in
Section~\ref{sec:higgs_discovery} from the $\min \chi^2$ value.  

All this is only true in the Gaussian limit, which means that we cannot use
it in our Higgs couplings measurement. Some of the channels involved
have a very small event count in the signal or in the background, many
uncertainties are heavily correlated, and we have very little control
over the form of systematic or theoretical uncertainties.\bigskip

In Section~\ref{sec:sim_errors} we construct the general form of a
$\chi^2$ variable including full correlations and without making any
assumptions on the form of uncertainties,
\begin{alignat}{5}
\chi^2(\vec{m}) \longrightarrow 
-2 \log \lumi(\vec{m}) = \vec{\chi}_T C^{-1} \vec{\chi} \; ,
\end{alignat}
with the definition of $\vec{\chi}$ and $C$ given in
Eq.\eqref{eq:sim_flat_errors}. This variable, evaluated at a model
parameter point $\vec{m}$, is the \underline{log-likelihood}. The name
likelihood implies that it is some kind of probability, but evaluated
over model parameter space and not over experimental outcomes. This
means that the normalization of a likelihood is only defined once we
agree on an integration measure in model space. Only in the Bayesian
approach we ever do that.
Here, we construct a completely exclusive
log-likelihood map over the $n_\text{mod}$-dimensional model parameter
space and then reduce the dimensionality one--by--one to show profile
likelihoods for individual model parameters or their
correlations. We will give details on this procedure in Section~\ref{sec:sim_errors}. 
Error bars on the individual couplings can be extracted
through toy measurements. These are assumed measurements which we generate
to trace the individual uncertainty distribution for each channel and
which define a distribution of best fit values. Typically $10^3$ toy
measurements give us a sufficiently precise distribution around the
best--fitting point to construct the model space volume which contains
68\% of the toy measurement distribution below and above the measured
central value.\bigskip

Technically, the log-likelihood can be extracted as a
\underline{Markov chain}\index{Markov chain}. This is a set of parameter points which
represent their log-likelihood value in their density. The
construction of such a Markov chain is simple:
\begin{enumerate}
\item start with any model parameter point $\vec{m}_0$ and compute
  $\lumi(\vec{m}_0)$.
\item random generate a second model parameter point $\vec{m}_1$
  according to some suggestion probability. This probability cannot
  have any memory (detailed balance), should be peaked for $\vec{m}_1
  \sim \vec{m}_0$, at the same time give a decent probability to move
  through the parameter space, and does not have to be symmetric
  $q(\vec{m}_0 \to \vec{m}_1) \ne q(\vec{m}_1 \to \vec{m}_0)$, We
  often use a Breit--Wigner or Cauchy distribution defined in
  Section~\ref{sec:qcd_bw} with a width of 1\% of the entire parameter
  range.
\item accept new point as the next point in the Markov chain if $\log
  \lumi(\vec{m}_1) > \log \lumi(\vec{m}_0)$. Otherwise,
  accept with reduced probability $\log \lumi(\vec{m}_1)/\log
  \lumi(\vec{m}_0)$.
\item stop once the chain is sufficiently long to cover at least part
  of the parameter space. Obviously, we can combine several different
  Markov chains
\end{enumerate}
There are ways to improve such a Markov chain. First, for a Markov
chain which carries information only about the log-likelihood itself,
\ie used to estimate the same property we define it through, we
can keep the value $\lumi(\vec{m})$ as a weight to each
point. Weighted Markov chains improve the convergence of the final
result. Secondly, we can slowly focus on regions of the relevant
parameter regions with larger $\lumi$ values. Following the
general idea of simulated annealing\index{simulated annealing} we can use cooled Markov chains
which include two stages: in the early stage the Markov chain rapidly
covers large fractions of the model space while in a second stage it
zooms in on the regions with the largest likelihoods. Technically, we
change the constant acceptance criterion relative to a linear random
number $r$ to a varying
\begin{alignat}{5}
\frac{\log \lumi(\vec{m}_1)}{\log \lumi(\vec{m}_0)} > r
\qqquad \longrightarrow \qqquad
\left( \frac{\log \lumi(\vec{m}_1)}{\log \lumi(\vec{m}_0)} \right)^{j/10} > r 
\qquad \text{for} \quad j = 1...10 \; .
\end{alignat}
Because the ratio of log-likelihoods is smaller than unity, for $j=1$
the weighted ratio becomes larger and the new point is more likely to
be accepted. For larger values of $j$ the weighted ratio becomes
smaller, which means that the Markov chain will stay more and more
localized and mostly move to another point if that one is more
likely. Obviously, cooling Markov chains are only reliable if we run
several chains to check if they really cover all interesting
structures in model space. In the case of the Higgs analysis we use
$\ope(10)$ chains with $10^6$ parameter points each. These
$10^7$ likelihood values define our completely exclusive likelihood
map.\bigskip

With this technical background we can continue with our Higgs
couplings analysis.  From Eq.\eqref{eq:higgs_lag_coup} we know that in
our case the model parameters $\vec{m}$ will be the Higgs couplings
$g_x$ or their deviations from the Standard Model values
$\Delta_x$ as defined in Eq.\eqref{eq:higgs_defcoup1}.
The observables will be the event rates in the Higgs
search channels shown in Table~\ref{tab:higgs_coup}. The errors on
signal and background rates include statistical, systematic
(correlated), and theoretical (correlated) sources. The
\underline{SFitter} version of the Higgs couplings analysis proceeds
in a series of independent steps:
\begin{enumerate}
\item construct an exclusive likelihood map over the Higgs couplings
  model space.
\item deflate the parameter space to profile likelihoods for
  one-dimensional and two-dimensional distributions of couplings. This
  gives us a \underline{global picture} of the Higgs couplings space.
\item determine the best--fitting parameter point with high
  resolution. This means starting from the best point in the Markov
  chain and using a minimization algorithm on $\log \lumi(\vec{m})$.
\item determine the 68\% uncertainty band on each extracted couplings
  from toy measurements, defining a \underline{local picture} of the
  Higgs couplings.
\end{enumerate}
The benefits of this approach as well as the impressive progress in
the experimental Higgs analyses can be nicely illustrated when we look
at the coupling determination from 2011 and early 2012 data.\bigskip

\begin{figure}[t]
\begin{center}
\includegraphics[width=0.35\hsize]{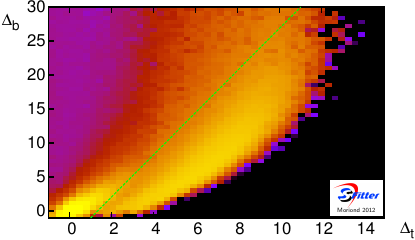}
\hspace*{0.1\textwidth}
\includegraphics[width=0.35\hsize]{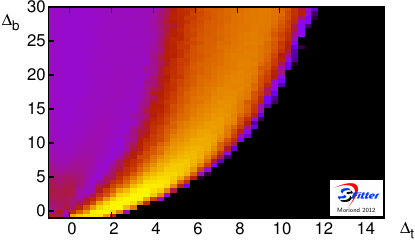}
\end{center}
\caption{Correlations of the top and bottom Yukawas extracted from the
  2011 data set with a center--of--mass energy of 7~TeV. The left panel
  shows the expected likelihood map in the Standard Model, the right
  panel shows the observed one. Figure from
  Ref.~\cite{sfitter_higgs7}.}
\label{fig:higgs_coup1}
\end{figure}

Based on 2011 data the number of measurements published by ATLAS and
CMS and shown in Moriond was relatively small. As listed in
Table~\ref{tab:higgs_coup} there should have been measurements
including a Higgs coupling to gluons and photons through effective
operators, plus couplings to $W$, $Z$, and tau leptons at the tree
level. Dealing with relatively early data we assume that the
higher--dimensional couplings are mediated by Standard Model top and
$W$ loops, with at best small corrections due to additional
states. Before we determine the individual couplings we can look at
the global picture, to make sure that everything looks roughly like
expected. In the left panel of Figure~\ref{fig:higgs_coup1} we show
the correlation between the top and bottom Yukawa couplings expected
for the 7~TeV run. In the $\Delta_{b,t} = 0...1$ range we see the
Standard Model coupling range. 

For large $\Delta_{b,t} \sim 5...10$ another solution appears. Its
main features is the simultaneous increase of both quark Yukawas. This
behavior can be explained by the indirect handles we have on each of
them. The top Yukawa is mostly measured through the effective
gluon--Higgs coupling while the bottom Yukawa enters the total
width. To keep for example the inclusive $H \to ZZ$ rate constant,
both couplings have to increase at roughly the same rate. For a
constant inclusive $H \to \gamma \gamma$ rate the discussion in
Section~\ref{sec:higgs_gamma} we also need to increase $\Delta_W$ at
the same rate. However, such an increase would be visible in the
inclusive and weak boson fusion $H \to WW$ channels. What we see in
Figure~\ref{fig:higgs_coup1} is that starting from Standard Model
values we expect all three couplings to increase. At some point
$\Delta_W$ hits the limits from other channels, so instead of further
increasing it switches back to its Standard Model value, $\Delta_b$
follows into the same direction, but $\Delta_t$ takes over the
increased effective photon-Higgs coupling with a different sign of
this effective couplings. From that point on $\Delta_b \sim \Delta_t$
can grow again. If we want to avoid the theoretical problem of a
hugely non--perturbative top Yukawa we can limit our couplings
extraction to the separable Standard-Model-like solution, as indicated
by the green line.

The problem with the same distribution extracted from 2011 data is
that this separations does not exist. This can be traced back to
essentially missing evidence for a Higgs boson decaying to $WW$ or to
$\tau\tau$. If in the argument above we remove the constraints from
visible and constraining $WW$ channels the two Yukawa couplings can
increase from $\Delta_{b,t} =0$ to huge values. The Standard Model and
the large couplings solutions are blended together. The global picture
tells us that we should expect any reasonable Yukawa coupling
measurements from 2011 data even if we allow for an indirect
determination from the higher--dimensional couplings.\bigskip

In the left panel of Figure~\ref{fig:higgs_coup2} we show the results
from the local analysis and see exactly what we expect from the global
picture: a universal Higgs coupling modification
\begin{alignat}{5}
\Delta_x \equiv \Delta_H \qquad \forall x
\label{eq:higgs_formfac}
\end{alignat}
is determined exactly as we expect from a set of measurements in agreement
with the Standard Model. In contrast, $\Delta_W
\sim \Delta_\tau \sim -1$, which means that there is no evidence for
such a coupling in the 2011 data set. The Higgs gauge coupling to $Z$
bosons is roughly what one would have expected, while the quark
Yukawas are very poorly determined.\bigskip

In the right panel of Figure~\ref{fig:higgs_coup2} we repeat the same
analysis with all data from the discovery papers published in
the Summer of 2012. As mentioned in Section~\ref{sec:higgs_discovery},
for ATLAS this includes all results presented in the talks on 4th of
July, plus an improved $H \to WW$ analysis.

\begin{figure}[t]
\begin{center}
\includegraphics[width=0.40\hsize]{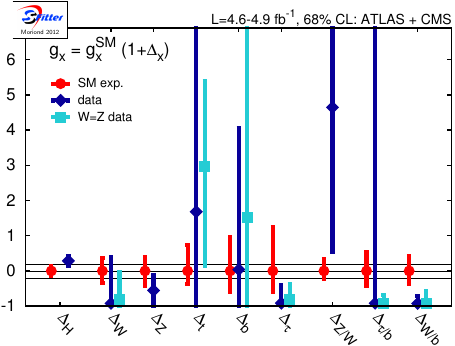}
\hspace*{0.1\textwidth}
\includegraphics[width=0.40\hsize]{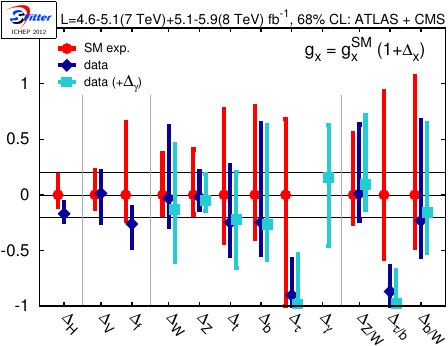}
\end{center}
\caption{Left: Higgs couplings extracted from the 2011 run at
  7~TeV. Right: couplings extracted from all data published in the
  Higgs discovery papers by ATLAS and CMS.  Figures from
  Refs.~\cite{sfitter_higgs7,sfitter_higgs8}.}
\label{fig:higgs_coup2}
\end{figure}

First, the measurements of a common Higgs form factor for all
couplings defined in Eq.\eqref{eq:higgs_formfac} has improved to a
20\% level, in very good agreement with the Standard Model. If we want
to separate this tree level form factor into a universal bosonic and a
universal fermionic coupling modification we see that both of them are
determined well. All these measurements are within our expectations
from the Standard Model for the central values as well as for the
uncertainties. The full set of couplings we can determine either
limiting the effective Higgs--photon couplings to Standard Model loops
or allowing for an additional contribution $\Delta_\gamma$ defined in
Eq.\eqref{eq:higgs_defcoup2}. It turns out that in the absence of any
direct quark Yukawa measurements we have no sensitivity to an
additional Higgs--gluon contribution.\bigskip

The most important result of the 2011/2012 coupling analysis is that
the Higgs couplings are typically determined at the 20\% to 50\%
level. The only coupling measurement still missing entirely is
$\Delta_\tau$. Allowing for a free Higgs--photon coupling does not
affect the other measurements significantly. The central value of
$\Delta_W$ decreases just sightly, allowing for a barely non-zero
central value of $\Delta_\gamma$. The bottom line that there is no
anomaly in the Higgs coupling to photons is in apparent disagreement
with other analyses. The difference is that in the SFitter results
shown here all Higgs couplings are allowed to vary in parallel. If the
event rate for $pp \to H \to \gamma \gamma$ is slightly high,
we do not simply translate this effect into a positive
value of $\Delta_\gamma$. Instead, we allow the top and $W$ couplings
to absorb some of this effect within their uncertainties from other
measurements. Only the part of the $\gamma \gamma$ anomaly which is
not compatible with the other Higgs channels then contributes to the
$\Delta_\gamma$ measurement shown in
Figure~\ref{fig:higgs_coup2}.\bigskip

Without showing the numerical outcome we state that the quality of the
fit, \ie the log-likelihood value in the best--fitting parameter point,
at this stage includes no useful information. Essentially any model of
Higgs couplings, with the exception of a chiral 4th generation, fits
the data equally well.  As indicated in Table~\ref{tab:higgs_coup} we
expect significant improvements of the Higgs coupling measurements
from the 2015 LHC run and beyond. This will mostly affect the highly
sensitive weak boson fusion signatures and a direct measurement of the
quark Yukawas based on fat jet techniques described in
Section~\ref{sec:sim_fatjet}. Any kind of measurement at the per-cent
level or better makes a very good case for a linear collider, running
at enough energy to study the $e^+ e^- \to ZH$, $t\bar{t}$ and
$t\bar{t}H$ production channels.

\subsubsection{Higgs quantum numbers}
\label{sec:higgs_cp}

One question for the LHC is: once we see a Higgs boson at the LHC, how
do we test its quantum numbers?  One such quantum number is its spin,
which in the ideal case we want to determine channel by channel in the
Higgs observation. Three questions people usually ask really mean the
same thing: 
\begin{itemize}
\item[--] what is the spin of the Higgs boson?  
\item[--] is the new resonance responsible for electroweak symmetry breaking?  
\item[--] what is the form of the Higgs operators in the Lagrangian? 
\end{itemize}
Given that the discovery of the
Higgs boson is an impressive confirmation that fundamental
interactions are described by quantum field theory, specifically by
gauge theories, the last version of the question is the most
appropriate. Lagrangians are what describes fields and interactions in
such a theory, and what we need to determine experimentally is the
Lagrangian of the observed Higgs boson. In that sense, the coupling
measurement described in the last section assumes the Higgs Lagrangian
of the Standard Model and measures the couplings of the corresponding
Higgs interaction terms.\bigskip

As far as the spin of the new particle is concerned, the spin-one case
is easily closed: following the Landau--Yang theorem a spin-one Higgs
boson would not couple to two photons, so in the photon decay channel
we only need to look at angular distributions for example in weak
boson fusion to distinguish spin zero from spin two.

Once we know that we are talking about a scalar field there are a few
options, linked to the CP properties of the new particle. The part of
the Higgs Lagrangian we are most interested in is the coupling to the
massive gauge bosons. In the Standard Model, the fundamental CP-even
Higgs boson couples to the $W$ and $Z$ bosons proportional to
$g^{\mu\nu}$. For general \underline{CP-even} and
\underline{CP-odd}\index{Higgs coupling!dimension-6 CP basis} Higgs
bosons there are two more ways to couple to $W$ bosons
using gauge invariant dimension-6 operators. If inside the dimension-6
operators we replace $(\phi^\dagger \phi)$ by the linear Higgs terms
we are interested in, we arrive at the dimension-5 Lagrangian
\begin{alignat}{5}
\lag \supset
  - g m_W \,H \, W_\mu W^\mu 
  -\frac{g_\text{D5}^+ v}{\Lambda^2} \; H \, W_{\mu\nu} W^{\mu\nu}
  -\frac{g_\text{D5}^- v}{\Lambda^2} \; A \, W_{\mu\nu} \widetilde W^{\mu\nu} \; .
\end{alignat}
In this notation $H$ is the scalar Higgs boson, while $A$ is a
pseudo-scalar. The coupling to $Z$ bosons is completely analogous to
the $W$ case, where $W^{\mu\nu}$ indicates the field strength tensor
and $\widetilde V^{\mu\nu}$ its dual. This set of gauge--invariant
terms in the Lagrangian can be translated into Feynman rules for the
Higgs coupling to massive gauge bosons. Their tensor structures are
\begin{alignat}{5}
\ope^+_\text{SM} = g^{\mu \nu}
  \qqquad \qqquad 
\ope^+_\text{D5} = g^{\mu\nu} - \frac{p_1^\mu p_2^\nu}{p_1 p_2}
  \qqquad \qqquad 
\ope^-_\text{D5} = \epsilon_{\mu\nu\rho\sigma} \; p_1^\rho p_2^\sigma \; .
\label{eq:higgs_coup}
\end{alignat}
These are the only gauge invariant dimension-5 couplings of two gauge
bosons to a (pseudo-) scalar field. The second one appears in the
effective one-loop gluon--gluon--Higgs coupling in
Eqs.\eqref{eq:higgs_eff1} and~\eqref{eq:higgs_eff2}. This second
tensor is not orthogonal to $g^{\mu\nu}$, but we can replace it with
The any linear combination with $g^{\mu\nu}$.  However, if we trust our
description in terms of a Lagrangian we obviously prefer the
gauge--invariant basis choice.\bigskip

The traditional observables reflecting the coupling structure of a
massive state decaying to two weak gauge bosons are the
\underline{Cabibbo--Maksymowicz--Dell'Aquila--Nelson angles}. They are
about the same age as the Higgs boson. We define them in the fully
reconstructable decay $X\to ZZ\to e^+ e^- \mu^+ \mu^-$. We already
know that one of the two $Z$ bosons will be far off its mass shell,
which does not affect the analysis of the decay angles. The four
lepton momenta reconstructing the Higgs--like state $X$ are given by
\begin{alignat}{5}
p_X = p_{Z_e}+p_{Z_\mu} 
\qqquad 
p_{Z_e} = p_{e^-}+p_{e^+} 
\qqquad 
p_{Z_\mu} = p_{\mu^-}+p_{\mu^+} \; .
\end{alignat}
Each of these momenta as well as the beam direction we can boost into
the $X$ rest frame and the two $Z_{e,\mu}$ rest frames, defining the
corresponding three-momenta $\hat{p}_i$.  The spin and CP angles are
then defined as
\begin{alignat}{5}
&\cos \theta_e =   
  \hat{p}_{e^-} \cdot\hat{p}_{Z_\mu} \Big|_{Z_e} 
&\qquad 
&\cos \theta_\mu =   
  \hat{p}_{\mu^-} \cdot\hat{p}_{Z_e} \Big|_{Z_\mu} 
\qqqquad 
\cos \theta^* = 
  \hat{p}_{Z_e} \cdot \hat{p}_\text{beam} \Big|_X  \notag \\
&\cos \phi_e = 
  (\hat{p}_\text{beam} \times \hat{p}_{Z_\mu}) \cdot (\hat{p}_{Z_\mu} \times \hat{p}_{e^-}) \Big|_{Z_e} 
  &\qquad
&\cos \Delta \phi = 
  (\hat{p}_{e^-} \times \hat{p}_{e^+}) \cdot (\hat{p}_{\mu^-} \times \hat{p}_{\mu^+}) \Big|_X \; .
\label{eq:angles_zz}
\end{alignat}
The index indicates the rest frame where the angles are defined. To
distinguish the different spin-zero hypotheses the angular difference
$\Delta \phi$ is most useful.  Looking at each of the decaying $Z$
bosons defining a plane opened by the two lepton momenta it is the
angle between these two planes in the Higgs rest frame is $\Delta \phi$. Its
distribution can be written as
\begin{alignat}{5}
\frac{d\sigma}{d \Delta\phi} \propto 1+a\cos \Delta \phi + b\cos (2 \Delta \phi) \; .
\end{alignat}
For the CP-odd Higgs coupling to $W^{\mu\nu}\widetilde{W}_{\mu\nu}$ we
find $a=0$ and $b=1/4$, while for the CP-even Higgs coupling
$g^{\mu\nu}$ we find $a>1/4$ depending on $m_H$. Some example
distributions for the decay planes angle we show in
Figure~\ref{fig:higgs_cp}.\bigskip

This method only works if we observe the decay $H\rightarrow ZZ
\rightarrow 4 \ell$ with a good signal--to--background ratio $S/B$.
We can derive an
alternative observable from studying its Feynman
diagrams: starting from the $H\rightarrow
ZZ\rightarrow 4 \ell$  decay topology we can switch two fermion legs from the final
state into the initial state
\begin{equation*}
\parbox{30mm}{
\begin{fmfgraph*}(80,50)
 \fmfset{arrow_len}{2mm}
 \fmfright{out1,in1,in2,out3}
 \fmf{fermion,width=0.5,tension=.5}{in1,v1}
 \fmf{fermion,width=0.5,tension=.5}{in2,v2}
 \fmf{fermion,width=0.5,tension=.5}{v1,out1}
 \fmf{fermion,width=0.5,tension=.5}{v2,out3}
 \fmf{photon,width=0.5}{v1,v3}
 \fmf{photon,width=0.5}{v2,v3}
 \fmf{dashes,width=0.5,tension=2}{v3,out2}
 \fmfleft{out2}
\end{fmfgraph*}
} \qquad \longrightarrow \qquad 
\parbox{30mm}{
\begin{fmfgraph*}(80,50)
 \fmfset{arrow_len}{2mm}
 \fmfright{in1,in2}
 \fmf{fermion,width=0.5,tension=1.2}{in1,v1}
 \fmf{fermion,width=0.5,tension=1.2}{in2,v2}
 \fmf{fermion,width=0.5,tension=.8}{v1,out1}
 \fmf{fermion,width=0.5,tension=.8}{v2,out3}
 \fmf{photon,width=0.5}{v1,v3}
 \fmf{photon,width=0.5}{v2,v3}
 \fmf{dashes,width=0.5,tension=.8}{v3,out2}
 \fmfleft{out1,out2,out3}
\end{fmfgraph*}
}
\end{equation*}
and read the diagram right--to--left. This gives us the Feynman diagram
for weak boson fusion Higgs production.  The angle between the decay
planes gets linked to the angular correlation of two forward the
tagging jets.  Its advantage is that it gives us a production-side
correlation, independent of the Higgs decay signature.

Going back to the transverse momentum spectra for the tagging
jets\index{tagging jet} shown in Eq.\eqref{eq:higgs_ptj} we already
know one way of telling apart these couplings: the dimension-3 $WWH$
coupling proportional to $g^{\mu \nu}$ comes out of the Higgs
potential as a side product of the $W$ mass term, \ie it couples the
longitudinal Goldstone modes in the $W$ boson to the Higgs. In
contrast, the CP-even dimension-6 operator is proportional to the
transverse tensor\index{transverse tensor}, which means it couples the
transverse $W$ polarizations to the Higgs and therefore produces a
harder $p_T$ spectrum of the tagging jets. The problem with such an
observation is that in the absence of a unitarizing Higgs scalar in
$WW$ scattering we should expect our theory to generally include
momentum dependent form factors instead. Any observable with units of
energy will become sensitive to these form factors.\bigskip

\begin{figure}[t]
 \includegraphics[width=0.3\textwidth]{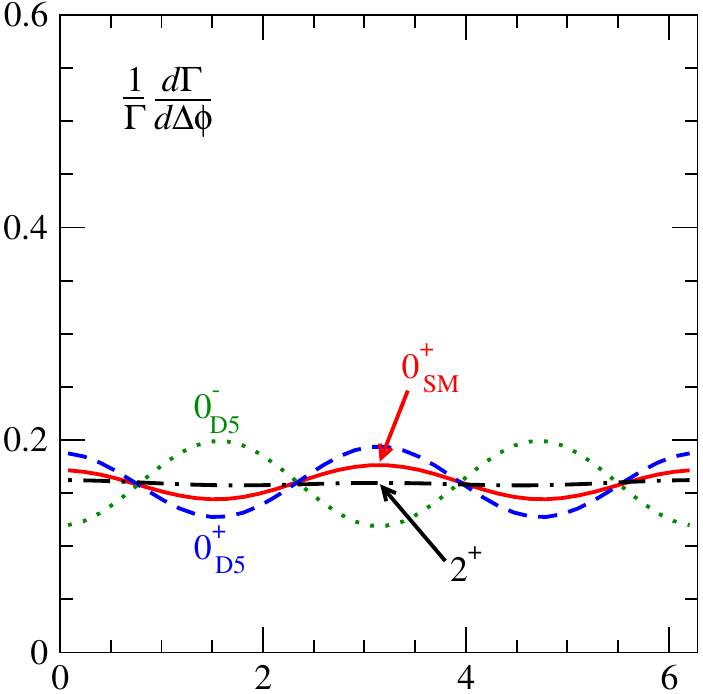}
 \hfill
 \includegraphics[width=0.3\textwidth]{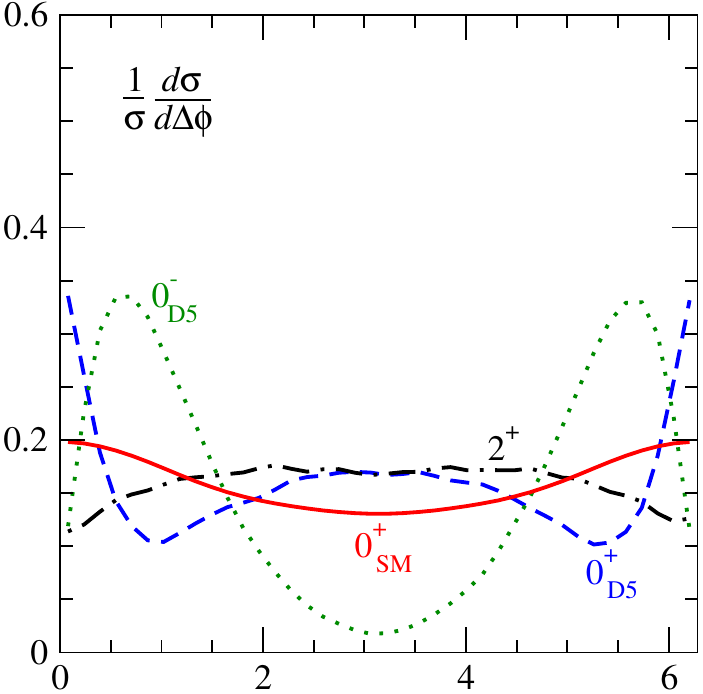}
 \hfill
 \includegraphics[width=0.3\textwidth]{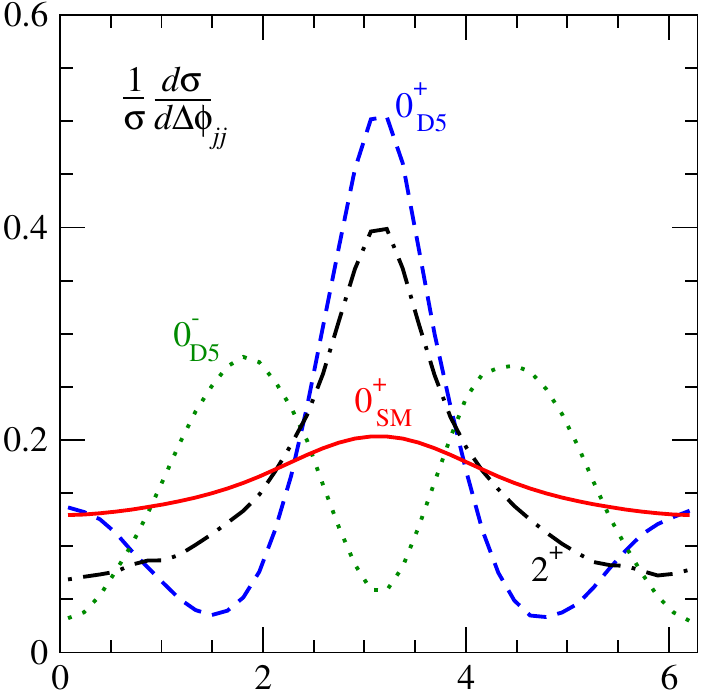}
\caption{$\Delta\phi$ distributions for $X\to ZZ$ events (left), weak boson fusion
  production in the Breit frame (center), and in the
  laboratory frame (right). Figures from
  Refs.~\cite{higgs_coupl,higgs_coupl_2}.}
\label{fig:higgs_cp}
\end{figure}

Sticking to angular observables, along the lines of
Eq.\eqref{eq:angles_zz}, we can first translate all decay angles into
the weak boson fusion topology. The problem in this approach are the
rest frames: a $W$ boson in the $t$-channel is space--like, implying $p_W^2
\equiv t < 0$. This means that we cannot boost into its rest
frame. What we can do is boost into the so-called \underline{Breit
  frame}\index{Breit frame}.  It is defined as the frame in
which the momentum of the $t$-channel particle only has space--like
entries and can be reached through a conventional boost. 

Writing the weak boson fusion momenta as $q_1 q_2 \to j_1 j_2 \, (X
\to d \bar{d} )$ we can define a modified version of the five angles
in Eq.\eqref{eq:angles_zz}, namely
\begin{alignat}{5}
&\cos \theta_1 =   
   \hat{p}_{j_1} \cdot\hat{p}_{V_2} \Big|_{V_1 \text{Breit}} 
&\qquad 
&\cos \theta_2 =   
   \hat{p}_{j_2} \cdot\hat{p}_{V_1} \Big|_{V_2 \text{Breit}} 
\qqqquad 
\cos \theta^* = 
  \hat{p}_{V_1} \cdot \hat{p}_{d} \Big|_X  \notag \\
&\cos \phi_1 = 
  (\hat{p}_{V_2} \times \hat{p}_{d}) \cdot (\hat{p}_{V_2} \times \hat{p}_{j_1}) \Big|_{V_1 \text{Breit}} 
  &\qquad
&\cos \Delta \phi = 
  (\hat{p}_{q_1} \times \hat{p}_{j_1}) \cdot (\hat{p}_{q_2} \times \hat{p}_{j_2}) \Big|_X \; .
\label{eq:angles_wbf}
\end{alignat}
In addition, we define the angle
$\phi_+\equiv 2\phi_1 +\Delta \phi $ which typically shows a
modulation for spin-two resonances. In Figure~\ref{fig:higgs_cp} we see
how $\Delta \phi$ in the Breit frame is closely related to the angle
between the $Z$ decay planes.\bigskip

In general, at hadron colliders we describe events using (pseudo-)
rapidities and azimuthal angles, suggesting to use the differences
$\left\{ \Delta \eta_{mn}, \Delta \phi_{mn} \right\}$ for $m,n =
j_{1,2}, X, d, \bar{d}$ to study properties of a Higgs--like resonance.

A very useful observable is the \underline{azimuthal
  angle}\index{azimuthal angle} between the two tagging jets, \ie the
angle separating the two jets in the transverse plane.  We can again
link it to the angle between the two $Z$ decay planes in $X \to ZZ$
decays: for weak boson fusion it is defined as $\cos \Delta \phi
= (\hat{p}_{q_1} \times \hat{p}_{j_1}) \cdot (\hat{p}_{q_2} \times
\hat{p}_{j_2})$ in the Higgs candidate's rest frame, as shown in Eq.\eqref{eq:angles_wbf}.  
We can links this rest frame to the laboratory frame through a 
boost with a modest shift in the transverse direction.
In the laboratory frame each cross product $(\hat{p}_q \times \hat{p}_j)$
then resides in the azimuthal plane. 
The difference $\Delta \phi$ is 
nothing but the azimuthal angle between two
vectors which are each orthogonal to one of the tagging jet
direction. This is the same as the azimuthal angle between the two
tagging jets themselves, $\Delta \phi_{jj}$.

In Figure~\ref{fig:higgs_cp} we finally show this azimuthal angle
between the tagging jets, for all three Higgs coupling operators
defined in Eq.\eqref{eq:higgs_coup} and a sample spin-two coupling. The
Standard Model operator predicts an essentially flat behavior, while
the other two show a distinctly different modulation.  The CP-odd
$\epsilon^{\mu\nu\rho\sigma}$ coupling vanishes once two momenta
contracted by the Levi--Civita tensor are equal.\index{Levi--Civita tensor} This explains the
zeros at $\phi=0, \pi$, where the two transverse jet momenta are no
more linearly independent. To explain the opposite shape of the
CP-even Higgs we use the low transverse momentum limit of the tagging
jets\index{tagging jet}. In that case the Higgs production matrix
element becomes proportional to the scalar product $(p_T^\text{tag 1}
p_T^\text{tag 2})$ which vanishes for a relative angle $\Delta \phi =
\pi/2$.\bigskip

In addition to the decay plane angle or the azimuthal angle $\Delta
\phi$ there is a large number of angular observables we can use to
study the quantum numbers. The only thing we have to make sure is that
we do not cut on such variables when extracting the Higgs signal from
the backgrounds. Examples of such cuts is the angle between the two
leptons in the $H \to W^+ W^-$ decay or the rapidity difference
between the two forward tagging jets in weak boson fusion. In those
cases the determination of the Higgs operators in the Lagrangian
requires additional information. Also, we would want to test if a
mixture of operators is responsible for the observed Higgs--like
resonance. In Eq.\eqref{eq:higgs_coup} the higher--dimensional CP-even
operator $\ope^+_\text{D5}$ should exist in the Standard Model,
but it is too small to be observed. As mentioned above, this question
about the structure of the Higgs Lagrangian we should really answer
before we measure the Higgs couplings as prefactors to the appropriate
operators, following Section~\ref{sec:higgs_couplings}.

\subsubsection{Higgs self coupling}
\label{sec:higgs_self}

If we should ever observe a scalar particle at the LHC, the crucial
question will be if this is indeed our Higgs boson arising from
electroweak symmetry breaking. In other words, does this observed
scalar field have a \underline{potential} with a minimum at a
non--vanishing vacuum expectation value?

Of course we could argue that a dimension-3 Higgs coupling to massive
$W$ bosons really is a Higgs--Goldstone self coupling, so we see it by
finding a Higgs in weak boson fusion. On the other hand, it would be
far more convincing to actually measure the self coupling of the
actual Higgs field. This trilinear coupling we can probe at the LHC
studying Higgs pair production, for example in gluon fusion via the usual
top quark loop
\begin{equation*}
\parbox{35mm}{
\begin{fmfgraph*}(100,60)
 \fmfset{arrow_len}{2mm}
 \fmfleft{in1,in2}
 \fmf{gluon,width=0.5}{in1,v1}
 \fmf{gluon,width=0.5}{in2,v2}
 \fmf{fermion,width=0.5,tension=0.5}{v1,v3,v4,v2,v1}
 \fmf{dashes,width=0.5}{v3,out1}
 \fmf{dashes,width=0.5}{v4,out2}
 \fmfright{out1,out2}
\end{fmfgraph*}
} \quad + \quad 
\parbox{35mm}{
\begin{fmfgraph*}(100,60)
 \fmfset{arrow_len}{2mm}
 \fmfleft{in1,in2}
 \fmf{gluon,width=0.5}{in1,v1}
 \fmf{gluon,width=0.5}{in2,v2}
 \fmf{fermion,width=0.5,tension=0.5}{v1,v3,v2,v1}
 \fmf{dashes,width=0.5}{v3,vh}
 \fmf{dashes,width=0.5}{vh,out1}
 \fmf{dashes,width=0.5}{vh,out2}
 \fmfdot{vh}
 \fmfright{out1,out2}
\end{fmfgraph*}
}
\end{equation*}
\bigskip

Following exactly the same argument as presented in
Section~\ref{sec:higgs_gluon} we can derive the two tensor structures
contributing to Higgs pair production for transverse gluons
\begin{alignat}{5}
\sqrt{2} P_T^{\mu \nu} &= 
  g^{\mu \nu} 
  - \frac{k_1^\nu k_2^\mu}{(k_1 k_2)}
\\
\sqrt{2} P_2^{\mu \nu} &= 
  g^{\mu \nu} 
  + \frac{k_3^2 k_1^\nu k_2^\mu}{k_T^2 (k_1 k_2)}
  - \frac{2 (k_2 k_3) k_1^\nu k_3^\mu}{k_T^2 (k_1 k_2)}
  - \frac{2 (k_1 k_3) k_2^\mu k_3^\nu}{k_T^2 (k_1 k_2)}
  + \frac{k_3^\nu k_3^\mu}{k_T^2}
 \qquad \text{with} \quad k_T^2 = 2 \frac{(k_1 k_3)(k_2 k_3)}{(k_1 k_2)} - k_3^2
\; . \notag 
\end{alignat}
The third momentum is one of the two Higgs momenta, so $k_3^2 =
m_H^2$. The two tensors are orthonormal, which
means $P_T^2 = \one, P_2^2 = \one$ and $P_T \cdot P_2 = 0$. The second
tensor structure is missing in single Higgs production, so it
only appears in the continuum (box) diagram and turns out to be
numerically sub-leading over most of the relevant phase space.\bigskip

\begin{figure}[t]
\begin{center}
\includegraphics[width=0.43\hsize]{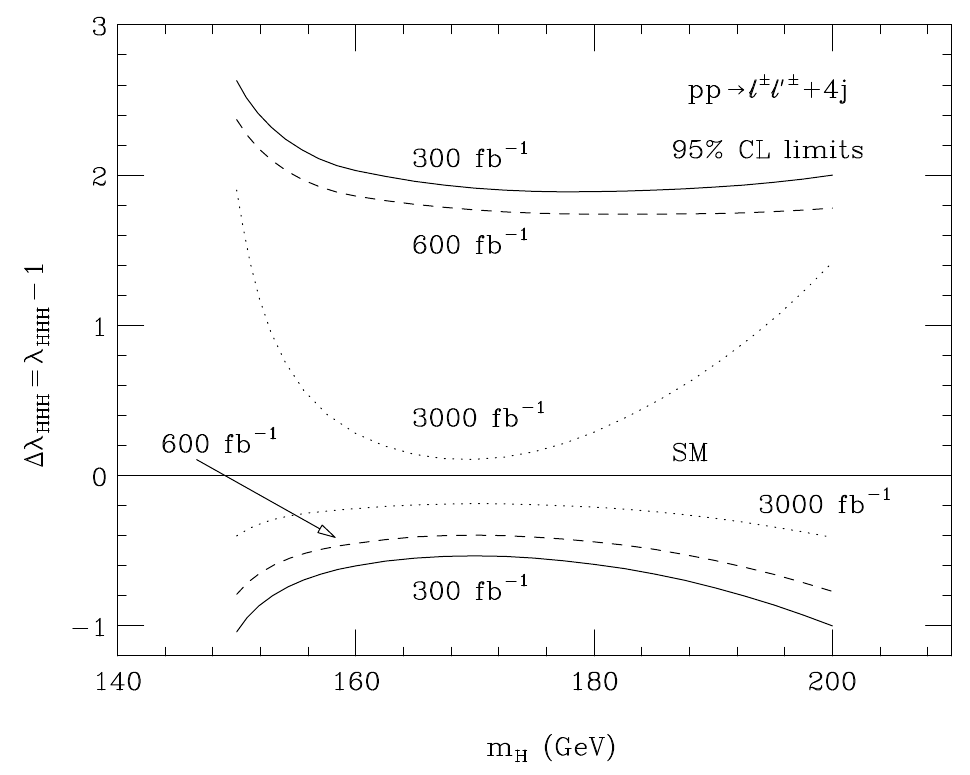}
\end{center}
\caption{The (parton--level) sensitivity limits for Higgs pair
  production and a measurement of the Higgs self coupling. The
  analysis is based on the decay $HH \to WW$. Figure from
  Ref.~\cite{Baur:2002qd}.}
\label{fig:higgs_hh}
\end{figure}

From Section~\ref{sec:higgs_low} on the effective Higgs--gluon coupling
we know that in the low energy limit we can compute the leading form
factors associated with the triangle and box diagrams, both
multiplying the transverse tensor $g^{\mu\nu}- k_1^\mu k_2^\nu/(k_1
k_2)$ for the incoming gluons
\begin{alignat}{5}
 g_{ggH} = -g_{ggHH} =
 -i \frac{\alpha_s}{12 \pi} + \ope \left(\frac{m_H^2}{4m_t^2}\right) \; .
\end{alignat}
Close to the production threshold $s \sim (2m_H)^2$ the leading
contribution to the loop--induced production cross section for
$gg\rightarrow HH$ involving the two Feynman diagrams above and the
Higgs self coupling derived in Section~\ref{sec:higgs_pot} is then
proportional to
\begin{alignat}{5}
\boxed{
\left[ 3m_H^2 \; \frac{g_{ggH}}{s-m_H^2}
      +g_{ggHH} \right]^2
=
g_{ggH}^2 \;
\left[ 3m_H^2 \; \frac{1}{s-m_H^2}
      -1 \right]^2
\sim
g_{ggH}^2 \;
\left[ 3m_H^2 \; \frac{1}{3m_H^2}
      -1 \right]^2
\rightarrow 0 
} \; ,
\label{eq:higgs_pair}
\end{alignat}
so the triangle diagram and the box diagram cancel.

In this argument we assume that the Higgs self coupling in the
Standard Model is proportional to $m_H$.  To see deviations from this
self coupling in the first term of Eq.\eqref{eq:higgs_pair} we can
look at something like the $m_{HH}$ distribution and measure the
threshold behavior. In the \underline{absence of any self
  coupling}\index{Higgs coupling!self coupling} this threshold
cancellation of the self coupling contribution with the continuum
should be absent as well. The threshold contribution to Higgs pair
production would be huge.  This way a shape analysis of the threshold
behavior will allow us to experimentally exclude the case of
$\lambda_{HHH}=0$ which predicts an unobserved large enhancement of
the production cross section at threshold. At this point it is still
under study if such a measurement will work at the LHC and what kind of
LHC luminosities would be required.\bigskip

As for all Higgs signatures we can go through the different Higgs pair
signatures to check which of them might work. All estimates
concerning detector performance have to be taken with a grain of salt,
because this measurement will only be possible after a significant
luminosity upgrade of the LHC, with integrated luminosities of
several $1000~\ifb$ of data. This might affect identification
efficiencies as well as the invariant (Higgs) mass reconstruction:
\begin{itemize}
\item[--] $gg \to HH \to b\bar{b} \; b\bar{b}$ is hopeless because of the
  overwhelming QCD backgrounds.
\item[--] $gg \to HH \to b\bar{b} \; W^+ W^-$ has a sizeable rate, but
  the irreducible background is $t\bar{t}$ production. In the very
  unlikely case that this background can be reduced to an acceptable
  level, the channel might work.
\item[--] $gg \to HH \to b\bar{b} \; \tau^+ \tau^-$ might or might not
  work. The key to this analysis is the reconstruction of the
  invariant masses of the bottom and tau pairs. Subjet methods along
  the lines of the Higgs tagger introduced in
  Section~\ref{sec:sim_fatjet} might help, if they survive the pile--up at
  this luminosity.
\item[--] $gg \to HH \to b\bar{b} \; \gamma \gamma$ should benefit
  from the excellent $m_{\gamma \gamma}$ reconstruction which ATLAS
  and CMS have already shown in the Higgs discovery. The backgrounds
  are not huge, either.
\item[--] $gg \to HH \to W^+ W^- \; W^+ W^-$ used to be the best
  channel for Higgs masses in the 160~GeV range. For lower masses it
  will be less promising. The most dangerous background is
  $t\bar{t}$+jets, which at least requires a very careful study.
\end{itemize}
Other channels should be tested, but at the least they will suffer
from small rates once we include the branching ratios.  While there
exist quite a number of studies for all of the channels listed above,
none of them has yet been shown to work for a 125~GeV Higgs boson. While
Nature's choice of Higgs mass is excellent when we are interested in
measuring as many Higgs couplings to Standard Model particles as
possible, it clearly suggests that we not look for the Higgs self
coupling.

\subsection{Alternatives and extensions}
\label{sec:higgs_alternatives}

The Higgs mechanism as discussed up to here is strictly speaking
missing two aspects: a reason why the Higgs field exists and formal
consistency of such a scalar field. In the following two sections we
will present two example models: first, we will show how technicolor
avoids introducing a new fundamental scalar field and instead relies
on a dynamic breaking of the electroweak gauge symmetry. The problem
is that technicolor does not predict the light scalar which ATLAS and
CMS discovered in the Summer of 2012. In addition, we will briefly
introduce the hierarchy problem, or the question why the Higgs is
light after including loop corrections to the Higgs mass. Little Higgs
models are one example for models stabilizing the Higgs mass,
supersymmetry is another.

\subsubsection{Technicolor}
\label{sec:higgs_technicolor}

Technicolor is an alternative way to break the electroweak symmetry
and create masses for gauge bosons essentially using a non--linear
sigma model, as introduced in Section~\ref{sec:higgs_sigma}.  There, we
give the scalar field $\phi$ a vacuum expectation value $v$ through a
potential, which is basically the Higgs mechanism. However, we know
another way to break (chiral) symmetries through
\underline{condensates} --- realized in QCD.  So let us review very
few aspects of QCD which we will need later.\bigskip

First, we illustrate why an asymptotically free theory like QCD is a
good model to explain electroweak symmetry breaking. This is what will
guide us to technicolor as a mechanism to break the electroweak gauge
symmetry.
As we will see in Section~\ref{sec:qcd_run_coup} the inherent mass scale
of QCD is $\lqcd \sim 200$~MeV. It describes the scale below which the
running QCD coupling constant $\alpha_s = g_s^2/(4 \pi)$ becomes
large, which means that perturbation theory in $\alpha_s$ breaks down and quarks
and gluons stop being QCD's physical degrees of freedom. 
This reflects the fact that QCD is not scale invariant. We
introduce a renormalization scale in our perturbative expansion. The
running of a dimensionless coupling constant can be translated into an
inherent mass scale. This mass scale characterizes the theory
in the sense that $\alpha_s(p^2 = \lqcd^2) \sim 1$; for scales
below $\lqcd$ the theory will become strongly interacting and hit 
its Landau pole\index{Landau pole}. Note that
first of all this scale could not appear if for some reason $\beta
\simeq 0$ and that it secondly does not depend on any mass scale in
the theory. This phenomenon of a logarithmically running coupling
introducing a mass scale in the theory is called
\underline{dimensional transmutation}. 
If at a high scale we start from a strong coupling in the $10^{-2}
\cdots 10^{-1}$ range the QCD scale will arrive at its known value
without any tuning.\bigskip

The symmetry QCD breaks is the chiral symmetry.  Just including the
quark doublets and the covariant derivative describing the $qqg$
interaction the QCD Lagrangian reads
\begin{equation}
 \lag_\text{QCD} \supset  \Psib_L \; i \Dslash \Psi_L
                       + \Psib_R \; i \Dslash \Psi_R
\label{eq:higgs_lag_qcd}
\end{equation}
From Eqs.\eqref{eq:higgs_chiral_mass} and~\eqref{eq:higgs_chiral_kin}
we know that the Lagrangian in Eq.\eqref{eq:higgs_lag_qcd} it is
symmetric under a chiral--type $SU(2)_L \times SU(2)_R$
transformation. Quark masses are not allowed, the chiral symmetry acts
as a custodial symmetry for the tiny quark masses we measure for
example for the valence quarks $u,d$. 
In Section~\ref{sec:qcd_run_coup} we will
elaborate more on the beta function of QCD. It is defined as
\begin{equation}
\frac{d \alpha}{d \log p^2} = 
\beta_\text{QCD} 
= - \frac{\alpha_s^2}{4\pi} 
      \left( \frac{11}{3} N_c - \frac{2}{3} n_f \right) 
\propto N_c \; ,
\end{equation}
where the $N_c$ scaling is only true in the pure Yang--Mills theory,
but gives asymmetric freedom due to $\beta_\text{QCD} < 0$. Towards
small energies the running strong coupling develops a Landau poles at
$\lqcd$. Because QCD is asymptotically free, at energies below $\lqcd$
the essentially massless quarks form condensates, which means two-quark
operators will develop a vacuum expectation value $\left< \Psib \Psi
\right>$.  This operator spontaneously breaks the $SU(2)_L \times
SU(2)_R$ symmetry into the (diagonal) $SU(2)$ of isospin.
This allows us to write down massive constituents which are the
different composite color--singlet mesons and baryons become the
relevant physical degrees of freedom.  Their masses are of the order
of the nucleon masses $m_\text{nucleon} \sim 1$~GeV.  The only
remaining massless particles are the Goldstone bosons from the
breaking of $SU(2)_L \times SU(2)_R$, the pions. Their masses are not
strictly zero, because the valence quarks do have a small mass of a
few MeV.  Their coupling strength (or decay rate) is governed by
$f_\pi$. It is defined via
\begin{equation}
\left< 0 | j_\mu^5 | \pi \right> = i f_\pi p_\mu \; , 
\label{eq:higgs_fpi}
\end{equation}
and parameterizes the breaking of the chiral symmetry via breaking
the axialvector--like $U(1)_A$.  The axial current can be computed as
$j_\mu^5 = \delta \lag /\delta (\p_\mu \pi)$ and in the $SU(2)$ basis
reads $j_\mu^5 = \psib \gamma_\mu \tau \psi/2$. From the measured
decays of the light color--singlet QCD pion into two leptons we know
that $f_\pi \sim 100$~MeV.\bigskip

To generalize them to technicolor we write the QCD observables in terms
of two QCD model parameters: the size of the gauge group, $N_c$, and
the scale at which the asymptotically free theory becomes strongly
interacting, $\lqcd$. It is hard to compute observables like
$m_\text{nucleon}$ or $f_\pi$ as a function of $N_c$ and
$\lqcd$. Instead, we derive simple scaling rules.

The $\lqcd$ dependence simply follows from the mass dimension which
for the vacuum expectation value is given by the mass dimension 3/2 of
each fermion field,
\begin{equation}
f_\pi \sim \lqcd 
  \qquad \qquad
\left< \overline{Q} Q \right> \sim \lqcd^3
  \qquad \qquad
m_\text{fermion} \sim \; \lqcd \; .
\label{eq:higgs_tcscaling1}
\end{equation}
The $N_c$ dependence of $f_\pi$ can be guessed from color factors: the
pion decay rate is by definition proportional to $f_\pi^2$. Leaving aside 
the strongly interacting complications parameterized by the appearance of $f_\pi$,
the
Feynman diagrams for this decay are
the same as for the Drell--Yan process $q \bar{q} \to \gamma, Z$. The
color structure of this process leads to an explicit factor of $N_c$,
to be combined with an averaging factor of $1/N_c$ for each of the
incoming quarks. Together, the pion decay rate is proportional to
$f_\pi^2/N_c$. We therefore postulate the pion decay constant to
scale like $f_\pi \sim \sqrt{N_c}$.  The vev--operator represents two
quarks exchanging a gluon at energy scales small enough for $\alpha_s$
to become large.  The color factor without any averaging over initial
states) simply sums over all colors states for the color--singlet
condensate, so it is proportional to $N_c$.  The fermion masses have
nothing to do with color states and hence should not depend on the
number of colors.  For details you should ask a lattice gauge
theorist, but we already get the idea how we can construct our
\underline{high--scale version of QCD} through
\begin{equation}
f_\pi \sim \sqrt{N_c} \; \lqcd 
  \qquad \qquad
\left< \overline{Q} Q \right> \sim N_c \; \lqcd^3
  \qquad \qquad
m_\text{fermion} \sim \; \lqcd \; .
\label{eq:higgs_tcscaling2}
\end{equation}
These scaling laws will allow us to make predictions for technicolor,
in spite of the fact that we cannot compute any of these observables
perturbatively.\bigskip

Let us work out the idea that a mechanism just like QCD condensates
could be the underlying theory of the non--linear $\sigma$ model.  In
contrast to QCD we now have a gauged custodial symmetry forbidding
weak gauge boson masses. The longitudinal modes of the massive $W$ and
$Z$ bosons come from the Goldstone modes of the
condensate's symmetry breaking called technipions. The corresponding mass
scale would have to be 
\begin{equation}
\ftc \sim v =246~\text{GeV} \; . 
\label{eq:higgs_def_ltc}
\end{equation}
Fermion masses we postpone to the next section --- in the 1970s, when
technicolor was developed, the top quark was not yet known. All
known fermions had masses of the order of GeV or much less, so they
were to a good approximation massless compared to the gauge bosons.

To induce $W$ and $Z$ masses we write down the non--linear sigma model
in its $SU(2)$ version. In Section~\ref{sec:higgs_higgsboson} we re-write
the linear sigma model using the Higgs field. Omitting the Goldstone
modes Eq.\eqref{eq:higgs_define} and Eq.\eqref{eq:def_phi} read
\begin{equation}
\s = \left( 1 + \frac{H}{v} \right) \one 
\qqqquad 
\phi = \frac{1}{\sqrt{2}}  
        \begin{pmatrix} 0 \\ v+H \end{pmatrix} 
= \frac{1}{\sqrt{2}}  
        \begin{pmatrix} 0 \\ v \s \end{pmatrix} \; .
\end{equation}
In the non--linear sigma model, defined in
Eq.\eqref{eq:higgs_goldstone_exp}, we replace $v$ by $\ftc$ and find
\begin{equation}
 \phi = \frac{1}{\sqrt{2}} \; 
        \begin{pmatrix} 0 \\ \ftc \s 
               \end{pmatrix}
 = \frac{1}{\sqrt{2}} \; 
        e^{- i (\pi \cdot \tau)/\ftc} \;
        \begin{pmatrix} 0 \\ \ftc
               \end{pmatrix}
      = \frac{1}{\sqrt{2}} \;
        \begin{pmatrix} 0 \\
                \ftc - i (\pi \cdot \tau) + \ope \left( \frac{1}{\ftc} \right)
               \end{pmatrix}
\label{eq:higgs_phi_tc}
\end{equation}
As basis vectors we use the three Pauli matrices\index{Pauli matrices} $\{\tau_a,\tau_b \}=2
\delta_{ab}$ which according to Eq.\eqref{eq:paulimat} fulfill $(\tau
\cdot \pi_1) \; (\tau \cdot \pi_2) = (\pi_1 \cdot \pi_2)$.  The
$SU(2)$-covariant derivative in the charge basis of the Pauli matrices
defined in Eq.\eqref{eq:cov_der} gives, when to simplify the formulas
we for a moment forget about the $U(1)_Y$ contribution and only keep
the non-zero upper entry in Eq.\eqref{eq:higgs_phi_tc}:
\begin{alignat}{5}
i D^\mu \phi \Bigg|_\text{lower}
&= \left[ i \p^\mu - \frac{g}{2} \, (\tau \cdot W^\mu)
                \right] \, 
                \frac{1}{\sqrt{2}}  
                \left[ \ftc - i (\tau \cdot \pi) + \ope \left( \frac{1}{\ftc} \right)
                \right] \notag \\
             &= \frac{1}{\sqrt{2}} \,
                \left[   \p^\mu (\tau \cdot \pi)
                       - \frac{g \ftc}{2} (\tau \cdot W^\mu)
                       + \ope(\ftc^0) 
                \right] \notag \\
(D_\mu \phi)^\dagger (D^\mu \phi) &=
                - \frac{1}{2} \,
                \left[   \p_\mu (\tau \pi)
                       - \frac{g \ftc}{2} (\tau \cdot W_\mu)
                       + \ope(\ftc^0) 
                \right] \;
                \left[   \p^\mu (\tau \pi)
                       - \frac{g \ftc}{2} (\tau \cdot W^\mu)
                       + \ope(\ftc^0) 
                \right] \notag \\
             & \supset
           - \frac{1}{2} (\p \pi)^2
           + \frac{g \ftc}{2} \, ( W_\mu \cdot (\p^\mu \pi) )
           + \ope(\ftc^0) \; .
\end{alignat}
If we also include the generator of the hypercharge $U(1)$ we find a
mixing term between the technipions and the $SU(2)$ gauge bosons
\begin{equation}
\boxed{
\lag \supset 
  \frac{g \ftc}{2} \, W_\mu^+ \p^\mu \pi^-
+ \frac{g \ftc}{2} \, W_\mu^- \p^\mu \pi^+
+ \ftc \, \left( \frac{g}{2} \, W_\mu^3 + \frac{g'}{2} \, B_\mu \right)\,
                             \p^\mu \pi^0
} \; .
\label{eq:higgs_mwz_tc}
\end{equation}
This is precisely the mixing term from the massive--photon example of
Eq.\eqref{eq:massive_photon} which we need to absorb the Goldstone
modes into the massive vector bosons.

We have strictly speaking not shown that the $\ftc$ appearing in the
scalar field $\phi$ is really the correctly normalized 
decay constant of the technipions and there is a lot of
confusion about factors $\sqrt{2}$ in the literature which we will
ignore in this sketchy argument. Nevertheless, if we assume the
correct normalization the massive Lagrangian
Eq.\eqref{eq:higgs_mwz_tc} with a mixing term proportional to $g
\ftc/2$ generates $m_W = g \ftc/2$. We know from
Eq.\eqref{eq:betaprime2} that $m_W = g v/2$, so electroweak symmetry
breaking might well be a scaled-up version of $f_\pi$.\bigskip

Once we are convinced that we can scale up QCD and break
electroweak symmetry we want to check what kind of predictions for the
electroweak observables come out of technicolor.  
We can study this scaling in the general case, where technicolor involves a gauge
group $SU(N_T)$ instead of $SU(N_c)$ and we have $N_D$ left handed
fermion doublets in the fundamental representation of $SU(N_T)$. To be
able to write down Dirac masses for the fermions at the end of the day
we also need ($2 N_D$) right handed fermion singlets. From the case of
more than one Higgs field contributing to $v$ we know that if we have
$N_D$ separate condensates their squares have to add up to $g^2 v^2$;
for equal vacuum expectation values they scale like $v \sim
\sqrt{N_D} \ftc$.  The scaling rules of Eq.\eqref{eq:higgs_tcscaling2}
then give
\begin{equation}
\frac{v}{\sqrt{N_D}} =
\ftc \sim \sqrt{\frac{N_T}{N_c}} \, \frac{\ltc}{\lqcd} \; f_\pi 
\qquad 
\Leftrightarrow 
\qquad
\frac{\ltc}{\lqcd} 
\sim \frac{v}{f_\pi} \; \sqrt{\frac{N_c}{N_D N_T}} 
\sim \frac{246~\gev}{130~\mev} \; \sqrt{\frac{N_c}{N_D N_T}} \; .
\label{eq:higgs_tcratio}
\end{equation} 
\bigskip

One simple example for this technicolor setup is the
\underline{Susskind--Weinberg model}. Its gauge group is $SU(N_T)
\times SU(3)_c \times SU(2)_L \times U(1)_Y$. As matter fields forming
the condensate which in turn breaks the electroweak symmetry we
include one doublet ($N_D=1$) of charged color--singlet technifermions
$(u^T,d^T)_{L,R}$.  In some ways this doublet and the two singlets
look like a fourth generation of chiral fermions, but with different
charges under all Standard Model gauge groups: for example, their
hypercharges $Y$ need to be chosen such that gauge anomalies do not
occur and we do not have to worry about non--perturbatively breaking
any symmetries, namely $Y=0$ for the left handed doublet and $Y=\pm
1$ for $u^T_R$ and $d^T_R$.  The usual formula $q = \tau_3/2 + Y/2$
then defines the electric charges $\pm 1/2$ for the heavy fermions
$u^T$ and $d^T$.

The additional $SU(N_T)$ gauge group gives us a running gauge coupling
which becomes large at the scale $\ltc$. Its beta function is modelled
after the QCD case
\begin{equation}
 \beta_\text{QCD} = - \frac{\alpha_s^2}{4\pi}
 \left( \frac{11}{3} N_c - \frac{2}{3} n_f \right)
 \qquad \qquad \qquad
 \beta_\text{T}   = - \frac{\alpha_T^2}{4\pi}
 \left( \frac{11}{3} N_T - \frac{4}{3} N_D \right) \; ,
\end{equation}
keeping in mind that $N_D$ counts the doublets, while $n_f$ counts the
number of flavors at the GUT scale. This relation holds for a simple
model, where quarks are only charged under $SU(3)_c$ and techniquarks
are only charged under $SU(N_T)$. Of course, both of them can carry
weak charges.  As a high--scale boundary condition we can for example
choose $\alpha_s(M_\text{GUT}) = \alpha_T(M_\text{GUT})$.  Using
Eq.\eqref{eq:run_alphas4} for $\lqcd$ we find
\begin{alignat}{5}
\frac{\ltc^2}{\lqcd^2} &= 
 \exp \left[+\frac{\alpha_T(m_\text{GUT})}{\beta_\text{T}}  
      \right] \;
 \exp \left[-\frac{\alpha_s(m_\text{GUT})}{\beta_\text{QCD}} 
      \right] \notag \\
&= 
 \exp \left[ \alpha_s(m_\text{GUT})
             \left( \frac{1}{\beta_\text{T}} - \frac{1}{\beta_\text{QCD}}
             \right)
      \right] \notag \\
&= 
 \exp \left[ - \frac{4 \pi}{\alpha_s(m_\text{GUT})}
             \left( \frac{1}{\dfrac{11}{3} N_T - \dfrac{4}{3} N_D} - \frac{1}{11-4}
             \right)
      \right] \; .
\end{alignat}
Such a GUT-inspired model based on an $SU(4)$ gauge group with
$\alpha_s(M_\text{GUT}) \sim 1/30$ and $N_D=1$ does not reproduce the
scale ratio required by Eq.\eqref{eq:higgs_tcratio}. However, for
example choosing $N_T = N_D = 4$ predicts $\ltc/\lqcd \sim 830$ and
with it exactly the measured ratio of $v/f_\pi$.\bigskip

At this stage, our fermion construction has two global chiral
symmetries $SU(2) \times SU(2)$ and $U(1) \times U(1)$ protecting the
technifermions from getting massive, which we will of course break
together with the local weak $SU(2)_L \times U(1)_Y$ symmetry. Details
about fermion masses we postpone to the next sections. Let us instead
briefly look at the \underline{particle spectrum} of our minimal
model:
\begin{itemize}
\item[--] techniquarks: from the scaling rules we know that the
  techniquark masses will be of the order $\ltc$ as give above.
  Numerically, the factor $\ltc/\lqcd \sim 800$ pushes the usual quark
  constituent masses to around $700$~GeV for the minimal model with
  $N_T=4$ and $N_D=1$. Because of the $SU(N_T)$ gauge symmetry there
  should exist four--techniquark bound states (technibaryons) which
  are stable due to the asymptotic freedom of the $SU(N_T)$
  symmetry. Those are not preferred by standard cosmology, so we
  should find ways to let them decay.

\item[--] Goldstone modes: from the breaking of the global chiral
  $SU(2) \times SU(2)$ and the $U(1) \times U(1)$ we will have four
  Goldstone modes. The three $SU(2)$ Goldstones are massless
  technipions, following our QCD analogy. Because we gauge the
  remaining Standard Model subgroup $SU(2)_L$, they become the
  longitudinal polarizations of the $W$ and $Z$ boson, after all this
  is the entire idea behind this construction. The remaining $U(1)$
  Goldstone mode also has an equivalent in QCD ($\eta'$), and its
  technicolor counter part acquires a mass though non--perturbative
  instanton breaking. Its mass can be estimates to $\sim 2$~TeV, so we
  are out of trouble.

\item[--] more exotic states: just like in QCD we will have a whole
  zoo of additional technicolor vector mesons and heavy resonances,
  but all we need to know about them is that they are heavy (and
  therefore not a problem for example for cosmology).
\end{itemize}
Before we move on, let us put ourselves into the shoes of the
technicolor proponents in the 70s. They knew how QCD gives masses to
protons, and the Higgs mechanism has nothing to do with it. Just
copying the appealing idea of dimensional transmutation without any
hierarchy problem they explained the measured $W$ and $Z$
masses. And just like in QCD, the masses of the four light quarks and
the leptons are well below a GeV and could be anything, but not linked
to weak--scale physics. Then, people found the massive bottom quark
and the even more massive top quark and it became clear that at least
the top mass was very relevant to the weak scale.  In this section we
will very briefly discuss how this challenge to technicolor basically
removed it from the list of models people take seriously --- until
extra dimensions came and brought it back to the mainstream.\bigskip

\underline{Extended technicolor} is a version of the original idea of
technicolor which attempts to solve two problems: create fermion
masses for three generations of quarks and leptons and let the heavy
techniquarks decay, to avoid stable technibaryons. From the
introduction we in principle know how to obtain a fermion mass from
Yukawa couplings, but to write down the Yukawa coupling to the sigma
field or to the TC condensate we need to write down some
Standard Model and technifermion operators. This is what ETC offers a
framework for.\bigskip

First, we need to introduce some kind of multiplets of matter
fermions. Just as before, the techniquarks, like all matter particles
have $SU(2)_L$ and $U(1)_Y$ or even $SU(2)_R$ quantum numbers.
However, there is no reason for them all to have a $SU(3)_c$ charge,
because we would prefer not to change $\beta_\text{QCD}$ too much.
Similarly, the Standard Model particles do not have a $SU(N_T)$
charge. This means we can write matter multiplets with explicitly
assigned color and technicolor charges as
\begin{equation}
\left( \;
    Q^T_{a=1..N_T}, \;
    Q^{(1)}_{j=1,...,N_c}, \;
    Q^{(2)}_{j=1,...,N_c}, \;
    Q^{(3)}_{j=1,...,N_c}, \;
    L^{(1)}, \;
    L^{(2)}, \;
    L^{(3)} \;
\right) \; .
\end{equation}
These multiplets replace the usual $SU(2)_L$ and $SU(2)_R$ singlets
and doublets in the Standard Model. The upper indices denote the
generation, the lower indices count the $N_T$ and $N_c$ fundamental
representations. In the minimal model with $N_T=4$ this multiplet has
$4+3+3+3+1+1+1=16$ entries.  In other words, we have embedded
$SU(N_T)$ and $SU(N_c)$ in a local gauge group $SU(16)$. If without
further discussion we also extend the Standard Model group by a
$SU(2)_R$ gauge group, the complete ETC symmetry group is $SU(16)
\times SU(2)_L \times SU(2)_R$, where we omit the additional
$U(1)_{B-L}$ throughout the discussion.

A technicolor condensate will now break $SU(2)_L \times SU(2)_R$,
while leaving $SU(3)_c$ untouched. If we think of the generators of
the ETC gauge group as ($16 \times 16$) matrices we can put a ($4
\times 4$) block of $SU(N_T)$ in the upper left corner and then three
($3 \times 3$) copies of $SU(N_c)$ on the diagonal. The last three
rows/columns can be the unit matrix. Once we break $SU(16)_\text{ETC}$
to $SU(N_T)$ and the Standard Model gauge groups, the Goldstone modes
corresponding to the broken generators obtain masses of the order of
$\letc$. This breaking should on the way produce the correct fermion
masses. The remaining $SU(N_T) \times SU(2)_L \times U(1)_Y$ will then
break the electroweak symmetry through a $SU(N_T)$ condensate and
create the measured $W$ and $Z$ masses as described in the last
section.\bigskip

In this construction we will have ETC gauge bosons which for example
in the quark sector form currents of the kind ($\overline{Q^T}
\gamma_\mu \, \tetc \, Q^T$), ($\overline{Q^T} \gamma_\mu \, \tetc \,
Q$), or ($\overline{Q} \gamma_\mu \, \tetc \, Q$). Here, $\tetc$ stands
for the $SU(16)_\text{ETC}$ generators. The multiplets $Q^T$ and $Q$
replace the $SU(2)_{L,R}$ singlet and doublets.
Below the the ETC breaking scale $\letc$ these currents become
four-fermion interactions, just like a Fermi interaction in the
electroweak theory,
\begin{equation}
\frac{(\overline{Q^T} \gamma_\mu \, \tetc^a \, Q^T) \; 
      (\overline{Q^T} \gamma^\mu \, \tetc^b \, Q^T)}{\letc^2}
\qquad 
\frac{(\overline{Q^T} \gamma_\mu \, \tetc^a \, Q) \; 
      (\overline{Q} \gamma^\mu \, \tetc^b \, Q^T)}{\letc^2}
\qquad 
\frac{(\overline{Q} \gamma_\mu \, \tetc^a \, Q) \; 
      (\overline{Q} \gamma^\mu \, \tetc^b \, Q)}{\letc^2} \; .
\label{eq:higgs_etc_ops1}
\end{equation}
The mass scale in this effective theory can be linked to the mass of
the ETC gauge bosons and their gauge coupling and should be of the
order $1/\letc \sim g_\text{ETC}/M_\text{ETC}$.  Let us see what this
kind of interaction predicts at energy scales below $\letc$ or
around the weak scale. Because
currents are hard to interpret, we Fierz--rearrange these
operators and then pick out three relevant classes of scalar
operators.\bigskip

Let us briefly recall this \underline{Fierz transformation}.\index{Fierz transformation}  The complete set
of four-fermion interactions is given by the structures
\begin{equation}
\lag \supset \left(\psib A_j \psi \right) \; 
             \left(\psib A^j \psi \right)
\qquad \qquad \text{with} \qquad \qquad
A_j = \one, \; \gamma_5, \;
      \gamma_\mu, \; \gamma_5 \gamma_\mu, \; \sigma_{\mu\nu} \; .
\end{equation}
The multi--index $j$ implies summing over all open indices in the
diagonal combination $A_j A^j$. These five types of ($4 \times 4$)
matrices form a basis of all real ($4 \times 4$) matrices which can
occur in the Lagrangian. 
If we now specify the spinors and cross them in this interaction we
should be able to write the new crossed (1,4,3,2) scalar combination
(or any new term, for that matter) as a linear combination of
the basis elements ordered as (1,2,3,4):
\begin{equation}
\left( \psib_1 A_i \psi_4 \right) \;
\left( \psib_3 A_i \psi_2 \right) =
\sum_j C_{ij}
\left( \psib_1 A_j \psi_2 \right) \;
\left( \psib_3 A_j \psi_4 \right) \; .
\end{equation}
In this notation we ignore the normal--ordering of the spinors in
the Lagrangian. It is easy to show that $C \cdot C = \one$.  The
coefficients $C_{ij}$ we list for completeness reasons:
\begin{equation}
\begin{array}{|l|rrrrr|}
\hline
                    & \one & \gamma_5 & \gamma_\mu & \gamma_5 \gamma_\mu & \sigma_{\mu\nu} \\ \hline
\one                & -1/4 & -1/4     & -1/4       &   1/4               & -1/8 \\           
\gamma_5            & -1/4 & -1/4     &  1/4       &  -1/4               & -1/8 \\     
\gamma_\mu          & -1   &  1       &  1/2       &   1/2               &  0   \\
\gamma_5 \gamma_\mu &  1   & -1       &  1/2       &   1/2               &  0   \\
\sigma_{\mu\nu}     & -3   & -3       &  1/2       &   0                 &  1/2 \\ \hline
\end{array}
\end{equation}
\bigskip

Applying a Fierz transformation to the three quark--techniquark
\underline{four-fermion operators} given in
Eq.\eqref{eq:higgs_etc_ops2} we obtain scalar ($A = \one$) operators,
\begin{equation}
\frac{(\overline{Q^T} \, \tetc^a \, Q^T) \; 
      (\overline{Q^T} \, \tetc^b \, Q^T)}{\letc^2}
\qquad \qquad 
\frac{(\overline{Q_L^T} \, \tetc^a \, Q_R^T) \; 
      (\overline{Q_R}   \, \tetc^b \, Q_L)}{\letc^2}
\qquad \qquad 
\frac{(\overline{Q_L} \, \tetc^a \, Q_R) \; 
      (\overline{Q_R} \, \tetc^b \, Q_L)}{\letc^2} \; .
\label{eq:higgs_etc_ops2}
\end{equation}
In these examples we pick certain chiralities of the Standard Model
fields and the technifermions.  Let us go through these operators one
by one. While not all of these
operators will be our friends we need them to give masses to the
Standard Model fermions, so we have to live with the
constraints.
\begin{enumerate}
\item Once technicolor forms condensates of the kind $\langle
  \overline{Q^T} Q^T \rangle$ the first operator in
  Eq.\eqref{eq:higgs_etc_ops2} will give masses to technicolor
  generators. This happens through loops involving technicolor
  states. We only quote the result for example for technipions which
  receive masses of the order $m \sim N_T \ltc^2/\letc$. For $\ltc
  \sim v$ and $\letc \sim 3$~TeV this not a large mass value, but
  nevertheless this first scalar operator clearly is our friend.

\item
  The scaling rules in Eq.\eqref{eq:higgs_tcscaling2} require the
  condensate to be proportional to $N_T \ltc^3$, The scalar
  dimension-6 operator adds a factor to $1/\letc^2$, so dimensional
  analysis tells us that the resulting Standard Model fermion masses
  will be of the order
\begin{equation}
 \lag \supset \frac{N_T \ltc^3}{\letc^2} \;
              \overline{Q_L} q_R
\equiv m_Q \; \overline{Q_L} q_R
\quad \Leftrightarrow
\quad 
\letc \sim \sqrt{ \frac{N_T \ltc^3}{m_Q} } \sim 
\left\{%
\begin{array}{ll}
3.7~\tev    & \qquad m_Q = 1~\gev \\
300~\gev  & \qquad m_Q = 150~\gev 
\end{array}
\right.
\end{equation}
  for $N_T=4$ and $\ltc = 150$~GeV. 
  This operator appears to be our friend for light quarks, but it
  becomes problematic for the top quark, where $\letc \sim v$ comes
  out too small.

  The top mass operator can be fierzed into a left handed
  fermion--technifermion current $( \overline{Q^T_L} \gamma_\mu Q_L) (
  \overline{Q_L} \gamma^\mu Q^T_L)$.  Because of the custodial
  $SU(2)_L \times SU(2)_R$ symmetry, which will turn out crucial to
  avoid electroweak precision constraints, we can rotate the top
  quarks into bottom quarks,
\begin{equation}
\frac{g_\text{ETC}^2}{M_\text{ETC}^2}
\left( \overline{Q^T_L} \gamma_\mu b_L   \right) \;
\left( \overline{Q_L}   \gamma^\mu Q^T_L \right) \; \; .
\end{equation}
  This operator induces a coupling of a charged ETC gauge boson to
  $T_L b_L$ which induces a one-loop contribution to the
  \underline{decay $Z \to b\bar{b}$}.  It contributes to the effective
  $bbZ$ coupling at the order $v^2/\letc^2 \sim \ope(1)$,
  considerably too big for the LEP measurement of $R_b =
  \Gamma_Z(b\bar{b})/\Gamma_Z(\text{hadrons})$. Note that such a
  constraint will affect any theory which induces a top mass through a
  partner of the top quark and allows for a general set of fierzed
  operators corresponding to this mass term, not just extended
  technicolor.

\item The third operator in Eq.\eqref{eq:higgs_etc_ops2} does not
  include any techniquarks, but all combinations of four-quark
  couplings. In the Standard Model such operators are
  very strongly constrained, in particular when they involve different
  quark flavors. \underline{Flavor--changing neutral currents}
  essentially force operators like
\begin{equation}
\frac{1}{\letc^2} 
\left( \bar{s} \gamma^\mu d \right) \;
\left( \bar{s} \gamma_\mu d \right) \;
\qquad \qquad \qquad 
\frac{1}{\letc^2} 
\left( \bar{\mu} \gamma^\mu e \right) \;
\left( \bar{e} \gamma_\mu \mu \right) \;
\end{equation}
  to vanish.  The currently strongest constraints come from kaon
  physics, for example the mass splitting between the $K^0$ and the
  $\overline{K}^0$. Its limit $\Delta M_K \lesssim 3.5 \cdot
  10^{-12}$~MeV implies $M_\text{ETC}/(g_\text{ETC} \theta_{sd})
  \gtrsim 600$~TeV in terms of the Cabibbo angle $\theta_{sd}$.  We
  can translate this bound on $\letc$ into an upper bound on fermion
  masses we can construct in our minimal model. $\letc > 10^3$~TeV
  simply translates in a maximum fermion mass which we can explain in
  this model: $m \lesssim 4$~MeV for $\ltc \lesssim 1$~TeV.  This is
  obviously not good news, unless we find a flavor symmetry to protect
  us from unwanted dimension-6 operators.
\end{enumerate}
\bigskip

Let us collect all the evidence we have against technicolor, in
spite of its very appealing numerical analogy to QCD. First and most
importantly, it predicts no light Higgs resonance, but a zoo of heavy
techni-particles. Both of these predictions are in disagreement with
current LHC data. In the next section we will use Goldstone's
theorem to break this degeneracy and generate a single light Higgs 
scalar. This Goldstone protection of a single light state can be 
applied to many models, including technicolor and other strongly
interacting Higgs sectors.

In addition, technicolor is strongly constrained by electroweak
precision constraints described in Section~\ref{sec:higgs_custodial}.
If we introduce new particles with $SU(2)_L \times U(1)_Y$ quantum
numbers, all of these particles will contribute to gauge boson self
energies.  Contributions from different states largely add, as we can
see for example for the $S$ and $T$ parameters in
Eq.\eqref{eq:s_t_sm}.  In technicolor the singlet techniquarks will
each contribute as $\Delta S \sim N_T/(6 \pi) \sim 4/20$, assuming
$N_D=1$. More realistic models easily get to $\Delta S \sim \ope(1)$,
which is firmly ruled out, no matter what kind of $\Delta T$ we manage
to generate.  The way out of some technicolor problems is so-called
walking technicolor, which still does not predict a light narrow Higgs
resonance. Nevertheless, it is instructive to understand dynamic
electroweak symmetry breaking because we know that this mechanism is
realized elsewhere in Nature and might eventually enter our Higgs
sector in some non--trivial way. After all, we did not yet manage to answer
the question where the Higgs field and the Higgs potential come from.

\subsubsection{Hierarchy problem and the little Higgs} 
\label{sec:higgs_lh_intro}

Before we introduce the little Higgs mechanism of breaking electroweak
symmetry we first need to formulate a major problem with the Higgs
boson as a fundamental scalar.  Let us start by assuming that the
Standard Model is a renormalizable theory.  

At next--to--leading order, the bare leading order Higgs mass gets
corrected by loops involving all heavy Standard Model particles.  Even
within the Higgs sector alone we can for example compute the four-point
Higgs loop proportional to the coupling given in
Eq.\eqref{eq:higgs_selfcoup}, namely $-3 i m_H^2/v^2$.  Introducing a
cutoff scale $\Lambda$ and sending it to infinity is certainly a valid
physical regularization scheme. We can implement the cutoff using the
\underline{Pauli--Villars regularization}\index{Pauli--Villars regularization},
\begin{alignat}{5}
  \frac{1}{q^2-m_H^2} \quad \longrightarrow \quad 
  \frac{1}{q^2-m_H^2} - \frac{1}{q^2-\Lambda^2}
 =\left\{%
  \begin{array}{ll}
     \dfrac{1}{q^2-m^2} \qquad & q^2\ll\Lambda^2 \\
     \dfrac{1}{q^2} - \dfrac{1}{q^2}=0        & q^2\gg\Lambda^2 \; . \\
  \end{array}
  \right.
\end{alignat}
The one-loop integral mediated by the Higgs self coupling then reads,
modulo prefactors,
\begin{alignat}{5}
\frac{3 m_H^2}{v^2} 
\int^{\Lambda} \frac{d^4q}{(2\pi)^4} \frac{1}{q^2-m_H^2} 
\quad &\to \; \frac{3 m_H^2}{v^2} 
\int \frac{d^4q}{(2\pi)^4}
 \left( \frac{1}{q^2-m_H^2} - \frac{1}{q^2-\Lambda^2} \right) \notag\\
&= \frac{3 m_H^2}{v^2} \left( m_H^2-\Lambda^2 \right)
\int \frac{d^4q}{(2\pi)^4} 
 \frac{1}{(q^2-m_H^2)(q^2-\Lambda^2)}  \notag\\
&= \frac{3 m_H^2}{v^2} \left( m_H^2-\Lambda^2 \right) \frac{C}{16 \pi^2} \; ,
\label{eq:higgs_loop_pv}
\end{alignat}
where $C$ is a numerical constant coming from the computation of the
four-dimensional integral. When we separate a remaining factor $1/(16
\pi^2)$ from the integral measure $1/(2 \pi)^4$ it comes out of
order unity. The problem is that Eq.\eqref{eq:higgs_loop_pv}
contributes to the Higgs mass, as we will show in some detail in
Section~\ref{sec:qcd_bw}. This means  that we observe a divergent one-loop
contribution to the Higgs mass $\Delta m_H^2 \propto
\Lambda^2$. Including all heavy Standard Model loops we find a full
set of quadratically divergent corrections to the bare Higgs mass
$m_{H,0}$\index{Higgs mass!quadratic divergence}
\begin{equation}
\boxed{  m_H^2 = m_{H,0}^2
         +\frac{3g^2}{32\pi^2}\; \frac{\Lambda^2}{m_W^2}
          \left[ m_H^2+2m_W^2+m_Z^2-\frac{4}{3}m_t^2
          \right] }
\; .
\label{eq:higgs_quadratic}
\end{equation}
This form of the Higgs mass corrections has one interesting new
aspect, when we compare it to the known behavior of the fermion
masses: the loop corrections to the mass and hence the quadratic
divergence are not proportional to the Higgs mass. Unlike fermion
masses which are linked to an approximate chiral symmetry, a finite
Higgs mass can appear entirely at loop level. We have already
exploited this feature writing down the Coleman--Weinberg mechanism in
Section~\ref{sec:higgs_coleman}.

The naive solution $m_H^2+2m_W^2+m_Z^2- 4n_f m_t^2/3 = 0$, called
Veltman's condition, assumes that fermionic and bosonic loop
corrections are regularized the same way, which is not
realistic.\bigskip

Why is the quadratic divergence in Eq.\eqref{eq:higgs_quadratic} a
problem? Dimensional regularization using $n=4-2\epsilon$ space--time
dimensions does not distinguish between logarithmic and quadratic
divergences. And we know that all masses develop poles $1/\epsilon$
which reflects the fact that the bare masses in the Lagrangian have to
be renormalized. In that sense dimensional regularization is not a
solution to our problem, but an approach which does not even see
the issue.

In an effective theory approach there exist many physical scales at
which we need to add new effects to the Standard Model. This could for
example be a see-saw scale to generate right handed neutrinos or some
scale where the quark flavor parameters are generated. In such an
effective theory we should be able to use a cutoff and matching scheme
around the high mass scale $\Lambda$. Varying this matching scale
slightly should not make a big difference. However, the quadratic
divergence of the Higgs mass implies that we have to compensate for a
large matching scale dependence of an observable mass on the
ultraviolet side of the matching. In other words, we need to seriously
fine-tune the ultraviolet completion of the effective Standard Model.

Alternatively, we can argue that in the presence of any large energy
scale the Higgs mass wants to run to this high scale. This is only
true for a fundamental scalar particle, fermion masses only run
logarithmically. This means that while the Higgs mechanism only works
for a light Higgs mass around the electroweak scale, the Higgs
naturally escapes, if we let it. Keeping the Higgs mass stable in the
presence of a larger physical energy scale is called the
\underline{hierarchy problem}\index{hierarchy problem}.\bigskip

We can quantify the level of fine tuning, which would be required to
remove the huge next--to--leading order contributions using a counter
term,
\begin{equation}
m_{H,0}^2
+\frac{3g^2}{32\pi^2}\; \frac{\Lambda^2}{m_W^2}
  \left[ m_H^2+2m_W^2+m_Z^2-\frac{4}{3}m_t^2 \right]
- \delta m_H^2 
\really m_{H,0}^2 \; .
\label{eq:higgs_quadratic2}
\end{equation}
Assuming $\Lambda=10$~TeV the different Standard Model contributions require 
\begin{equation}
\delta m_H^2 = \left\{
\begin{array}{rlr}
   -\frac{\displaystyle 3}{\displaystyle 8\pi^2} \; \lambda_t^2 \; \Lambda^2
    & \sim -(2~\tev)^2 & \qquad \qquad t \; \text{loop} \\[2mm]
    \frac{\displaystyle 1}{\displaystyle 16\pi^2} \; g^2 \; \Lambda^2
    & \sim (100~\tev)^2 & W \; \text{loop} \\[2mm]
    \frac{\displaystyle 1}{\displaystyle 16\pi^2} \; \lambda^2 \; \Lambda^2
    & \sim (500~\tev)^2 & H \; \text{loop.} 
\end{array}
\right.
\end{equation}
Varying the cutoff scale $\Lambda$ we need to ensure 
\begin{equation}
m_H = m_{H,0}^2 - \delta m_H^2 +
\left\{
\begin{array}{rl}
    (-250+50+25) \; (125~\gev)^2 
      & \text{for} \; \Lambda=10~\text{TeV} \\
    (-25000+2500+1250) \; (125~\text{GeV})^2
      & \text{for} \; \Lambda=100~\text{TeV} \\
     ... 
\end{array}
\right. 
\end{equation}
While we need to emphasize that the hierarchy problem is a
mathematical or even esthetic problem of the Higgs sector, it might
guide us to a better understanding of the Higgs sector. The best--known
solution to the hierarchy problem is supersymmetry, with the modified
Higgs sector discussed in Section~\ref{sec:higgs_2hdm}. Alternatively,
extra space--time dimensions, flat or warped, offer a
solution. Finally, we will show how little Higgs models protect the
Higgs mass based on concepts related to Goldstone's theorem\index{Goldstone's theorem}.\bigskip

Trying to solve this hierarchy problem using broken symmetries will
lead us to \underline{little Higgs models}.  This mechanism of
stabilizing a small Higgs mass is based on Goldstone's theorem\index{Goldstone's theorem}: we
make the Higgs a Goldstone mode of a broken symmetry at some higher
scale. This way the Higgs mass is forbidden by a symmetry and cannot
diverge quadratically at large energy scales. More precisely, the
Higgs has to be a pseudo--Goldstone, so that we can write down a Higgs
mass and potential. This idea has been around for a long time, but
for decades people did not know how to construct such a symmetry.

Before we solve this problem via the little Higgs mechanism, let us
start by constructing an example symmetry which protects the Higgs
mass from quadratic divergences at one loop.\index{Higgs mass!quadratic divergence} We break a for now global
$SU(3)$ symmetry to $SU(2)_L$. The number of generators which are set
free when we break $SU(N) \to SU(N-1)$ is
\begin{equation}
 (N^2-1)^2-((N-1)^2-1)=2N-1 \; .
\label{eq:higgs_goldstone_count}
\end{equation}
The $SU(2)$ generators are the Pauli matrices given in
Eq.\eqref{eq:cov_der}. For $SU(3)$ the basis is given by the traceless
hermitian and unitary \underline{Gell--Mann
  matrices}\index{Gell--Mann matrices},
\begin{small}
\begin{alignat}{5}
&\lambda^1  = \begin{pmatrix} \tau^1& &0\\& &0\\0&0&0\\ \end{pmatrix}  \qquad 
 \lambda^2  = \begin{pmatrix} \tau^2& &0\\& &0\\0&0&0\\ \end{pmatrix}  \qquad 
 \lambda^3  = \begin{pmatrix} \tau^3& &0\\& &0\\0&0&0\\  \end{pmatrix}  \qquad 
\lambda^8   = \frac{1}{\sqrt{3}}
             \begin{pmatrix} \one& &0\\ &&0\\0&0&-2\\   \end{pmatrix}  \notag \\
&\lambda^4  = \begin{pmatrix} 0& &1\\ & &0\\1&0&0\\       \end{pmatrix}  \qquad \;\;
 \lambda^5  = \begin{pmatrix} 0& &-i\\ & &0\\i&0&0\\      \end{pmatrix}  \qquad 
 \lambda^6  = \begin{pmatrix} 0& &0\\ & &1\\0&1&0\\       \end{pmatrix}  \qquad \;\;
 \lambda^7  = \begin{pmatrix} 0& &0\\ & &-i\\0&i&0\\      \end{pmatrix} \; . 
\label{eq:higgs_gell_mann_mat}
\end{alignat}
\end{small} 
We can arrange all generators of $SU(3)$ which are not generators of
$SU(2)$, and hence turn into Goldstones, in the outside column and row
of the $3\times 3$ matrix
\begin{equation}
\begin{pmatrix} SU(2)& & w_1\\ && w_2\\ w_1^*& w_2^*& w_0 \end{pmatrix}
\equiv
\begin{pmatrix} SU(2) & \phi \\ \phi^\dag  & w_0 \\ \end{pmatrix} \; .
\label{eq:higgs_su3_su2}
\end{equation}
The entry $w_0$ is fixed by the requirement that the matrix has to be
traceless when we include $\one$ as the fourth $SU(2)$ matrix in the
top--left corner. The corresponding field is an $SU(2)$ singlet and
can be ignored for now.  

We now assume that the $SU(2)_L$ doublet $\phi$ formed by the broken
$SU(3)$ generators is the Standard Model Higgs doublet. Normalization
factors $1/\sqrt{2}$ we omit in this section.  The Higgs can then only
acquire a mass at the electroweak scale, where $SU(2)_L$ is
broken. Based on Eq.\eqref{eq:higgs_su3_su2} we define a sigma field
as in Eq.\eqref{eq:higgs_goldstone_exp}. The only difference to 
Eq.\eqref{eq:higgs_goldstone_exp} is that $\s$ is now a triplet and
includes a symmetry breaking scale $f > v$
\begin{alignat}{2}
 \s &= \exp \left[-\frac{i}{f}\begin{pmatrix} 0_{2\times2}&\phi\\\phi^\dag  & 0\\ 
                                \end{pmatrix} \right] \;
              \begin{pmatrix} 0_2 \\ f\\
              \end{pmatrix}  \notag\\
      &= \left[ \one 
        -\frac{i}{f} \begin{pmatrix} 0&\phi \\ \phi^\dag  & 0\\
                                  \end{pmatrix}
        -\frac{1}{2} \left( \frac{-1}{f} \right)^2
                       \begin{pmatrix} 0&\phi \\ \phi^\dag  & 0\\
                             \end{pmatrix} \;
                       \begin{pmatrix} 0&\phi \\ \phi^\dag  & 0\\
                             \end{pmatrix} 
        + \ope \left( \frac{1}{f^3} \right) 
        \right] \;  \begin{pmatrix}0\\f\\\end{pmatrix} \notag \\
      &= \begin{pmatrix}0\\f\\\end{pmatrix}
        -\begin{pmatrix}i\phi \\ 0 \\ \end{pmatrix}
        -\frac{1}{2f^2} \begin{pmatrix}0\\\phi^\dag \phi f\\
                              \end{pmatrix} 
        + \ope \left( \frac{1}{f^3} \right) \notag \\
      &= \begin{pmatrix}0\\f\\\end{pmatrix}
        -\begin{pmatrix}i\phi \\ \phi^\dag \phi/(2f)\\\end{pmatrix} 
        + \ope \left( \frac{1}{f^3} \right) \; .
\label{eq:higgs_def_su3}
\end{alignat}
Only in the first line we indicate which of the zeros in the $3 \times
3$ matrix is a $2 \times 2$ sub-matrix. This is easy to keep track of
if we remember that the Higgs field $\phi$ is a doublet, while $\phi^\dagger
\phi$ is a scalar number. The kinetic term for the triplet field $\s$
becomes
\begin{alignat}{5}
  |\partial_\mu \s|^2
  & 
    = (i\partial_\mu \phi^*)_i (-i\partial^\mu \phi)_i
     + \frac{1}{4f^2}(\partial_\mu \phi^\dag  \phi) (\partial^\mu \phi^\dag \phi) 
  &= |\partial_\mu \phi|^2
     \left( 1 + \frac{\phi^\dag \phi}{f^2} \right) \; ,
\label{eq:higgs_strong_self}
\end{alignat}
where we skip the non--trivial intermediate steps.  The second term in
the brackets includes two Higgs fields which we can link to a
propagator, generating a one-loop correction to the Higgs
propagator. We know that our theory is an effective non--renormalizable
field theory, so we can apply a cutoff to the divergent loop
diagram. The result we already know from
Eq.\eqref{eq:higgs_loop_pv}. Ignoring the constant $C$ but keeping the
factor $1/(4\pi)^2$ from the integral measure we can write down the
condition under which the second term in
Eq.\eqref{eq:higgs_strong_self} does not dominate the tree level
propagator,
\begin{equation}
\boxed{   \frac{\Lambda^2}{(4\pi)^2 f^2} \lesssim 1 } 
\qquad \Leftrightarrow \qquad 
\Lambda \lesssim 10 \times f \; .
\label{eq:littleh_highscale}
\end{equation}
Once the loop--induced effect exceeds the tree level propagator at high
energies we consider the theory strongly interacting. Our
perturbative picture of the little Higgs theory breaks down. Without
even writing out a model for a Higgs mass protected by Goldstone's
theorem we already know that its ultraviolet completion will not be
perturbative and hence not predictive, and that it's range of
validity will be rather limited. Accepting these limitations we now
introduce a coupling to the $SU(2)$ gauge bosons and see what happens
to the Higgs mass. Of course, from the discussion of Goldstone's
theorem in Section~\ref{sec:higgs_ewsb} we already know that we will
not be able to generate the Higgs mass or potential in a
straightforward way, but it is constructive to see the problems which
will arise.\bigskip

As a \underline{first attempt} we simply add $g \; ( W^\mu
\cdot \tau)$ as part of the covariant derivative to the kinetic
term. In other words, we gauge the $SU(2)$ subgroup of the global
$SU(3)$ group. This automatically creates a four-point coupling of the
kind $g^2 |\vec{W}_\mu \phi|^2$.  As we did for
Eq.\eqref{eq:higgs_strong_self} we combine the two $W$ bosons to a
propagator and generate a one--loop Higgs mass term of the kind
\begin{equation}
  \lag \supset \frac{g^2 \Lambda^2}{(4\pi)^2} \; \phi^\dag \phi \; .
\label{eq:higgs_little_wrong1}
\end{equation}
This term gives the quadratically divergent Higgs mass we know from
Eq.\eqref{eq:higgs_quadratic}. Our ansatz does not solve or even
alleviate the hierarchy problem, so we discard it. What we learn from
it is that we cannot just write down the Standard Model $SU(2)_L$
gauge sector and expect the hierarchy problem to vanish.\bigskip

In a \underline{second attempt} we therefore write the same
interaction in terms of the triplet field $\s$, just leaving the third
entry in the gauge--boson matrix empty,
\begin{equation}
 g^2 \left|\begin{pmatrix} (W_\mu \cdot \tau) &0 \\ 0&0 \\
           \end{pmatrix} \s
     \right|^2 \; .
\end{equation}
We can again square this interaction term contributing to the
Higgs mass and find schematically
\begin{equation}
  g^2 \s^\dag \begin{pmatrix} \one &0\\0&0\\\end{pmatrix} \; 
               \begin{pmatrix} \one &0\\0&0\\\end{pmatrix} \s
\sim g^2 \s^\dag \begin{pmatrix} \one &0\\0&0\\\end{pmatrix} \s
\sim g^2 \; \phi^\dag \phi \; ,
\end{equation}
so the self energy contribution with the two $W$ fields linked now
reads
\begin{equation}
 \lag \supset \frac{g^2 \Lambda^2}{(4\pi)^2} \; \s^\dag  
                  \begin{pmatrix} \one &0\\0&0\\\end{pmatrix}
                  \; \s
= \frac{g^2 \Lambda^2}{(4\pi)^2} \; \phi^\dag \phi \; .
\label{eq:higgs_little_wrong2}
\end{equation}
This is precisely Eq.\eqref{eq:higgs_little_wrong1} and leads us to
also discard this second attempt. This outcome is not surprising
because we really only write the same thing in two different
notations, either using $\s^\dag \s$ or $\phi^\dag \phi$. Embedding
the $SU(2)_L$ gauge sector into a $SU(3)$ structure can only improve
our situation when the $SU(3)$ gauge group actually extends beyond
$SU(2)_L$.\bigskip

Learning from these two failed attempts we can go for a
\underline{third attempt}, where we add a proper covariant derivative
including all $SU(3)$ degrees of freedom.  Closing all of them
into loops we obtain in a proper basis
\begin{equation}
\lag\supset \frac{g^2 \Lambda^2}{(4\pi)^2} \; \s^\dag \; \one \; \s
 = \frac{g^2 \Lambda^2}{(4\pi)^2} \; f^2 \; .
\label{eq:higgs_little_wrong3}
\end{equation}
This is no contribution to the Higgs mass because the now massive
$SU(3)$ gauge bosons ate the Goldstones altogether.  According to
Eq.\eqref{eq:higgs_goldstone_count} the numbers match for the
breaking of $SU(N)$ to $SU(N-1)$. However, this attempt brings us
closer to solving Higgs--Goldstone problem. We are stuck between
either including only the $SU(2)_L$ covariant derivative and finding
quadratic divergences or including the $SU(3)$ covariant derivative
and turning the Higgs into a Goldstone mode which gives a mass of
scale $f$ to the heavy gauge bosons. What we need is a mix of an
extended $SU(3)$ gauge sector and a global symmetry where the
Goldstone modes are not eaten.\bigskip

\begin{figure}[t]
\begin{center}
\includegraphics[width=0.15\textwidth]{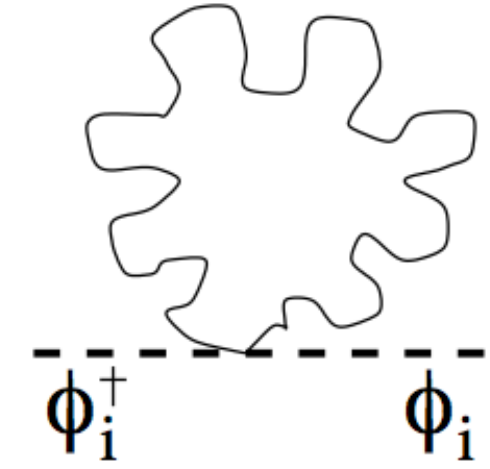}
\hspace*{0.06\textwidth}
\includegraphics[width=0.17\textwidth]{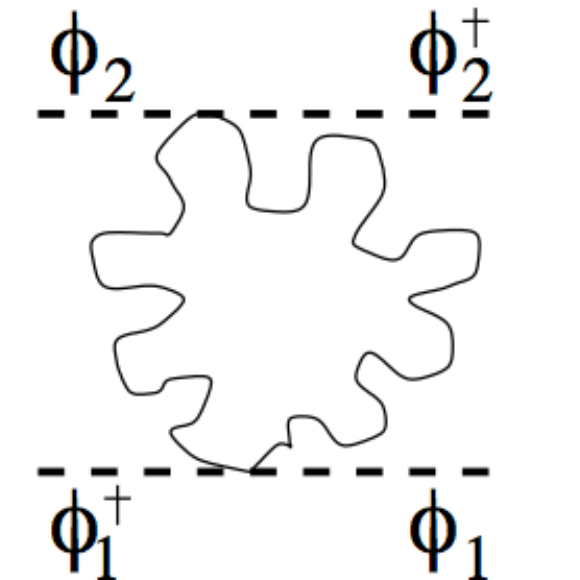}
\end{center}
\caption{Feynman diagrams contributing to the Higgs mass in
  little Higgs models. Beautiful figure from Ref.~\cite{schmaltz}.
\label{fig:littleh_feyn}}
\end{figure}

In our \underline{fourth and correct attempt} we find a way our
of this dilemma by using two independent sets of $SU(3)$
generators. We break them to our $SU(2)_L$ gauge group through a
combination of spontaneous and explicit breaking. This way will get
eaten Goldstones which make the $SU(3)$ gauge bosons heavy and at the
same time uneaten Goldstones which can form our Higgs, provided we
only gauge one $SU(3)$ gauge group.  Naively, we will be able to
distribute $8+8-3=13$ Goldstones this way.  However, we have have to
be careful not to double count three of them in the case where we
identify both $SU(2)$ subgroups of the two original $SU(3)$ groups; in
this case we are down to ten Goldstone modes. The art will be to
arrange the spontaneous and hard symmetry breakings into a workable
model.\bigskip

First, we write each of the set of $SU(3)$ generators the same way as
shown in Eq.\eqref{eq:higgs_def_su3} and identify those degrees of
freedom which we hope will include the Higgs field
\begin{equation}
 \s_j = \exp \left[- \frac{i}{f} 
                     \begin{pmatrix}0_{2 \times 2}&\phi_j\\
                                            \phi_j^\dag &0\\\end{pmatrix}
               \right] 
               \begin{pmatrix}0_2\\f\\\end{pmatrix}
               \qqqquad j=1,2 \; .
\label{eq:littleh_basic}
\end{equation}
For simplicity we set $f_1\equiv f_2\equiv f$. Each of the two $\s$
fields couples to the one set of $SU(3)$ gauge bosons through the
usual covariant derivative
\begin{equation}
  \lag \supset |D_\mu \s_1|^2 + |D_\mu \s_2|^2
       \supset g_1^2 \, |W_\mu \s_1|^2 + g_2^2 \, |W_\mu \s_2|^2 \; .
\end{equation}
The gauge boson fields we can linked to form propagators in loop
diagrams of the kind shown in the left panel of
Figure~\ref{fig:littleh_feyn}. From our attempt number three we 
know that for a universal coupling $g_1=g_2=g$ these diagrams give us
\begin{equation}
\lag \supset
  \frac{\Lambda^2}{(4\pi)^2} \; 
  \left(  g_1^2 \, \s^\dag _1\s_1 
        + g_2^2 \, \s^\dag _2\s_2 \right)
= \frac{2 g^2 \Lambda^2}{(4\pi)^2} \; f^2 \; .
\label{eq:littleh_symmetric}
\end{equation}
However, these are not the only loop diagrams we can generate with two
sets of Goldstones. For example, we can write diagrams like the one in
the right panel of Figure~\ref{fig:littleh_feyn}, where we couple
$\s_1$ to $\s_2$ directly through a gauge--boson loop. Counting powers
of momentum we can guess that it will only be logarithmically
divergent, so its contribution to the Lagrangian should be of the kind
\begin{equation}
\lag \supset
 \frac{g_1^2 g_2^2}{(4\pi)^2} \; \log \frac{\Lambda^2}{\mu^2} \;
 |\s^\dag _1\s_2|^2 \; ,
\label{eq:littleh_phi12}
\end{equation}
including a free renormalization scale $\mu$.  The combination
$\s_1^\dag \s_2$ is a scalar. It is indeed gauge--invariant only under
the diagonal subgroup of $SU(3)_1 \times SU(3)_2$, just like we
discuss in case of the custodial $SU(2)_L \times SU(2)_R$ in
Section~\ref{sec:higgs_custodial}.  The combined gauge interactions
$g_1$ and $g_2$ break the large symmetry group $SU(3) \times SU(3)$ of
Eq.\eqref{eq:littleh_symmetric} to their diagonal subgroup
$SU(3)_\text{diag}$. This happens as hard symmetry breaking via a
loop--induced term in the Lagrangian.\bigskip

Next, we translate Eq.\eqref{eq:littleh_phi12} into the Higgs
fields $\phi_j$ and see if it gives them a mass. This is easier if we
re-organize the $\phi_j$ in a more symmetric manner; if we shift $\phi_j \to
G \pm \phi$ the Goldstone modes are $SU(3)$ rotations common to
$\s_1$ and $\s_2$ and lend longitudinal degrees of freedom to the
massive gauge bosons of the gauged $SU(3)$ group.  For the supposed
Higgs mass term in the Lagrangian we find to leading order and neglecting
commutators
\begin{alignat}{6}
\s^\dag _1\s_2 
& =      \begin{pmatrix} 0 & f\\ \end{pmatrix}
e^{ \frac{i}{f} (\phi_1 \cdot \lambda)} \;
e^{-\frac{i}{f} (\phi_2 \cdot \lambda)}
         \begin{pmatrix} 0 \\ f\\ \end{pmatrix}
\notag \\
& = 
  \begin{pmatrix} 0 & f\\ \end{pmatrix} 
  e^{ \frac{i}{f} (\phi \cdot \lambda)} 
  e^{ \frac{i}{f} (G \cdot \lambda)}
  e^{-\frac{i}{f} (G \cdot \lambda)}
  e^{ \frac{i}{f} (\phi \cdot \lambda)}
  \begin{pmatrix} 0\\ f\\ \end{pmatrix} 
  + \text{commutator terms} \notag \\
& \simeq  
  \begin{pmatrix}0& f\\ \end{pmatrix}
  e^{\frac{2i}{f} (\phi \cdot \lambda)} 
  \begin{pmatrix}0\\ f\\ \end{pmatrix} \notag \\
& =  
  \begin{pmatrix}0& f\\ \end{pmatrix}
  \left[ \one 
        + \frac{2i}{f}
          \begin{pmatrix}0 & \phi\\ \phi^\dag & 0\\ \end{pmatrix}
        + \frac{1}{2}
          \left(\frac{2i}{f}\right)^2
          \begin{pmatrix}\phi\phi^\dag  & 0\\ 
                                 0& \phi^\dag \phi\\ \end{pmatrix} \right. \notag \\
& \phantom{haaalloooooo}  \left.      + \frac{1}{6}
          \left(\frac{2i}{f}\right)^3
          \begin{pmatrix}0 & \phi\phi^\dag \phi \\ 
                         \phi^\dag \phi \phi^\dag & 0\\ \end{pmatrix}
        + \frac{1}{24}
          \left(\frac{2i}{f}\right)^4
          \begin{pmatrix} (\phi\phi^\dag)^2  & 0\\ 
                                 0& (\phi^\dag \phi)^2 \\ \end{pmatrix}
        + \ope \left( \frac{1}{f^5} \right) 
  \right]
  \begin{pmatrix}0\\ f\\ \end{pmatrix} \notag \\
& = f^2  
   -\frac{2}{f^2} \begin{pmatrix}0& f\\ \end{pmatrix}
    \begin{pmatrix}0\\ \phi^\dag \phi f\\ \end{pmatrix}
   +\frac{2}{3 f^4} \begin{pmatrix}0& f\\ \end{pmatrix}
    \begin{pmatrix}0\\ (\phi^\dag \phi)^2 f\\ \end{pmatrix}
   +\ope \left( \frac{1}{f^6} \right) \notag \\
& = f^2 
   - 2\phi^\dag \phi
   + \frac{2}{3 f^2} (\phi^\dag \phi)^2
   + \ope \left( \frac{1}{f^4} \right) \; .
\end{alignat}
After squaring this expression we find
\begin{alignat}{5}
|\s^\dag _1\s_2|^2 \Bigg|_\text{gauge} & = f^4  - 4f^2 \phi^\dag \phi 
   + \frac{16}{3} (\phi^\dag \phi )^2 
   + \ope \left( \frac{1}{f^2} \right) \; .
\label{eq:littleh_sigma12}
\end{alignat}
This combination of spontaneous symmetry breaking of the two $SU(3)$
symmetries at the scale $f$ and explicit hard breaking to the diagonal
$SU(3)$ the pseudo--Goldstone field $\phi$ develops a mass and a
potential as powers of $|\s_1^\dag \s_2|$. The mass scales for
spontaneous symmetry breaking, $f$, and the hard breaking scale in
Eq.\eqref{eq:littleh_sigma12} are linked by loop effects.  For
example, its mass term just combining the two above formulae reads
\begin{equation}
 \boxed{\lag \supset 
 -\frac{g_1^2 g_2^2 f^2}{(2\pi)^2} \; \log\frac{\Lambda^2}{\mu^2} \; \phi^\dag \phi} 
\qqquad \Leftrightarrow \qqquad 
\boxed{m_H \sim \frac{g^2 f}{2 \pi} 
    \gtrsim \frac{g^2\Lambda}{8 \pi^2} 
    \sim \frac{\Lambda}{100}} \; .
\label{eq:littleh_range}
\end{equation}
This relation points to a \underline{new physics energy scale} $f \sim 1$~TeV.
Following the constraint given by Eq.\eqref{eq:littleh_highscale} we
do not expect $\log \Lambda/\mu$ to give a contribution to the Higgs
mass beyond a factor of $\ope(1)$. While this relation of scales indicates a
suppression of $g^2$ instead of $g$, we do not collect additional
factors $1/(4 \pi)$, because we are still looking at one--loop
diagrams.\bigskip

The mechanism described above is called \underline{collective symmetry
  breaking}.  It is a convoluted way of spontaneously and explicitly
breaking a global symmetry $SU(3)_1 \times SU(3)_2$ to our $SU(2)_L$,
the latter by introducing gauge or Yukawa coupling terms in the
Lagrangian.  Of the two sets of Goldstones arising in the spontaneous
breaking of each $SU(3)_{1,2} \to SU(2)_L$, denoted as $\phi_1$ and
$\phi_2$, we use $(\phi_1+\phi_2)/2$ to give the gauge bosons of one
of the broken $SU(3)$ groups a mass around $f$. The remaining
Goldstones $\phi=(\phi_1-\phi_2)/2$ at this stage remain
massless. They turn into pseudo--Goldstones and acquire a mass as well
as a potential in the explicit breaking of the global $SU(3)_1 \times
SU(3)_2$ symmetry into the gauged $SU(3)_\text{diag}$.\bigskip

The reason why this symmetry breaking is called `collective' is that
we need to break two symmetries explicitly to allow for mass and
potential terms for the pseudo--Goldstone. Only breaking one of them
leaves the respective other one as a global symmetry under which the
Higgs fields transforms non--linearly.  Because the original global
symmetry group is explicitly broken the Higgs will develop mass and
potential terms at the scale $f$, but doubly loop suppressed either
via gauge boson or via fermion loops.  This translates into a double
Higgs mass suppression $g_1 g_2$ relative to $f$.
Equation\eqref{eq:littleh_range} tells us that we can write down a
perturbative theory which is valid from $v \sim m_H$ to an ultraviolet
cutoff around $100 \times m_H$.

\subsubsection{Little Higgs Models}
\label{sec:higgs_lh_models}

Collective symmetry breaking can be implemented in a wide variety of
models. Our first example is based on the smallest useful
extension of $SU(2)_L$, namely $SU(3)$.  For decades people tried to
implement a Goldstone Higgs in this symmetry structure and learned
that to protect the Higgs mass a single broken $SU(3)$ symmetry is not
sufficient. For the simplest little Higgs model or \underline{Schmaltz
  model} we instead postulate a global $SU(3) \times SU(3)$ symmetry
and break it down to $SU(2)_L$ the way we introduce it in
Section~\ref{sec:higgs_lh_intro}. We can then express all mass scales
in terms of the symmetry--breaking scale $f$. Starting from the
ultraviolet the basic structure of our model in terms of its particle
content in the gauge sector is
\begin{itemize}
\item[--] for $E > 4\pi f$ our effective theory in $E/f$ breaks down,
  so our theory is strongly interacting and/or needs a ultraviolet
  completion.
\item[--] below that, the effective Lagrangian obeys a global and
  partly gauged $SU(3)_1 \times SU(3)_2$ symmetry with two gauge
  couplings $g_{1,2}$.  Both couplings are attached to one set of
  $SU(3)$ gauge bosons, containing three $SU(2)$ gauge bosons plus
  complex fields $W'_{\pm},W'_0$ with hypercharge $1/2$ and a singlet
  $Z'$.
\item[--] through loop effects the combined gauge couplings explicitly
  break $SU(3)_1 \times SU(3)_2 \to SU(3)_\text{diag}$. The related
  Goldstone modes give masses of the order $g f$ to the heavy $SU(3)$
  gauge bosons.
\item[--] the other five broken generators of $SU(3)_1 \times SU(3)_2$
  become Goldstone modes and the Standard Model Higgs doublet. Terms
  like $\s^\dag_1\s_2$ give rise to a Higgs mass of the order $g^2
  f/(2\pi)$.
\item[--] to introduce hypercharge $U(1)_Y$ we have to postulate
  another $U(1)_X$, which includes a heavy gauge boson mixing with the
  $SU(3)/SU(2)$ and the $SU(2)$ gauge bosons, to produce $\gamma, Z,
  Z'$. This will turn into a problem, because we lose custodial
  symmetry. For our discussion we ignore the $U(1)$ gauge bosons.
\end{itemize}
Until now we have not discussed any fermionic aspects of the little
Higgs setup.
However, to remove the leading quadratic divergence in
Eq.\eqref{eq:higgs_quadratic} we obviously need to modify the
\underline{fermion sector} as well.  For this purpose we enlarge the
$SU(2)$ heavy--quark doublet $Q$ to an $SU(3)$ triplet $\Psi=(t,b,T)
\equiv (Q,T)$.  The Yukawa couplings look like $\lambda \s^\dag \Psi
t^c$, in analogy to the Standard Model, but with two right handed top
singlets $t_j^c$ which will combine to the Standard--Model and a heavy
right handed top. We can compute this in terms of the physical
fields,
\begin{alignat}{4}
 \s_j^\dag\Psi & 
   = \begin{pmatrix}0& f\\ \end{pmatrix} 
     \exp \left[\frac{i}{f}
                      \begin{pmatrix} 0 &\phi\\\phi^\dag& 0 \\ \end{pmatrix}
                \right]
           \begin{pmatrix}Q\\T\\\end{pmatrix} \notag \\
 & = \begin{pmatrix}0& f\\ \end{pmatrix} 
     \left[\one + \frac{i}{f}
                         \begin{pmatrix} 0 &\phi\\\phi^\dag& 0 \\\end{pmatrix}
                      + \frac{1}{2} \left( \frac{i}{f} \right)^2
                         \begin{pmatrix} \phi\phi^\dag& 0 \\
                                         0  &\phi^\dag \phi\\\end{pmatrix}
                      + \ope \left( \frac{1}{f^3} \right)  \right]
           \begin{pmatrix}Q\\T\\\end{pmatrix} \notag \\
 & = \begin{pmatrix}0& f\\ \end{pmatrix} 
     \left[ \begin{pmatrix} Q\\T\\\end{pmatrix}
                       + \frac{i}{f} 
                           \begin{pmatrix}\phi T\\
                                             \phi^\dag Q\\\end{pmatrix}
                      - \frac{1}{2f^2} 
                           \begin{pmatrix} \phi\phi^\dag Q\\
                                                 \phi^\dag \phi T\\\end{pmatrix}
                      + \ope \left( \frac{1}{f^3} \right) \right]\notag \\
 & = f T + i\phi^\dag Q
    - \frac{1}{2f} \; \phi^\dag \phi T
    + \ope \left( \frac{1}{f^2} \right) \; .
\end{alignat}
Combining the two Yukawas with the simplification
$\lambda_1=\lambda_2=\lambda$ gives us the leading terms
\begin{equation}
  \lag \supset \lambda f \; \left( 1 - \frac{1}{2f^2} \; \phi^\dag \phi
                            \right) T T^c
               + \lambda \; \phi^\dag Q t^c \; ,
\label{eq:littleh_topcoups}
\end{equation}
where we define the SM top quark as $t_2^c-t_1^2=-i\sqrt{2}t^c$ and
its orthogonal partner $t_1^c+t_2^c=\sqrt{2}T^c$.\bigskip

According to Eq.\eqref{eq:littleh_topcoups} both top quarks contribute
to the Higgs propagator and Higgs mass corrections, the Standard Model
top through the usual $ttH$ Yukawa coupling and the new heavy top
particle through a four-point $TTHH$ interaction.  For the Feynman
rule we will need to include an additional factor 2 in the $TTHH$
coupling, stemming from the two permutations of the Higgs fields. The
question becomes how these two diagrams cancel.

The scalar integrals involved we know; generally omitting a factor
$1/(4\pi)^2$ the two-point function from the Standard Model top loop
has a quadratic pole $B(0;m,m) \sim (\Lambda/m)^2$.  Adding two
fermion propagators with mass $m_t$ and two Yukawa couplings $\lambda$
gives a combined prefactor $i^4 \, \lambda^2 \, \Lambda^2= \lambda^2
\, \Lambda^2$.  The heavy top diagram gives a one-point function with
the pole $A(m_T) \sim \Lambda^2$.  Adding one fermion propagator with
mass $m_T$ and the coupling $\lambda/f$ yields $i^2 \lambda/f \, m_T
\, \Lambda^2 = - \lambda^2 \Lambda^2$. This hand--waving estimate
illustrates how these two top quarks cancel each other's quadratic
divergence for the Higgs mass.\index{Higgs mass!quadratic divergence}
If we do this calculation more carefully, we find that for an
$SU(3)$-invariant regulator the quadratic divergences cancel, and
terms proportional to $\log m_t/m_T$ remain. 
Instead keeping the two $\lambda_j$ separated we would find
\begin{alignat}{2}
 m_T &= \sqrt{ \lambda_1^2 f_1^2 + \lambda_2^2 f_2^2}
     \sim \text{max}_j (\lambda_j f_j ) 
\qqqquad
\lambda_t &= \lambda_1 \lambda_2 \; 
             \frac{\sqrt{f_1^2+f_2^2}}{m_T} \; .
\label{eq:littleh_topyukawa}
\end{alignat}
Following the same logic as for the gauge boson loop shown in
Figure~\ref{fig:littleh_feyn} the combination of $\lambda_1$ and
$\lambda_2$ breaks $SU(3)_1 \times SU(3)_2 \to SU(3)_\text{diag}$
explicitly.  This turns the Higgs into a pseudo--Goldstone and allows
contributions proportional to $\lambda_1 \lambda_2$ in the Higgs mass
and potential.\bigskip

To arrive at the Standard Model in the infrared limit we need to
generate a \underline{Higgs potential}\index{Higgs potential} $V = \mu^2 |\phi|^2 + \lambda
|\phi|^2$.  The two parameters are related via $\mu^2 = -\lambda
v^2$. We already know that gauge boson loops generate such a
potential, as shown in Eq.\eqref{eq:littleh_sigma12}. Similarly,
fermion loops in the Schmaltz model give
\begin{alignat}{2}
 |\s_1^\dag \s_2|^2 \Bigg|_\text{fermion} & 
   =   f^4
     - 4f^2 \phi^\dag \phi
     + \frac{14}{3} \; (\phi^\dag \phi)^2
     + \ope \left( \frac{1}{f^2} \right) 
   \equiv f^4 + \mu^2 \phi^\dag \phi
        + \lambda(\phi^\dag \phi)^2 
        + \ope \left( \frac{1}{f^2} \right) \notag \\
& \Rightarrow \qqquad
  \left| \frac{\mu^2}{\lambda} \right|_\text{fermion}
 +\left| \frac{\mu^2}{\lambda} \right|_\text{gauge}
  \sim \frac{12 f^2}{14} + \frac{12 f^2}{16}
  \gg v^2 \; .
\label{eq:littleh_selfcoup}
\end{alignat}
The self coupling $\lambda$ is too small to give us anything like the
Standard Model Higgs potential.  There is no easy cure to this, but we
can resort to ad--hoc introducing a tree level $\mu$ parameter with
the proper size.
\begin{equation}
  \lag \supset \mu^2 \s_1^\dag\s_2
       = \mu^2 \left[  f^2 
                     - 2 \; \phi^\dag \phi
                     + \ope \left(\frac{1}{f^2}\right)  
               \right] \; .
\end{equation}
Roughly $\mu\sim v$ brings the Higgs potential terms to the correct
value. Ironically, this ad-hoc terms gives the model its alternative
name \underline{$\mu$ model}.  As a side remark, such a term also
breaks the $U(1)$ symmetry linked to the 8th $SU(3)$ generators and
gives the corresponding Goldstone a mass of the order $v$.

Now we can summarize the particle content of this first little Higgs
model.  Apart from the Standard Model particles and a protected light
Higgs we find the new particle spectrum, still not including the
$U(1)$ structure
\begin{alignat}{4}
\text{$SU(3)$ gauge bosons $W'^\pm,W'^0$} \qqquad & \text{with} 
  \quad m_{W'} = \ope(g f) \notag \\ 
\text{singlet $Z'$}                      \qqquad         &\text{with} 
  \quad m_{Z'} = \ope(g f) \notag \\ 
\text{heavy top $T$}                     \qqquad         &\text{with}
  \quad m_T    = \ope(\lambda_t f) \; .
\end{alignat}
\bigskip

The Schmaltz model discussed above has a few disadvantages. Among them
is the missing $U(1)$ gauge group and the need for an ad-hoc $\mu$
term. From what we know about sigma models and collective symmetry
breaking we can construct a second economic little--Higgs model, the
\underline{littlest Higgs} model.  This time, we embed two gauge
symmetries which overlap by the Standard Model Higgs doublet into one
sigma field: it includes two copies of $SU(2)$ as part of the large
global $SU(5)$ symmetry group. This will also have space to the $U(1)$
gauge group. Our enlarged symmetry group now has $(5^2-1)=24$
generators $T$ in the usual adjoint representation $SU(5)$.  The Pauli
matrices\index{Pauli matrices} as one set of $SU(2)$ generators are arranged in the $5
\times 5$ matrix similarly to Eq.\eqref{eq:higgs_su3_su2}
\begin{equation}
          \begin{pmatrix} -SU(2)^* & 0_{2 \times 3}\\
                                       0_{3 \times 2} & 0_{3 \times 3}\\ \end{pmatrix} 
  \qqqquad
          \begin{pmatrix} 0_{3 \times 3}  & 0_{2 \times 3}\\   
                              0_{3 \times 2} & SU(2)\\ \end{pmatrix} \; .
\label{eq:littlesth_su2l}
\end{equation}
This $SU(2)$ symmetry will need to stay unbroken when we break
$SU(5)$. This double appearance looks like a double counting, so we
will will eventually get rid of half of these generators. Moreover,
these unbroken $SU(2)$ generators only use half of the number of
degrees of freedom which $SU(5)$ offers in each of these $2 \times 2$
sub-matrices.\bigskip

Next, we need to identify those broken Goldstone modes $\hat{T}$ which
form the Higgs field. In the $SU(5)$ generator matrix the Higgs
doublet has to be arranged such that neither set of $SU(2)$ generators
includes $\phi$, so when we break $SU(5)$ into the the $SU(2)$
subgroups the Higgs always stays a (pseudo--) Goldstone. We construct
a pattern similar to Eq.\eqref{eq:littleh_basic} in the Schmaltz model
when we only consider the upper left or lower right $3 \times 3$
matrices of the full $SU(5)$ group,
\begin{equation}
 \s = \exp \left[ - \frac{i}{f} (w \cdot \hat{T} \right]
      \langle\s\rangle 
\qquad \text{with} \qquad
 (w \cdot \hat{T} =
                        \begin{pmatrix} 0 & \phi^* & 0 \\
                                        \phi^T & 0 & \phi^\dag \\
                                         0 & \phi & 0         \\ 
                              \end{pmatrix} \; .
\label{eq:littlesth_goldstones1}
\end{equation}
Again, the four-fold appearance looks like we introduced too many
Higgs fields, but the symmetry structure of the broken $SU(5)$ will
ensure that they really all are the same field.\bigskip

Spontaneously breaking the global $SU(5)$ symmetry by an appropriate
vacuum expectation value $\langle \s \rangle$ will eventually allow
the $\phi$ doublet to develop a potential, including a mass and a self
coupling.  The Standard--Model $SU(2)_L$ generators should not be
affected. We try
\begin{equation}
  \langle \s\rangle =
     \begin{pmatrix} 0 & 0 & \one_{2\times 2} \\
                     0 & 1 & 0 \\
                     \one_{2\times 2} & 0 & 0
           \end{pmatrix} \; .
\label{eq:littlesth_vev}
\end{equation}
This vacuum expectation value obviously breaks our global $SU(5)$
symmetry.  What remains in the $\langle \s \rangle$ background is an
$SO(5)$ symmetry, generated by the antisymmetric tensor with
$(4+3+2+1)=10$ entries.  This way 14 of the original 24 generators are
broken and the multiple appearance of some of the Goldstone fields in
Eq.\eqref{eq:littlesth_su2l} and Eq.\eqref{eq:littlesth_goldstones1}
is explained.
Using commutation relations we can show that the Standard--Model
$SU(2)_L$ generators in Eq.\eqref{eq:littlesth_su2l} are indeed
unbroken.  The corresponding unbroken $U(1)$ generators are the
equally symmetric diagonals diag$(-3,-3,2,2,2)/10$ and
diag$(-2,-2,-2,3,3)/10$.\bigskip

To compute the spectrum of the littlest Higgs model based on breaking
$SU(5)\to SO(5)$ through the vacuum expectation value shown in
Eq.\eqref{eq:littlesth_vev} we generalize
Eq.\eqref{eq:littlesth_goldstones1} to include the complete set of
Goldstones associated with the broken generators,
\begin{equation}
  (w \cdot \hat{T}) =
      \begin{pmatrix}
            \chi_{2\times2} & \phi^* & \kappa^\dag_{2\times2} \\
            \phi^T    & 0            & \phi^\dag      \\
            \kappa_{2\times2} & \phi  & \chi_{2\times 2}^T   \\
            \end{pmatrix}
    + \frac{\eta}{2\sqrt{5}}
            \begin{pmatrix} \one & 0 & 0 \\
                               0 &-4 & 0 \\
                               0 & 0 & \one \\
                  \end{pmatrix} \; .
\label{eq:littlesth_goldstones2}
\end{equation}
The form reflects the commutation property $\langle \s \rangle \;
\hat{T}^T = \hat{T} \; \langle \s \rangle$, which links opposite
corners of $w \cdot \hat{T}$. The $\chi$ field differs from the
unbroken $SU(2)$ generators in Eq.\eqref{eq:littlesth_su2l} in the
relative sign between the two appearances. They can be shown to also
form hermitian traceless $2\times 2$ matrices, which means that they can be written
as a second triplet of $SU(2)$ fields. The combination of broken
generators $\chi$ and $\chi^T$ and unbroken $SU(2)_L$ generators in
Eq.\eqref{eq:littlesth_su2l} account for all degrees of freedom in
those sub-matrices.  The $2 \times 2$ matrix $\kappa$ is not
traceless, but complex symmetric, so instead of another set of $SU(2)$
gauge bosons they form a complex scalar triplet.  The complex doublet
$\phi$ will become the Standard Model Higgs doublet, and $\eta$ is the
usual real singlet. Together, these field indeed correspond to $3_\chi
+6_\kappa +4_\phi +1_\eta =14$ broken Goldstones.

Unless something happens the fields linked to the broken
generators $\hat{T}$ can either turn into gauge boson mass terms of
the order $f$ or stay massless.  In particular, $\chi$ will make one
of the two sets of $SU(2)$ gauge bosons $W^{'\pm},W^{'0}$ heavy, while
$\eta$ gets eaten by the $B'$ field.\bigskip

The trick in this littlest Higgs model is to mix the two $SU(2)$
groups in two opposite corners of the sigma field in
Eq.\eqref{eq:littlesth_su2l} and
Eq.\eqref{eq:littlesth_goldstones2}. We can for example introduce two
gauge couplings $g_1$ and $g_2$, one for each corner. By construction,
the combination with a relative minus sign between the upper--left and
lower--right fields stays unbroken after spontaneously breaking $SU(5)
\to SO(5)$, so this linear combination will give the Standard Model
gauge bosons. Its orthogonal combination $\chi$ will become a set of
heavy $W'$ states. 
Introducing an $SU(2)$ mixing angle $\tan \theta = g_2/g_1$ we can
define a rotation from the $SU(2)_1 \times SU(2)_2$ interaction basis
to the mass basis after spontaneous symmetry breaking. Exactly the
same works for the $B$ fields corresponding to $U(1)_Y$,
\begin{equation}
  \begin{pmatrix}W^{'a}\\W^a\\\end{pmatrix}
  = \begin{pmatrix}-\cos\theta&\sin \theta\\\sin\theta&\cos\theta\\
          \end{pmatrix}
    \begin{pmatrix}W_1^a\\W_2^a\\\end{pmatrix} 
          \qquad \qquad \qquad
  \begin{pmatrix}B' \\B \\\end{pmatrix}
  = \begin{pmatrix}-\cos\theta'&\sin\theta'\\\sin\theta'&\cos\theta'\\
          \end{pmatrix}
    \begin{pmatrix}B_1^a\\B_2^a\\\end{pmatrix} \; .
\end{equation}
Including all factors the heavy gauge bosons acquire the masses
\begin{equation}
  m_{W'}=\frac{gf}{\sin (2\theta)}
  \qquad \qquad \qquad
  m_{B'}=\frac{g'f}{\sqrt{5}\sin (2\theta')} \; .
\end{equation}
In our discussion of little Higgs models factors of two never really
matter. However, in the case of $m_{B'}$ we see that for $f$ in the
TeV range the heavy $U(1)$ gauge boson is predicted to be very light,
making the model experimentally very vulnerable.\bigskip

Protecting the Higgs mass from quadratic divergences in the gauge
sector of the littlest--Higgs model works similar to the Schmaltz
model. Each of the two sets of $SU(2)_L$ generators in
Eq.\eqref{eq:littlesth_su2l} corresponds to a $2 \times 2$ sub-matrix
in one of the corners of the $SU(5)$ sigma field. If we break the
global $SU(5)$ down to one of the two $SU(2)$ groups the Higgs doublet
will be a broken generator of the global $SU(5)$ and therefore be
massless. Unlike in the $SU(3) \times SU(3)$ setup the finite Higgs
mass in the littlest Higgs model is not induced by gauge boson or
fermion loops. It appears once we integrate out the heavy field
$\kappa$ in the presence of $g_1$ and $g_2$, communicated to the Higgs
field $\phi$ via the \underline{Coleman--Weinberg} mechanism
introduced in Section~\ref{sec:higgs_coleman}. The resulting quartic
Higgs term, only taking into account the $SU(2)$ couplings becomes
\begin{equation}
\lag \supset - c \frac{g_1^2 g_2^2}{g_1^2 + g_2^2} |\phi^\dag \phi|^2 \; ,
\end{equation}
with an order-one constant $c$. Unlike in the Schmaltz model this
value does not have to be too large; in the Coleman--Weinberg model we
typically find 
\begin{equation}
\frac{m_H^2}{\lambda} \sim \left( \frac{m_\kappa}{g} \right)^2 \; ,
\label{eq:littlesth_selfcoup}
\end{equation}
which we need to adjust to stay below the order $f^2$ which
Eq.\eqref{eq:littleh_selfcoup} gives for the Schmaltz model.\bigskip

To protect the Higgs mass against the top loop we again extend the
$SU(2)_L$ quark doublet to the triplet $\Psi=(t, b ,T)$ and add a
right handed singlet $t'^c$. Because we expect mixing between the two
top singlets which will give us the Standard--Model and a heavy top
quark we write two general Yukawa couplings for the Standard Model
doublet and the additional heavy states. The first is mediated by the
$\s$ field as
\begin{equation}
  \lag \supset
  \lambda_1 \; f \; \epsilon_{ijk} \Psi_i \; \s_{j4} \s_{k5} \; t_1^c
 +\lambda_2 \; f \; T t_2^c \; .
\end{equation}
This form uses the $2\times 3$ triplets from the upper--right corner of
the Goldstone matrix in Eq.\eqref{eq:littlesth_goldstones2}
\begin{equation}
 \s_{jm} = \begin{pmatrix}\kappa^\dag\\ \phi^\dag  \\
                     \end{pmatrix}
 \qquad \qquad
  j=1,2,3 \qquad m=4,5 \; .
\end{equation}
These triplets represent the $SU(3)$ sub-matrix of the $SU(5)$
generators which requires the Higgs mass to be zero.  This means that
if we set $\lambda_2=0$ this Yukawa coupling as an anti--symmetric
combination of three triplets is $SU(3)$ symmetric.  The top--induced
contributions to the Higgs mass will be proportional to $\lambda_1^2
\lambda_2^2$ and quadratic divergences are forbidden at one loop.

The two heavy quarks mix to the SM top quark and an additional heavy
top with mass
$m_T = \sqrt{\lambda_1^2+\lambda_2^2} \, f$
where as before we assume $f=f_1=f_2$. The actual top--Higgs coupling
are of the order $\min \lambda_j$ for the three-point Higgs coupling
to the Standard Model top and $\lambda/f$ for the four-point Higgs
coupling to a pair of heavy tops.\bigskip

Looking at the complete set of $SU(5)$ generators in
Eq.\eqref{eq:littlesth_goldstones2} we can collect the heavy spectrum of
the littlest Higgs model,
\begin{alignat}{3}
 \text{$SU(2)_L \times U(1)_Y$ gauge bosons $B',W'^{\pm},Z'$} \qquad & \text{with}
   \quad m_{B',W',Z'} = \ope(g f) \notag \\ 
 \text{heavy `Higgs' triplet} \;
 \kappa=\begin{pmatrix} \kappa^{++} & \kappa^+ \\ 
                        \kappa^+ & \kappa^0\\
        \end{pmatrix}      \qquad & \text{with}
   \quad m_\kappa = \ope(g f) \notag \\ 
 \text{heavy top $T$}                                    \qquad & \text{with}
   \quad m_T = \ope(\lambda f) \; .
\end{alignat}
From the $B'$ and $\kappa$ fields we expect a serious violation of the custodial
$SU(2)$ symmetry.  This means that electroweak precision data forces
us to choose $f$ unusually large, in conflict with the requirement in
Eq.\eqref{eq:littlesth_selfcoup}. Moreover, the Higgs triplet should
not become too heavy to maintain the correct relative size of the
Higgs mass and the Higgs self coupling.  Such a Higgs triplet with a
doubly charged Higgs boson has a smoking--gun signature at the LHC,
namely its production in weak boson fusion $uu \to dd W^+W^+ \to dd
H^{++}$.\bigskip

While the littlest Higgs setup solves some of the issues of the
Schmaltz model, it definitely has its problems linked to the Higgs
triplet and the heavy $U(1)$ gauge boson. To not violate the custodial
symmetry too badly it would be great to introduce some kind of $Z_2$
symmetry which allows only two heavy new particles in any vertex. The
same symmetry would give us a stable lightest new 
particle as a weakly interacting dark matter candidate.  All we need to to is define a
quantum number with one value for all weak--scale Standard Model
particles and another value for all particles with masses around $f$.
Such a parity will be called \underline{$T$ parity}.\bigskip

For the littlest Higgs, we would like to separate the additional heavy
states, including the $SU(2)$ doublet, from our Standard Model Higgs
and gauge bosons.  The symmetry we want to introduce should multiply
all heavy entries in the sigma field of
Eq.\eqref{eq:littlesth_goldstones2} by a factor $(-1)$. By
multiplying out the matrices we show that there exists a matrix
$\Omega$ such that
\begin{equation}
(w \cdot \hat{T}) \to \Omega^{-1} \, \left( w \cdot \hat{T} \right) \, \Omega \; .
\end{equation}
One matrix $\Omega$ for which this works is
\begin{alignat}{4}
\begin{pmatrix}\chi & \phi^*&\kappa^\dag\\
                                                 \phi^T& 0  & \phi^\dag  \\ 
                                                \kappa& \phi  &\chi^T\\
                             \end{pmatrix}
                     &+\frac{\eta}{2 \sqrt{5}} \begin{pmatrix}\one&0&0\\0&-4&0\\0&0&\one\\
                                 \end{pmatrix} \notag \\
      & \to  i^2 \begin{pmatrix}\one&0&0\\0&-1&0\\0&0&\one\\
                             \end{pmatrix}
                       \left[
                             \begin{pmatrix}\chi & \phi^*&\kappa^\dag\\
                                                       \phi^T& 0  & \phi^\dag  \\ 
                                                      \kappa& \phi  &\chi^T   \\
                                   \end{pmatrix}
                             +\frac{\eta}{2 \sqrt{5}}
                             \begin{pmatrix}\one&0&0\\0&-4&0\\0&0&\one\\
                                   \end{pmatrix}
                       \right]
                       \begin{pmatrix}\one&0&0\\0&-1&0\\0&0&\one\\    
                             \end{pmatrix} \notag \\
                  &= - \begin{pmatrix}\one&0&0\\0&-1&0\\0&0&\one\\
                             \end{pmatrix}
                       \left[ 
                             \begin{pmatrix}\chi & -\phi^*&\kappa^\dag\\
                                                       \phi^T& 0   & \phi^\dag  \\ 
                                                      \kappa& -\phi  &\chi^T   \\
                                   \end{pmatrix}
                             +\frac{\eta}{2 \sqrt{5}}
                             \begin{pmatrix}\one&0&0\\0&4&0\\0&0&\one\\
                                   \end{pmatrix}
                       \right] \notag \\
                  &= - \begin{pmatrix}\chi & -\phi^*&\kappa^\dag\\
                                                -\phi^T& 0   & -\phi^\dag \\ 
                                                \kappa& -\phi  &\chi^T   \\
                             \end{pmatrix}
                     - \frac{\eta}{2 \sqrt{5}}
                       \begin{pmatrix}\one&0&0\\0&-4&0\\0&0&\one\\
                             \end{pmatrix} \notag \\
                  &=   \begin{pmatrix}-\chi & \phi^*&-\kappa^\dag\\
                                                  \phi^T& 0  & \phi^\dag  \\ 
                                                -\kappa& \phi  &-\chi^T  \\
                             \end{pmatrix}
                     -\frac{\eta}{2 \sqrt{5}}
                       \begin{pmatrix}\one&0&0\\0&-4&0\\0&0&\one\\
                             \end{pmatrix} \; .
\end{alignat}
This symmetry works perfectly for the additional gauge bosons,
including the heavy scalars $\kappa$. For the massive twins of the
$SU(2)_L$ gauge bosons we rely on the fact that in the special case
$g_1 = g_2$ the Lagrangian involving $D_\mu \s$ is symmetric under the
exchange of the two $SU(2) \times U(1)$ groups. The eigenstates we can
choose as $W_\pm = (W_1 \pm W_2)/\sqrt{2}$ and the same for the $B$
fields.  Of these two $W_+,B_+$ are Standard Model gauge bosons, while
$W_-,B_-$ are heavy. Exchanging the indices (1$\leftrightarrow$2) is
an even transformation for $W_+$, while it is odd for $W_-$, again
just as we want.

A problem arises when we assign such a quantum number to the heavy
tops which form part of a triplet extending the usual Standard Model
quark doublets. Getting worse, we have to be very careful to then
implement $T$ parity specifically taking care that it is not broken by
anomalies. At this point, it turns our that we have to introduce
additional fermions and the model rapidly loses its concise structure
as the price for a better agreement with electroweak precision
constraints.\bigskip

In summary, it is fair to say that collective symmetry breaking is an
attractive idea, based on a fundamental property like Goldstone's
theorem\index{Goldstone's theorem}. Already at the very beginning we
notice that its ultraviolet completion will be strongly interacting,
which some theorists would consider not attractive. Certainly, it is
not clear how the measurement of an approximate gauge coupling
unification would fit into such a picture. The same holds true for the
fixed point arguments which we present in
Section~\ref{sec:higgs_fixedpoint}. What is more worrisome is that it
appears to be hard to implement collective symmetry breaking in a
compact model which is not obviously ruled out or inconsistent. This
might well be a sign that protecting the Higgs mass through a
pseudo--Goldstone property is not what is realized in Nature.

\subsection{Higgs inflation} 
\label{sec:higgs_inflation}

Going beyond the weak scale and any energy scale we will probe with
the LHC we can ask another question after discovering the first
fundamental scalar particle in the Standard Model: can a
Standard--Model--like Higgs boson be the scalar particle responsible
for inflation?\bigskip

One of the most pressing problems in cosmology is the question, why
the cosmic microwave background radiation is so homogeneous while
based on the usual evolution of the Universe different regions cannot
be causally connected. A solution to this problem is to postulate an
era of exponentially accelerated expansion of the Universe which
would allow all these regions to actually know about each other.

We can trigger inflation through a scalar field $\chi$ in a potential
$U(\chi)$. In the beginning, this field is located far away from its
stable vacuum state. While moving towards the minimum of its potential
it releases energy. Slow roll inflation can be linked to two physical
conditions: on the one hand we require that in the equation of motion
for the inflaton field $\chi$ we can neglect the kinetic term, which
means
\begin{alignat}{5}
0 = \frac{\p^2 \chi}{\p t^2} + 3 \hat{H} \frac{\p \chi}{\p t} 
    + \frac{d U}{d \chi} 
\simeq 3 \hat{H} \frac{\p \chi}{\p t} + \frac{d U}{d \chi} \; .
\label{eq:inflation_eom}
\end{alignat}
The second term is proportional to $\p \chi/\p t$ and therefore a
friction term, called Hubble friction. To not confuse it with the
Higgs field we denote the Hubble constant as $\hat{H}$. In the absence
of the kinetic term the equation of state for the inflaton field
$\chi$ can behave like $w \equiv p/\rho <-1/3$. Given the pressure $p$
and the energy density $\rho$ this is the condition for inflation.
Negative values for $w$ arise when the potential $U$ dominates the
energy of the inflaton and the change of the field value $\chi$ with
time it small.  Equivalently, we can require two parameters which
describe the variation of the potential $U(\chi)$ to be small,
\begin{alignat}{5}
\epsilon = \frac{M_\text{Planck}^2}{2} \; 
\left( \frac{1}{U} \; \frac{dU}{d\chi}
\right)^2 \ll 1 
\qqqquad          
\left| \eta \right| = M_\text{Planck}^2
\left| \frac{1}{U} \; \frac{d^2 U}{d \chi^2} \right| \ll 1 \; .
\label{eq:inflation_slowroll}
\end{alignat}
The powers of $M_\text{Planck}$ give the correct units.  The two slow
roll conditions in Eq.\eqref{eq:inflation_eom} and
Eq.\eqref{eq:inflation_slowroll} are equivalent, which means that for
Higgs inflation we need to compute $U(\chi)$ with the appropriate
inflaton field and test the conditions given in
Eq.\eqref{eq:inflation_slowroll}.\bigskip

The starting point of any field theoretical explanation of inflation
is the Einstein--Hilbert action for gravity,
\begin{equation}
\boxed{
S = - \int d^4x \; \sqrt{-g} \; \frac{M_\text{Planck}^2}{2} R 
} \; ,
\label{eq:inflation_eh}
\end{equation}
with the Planck mass $M_\text{Planck}$ as the only free parameter, the
Ricci scalar with the graviton field, and no interactions between the
gravitational sector and matter. In addition, we neglect the
cosmological constant. If we combine the Higgs and gravitational
sectors the minimal coupling of the two sectors is generated through
the gravitational coupling to the energy--momentum tensor including
all Standard Model particles. It turns out that for Higgs inflation we
need an additional \underline{non--minimal coupling} between the two
sectors, so we start with the ansatz
\begin{alignat}{5}
S_J(H) 
&= \int d^4x \; \sqrt{-g}
\left[ - \frac{M^2}{2} R - \xi \frac{(v+H)^2}{2} R
       + \frac{1}{2} \p_\mu H \p^\mu H
       - \frac{\lambda}{4} ((v+H)^2 - v^2)^2 
\right] \notag \\
&= \int d^4x \; \sqrt{-g}
\left[ - \frac{M^2}{2} R - \xi \frac{(v+H)^2}{2} R
       + \frac{1}{2} \p_\mu H \p^\mu H
       - \frac{\lambda}{4} (v+H)^4
       + \frac{\lambda v^2}{2} (v+H)^2 
       + \text{const}
\right] \; .
\label{eq:inflation_s_jordan}
\end{alignat}
This form of the Higgs potential in the second line corresponds
Eq.\eqref{eq:higgs_pot} after inserting $\mu^2 =
-\lambda v^2 $ according to Eq.\eqref{eq:higgs_d4}.
The form of the \underline{Einstein--Hilbert action} suggests that
first of all the fundamental Planck mass $M$ is replaced by an
effective, observed Planck mass $M_\text{Planck}$ in a scalar field
background,
\begin{equation}
M_\text{Planck}^2 = M^2 + \xi (v + H)^2 \; .
\label{eq:inflation_s_jordan2}
\end{equation}
First, we assume $\xi v^2 \ll M^2$. For the specific case of Higgs
inflation we in addition postulate $\xi \gg 1$, but with the original
hierarchy still intact. The hierarchy between $\xi H$ and $M$ will be
discussed later.\bigskip

The action in the Jordan frame given by
Eq.\eqref{eq:inflation_s_jordan} with the identification
Eq.\eqref{eq:inflation_s_jordan2} is a little cumbersome to treat
gravity problems. We can decouple the gravitational and Higgs sectors
via a field re-definition into the Einstein frame and quote the result
as
\begin{alignat}{5}
S_E(\chi) 
&= \int d^4x \; \sqrt{-\hat{g}}
\left[ - \frac{M_\text{Planck}^2}{2} R 
       + \frac{1}{2} \p_\mu \chi \p^\mu \chi
       - \underbrace{\frac{\lambda}{4} \; \frac{M_\text{Planck}^4}{(M^2 + \xi (v+H(\chi))^2)^2} ((v+H(\chi))^2 - v^2)^2}_{\equiv U(\chi)}
\right] \notag \\
& \hat{g}_{\mu \nu} = \frac{M^2 + \xi (v+H)^2}{M_\text{Planck}^2} \; 
                     g_{\mu \nu} \notag \\
& \frac{d \chi}{dH} = 
\left[ 
\frac{1 + (\xi + 6 \xi^2) \dfrac{(v+H)^2}{M_\text{Planck}^2}}
     {\left(1 + \xi \dfrac{(v+H)^2}{M_\text{Planck}^2} \right)^2}
\right]^{1/2} \; .
\label{eq:inflation_s_einstein}
\end{alignat}
The original Higgs potential in terms of $H$ is replaced by the
inflaton--Higgs scalar $\chi$ and its combined potential
$U(\chi)$. The question is if this scalar field $\chi$ can explain
inflation. After studying some basic features of this theory our main
task will be to determine the value of $\xi$ which would make such a
model theoretically and experimentally feasible.\bigskip

If we are interested in the evolution of the early universe the
condition $\xi v^2 \ll M^2$ simplifies the above equations, but it
does not imply anything for the hierarchy between the Higgs field
values $H$ and the fundamental Planck scale $M$.  First, in the limit
$\xi H^2, \xi^2 H^2 \ll M_\text{Planck}^2 \sim M^2$ we can solve the
relation between the Higgs field $H$ and its re-scaled counter part
$\chi$,
\begin{alignat}{5}
\frac{d \chi}{dH} = 
\left[ 1 
+ \ope \left(\frac{\xi H^2}{M_\text{Planck}^2}\right) 
+ \ope \left(\frac{\xi^2 H^2}{M_\text{Planck}^2}\right) 
\right]^{1/2} 
\simeq 1 
\qquad \Leftrightarrow \qquad 
\chi \simeq H \; .
\end{alignat}
In this derivation we already see that we have to deal with two
additional energy scales, $M_\text{Planck}/\sqrt{\xi}$ and
$M_\text{Planck}/\xi$.  In this limit the potential for the re-scaled
Higgs field $\chi$ in the Einstein frame becomes
\begin{alignat}{5}
U(\chi) = 
\frac{\lambda}{4} \; \frac{M_\text{Planck}^4}{(M^2 + \xi (v+H)^2)^2} \; ((v+H)^2 - v^2)^2 
&\simeq 
\frac{\lambda}{4} \; ((v+\chi)^2 - v^2)^2 \; .
\label{eq:inflation_pot_low}
\end{alignat}
This is exactly the usual Higgs potential at low energies.\bigskip

The opposite limit is $\xi H^2 \gg M_\text{Planck}^2 \sim M^2 \gg \xi
v^2$. If we avoid Higgs field strength exceeding the Planck scale this
condition implicitly assume $\xi \gg 1$ and therefore $\xi^2 H \gg
\xi H^2$. We find
\begin{alignat}{5}
\frac{d \chi}{dH} 
&= 
\left[ 
\frac{6 \xi^2 H^2/M_\text{Planck}^2}{\xi^2 H^4/M_\text{Planck}^4}
+ \ope \left(\frac{M_\text{Planck}^2}{\xi H^2}\right) 
\right]^{1/2}
= 
\left[ 
\frac{6 M_\text{Planck}^2}{H^2}
+ \ope \left(\frac{M_\text{Planck}^2}{\xi H^2}\right) 
\right]^{1/2}
\simeq 
\frac{\sqrt{6} M_\text{Planck}}{H}
\notag \\
\Leftrightarrow \qqquad 
\chi &\simeq \sqrt{\frac{3}{2}} M_\text{Planck} \log \frac{\xi H^2}{M_\text{Planck}^2} 
\qqquad \Leftrightarrow \qqquad 
H^2 \simeq \frac{M_\text{Planck}^2}{\xi} \; 
      e^{\sqrt{2} \chi/(\sqrt{3} M_\text{Planck})} 
 \; .
\end{alignat}
The integration constants in this result are chosen appropriately. We
can use this relation to compute the leading terms in the scalar
potential for $\chi$ in the Einstein frame and for large Higgs field
values,
\begin{alignat}{5}
&\; U(\chi) \simeq 
\frac{\lambda}{4} \; \frac{M_\text{Planck}^4}{(M^2 + \xi H^2)^2} \; H^4
\simeq 
\frac{\lambda}{4} \; \dfrac{M_\text{Planck}^4 H^4}{\xi^2 H^4 \left( 1 + \dfrac{M^2}{\xi H^2}\right)^2} 
\simeq 
\frac{\lambda}{4} \; \frac{M_\text{Planck}^4}{\xi^2} 
\left( 1 - 2 \frac{M_\text{Planck}^2}{\xi H^2} \right) \notag \\
&\boxed{ U(\chi) \simeq 
\frac{\lambda}{4} \; \frac{M_\text{Planck}^4}{\xi^2} 
\left( 1 - 2 e^{-\frac{\sqrt{2}\chi}{\sqrt{3}M_\text{Planck}}} \right) } \; .
\label{eq:inflation_pot_high}
\end{alignat}
We show $U(\chi)$ in Figure~\ref{fig:inflation}. Following
Eq.\eqref{eq:inflation_pot_low} and Eq.\eqref{eq:inflation_pot_high}
it resembles the usual Higgs potential at small values of $H$ and
$\chi$ and becomes flat at large field values. This means that we can
indeed use the Higgs scalar as the inflaton, but we need to see what
the slow roll conditions from Eq.~\eqref{eq:inflation_slowroll} tell
us about the model parameter $\xi$ in the action introduced in
Eq.\eqref{eq:inflation_s_jordan}. Following the discussion in
Section~\ref{sec:higgs_coleman} we can compute additional
contributions to $U$ from all Standard Model fields, but for our
purpose the leading behavior is fine.\bigskip

\begin{figure}[t]
\begin{center}
\includegraphics[width=0.50\textwidth]{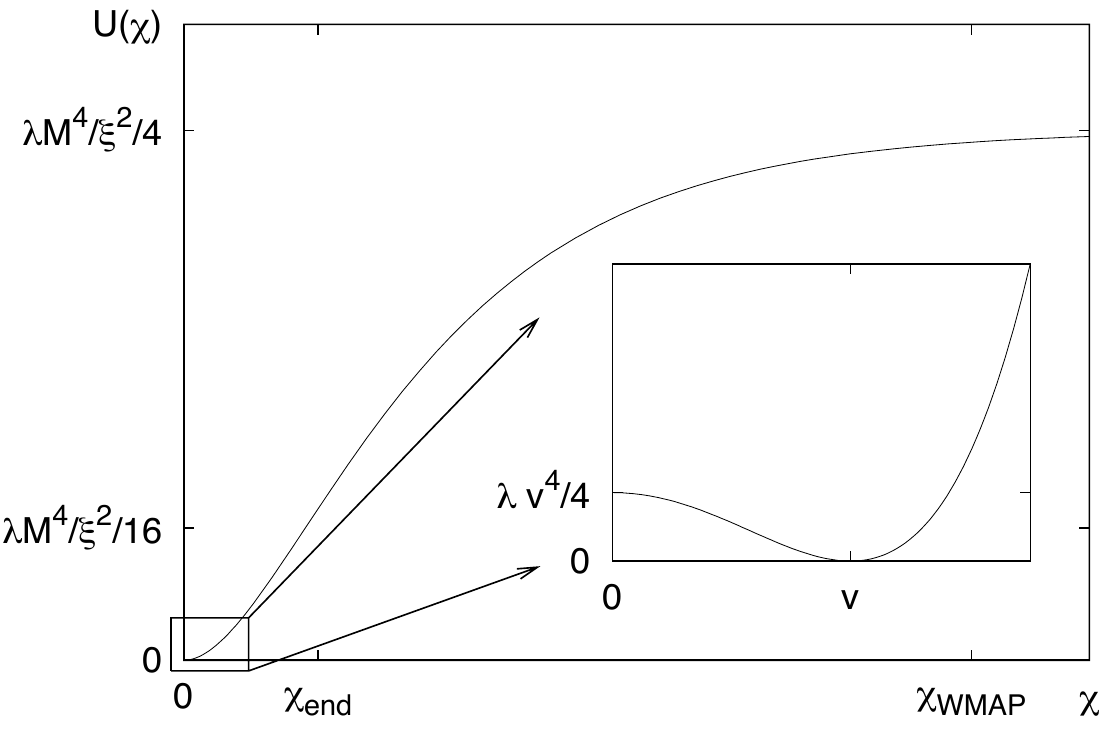}
\end{center}
\caption{Potential $U(\chi)$ as a function of the $\chi$.
  Figure from Ref.~\cite{Bezrukov:2013fka}.
\label{fig:inflation}}
\end{figure}

For this test we need to compute the first derivative of the potential
$U$ in the limit of large field values, because this is where we
expect the field $\chi$ to act as the inflaton. For the first slow
roll parameter $\epsilon$ we find
\begin{alignat}{5}
\frac{d U}{d\chi}
&= - \frac{\lambda}{2} \; \frac{M_\text{Planck}^4}{\xi^2} \; 
\frac{d}{d\chi} \; 
 e^{-\frac{\sqrt{2}\chi}{\sqrt{3}M_\text{Planck}}} 
= \frac{\lambda}{\sqrt{6}} \; \frac{M_\text{Planck}^3}{\xi^2} \; 
 e^{-\frac{\sqrt{2}\chi}{\sqrt{3}M_\text{Planck}}} 
= \frac{\lambda}{\sqrt{6}} \; \frac{M_\text{Planck}^5}{\xi^3 H^2} \notag \\
\epsilon&= 
\frac{M_\text{Planck}^2}{2} \; 
\left( \frac{\lambda^2}{6} \; \frac{M_\text{Planck}^{10}}{\xi^6 H^4} \right) \; 
\left( \frac{16}{\lambda^2} \; \frac{\xi^4}{M_\text{Planck}^8} \right)
= \frac{4}{3} \; \frac{M_\text{Planck}^4}{\xi^2 H^4} \ll 1
\qquad \Leftrightarrow \qquad H \gg \frac{M_\text{Planck}}{\sqrt{\xi}} \; .
\label{eq:inflation_epsilon}
\end{alignat}
Similarly, the second parameter $\eta$ comes out as
\begin{alignat}{5}
\frac{d^2 U}{d\chi^2}
&= \frac{\lambda}{\sqrt{6}} \; \frac{M_\text{Planck}^3}{\xi^2} \; 
 \frac{d}{d\chi} \; e^{-\frac{\sqrt{2}\chi}{\sqrt{3}M_\text{Planck}}} 
= - \frac{\lambda}{3} \; \frac{M_\text{Planck}^2}{\xi^2} \; 
 e^{-\frac{\sqrt{2}\chi}{\sqrt{3}M_\text{Planck}}} 
= - \frac{\lambda}{3} \; \frac{M_\text{Planck}^4}{\xi^3 H^2} \notag \\
|\eta|
&= M_\text{Planck}^2 \; 
\left( \frac{\lambda}{3} \; \frac{M_\text{Planck}^4}{\xi^3 H^2} \right) \;
\left( \frac{4}{\lambda} \; \frac{\xi^2}{M_\text{Planck}^4} \right)
= \frac{4}{3} \; \frac{M_\text{Planck}^2}{\xi H^2} \ll 1 
\qquad \Leftrightarrow \qquad H \gg \frac{M_\text{Planck}}{\sqrt{\xi}} \; .
\label{eq:inflation_eta}
\end{alignat}
Both slow roll conditions are identical and correspond to the original
condition we used to compute the potential $U(\chi)$ in the limit of
large field values. The potential $U(\chi)$ leads to slow roll
inflation, and since with time the Higgs field value becomes smaller
inflation ends for $H \sim M_\text{Planck}/\sqrt{\xi}$ and
correspondingly $\chi \sim M_\text{Planck}$.\bigskip

After confirming that Higgs inflation has all the correct theoretical
features we can confront it with data, most specifically with the
different measurements of the cosmic microwave background. Without
derivation we quote that during inflation the visible CMB modes
crossed the horizon at field values $H \simeq 9
M_\text{Planck}/\sqrt{\xi}$, shortly before the condition $H \simeq
M_\text{Planck}/\sqrt{\xi}$ for the end of inflation is finally reached.
Experimentally, the normalization of these observed CMB modes requires
\begin{alignat}{5}
\left( \frac{M_\text{Planck}}{36} \right)^4 \simeq \frac{U}{\epsilon} 
&= \frac{\lambda}{4} \; \frac{M_\text{Planck}^4}{\xi^2} \;
\frac{4}{3} \; \frac{\xi^2 H^4}{M_\text{Planck}^4} 
 = \frac{\lambda}{3} \; H^4 \; 
\qquad 
&&\text{using Eq.\eqref{eq:inflation_pot_high} and Eq.\eqref{eq:inflation_epsilon}}
\notag \\
&\simeq \frac{\lambda}{3} \; 
\left( \frac{9 M_\text{Planck}}{\sqrt{\xi}} \right)^4 \; 
&&\text{from CMB measurement}
\notag \\
\Leftrightarrow \qquad 
\lambda &=\frac{3}{324^4} \;  \xi^2 
\qquad \Leftrightarrow \qquad 
\boxed{\xi \simeq 60000 \; \sqrt{\lambda}} \; .
\end{alignat}
While at the electroweak scale we know that $\lambda \sim 1/8$ it
becomes much smaller at high energies, so the actual value of $\xi$ is
tricky to extract with the Higgs data available right now. However,
this measurement of $\xi$ is in agreement with our original
assumption. Additional measurements of the cosmic microwave background
are in agreement with this setup.  In particular, the Hubble scale
$\sqrt{\lambda} M_\text{Planck}/\xi$ is small enough to avoid
non--Gaussianities in the cosmic microwave background. This short
introduction leaves us with a few open questions:
\begin{enumerate}
\item Higgs inflation is theoretically and experimentally feasible
\item the coupling strength $\xi \gg 1$ needs a good explanation
\item adding the $\xi H^2 R$ coupling leads to unitarity violation in
  $HH \to HH$ scattering, so it requires an unknown ultraviolet
  completion
\item the cosmic microwave background decouples around $H \simeq
  M_\text{Planck}/10 \gg v$, so slow roll inflation and electroweak
  symmetry breaking are really described by separate potentials written as
  one function $U(\chi)$
\item should the Higgs potential become unstable at high energies,
  the finite life time of the Universe can be used to save the
  Standard Model, but Higgs inflation will break down in the
  presence of an alternative global minimum.
\end{enumerate}

\subsection{Further reading} 

At this point we are done with our brief review of the Higgs sector in
the Standard Model and of contemporary Higgs phenomenology. From the
discussions it should be clear that we have to move on to QCD, to
understand the strengths and weaknesses of these searches and what
distinguishes a good from a bad search channel.

Before moving on we mention a few papers where you can learn more
about Higgs phenomenology at the LHC.  Luckily, for Higgs searches
there are several very complete and useful reviews available:

\begin{itemize}
\item[--] You can look at the original articles by
  Higgs~\cite{Higgs:1964ia,Higgs:1964pj} or Brout and
  Englert~\cite{Englert:1964et}, but they are rather short and not
  really on top of the phenomenological aspects of the topic. Other
  papers for example by Guralnik, Hagen, Kibble~\cite{Guralnik:1964eu}
  tend to be harder to read for phenomenologically interested
  students.

\item[--] Wolfgang Kilian's book~\cite{Kilian:2003pc} on the effective
  field theory approach to the Higgs mechanism is what the
  corresponding sections in these lecture notes are largely based on
  it. The underlying symmetry structure, including the custodial
  symmetry is nicely described in Scott Willenbrock's TASI
  lectures~\cite{Willenbrock:2004hu}.

\item[--] If you are interested in a comprehensive overview of Higgs
  physics as an effective field theory with a special focus on
  higher--dimensional operators in linear and non--linear sigma models
  there is a really useful paper from Spain~\cite{Brivio:2013pma}.

\item[--] Abdelhak Djouadi's compilation on `absolutely everything we
  know about Higgs phenomenology' is indeed exhaustive. It has two
  volumes, one on the Standard Model~\cite{Djouadi:2005gi} and another
  one on its minimal supersymmetric extension~\cite{Djouadi:2005gj}

\item[--] For more experimental details you might want to have a look
  at Karl Jakobs' and Volker B\"uscher's review of LHC Higgs
  physics~\cite{Buescher:2005re}.
 
\item[--] As a theory view on LHC Higgs physics, mostly focused on
  gluon fusion production and its QCD aspects, there is Michael
  Spira's classic~\cite{Spira:1997dg}. This is where you can find more
  information on the low energy theorem. Michael and his collaborators
  also published a set of lecture notes on Higgs
  physics~\cite{GomezBock:2007hp}. 

\item[--] As always, there is a TASI lecture on the topic. TASI
  lecture notes are generally what you should look for when you are
  interested in an area of high energy physics. Dave Rainwater did not
  only write his thesis on Higgs searches in weak boson
  fusion~\cite{Rainwater:1999gg}, he also left us all he knows about
  Higgs phenomenology at the LHC in his TASI
  notes~\cite{Rainwater:2007cp}.

\item[--] Tao Han wrote a very comprehensible set of TASI lecture
  notes on basic LHC phenomenology, in case you need to catch up on
  this~\cite{Han:2005mu}.

\item[--] For some information on electroweak precision data and the
  $\rho$ parameter, there are James Wells' TASI
  lectures~\cite{Wells:2005vk}.

\item[--] If you are interested in Higgs production in association
  with a $W$ or $Z$ boson and the way to observe boosted $H \to b
  \bar{b}$ decays you need to read the original
  paper~\cite{Butterworth:2008iy}. The same is true for the
  $t\bar{t}H$ analysis.

\item[--] For cut rules and scalar integrals the best writeup I know
  is Wim Beenakker's PhD thesis. Unfortunately, I am not sure where
  to get it from except from the author by request.

\item[--] If you are getting interested in fixed points in RGE
  analyses you can look at Christoph Wetterich's original
  paper~\cite{christof} or a nice introductory review by Barbara
  Schrempp and Michael Wimmer~\cite{Schrempp:1996fb}.

\item[--] My discussion on technicolor largely follows the extensive
  review by Chris Hill and Elisabeth Simmons~\cite{Hill:2002ap}.

\item[--] A really nice writeup which my little Higgs discussion is
  based on is Martin Schmaltz' and David Tucker--Smith's review
  article~\cite{schmaltz}.

\item[--] For more information on Higgs inflation you can start with a
  nice set of TASI lectures on inflation~\cite{Baumann:2009ds} and
  then dive into a specific review of Higgs inflation by one of its
  inventors~\cite{Bezrukov:2013fka}.

\end{itemize}

\newpage

\section{QCD}
\label{sec:qcd}

Just as Section~\ref{sec:higgs} is not meant to be a complete
introduction to electroweak symmetry breaking but is aimed at
introducing the aspects of Higgs physics most relevant to the LHC this
section cannot cover the entire field of QCD. Instead, we will focus
on QCD as it impacts LHC physics, like for example the Higgs searches
discussed in the first part of the lecture.

In Section~\ref{sec:qcd_dy} we will introduce the most important
process at the LHC, the Drell--Yan process or lepton pair
production. This process will lead us through all of the introduction
into QCD. Ultraviolet divergences and renormalization we will only
mention in passing, to get an idea how much of the treatment of
ultraviolet and infrared divergences works the same way. After
discussing in detail infrared divergences in Sections~\ref{sec:qcd_ir}
to~\ref{sec:qcd_shower} we will spend some time on modern approaches
on combining QCD matrix element calculations at leading order and
next--to--leading order in perturbative QCD with parton showers. This
last part is fairly close to current research with the technical
details changing rapidly. Therefore, we will rely on toy models to
illustrate the different approaches.

\subsection{Drell--Yan process}
\label{sec:qcd_dy}

Most text books on QCD start from a very simple class of QCD
processes, called deep inelastic scattering. These are processes with
the HERA initial state $e^\pm p$. The problem with this approach is
that in the LHC era we would
like to instead understand processes of the kind $pp \to W$+jets,
$pp \to H$+jets, $pp \to t\bar{t}+$jets, or the production of new
particles with or without jets. These kind of signal and background
processes and their relevance in an LHC analysis we already mentioned
in Section~\ref{sec:higgs_gf_lhc}.\bigskip

From a QCD perspective such processes are very complex, so we need to step
back a little and start with a simple question: we know how to compute
the production rate and distributions for photon or $Z$ production
for example at LEP, $e^+e^- \to \gamma,Z \to \ell^+ \ell^-$. What is
then the production rate for the same final state at the LHC, how do we account
for quarks inside the protons, and what are the best--suited kinematic
variables to use at a hadron collider? 

\subsubsection{Gauge boson production}
\label{sec:qcd_dy_r}

The simplest question we can ask at the LHC is: how do we compute the
production of a single weak gauge boson?  This process we refer to as the
\underline{Drell--Yan production} process, in spite of producing neither Drell
nor Yan at the LHC.  In our first attempts we will explicitly not care
about additional jets, so if we assume the proton to consist of quarks
and gluons and simply compute the process $q \bar{q} \to \gamma,Z$
under the assumption that the quarks are partons inside
protons. Gluons do not couple to electroweak gauge bosons, so we only
have to consider valence quark vs sea antiquark scattering in the
initial state. Modulo the $SU(2)_L$ and $U(1)_Y$ charges which
describe the $Zf \bar{f}$ and $\gamma \bar{f} f$ couplings in the
Feynman rules\index{chiral projectors}
\begin{alignat}{5}
\boxed{- i \gamma^\mu \left( \ell \prol + r \pror \right)}
 \qquad \qquad \text{with} \quad 
 \ell &= \frac{e}{s_w c_w} \; \left( T_3 - 2 Q s_w^2 \right)
 \qquad \quad
 r = \ell \Big|_{T_3=0}
 \qquad \qquad &&(Zf \bar{f})
\notag \\
 \ell &= r = Q e
 &&(\gamma f \bar{f}) \; ,
\label{eq:dy_feyn}
\end{alignat}
with $T_3 = \pm 1/2$,
the matrix element and the squared matrix element for the partonic
process 
\begin{alignat}{5}
q \bar{q} \to \gamma,Z
\label{eq:qcd_dy1}
\end{alignat}
will be the same as the corresponding matrix element squared for $e^+
e^- \to \gamma,Z$, with an additional color factor.  The general
amplitude for \underline{massless fermions} is
\begin{alignat}{5}
\mat &= - i \bar{v}(k_2) \gamma^\mu \left( \ell \prol + r \pror \right) u(k_1)
        \epsilon_\mu \; .
\label{eq:dy_me1}
\end{alignat}
At the LHC massless fermions are a good approximation for all
particles except for the top quark. For the bottom quark we need to be
careful with some aspects of this approximation, but the first two
generations of quarks and all leptons are usually assumed to be massless
in LHC simulations. Once we will arrive at infrared divergences in LHC
cross sections we will specifically discuss ways of regulating them
without introducing masses.

Squaring the matrix element in Eq.\eqref{eq:dy_me1} means adding the
same structure once again, just walking through the Feynman diagram in the opposite
direction. Luckily, we do not have to care about factors of ($-i$)
since we are only interested in the absolute value squared. Because
the chiral projectors $\prolr = (\one \mp \gamma_5)/2$ defined in Eq.\eqref{eq:def_prolr} are real and
$\gamma_5^T=\gamma_5$ is symmetric, the left and right handed gauge
boson vertices described by the Feynman rules in Eq.\eqref{eq:dy_feyn} do not change under transposition.
For the production of a massive $Z$
boson on or off its mass shell we obtain
\begin{alignat}{5}
\matx &= \sum_\text{spin,pol,color} 
          \bar{u}(k_1) \gamma^\nu \left( \ell \prol + r \pror \right)  v(k_2)
   \;\;   \bar{v}(k_2) \gamma^\mu \left( \ell \prol + r \pror \right)  u(k_1)
   \;\;   \epsilon_\mu \epsilon^*_\nu
  && \text{incoming (anti-) quark} \; k_{1,2} 
  \notag \\
      &= N_c \tr \left[ \slashchar{k}_1
          \gamma^\nu \left( \ell \prol + r \pror \right) \slashchar{k}_2
          \gamma^\mu \left( \ell \prol + r \pror \right) \right] 
       \; \left( - g_{\mu \nu} + \frac{q_\mu q_\nu}{m_Z^2} \right)
  \qqquad &&\text{unitary gauge with} \; q = -k_1 - k_2
  \notag \\
      &= N_c \tr \left[  \slashchar{k}_1 \gamma^\nu
          \left( \ell \prol + r \pror \right) 
          \left( \ell \prol + r \pror \right) \slashchar{k}_2 \gamma^\mu \right] 
       \; \left( - g_{\mu \nu} + \frac{q_\mu q_\nu}{m_Z^2} \right)
  \qqquad &&\text{with} \, \{ \gamma_\mu,\gamma_5 \} = 0 
  \notag \\
      &= N_c \tr \left[  \slashchar{k}_1 \gamma^\nu
          \left( \ell^2 \frac{\one}{2} + r^2 \frac{\one}{2} \right) 
          \slashchar{k}_2 \gamma^\mu \right] 
       \; \left( - g_{\mu \nu} + \frac{q_\mu q_\nu}{m_Z^2} \right)
  \qqquad &&\text{symmetric polarization sum}
  \notag \\
      &= \frac{N_c}{2} \left( \ell^2 + r^2 \right) \tr \left[  
          \slashchar{k}_1 \gamma^\nu \slashchar{k}_2
          \gamma^\mu  \right] 
       \; \left( - g_{\mu \nu} + \frac{q_\mu q_\nu}{m_Z^2} \right)
  \notag \\
      &= 2 N_c \left( \ell^2 + r^2 \right) \left[  
         k_1^\mu k_2^\nu + k_1^\nu k_2^\mu - (k_1 k_2) g^{\mu \nu} \right] 
       \; \left( - g_{\mu \nu} + \frac{q_\mu q_\nu}{m_Z^2} \right)
  \notag \\
      &= 2 N_c \left( \ell^2 + r^2 \right) \left[  
         - 2 (k_1 k_2) + 4 (k_1 k_2)
         + 2 \frac{(-k_1 k_2)^2}{m_Z^2}
         - \frac{(k_1 k_2)q^2}{m_Z^2} \right] 
  \qqquad &&\text{with} \, (q k_1) = - (k_1 k_2) 
  \notag \\
      &= 2 N_c \left( \ell^2 + r^2 \right) \left[  
           2 (k_1 k_2) 
         + \frac{q^4}{2m_Z^2}
         - \frac{q^4}{2m_Z^2} \right] 
  \qqquad &&\text{with} \, q^2 = (k_1 + k_2)^2 
  \notag \\
      &= 2 N_c \left( \ell^2 + r^2 \right) q^2
\label{eq:dy_me2}
\end{alignat}
The color factor $N_c$ accounts for the number of $SU(3)$ states which
can be combined to form a color singlet like the $Z$.

An interesting aspect coming out of our calculation is that the
$1/m_Z$-dependent terms in the polarization sum do not contribute ---
as far as the matrix element squared is concerned the $Z$ boson could
as well be transverse. This reflects the fact that the Goldstone modes
do not couple to massless fermions, just like the Higgs boson.  This
means that not only the matrix element squared for the on--shell $Z$
case corresponds to $q^2 = m_Z^2$ but also that the on--shell photon
case is given by $q^2 \to 0$. The apparently vanishing matrix element
in this limit has to be combined with the phase space definition to
give a finite result.\bigskip

What is still missing is an averaging factor for initial--state spins
and colors, only the sum is included in Eq.\eqref{eq:dy_me2}.  For
incoming electrons as well as incoming quarks this factor $K_{ij}$
includes $1/4$ for the spins.  Since we do not observe color in the
initial state, and the color structure of the incoming $q \bar{q}$ pair
has no impact on the $Z$--production matrix element, we also average
over the color. This gives us another factor $1/N_c^2$ for the averaged
matrix element, which altogether becomes
\begin{alignat}{5}
K_{ij} = \frac{1}{4 N_c^2} \; .
\end{alignat}
In spite of our specific case in Eq.\eqref{eq:dy_me2} looking that way,
matrix elements we compute from our Feynman rules are not
automatically numbers with mass unit zero.\bigskip

If for the partonic invariant mass of the two quarks we introduce the
\underline{Mandelstam variable}\index{Mandelstam variable} $s = (k_2+k_2)^2 = 2 (k_1 k_2)$, so 
momentum conservation for on--shell $Z$ production implies $s = q^2 = m_Z^2$. 
In four space--time
dimensions (this detail will become important later) we can compute a
\underline{total cross section}\index{cross section!total $2 \to 1$}
from the matrix element squared, for example as given in
Eq.\eqref{eq:dy_me2}, as
\begin{alignat}{5}
\boxed{
 s \; \frac{d \sigma}{d y} \Bigg|_{2 \to 1}  = \frac{\pi}{(4 \pi)^2} \; K_{ij} \;
                               \left( 1 - \tau \right) \; \matx
}
\qqqquad 
\tau = \frac{m_Z^2}{s} \qqquad 
y = \frac{1 - \cos \theta}{2} \; .
\label{eq:qcd_two_to_one}
\end{alignat}
The scattering angle $\theta$ enters through the definition of $y = 0...1$.  The mass of
the final state appears in $\tau$, with $\tau =0$ for a massless
photon. It would be replaced to include $m_W$ or the Higgs mass or the
mass of a Kaluza--Klein graviton if needed. At the production threshold
of an on--shell particle the phase space opens in the limit $\tau \to
1$, slowly increasing the cross section above threshold $\tau <
1$.\bigskip

We know that such a heavy gauge boson we do not actually
observe at colliders.  What we should really calculate is the
production for example of a pair of fermions through an $s$-channel
$Z$ and $\gamma$, where the $Z$ might or might not be on its mass
shell. The matrix element for this process we can derive from the same
Feynman rules in Eq.\eqref{eq:dy_feyn}, now for an incoming fermion
$k_1$, incoming anti--fermion $k_2$, outgoing fermion $p_1$ and
outgoing anti--fermion $p_2$. To make it easy to switch particles
between initial and final states, we can define all momenta as
incoming, so momentum conservation means $k_1+k_2+p_1+p_2=0$. The
additional \underline{Mandelstam variables}\index{Mandelstam variable}
we need to describe this ($2 \to 2$) process are $t=(k_1+p_1)^2<0$ and
$u=(k_1+p_2)^2<0$, as usually with $s+t+u=0$ for massless final--state
particles. The ($2 \to 2$) matrix element for the two sets of incoming
and outgoing fermions becomes 
\begin{alignat}{5}
\mat = (- i)^2 
     \; 
     \bar{u}(p_1) \gamma^\nu \left( \ell' \prol + r' \pror \right) v(p_2)
     \; \; 
     \bar{v}(k_2) \gamma^\mu \left( \ell \prol + r \pror \right) u(k_1)
     \; \frac{i}{q^2-m_Z^2} \left( - g_{\mu \nu} + \frac{q_\mu q_\nu}{m_Z^2} \right) \; .
\label{eq:dy_me3}
\end{alignat}
The coupling to the gauge bosons are $\ell$ and $r$ for the incoming
quarks and $\ell'$ and $r'$ for the outgoing leptons. The chiral projectors 
are defined in Eq.\eqref{eq:def_prolr}. When we combine
the four different spinors and their momenta correctly the matrix
element squared factorizes into twice the trace we have computed
before. The corresponding picture is two fermion currents interacting
with each other through a gauge boson. All we have to do is combine
the traces properly. If the incoming trace in the matrix element and
its conjugate includes the indices $\mu$ and $\rho$ and the outgoing
trace the indices $\nu$ and $\sigma$, the $Z$ bosons link $\mu$ and
$\nu$ as well as $\rho$ and $\sigma$.

To make the results a little more compact we compute this process for
a \underline{massless photon} instead of the $Z$ boson, \ie for the
physical scenario where the initial--state fermions do not have enough
energy to excite the intermediate $Z$ boson. The specific features of
an intermediate massive $Z$ boson we postpone to
Section~\ref{sec:qcd_bw}.  The assumption of a massless photon
simplifies the couplings to $(\ell^2 + r^2) = 2Q^2 e^2$ and the
polarization sums to $-g_{\mu \nu}$ and $-g_{\rho \sigma}$:
\begin{alignat}{5}
\matx &= 4 N_c \; (2 Q^2 e^2) \; (2 {Q'}^2 e^2) \; \frac{1}{q^4} \;
   \left[ k_1^\mu k_2^\rho + k_1^\rho k_2^\mu - (k_1 k_2) g^{\mu \rho} \right] \;
   \left( - g_{\mu \nu} \right) \;
   \left[ p_1^\nu p_2^\sigma + p_1^\sigma p_2^\nu - (p_1 p_2) g^{\nu \sigma} \right] \;
   \left( - g_{\rho \sigma} \right)
 \notag \\
     &= 16 N_c \; Q^2 {Q'}^2 e^4 \; \frac{1}{q^4} \;
   \left[ k_1^\mu k_{2 \sigma} + k_{1 \sigma} k_2^\mu - (k_1 k_2) g_\sigma^\mu \right] \;
   \left[ p_{1 \mu} p_2^\sigma + p_1^\sigma p_{2 \mu} - (p_1 p_2) g^\sigma_\mu \right]
 \notag \\
     &= 16 N_c \; Q^2 {Q'}^2 e^4 \; \frac{1}{q^4} \;
   \left[  2 (k_1 p_1) (k_2 p_2) + 2 (k_1 p_2) (k_2 p_1) - 2 (k_1 k_2) (p_1 p_2)  
         - 2 (k_1 k_2) (p_1 p_2) + 4 (k_1 k_2) (p_1 p_2)  \right]
 \notag \\
     &= 32 N_c \; Q^2 {Q'}^2 e^4 \; \frac{1}{q^4} \;
   \left[  (k_1 p_1) (k_2 p_2) + (k_1 p_2) (k_2 p_1) \right]
 \notag \\
     &= 32 N_c \; Q^2 {Q'}^2 e^4 \; \frac{1}{s^2} \;
   \left[  \frac{t^2}{4} + \frac{u^2}{4} \right]
 \notag \\
     &= 8 N_c \; Q^2 {Q'}^2 e^4 \; \frac{1}{s^2} \; \left[ s^2 +2st + 2 t^2 \right]
 \notag \\
     &= 8 N_c \; Q^2 {Q'}^2 e^4 \; \left[ 1 + 2 \frac{t}{s} + 2 \frac{t^2}{s^2} \right] \; .
\label{eq:dy_me4}
\end{alignat}
We can briefly check if this number is indeed positive, using the
definition of the Mandelstam variable $t$ for massless external
particles in terms of the polar angle $t = s (-1 + \cos \theta)/2 = -s
\cdots 0$: the upper phase space boundary $t=0$ inserted into the
brackets in Eq.\eqref{eq:dy_me4} gives $[ \cdots ]=1$, just as the
lower boundary $t=-s$ with $[ \cdots ]=1-2+2=1$. For the
central value $t=-s/2$ the minimum value of the brackets is $[\cdots ]
= 1 -1 + 0.5 = 0.5$. 

The azimuthal angle $\phi$ plays no role at colliders, unless you
want to compute gravitational effects on Higgs production at ATLAS and
CMS.  Any LHC Monte Carlo will either random-generate a reference
angle $\phi$ for the partonic process or pick one and keep it fixed.

The two-particle phase space integration for massless particles then gives
us\index{cross section!total $2 \to 2$}
\begin{alignat}{5}
\boxed{
  s^2 \frac{d \sigma}{d t} \Bigg|_{2 \to 2} = 
 \frac{\pi}{(4 \pi)^2} \; K_{ij} \;  \matx
}
\qqqquad 
t = \frac{s}{2} \left( -1 + \cos \theta \right) \; .
\label{eq:qcd_two_to_two}
\end{alignat}
For our Drell--Yan process we then find the differential cross section
in four space--time dimensions, using $\alpha = e^2/(4 \pi)$
\begin{alignat}{5}
\frac{d \sigma}{d t} 
 &= \frac{1}{s^2} \frac{\pi}{(4 \pi)^2} \; \frac{1}{4 N_c} \; 
    8 \; Q^2 {Q'}^2 (4 \pi \alpha)^2 \; \left( 1 + 2 \frac{t}{s} + 2 \frac{t^2}{s^2} \right)
  \notag \\
 &= \frac{1}{s^2} \frac{2 \pi\alpha^2}{N_c} \; 
     Q^2 {Q'}^2 \; \left( 1 + 2 \frac{t}{s} + 2 \frac{t^2}{s^2} \right) \; ,
\end{alignat}
which we can integrate over the polar angle or the Mandelstam variable
$t$ to compute the total cross section
\begin{alignat}{5}
\sigma &=  \frac{1}{s^2} \frac{2 \pi\alpha^2}{N_c} \; 
     Q^2 {Q'}^2 \int_{-s}^0 dt \; \left( 1 + 2 \frac{t}{s} + 2 \frac{t^2}{s^2} \right)
  \notag \\
       &=  \frac{1}{s^2} \frac{2 \pi\alpha^2}{N_c} \; 
     Q^2 {Q'}^2 \left[ t + \frac{t^2}{s} + \frac{2t^3}{3s^2} \right]_{-s}^0
  \notag \\
       &=  \frac{1}{s^2} \frac{2 \pi\alpha^2}{N_c} \; 
     Q^2 {Q'}^2 \left( s - \frac{s^2}{s} + \frac{2s^3}{3s^2} \right)
  \notag \\
       &=  \frac{1}{s} \frac{2 \pi\alpha^2}{N_c} \; 
     Q^2 {Q'}^2 \; \frac{2}{3}
\qqquad \Rightarrow \qqquad 
\boxed{ 
\sigma(q\bar{q} \to \ell^+ \ell^-) \Bigg|_\text{QED}= \frac{4 \pi\alpha^2}{3 N_c s} \; Q_\ell^2 Q_q^2
}
\label{eq:dy_sigma1}
\end{alignat}
\bigskip

As a side remark --- in the history of QCD, the same process but
read \underline{right--to--left} played a crucial role, namely the
production rate of quarks in $e^+ e^-$ scattering. For small enough
energies we can neglect the $Z$ exchange contribution. At leading
order we can then compute the corresponding production cross sections
for muon pairs and for quark pairs in $e^+ e^-$ collisions. Moving the
quarks into the final state means that we do not average of the color
in the initial state, but sum over all possible color combinations,
which in Eq.\eqref{eq:qcd_two_to_two} gives us an averaging factor
$K_{ij} = 1/4$. Everything else stays the same as for the Drell--Yan
process
\begin{alignat}{5}
\boxed{
R \equiv \frac{\sigma(e^+ e^- \to \text{hadrons})}{\sigma(e^+ e^- \to \ell^+ \ell^-)}
} 
 = \dfrac{\sum_\text{quarks} \dfrac{4 \pi\alpha^2 N_c}{3 s} \; Q_e^2 Q_q^2}
         {\dfrac{4 \pi\alpha^2 }{3 s} \; Q_e^2 Q_\ell^2}
  = N_c \left( 3 \frac{1}{9} + 2 \frac{4}{9} \right) = \frac{11 N_c}{9} \; ,
\label{eq:qcd_r}
\end{alignat}
\index{R ratio}for example for five quark flavors where the top quark
is too heavy to be produced at the given $e^+ e^-$ collider
energy. For those interested in the details we did take one short cut:
hadrons are also produced in the hadronic decays of $e^+ e^- \to
\tau^+ \tau^-$ which we strictly speaking need to subtract. This way,
$R$ as a function of the collider energy is a beautiful measurement of
the weak and color charges of the quarks in QCD.

\subsubsection{Massive intermediate states}
\label{sec:qcd_bw}

At hadron colliders we cannot tune the energies of the incoming
partons. This means that for any particle we will always observe a mix
of on--shell and off--shell production, depending on the structure of
the matrix element and the distribution of partons inside the proton.
For hadron collider analyses this has profound consequences: unlike at
an $e^+ e^-$ collider we have to base all measurements on
reconstructed final--state particles. For studies of particles decaying
to jets this generally limits the possible precision at hadron
colliders to energy scales above $\lqcd$.  Therefore, before we move
on to describing incoming quarks inside protons we should briefly
consider the second Feynman diagram contributing to the Drell--Yan
production rate in Eq.\eqref{eq:dy_sigma1}, the on--shell or off--shell
$Z$ boson
\begin{alignat}{5}
 |\mat|^2 = |\mat_\gamma + \mat_Z |^2
          = |\mat_\gamma |^2 + |\mat_Z |^2 + 2 \, \text{Re} \, \mat_Z \mat_\gamma \; .
\end{alignat}
Interference occurs in phase space regions where for both
intermediate states the invariant masses of the muon pair are the
same. 

For the photon the on--shell pole is not a problem. It has zero
mass, which means that we hit the pole $1/q^2$ in the matrix element
squared only in the limit of zero incoming energy. Strictly speaking
we never hit it, because the energy of the incoming particles has to
be large enough to produce the final--state particles with their tiny
but finite masses and with some kind of momentum driving them through
the detector.

A problem arises when we consider the intermediate $Z$ boson. In that
case, the propagator contributes as $\matx \propto 1/(s-m_Z^2)^2$
which diverges on the mass shell.
Before we can ask what such a pole means for LHC simulations we have
to recall how we deal with it in field theory.  There, we encounter
the same issue when we solve for example the \underline{Klein--Gordon
equation}\index{Klein--Gordon equation}. The Green function for a field obeying this equation is the
inverse of the Klein--Gordon operator
\begin{alignat}{5}
\left( \Box + m^2 \right) G(x-x') = \delta^4 (x-x') \; .
\end{alignat}
Fourier transforming $G(x-x')$ into momentum space we find
\begin{alignat}{5}
G(x-x') &= \int \frac{d^4 q}{(2 \pi)^4} \; e^{-i q \cdot (x-x')} \tilde{G}(q)
 \notag \\
\left( \Box + m^2 \right) G(x-x') &= \int \frac{d^4 q}{(2 \pi)^4} \; 
  \left( \Box + m^2 \right) e^{-i q \cdot (x-x')} \; \tilde{G}(q)
 \notag \\
 &= \int \frac{d^4 q}{(2 \pi)^4} \; 
  \left( (i q)^2 + m^2 \right) e^{-i q \cdot (x-x')} \; \tilde{G}(q) 
 \notag \\
 &= \int \frac{d^4 q}{(2 \pi)^4} \; 
  e^{-i q \cdot (x-x')} \; \left( -q^2 + m^2 \right) \tilde{G}(q) \; \really \;
  \delta^4 (x-x') = \int \frac{d^4 q}{(2 \pi)^4} \; e^{-i q \cdot (x-x')}
 \notag \\
& \qquad \Leftrightarrow  \qquad ( -q^2 + m^2 ) \, \tilde{G}(q) = 1
\qquad \Leftrightarrow \qquad \tilde{G}(q) = - \frac{1}{q^2 - m^2} \; .
\label{eq:fourier_green}
\end{alignat}
The problem with the Green function in momentum space is that as an
inverse it is not defined for $q^2 = m^2$. We usually avoid this
problem by slightly shifting this pole following the \underline{Feynman $i
\epsilon$}\index{propagator!Feynman $i \epsilon$} prescription to $m^2 \to m^2 - i \epsilon$, or equivalently
deforming our integration contours appropriately. The sign of this
infinitesimal shift we need to understand because it will become
relevant for phenomenology when we introduce an actual finite decay
width of intermediate states.

In the Feynman $i \epsilon$ prescription the sign is crucial to
correctly complete the $q_0$ integration of the Fourier transform in
the complex plane
\begin{alignat}{5}
  \int_{-\infty}^\infty d q_0 \; 
  \frac{e^{-i q_0 x_0}}{q^2-m^2+i \epsilon}
&= \left( \theta(x_0) + \theta(-x_0) \right) \; \int_{-\infty}^\infty  d q_0 \; 
  \frac{e^{-i q_0 x_0}}{q_0^2- (\omega^2-i \epsilon )}
 \qqquad \text{with} \; \omega^2 = \vec{q}^2 + m^2
\\
&= \left( \theta(x_0) + \theta(-x_0) \right) \; \int_{-\infty}^\infty  d q_0 \; 
  \frac{e^{-i q_0 x_0}}
       {(q_0-\sqrt{\omega^2-i \epsilon})(q_0+\sqrt{\omega^2-i \epsilon})}
 \notag \\
&= \left( \theta(x_0) \oint_{C_2} + \theta(-x_0) \oint_{C_1} \right) dq_0
  \frac{e^{-i q_0 x_0}}
       {(q_0 -\omega (1-i \epsilon'))(q_0 + \omega (1-i \epsilon'))}
 \quad \text{with} \; \epsilon' = \frac{\epsilon}{2 \omega^2} \notag 
\end{alignat}
In the last step we have closed the integration contour along the real
$q_0$ axis in the complex $q_0$ plane. Because the integrand has to
vanish for large $q_0$, we have to make sure the exponent $-i x_0 \, i
\text{Im} \, q_0 = x_0 \text{Im} \, q_0$ is negative. For $x_0>0$ this
means $\text{Im} q_0 <0$ and vice versa. This argument forces $C_1$ to
close for positive and $C_2$ for negative imaginary parts in the
complex $q_0$ plane.

The contour integrals we can solve using Cauchy's formula\index{Cauchy integral}, keeping in
mind that the integrand has two poles at $q_0 = \pm \omega (1 - i
\epsilon')$. They lie in the upper (lower) half plane for negative
(positive) real parts of $q_0$. The contour $C_1$ through the upper
half plane includes the pole at $q_0 \sim -\omega$ while the contour
$C_2$ includes the pole at $q_0 \sim \omega$, all assuming $\omega >
0$:
\begin{alignat}{5}
  \int_{-\infty}^\infty d q_0 \; 
  \frac{e^{-i q_0 x_0}}{q^2-m^2+i \epsilon}
&= 2 \pi i \left[
   \theta(x_0) \frac{(-1) e^{-i \omega x_0}}
       {\omega + \omega (1-i \epsilon')}
 + \theta(-x_0) \frac{e^{i \omega x_0}}
       {-\omega - \omega (1-i \epsilon')} \right]
\notag \\
&\stackrel{\epsilon' \to 0}{=} - i \frac{\pi}{\omega} \left[
   \theta(x_0) e^{-i \omega x_0}
 + \theta(-x_0) e^{i \omega x_0} \right] \; .
\label{eq:prop_final}
\end{alignat}
The factor $(-1)$ in the $C_2$ integration arises because Cauchy's
integration formula requires us to integrate counter--clockwise, while
going from negative to positive $\text{Re} q_0$ the contour $C_2$ is
defined clockwise. Using this result we can complete the
four-dimensional Fourier transform from Eq.\eqref{eq:fourier_green}
\begin{alignat}{5}
G(x) 
 &=
  \int d^4 q \; e^{-i (q \cdot x)} \tilde{G}(q)  \notag \\
 &= 
  \int d^4 q \; \frac{e^{-i (q \cdot x)}}{q^2-m^2+i \epsilon} \notag \\
 &= -i \pi 
    \int d^3 \vec{q} \; e^{i \vec{q} \vec{x}} \; \frac{1}{\omega} \;
    \left[ \theta(x_0) e^{-i \omega x_0} + \theta(-x_0) e^{i \omega x_0} \right]
 \notag \\
 &= -i \pi 
    \int d^4 q \; e^{i \vec{q} \vec{x}} \; \frac{1}{\omega} \;
    \left[ \theta(x_0)   e^{-i q_0 x_0} \delta(q_0 - \omega) 
         + \theta(-x_0)  e^{-i q_0 x_0} \delta(q_o + \omega) \right] 
 \notag \\
 &= -i \pi 
    \int d^4 q \; e^{-i (q \cdot x)} \; \frac{1}{\omega} \;
    \left[ \theta(x_0)   \delta(\omega - q_0) 
         + \theta(-x_0)  \delta(\omega + q_0) \right] 
 \qqquad \text{with} \; \delta(x) = \delta(-x) 
 \notag \\
 &= -i \pi 
    \int d^4 q \; e^{-i (q \cdot x)} \; \frac{1}{\omega} \; 2 \omega \;
    \left[ \theta(x_0)   \delta(\omega^2 - q_0^2) 
         + \theta(-x_0)  \delta(\omega^2 - q_0^2) \right] 
 \notag \\
 &= -2 \pi i  
    \int d^4 q \; e^{-i (q \cdot x)} \; 
    \left[ \theta(x_0)  
         + \theta(-x_0) \right]  \; \delta(q_0^2 - \omega^2) 
 \notag \\
 &= -2 \pi i  
    \int d^4 q \; e^{-i (q \cdot x)} \; 
    \left[ \theta(x_0)  
         + \theta(-x_0) \right]  \; \delta(q^2 - m^2) 
 \qqquad \text{with} \; q_0^2 - \omega^2 = q^2 - m^2  \; .
\label{eq:qcd_feynmanepsilon}
\end{alignat}
This is exactly the usual decomposition of the propagator function
$\Delta_F(x) = \theta(x_0) \Delta^+(x) + \theta(-x_0) \Delta^-(x)$
into positive and negative energy contributions.\bigskip

Let us briefly recapitulate what would have happened if we instead had
chosen the Feynman parameter $\epsilon < 0$. We summarize all steps leading to the
propagator function in Eq.\eqref{eq:qcd_feynmanepsilon} 
in Table~\ref{tab:qcd_feynmanepsilon}. For the wrong sign of $i
\epsilon$ the two poles in the complex $q_0$ plane would be mirrored
by the real axis. The solution with $\text{Re} \, q_0 > 0$ would sit
in the quadrant with $\text{Im} \, q_0>0$ and the second pole at a
negative real and imaginary part. To be able to close the integration
path in the upper half plane in the mathematically positive direction
the real pole would have to be matched up with $\theta(-x_0)$. The
residue in the Cauchy integral would now include a factor $+1/(2
\omega)$. At the end, the two poles would give the same result as for
the correct sign of $i \epsilon$, except with a wrong over--all
sign.\bigskip

\begin{table}[t]
\begin{tabular}{l|ll|ll}
& 
\multicolumn{2}{c}{$\dfrac{1}{q^2 - m^2 + i \epsilon}$} &
\multicolumn{2}{c}{$\dfrac{1}{q^2 - m^2 - i \epsilon}$} \\[2mm]
\hline
pole &
  $q_0 =  \omega (1 - i \epsilon)$ &
  $q_0 = -\omega (1 - i \epsilon)$ &
  $q_0 =  \omega (1 + i \epsilon)$ &
  $q_0 = -\omega (1 + i \epsilon)$   \\[2mm]
\hline
complex quadrant & 
  $(+,-)$ &
  $(-,+)$ &
  $(+,+)$ &
  $(-,-)$  \\[2mm]
convergence: $x_0 \text{Im} q_0 <0$ & 
  $x_0>0$ &
  $x_0<0$ &
  $x_0<0$ &
  $x_0>0$  \\[2mm]
part of real axis &
  $\theta (x_0) $ & 
  $\theta (-x_0) $ & 
  $\theta (-x_0) $ & 
  $\theta (x_0) $  \\[2mm]
\hline
closed contour &
  $\text{Im} \, q_0 < 0$ &
  $\text{Im} \, q_0 > 0$ &
  $\text{Im} \, q_0 > 0$ &
  $\text{Im} \, q_0 < 0$  \\[2mm]
direction of contour &
  $-1$ &
  $+1$ &
  $+1$ &
  $-1$  \\[2mm]
residue &
  $+\dfrac{1}{2\omega}$ & 
  $-\dfrac{1}{2\omega}$ & 
  $+\dfrac{1}{2\omega}$ & 
  $-\dfrac{1}{2\omega}$ \\[2mm]
Fourier exponent &
  $e^{-i \omega x_0}$ &
  $e^{+i \omega x_0}$ &
  $e^{+i \omega x_0}$ &
  $e^{-i \omega x_0}$  \\[2mm]
\hline
all combined &
  $-\dfrac{e^{-i \omega x_0}}{2 \omega} \, \theta(x_0)$  &
  $-\dfrac{e^{+i \omega x_0}}{2 \omega} \, \theta(-x_0)$ &
  $+\dfrac{e^{+i \omega x_0}}{2 \omega} \, \theta(-x_0)$ &
  $+\dfrac{e^{-i \omega x_0}}{2 \omega} \, \theta(x_0)$ 
\end{tabular}
\caption{Contributions to the propagator function
  Eq.\eqref{eq:qcd_feynmanepsilon} for both signs of $i \epsilon$.}
\label{tab:qcd_feynmanepsilon}
\end{table}

When we are interested in the kinematic distributions of on--shell
massive states the situation is a little different. Measurements of
differential distributions for example at LEP include information on
the \underline{physical width}\index{particle width} of the decaying
particle, which means we cannot simply apply the Feynman $i \epsilon$
prescription as if we were dealing with an asymptotic stable
state. From the same couplings governing the $Z$ decay, the $Z$
propagator receives corrections, for example including fermion loops:
\begin{center}
\begin{fmfgraph*}(80,40)
 \fmfset{arrow_len}{2mm}
 \fmfleft{in1}
 \fmf{photon,width=0.5}{in1,v1}
 \fmf{fermion,width=0.5,left,tension=0.3}{v1,v2,v1}
 \fmf{photon,width=0.5}{v2,out1}
 \fmfright{out1}
\end{fmfgraph*}
\end{center}
Such one-particle irreducible diagrams can occur in the same
propagator repeatedly. Schematically written as a scalar they are of
the form
\begin{alignat}{5}
  \frac{i}{q^2-m_0^2+i \epsilon} 
+&\frac{i}{q^2-m_0^2+i \epsilon} (-i M^2) \frac{i}{q^2-m_0^2+i \epsilon} 
\notag \\
+& \frac{i}{q^2-m_0^2+i \epsilon} (-i M^2) \frac{i}{q^2-m_0^2+i \epsilon} (-i M^2) \frac{i}{q^2-m_0^2+i \epsilon} + \cdots
 \notag \\
=& \frac{i}{q^2-m_0^2+i \epsilon} \sum_{n=0} \left( \frac{M^2}{q^2-m_0^2+i \epsilon} \right)^n
 \notag \\
=& \frac{i}{q^2-m_0^2+i \epsilon} \; \frac{1}{1 - \dfrac{M^2}{q^2-m_0^2+i \epsilon}}
\qqqquad \text{summing the geometric series}
 \notag \\
=& \frac{i}{q^2-m_0^2+i \epsilon-M^2} \; .
\label{eq:prop_sum}
\end{alignat}
We denote the loop as $M^2$ for reasons which will become
obvious later.  Requiring that the residue of the
propagator\index{propagator!residue} be unity at the pole we
\underline{renormalize}\index{renormalization!mass}\index{renormalization!wave function} the wave function and the mass in the corresponding
process. For example for a massive scalar or gauge boson with a real
correction $M^2(q^2)$ this reads
\begin{alignat}{5}
\boxed{
 \frac{i}{q^2-m_0^2-M^2(q^2)}
 = \frac{iZ}{q^2-m^2} \qqquad \text{for} \; q^2 \sim m^2
} \; ,
\end{alignat}
including a renormalized mass $m$ and a wave function renormalization
constant $Z$.
\bigskip 

The important step in our argument is that in analogy to the effective
$ggH$ coupling discussed in Section~\ref{sec:higgs_gluon} the one-loop
correction $M^2$ depends on the momentum flowing through the
propagator. Above a certain threshold it can develop an imaginary part
because the momentum flowing through the diagram is large enough to
produce on--shell states in the loop. Just as for the $ggH$ coupling
such \underline{absorptive parts}\index{absorptive integral} appear
when a real decay like $Z \to \ell^+ \ell^-$ becomes kinematically
allowed. After splitting $M^2(q^2)$ into its real and imaginary parts
we know what to do with the real part: the solution to $q^2 - m_0^2 -
\text{Re} M^2(q^2) \really 0$ defines the renormalized particle mass
$q^2 = m^2$ and the wave function renormalization $Z$. The imaginary
part looks like the Feynman $i \epsilon$ term discussed before
\begin{alignat}{5}
 \frac{i}{q^2-m_0^2+i \epsilon-\text{Re} M^2(q^2)-i \text{Im} M^2}
 &= \frac{iZ}{q^2-m^2+i \epsilon-i Z \text{Im} M^2} \notag \\
 &\equiv \frac{iZ}{q^2-m^2+i m \Gamma} 
\qquad \Leftrightarrow \qquad
\Gamma = - \frac{Z}{m} \, \text{Im} \, M^2(q^2=m^2) \; ,
\end{alignat}
for $\epsilon \to 0$ and finite $\Gamma \ne 0$. We can illustrate the link
between the element squared $M^2$ of a self energy and the partial
width by remembering one way to compute scalar
integrals or one-loop amplitudes by gluing them together using
tree level amplitudes. Schematically written, the Cuskosky cutting rule discussed in
Section~\ref{sec:higgs_gluon} tells us $\text{Im}
\; M^2 \sim M^2 |_\text{cut} \equiv \Gamma$. Cutting the
one-loop bubble diagram at the one possible place is nothing but
squaring the two tree level matrix element for the decay $Z \to \ell^+
\ell^-$. One thing that we need to keep track of, apart from the
additional factor $m$ due to dimensional analysis, is the sign of the
$i m \Gamma$ term which just like the $i \epsilon$ prescription is
fixed by causality. \bigskip

Going back to the Drell--Yan process $q \bar{q} \to \ell^+ \ell^-$ we now
know that for massive unstable particles the Feynman epsilon which we
need to define the Green function for internal states acquires a
finite value, proportional to the total width of the unstable
particle.  This definition of a propagator of an unstable particle in
the $s$-channel is what we need for the second Feynman diagram contributing to the
Drell--Yan process: $q\bar{q} \to Z^* \to \ell^+ \ell^-$.  The resulting
shape of the propagator squared is a \underline{Breit--Wigner
  propagator}\index{particle width!Breit--Wigner propagator}
\begin{alignat}{5}
\boxed{
\sigma(q\bar{q} \to Z \to \ell^+ \ell^-) 
\propto \left| \frac{1}{s-m_Z^2+i m_Z \Gamma_Z}  \right|^2
= \frac{1}{(s-m_Z^2)^2 + m_Z^2 \Gamma_Z^2}
} \; .
\end{alignat}
When taking everything into account, the $(2 \to 2)$ production cross
section also includes the squared matrix element for the decay $Z \to \ell^+
\ell^-$ in the numerator. In the \underline{narrow width
  approximation}\index{particle width!narrow width approximation}, the $(2 \to 2)$ matrix element factorizes into the production
process times the branching ratio for $Z \to \ell^+ \ell^-$, simply by
definition of the Breit--Wigner or Lorentz or Cauchy distribution
\begin{alignat}{5}
\lim_{\Gamma \to 0} \; 
\frac{\Gamma_{Z, \ell \ell}}{(s-m_Z^2)^2 + m_Z^2 \Gamma_{Z,\text{tot}}^2}
= \Gamma_{Z, \ell \ell} \frac{\pi}{\Gamma_{Z, \text{tot}}} \delta(s-m_Z^2)
\equiv \pi \; \br(Z \to \ell \ell) \; \delta(s-m_Z^2) \; .
\label{eq:def_bw}
\end{alignat}
The additional factor $\pi$ will be absorbed in the different
one-particle and two-particle phase space definitions. We immediately
see that this narrow width approximation is only exact for scalar
particles. It does not keep information about the structure of the
matrix element, \eg when a non--trivial structure of the numerator
gives us the spin and angular correlations between the production and
decay processes.

Because of the $\gamma$-$Z$ interference we will always simulate
lepton pair production using the full on--shell and off--shell ($2 \to
2$) process. For example for top pair production with three-body
decays this is less clear. Sometimes, we will simulate the
production and the decay independently and rely on the limit
$\Gamma \to 0$. In that case it makes sense to nevertheless require a
Breit--Wigner shape for the momenta of the supposedly on--shell top
quarks. For top mass measurements we do, however, have to
take into account off--shell effects and QCD effects linking the decay
and production sides of the full Feynman diagrams.\bigskip

Equation~\eqref{eq:def_bw} uses a mathematical relation we might want to
remember for life, and that is the definition of the one-dimensional
\underline{Dirac delta distribution}\index{Dirac delta distribution} in three ways and including all
factors of 2 and $\pi$
\begin{alignat}{5}
\delta(x) = \int \frac{d q}{2 \pi} \; e^{-i x q} 
          = \lim_{\sigma \to 0} \; \frac{1}{\sigma \sqrt{\pi}} \, e^{-x^2/\sigma^2}
          = \lim_{\Gamma \to 0} \; \frac{1}{\pi} \, \frac{\Gamma}{x^2 + \Gamma^2} \; .
\end{alignat}
The second distribution is a \underline{Gaussian}\index{Gaussian distribution} and the third one
we would refer to as a \underline{Breit--Wigner}\index{propagator!Breit--Wigner propagator} shape while most other
people call it a Cauchy distribution\index{Cauchy distribution}.\bigskip

Now, we know  everything necessary to compute all Feynman
diagrams contributing to muon pair production at a hadron
collider. Strictly speaking, the two amplitudes interfere, so we end
up with three distinct contributions: $\gamma$ exchange, $Z$ exchange
and the $\gamma -Z$ interference terms. They have the properties
\begin{itemize}
\item[--] for small energies the $\gamma$ contribution dominates and 
          can be linked to the $R$ parameter.
\item[--] on the $Z$ pole the rate is regularized by the $Z$
  width and $Z$ contribution dominates over the photon.
\item[--] in the tails of the Breit--Wigner distribution we expect $Z -
  \gamma$ interference. For $m_{\ell \ell} > 120$~GeV the $\gamma$ and
  $Z$ contributions at the LHC are roughly equal in size.
\item[--] for large energies we are again dominated by the photon
  channel.
\item[--] quantum effects allow unstable particles like the $Z$ 
  to decay off--shell, defining a Breit--Wigner propagator.
\item[--] in the limit of vanishing width the $Z$ contribution
  factorizes into $\sigma \cdot \br$.
\end{itemize}
%

\subsubsection{Parton densities}
\label{sec:qcd_pdf}

At the end of Section~\ref{sec:qcd_dy_r} the discussion of different
energy regimes for $R$ experimentally makes sense --- at an $e^+ e^-$
collider we can tune the energy of the initial state\index{R ratio}. At hadron
colliders the situation is very different. The energy distribution of
incoming quarks as parts of the colliding protons has to be taken into
account. We first assume that quarks move collinearly with the
surrounding proton such that at the LHC incoming partons have zero
$p_T$. Under that condition we can define a probability distribution for finding a parton
just depending on the respective fraction of the proton's momentum.
For this momentum fraction $x = 0\cdots 1$ the \underline{parton
  density function}\index{parton densities} (pdf) is written as $f_i(x)$, where $i$ denotes the
different partons in the proton, for our purposes $u,d,c,s,g$ and,
depending on the details, $b$. All incoming partons we assume to be
massless.

In contrast to so-called structure functions a pdf is not an
observable. It is a distribution in the mathematical sense, which
means it has to produce reasonable results when we integrate it
together with a test function. Different parton densities have very
different behavior --- for the valence quarks ($uud$) they peak
somewhere around $x \lesssim 1/3$, while the gluon pdf is small at $x
\sim 1$ and grows very rapidly towards small $x$. For some typical
part of the relevant parameter space ($x = 10^{-3} \cdots 10^{-1}$) the gluon density
roughly scales like $f_g(x) \propto x^{-2}$. Towards smaller $x$
values it becomes even steeper. This steep gluon distribution was
initially not expected and means that for small enough $x$ LHC
processes will dominantly be gluon fusion processes.\bigskip

While we cannot actually compute parton distribution functions
$f_i(x)$ as a function of the momentum fraction $x$ there are a few
predictions we can make based on symmetries and properties of the
hadrons. Such arguments for example lead to \underline{sum
  rules}\index{QCD sum rules}: \bigskip

The parton distributions inside an antiproton are linked to those
inside a proton through the CP symmetry, which is an exact symmetry of
QCD. Therefore, we know that
\begin{alignat}{5}
f^{\bar{p}}_q(x) = f_{\bar{q}}(x) \qqqquad 
f^{\bar{p}}_{\bar{q}}(x) = f_q(x) \qqqquad 
f^{\bar{p}}_g(x) = f_g(x) 
\end{alignat}
for all values of $x$.

If the proton consists of three valence quarks $uud$, plus quantum
fluctuations from the vacuum which can either involve gluons or
quark--antiquark pairs, the contribution from the sea quarks has to be
symmetric in quarks and antiquarks. The expectation values for the
signed numbers of up and down quarks inside a proton have to fulfill
\begin{alignat}{5}
\langle N_u \rangle = 
\int_0^1 \, dx \; \left( f_u(x) - f_{\bar{u}}(x) \right) = 2
\qqqquad
\langle N_d \rangle = 
\int_0^1 \, dx \; \left( f_d(x) - f_{\bar{d}}(x) \right) = 1 \; .
\end{alignat}

Similarly, the total momentum of the proton has to consist of sum of
all parton momenta. We can write this as the expectation value of
$\sum x_i$
\begin{alignat}{5}
\langle \sum x_i \rangle = 
\int_0^1 \, dx \; x \; \left( \sum_q f_q(x) 
                            + \sum_{\bar{q}} f_{\bar{q}}(x) 
                            + f_g(x) \right) 
= 1
\end{alignat}
What makes this prediction interesting is that we can compute the same
sum only taking into account the measured quark and antiquark parton
densities. We find that the momentum sum rule only comes to 1/2.  Half
of the proton momentum is then carried by gluons.\bigskip

Given the correct definition and normalization of the pdf we can now
compute the \underline{hadronic cross section}\index{cross section!hadronic} from its partonic
counterpart, like the QED result in Eq.\eqref{eq:dy_sigma1}, as
\begin{alignat}{5}
\boxed{
\sigma_\text{tot} = \int_0^1 dx_1 \int_0^1 dx_2 \;
                   \sum_{ij} f_i(x_1) \, f_j(x_2) \;
                   \hat{\sigma}_{ij}(x_1 x_2 S)
} \; ,
\label{eq:qcd_sigtot}
\end{alignat}
where $i,j$ are the incoming partons with the momentum factions
$x_{i,j}$. The partonic energy of the scattering process is $s=x_1 x_2
S$ with the LHC proton energy of eventually $\sqrt{S}=14$~TeV. The
partonic cross section $\hat{\sigma}$ corresponds to the cross
sections $\sigma$ computed for example in Eq.\eqref{eq:dy_sigma1}. It
has to include all the necessary $\theta$ and $\delta$ functions for
energy--momentum conservation. When we express a general $n$--particle
cross section $\hat{\sigma}$ including the phase space integration,
the $x_i$ integrations and the phase space integrations can of course
be interchanged, but Jacobians will make life hard.
In Section~\ref{sec:qcd_mc} we will discuss an easier
way to compute kinematic distributions instead of from the fully
integrated total rate in Eq.\eqref{eq:qcd_sigtot}.

\subsubsection{Hadron collider kinematics}
\label{sec:qcd_kinematics}

Hadron colliders have a particular kinematic feature in that event by
event we do not know the longitudinal velocity of the initial state,
\ie the relative longitudinal boost from the laboratory frame to the
partonic center of mass. This sensitivity to longitudinal boosts is
reflected in the choice of kinematic variables. The first thing we 
consider is the projection of all momenta onto the transverse
plane. These transverse components are trivially invariant under 
longitudinal boosts because the two are orthogonal to each other.

In addition, for the production of a single electroweak gauge boson we
remember that the produced particle does not have any momentum
transverse to the beam direction. This reflects the fact that the
incoming quarks are collinear with the protons and hence have zero
transverse momentum. Such a gauge boson not recoiling against anything
else cannot develop a finite transverse momentum. Of course, once we
decay this gauge boson for example into a pair of muons, each muon
will have transverse momentum, only their vector sum will be zero:
\begin{alignat}{5}
\sum_\text{final state} \vec{p}_{T,j} = \vec{0} \; .
\end{alignat}
This is a relation between two-dimensional, not three dimensional
vectors. For more than one particle in the final state we define an
azimuthal angle in the transverse plane transverse.
While differences of azimuthal angles are
observables, the over--all angle is a symmetry of the detector as well
as of our physics.\bigskip

In addition to the transverse plane we need to parameterize the
longitudinal momenta in a way which makes it easy to implement
longitudinal boosts. In Eq.\eqref{eq:qcd_sigtot} we integrate over the
two momentum fractions $x_{1,2}$ and can at best determine their
product $x_1 x_2 = s/S$ from the final--state kinematics. Our task is
to replace both, $x_1$ and $x_2$ with a more physical variable which
should be well behaved under longitudinal boosts. 

A longitudinal boost for example from the rest frame of a massive
particle reads
\begin{alignat}{5}
\begin{pmatrix} E \\ p_L \end{pmatrix} 
&= 
\exp \left[ y \begin{pmatrix} 0 & 1 \\ 1 & 0 \end{pmatrix} \right]
    \begin{pmatrix} m \\ 0 \end{pmatrix} 
\notag \\
&= 
\left[ \one 
     + y \begin{pmatrix} 0 & 1 \\ 1 & 0 \end{pmatrix} 
     + \frac{y^2}{2} \one  
     + \frac{y^3}{6} \begin{pmatrix} 0 & 1 \\ 1 & 0 \end{pmatrix} \cdots \right]
    \begin{pmatrix} m \\ 0 \end{pmatrix} 
\notag \\
&= 
\left[ \one \; \sum_\text{$j$ even} \frac{y^j}{j!} 
     + \begin{pmatrix} 0 & 1 \\ 1 & 0 \end{pmatrix} \; \sum_\text{$j$ odd} \frac{y^j}{j!} 
\right]
    \begin{pmatrix} m \\ 0 \end{pmatrix} 
\notag \\
&= 
\left[ \one \; \cosh y
     + \begin{pmatrix} 0 & 1 \\ 1 & 0 \end{pmatrix} \; \sinh y 
\right]
    \begin{pmatrix} m \\ 0 \end{pmatrix} 
= m \begin{pmatrix} \cosh y \\ \sinh y \end{pmatrix} \; .
\label{eq:qcd_rap1}
\end{alignat}
We can re-write the \underline{rapidity $y$}\index{rapidity}
defined above in a way which allows us to compute it from the
four-momentum for example in the LHC lab frame
\begin{alignat}{5}
\frac{1}{2} \log \frac{E+p_L}{E-p_L}
= \frac{1}{2} \log \frac{\cosh y + \sinh y}{\cosh y - \sinh y}
= \frac{1}{2} \log \frac{e^y}{e^{- y}} = y \; .
\label{eq:qcd_rap2}
\end{alignat}
We can explicitly check that the rapidity is indeed additive
by applying a second longitudinal boost to $(E, p_L)$ in
Eq.\eqref{eq:qcd_rap1}
\begin{alignat}{5}
    \begin{pmatrix} E' \\ p_L' \end{pmatrix} 
&=
\exp \left[ y' \begin{pmatrix} 0 & 1 \\ 1 & 0 \end{pmatrix} \right]
    \begin{pmatrix} E \\ p_L \end{pmatrix} 
= 
\left[ \one \; \cosh y'
     + \begin{pmatrix} 0 & 1 \\ 1 & 0 \end{pmatrix} \; \sinh y' 
\right]
    \begin{pmatrix} E \\ p_L \end{pmatrix} \notag \\
&=   \begin{pmatrix} E \cosh y' + p_L \sinh y' \\ 
                    p_L \cosh y' + E \sinh y' \end{pmatrix} \; ,
\end{alignat}
which gives for the combined rapidity, following its extraction in
Eq.\eqref{eq:qcd_rap2}
\begin{alignat}{5}
\frac{1}{2} \log \frac{E'+p_L'}{E'-p_L'}
= \frac{1}{2} \log \frac{(E + p_L) (\cosh y' + \sinh y')}
                        {(E - p_L) (\cosh y' - \sinh y')}
= \frac{1}{2} \log \frac{E+p_L}{E-p_L} + y'  
= y + y' \; .
\end{alignat}
Two successive boosts with rapidities $y$ and $y'$ can be 
combined into a single boost by $y+y'$.
This combination of several longitudinal boosts is important in the
case of massless particles. They do not have a rest frame, which means
we can only boost them from one finite-momentum frame to the other.
For such massless particles we can simplify the formula for the
rapidity Eq.\eqref{eq:qcd_rap2}, in
terms of the polar angle $\theta$. We use that for
massless particles $E = |\vec{p}|$, giving us
\begin{alignat}{5}
y = 
\frac{1}{2} \log \frac{E+p_L}{E-p_L} =
\frac{1}{2} \log \frac{|\vec{p}|+p_L}{|\vec{p}|-p_L} = 
\frac{1}{2} \log \frac{1 + \cos \theta}{1 - \cos \theta} = 
\frac{1}{2} \log \dfrac{1}{\tan^2 \dfrac{\theta}{2}} = 
- \log \tan \frac{\theta}{2} \equiv \eta
\label{eq:qcd_pseudorap}
\end{alignat}
This \underline{pseudo-rapidity}\index{pseudo-rapidity} $\eta$ is more
handy, but coincides with the actual rapidity only for massless
particles. To get an idea about the experimental setup at the LHC ---
in CMS and ATLAS we can observe different particles to polar angles of
between 10 and 1.3 degrees, corresponding to maximum pseudo-rapidities
of 2.5 to 4.5. Because this is numerically about the same range as the
range of the \underline{azimuthal angle}\index{azimuthal angle} $[0,
  \pi]$ we define a distance measure inside the detector
\begin{alignat}{5}
(\Delta R)^2 &= ( \Delta y)^2 + ( \Delta \phi )^2 
\notag \\
&= ( \Delta \eta)^2 + ( \Delta \phi )^2  
\qquad && \text{massless particles}
\notag \\
&= \left( \log \dfrac{\tan \dfrac{\theta + \Delta \theta}{2}}{\dfrac{\theta}{2}}
   \right)^2 + ( \Delta \phi )^2  
\notag \\
&= \frac{(\Delta \theta)^2}{\sin^2 \theta} + ( \Delta \phi )^2  + 
   \ope( (\Delta \theta)^3) 
\label{eq:eta_phi}
\end{alignat}
The angle $\theta$ is the polar angle of one of the two particles
considered and in our leading approximation can be chosen as each of
them without changing Eq.\eqref{eq:eta_phi}.\bigskip

Still for the case of single gauge boson production we can
express the final--state kinematics in terms of two parameters, the
invariant mass of the final--state particle $q^2$ and its rapidity. 
We already know that the
transverse momentum of a single particle in the final state is zero. The two incoming and
approximately massless protons have the momenta
\begin{alignat}{5}
p_1 = (E,0,0,E) \qqqquad p_2 = (E,0,0,-E) \qqqquad S = (2E)^2 \; .
\end{alignat}
For the momentum of the final--state gauge boson in terms
of the parton momentum fractions this means in combination with
Eq.\eqref{eq:qcd_rap1}
\begin{alignat}{5}
q = x_1 p_1 + x_2 p_2 = E \begin{pmatrix} x_1+x_2 \\ 0 \\ 0 \\ x_1-x_2 \end{pmatrix}
  &\really \sqrt{q^2} \begin{pmatrix} \cosh y \\ 0 \\ 0 \\ \sinh y  \end{pmatrix}
  = 2 E \, \sqrt{x_1 x_2} \, \begin{pmatrix} \cosh y \\ 0 \\ 0 \\ \sinh y  \end{pmatrix}
\notag \\
\Leftrightarrow \qquad
  \cosh y &= \frac{x_1 + x_2}{2 \sqrt{x_1 x_2}}
           = \frac{1}{2} \; \left( \sqrt{\frac{x_1}{x_2}} + \sqrt{\frac{x_2}{x_1}} \right)
\qquad \Leftrightarrow \qquad
      e^y =  \sqrt{\frac{x_1}{x_2}} \; .
\end{alignat}
This result can be combined with $x_1 x_2 = q^2/S$ to obtain
\begin{alignat}{5}
x_1 = \sqrt{\frac{q^2}{S}} \, e^y \qqqquad
x_2 =  \sqrt{\frac{q^2}{S}} \, e^{-y} \; .
\end{alignat}
\bigskip

These relations allow us to for example compute the hadronic
\underline{total cross section} for lepton pair production in QED
\begin{alignat}{5}
\boxed{
\sigma(pp \to \ell^+ \ell^-) \Bigg|_\text{QED}
= \frac{4 \pi\alpha^2  Q_\ell^2}{3 N_c}
  \int_0^1 dx_1 dx_2 \; \sum_j \; Q_j^2 \; f_j(x_1) \, f_{\bar{j}}(x_2) \; \frac{1}{q^2}
} \; ,
\label{eq:qcd_dy_naive}
\end{alignat}
instead in terms of the hadronic phase space variables $x_{1,2}$ in terms of the
kinematic final--state observables $q^2$ and $y$. Remember that the 
partonic or quark--antiquark cross section
$\hat{\sigma}$ is already integrated over the (symmetric)
azimuthal angle $\phi$ and the polar angle Mandelstam variable
$t$. The transverse momentum of the two leptons is therefore fixed by
momentum conservation.

The Jacobian for this change of variables reads
\begin{alignat}{5}
\frac{\partial (q^2,y)}{\partial (x_1,x_2)} 
= \begin{vmatrix} x_2 S & x_1 S \\ 1/(2 x_1) & -1/(2 x_2) \end{vmatrix}
 = S = \frac{q^2}{x_1 x_2} \; ,
\end{alignat}
which inserted into Eq.\eqref{eq:qcd_dy_naive} gives us
\begin{alignat}{5}
\sigma(pp \to \ell^+ \ell^-) \Bigg|_\text{QED}
&= \frac{4 \pi\alpha^2  Q_\ell^2}{3 N_c}
  \int dq^2 dy \; \frac{x_1 x_2}{q^2} \; \frac{1}{q^2} \; \sum_j Q_j^2 \; f_j(x_1) \, f_{\bar{j}}(x_2)
\notag \\
&= \frac{4 \pi\alpha^2  Q_\ell^2}{3 N_c}
  \int dq^2 dy \; \frac{1}{q^4} \; \sum_j \; Q_j^2 \; x_1 f_j(x_1) \, x_2 f_{\bar{j}}(x_2) \; .
\label{eq:qcd_diff_cxn}
\end{alignat}
In contrast to the original form of the integration over the hadronic
phase space this form reflects the kinematic observables. For the
Drell--Yan process at leading order the $q^2$ distribution is the same
as $m_{\ell \ell}^2$, one of the most interesting distributions to
study because of different contributions from the photon, the $Z$
boson, or extra dimensional gravitons. On the other hand, the rapidity
integral still suffers from the fact that at hadron colliders we do
not know the longitudinal kinematics of the initial state and
therefore have to integrate over it.

\subsubsection{Phase space integration} 
\label{sec:qcd_mc}

In the previous example we have computed the simple two-dimensional
distribution, by leaving out the double integration in
Eq.\eqref{eq:qcd_diff_cxn}
\begin{alignat}{5}
\frac{d \sigma(pp \to \ell^+ \ell^-)}{dq^2 dy} \Bigg|_\text{QED}
= \frac{4 \pi\alpha^2  Q_\ell^2}{3 N_c q^4}
  \; \sum_j \; Q_j^2 \; x_1 f_j(x_1) \, x_2 f_{\bar{j}}(x_2) \; .
\end{alignat}
We can numerically evaluate this expression and compare it to
experiment. However, the rapidity $y$ and the momentum transfer $q^2$
of the $\ell^+ \ell^-$ pair are by no means the only distribution we would like to look
at. Moreover, we have to integrate numerically over the parton
densities $f(x)$, so we will have to rely on numerical integration
tools no matter what we are doing. Looking at a simple ($2 \to 2$)
process we can write the total cross section as
\begin{alignat}{5}
\sigma_\text{tot} = \int d \phi 
                   \int d \cos \theta
                   \int d x_1
                   \int d x_2 \;
                   F_\text{PS} 
                   \matx
                 = \int_0^1 dy_1 \cdots dy_4 \; 
                   J_\text{PS}(\vec{y}) 
                   \matx \; ,
\label{eq:ps_integ1}
\end{alignat}
with an appropriate function $F_\text{PS}$.  In the second step we
have re-written the phase space integral as an integral over the
four-dimensional unit cube, implicitly defining the appropriate
Jacobian.  Like any integral we can numerically evaluate this phase
space integral by binning the variable we integrate over
\begin{alignat}{5}
\int_0^1 dy \; f(y) 
 \quad \longrightarrow \quad 
 \sum_j (\Delta y)_j f(y_j)
 \sim \Delta y \sum_j f(y_j) \; .
\label{eq:ps_int}
\end{alignat}
Without any loss of generality we assume that the integration
boundaries are $0 \cdots 1$. We can
divide the integration variable $y$ into a discrete set of points $y_j$, for example equidistant
in $y$ or as a chain of random numbers $y_j \epsilon
[0,1]$. In the latter case we need to keep track of the bin widths
$(\Delta y)_j$. When we extend the integral to $d$ dimensions we can
in principle divide each axis into bins and compute the functional
values for this grid. For not equidistant bins generated by random
numbers we again keep track of the associated phase space volume for
each random number vector. Once we know these phase space weights
for each phase space point there is no reason to consider the set of
random numbers as in any way linked to the $d$ axes. 
All we need is a chain of random points with an associated
phase space weight and their transition matrix element, to integrate
over the phase space in complete analogy to
Eq.\eqref{eq:ps_int}.\bigskip

The obvious question is how such random numbers can be chosen in a
smart way. However, before we discuss how to best evaluate such an integral
numerically, let us first illustrate how this integral is much more
useful than just to provide the total cross section. If we are
interested in the distribution of an observable, like for example the
distribution of the transverse momentum of a muon in the Drell--Yan
process, we need to compute $d \sigma/d p_T$ as a function of
$p_T$. In terms of Eq.\eqref{eq:ps_integ1} any physical $y_1$
distribution is given by
\begin{alignat}{5}
\sigma &= \int dy_1 \cdots dy_d \; f(\vec{y}) 
        = \int dy_1 \; \frac{d\sigma}{dy_1} \notag \\
\frac{d\sigma}{dy_1} \Bigg|_{y_1^0} &= \int dy_2 \cdots dy_d \; f(y_1^0)
                             = \int dy_1 \cdots dy_d \; f(\vec{y}) 
                                      \; \delta(y_1-y_1^0) \; .
\label{eq:ps_integ2}
\end{alignat}
We can compute this distribution numerically in two ways: one way
corresponds to the first line in Eq.\eqref{eq:ps_integ2} and means
evaluating the $y_2 \cdots y_d$ integrations and leaving out
the $y_1$ integration. The result will be a function of $y_1$ which we
then evaluate at different points $y_1^0$.  

The second and much more efficient option corresponds to the second
line of Eq.\eqref{eq:ps_integ2}, with the delta distribution defined
for discretized $y_1$. First, we define an array with the size given
by the number of bins in the $y_1$ integration. Then, for each $y_1$
value of the complete $y_1 \cdots y_d$ integration we decide where the
value $y_1$ goes in this array and add $f(\vec{y})$ to the
corresponding column. Finally, we print these columns as a function of
$y_1$ to see the distribution. This set of columns is referred to as a
\underline{histogram}\index{histogram} and can be produced using
publicly available software. This histogram approach does not sound
like much, but imagine we want to compute a distribution
$d\sigma/dp_T$, where $p_T(\vec{y})$ is a complicated function of the
integration variables and kinematic phase space cuts. We then simply evaluate
\begin{alignat}{5}
\frac{d\sigma}{dp_T} = \int dy_1 \cdots dy_d \; f(\vec{y})  
                                      \; \delta \left( p_T(\vec{y}-p_T^0) 
                                               \right) 
\end{alignat}
numerically and read off the $p_T$ distribution as a side product of
the calculation of the total rate.  Histograms mean that computing a
total cross section numerically we can trivially extract all
distributions in the same process.\bigskip

The procedure outlined above has an interesting interpretation.
Imagine we do the entire phase space integration numerically. Just
like computing the interesting observables we can compute the momenta
of all external particles. These momenta are not all independent,
because of energy--momentum conservation, but this can be taken care
of. The tool which translates the vector of integration variables
$\vec{y}$ into the external momenta is called a \underline{phase space
  generator}\index{phase space!phase space generator}.  Because the
phase space is not uniquely defined in terms of the integration
variables, the phase space generator also returns the Jacobian
$J_\text{PS}$, called the phase space weight. If we think of the
integration as an integration over the unit cube, this weight needs to
be combined with the matrix element squared $\matx$. Once we compute
the unique phase space configuration $(k_1, k_2, p_1 \cdots)_j$
corresponding to the vector $\vec{y}_j$, the combined weight $W =
J_\text{PS} \matx$ is the probability that this configuration will
appear at the LHC. This means we do not only integrate over the phase
space, we really simulate LHC events.  The only complication is that
the probability of a given configuration is not only given by the
frequency with which it appears, but also by the explicit weight. So
when we run our numerical integration through the phase space
generator and histogram all the distributions we are interested in we
generate \underline{weighted events}\index{event generation!weighted events}. These events, which consist of the momenta of all external particles and
the weight $W$, we can for example store in a big file.\bigskip

This simulation is not yet what experimentalists want --- they want to
represent the probability of a certain configuration appearing only by
its frequency and not by an additional event weight. Experimentally measured events do not come with a
variable weight, either they are recorded or they are not. This means
we have to unweight the events by translating the event weight into
frequency. 

There are two ways to do that. On the one hand, we can look at the
minimum event weight and express all other events in relative
probability to this event. Translating this relative event weight into
a frequency means replacing an event with the relative weight
$W_j/W_\text{min}$ by $W_j/W_\text{min}$ unit--weight events in the
same phase space point. The problem with this method is that we are
really dealing with a binned phase space, so we would not know how to
distribute these events in and around the given bin.  

Alternatively, we can translate the weight of each event into a
probability to keep it or drop it. Because such a probability has to
be limited from above we start from to the maximum weight
$W_\text{max}$ and compute the ratio $W_j/W_\text{max} \epsilon [0,1]$
for each event. We then generate a flat random number $r \epsilon
[0,1]$ and only keep an event if $r < W_j/W_\text{max}$. This way, we
keep an event with a large weight $W_j/W_\text{max} \sim 1$ for almost
all values of $r$, while events with small weights are more likely to
drop out.  The challenge in this translation is that we always lose
events. If it was not for the experimentalists we would hardly use
such \underline{unweighted events}
\index{event generation!unweighted events}, but they have good reasons 
to want such unweighted events
which feed best through detector simulations.

The last comment is that if the phase space configuration $(k_1, k_2,
p_1 \cdots )_j$ can be measured, its weight $W_j$ better be
positive. This is not trivial once we go beyond leading order.  There,
we need to add several contributions to produce a physical event, like
for example different $n$--particle final states. There is no
guarantee for each of them to be positive. Instead, we ensure that
after adding up all contributions and after integrating over any kind
of unphysical degrees of freedom we might have introduced, the
probability of a physics configuration is positive. From this point of
view negative values for parton densities $f(x)<0$ are in principle
not problematic, as long as we always keep a positive hadronic rate
$d\sigma_{pp \to X}>0$.\bigskip

Going back to the numerical phase space integration for many
particles, it faces two problems. First, the partonic phase space for
$n$ on--shell particles in the final state has $3(n+2)-3$
dimensions. If we divide each of these directions in 100 bins, the
number of phase space points we need to evaluate for a $(2 \to 4)$
process is $100^{15}=10^{30}$, which is not realistic.

To integrate over a large number of dimensions we use \underline{Monte
  Carlo integration}\index{Monte Carlo!Monte Carlo integration}. In
this approach we first replace the binned directions in phase space by
a chain of random numbers $Y_j$, which can be organized in any number
of dimensions. In one dimension it replaces the equidistant bins in
the direction $y$. Because the distance between these new random
numbers is not constant, each random number will come with yet 
another weight.  The probability of finding $Y_j \epsilon [y,
  y+dy]$ is given by a smartly chosen function $p_Y(y)$. Integrating a
function $g(y)$ now returns an expectation value of $g$ evaluated over
the chain $Y_j$,
\begin{alignat}{5}
\langle g(Y) \rangle
= \int_0^1 dy \; p_Y(y) \; g(y) 
\quad \longrightarrow \quad 
\frac{1}{N_Y} \sum_{j=1}^{N_Y} g(Y_j) \; .
\label{eq:vegas1}
\end{alignat}
First of all, we can 
immediately generalize this approach to any number of $d$ dimensions,
just by organizing the random numbers $Y_j$ in one large chain instead
of a $d$-dimensional array. Second, in Eq.\eqref{eq:ps_integ1} we are
interested in an integral over $g$, which means that we should rewrite
the integration as
\begin{alignat}{5}
\int_0^1 d^d y \; f(y) 
  = \int_0^1 d^d y \; \frac{f(y)}{p_Y(y)} \; p_Y(y)
  = \left< \frac{f(Y)}{p_Y(Y)} \right> 
\; \rightarrow \;
\frac{1}{N_Y} \sum_j \frac{f(Y_j)}{p_Y(Y_j)} \; .
\end{alignat}
To compute the integral we now average over all values of $f/p_Y$
along the random number chain $Y_j$. In the ideal case where we
exactly know the form of the integrand and can map it into our random
numbers, the error of the numerical integration will be zero. So what
we have to find is a way to encode $f(Y_j)$ into $p_Y(Y_j)$. This task
is called \underline{importance sampling}\index{Monte Carlo!importance sampling} and you can find some documentation for example on the
standard implementation VEGAS to look at the details.

Technically, VEGAS will call the function which
computes the weight $W = J_\text{PS} \matx$ for a number of phase
space points and average over these points, but including another
weight factor $W_\text{MC}$ representing the importance sampling. If
we want to extract distributions via histograms we have to 
add the total weight $W = W_\text{MC} J_\text{PS} \matx$ to the
columns. \bigskip

The second numerical challenge is that the matrix elements for
interesting processes are by no means flat. We would therefore like to
help our adaptive or importance sampling Monte Carlo by defining the
integration variables such that the integrand becomes as flat as
possible. For example for the integration over the partonic momentum
fraction we know that the integrand usual falls off as
$1/x$. In that situation we can substitute
\begin{alignat}{5}
\int_\delta dx \; \frac{C}{x} = 
\int_{\log \delta} d \log x \; \left( \frac{d \log x}{dx} \right)^{-1} \; 
                   \frac{C}{x} =
\int_{\log \delta} d \log x \; C \; ,
\end{alignat}
to obtain a flat integrand. There exists an even more
impressive and relevant example: intermediate particles with
Breit--Wigner propagators squared are particularly painful to integrate over the
momentum $s = p^2$ flowing through it
\begin{alignat}{5}
P(s,m) = \frac{1}{(s-m^2)^2 + m^2 \Gamma^2} \; .
\end{alignat}
For example, a Standard Model Higgs boson with a mass of 126~GeV has
a width around $0.005$~GeV, which means that the integration over the
invariant mass of the Higgs decay products $\sqrt{s}$ requires a
relative resolution of $10^{-5}$. Since this is unlikely to be
achievable what we should really do is find a substitution which
produces the inverse Breit--Wigner as a Jacobian and leads to a flat
integrand --- et voil{\'a}\index{particle width!Breit--Wigner propagator}
\begin{alignat}{5}
\int ds \; \frac{C}{(s-m^2)^2 + m^2 \Gamma^2} 
&= \int dz \; \left( \frac{dz}{ds} \right)^{-1} 
              \frac{C}{(s-m^2)^2 + m^2 \Gamma^2} 
\notag \\
&= \int dz \; \frac{(s-m^2)^2 + m^2 \Gamma^2}{m \Gamma} \;
              \frac{C}{(s-m^2)^2 + m^2 \Gamma^2} 
\notag \\
&= \frac{1}{m \Gamma} \int dz \; C 
\qquad \text{with} \qquad \tan z = \frac{s - m^2}{m \Gamma}  \; .
\end{alignat}
This is the most useful \underline{phase space mapping}\index{phase space!phase space mapping} in LHC physics.  Of
course, any adaptive Monte Carlo will eventually converge on such an
integrand, but a well--chosen set of integration parameters will speed
up simulations very significantly.

\subsection{Ultraviolet divergences}
\label{sec:qcd_uv}

From general field theory we know that when we are interested for
example in cross section prediction with higher precision we need to
compute further terms in its perturbative series in $\alpha_s$. This
computation will lead to ultraviolet divergences which can be absorbed
into counter terms for any parameter in the Lagrangian. The crucial
feature is that for a renormalizable theory like our Standard Model
the number of counter terms is finite, which
means once we know all parameters including their counter terms our
theory becomes predictive.

In Section~\ref{sec:qcd_ir} we will see that in QCD processes we also
encounter another kind of divergences. They arise from the infrared
momentum regime. Infrared divergences is what this lecture is really
going to be about, but before dealing with them it is very instructive to
see what happens to the much better understood ultraviolet divergences. In
Section~\ref{sec:qcd_counter_terms} we will review how such
ultraviolet divergences arise and how they are removed. In
Section~\ref{sec:qcd_run_coup} we will review how running parameters
appear in this procedure, \ie how scale dependence is linked to the
appearance of divergences. Finally, in
Section~\ref{sec:qcd_scale_logs} we will interpret the use of running
parameters physically and see that in perturbation theory they resum
classes of logarithms to all orders in perturbation theory. Later in
Section~\ref{sec:qcd_ir} we will follow exactly the same steps for
infrared divergences and develop some crucial features of hadron
collider physics.

\subsubsection{Counter terms} 
\label{sec:qcd_counter_terms}

Renormalization as the proper treatment of ultraviolet divergences
is one of the most important things to understand about field
theories; you can find more detailed discussions in any book on
advanced field theory. The particular aspect of renormalization which
will guide us through this section is the appearance of the
renormalization scale.

In perturbation theory, scales automatically arise from the
regularization of infrared or ultraviolet divergences. We can see this
by writing down a simple scalar loop integral, with to two
virtual scalar propagators with masses $m_{1,2}$ and an external momentum
$p$ flowing through a diagram, similar to those summed in
Section~\ref{sec:qcd_bw}

\begin{alignat}{5}
  B(p^2;m_1,m_2) \equiv
  \int \frac{d^4q}{16 \pi^2} \; \frac{1}{q^2-m_1^2} \frac{1}{(q+p)^2-m_2^2} \; .
\label{eq:twopoint}
\end{alignat}
Such two-point functions appear for example in the gluon self energy
with virtual gluons, with massless ghost scalars, with a Dirac trace in the numerator
for quarks, and with massive scalars for supersymmetric scalar
quarks\index{supersymmetry}. In those cases the two masses are
identical $m_1 = m_2$. The integration measure $1/(16 \pi^2)$ is
dictated by the Feynman rule for the integration over loop
momenta.  Counting powers of $q$ in Eq.\eqref{eq:twopoint} we see that
the integrand is not suppressed by powers of $1/q$ in the ultraviolet,
so it is logarithmically divergent and we have to regularize
it. Regularizing means expressing the divergence in a well--defined
manner or scheme, allowing us to get rid of it by
renormalization.\bigskip

One regularization scheme is to introduce a cutoff into the momentum
integral $\Lambda$, for example through the so-called Pauli---Villars
regularization\index{Pauli--Villars regularization}. Because the ultraviolet behavior of the integrand or
integral cannot depend on any parameter living at a small energy
scales, the parameterization of the ultraviolet divergence in
Eq.\eqref{eq:twopoint} cannot involve the mass $m$ or the external
momentum $p^2$. The scalar two-point function has mass dimension zero,
so its divergence has to be proportional to $\log (\Lambda/\mu_R)$
with a dimensionless prefactor and some scale $\mu_R^2$ which is an
artifact of the regularization of such a Feynman diagram.\bigskip

A more elegant regularization scheme is \underline{dimensional
  regularization}\index{dimensional regularization}. It is designed
not to break gauge invariance and naively seems to not introduce a
mass scale $\mu_R$.  When we shift the momentum integration from 4 to
$4 - 2\epsilon$ dimensions and use analytic continuation in the number
of space--time dimensions to renormalize the theory, a
\underline{renormalization scale}\index{scales!renormalization scale}
$\mu_R$ nevertheless appears once we ensure the two-point function and with
it observables like cross sections keep their correct mass dimension
\begin{alignat}{5}
  \int \frac{d^4q}{16 \pi^2} \cdots \longrightarrow
  \mu_R^{2\epsilon} \;  \int \frac{d^{4-2 \epsilon}q}{16 \pi^2} \cdots
  =
  \frac{i \mu_R^{2\epsilon}}{(4 \pi)^2} \;
                      \left[ \frac{C_{-1}}{\epsilon} 
                           + C_0 
                           + C_1 \, \epsilon + \ope(\epsilon^2)
                      \right] \; .
\label{eq:uv_poles}
\end{alignat}
At the end, the scale $\mu_R$ might become irrelevant and drop out
after renormalization and analytic continuation, but to be on the save
side we keep it.  The constants $C_i$ in the series in $1/\epsilon$
depend on the loop integral we are considering.  To regularize the
ultraviolet divergence we assume $\epsilon>0$ and find mathematically
well defined poles $1/\epsilon$. Defining scalar integrals with the
integration measure $1/(i \pi^2)$ will make for example $C_{-1}$ come
out as of the order $\ope(1)$. This is the reason we usually find
factors $1/(4 \pi)^2 = \pi^2/(2 \pi)^4$ in front of the loop
integrals.

The poles in $1/\epsilon$ will cancel with the universal
\underline{counter terms} once we renormalize the theory. Counter
terms we include by shifting parameters in the Lagrangian and the leading
order matrix element. They cancel the poles in the combined leading
order and virtual one-loop prediction
\begin{alignat}{5}
\left| \mat_\text{LO}(g) + \mat_\text{virt} \right|^2 
&= \left| \mat_\text{LO}(g) \right|^2 
 + 2 \, \text{Re} \; \mat_\text{LO}(g) \mat_\text{virt} + \cdots \notag \\
&\to \left| \mat_\text{LO}(g+\delta g) \right|^2 
 + 2 \, \text{Re} \; \mat_\text{LO}(g) \mat_\text{virt} 
 + \cdots \notag \\
\text{with} \qquad  
g &\to g^\text{bare} = 
g + \delta g 
\qquad \text{and} \quad 
\delta g \propto \alpha_s/\epsilon \; .
\label{eq:renorm}
\end{alignat}
\index{renormalization!strong coupling}The dots indicate higher orders in $\alpha_s$, for example
absorbing the $\delta g$ corrections in the leading order and virtual
interference. As we can see in Eq.\eqref{eq:renorm} the counter terms
do not come with a factor $\mu_R^{2 \epsilon}$ in front. Therefore,
while the poles $1/\epsilon$ cancel just fine, the scale factor
$\mu_R^{2 \epsilon}$ will not be matched between the actual
ultraviolet divergence and the counter term. 

We can keep track of the renormalization scale best by expanding the
prefactor of the regularized but not yet renormalized integral in
Eq.\eqref{eq:uv_poles} in a Taylor series in $\epsilon$, no question
asked about convergence radii
\begin{alignat}{5}
 \mu_R^{2\epsilon} \; 
  \left[ \frac{C_{-1}}{\epsilon} + C_0 + \ope(\epsilon) \right] 
 &= 
 e^{2 \epsilon \log \mu_R} \;
  \left[ \frac{C_{-1}}{\epsilon} + C_0 + \ope(\epsilon) \right] 
  \notag \\ 
 &= \left[ 1 + 2 \epsilon \log \mu_R + \ope(\epsilon^2) \right] \;
  \left[ \frac{C_{-1}}{\epsilon} + C_0 + \ope(\epsilon) \right]
  \notag \\ 
 &= \frac{C_{-1}}{\epsilon} 
   + C_0
   + C_{-1} \, \log \mu_R^2 
   +   \ope(\epsilon) \notag \\
 &\to \frac{C_{-1}}{\epsilon} 
   + C_0
   + C_{-1} \, \log \frac{\mu_R^2}{M^2}
   +   \ope(\epsilon) \; .
\label{eq:qcd_mu_taylor}
\end{alignat}
In the last step we correct by hand for the fact that $\log
\mu_R^2$ with a mass dimension inside the logarithm cannot appear in
our calculations. From somewhere else in our calculation the
logarithm will be matched with a $\log M^2$ where $M^2$ is
the typical mass or energy scale in our process. This little argument
shows that also in dimensional regularization we introduce a mass
scale $\mu_R$ which appears as $\log (\mu_R^2/M^2)$ in the
renormalized expression for our observables. There is no way of
removing ultraviolet divergences without introducing some kind of
renormalization scale.

In Eq.\eqref{eq:qcd_mu_taylor} there appear two contributions to a
given observable, the expected $C_0$ and the renormalization--induced
$C_{-1}$. Because the factors $C_{-1}$ are linked to the counter terms
in the theory we can often guess them without actually computing the
loop integral, which is very useful in cases where they numerically
dominate.\bigskip

Counter terms as they schematically appear in Eq.\eqref{eq:renorm} are
not uniquely defined. They need to include a given divergence to
return finite observables, but we are free to add any finite
contribution we want.  This opens many ways to define a counter term
for example based on physical processes where counter terms do not
only cancel the pole but also finite contributions at a given order in 
perturbation theory. Needless to say, such schemes do not automatically work
universally. An example for such a \underline{physical renormalization
  scheme} is the on--shell scheme for masses, where we define a counter
term such that external on--shell particles do not receive any
corrections to their masses. For the top mass this means that 
we replace the leading order mass with
the bare mass, for which we then insert the expression in terms of the
renormalized mass and the counter term
\begin{alignat}{5}
m_t^\text{bare} &= m_t + \delta m_t \notag \\
     &= m_t + m_t \frac{\alpha_s C_F}{4 \pi} 
                 \left( 3 \left(-\frac{1}{\epsilon} + \gamma_E
                                -\log (4 \pi) - \log \frac{\mu_R^2}{M^2}
                          \right)
                       -4 + 3 \log \frac{m_t^2}{M^2}
                 \right) \notag \\
     &\equiv m_t + m_t \frac{\alpha_s C_F}{4 \pi} 
                 \left( - \frac{3}{\tilde \epsilon}
                       -4 + 3 \log \frac{m_t^2}{M^2}
                 \right)
\qqquad \Leftrightarrow \qqquad
\frac{1}{\tilde{\epsilon}\left(\dfrac{\mu_R}{M}\right)} \equiv 
\frac{1}{\epsilon} - \gamma_E + \log \frac{4 \pi \mu_R^2}{M^2} \; ,
\label{eq:top_counter}
\end{alignat}
with the color factor $C_F = (N^2-1)/(2N)$.
\index{renormalization!top quark mass}
The convenient scale dependent pole $1/\tilde{\epsilon}$ 
includes the universal additional terms like the Euler gamma
function and the scaling logarithm.  This logarithm is the big problem
in this universality argument, since we need to introduce the
arbitrary energy scale $M$ to separate the universal logarithm of the
renormalization scale and the parameter-dependent logarithm of the
physical process.

A theoretical problem with this \underline{on--shell renormalization
  scheme} is that it is not gauge invariant. On the other hand, it
describes for example the kinematic features of top pair production at
hadron colliders in a stable perturbation series. This means that once
we define a more appropriate scheme for heavy particle masses in
collider production mechanisms it better be numerically close to the
pole mass. For the computation of total cross sections at hadron
colliders or the production thresholds at $e^+ e^-$ colliders the pole
mass is not well suited at all, but as we will see in
Section~\ref{sec:sim} this is not where we expect to
measure particle masses at the LHC, so we should do fine with something very
similar to the pole mass.\bigskip

Another example for a process dependent renormalization scheme is the
mixing of $\gamma$ and $Z$ propagators. There we choose the counter
term of the weak mixing angle such that an on--shell $Z$ boson cannot
oscillate into a photon, and vice versa. We can generalize this scheme
for mixing scalars as they for example appear in supersymmetry, but it
is not gauge invariant with respect to the weak gauge symmetries
of the Standard Model either. For QCD
corrections, on the other hand, it is the most convenient scheme
keeping all exchange symmetries of the two scalars.\bigskip

To finalize this discussion of process dependent mass renormalization
we quote the result for a scalar supersymmetric quark, a squark, where
in the on--shell scheme we find\index{renormalization!squark mass}
\begin{alignat}{5}
m_{\tilde q}^\text{bare} &= m_{\tilde q} + \delta m_{\tilde q} \notag \\
     &= m_{\tilde q} + m_{\tilde q} \frac{\alpha_s C_F}{4 \pi} 
                 \left( - \frac{2r}{\tilde \epsilon}
                       - 1
                       - 3 r 
                       - \left( 1 - 2 r \right) \log r 
                       - \left( 1 - r \right)^2
                         \log \left| \frac{1}{r} - 1 \right|
                       - 2 r \log \frac{m_{\tilde q}^2}{M^2}
                 \right) \; .
\label{eq:counter_msq1}
\end{alignat}
with $r = m_{\tilde g}^2/m_{\tilde q}^2$. The interesting aspect of
this squark mass counter term is that it also depends on the gluino
mass, not just the squark mass itself. The reason why QCD counter
terms tend to depend only on the renormalized quantity itself is that
the gluon is massless. In the limit of vanishing gluino contribution
the squark mass counter term is again only proportional to the
squark mass itself
\begin{alignat}{5}
m_{\tilde q}^\text{bare} \Bigg|_{m_{\tilde g} = 0} 
= m_{\tilde q} + \delta m_{\tilde q} 
= m_{\tilde q} + m_{\tilde q} \frac{\alpha_s C_F}{4 \pi} 
\left( - \frac{1}{\tilde \epsilon}
       - 3 
       + \log \frac{m_{\tilde q}^2}{M^2}
\right) \; .
\label{eq:counter_msq2}
\end{alignat}
Taking the limit of Eq.\eqref{eq:counter_msq1} to derive
Eq.\eqref{eq:counter_msq2} is computationally not trivial,
though.\bigskip

One common feature of all mass counter terms listed above is
$\delta m \propto m$, which means that we actually encounter a
multiplicative renormalization
\begin{alignat}{5}
m^\text{bare} = Z_m \, m
             = \left( 1 + \delta Z_m \right) m
             = \left( 1 + \frac{\delta m}{m} \right) m
             = m + \delta m \; ,
\end{alignat}
with $\delta Z_m = \delta m/m$ linking the two ways of writing the mass
counter term.  This form implies that particles with zero mass will
not obtain a finite mass through renormalization. If we remember
that chiral symmetry protects a Lagrangian from acquiring fermion
masses this means that on--shell renormalization does not break this
symmetry. A massless theory cannot become massive by mass
renormalization. Regularization and renormalization schemes which do
not break symmetries of the Lagrangian are ideal.\bigskip

When we introduce counter terms in general field theory we usually
choose a slightly more model independent scheme --- we define a
renormalization point. This is the energy scale at which the counter
terms cancels all higher order contributions, divergent as well as
finite. The best known example is the electric charge which we
renormalize in the \underline{Thomson
  limit}\index{renormalization!Thomson limit} of zero momentum
transfer through the photon propagator
\begin{alignat}{5}
e \to e^\text{bare} = e + \delta e \; .
\end{alignat}
\bigskip

Looking back at $\delta m_t$ as defined in Eq.\eqref{eq:top_counter}
we also see a way to define a completely general counter term: if
dimensional regularization, \ie the introduction of $4 - 2\epsilon$
dimensions does not break any of the symmetries of our Lagrangian,
like Lorentz symmetry or gauge symmetries, we can simply subtract
the ultraviolet pole and nothing else. The only question is: do we
subtract $1/\epsilon$ in the MS scheme or do we subtract $1/\tilde
\epsilon$ in the \underline{$\msbar$
  scheme}\index{renormalization!$\msbar$ scheme}.  In the $\msbar$
scheme the counter term is then scale dependent.

Carefully counting, there are three scales present in such a
scheme. First, there is the physical scale in the process. In our case
of a top self energy this is for example the top mass $m_t$ appearing
in the matrix element for the process $pp \to t\bar{t}$. Next, there
is the renormalization scale $\mu_R$, a reference scale which is part
of the definition of any counter term. And last but not least, there
is the scale $M$ separating the counter term from the process
dependent result, which we can choose however we want, but which as we
will see implies a running of the counter term. The role of this scale
$M$ will become clear when we go through the example of the running
strong coupling $\alpha_s$. Of course, we would prefer to choose all
three scales the same, but in a complex physical process this might
not always be possible. For example, any massive $(2 \to 3)$
production process naturally involves several external physical
scales.\bigskip

Just a side remark for completeness: a one loop integral which has no
intrinsic mass scale is the two-point function with zero mass in the
loop and zero momentum flowing through the integral:
$B(p^2=0;0,0)$. It appears for example in the self energy corrections
of external quarks and gluons.  Based on dimensional arguments
this integral has to vanish altogether. On the other hand, we know
that like any massive two-point function it has to be ultraviolet
divergent $B \sim 1/\epsilon_\text{UV}$ because setting all internal
and external mass scales to zero is nothing special from an
ultraviolet point of view.  This can only work if the scalar integral
also has an infrared divergence appearing in dimensional
regularization.  We can then write the entire massless two-point
function as 
\begin{alignat}{5}
B(p^2=0;0,0) 
= \int \frac{d^4q}{16 \pi^2} \; \frac{1}{q^2} \frac{1}{(q+p)^2} 
= \frac{i \pi^2}{16\pi^2} \, 
  \left( \frac{1}{\epsilon_\text{UV}} - \frac{1}{\epsilon_\text{IR}} \right) \; ,
\end{alignat}
keeping track of the divergent contributions from the infrared and the
ultraviolet regimes. For this particular integral they precisely
cancel, so the result for $B(0;0,0)$ is zero, but setting it to zero
too early will spoil any ultraviolet and infrared finiteness test.
Treating the two divergences strictly
separately and dealing with them one after the other also ensures that
for ultraviolet divergences we can choose $\epsilon >0$ while for
infrared divergences we require $\epsilon <0$.

\subsubsection{Running strong coupling}
\label{sec:qcd_run_coup}

To get an idea what these different scales which appear in the process
of renormalization mean let us compute such a scale dependent
parameter, namely the \underline{running strong coupling}\index{running coupling} 
$\alpha_s(\mu_R^2)$. The Drell--Yan process is one of the very few
relevant processes at hadron colliders where the strong coupling does
not appear at tree level, so we cannot use it as our toy process this
time.  Another simple process where we can study this coupling is
bottom pair production at the LHC, where at some energy range we will
be dominated by valence quarks: $q \bar{q} \to b \bar{b}$.  The only
Feynman diagram is an $s$-channel off--shell gluon with a momentum
flow $p^2 \equiv s$.
\begin{center}
\begin{fmfgraph*}(80,60)
 \fmfset{arrow_len}{2mm}
 \fmfleft{in1,in2}
 \fmf{fermion,width=0.5}{in1,v1}
 \fmf{fermion,width=0.5}{v1,in2}
 \fmf{fermion,width=0.5}{out1,v2}
 \fmf{fermion,width=0.5}{v2,out2}
 \fmf{gluon,width=0.5}{v1,v2}
 \fmfright{out1,out2}
\end{fmfgraph*}
\end{center}
At next--to--leading order this gluon propagator will be corrected by
self energy loops, where the gluon splits into two quarks or gluons
and re-combines before it produces the two final--state bottoms. Let us
for now assume that all quarks are massless. The Feynman diagrams for
the gluon self energy include a quark look, a gluon loop, and the
ghost loop which removes the unphysical degrees of freedom of the
gluon inside the loop.
\begin{equation*}
\parbox{30mm}{
\begin{fmfgraph*}(80,60)
 \fmfset{arrow_len}{2mm}
 \fmfleft{in1}
 \fmf{gluon,width=0.5}{in1,v1}
 \fmf{fermion,width=0.5,left,tension=0.3}{v1,v2,v1}
 \fmf{gluon,width=0.5}{v2,out1}
 \fmfright{out1}
\end{fmfgraph*}
} \quad + \quad 
\parbox{30mm}{
\begin{fmfgraph*}(80,60)
 \fmfset{arrow_len}{2mm}
 \fmfleft{in1}
 \fmf{gluon,width=0.5}{in1,v1}
 \fmf{gluon,width=0.5,right,tension=0.6}{v1,v2,v1}
 \fmf{gluon,width=0.5}{v2,out1}
 \fmfright{out1}
\end{fmfgraph*}
} \quad + \quad 
\parbox{30mm}{
\begin{fmfgraph*}(80,60)
 \fmfset{arrow_len}{2mm}
 \fmfleft{in1}
 \fmf{gluon,width=0.5}{in1,v1}
 \fmf{dashes,width=0.5,right,tension=0.3}{v1,v2,v1}
 \fmf{gluon,width=0.5}{v2,out1}
 \fmfright{out1}
\end{fmfgraph*}
}
\end{equation*}
The gluon self energy correction or \underline{vacuum polarization},
as propagator corrections to gauge bosons are usually labelled, will
be a scalar. This way, all fermion lines close in the Feynman diagram and the Dirac trace is
computed inside the loop. In color space the self energy will
(hopefully) be diagonal, just like the gluon propagator itself, so we
can ignore the color indices for now. In unitary gauge the gluon
propagator\index{propagator!gluon} is proportional to the transverse
tensor $T^{\mu \nu} = g^{\mu\nu} - p^\nu p^\mu/p^2$. As mentioned in
the context of the effective gluon--Higgs coupling, the same should be
true for the gluon self energy, which we therefore write as $\Pi^{\mu
  \nu} \equiv \Pi \, T^{\mu \nu}$. Unlike for two different external
momenta $k_1 \ne k_2$ shown in Eq.\eqref{eq:higgs_transverse} the case
with only one external momentum gives us the useful simple relations
\begin{alignat}{5}
T^{\mu \nu} g_\nu^\rho &= 
\left( g^{\mu \nu} - \frac{p^\mu p^\nu}{p^2} \right) g_\nu^\rho
= T^{\mu \rho} 
\notag \\
T^{\mu \nu} T_\nu^\rho &= 
\left( g^{\mu \nu} - \frac{p^\mu p^\nu}{p^2} \right) \; 
\left( g_\nu^\rho - \frac{p_\nu p^\rho}{p^2} \right) 
= g^{\mu \rho} - 2 \frac{p^\mu p^\rho}{p^2} + p^2 \frac{p^\mu p^\rho}{p^4}
= T^{\mu \rho} \; . 
\label{eq:gluon_transverse}
\end{alignat}
Including the gluon, quark, and ghost loops the regularized gluon self
energy with a momentum flow $p^2$ through the propagator reads
\begin{alignat}{5}
- \frac{1}{p^2} \; \Pi\left( \frac{\mu_R^2}{p^2} \right) 
    &=&&     \frac{\alpha_s}{4 \pi} \;
             \left( - \frac{1}{\tilde \epsilon} 
                    + \log \frac{p^2}{M^2} \right) \;
             \left( \frac{13}{6} N_c - \frac{2}{3} n_f \right) 
             + \ope(\log m_t^2 ) \notag \\
    &\equiv \;&& 
            \alpha_s \;
             \left( - \frac{1}{\tilde \epsilon} 
                    + \log \frac{p^2}{M^2} \right) \;
             b_0
             + \ope(\log m_t^2 ) \notag \\
   &&&\text{with} \qquad
  \boxed{b_0 = \frac{1}{4 \pi} \left( \frac{11}{3} N_c - \frac{2}{3} n_f
                                   \right) } \; .
\label{eq:qcd_bzero}
\end{alignat}
The minus sign arises from the factors $i$ in the propagators, as
shown in Eq.\eqref{eq:prop_sum}.  The number of fermions coupling to
the gluons is $n_f$. From the comments on $B(p^2;0,0)$ we could guess
that the loop integrals will only give a logarithm $\log p^2$ which is
then matched by the logarithm $\log M^2$ implicitly included in the
definition of $\tilde \epsilon$. The factor $b_0$ arises from the
one-loop corrections to the gluon self energy, \ie from diagrams which
include one additional factor $\alpha_s$. Strictly speaking, this form
is the first term in a \underline{perturbative series}\index{QCD perturbative series} in the strong coupling $\alpha_s = g_s^2/(4
\pi)$. Later on, we will indicate where additional higher order
corrections would enter. For later we keep in mind that for a
sufficiently small number of quark generations the sign of $b_0$ is
positive.

In the second step of Eq.\eqref{eq:qcd_bzero} we have sneaked in
additional contributions to the renormalization of the strong coupling
from the other one-loop diagrams in the process, replacing the factor
13/6 by a factor 11/3. This is related to the fact that there are
actually three types of divergent virtual gluon diagrams in the
physical process $q \bar{q} \to b \bar{b}$: the external quark self
energies with renormalization factors $Z_f^{1/2}$, the internal gluon self
energy $Z_A$, and the vertex corrections $Z_{Aff}$. The only physical
parameters we can renormalize in this process are the strong coupling
and, if finite, the bottom mass. Wave function renormalization
constants are not physical, but vertex renormalization terms are. The entire divergence in our $q \bar{q}
\to b \bar{b}$ process which needs to be absorbed in the strong
coupling $Z_g$ is given by the combination
\begin{alignat}{5}
Z_{Aff} = Z_g Z_A^{1/2} Z_f
\qqqquad \Leftrightarrow \qqqquad
\frac{Z_{Aff}}{Z_A^{1/2} Z_f} \equiv Z_g \; .
\label{eq:qcd_wf_renorm}
\end{alignat}
We can check this definition of $Z_g$ by comparing all vertices in
which the strong coupling $g_s$ appears, namely the gluon coupling to
quarks, ghosts as well as the triple and quartic gluon vertex. All of
them need to have the same divergence structure
\begin{alignat}{5}
\frac{Z_{Aff}}{Z_A^{1/2} Z_f} 
\really \frac{Z_{A \eta \eta}}{Z_A^{1/2} Z_\eta} 
\really \frac{Z_{3A}}{Z_A^{3/2}} 
\really \sqrt{ \frac{Z_{4A}}{Z_A^2} } \; .
\end{alignat}
If we had done the same calculation in QED and looked for a
running electric charge, we would have found that the vacuum
polarization diagrams for the photon do account for the entire counter
term of the electric charge. The other two renormalization constants
$Z_{Aff}$ and $Z_f$ cancel because of gauge invariance.\bigskip

In contrast to QED, the strong coupling diverges in the Thomson limit
because QCD is confined towards large distances and weakly coupled at
small distances. Lacking a well enough motivated reference point we
are lead to renormalize $\alpha_s$ in the $\msbar$ scheme. From
Eq.\eqref{eq:qcd_bzero} we know that the ultraviolet pole which needs
to be cancelled by the counter term is proportional to the function
$b_0$
\begin{alignat}{5}
&& g_s^\text{bare}   &= Z_g g_s 
                = \left( 1 + \delta Z_g \right) g_s 
                = \left( 1 + \frac{\delta g_s}{g_s} \right) g_s \notag \\
&\Rightarrow& \qquad
(g_s^2)^\text{bare} &= ( Z_g g_s )^2 
                = \left( 1 + \frac{\delta g_s}{g_s} \right)^2 g_s^2
                = \left( 1 + 2 \frac{\delta g_s}{g_s} \right) g_s^2 
                = \left( 1 + \frac{\delta g_s^2}{g_s^2} \right) g_s^2 \notag \\
&\Rightarrow& \qquad
\alpha_s^\text{bare} &= 
        \left( 1 + \frac{\delta \alpha_s}{\alpha_s} \right) \alpha_s 
        \really \left( 1 
                  - \frac{\Pi}{p^2} \Bigg|_\text{pole} \right) \alpha_s(M^2) 
\stackrel{\text{Eq.\eqref{eq:qcd_bzero}}}{=}
          \left( 1 
                  - \frac{\alpha_s}{\tilde \epsilon \left( \dfrac{\mu_R}{M} \right) } \, b_0 \right) \alpha_s(M^2) \; .
\label{eq:qcd_shift_alpha}
\end{alignat}
Only in the last step we have explicitly included the scale dependence
of the counter term. Because the bare coupling does not depend on any
scales, this means that $\alpha_s$ depends on the unphysical scale
$M$. Similar to the top mass renormalization scheme we can switch to a
more physical scheme for the strong coupling as well: we can absorb
also the finite contributions of $\Pi(\mu_R^2/p^2)$ into the strong
coupling by simply identifying $M^2 = p^2$. Based again
on Eq.\eqref{eq:qcd_bzero} this implies
\begin{alignat}{5}
\alpha_s^\text{bare} 
= \alpha_s(p^2) \left( 1 - \frac{\alpha_s(p^2) b_0}{\tilde \epsilon} 
                      + \alpha_s(p^2) b_0 \log \frac{p^2}{M^2} \right) \; .
\label{eq:run_alphas1}
\end{alignat}
On the right hand side $\alpha_s$ is consistently evaluated as a
function of the physical scale $p^2$.  This formula defines a running
coupling $\alpha_s(p^2)$, because the definition of the coupling now
has to account for a possible shift between the original argument
$p^2$ and the scale $M^2$ coming out of the $\msbar$ scheme. Since
according to Eqs.\eqref{eq:qcd_shift_alpha} and~\eqref{eq:run_alphas1}
the bare strong coupling can be expressed in terms of $\alpha_s(M^2)$
as well as in terms of $\alpha_s(p^2)$ we can link the two scales
through
\begin{alignat}{5}
\alpha_s(M^2) 
&= \alpha_s(p^2) + \alpha_s^2(p^2) b_0 \log \frac{p^2}{M^2} 
 = \alpha_s(p^2) \left( 1 + \alpha_s(p^2) b_0 \log \frac{p^2}{M^2} \right)
\notag \\
\Leftrightarrow \qquad 
\frac{d \alpha_s(p^2)}{d \log p^2} &= - \alpha_s^2(p^2) b_0 + \ope(\alpha_s^3) \; .
\label{eq:run_alphas2}
\end{alignat}
To the given loop order the argument of the strong coupling squared in
this formula can be neglected --- its effect is of higher order. We
nevertheless keep the argument as a higher order effect to later distinguish
different approaches to the running coupling. From
Eq.\eqref{eq:qcd_bzero} we know that $b_0 >0$, which means that towards
larger scales the strong coupling has a negative slope.
The ultraviolet limit of the strong coupling is zero. This makes
QCD an \underline{asymptotically free}\index{asymptotic freedom} theory. We can compute the
function $b_0$ in general models by simply adding all contributions of
strongly interacting particles in this loop
\begin{alignat}{5}
b_0 = - \frac{1}{12 \pi} \; \sum_\text{colored states} D_j \; T_{R,j} \; ,
\end{alignat}
where we need to know some kind of counting factor $D_j$ which is -11
for a vector boson (gluon), +4 for a Dirac fermion (quark), +2 for a
Majorana fermion (gluino), +1 for a complex scalar (squark) and +1/2
for a real scalar. Note that this sign is not given by the fermionic
or bosonic nature of the particle in the loop.  The color charges are
$T_R=1/2$ for the fundamental representation of $SU(3)$ and $C_A =
N_c$ for the adjoint representation. The masses of the loop particles
are not relevant in this approximation because we are only interested
in the ultraviolet regime of QCD where all particles can be regarded
massless. When we really model the running of $\alpha_s$ we need to
take into account threshold effects of heavy particles, because
particles can only contribute to the running of $\alpha_s$ at scales
above their mass scale. This is why the $R$ ratio computed in
Eq.\eqref{eq:qcd_r} is so interesting once we vary the energy of the
incoming electron--positron pair.\bigskip

We can do even better than this fixed order in perturbation theory:
while the correction to $\alpha_s$ in Eq.\eqref{eq:run_alphas1} is
perturbatively suppressed by the usual factor $\alpha_s/(4 \pi)$ it
includes a logarithm of a ratio of scales which does not need to be
small. Instead of simply including these gluon self energy corrections
at a given order in perturbation theory we can instead include chains
of one-loop diagrams with $\Pi$ appearing many times in the off--shell
gluon propagator. This series of Feynman diagrams is identical to the
one we sum for the mass renormalization in
Eq.\eqref{eq:prop_sum}. It means we replace the off--shell
gluon propagator by
\begin{alignat}{5}
\frac{T^{\mu \nu}}{p^2}
\rightarrow &
  \frac{T^{\mu \nu}}{p^2}
 + \left ( \frac{T}{p^2} \cdot (-T \, \Pi) \cdot \frac{T}{p^2} 
    \right)^{\mu \nu} \notag \\
 & \hspace*{20pt}
 + \left( \frac{T}{p^2} \cdot (-T \, \Pi) \cdot \frac{T}{p^2} 
                        \cdot (-T \, \Pi) \cdot \frac{T}{p^2} 
   \right)^{\mu \nu} + \cdots \notag \\
= & \;
  \frac{T^{\mu \nu}}{p^2} \;
  \sum_{j=0}^\infty \left( -\frac{\Pi}{p^2} \right)^j
= \frac{T^{\mu \nu}}{p^2} \; \frac{1}{1 + \Pi/p^2} \; ,
\end{alignat}
schematically written without the factors $i$.
To avoid indices we abbreviate $T^{\mu \nu} T_\nu^\rho = T \cdot T$
which make sense because of $(T \cdot T \cdot T)^{\mu \nu} = T^{\mu
  \rho} T^\sigma_\rho T_\sigma^\nu = T^{\mu \nu}$. This resummation
of the logarithm which appears in the next--to--leading order
corrections to $\alpha_s$ moves the finite shift in $\alpha_s$ shown
in Eqs.\eqref{eq:qcd_bzero} and~\eqref{eq:run_alphas1} into the
denominator, while we assume that the pole will be properly taken care
off in any of the schemes we discuss
\begin{alignat}{5}
 \alpha_s^\text{bare} =
 \alpha_s(M^2)
 - \frac{\alpha_s^2 b_0}{\tilde{\epsilon}} 
 \equiv \frac{\alpha_s(p^2)}{1 - \alpha_s(p^2) \;
                 b_0 \; \log \dfrac{p^2}{M^2}} 
 - \frac{\alpha_s^2 b_0}{\tilde{\epsilon}} \; .
\label{eq:run_alphas2a}
\end{alignat}
Just as in the case without resummation, we can use this complete
formula to relate the values of $\alpha_s$ at two reference
points, \ie we consider it a \underline{renormalization group
  equation} (RGE) which evolves physical parameters from one scale to
another in analogy to the fixed order version in Eq.\eqref{eq:run_alphas2}
\begin{alignat}{5}
 \frac{1}{\alpha_s(M^2)} &=
                 \frac{1}{\alpha_s(p^2)}
                 \left( 1
               - \alpha_s(p^2) \;
                 b_0 \;
                 \log \frac{p^2}{M^2} 
                 \right)
               = \frac{1}{\alpha_s(p^2)}
               - b_0 \;
                 \log \frac{p^2}{M^2} 
               + \ope(\alpha_s) \; .
\label{eq:run_alphas3}
\end{alignat}
\index{renormalization group equation!strong coupling} The factor
$\alpha_s$ inside the parentheses we can again evaluate at either of
the two scales, the difference is a higher order effect. If we keep it
at $p^2$ we see that the expression in Eq.\eqref{eq:run_alphas3} is
different from the un-resummed version in Eq.\eqref{eq:run_alphas1}.
If we ignore this higher order effect the two formulas become
equivalent after switching $p^2$ and $M^2$. Resumming the vacuum
expectation bubbles only differs from the un-resummed result once we
include some next--to--leading order contribution.  When we
differentiate $\alpha_s(p^2)$ with respect to the momentum transfer
$p^2$ we find, using the relation $d/d x (1/\alpha_s) = - 1/\alpha_s^2
\; d \alpha_s/dx$
\begin{alignat}{5}
 \frac{1}{\alpha_s} \; \frac{d \alpha_s}{d \log p^2} 
 = - \alpha_s \frac{d}{d \log p^2} \; \frac{1}{\alpha_s}
 = - \alpha_s \, b_0 + \ope(\alpha_s^2)
 \qquad \text{or} \qquad
\boxed{ p^2 \frac{d \alpha_s}{d p^2} 
 \equiv \frac{d \alpha_s}{d \log p^2} 
 = \beta = -\alpha_s^2 \sum_{n=0} b_n \alpha_s^n } \; .
\label{eq:run_alphas4}
\end{alignat}
This is the famous running of the strong coupling constant including
all higher order terms $b_n$.\bigskip

In the running of the strong coupling constant we relate the different
values of $\alpha_s$ through multiplicative factors of the kind
\begin{alignat}{5}
\left( 
1 \pm \alpha_s(p^2) b_0 \log \frac{p^2}{M^2} 
\right) \; .
\end{alignat}
Such factors appear in the un-resummed computation of
Eq.\eqref{eq:run_alphas2} as well as in Eq.\eqref{eq:run_alphas2a}
after resummation. Because they are multiplicative, these factors can
move into the denominator, where we need to ensure that they do not
vanish.  Dependent on the sign of $b_0$ this becomes a problem for
large scale ratios $|\alpha_s \log p^2/M^2| > 1$, where it leads to the Landau
pole\index{Landau pole}. We discuss it in detail for the Higgs self coupling in
Section~\ref{sec:higgs_rge}. For the strong coupling with $b_0 > 0$
and large coupling values at small scales $p^2 \ll M^2$ the
combination $(1 + \alpha_s b_0 \log p^2/M^2)$ can indeed vanish and
become a problem. For the opposite case of a coupling with $b_0 < 0$
and large coupling values at large scales $p^2 \gg M^2$ the same
combination $(1 + \alpha_s b_0 \log p^2/M^2)$ crosses zero.

It is customary to replace the renormalization point of $\alpha_s$ in
Eq.\eqref{eq:run_alphas2a} with a reference scale defined by the
Landau pole.  At one loop order this reads
\begin{alignat}{5}
  1 + \alpha_s \;
      b_0 \; \log \frac{\lqcd^2}{M^2}  &\really 0 
  \qquad \Leftrightarrow \qquad 
  \log \frac{\lqcd^2}{M^2} = - \frac{1}{\alpha_s(M^2) b_0}
  \qquad \Leftrightarrow \qquad 
  \log \frac{p^2}{M^2} = \log \frac{p^2}{\lqcd^2}
                       - \frac{1}{\alpha_s(M^2) b_0} \notag \\
  \frac{1}{\alpha_s(p^2)} & 
  \stackrel{\text{Eq.\eqref{eq:run_alphas3}}}{=}
                 \frac{1}{\alpha_s(M^2)}
               + b_0 \; \log \frac{p^2}{M^2} \\
               &= \frac{1}{\alpha_s(M^2)} \;
                 + b_0 \log \frac{p^2}{\lqcd^2} 
                 - \frac{1}{\alpha_s(M^2)}
               = b_0
                 \log \frac{p^2}{\lqcd^2} 
  \qquad \Leftrightarrow \qquad 
  \boxed{\alpha_s(p^2) =  \frac{1}{b_0 \log \dfrac{p^2}{\lqcd^2}}} \; .\notag
\end{alignat}
This scheme can be generalized to any order in perturbative QCD and is
not that different from the Thomson limit renormalization scheme of
QED, except that with the introduction of $\lqcd$ we are choosing a
reference point which is particularly hard to compute
perturbatively. One thing that is interesting in the way we introduce
$\lqcd$ is the fact that we introduce a scale into our theory
without ever setting it. All we did was renormalize a coupling which
becomes strong at large energies and search for the mass scale of this
strong interaction. This trick is called \underline{dimensional
  transmutation}\index{dimensional transmutation}.\bigskip

In terms of language, there is a little bit of \underline{confusion}
between field theorists and phenomenologists: up to now we have
introduced the renormalization scale $\mu_R$ as the 
renormalization point, for example of the strong coupling
constant.  In the $\msbar$ scheme, the subtraction of
$1/\tilde{\epsilon}$ shifts the scale dependence of the strong
coupling to $M^2$ and moves the logarithm $\log
M^2/\Lambda^2_\text{QCD}$ into the definition of the renormalized
parameter. This is what we will from now on call the renormalization
scale in the phenomenological sense, \ie the argument we evaluate
$\alpha_s$ at.  Throughout this section we will keep the symbol $M$
for this renormalization scale in the $\msbar$ scheme, but from
Section~\ref{sec:qcd_ir} on we will shift back to $\mu_R$ instead of
$M$ as the argument of the running coupling, to be consistent with the literature.

\subsubsection{Resumming scaling logarithms}
\label{sec:qcd_scale_logs}

In the last Section~\ref{sec:qcd_run_coup} we have introduced the
running strong coupling in a fairly abstract manner. For example, we
did not link the resummation of diagrams and the running of $\alpha_s$
in Eqs.\eqref{eq:run_alphas2} and~\eqref{eq:run_alphas4} to
physics. In what way does the resummation of the one-loop diagrams for
the $s$-channel gluon improve our prediction of the bottom pair
production rate at the LHC?\bigskip

To illustrate those effects we best look at a simple observable which
depends on just one energy scale $p^2$. The first observable
coming to mind is again the Drell--Yan cross section $\sigma(q \bar{q}
\to \mu^+ \mu^-)$, but since we are not really sure what to do with
the parton densities which are included in the actual hadronic
observable, we better use an observable at an $e^+ e^-$ collider.
Something that will work and includes $\alpha_s$ at least in the
one-loop corrections is the $R$ parameter defined in Eq.\eqref{eq:qcd_r}\index{R ratio}
\begin{alignat}{5}
R = \frac{\sigma(e^+ e^- \to \text{hadrons})}{\sigma(e^+ e^- \to \mu^+ \mu^-)}
  = N_c \sum_\text{quarks} Q_q^2 = \frac{11 N_c}{9} \; .
\label{eq:qcd_r2}
\end{alignat}
The numerical value at leading order assumes five quarks.
Including higher order corrections we can express the result in a
power series in the renormalized strong coupling $\alpha_s$.  In the
$\msbar$ scheme we subtract $1/\tilde \epsilon(\mu_R/M)$ and in
general include a scale dependence on $M$ in the individual prefactors
$r_n$
\begin{alignat}{5}
R \left( \frac{p^2}{M^2}, \alpha_s \right)
= \sum_{n=0} \; r_n\left( \frac{p^2}{M^2} \right) 
       \; \alpha_s^n(M^2)
\qqqquad r_0 = \frac{11 N_c}{9} \; .
\label{eq:r_pert}
\end{alignat}
The $r_n$ we can assume to be dimensionless ---
if they are not, we can scale $R$ appropriately using $p^2$. This
implies that the $r_n$ only depend on ratios of two scales, the
externally fixed $p^2$ on the one hand and the artificial $M^2$ on the
other.\bigskip

At the same time we know that $R$ is an observable, which means that
including all orders in perturbation theory it cannot depend on any
artificial scale choice $M$. Writing this dependence as a total
derivative and setting it to zero we find an equation which would be
called a \underline{Callan--Symanzik equation}\index{Callan--Symanzik equation} if instead of the
running coupling we had included a running mass
\begin{alignat}{5}
0 &\really M^2 \frac{d}{d M^2} R \left( \frac{p^2}{M^2}, \alpha_s(M^2) \right)
   = M^2 \left[  \frac{\p}{\p M^2} 
                 + \frac{\p \alpha_s}{\p M^2} \, \frac{\p}{\p \alpha_s}
           \right] 
     R \left( \frac{p^2}{M^2}, \alpha_s \right)
\notag \\
  &= \left[  M^2 \frac{\p}{\p M^2} 
           + \beta \, \frac{\p}{\p \alpha_s}
     \right] \; 
     \sum_{n=0} \; r_n\left( \frac{p^2}{M^2} \right) \; \alpha_s^n
\notag \\
  &= \sum_{n=1}
     M^2 \frac{\p r_n}{\p M^2} \, \alpha_s^n
   + \sum_{n=1}
     \beta \, r_n \, n \alpha_s^{n-1}
   &&\text{with} \quad r_0 = \frac{11 N_c}{9} = \text{const}
\notag \\
  &= M^2 \sum_{n=1}
     \frac{\p r_n}{\p M^2} \, \alpha_s^n
   - \sum_{n=1} \sum_{m=0} 
     n r_n \, \alpha_s^{n+m+1} \, b_m
   &&\text{with} \quad
   \beta = -\alpha_s^2 \sum_{m=0} b_m \alpha_s^m
\notag \\
  &= M^2 \frac{\p r_1}{\p M^2} \, \alpha_s
   + \left( M^2 \frac{\p r_2}{\p M^2} 
          - r_1 b_0
     \right) \, \alpha_s^2
   + \left( M^2 \frac{\p r_3}{\p M^2} 
          - 2 r_2 b_0 - r_1 b_1 
     \right) \, \alpha_s^3
   &&+ \ope(\alpha_s^4) \; .
\label{eq:callan_sym}
\end{alignat}
In the second line we have to remember that the $M$ dependence of
$\alpha_s$ is already included in the appearance of $\beta$, so
$\alpha_s$ should be considered a variable by itself.  This
perturbative series in $\alpha_s$ has to vanish in each order of
perturbation theory. The non--trivial structure, namely the mix of
$r_n$ derivatives and the perturbative terms in the $\beta$ function
we can read off the $\alpha_s^3$ term in Eq.\eqref{eq:callan_sym}:
first, we have the appropriate NNNLO corrections $r_3$. Next, we have
one loop in the gluon propagator $b_0$ and two loops for example in
the vertex $r_2$. And finally, we need the two-loop diagram for the
gluon propagator $b_1$ and a one-loop vertex correction $r_1$. The
kind--of--Callan--Symanzik equation Eq.\eqref{eq:callan_sym} requires 
\begin{alignat}{5}
 \frac{\p r_1}{\p \log M^2/p^2} &= 0 \notag \\
 \frac{\p r_2}{\p \log M^2/p^2} &= r_1 b_0 \notag \\
 \frac{\p r_3}{\p \log M^2/p^2} &= r_1 b_1 + 2 r_2(M^2) b_0 \notag \\
\cdots
\end{alignat}
The dependence on the argument $M^2$ vanishes for $r_0$ and $r_1$.
Keeping in mind that there will be integration constants $c_n$
and that another, in our case, unique
momentum scale $p^2$ has to cancel the mass units inside $\log M^2$ we
find
\begin{alignat}{5}
r_0 &= c_0 = \frac{11 N_c}{9}
\notag \\
r_1 &= c_1 
\notag \\
r_2 &= c_2 + r_1 b_0 \log \frac{M^2}{p^2} 
     = c_2 + c_1 b_0 \log \frac{M^2}{p^2} 
\notag \\
r_3 
    &= \int d \log \frac{{M'}^2}{p^2}
    \left( c_1 b_1 + 2 \left( c_2 + c_1 b_0 \log \frac{{M'}^2}{p^2} \right) b_0 \right) 
\notag \\
    &= c_3 
    + \left( c_1 b_1 
           + 2 c_2 b_0 \right) \log \frac{M^2}{p^2} 
    + c_1 b_0^2 \log^2 \frac{M^2}{p^2} 
\notag \\
\cdots
\label{eq:qcd_r_c}
\end{alignat}
This chain of $r_n$ values looks like we should interpret the apparent
fixed-order perturbative series for $R$ in Eq.\eqref{eq:r_pert} as a
series which implicitly includes terms of the order $\log^{n-1}
M^2/p^2$ in each $r_n$. They can become problematic if this logarithm
becomes large enough to spoil the fast convergence in terms of
$\alpha_s \sim 0.1$, evaluating the observable $R$ at scales far
away from the scale choice for the strong coupling constant
$M$.\bigskip

Instead of the series in $r_n$ we can use the conditions in
Eq.\eqref{eq:qcd_r_c} to express $R$ in terms of the $c_n$ and collect
the logarithms appearing with each $c_n$. The geometric series we then
resum to
\begin{alignat}{5}
R = \sum_n r_n \left( \frac{p^2}{M^2} \right) \; \alpha_s^n(M^2) 
  =&  \; c_0 
     + c_1 \left( 1 + \alpha_s(M^2) b_0 \log \frac{M^2}{p^2} 
                    + \alpha_s^2(M^2) b_0^2 \log^2 \frac{M^2}{p^2} +\cdots
           \right) \alpha_s(M^2) 
\notag \\
 &\hspace*{5mm} 
     + c_2 \left( 1 + 2 \alpha_s(M^2) b_0 \log \frac{M^2}{p^2} +\cdots
           \right) \alpha_s^2(M^2) + \cdots
\notag \\
  =&  \; c_0 
     + c_1 \frac{\alpha_s(M^2)}
                {1 - \alpha_s(M^2) b_0 \log \dfrac{M^2}{p^2}} 
     + c_2 \left( 
           \frac{\alpha_s(M^2)}
                {1 - \alpha_s(M^2) b_0 \log \dfrac{M^2}{p^2}}
           \right)^2 + \cdots
\notag \\
  \equiv& \sum c_n \; \alpha_s^n(p^2) \; .
\label{eq:r_in_c}
\end{alignat}
In the original ansatz $\alpha_s$ is always evaluated at the
scale $M^2$.  In the last step we use Eq.\eqref{eq:run_alphas3} with
flipped arguments $p^2$ and $M^2$, derived from the resummation of the
vacuum polarization bubbles. In contrast to the $r_n$ integration
constants the $c_n$ are by definition independent of $p^2/M^2$ and
therefore more suitable as a perturbative series in the presence of
potentially large logarithms. Note that the un-resummed version of
the running coupling in Eq.\eqref{eq:run_alphas1} would not give the
correct result, so Eq.\eqref{eq:r_in_c} only holds for resummed vacuum
polarization bubbles.\bigskip

This re-organization of the perturbation series for $R$ can be
interpreted as \underline{resumming all
  logarithms}\index{resummation!scaling logarithms} of the kind $\log
M^2/p^2$ in the new organization of the perturbative series and
absorbing them into the running strong coupling evaluated at the scale
$p^2$. All scale dependence in the perturbative series for the
dimensionless observable $R$ is moved into $\alpha_s$, so possibly
large logarithms $\log M^2/p^2$ have disappeared.  In
Eq.\eqref{eq:r_in_c} we also see that this series in $c_n$ will never
lead to a scale-invariant result when we include a finite order in
perturbation theory. Some higher--order factors $c_n$ are known, for
example inserting $N_c = 3$ and five quark flavors just as we assume
in Eq.\eqref{eq:qcd_r2}
\begin{alignat}{5}
R = \frac{11}{3} \; 
\left( 1 + \frac{\alpha_s(p^2)}{\pi}
         + 1.4 \left( \frac{\alpha_s(p^2)}{\pi} \right)^2
         - 12 \left( \frac{\alpha_s(p^2)}{\pi} \right)^3
         + \ope\left( \frac{\alpha_s(p^2)}{\pi} \right)^4 
\right) \; .
\end{alignat}
This alternating series with increasing perturbative prefactors seems
to indicate the asymptotic instead of convergent behavior of
perturbative QCD. At the bottom mass scale the relevant coupling
factor is only $\alpha_s(m_b^2)/\pi \sim 1/14$, so a further increase
of the $c_n$ would become dangerous. However, a detailed look into the
calculation shows that the dominant contributions to $c_n$ arise from
the analytic continuation of logarithms, which are large finite terms for
example from $\text{Re} (\log^2 (-E^2)) = \log^2 E^2 + \pi^2$. In the
literature such $\pi^2$ terms arising from the analytic continuation
of loop integrals are often phrased in terms of $\zeta_2 =
\pi^2/6$.\bigskip

Before moving on we collect the logic of the argument given in this
section: when we regularize an ultraviolet divergence we automatically
introduce a reference scale $\mu_R$. Naively, this could be an
ultraviolet cutoff scale, but even the seemingly scale invariant
dimensional regularization in the conformal limit of our field theory
cannot avoid the introduction of a scale. There are several ways of
dealing with such a scale: first, we can renormalize our parameter at
a reference point.  Secondly, we can define a running parameter and this way
absorb the scale logarithm into the $\msbar$ counter term. In that
case introducing $\lqcd$ leaves us with a compact form of the running
coupling $\alpha_s(M^2;\lqcd)$.

Strictly speaking, at each order in perturbation theory the scale
dependence should vanish together with the ultraviolet poles, as long as there
is only one scale affecting a given observable. However, defining the
running strong coupling we sum one-loop vacuum polarization
graphs. Even when we compute an observable at a given loop order, we
implicitly include higher order contributions. They lead to a
dependence of our perturbative result on the artificial scale $M^2$,
which phenomenologists refer to as renormalization scale dependence.

Using the $R$ ratio we see what our definition of
the running coupling means in terms of resumming logarithms:
reorganizing our perturbative series to get rid of the ultraviolet
divergence $\alpha_s(p^2)$ resums the scale logarithms $\log
p^2/M^2$ to all orders in perturbation theory. We will need this
picture once we introduce infrared divergences in the
following section.

\subsection{Infrared divergences}
\label{sec:qcd_ir}

After this brief excursion into ultraviolet divergences and
renormalization we can return to the original example, the Drell--Yan
process.  Last, we wrote down the hadronic cross sections in terms of
parton distributions at leading order in
Eq.\eqref{eq:qcd_dy_naive}. At this stage parton distributions (pdfs)
in the proton are only functions of the collinear momentum fraction of
the partons inside the proton about which from a theory point of view
we only know a set of sum rules.

The perturbative question we need to ask for $\mu^+ \mu^-$ production
at the LHC is: what happens if together with the two leptons we
produce additional jets which for one reason or another we do not
observe in the detector. Such jets could for example come from the
radiation of a gluon from the initial--state quarks. In
Section~\ref{sec:qcd_single_jet} we will study the kinematics of
radiating such jets and specify the infrared divergences this leads
to. In Sections~\ref{sec:qcd_splitting} and~\ref{sec:qcd_dglap} we
will show that these divergences have a generic structure and can be
absorbed into a re-definition of the parton densities, similar to an
ultraviolet renormalization of a Lagrangian parameter. In
Sections~\ref{sec:qcd_solve_dglap} and~\ref{sec:qcd_resum_collinear} we
will again follow the example of the ultraviolet divergences and
specify what absorbing these divergences means in terms logarithms
appearing in QCD calculations.

Throughout this writeup we will use the terms \underline{jets and final
  state partons}\index{jet} synonymously. This is not really correct once we
include jet algorithms and hadronization. On the other hand, in
Section~\ref{sec:sim_fatjet} we will see that the purpose of a jet
algorithm is to take us from some kind of energy deposition in the
calorimeter to the parton radiated in the hard process. The two should 
therefore be closely related.

\subsubsection{Single jet radiation} 
\label{sec:qcd_single_jet}

Let us get back to the radiation of additional partons in the
Drell--Yan process. We can start for example by computing
the cross section for the partonic process $q \bar{q} \to Z
g$. However, this partonic process involves renormalization of
ultraviolet divergences as well as loop diagrams which we have to 
include before we can say anything reasonable, \ie ultraviolet and
infrared finite.

To make life easier and still learn about the structure of collinear
infrared divergences\index{collinear divergence} we instead look at the
crossed process 

\begin{center}
\begin{fmfgraph*}(80,50)
 \fmfset{arrow_len}{2mm}
 \fmfleft{in1,in2}
 \fmf{fermion,width=0.5,label=$q$}{in1,v1}
 \fmf{fermion,width=0.5}{v1,v2}
 \fmf{fermion,width=0.5}{v2,out2}
 \fmf{photon,width=0.5,lab.side=left,label=$Z$,tension=1}{v1,out1}
 \fmf{gluon,width=0.5,lab.side=right,label=$g$,tension=1}{in2,v2}
 \fmfright{out1,out2}
\end{fmfgraph*}
\end{center}

It should behave similar to any other $(2 \to 2)$ jet radiation,
except that it has a different incoming state than the leading order
Drell--Yan process and hence does not involve virtual
corrections. This means we do not have to deal with ultraviolet
divergences and renormalization, and can concentrate on parton or jet
radiation from the initial state. Moreover, let us go back to $Z$
production instead of a photon, to avoid confusion with additional massless
particles in the final state.\bigskip

The amplitude for this $(2 \to 2)$ process is --- modulo charges and
averaging factors, but including all Mandelstam variables
\begin{alignat}{5}
 \matx  \sim          - \frac{t}{s}
                      - \frac{s^2 -2 m_Z^2 (s + t - m_Z^2)}{st} \; .
\label{eq:qcd_twototwo}
\end{alignat}
As discussed in Section~\ref{sec:qcd_dy_r},
the Mandelstam variable $t$ for one massless final--state particle can
be expressed as $t = -s (1-\tau) y$ in terms of the rescaled
gluon emission angle $y=(1 - \cos \theta)/2$ and $\tau =
m_Z^2/s$. Similarly, we obtain $u = -s (1-\tau) (1-y)$, so as a first
check we can confirm that $t+u=-s(1-\tau) = -s+m_Z^2$. The collinear
limit\index{collinear limit} when the gluon is radiated in the beam
direction is given by $y \to 0$, corresponding to negative $t \to 0$
with finite $u=-s+m_Z^2$. In this limit the matrix element can also be written as
\begin{alignat}{5}
 \matx \sim \frac{s^2 - 2 s m_Z^2 + 2 m_Z^4}{s(s-m_Z^2)} \; \frac{1}{y}
                 + \ope(y) \; .
\label{eq:dy_div1}
\end{alignat}
This expression is divergent for collinear gluon radiation or
gluon splitting, \ie for small angles $y$. We can translate this $1/y$
divergence for example into the transverse momentum of the gluon or
$Z$ 
\begin{alignat}{5}
s p_T^2 
= t u 
= s^2 (1 - \tau)^2 \; y (1-y) 
= (s-m_Z^2)^2 y + \ope(y^2)
\end{alignat}
In the collinear limit our matrix element squared in
Eq.\eqref{eq:dy_div1} becomes
\begin{alignat}{5}
\boxed{
 \matx \sim \frac{s^2 - 2 s m_Z^2 + 2 m_Z^4}{s^2} \;
                   \frac{(s-m_Z^2)}{p_T^2}
                 + \ope(p_T^0) 
} \; .
\label{eq:dy_div2}
\end{alignat}
The matrix element for the tree level process $q g \to Z q$ has a
leading divergence proportional to $1/p_T^2$. To compute the total
cross section for this process we need to integrate the matrix element
over the entire two-particle phase space. Starting from
Eq.\eqref{eq:qcd_diff_cxn} and using the appropriate Jacobian this
integration can be written in terms of the reduced angle $y$.
Approximating the matrix element as $C'/y$ or $C/p_T^2$, we then 
integrate
\begin{alignat}{5}
  \int_{y^\text{min}}^{y^\text{max}} d y \frac{C'}{y}
= \int_{p_T^\text{min}}^{p_T^\text{max}} d p_T^2 \frac{C}{p_T^2}
= \; 2 \int_{p_T^\text{min}}^{p_T^\text{max}} d p_T \; p_T \; \frac{C}{p_T^2}
\simeq& \; 2 C \int_{p_T^\text{min}}^{p_T^\text{max}} d p_T \frac{1}{p_T}
= 2 C \; \log \frac{p_T^\text{max}}{p_T^\text{min}}
\label{eq:qcd_collinear}
\end{alignat}
The form $C/p_T^2$ for the matrix element is of course only valid in
the collinear limit; in the non--collinear phase space $C$ is not a
constant.  However, Eq.\eqref{eq:qcd_collinear} describes well the
collinear divergence arising from quark radiation at the LHC.\bigskip

Next, we follow the same strategy as for the ultraviolet
divergence.  First, we regularize the divergence for example using
dimensional regularization. Then, we find a well--defined way to get
rid of it. Dimensional regularization means writing the two-particle
phase space in $n=4-2 \epsilon$ dimensions. Just for reference, the
complete formula in terms of the angular variable $y$ reads
\begin{alignat}{5}
 s \; \frac{d \sigma}{d y} =
  \frac{\pi (4 \pi)^{-2+\epsilon}}{\Gamma(1-\epsilon)} \;
  \left( \frac{\mu_F^2}{m_Z^2} \right)^\epsilon \;
  \frac{\tau^\epsilon (1-\tau)^{1-2 \epsilon}}
       {y^\epsilon (1-y)^\epsilon} \matx
 \sim 
  \left( \frac{\mu_F^2}{m_Z^2} \right)^\epsilon \;
  \frac{\matx}{y^\epsilon (1-y)^\epsilon} \; .
\label{eq:qcd_phase}
\end{alignat}
In the second step we only keep the factors we are interested in. The
additional factor $1/y^\epsilon$ regularizes the integral at $y \to
0$, as long as $\epsilon<0$ by slightly increasing the suppression of
the integrand in the infrared regime. This means that for infrared
divergences we can as well choose $n = 4 + 2 \epsilon$ space--time
dimensions with $\epsilon > 0$. After integrating the leading
collinear divergence $1/y^{1+\epsilon}$ we are left with a pole
$1/(-\epsilon)$. This regularization procedure is symmetric in $y
\leftrightarrow (1-y)$. What is important to notice is again the
appearance of a scale $\mu_F^{2 \epsilon}$ with the $n$-dimensional
integral. This scale arises from the infrared regularization of the
phase space integral and is referred to as \underline{factorization
  scale}\index{scales!factorization scale}. The actual removal of the
infrared pole --- corresponding to the renormalization in the
ultraviolet case --- is called \underline{mass
  factorization}\index{mass factorization} and works exactly the same
way as renormalizing a parameter: in a well--defined scheme we simply
subtract the pole from the fixed-order matrix element squared.

\subsubsection{Parton splitting}
\label{sec:qcd_splitting}

\begin{figure}[t]
\begin{center}
\includegraphics[width=0.25\hsize]{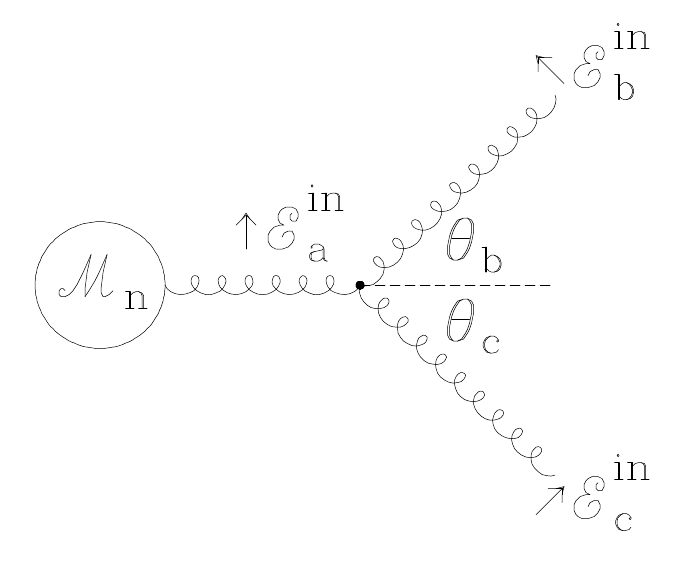}
\end{center}
\caption{Splitting of one gluon into two gluons.  Figure from
  Ref.~\cite{Ellis:1991qj}.}
\label{fig:qcd_gsplit}
\end{figure}

From the discussion of the process $q g \to Z q$ we can at least hope
that after taking care of all other infrared and ultraviolet
divergences the collinear structure of the process $q \bar{q} \to Z g$
will be similar. In this section we will show that we can indeed write
all collinear divergences in a universal form, independent of the hard
process which we choose as the Drell--Yan process. In the
collinear limit, the radiation of additional partons or the splitting
into additional partons will be described by universal
\underline{splitting functions}\index{splitting kernel}.\bigskip

Infrared divergences occur for massless particles in the initial or
final state, so we need to go through all ways incoming or outgoing
gluons and quark can split into each other. The description of the
factorized phase space, with which we will start, is common to all
these different channels. The first and at the LHC most important case
is the splitting of one gluon into two, shown in
Figure~\ref{fig:qcd_gsplit}. The two daughter gluons are close
to mass shell while the mother has to have a finite positive invariant
mass $p_a^2 \gg p_b^2, p_c^2$. We again assign the direction of the
momenta as $p_a = - p_b - p_c$, which means we have to take care of
minus signs in the particle energies. The kinematics of this
approximately collinear process we can describe in terms of the energy
fractions $z$ and $1-z$ defined as
\begin{alignat}{5}
z = \frac{|E_b|}{|E_a|} = 1 - \frac{|E_c|}{|E_a|}
\qqquad 
p_a^2 &= (-p_b -p_c)^2
       = 2 (p_b p_c)  
       = 2 z (1-z) (1 - \cos \theta ) E_a^2
       = z (1-z) E_a^2 \theta^2 + \ope(\theta^4) 
\notag \\
\Leftrightarrow \qquad
\theta &\equiv \theta_b + \theta_c 
         \simeq \frac{1}{|E_a|} \; \sqrt{ \frac{p_a^2}{z (1-z)} } \; ,
\label{eq:qcd_pre_sudakov}
\end{alignat}
in the collinear limit and in terms of the opening angle $\theta$
between $\vec{p}_b$ and $\vec{p}_c$. Because $p_a^2 >0$ we call this
final--state splitting configuration \underline{time--like
  branching}\index{splitting!time--like branching}. For this
configuration we can write down the so-called \underline{Sudakov
  decomposition}\index{phase space!Sudakov decomposition} of the
four-momenta
\begin{alignat}{5}
- p_a = p_b + p_c =
  \left( - z p_a + \beta n + p_T \right) \; + \;
  \left( - (1-z) p_a - \beta n - p_T \right) \; .
\label{eq:qcd_sudakovdec}
\end{alignat}
It defines an arbitrary unit four-vector $n$, a component orthogonal
to the mother momentum and $n$, \ie $p_a$ $(p_a p_T) = 0 = (n p_T)$,
and a free factor $\beta$. This way, we can specify $n$ such that it
defines the direction of the $p_b$--$p_c$ decay plane. In this
decomposition we can set only one invariant mass to zero, for example
that of a radiated gluon $p_c^2=0$. The second final state will have a
finite invariant mass $p_b^2 \neq 0$.\bigskip

As specific choice for the three reference four-vectors is
\begin{alignat}{5}
p_a = \begin{pmatrix}  |E_a| \\ 0 \\ 0 \\ p_{a,3} \end{pmatrix}
    = |E_a| \begin{pmatrix}  1 \\ 0 \\ 0 \\ 1 + \ope(\theta) \end{pmatrix}
\qqquad
n = \begin{pmatrix}  1 \\ 0 \\ 0 \\ -1 \end{pmatrix}
\qqquad
p_T = \begin{pmatrix}  0 \\ p_{T,1} \\ p_{T,2} \\ 0 \end{pmatrix} \; .
\label{eq:sudakov_momenta}
\end{alignat}
Relative to $\vec{p}_a$ we can split the opening angle $\theta$ for
massless partons according to Figure~\ref{fig:qcd_gsplit}
\begin{alignat}{5}
\theta = \theta_b + \theta_c
\qquad \text{and} \qquad
\frac{\theta_b}{\theta_c}
= \frac{p_T}{|E_b|} \left( \frac{p_T}{|E_c|} \right)^{-1}
= \frac{1-z}{z} 
\qquad \Leftrightarrow \qquad 
\theta = \frac{\theta_b}{1-z}
       = \frac{\theta_c}{z} \; .
\label{eq:sudakov_pt}
\end{alignat}
The momentum choice in Eq.\eqref{eq:sudakov_momenta} has the
additional feature that $n^2=0$, which allows us to extract $\beta$ from
the momentum parameterization shown in Eq.\eqref{eq:qcd_sudakovdec}
and the additional condition that $p_c^2=0$
\begin{alignat}{5}
p_c^2 &= \left( -(1-z) p_a - \beta n - p_T 
        \right)^2 \notag \\
      &= (1-z)^2 p_a^2 + p_T^2 + 2 \beta (1-z) (n p_a) \notag \\
      &= (1-z)^2 p_a^2 + p_T^2 + 4 \beta (1-z) |E_a| (1 + \ope(\theta)) \really 0 
   \qquad \Leftrightarrow \qquad 
   \beta &\simeq - \frac{p_T^2 + (1-z)^2 p_a^2}{4 (1-z) |E_a|} \; .
\label{eq:qcd_sudakovdec_solve}
\end{alignat}
\bigskip

Using this specific phase space parameterization we can divide an
$(n+1)$-particle process into an $n$-particle process and a \underline{splitting
process} of quarks and gluons. First, this requires us to
split the $(n+1)$-particle phase space alone into an $n$-particle
phase space and the collinear splitting.\index{splitting!phase space} The general $(n+1)$-particle
phase space separating off the $n$-particle contribution
\begin{alignat}{5}
d \Phi_{n+1} 
&= \cdots 
          \frac{d^3 \vec{p}_b}{2 (2 \pi)^3 |E_b|} \; 
          \frac{d^3 \vec{p}_c}{2 (2 \pi)^3 |E_c|} \;
 = \cdots 
          \frac{d^3 \vec{p}_a}{2 (2 \pi)^3 |E_a|}  \; 
          \frac{d^3 \vec{p}_c}{2 (2 \pi)^3 |E_c|} \frac{|E_a|}{|E_b|} 
  \qqquad &&\text{at fixed $p_a$}
  \notag \\
&= d \Phi_n \;
   \frac{d p_{c,3} d p_T p_T d \phi}{2 (2 \pi)^3 |E_c|} \; \frac{1}{z} 
   \notag \\
&= d \Phi_n \;
   \frac{d p_{c,3} d p_T^2 d \phi}{4 (2 \pi)^3 |E_c|} \; \frac{1}{z} 
\label{eq:ps_separate2}
\end{alignat}
is best expressed in terms of the energy fraction $z$ and the
azimuthal angle\index{azimuthal angle} $\phi$.  In other words,
separating the $(n+1)$-particle space into an $n$-particle phase space
and a $(1 \to 2)$ splitting phase space is possible without any
approximation, and all we have to take care of is the correct
prefactors in the new parameterization.\bigskip

Our next task is to translate the phase space parameters $p_{c,3}$ and
$p_T^2$ appearing in Eq.\eqref{eq:ps_separate2} into $z$ and $p_a^2$.
Starting from Eq.\eqref{eq:qcd_sudakovdec} for $p_{c,3}$ with the
third components of $p_a$ and $p_T$ given by
Eq.\eqref{eq:sudakov_momenta} we insert $\beta$ from
Eq.\eqref{eq:qcd_sudakovdec_solve} and obtain
\begin{alignat}{5}
\frac{d p_{c,3}}{d z} 
&= \frac{d}{d z} \left[ -(1-z) |E_a| (1 + \ope(\theta)) + \beta \right] 
= \frac{d}{d z} \left[ -(1-z) |E_a| (1 + \ope(\theta)) - \frac{p_T^2+(1-z)^2 p_a^2}{4 (1-z) |E_a|} \right] \notag \\
&= |E_a| (1 + \ope(\theta))
   - \frac{p_T^2}{4 (1-z)^2 E_a} 
   + \frac{p_a^2}{4 |E_a|} \notag \\
&= \frac{|E_c|}{1-z} (1 + \ope(\theta))
   - \frac{\theta^2 z^2 E_c^2}{4 (1-z)^2 E_a} 
   + \frac{z (1-z) E_a^2 \theta^2 + \ope(\theta^4)}{4 |E_a|} 
\qqquad \text{using Eq.\eqref{eq:qcd_pre_sudakov} and Eq.\eqref{eq:sudakov_pt}}
\notag \\
&=  \frac{|E_c|}{1-z} + \ope(\theta)
   \qquad \Leftrightarrow \qquad 
   \frac{d p_{c,3}}{|E_c|} \simeq \frac{dz}{1-z} \; .
\end{alignat}
In addition to substituting $d p_{c,3}$ by $d z$ in
Eq.\eqref{eq:ps_separate2} we also replace $d p_T^2$ with $d p_a^2$
according to
\begin{alignat}{5}
\frac{p_T^2}{p_a^2} = 
\frac{E_b^2 \theta_b^2}{z (1-z) E_a^2 \theta^2} = 
\frac{z^2 E_a^2 (1-z)^2 \theta^2}{z (1-z) E_a^2 \theta^2} = z (1-z) 
\quad \Leftrightarrow \quad
d p_T^2 = z (1-z) dp_a^2 \; .
\end{alignat}
This gives us the final result for the separated \underline{collinear
  phase space}
\begin{alignat}{5}
\boxed{
d \Phi_{n+1} = 
d \Phi_n \;   \frac{d z \, d p_a^2 \, d \phi}{4 (2 \pi)^3} =
d \Phi_n \;   \frac{d z \, d p_a^2}{4 (2 \pi)^2} 
} \; , 
\label{eq:coll_phasespace}
\end{alignat}
where in the second step we assume an azimuthal symmetry.\bigskip

Adding the transition matrix elements to this factorization of the
phase space and ignoring the initial--state flux factor which is common
to both processes we can now postulate a full factorization for one collinear emission and
in the collinear approximation
\begin{alignat}{5}
d \sigma_{n+1} 
&= \overline{|\mat_{n+1}|^2} \; d \Phi_{n+1} 
  \notag \\
&=  \overline{|\mat_{n+1}|^2} \; d \Phi_n \frac{d p_a^2 \, dz}{4(2 \pi)^2} \;
    \left( 1 + \ope(\theta) \right)
  \notag \\
&\simeq  \frac{2 g_s^2}{p_a^2} \; \hat{P}(z) \; \overline{|\mat_n|^2} \; d \Phi_n \frac{d p_a^2 \, dz}{16 \pi^2} 
        \qquad \text{assuming} \quad 
\boxed{
\overline{|\mat_{n+1}|^2} \simeq \frac{2 g_s^2}{p_a^2} \; \hat{P}(z) \; \overline{|\mat_n|^2}
} \; .
\label{eq:qcd_phase_space}
\end{alignat}
This last step is an assumption. We will proceed to show it step by
step by constructing the appropriate \underline{splitting kernels}
$\hat{P}(z)$\index{splitting kernel} for all different quark and gluon
configurations. If Eq.\eqref{eq:qcd_phase_space} holds true this means that
we can compute the $(n+1)$ particle amplitude squared from the
$n$-particle case convoluted with the appropriate splitting kernel.
Using $d \sigma_n \sim \overline{|\mat_n|^2} \; d \Phi_n$ and $g_s^2 =
4 \pi \alpha_s$ we can write this relation in its most common
form\index{factorization}
\begin{alignat}{5}
\boxed{\sigma_{n+1}
 \simeq \int \sigma_n \; 
   \frac{d p_a^2}{p_a^2} dz \; 
   \frac{\alpha_s}{2 \pi} \; \hat{P}(z)
} \; .
\label{eq:qcd_factorize}
\end{alignat}
Reminding ourselves that relations of the kind
$\overline{|\mat_{n+1}|^2} = p \overline{|\mat_{n}|^2}$ can typically
be summed, for example for the case of successive soft photon
radiation in QED, we see that Eq.\eqref{eq:qcd_factorize} is not the
final answer.  It does not include the necessary phase space factor
$1/n!$ from identical bosons in the final state which leads to the
simple exponentiation.\bigskip

As the first parton splitting in QCD we study a \underline{gluon
  splitting into two gluons}, shown in Figure~\ref{fig:qcd_gsplit}.
To compute its transition amplitude we write down all gluon momenta
and polarizations in a specific frame. With respect to the scattering
plane opened by $\vec{p}_b$ and $\vec{p}_c$ all three gluons have two
transverse polarizations, one in the plane, $\epsilon^\|$, and one
perpendicular to it, $\epsilon^\perp$. In the limit of small
scattering angles, the three parallel as well as the three
perpendicular polarization vectors are aligned. The perpendicular
polarizations are also orthogonal to all three gluon momenta. The
physical transverse polarizations in the plane are orthogonal to their
corresponding momenta and only approximately orthogonal to the other
momenta. Altogether, this means for the three-vectors $\epsilon^\|$
and $\epsilon^\perp$
\begin{alignat}{5}
( \epsilon_i^\| \epsilon_j^\| ) = -1 + \ope(\theta) \qquad
( \epsilon_i^\perp \epsilon_j^\perp ) = -1 \qquad
( \epsilon_i^\perp \epsilon_j^\| ) = 0 \qqquad
( \epsilon_i^\perp p_j ) = 0 \qquad
( \epsilon_j^\| p_j ) = 0\; ,
\label{eq:pol_vec1}
\end{alignat}
with general $i \ne j$. For $i = j$ we find exactly one and zero.  Using
these kinematic relations we can tackle the splitting amplitude $g \to
gg$. It is proportional to the vertex $V_{ggg}$ which in terms of all
incoming momenta reads
\begin{alignat}{5}
V_{ggg} 
&= i g_s f^{abc} \; \epsilon_a^\alpha \epsilon_b^\beta \epsilon_c^\gamma \;
    \left[  g_{\alpha \beta} (p_a - p_b)_\gamma
          + g_{\beta \gamma} (p_b - p_c)_\alpha
          + g_{\gamma \alpha} (p_c - p_a)_\beta
    \right]
\notag \\
&= i g_s f^{abc} \; \epsilon_a^\alpha \epsilon_b^\beta \epsilon_c^\gamma \;
    \left[  g_{\alpha \beta} (-p_c - 2 p_b)_\gamma
          + g_{\beta \gamma} (p_b - p_c)_\alpha
          + g_{\gamma \alpha} (2 p_c + p_b)_\beta
    \right]
\qquad && \text{with} \; p_a = - p_b - p_c 
\notag \\
&= i g_s f^{abc} \; 
    \left[  -2 (\epsilon_a \epsilon_b) (\epsilon_c p_b)
          + (\epsilon_b \epsilon_c) (\epsilon_a p_b) 
          - (\epsilon_b \epsilon_c) (\epsilon_a p_c) 
          + 2 (\epsilon_c \epsilon_a) (\epsilon_b p_c) 
    \right]
       && \text{with} \; (\epsilon_j p_j) = 0 
\notag \\
&= - 2 i g_s f^{abc} \; 
    \left[ (\epsilon_a \epsilon_b) (\epsilon_c p_b)
           -(\epsilon_b \epsilon_c) (\epsilon_a p_b) 
           -(\epsilon_c \epsilon_a) (\epsilon_b p_c) 
    \right]
       && \text{with} \; (\epsilon_a p_c) = - (\epsilon_a p_b) \notag \\
&= - 2 i g_s f^{abc} \; 
    \left[ (\epsilon_a \epsilon_b) (\epsilon_c^\parallel p_b)
           -(\epsilon_b \epsilon_c) (\epsilon_a^\parallel p_b) 
           -(\epsilon_c \epsilon_a) (\epsilon_b^\parallel p_c) 
    \right]
       && \text{with} \; (\epsilon_i^\perp p_j) = 0 \; .
\label{eq:gluon_split1}
\end{alignat}
Squaring the splitting matrix element to compute the $(n+1)$ and $n$
particle matrix elements squared for the unpolarized case gives us
\begin{alignat}{5}
\overline{| \mat_{n+1} |^2} 
&= \frac{1}{2} \; \left( \frac{1}{p_a^2} \right)^2  \;
   4 g_s^2 \; 
   \frac{1}{N_c^2-1} \;
   \frac{1}{N_a} \;
   \sum_\text{color,pols} 
   \left[ \sum_\text{3 terms} \pm
          f^{abc} \;
          (\epsilon \cdot \epsilon) (\epsilon \cdot p) \right]^2 \;
   \overline{| \mat_n |^2}
      \notag \\
&= \frac{2 g_s^2}{p_a^4} \;
   \frac{1}{2 (N_c^2-1)} \;
   \sum_\text{color,pols} 
   \left[ \sum_\text{3 terms} \pm
          f^{abc} \;
          (\epsilon \cdot \epsilon) (\epsilon \cdot p) \right]^2 \;
   \overline{| \mat_n |^2} \; ,
\label{eq:gluon_split1b}
\end{alignat}
where the sums runs over all color and polarizations and over the
three terms in the brackets of Eq.\eqref{eq:gluon_split1}.  The factor
$1/2$ in the first line takes into account that for two final--state
gluons the $(n+1)$-particle phase space is only half its usual size.
Because we compute the color factor and spin sum for the decay of
gluon $a$ the formula includes averaging factors for the color
$(N_c^2-1)$ and the polarization $N_a = 2$ of the mother particle.

Inside the color and polarization sum each term $(\epsilon \cdot
\epsilon)(\epsilon \cdot p)$ is symmetric in two indices but gets
multiplied with the anti--symmetric color factor. This means that the
final result will only be finite if we square each term individually
as a product of two symmetric and two anti--symmetric terms. In other
words, the sum over the external gluons becomes an incoherent
polarization sum,
\begin{alignat}{5}
\overline{| \mat_{n+1} |^2} 
&= \frac{2 g_s^2}{p_a^2}  \;
   \frac{N_c}{2} \; 
   \sum_\text{pols} 
   \left[ \sum_\text{3 terms} \frac{(\epsilon \cdot \epsilon)^2 (\epsilon \cdot p)^2}{p_a^2}  \right] \;
   \overline{| \mat_n |^2} \; ,
\label{eq:gluon_split1c}
\end{alignat}
using $f^{abc} f^{abd} \delta^{cd} = N_c \delta^{cd} \delta^{cd} = N_c
(N_c^2-1)$.\bigskip

Going through all possible combinations we know what can contribute
inside the brackets of Eq.\eqref{eq:gluon_split1}: $(\epsilon^\|_a
\epsilon^\|_b)$ as well as $(\epsilon^\perp_a \epsilon^\perp_b)$ can
be combined with $(\epsilon^\|_c p_b)$; $(\epsilon^\|_b
\epsilon^\|_c)$ or $(\epsilon^\perp_b \epsilon^\perp_c)$ with
$(\epsilon^\|_a p_b)$; and last but not least we can combine
$(\epsilon^\|_a \epsilon^\|_c)$ and $(\epsilon^\perp_a
\epsilon^\perp_c)$ with $(\epsilon^\|_b p_c)$.  The finite
combinations between polarization vectors and momenta which we appear
in Eq.\eqref{eq:gluon_split1c} are, in terms of $z$, $E_a$, and
$\theta$
\begin{alignat}{5}
( \epsilon_c^\| p_b ) 
&= - E_b \cos \angle (\vec{\epsilon}_c^\|,\vec{p}_b )
 = - E_b \cos \left( \frac{\pi}{2} - \theta \right)
 = - E_b \sin \theta
 \simeq - E_b \theta
 = - z E_a \theta  
  \notag \\
( \epsilon_a^\| p_b ) 
&= - E_b \cos \angle (\vec{\epsilon}_a^\|,\vec{p}_b )
 = - E_b \cos \left( \frac{\pi}{2} - \theta_b \right)
 = - E_b \sin \theta_b
 \simeq - E_b \theta_b
 = - z (1-z) E_a \theta
  \notag \\
( \epsilon_b^\| p_c ) 
&= - E_c \cos \angle (\vec{\epsilon}_b^\|,\vec{p}_c )
 = - E_c \cos \left( \frac{\pi}{2} - \theta \right)
 = - E_c \sin \theta
 \simeq - E_c \theta
 = - (1-z) E_a \theta
\; .
\label{eq:pol_vec2}
\end{alignat}
For the four non--zero combinations of gluon polarizations, the
splitting matrix elements ordered still the same way are
\begin{center}
\begin{tabular}{lll|r|c}
$\epsilon_a$ & $\epsilon_b$ & $\epsilon_c$ & 
 $\pm (\epsilon \cdot \epsilon) (\epsilon \cdot p)$ & 
 $\dfrac{(\epsilon \cdot \epsilon)^2 (\epsilon \cdot p)^2}{p_a^2}
 = \dfrac{(\epsilon \cdot \epsilon)^2 (\epsilon \cdot p)^2}{z(1-z)E_a^2\theta^2}$ \\ \hline
$\|$         & $\|$         & $\|$    &&\\
$\perp$      & $\perp$      & $\|$    & 
 \raisebox{1.5ex}{$(-1) (-z) E_a \theta$} &
 \raisebox{1.5ex}{$\dfrac{z}{1-z}$} \\ \hline
$\|$         & $\|$         & $\|$    &&\\
$\|$         & $\perp$      & $\perp$ & 
 \raisebox{1.2ex}{$-(-1) (-z)(1-z) E_a \theta$} &
 \raisebox{1.2ex}{$z (1-z)$} \\ \hline
$\|$         & $\|$         & $\|$    &&\\
$\perp$      & $\|$         & $\perp$ & 
 \raisebox{1.2ex}{$-(-1) (-1) (1-z) E_a \theta$} &
 \raisebox{1.2ex}{$\dfrac{1-z}{z}$} 
\end{tabular}
\end{center}
For the incoherent sum in Eq.\eqref{eq:gluon_split1c} we find
\begin{alignat}{5}
\overline{| \mat_{n+1} |^2} 
&= \frac{2 g_s^2}{p_a^2} \; \frac{N_c}{2} \; 2
   \left[ \frac{z}{1-z} + z (1-z) + \frac{1-z}{z}  \right] \;
   \overline{| \mat_n |^2}
      \notag \\
&\equiv \frac{2 g_s^2}{p_a^2} \; \hat{P}_{g  \leftarrow g}(z) \; \overline{| \mat_n |^2}
      \notag \\
 &\Leftrightarrow \qquad 
 \boxed{
 \hat{P}_{g \leftarrow g}(z) = C_A \left[ \frac{z}{1-z} + \frac{1-z}{z} + z (1-z) \right]
 } \; ,
\label{eq:qcd_pgg}
\end{alignat}
using $C_A = N_c$\index{splitting kernel!unsubtracted $\hat{P}_{g \leftarrow g}$}. The form of the splitting kernel is symmetric
when we exchange the two gluons $z$ and $(1-z)$. It diverges if either
of the gluons become soft. The notation $\hat{P}_{i \leftarrow j} \sim
\hat{P}_{ij}$ is inspired by a matrix notation which we can use to
multiply the splitting matrix from the right with the incoming parton
vector to get the final parton vector. Following the logic described
above, with this calculation we prove that the factorized form of the
$(n+1)$-particle matrix element squared in
Eq.\eqref{eq:qcd_phase_space} holds for gluons only.\bigskip

The same kind of splitting kernel we can compute for the splitting of
a \underline{gluon into two quarks} and the splitting of a
\underline{quark into a quark and a gluon}
\begin{alignat}{5}
g(p_a) \to q(p_b) + \bar{q}(p_c) 
\qquad \qquad \text{and} \qquad \qquad 
q(p_a) \to q(p_b) + g(p_c) 
\; .
\end{alignat}
Both splittings include the quark--quark--gluon vertex, coupling the
gluon current to the quark and antiquark spinors. For small angle
scattering we can write the spinors of the massless quark $u(p_b)$ and
the massless antiquark $v(p_c)$ in terms of two-component spinors
\begin{alignat}{5}
u(p)  &= \sqrt{E} \begin{pmatrix} \chi_\pm \\ \pm \chi_\pm \end{pmatrix}
  \qqqquad \text{with} \quad 
   &\chi_+ &= \begin{pmatrix} 1 \\ \theta/2 \end{pmatrix} 
     \qquad &&\text{(spin up)}  \notag \\
  &&\chi_- &= \begin{pmatrix} - \theta/2 \\ 1 \end{pmatrix}
            &&\text{(spin down)} \; .
\label{eq:spinor_u}
\end{alignat}
For the massless antiquark we need to replace $\theta \to -\theta$
and take into account the different relative spin-momentum directions
$(\sigma \hat{p}$, leading to the additional sign in the lower two
spinor entries. The antiquark spinors then become
\begin{alignat}{9}
v(p)  &= - i \sqrt{E} \begin{pmatrix} \mp \epsilon \chi_\pm \\ \epsilon \chi_\pm \end{pmatrix}
  \qqquad \text{with} \quad 
   &\chi_+ &= \begin{pmatrix} 1 \\ -\theta/2 \end{pmatrix} \qquad 
   &\epsilon \chi_+ &= \begin{pmatrix} -\theta/2 \\ -1 \end{pmatrix} 
     \qquad &&\text{(spin up)}  \notag \\
  &&\chi_- &= \begin{pmatrix} \theta/2 \\ 1 \end{pmatrix}  \qquad 
   &\epsilon \chi_- &= \begin{pmatrix} 1 \\ -\theta/2 \end{pmatrix}
            &&\text{(spin down)} \; .
\label{eq:spinor_v}
\end{alignat}
We again limit our calculations to the leading terms in the small
scattering angle $\theta$.  In addition to the fermion spinors, for
the coupling to a gluonic current we need the \underline{Dirac
  matrices}\index{Dirac matrices} which in the Dirac representation
are conveniently expressed in terms of the Pauli matrices defined in
Eq.\eqref{eq:cov_der}
\begin{alignat}{5}
\gamma^0 = \begin{pmatrix} \one & 0 \\ 0 & -\one \end{pmatrix}
\qqquad 
\gamma^j = \begin{pmatrix} 0 & \tau^j \\ -\tau^j & 0 \end{pmatrix}
\qquad \Rightarrow \qquad 
\gamma^0 \gamma^0 = \one
\qqquad 
\gamma^0 \gamma^j =  \begin{pmatrix} 0 & \tau^j \\ \tau^j & 0 \end{pmatrix}
\label{eq:dirac_matrices}
\end{alignat}
We are particularly interested in the combination $\gamma^0 \gamma^j$
because of the definition of the conjugated spinor $\bar{u} = u^T
\gamma^0$.\bigskip

In the notation introduced in Eq.\eqref{eq:qcd_pgg} we first compute
the splitting kernel $\hat{P}_{q \leftarrow g}$, sandwiching the $qqg$
vertex between an outgoing quark $\bar{u}_\pm(p_b)$ and an outgoing
antiquark $v_\pm(p_a)$ for all possible spin combinations. We start
with all four gluon polarizations, \ie all four gamma matrices,
between two spin-up quarks and their spinors written out in
Eqs.\eqref{eq:spinor_u} and~\eqref{eq:spinor_v}
\begin{alignat}{5}
\frac{\bar{u}_+(p_b) \gamma^0 v_-(p_c)}{-i \sqrt{E_b} \sqrt{E_c}}
&=
  \left( 1, \frac{\theta_b}{2}, 1, \frac{\theta_b}{2} \right) 
  \begin{pmatrix} 1&&& \\ &1&& \\ &&1& \\ &&&1 \end{pmatrix}
  \begin{pmatrix} 1 \\ -\theta_c/2 \\ 1 \\ -\theta_c/2 \end{pmatrix}
&=&  
  \left( 1, \frac{\theta_b}{2}, 1, \frac{\theta_b}{2} \right) 
  \begin{pmatrix} 1 \\ -\theta_c/2 \\ 1 \\ -\theta_c/2 \end{pmatrix}
 = 2
\notag \\
\frac{\bar{u}_+(p_b) \gamma^1 v_-(p_c)}{-i \sqrt{E_b} \sqrt{E_c}}
&= 
  \left( 1, \frac{\theta_b}{2}, 1, \frac{\theta_b}{2} \right) 
  \begin{pmatrix} &&&1 \\ &&1& \\ &1&& \\ 1&&& \end{pmatrix}
  \begin{pmatrix} 1 \\ -\theta_c/2 \\ 1 \\ -\theta_c/2 \end{pmatrix}
&=&  
  \left( 1, \frac{\theta_b}{2}, 1, \frac{\theta_b}{2} \right) 
  \begin{pmatrix} -\theta_c/2 \\ 1 \\ -\theta_c/2 \\ 1 \end{pmatrix}
 = \theta_b - \theta_c
\notag \\
\frac{\bar{u}_+(p_b) \gamma^2 v_-(p_c)}{-i \sqrt{E_b} \sqrt{E_c}}
&=
  \left( 1, \frac{\theta_b}{2}, 1, \frac{\theta_b}{2} \right) 
  \begin{pmatrix} &&&-i \\ &&i& \\ &-i&& \\ i&&& \end{pmatrix}
  \begin{pmatrix} 1 \\ -\theta_c/2 \\ 1 \\ -\theta_c/2 \end{pmatrix}
&=& 
  i \left( 1, \frac{\theta_b}{2}, 1, \frac{\theta_b}{2} \right) 
  \begin{pmatrix} \theta_c/2 \\ 1 \\ \theta_c/2 \\ 1 \end{pmatrix}
 = i ( \theta_b + \theta_c) 
\notag \\
\frac{\bar{u}_+(p_b) \gamma^3 v_-(p_c)}{-i \sqrt{E_b} \sqrt{E_c}}
&=
  \left( 1, \frac{\theta_b}{2}, 1, \frac{\theta_b}{2} \right) 
  \begin{pmatrix} &&1& \\ &&&-1 \\ 1&&& \\ &-1&& \end{pmatrix}
  \begin{pmatrix} 1 \\ -\theta_c/2 \\ 1 \\ -\theta_c/2 \end{pmatrix}
&=&  
  \left( 1, \frac{\theta_b}{2}, 1, \frac{\theta_b}{2} \right) 
  \begin{pmatrix} 1 \\ \theta_c/2 \\ 1 \\ \theta_c/2 \end{pmatrix}
 = 2 \; .
\end{alignat}
Somewhat surprisingly the unphysical scalar and longitudinal gluon
polarizations seem to contribute to this vertex. However, after adding
the two unphysical degrees of freedom they cancel because of the form
of our metric. Assuming transverse gluons we compute this vertex factor
also for the other diagonal spin combination
\begin{alignat}{5}
\frac{\bar{u}_-(p_b) \gamma^1 v_+(p_c)}{-i \sqrt{E_b} \sqrt{E_c}}
&=
  \left( -\frac{\theta_b}{2}, 1, \frac{\theta_b}{2}, -1 \right) 
  \begin{pmatrix} &&&1 \\ &&1& \\ &1&& \\ 1&&& \end{pmatrix}
  \begin{pmatrix} \theta_c/2 \\ 1 \\ -\theta_c/2 \\ -1 \end{pmatrix}
&=&  
  \left( -\frac{\theta_b}{2}, 1, \frac{\theta_b}{2}, -1 \right) 
  \begin{pmatrix} -1 \\ -\theta_c/2 \\ 1 \\ \theta_c/2 \end{pmatrix}
 = \theta_b - \theta_c
\notag \\
\frac{\bar{u}_-(p_b) \gamma^2 v_+(p_c)}{-i \sqrt{E_b} \sqrt{E_c}}
&=
  \left( -\frac{\theta_b}{2}, 1, \frac{\theta_b}{2}, -1 \right) 
  \begin{pmatrix} &&&-i \\ &&i& \\ &-i&& \\ i&&& \end{pmatrix}
  \begin{pmatrix} \theta_c/2 \\ 1 \\ -\theta_c/2 \\ -1 \end{pmatrix}
&=& 
  i \left( -\frac{\theta_b}{2}, 1, \frac{\theta_b}{2}, -1 \right) 
  \begin{pmatrix} 1 \\ -\theta_c/2 \\ -1 \\ \theta_c/2 \end{pmatrix}
 = -i (\theta_b + \theta_c) \; .
\end{alignat}
Before collecting the prefactors for this gluon--quark splitting, we
also need the same--spin case
\begin{alignat}{5}
\frac{\bar{u}_+(p_b) \gamma^1 v_+(p_c)}{-i \sqrt{E_b} \sqrt{E_c}}
&=
  \left( 1, \frac{\theta_b}{2}, 1, \frac{\theta_b}{2} \right) 
  \begin{pmatrix} &&&1 \\ &&1& \\ &1&& \\ 1&&& \end{pmatrix}
  \begin{pmatrix} \theta_c/2 \\ 1 \\ -\theta_c/2 \\ -1 \end{pmatrix}
&=&  
  \left( 1, \frac{\theta_b}{2}, 1, \frac{\theta_b}{2} \right) 
  \begin{pmatrix} -1 \\ -\theta_c/2 \\ 1 \\ \theta_c/2 \end{pmatrix}
 = 0
\notag \\
\frac{\bar{u}_+(p_b) \gamma^2 v_+(p_c)}{-i \sqrt{E_b} \sqrt{E_c}}
&=
  \left( 1, \frac{\theta_b}{2}, 1, \frac{\theta_b}{2} \right) 
  \begin{pmatrix} &&&-i \\ &&i& \\ &-i&& \\ i&&& \end{pmatrix}
  \begin{pmatrix} \theta_c/2 \\ 1 \\ -\theta_c/2 \\ -1 \end{pmatrix}
&=& 
  i \left( 1, \frac{\theta_b}{2}, 1, \frac{\theta_b}{2} \right) 
  \begin{pmatrix} 1 \\ -\theta_c/2 \\ -1 \\ \theta_c/2 \end{pmatrix}
 = 0 \; ,
\end{alignat}
which vanishes.  The gluon current can only couple to two fermions via
a spin flip. For massless fermions this means that the gluon splitting
into two quarks involves two quark spin cases, each of them coupling
to two transverse gluon polarizations.  Keeping track of all the
relevant factors our vertex function for the splitting $g \to q
\bar{q}$ becomes for each of the two quark spins
\begin{alignat}{5}
V_{qqg}
   & = -i g_s T^a \; \bar{u}_\pm(p_b) \gamma_\mu \epsilon_a^\mu v_\mp(p_c)
     \equiv -i g_s T^a \; \epsilon_a^j \; F_\pm^{(j)}
\qqquad \text{for} \quad j=1,2
\notag \\
&\frac{|F^{(1)}_+|^2}{p_a^2}
=\frac{|F^{(1)}_-|^2}{p_a^2}
   = \frac{E_b E_c ( \theta_b - \theta_c )^2}{p_a^2}
   = \frac{E_a^2 z (1-z) (1-z-z)^2 \theta^2}{E_a^2 z (1-z) \theta^2}
   = (1-2z)^2
\notag \\
&\frac{|F^{(2)}_+|^2}{p_a^2}
=\frac{|F^{(2)}_-|^2}{p_a^2}
   = \frac{E_b E_c ( \theta_b + \theta_c )^2}{p_a^2}
   = \frac{E_a^2 z (1-z) (1-z+z)^2 \theta^2}{E_a^2 z (1-z) \theta^2}
   = 1 \; .
\end{alignat}
We omit irrelevant factors $i$ and $(-1)$ which drop out once we
compute the absolute value squared. In complete analogy to the gluon
splitting case we can factorize the $(n+1)$-particle matrix element
into
\begin{alignat}{5}
\overline{| \mat_{n+1} |^2} 
&= \left( \frac{1}{p_a^2} \right)^2  \;
   g_s^2 \; \frac{\tr T^a T^a}{N_c^2-1} \;
   \frac{1}{N_a} \;
   \left[  |F^{(1)}_+|^2 + |F^{(1)}_-|^2 + |F^{(2)}_+|^2 + |F^{(2)}_-|^2 \right] \;
   \overline{| \mat_n |^2}
      \notag \\
&= \frac{g_s^2}{p_a^2} \; T_R \frac{N_c^2-1}{N_c^2-1} \;
   \left[  (1-2z)^2 + 1 \right] \;
   \overline{| \mat_n |^2}
   \qqquad \text{with} \; \tr T^a T^b = T_R \delta^{ab} \; 
           \text{and} \; N_a = 2
      \notag \\
&= \frac{2 g_s^2}{p_a^2} \; T_R \;
   \left[  z^2 + (1-z)^2 \right] \;
   \overline{| \mat_n |^2}
      \notag \\
&\equiv \frac{2 g_s^2}{p_a^2} \hat{P}_{q  \leftarrow g}(z) \;
        \overline{| \mat_n |^2}
      \notag \\
 &\Leftrightarrow \qquad 
 \boxed{
 \hat{P}_{q \leftarrow g}(z) = T_R \left[ z^2 + (1-z)^2 \right]
 } \; ,
\label{eq:qcd_pqg}
\end{alignat}
with $T_R = 1/2$\index{splitting kernel!unsubtracted $\hat{P}_{q  \leftarrow g}$}. In the first line we implicitly assume that the
internal quark propagator can be written as something like $u
\bar{u}/p_a^2$\index{propagator!spinor} and we only need to consider the denominator.
This splitting kernel is again symmetric in $z$ and $(1-z)$ because
QCD does not distinguish between the outgoing quark and the outgoing
antiquark.\bigskip

The third splitting we compute is gluon radiation off a quark, \ie
$q(p_a) \to q(p_b) + g(p_c)$, sandwiching the $qqg$ vertex between an
outgoing quark $\bar{u}_\pm(p_b)$ and an incoming quark
$u_\pm(p_a)$. From the splitting of a gluon into a quark--antiquark
pair we already know that we can limit our analysis to the physical
gluon polarizations and a spin flip in the quarks. Inserting the
spinors from Eq.\eqref{eq:spinor_u} and the two relevant gamma
matrices gives us
\begin{alignat}{5}
\frac{\bar{u}_+(p_b) \gamma^1 u_+(p_a)}{E}
&=
  \left( 1, \frac{\theta_b^*}{2}, 1, \frac{\theta_b^*}{2} \right) 
  \begin{pmatrix} &&&1 \\ &&1& \\ &1&& \\ 1&&& \end{pmatrix}
  \begin{pmatrix} 1 \\ \theta_a^*/2 \\ 1 \\ \theta_a^*/2 \end{pmatrix}
&=&  
  \left( 1, \frac{\theta_b^*}{2}, 1, \frac{\theta_b^*}{2} \right) 
  \begin{pmatrix} \theta_a^*/2 \\ 1 \\ \theta_a^*/2 \\ 1 \end{pmatrix}
 = \theta_a^* + \theta_b^*
\notag \\
\frac{\bar{u}_+(p_b) \gamma^2 u_+(p_a)}{E}
&=
  \left( 1, \frac{\theta_b^*}{2}, 1, \frac{\theta_b^*}{2} \right) 
  \begin{pmatrix} &&&-i \\ &&i& \\ &-i&& \\ i&&& \end{pmatrix}
  \begin{pmatrix} 1 \\ \theta_a^*/2 \\ 1 \\ \theta_a^*/2 \end{pmatrix}
&=& 
  i \left( 1, \frac{\theta_b^*}{2}, 1, \frac{\theta_b^*}{2} \right) 
  \begin{pmatrix} - \theta_a^*/2 \\ 1 \\ -\theta_a^*/2 \\ 1 \end{pmatrix}
 = i ( \theta_b^* - \theta_a^*) \; ,
\end{alignat}
with the angles $\theta_b^*$ and $\theta_a^*$ relative to the final
state gluon direction $\vec{p}_c$. Comparing to the situation shown in
Figure~\ref{fig:qcd_gsplit} for the angle relative to the scattered
gluon we now find $\theta_b^* = \theta$ while for the incoming quark
$\theta_a^* = - \theta_c = -z \theta$. The spin--down
case gives the same result, modulo a complex conjugation
\begin{alignat}{5}
\frac{\bar{u}_-(p_b) \gamma^1 u_-(p_a)}{\sqrt{E_b} \sqrt{E_a}}
&=
  \left( -\frac{\theta_b^*}{2}, 1, \frac{\theta_b^*}{2}, -1 \right) 
  \begin{pmatrix} &&&1 \\ &&1& \\ &1&& \\ 1&&& \end{pmatrix}
  \begin{pmatrix} -\theta_a^*/2 \\ 1 \\ \theta_a^*/2 \\ -1 \end{pmatrix}
&=&  
  \left( -\frac{\theta_b^*}{2}, 1, \frac{\theta_b^*}{2}, -1 \right) 
  \begin{pmatrix} -1 \\ \theta_a^*/2 \\ 1 \\ -\theta_a^*/2 \end{pmatrix}
 = \theta_a^* + \theta_b^*
\notag \\
\frac{\bar{u}_-(p_b) \gamma^2 u_-(p_a)}{\sqrt{E_b} \sqrt{E_a}}
&=
  \left( -\frac{\theta_b^*}{2}, 1, \frac{\theta_b^*}{2}, -1 \right) 
  \begin{pmatrix} &&&-i \\ &&i& \\ &-i&& \\ i&&& \end{pmatrix}
  \begin{pmatrix} -\theta_a^*/2 \\ 1 \\ \theta_a^*/2 \\ -1 \end{pmatrix}
&=& 
  i \left( -\frac{\theta_b^*}{2}, 1, \frac{\theta_b^*}{2}, -1 \right) 
  \begin{pmatrix} 1 \\ \theta_a^*/2 \\ -1 \\ -\theta_a^*/2 \end{pmatrix}
 = i ( \theta_a^* - \theta_b^* ) \; .
\end{alignat}
In terms of $\theta$ the two combinations of angles become $\theta_a^*
+ \theta_b^* = \theta (1-z)$ and $\theta_a^* - \theta_b^* = \theta
(-z-1)$. The vertex function for gluon radiation off a quark then reads
\begin{alignat}{5}
V_{qqg}
   & = -i g_s T^a \; \bar{u}_\pm(p_b) \gamma_\mu \epsilon_a^\mu u_\pm(p_c)
     \equiv -i g_s T^a \; \epsilon_a^j \; F_\pm^{(j)}
\qqquad \text{for} \quad j=1,2
\notag \\
&\frac{|F^{(1)}_+|^2}{p_a^2}
=\frac{|F^{(1)}_-|^2}{p_a^2}
   = \frac{E_a E_b (\theta_a^* + \theta_b^*)^2}{p_a^2}
   = \frac{E_a^2 z (z-1)^2 \theta^2}{E_a^2 z (1-z) \theta^2}
   = (1-z)
\notag \\
&\frac{|F^{(2)}|^2_+}{p_a^2}
=\frac{|F^{(2)}|^2_-}{p_a^2}
   = \frac{E_a E_b (\theta_b^* - \theta_a^*)^2}{p_a^2}
   = \frac{E_a^2 z (1+z)^2 \theta^2}{E_a^2 z (1-z) \theta^2}
   = \frac{(1+z)^2}{1-z} \; ,
\end{alignat}
again dropping irrelevant prefactors. The factorized matrix element
for this channel has the same form as Eq.\eqref{eq:qcd_pqg}, except
for the color averaging factor of the now incoming quark,
\begin{alignat}{5}
\overline{| \mat_{n+1} |^2} 
&= \left( \frac{1}{p_a^2} \right)^2  \;
   g_s^2 \; \frac{\tr T^a T^a}{N_c} \;
   \frac{1}{N_a} \;
   \left[  |F^{(1)}_+|^2 + |F^{(1)}_-|^2 + |F^{(2)}_+|^2 + |F^{(2)}_-|^2 \right] \;
   \overline{| \mat_n |^2}
      \notag \\
&= \frac{g_s^2}{p_a^2} \; \frac{N_c^2-1}{2N_c} \;
   \frac{(1+z)^2 + (1-z)^2}{1-z} \;
   \overline{| \mat_n |^2}
      \notag \\
&= \frac{2 g_s^2}{p_a^2} \; C_F \;
   \frac{1+z^2}{1-z} \;
   \overline{| \mat_n |^2}
      \notag \\
&\equiv \frac{2 g_s^2}{p_a^2} \hat{P}_{q  \leftarrow g}(z) \;
        \overline{| \mat_n |^2}
      \notag \\
 &\Leftrightarrow \qquad 
 \boxed{
 \hat{P}_{q \leftarrow q}(z) = C_F \frac{1+z^2}{1-z}
 } \; .
\label{eq:qcd_pqq}
\end{alignat}
The color factor for gluon radiation off a quark is $C_F = (N^2-1)/(2N)$.
The averaging factor $1/N_a=2$ now is the number of quark spins in the
intermediate state\index{splitting kernel!unsubtracted $\hat{P}_{q  \leftarrow q}$}. Just switching $z \leftrightarrow (1-z)$ we can
read off the kernel for a quark splitting written in terms of the
final--state gluon
\begin{alignat}{5}
 \boxed{
 \hat{P}_{g \leftarrow q}(z) = C_F \frac{1+(1-z)^2}{z}
 } \; .
\label{eq:qcd_pgq}
\end{alignat}
This result\index{splitting kernel!unsubtracted $\hat{P}_{g \leftarrow q}$} finalizes our calculation of \underline{all QCD splitting
  kernels} $\hat{P}_{i \leftarrow j}(z)$ between quarks and gluons
. As alluded to earlier, similar to ultraviolet divergences which get
removed by counter terms these splitting kernels are universal.
They do not depend on the hard $n$-particle matrix element which is
part of the original ($n+1$)-particle process. We show all four results
in Eqs.\eqref{eq:qcd_pgg}, \eqref{eq:qcd_pqg},
\eqref{eq:qcd_pqq}, and~\eqref{eq:qcd_pgq}. This means that by
construction of the kernels $\hat{P}$ we have shown that the collinear
factorization Eq.\eqref{eq:qcd_factorize} holds at this level in
perturbation theory.\index{factorization}\bigskip

Before using this splitting property to describe QCD effects at the
LHC we need to look at the splitting of partons in the initial state,
meaning $|p_a^2|, p_c^2 \ll |p_b^2|$ where $p_b$ is the momentum
entering the hard interaction. The difference to the final--state
splitting is that now we can consider the split parton momentum $p_b =
p_a - p_c$ as a $t$-channel diagram, so we already know $p_b^2 = t <0$
from our usual Mandelstam variables argument.  This
\underline{space--like splitting}\index{splitting!space--like branching}
version of Eq.\eqref{eq:qcd_sudakovdec} for $p_b^2$ gives us
\begin{alignat}{5}
t \equiv p_b^2 
&= (-z p_a + \beta n + p_T )^2 
\notag \\
&= p_T^2 - 2 z \beta (p_a n) 
      \qqquad &&\text{with} \; p_a^2 = n^2 = (p_a p_T) = (n p_T) = 0
\notag \\
&= p_T^2 + \frac{p_T^2 z}{1-z} 
      \qqquad &&\text{using Eq.\eqref{eq:qcd_sudakovdec_solve}}
\notag \\
&= \frac{p_T^2}{1-z}
 = -\frac{p_{T,1}^2+p_{T,2}^2}{1-z} <0 \; .
\label{eq:qcd_sudakovdec2}
\end{alignat}
The calculation of the splitting kernels and matrix elements is the
same as for the time--like case, with the one exception that for
splitting in the initial state the flow factor has to be evaluated at
the reduced partonic energy $E_b = z E_a$ and that the energy fraction
entering the parton density needs to be replaced by $x_b \to z
x_b$. The factorized matrix element for initial--state splitting then
reads just like Eq.\eqref{eq:qcd_factorize}
\begin{alignat}{5}
\sigma_{n+1}
 = \int \sigma_n \; 
   \frac{d t}{t} dz \; 
   \frac{\alpha_s}{2 \pi} \; \hat{P}(z) \; .
\label{eq:qcd_factorize2}
\end{alignat}
How to use this property to make statements about the quark and gluon
content in the proton will be the focus of the next section.

\subsubsection{DGLAP equation}
\label{sec:qcd_dglap}

We can use everything we now know about collinear parton splitting  
to describe incoming partons at hadron colliders. For example in $pp
\to Z$ production incoming partons inside the protons 
transform into each other via collinear splitting until they enter the
$Z$ production process as quarks.  Taking Eq.\eqref{eq:qcd_factorize2}
seriously, the parton density we insert into Eq.\eqref{eq:qcd_sigtot}
depends on two parameters, the final energy fraction and the
virtuality $f(x_n,-t_n)$. The second parameter $t$ is new 
compared to the purely probabilistic picture in
Eq.\eqref{eq:qcd_sigtot}. However, it cannot be neglected unless we 
convince ourselves that it is unphysical. As we will see later it
corresponds exactly to the artificial renormalization scale which
appears when we resum the scaling logarithms which appear in counter
terms.\bigskip

More quantitatively, we start with a quark inside the proton with an
energy fraction $x_0$, as it enters the hadronic phase space integral
shown in Section~\ref{sec:qcd_kinematics}. Since this quark is
confined inside the proton it can only have small transverse momentum,
which means its four-momentum squared $t_0$ is negative and its
absolute value $|t_0|$ is small. The variable $t$ we call
\underline{virtuality}\index{virtuality}. For the incoming
partons which if on--shell have $p^2=0$ it gives the distance to the
mass shell.  Let us simplify our kinematic argument by assuming that
there exists only one splitting, namely successive gluon radiation
off an incoming quark, where the outgoing gluons are not relevant
\begin{center}
\begin{fmfgraph*}(350,50)
 \fmfset{arrow_len}{2mm}
 \fmfleft{in1}
 \fmfbottom{in2}
 \fmf{fermion,width=0.5}{v6,in2}
 \fmf{fermion,width=0.5,lab.side=left,label=$(x_0,,t_0)$}{in1,v1}
 \fmf{fermion,width=0.5,lab.side=left,label=$(x_1,,t_1)$}{v1,v2}
 \fmf{fermion,width=0.5}{v2,v3}
 \fmf{dashes,width=0.5}{v3,v4}
 \fmf{fermion,width=0.5}{v4,v5}
 \fmf{fermion,width=0.5,lab.side=left,label=$(x_n,,t_n)$}{v5,v6}
 \fmf{photon,width=0.5,lab.side=left,tension=1.8}{v6,v7}
 \fmf{fermion,width=0.5,tension=1.5}{out1,v7}
 \fmf{fermion,width=0.5,tension=1.5}{v7,out2}
 \fmfright{out1,out2}
\end{fmfgraph*}
\end{center}
In that case each collinear gluon radiation will decrease the quark
energy $x_{j+1} < x_j$ and increase its virtuality
$|t_{j+1}| = -t_{j+1} > -t_j = |t_j|$ through its recoil.\bigskip

From the last section we know what the successive splitting means in
terms of splitting probabilities. We can describe how the parton
density $f(x,-t)$ evolves in the $(x-t)$ plane as depicted in
Figure~\ref{fig:qcd_tx}.  The starting point $(x_0,t_0)$ is at
least probabilistically given by the energy and kind of the hadron,
for example the proton.  For a given small virtuality $|t_0|$ we start
at some kind of fixed $x_0$ distribution. We then interpret each
branching as a step strictly downward in $x_j \to x_{j+1}$ where the
$t$ value we assign to this step is the ever increasing virtuality
$|t_{j+1}|$ after the branching.  Each splitting means a synchronous
shift in $x$ and $t$, so the actual path in the $(x-t)$ plane really
consists of discrete points.  The probability of such a splitting to
occur is given by $\hat{P}_{q \leftarrow q}(z) \equiv \hat{P}(z)$ as
it appears in Eq.\eqref{eq:qcd_factorize2}\index{splitting kernel}
\begin{alignat}{5}
\frac{\alpha_s}{2\pi} \; \hat{P}(z) \; \frac{dt}{t} \; dz \; .
\end{alignat}
In this picture we consider this probability a smooth function in $t$
and $z$. At the end of the path we will probe this evolved parton
density, where $x_n$ and $t_n$ enter the hard scattering
process and its energy--momentum conservation.\bigskip

\begin{figure}[t]
\begin{center}
\includegraphics[width=0.35\hsize]{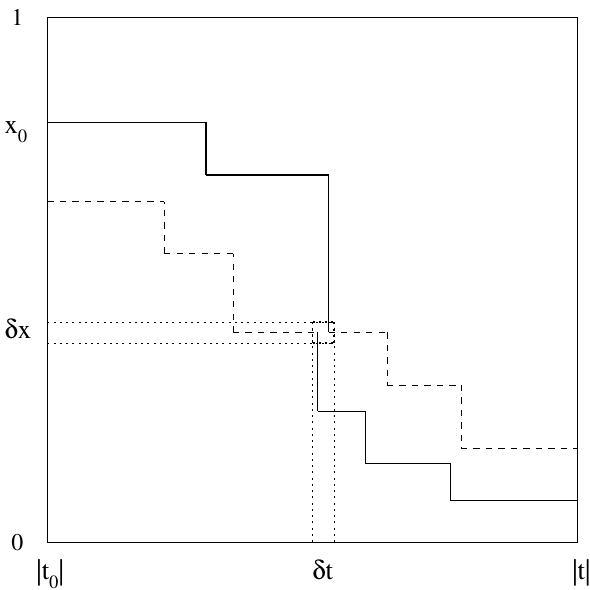}
\end{center}
\caption{Path of an incoming parton in the $(x-t)$ plane. Because we
  define $t$ as a negative number its axis is labelled $|t|$.}
\label{fig:qcd_tx}
\end{figure}

When we convert a partonic into a hadronic cross section numerically
we need to specify the probability of the parton density $f(x,-t)$
residing in an infinitesimal square $[x_j,x_j+\delta x]$ and, if this
second parameter has anything to do with physics, $[|t_j|,|t_j|+\delta
  t]$. Using our $(x,t)$ plane we compute the flows into this square
and out of this square, which together define the net shift in $f$ in
the sense of a differential equation, similar to the derivation of
Gauss' theorem for vector fields inside a surface
\begin{alignat}{5}
\delta f_\text{in} - \delta f_\text{out} = \delta f(x,-t) \; .
\end{alignat}
We compute the incoming and outgoing flows from the history of the
$(x,t)$ evolution. At this stage our picture becomes a little subtle;
the way we define the path between two splittings in
Figure~\ref{fig:qcd_tx} it can enter and leave the square either
vertically or horizontally. Because we do not consider the movement in
the $(x,t)$ plane continuous we can choose this direction as
vertical or horizontal. Because we want to arrive at a differential
equation in $t$ we choose the vertical drop, such that the area the
incoming and outgoing flows see is given by $\delta t$.  If we define
a splitting as such a vertical drop in $x$ at the target value
$t_{j+1}$ an incoming path hitting the square at some value $t$ can
come from any $x$ value above the square. Using this convention and
following the fat solid lines in Figure~\ref{fig:qcd_tx} the vertical
flow into (and out of) the square $(x,t)$ square is proportional to
$\delta t$
\begin{alignat}{5}
\delta f_\text{in}(-t) 
&= \delta t \; \left( \frac{\alpha_s \hat{P}}{2\pi t} 
                \otimes f \right)(x,-t)
= \frac{\delta t}{t} 
   \int_x^1 \frac{dz}{z} \; \frac{\alpha_s}{2\pi} \;
   \hat{P}(z) f\left(\frac{x}{z},-t\right) 
  \notag \\
&= \frac{\delta t}{t} 
   \int_0^1 \frac{dz}{z} \; \frac{\alpha_s}{2\pi} \;
   \hat{P}(z) f\left(\frac{x}{z},-t\right) 
   \qquad \text{assuming} \; f(x',-t)=0 \; \text{for} \; x' > 1 \; ,
\label{eq:flow_in}
\end{alignat}
where $\delta t$ is the size of the interval covered by the virtuality
value $t$. We use the definition of a \underline{convolution}\index{convolution}
\begin{alignat}{5}
(f \otimes g)(x) 
= \int_0^1 dx_1 dx_2 f(x_1) g(x_2) \; \delta(x - x_1 x_2)
= \int_0^1 \frac{dx_1}{x_1} f(x_1) g\left( \frac{x}{x_1} \right)
= \int_0^1 \frac{dx_2}{x_2} f\left( \frac{x}{x_2} \right) g(x_2) \; .
\end{alignat}
The outgoing flow we define in complete analogy, again leaving the
infinitesimal square vertically. Following the fat solid line in
Figure~\ref{fig:qcd_tx} it is also proportional to
$\delta t$
\begin{alignat}{5}
\delta f_\text{out}(-t) 
&= \delta t \; \int_0^1 dy \frac{\alpha_s \hat{P}(y)}{2 \pi t} \; f(x,-t)
 = \frac{\delta t}{t} f(x,-t) 
   \int_0^1 dy \; \frac{\alpha_s}{2\pi} \; \hat{P}(y)  \; .
\label{eq:flow_out}
\end{alignat}
The $y$ integration, unlike the $z$ integration for the incoming flow
is not a convolution. This integration appears because we do
not know the normalization of $\hat{P}(z)$ distribution which we
interpret as a probability. The reason why it is not a convolution is
that for the outgoing flow we know the starting condition and 
integrate over the final configurations; this aspect will become
important later.  Combining Eq.\eqref{eq:flow_in} and
Eq.\eqref{eq:flow_out} we can compute the change in the parton density
of the quarks as
\begin{alignat}{5}
\delta f(x,-t) 
&= \frac{\delta t}{t} \; \left[ 
   \int_0^1 \frac{dz}{z} \; \frac{\alpha_s}{2\pi} \; \hat{P}(z) \; 
            f\left(\frac{x}{z},-t\right) 
 - \int_0^1 dy  \; \frac{\alpha_s}{2\pi} \; \hat{P}(y) \; 
            f(x,-t) 
   \right]
 \notag \\
&= \frac{\delta t}{t}
   \int_0^1 \frac{dz}{z}  \; \frac{\alpha_s}{2\pi} \;
   \left[ \hat{P}(z) - \delta(1-z) \int_0^1 dy \hat{P}(y) 
   \right] \; f\left(\frac{x}{z},-t\right) 
 \notag \\
&\equiv \frac{\delta t}{t}
   \int_x^1 \frac{dz}{z}  \; \frac{\alpha_s}{2\pi} \;
   \hat{P}(z)_+ \; f\left(\frac{x}{z},-t\right) 
 \notag \\
\Leftrightarrow \qquad
\frac{\delta f(x,-t)}{\delta (-t)}
&= \frac{1}{(-t)}
   \int_x^1 \frac{dz}{z}  \; \frac{\alpha_s}{2\pi} \;
   \hat{P}(z)_+ \; f\left(\frac{x}{z},-t\right) \; ,
\label{eq:qcd_almost_dglap}
\end{alignat}
again assuming $f(x)=0$ for $x>1$, strictly speaking requiring
$\alpha_s$ to only depend on $t$ but not on $z$, and using the
specifically defined \underline{plus subtraction}\index{plus subtraction} scheme
\begin{alignat}{5}
 \boxed{F(z)_+ \equiv 
  F(z) - \delta(1-z)  \int_0^1 dy \; F(y) }
\qquad \text{or} \qquad 
 \int_0^1 dz \; \frac{f(z)}{(1-z)_+} =
  \int_0^1 dz \; \left( \frac{f(z)}{1-z} - \frac{f(1)}{1-z} \right) \; .
\label{eq:plus_subtraction}
\end{alignat}
For the second term we
choose $F(z) = 1/(1-z)$, multiply it with an arbitrary test
function $f(z)$ and integrate over $z$. 
In contrast to the original $z$ integral the
plus--subtracted integral is by definition finite in the limit $z \to
1$, where some of the splitting kernels diverge. For example, the quark
splitting kernel including the plus prescription becomes 
$C_F ((1+z^2)/(1-z))_+$. At this stage the plus prescription is simply
a convenient way of writing a complicated combination of splitting
kernels, but we will see that it also has a physics meaning.\bigskip

Next, we check that the plus prescription indeed acts as a
regularization technique for the parton densities. Obviously, the
integral over $f(z)/(1-z)$ is divergent at the boundary $z \to 1$,
which we know we can cure using \underline{dimensional
  regularization}\index{dimensional regularization}. The special case $f(z)=1$ illustrates how
  dimensional regularization of infrared divergences in the phase
  space integration Eq.\eqref{eq:qcd_phase} works
\begin{alignat}{5}
\int_0^1 dz \; \frac{1}{(1-z)^{1-\epsilon}}
= \int_0^1 dz \; \frac{1}{z^{1-\epsilon}}
= \frac{z^\epsilon}{\epsilon} \Bigg|_0^1
= \frac{1}{\epsilon} 
\qqqquad \text{with} \; \epsilon > 0 \; ,
\end{alignat}
for $4+2 \epsilon$ dimensions. This change in sign avoids the analytic
continuation of the usual value $n=4-2\epsilon$ to $\epsilon <
0$. The dimensionally regularized integral we can write as
\begin{alignat}{5}
\int_0^1 dz \; \frac{f(z)}{(1-z)^{1-\epsilon}}
=&  \int_0^1 dz \; \frac{f(z)-f(1)}{(1-z)^{1-\epsilon}}
 + f(1) \int_0^1 dz \; \frac{1}{(1-z)^{1-\epsilon}}
\notag \\
=&  \int_0^1 dz \; \frac{f(z)-f(1)}{1-z}
    \left( 1 + \ope(\epsilon) \right) 
 + \frac{f(1)}{\epsilon} 
\notag \\
=&  \int_0^1 dz \; \frac{f(z)}{(1-z)_+}
    \left( 1 + \ope(\epsilon) \right) 
 + \frac{f(1)}{\epsilon} 
\qqquad \text{by definition}
\notag \\
\Leftrightarrow \qquad 
  \int_0^1 dz \; \frac{f(z)}{(1-z)^{1-\epsilon}}
- \frac{f(1)}{\epsilon} 
=& \int_0^1 dz \; \frac{f(z)}{(1-z)_+}
    \left( 1 + \ope(\epsilon) \right) \; .
\label{eq:plus_dimreg}
\end{alignat}
The dimensionally regularized integral minus the pole, \ie the
finite part of the dimensionally regularized integral, is the same as
the plus--subtracted integral modulo terms of the order $\epsilon$.
The third line in Eq.\eqref{eq:plus_dimreg} shows that the difference
between a dimensionally regularized splitting kernel and a
plus--subtracted splitting kernel manifests itself as terms
proportional to $\delta(1-z)$. Physically, they represent
contributions to a soft--radiation phase space integral.\bigskip

Before we move on introducing a gluon density we can slightly
reformulate the splitting kernel $\hat{P}_{q \leftarrow q}$ in
Eq.\eqref{eq:qcd_pqq}.  If the plus prescription regularizes the pole
at $z \to 1$, what happens when we include the numerator of the
regularized function, \eg the quark splitting kernel?
The finite difference between these results is
\begin{alignat}{5}
\left( \frac{1+z^2}{1-z} \right)_+
- (1+z^2) \; \left( \frac{1}{1-z}\right)_+ 
&= 
 \frac{1+z^2}{1-z} 
 - \delta(1-z) \int_0^1 dy \; \frac{1+y^2}{1-y} 
 - \frac{1+z^2}{1-z} + \delta(1-z) \int_0^1 dy \; \frac{1+z^2}{1-y} 
\notag \\
& = 
 - \delta(1-z) \int_0^1 dy \; 
 \left( \frac{1+y^2}{1-y} - \frac{2}{1-y}  \right) 
\notag \\
& = 
   \delta(1-z) \int_0^1 dy \; 
   \frac{y^2-1}{y-1} 
  = 
   \delta(1-z) \int_0^1 dy \;  (y+1)
  = 
   \frac{3}{2} \delta(1-z) \; .
\end{alignat}
We can therefore write the quark's splitting kernel in two equivalent
ways
\begin{alignat}{5}
\boxed{
P_{q \leftarrow q}(z) = 
C_F \left( \frac{1+z^2}{1-z} \right)_+ = 
C_F \left[ \frac{1+z^2}{(1-z)_+} 
                               + \frac{3}{2} \delta(1-z) \right]
} \; .
\label{eq:plus_pqq}
\end{alignat}
\index{splitting kernel!subtracted $P_{q  \leftarrow q}$} \bigskip

The infinitesimal version of Eq.\eqref{eq:qcd_almost_dglap} is the
Dokshitzer--Gribov--Lipatov--Altarelli--Parisi or DGLAP
\underline{integro-differential equation}\index{DGLAP equation} which describes the scale
dependence of the quark parton density. As we already know quarks do
not only appear in $q \to q$ splitting, but also in gluon
splitting. Therefore, we generalize Eq.\eqref{eq:qcd_almost_dglap} to
include the full set of QCD partons, \ie quarks and gluons. This
generalization involves a sum over all allowed splittings and the
plus--subtracted splitting kernels. For the quark density on the left
hand side it is
\begin{alignat}{5}
\boxed{
\frac{d f_q(x,-t)}{d \log (-t)} 
= -t \; \frac{d f_q(x,-t)}{d (-t)} 
= \sum_{j=q,g} \int_x^1 \frac{dz}{z}  \; \frac{\alpha_s}{2\pi} \;
   P_{q \leftarrow j}(z) \; f_j\left(\frac{x}{z},-t\right) 
}
\qquad \text{with} \quad
P_{q \leftarrow j}(z) \equiv \hat{P}_{q \leftarrow j}(z)_+  \; .
\label{eq:qcd_dglap}
\end{alignat}
Going back to Eq.\eqref{eq:qcd_almost_dglap} we add all relevant
parton indices and splittings and arrive at
\begin{alignat}{5}
\delta f_q(x,-t) 
&= \frac{\delta t}{t} \; \left[ 
   \int_0^1 \frac{dz}{z} \; \frac{\alpha_s}{2\pi} \; 
            \hat{P}_{q \leftarrow q}(z) \; 
            f_q\left(\frac{x}{z},-t\right) 
 + \int_0^1 \frac{dz}{z} \; \frac{\alpha_s}{2\pi} \; 
            \hat{P}_{q \leftarrow g}(z) \; 
            f_g\left(\frac{x}{z},-t\right) \right.
\notag \\
& \hspace*{20pt} \left.
 - \int_0^1 dy  \; \frac{\alpha_s}{2\pi} \; 
            \hat{P}_{q \leftarrow q}(y) \; 
            f_q(x,-t) 
   \right] \; .
\label{eq:qcd_dglap_quark}
\end{alignat}
Of the three terms on the right hand side the first and the third
together define the plus--subtracted splitting kernel $P_{q \leftarrow
  q}(z)$, just following the argument above. The second term is a
proper convolution and the only term proportional to the gluon parton
density. Quarks can be produced in gluon splitting but cannot vanish
into it.  Therefore, we have to identify the second term with $P_{q \leftarrow g}$
in Eq.\eqref{eq:qcd_dglap} without adding a plus--regulator
\begin{alignat}{5}
\boxed{
P_{q \leftarrow g}(z) \equiv \hat{P}_{q \leftarrow g}(z) 
= T_R \left[ z^2 + (1-z)^2 \right]
}  \; .
\label{eq:plus_pqg}
\end{alignat}
\index{splitting kernel!subtracted $P_{q \leftarrow g}$} In principle,
the splitting kernel $\hat{P}_{g \leftarrow q}$\index{splitting kernel!subtracted $P_{g \leftarrow q}$} also generates a quark, in
addition to the final--state gluon.  However, comparing this to the
terms proportional to $\hat{P}_{q \leftarrow q}$ they both arise from
the same splitting, namely a quark density leaving the infinitesimal
square in the $(x-t)$ plane via the splitting $q \to qg$. Including
the additional $\hat{P}_{g \leftarrow q}(y)$ would be double counting
and should not appear, as the notation $g \leftarrow q$ already
suggests.\bigskip

The second QCD parton density we have to study is the \underline{gluon
density}. The incoming contribution to the infinitesimal square is
given by the sum of four splitting scenarios each leading to a gluon
with virtuality $-t_{j+1}$
\begin{alignat}{5}
\delta f_\text{in}(-t) 
&= \frac{\delta t}{t} 
   \int_0^1 \frac{dz}{z} \; \frac{\alpha_s}{2\pi} \;
   \left[ \hat{P}_{g \leftarrow g}(z) 
          \left( f_g\left(\frac{x}{z},-t\right) 
               + f_g\left(\frac{x}{1-z},-t\right) \right)
        + \hat{P}_{g \leftarrow q}(z) 
         \left( f_q\left(\frac{x}{z},-t\right) 
              + f_{\bar{q}} \left(\frac{x}{z},-t\right) \right)
   \right]
\notag \\
&= \frac{\delta t}{t} 
   \int_0^1 \frac{dz}{z} \; \frac{\alpha_s}{2\pi} \;
   \left[ 2 \hat{P}_{g \leftarrow g}(z) 
                f_g\left(\frac{x}{z},-t\right) 
        + \hat{P}_{g \leftarrow q}(z) 
         \left( f_q\left(\frac{x}{z},-t\right) 
              + f_{\bar{q}} \left(\frac{x}{z},-t\right) \right)
   \right]  \; ,
\end{alignat}
using $P_{g \leftarrow \bar{q}} = P_{g \leftarrow q}$ in the first
line and $P_{g \leftarrow g}(1-z) = P_{g \leftarrow g}(z)$ in the
second. To leave the volume element in the $(x,t)$ space a gluon can
either split into two gluons or radiate one of $n_f$ light-quark
flavors. Combining the incoming and outgoing flows we find\index{DGLAP
  equation}
\begin{alignat}{5}
\delta f_g(x,-t) 
=& \frac{\delta t}{t} 
   \int_0^1 \frac{dz}{z} \; \frac{\alpha_s}{2\pi} \;
   \left[ 2 \hat{P}_{g \leftarrow g}(z) 
                f_g\left(\frac{x}{z},-t\right) 
        + \hat{P}_{g \leftarrow q}(z) 
         \left( f_q\left(\frac{x}{z},-t\right) 
              + f_{\bar{q}} \left(\frac{x}{z},-t\right) \right)
   \right]
\notag \\
-& \frac{\delta t}{t} \; 
   \int_0^1 dy \; \frac{\alpha_s}{2\pi} \;
   \left[ \hat{P}_{g \leftarrow g}(y) 
        + n_f \hat{P}_{q \leftarrow g}(y) 
   \right] f_g(x,-t) 
\label{eq:qcd_dglap_gluon}
\end{alignat}
We have to evaluate the four terms in this expression one after the
other. Unlike in the quark case they do not immediately correspond to
regularizing the diagonal splitting kernel using the plus
prescription.

First, there exists a contribution to $\delta f_\text{in}$
proportional to $f_q$ or $f_{\bar{q}}$ which is not matched by the
outgoing flow. From the quark case we already know how to deal with
it.  For the corresponding splitting kernel there is no regularization
through the plus prescription needed, so we define
\begin{alignat}{5}
\boxed{
P_{g \leftarrow q}(z) \equiv
\hat{P}_{g \leftarrow q}(z) = C_F \frac{1+(1-z)^2}{z} 
} \; .
\label{eq:plus_pgq}
\end{alignat}
\index{splitting kernel!subtracted $P_{g \leftarrow q}$}This ensures
that the off-diagonal contribution to the gluon density is taken into
account when we extend Eq.\eqref{eq:qcd_dglap} to a combined
quark/antiquark and gluon form.
Hence, the structure of the DGLAP equation implies that the two
off-diagonal splitting kernels do not include any plus prescription
$\hat{P}_{i \leftarrow j} = P_{i \leftarrow j}$.  We could have
expected this because off-diagonal kernels are finite in the
soft limit, $z \to 1$. Applying a plus prescription would
only have modified the splitting kernels at the isolated
(zero-measure) point $y=1$ which for a finite value of the integrand
does not affect the integral on the right hand side of the DGLAP
equation.

Second, the $y$ integral describing the gluon splitting into a quark
pair we can compute directly,
\begin{alignat}{5}
 - \int_0^1 dy \; \frac{\alpha_s}{2\pi} \; n_f \; \hat{P}_{q \leftarrow g}(y) 
&= - \frac{\alpha_s}{2\pi} \; n_f \; T_R \; \int_0^1 dy \; \left[ 1 - 2y + 2 y^2 \right] 
\qqquad \text{using Eq.\eqref{eq:plus_pqg}}
\notag \\
&= - \frac{\alpha_s}{2\pi} \; n_f \; T_R \; \left[ y - y^2 + \frac{2 y^3}{3} \right]_0^1 \notag \\
&= - \frac{2}{3}\; \frac{\alpha_s}{2\pi} \; n_f \; T_R \; .
\label{eq:qcd_pgg_1}
\end{alignat}

Finally, the terms proportional to the purely gluonic splitting $P_{g
  \leftarrow g}$ appearing in Eq.\eqref{eq:qcd_dglap_gluon} require
some more work. The $y$ integral coming from the outgoing flow has to
consist of a finite term and a term we can use to define the plus
prescription for $\hat{P}_{g \leftarrow g}$. We can compute the
integral as
\begin{alignat}{5}
 - \int_0^1 dy \; \frac{\alpha_s}{2\pi} \; \hat{P}_{g \leftarrow g}(y) 
=& - \frac{\alpha_s}{2\pi} \; C_A \; \int_0^1 dy \; 
   \left[ \frac{y}{1-y} + \frac{1-y}{y} + y(1-y) \right]
\qqquad \text{using Eq.\eqref{eq:qcd_pgg}}
\notag \\
=& - \frac{\alpha_s}{2\pi} \; C_A \; \int_0^1 dy \; 
   \left[ \frac{2 y}{1-y} + y(1-y) \right]
\notag \\
=& - \frac{\alpha_s}{2\pi} \; C_A \; \int_0^1 dy \; 
   \left[ \frac{2 (y-1)}{1-y} + y(1-y) \right]
   - \frac{\alpha_s}{2\pi} \; C_A \; \int_0^1 dy \; \frac{2}{1-y}
\notag \\
=& - \frac{\alpha_s}{2\pi} \; C_A \; \int_0^1 dy \; 
   \left[ -2 + y - y^2 \right]
   - \frac{\alpha_s}{2\pi} \; 2 C_A \; \int_0^1 dz \; \frac{1}{1-z}
\notag \\
=& - \frac{\alpha_s}{2\pi} \; C_A \; 
   \left[ -2 + \frac{1}{2} - \frac{1}{3} \right]
  -\frac{\alpha_s}{2\pi} \; 2 C_A \; \int_0^1 dz \; \frac{1}{1-z}
\notag \\
=& \; \frac{\alpha_s}{2\pi} \; \frac{11}{6} \; C_A \; 
  - \frac{\alpha_s}{2\pi} \; 2 C_A \; \int_0^1 dz \; \frac{1}{1-z} \; .
\label{eq:qcd_pgg_2}
\end{alignat}
The second term in this result is what we need to replace the first term in the
splitting kernel of Eq.\eqref{eq:qcd_pgg} proportional to $1/(1-z)$ by
$1/(1-z)_+$. We can see this using $f(z) = z$ and correspondingly
$f(1) = 1$ in Eq.\eqref{eq:plus_subtraction}. The two finite terms in
Eq.\eqref{eq:qcd_pgg_1} and Eq.\eqref{eq:qcd_pgg_2} we have to include
in the definition of $\hat{P}_{g \leftarrow g}$ ad hoc. Because the
regularized splitting kernel appear inside a convolution the two
finite terms require an additional term $\delta(1-z)$. Collecting all
of them we arrive at
\begin{alignat}{5}
\boxed{
P_{g \leftarrow g}(z) = 2 C_A \;
                      \left( \frac{z}{(1-z)_+} + \frac{1-z}{z} + z(1-z) \right) 
                    + \frac{11}{6} \; C_A \; \delta(1-z) 
                    - \frac{2}{3}\; \; n_f \; T_R \; \delta(1-z)
} \; .
\end{alignat}
\index{splitting kernel!subtracted $P_{g  \leftarrow g}$}This result concludes our computation of all four regularized splitting
functions which appear in the DGLAP equation
Eq.\eqref{eq:qcd_dglap}.\bigskip

Before discussing and solving the DGLAP equation, let us briefly
recapitulate: for the full quark and gluon particle content of QCD we
have derived the DGLAP equation which describes a 
factorization scale dependence of the quark and gluon parton
densities. The universality of the splitting kernels is obvious from
the way we derive them --- no information on the $n$-particle process
ever enters the derivation. 

The DGLAP equation is formulated in terms of four splitting kernels of
gluons and quarks which are linked to the splitting probabilities, but
which for the DGLAP equation have to be regularized. With the help of
a plus--subtraction all kernels $P_{i \leftarrow j}(z)$ become finite,
including in the soft limit $z \to 1$.  However, splitting kernels are
only regularized when needed, so the finite off-diagonal quark--gluon
and gluon--quark splittings are unchanged.  This means the plus
prescription really acts as an infrared renormalization, moving
universal infrared divergences into the definition of the parton
densities. The original collinear divergence has vanished as well.

The only approximation we make in the computation of the splitting
kernels is that in the $y$ integrals we implicitly assume that the
running coupling $\alpha_s$ does not depend on the momentum fraction.
In its standard form and in terms of the factorization scale $\mu_F^2
\equiv -t$ the \underline{DGLAP equation}\index{DGLAP equation}\index{scales!factorization scale} reads
\begin{alignat}{5}
\boxed{
  \frac{d f_i(x,\mu_F)}{d \log \mu_F^2} 
  = \sum_j \; \int_x^1 \; \frac{dz}{z} \;
    \frac{\alpha_s}{2 \pi} \;
    P_{i \leftarrow j}(z) \;
    f_j\left(\frac{x}{z},\mu_F\right)
  =  \frac{\alpha_s}{2 \pi} \; \sum_j \;     
     \left( P_{i \leftarrow j} \otimes f_j \right) (x,\mu_F) 
}  \; .
\label{eq:qcd_dglap2}
\end{alignat}
%

\subsubsection{Parton densities}
\label{sec:qcd_solve_dglap}

Solving the integro-differential DGLAP equation
Eq.\eqref{eq:qcd_dglap2}\index{DGLAP equation!solutions} for the
parton densities is clearly beyond the scope of this
writeup. Nevertheless, we will sketch how we would approach this.
This will give us some information on the structure of its solutions
which we need to understand the physics of the DGLAP equation.

One simplification we can make in this illustration is to postulate
eigenvalues in parton space and solve the equation for them. This
gets rid of the sum over partons on the right hand side. One such
parton density is the \underline{non--singlet parton density}, defined
as the difference of two parton densities $\qns = (f_u -
f_{\bar{u}})$. Since gluons cannot distinguish between quarks and
antiquarks, the gluon contribution to their evolution cancels, at
least in the massless limit. This will be true at arbitrary loop
order, since flavor $SU(3)$ commutes with the QCD gauge group. The
corresponding DGLAP equation with leading order splitting kernels now
reads
\begin{alignat}{5}
  \frac{d \qns(x,\mu_F)}{d \log \mu_F^2} 
  = \int_x^1 \; \frac{dz}{z} \;
     \frac{\alpha_s}{2 \pi} \;
     P_{q \leftarrow q}(z) \;
     \qns\left( \frac{x}{z},\mu_F\right) 
  = \frac{\alpha_s}{2 \pi} \;
     \left( P_{q \leftarrow q} \otimes
     \qns \right)(x,\mu_F) \; .
\end{alignat}
To solve it we need a transformation which simplifies a
convolution, leading us to the \underline{Mellin transform}\index{Mellin transform}.  Starting from a
function $f(x)$ of a real variable $x$ we define the Mellin transform
into moment space $m$
\begin{alignat}{5}
\boxed{
\mat[f](m) \equiv \int_0^1 dx \, x^{m-1} f(x)
}
\qqqquad
f(x) = \frac{1}{2 \pi i} \int_{c - i \infty}^{c - i \infty} dm \;
        \frac{\mat[f](m)}{x^m} \; .
\end{alignat}
The integration contour for the inverse transformation lies to the
right of all singularities of the analytic continuation of
$\mat[f](m)$, which fixes the offset $c$. The Mellin transform of a
convolution is the product of the two Mellin transforms, which gives
us the transformed DGLAP equation
\begin{alignat}{5}
\mat[P_{q \leftarrow q} \otimes \qns](m) &=
\mat \left[\int_0^1 \frac{dz}{z} 
           P_{q \leftarrow q} \left( \frac{x}{z} \right) \qns(z) \right](m) =
\mat[P_{q \leftarrow q}](m) \; \mat[\qns](m,\mu_F) 
\notag \\
  \frac{d \mat[\qns](m,\mu_F)}{d \log \mu_F^2} 
  &= \frac{\alpha_s}{2 \pi} \;
     \mat[P_{q \leftarrow q}](m) \; \mat[\qns](m,\mu_F) \; ,
\label{eq:dglap_mellin}
\end{alignat}
and its solution
\begin{alignat}{5}
  \mat[\qns](m,\mu_F)
  &= \mat[\qns](m,\mu_{F,0} \;
     \exp \left( \frac{\alpha_s}{2 \pi} \;
                 \mat[P_{q \leftarrow q}](m) 
                 \log \frac{\mu_F^2}{\mu_{F,0}^2}
          \right)
\notag \\
  &= \mat[\qns](m,\mu_{F,0} \;
     \left( \frac{\mu_F^2}{\mu_{F,0}^2}
     \right)^{\frac{\alpha_s}{2 \pi} \mat[P_{q \leftarrow q}](m)}
\notag \\
  &\equiv \mat[\qns](m,\mu_{F,0} \;
     \left( \frac{\mu_F^2}{\mu_{F,0}^2}
     \right)^{\frac{\alpha_s}{2 \pi} \gamma(m)} \; ,
\label{eq:def_anodim}
\end{alignat}
defining $\gamma(m) = \mat[P](m)$.\bigskip

The solution given by Eq.\eqref{eq:def_anodim} still has the
complication that it includes $\mu_F$ and $\alpha_s$ as two free
parameters.  To simplify this form we can include $\alpha_s(\mu_R^2)$
in the running of the DGLAP equation and identify the renormalization
scale $\mu_R$ of the strong coupling with the factorization scale
\underline{$\mu_F = \mu_R \equiv \mu$}\index{scales!factorization scale}\index{scales!renormalization scale}. This allows us to
replace $\log \mu^2$ in the DGLAP equation by $\alpha_s$, including
the leading order Jacobian. This is clearly correct for all one-scale
problems where we have no freedom to choose either of the two
scales. We find
\begin{alignat}{5}
  \frac{d}{d \log \mu^2} 
= \frac{d \log \alpha_s}{d \log \mu^2} \; 
  \frac{d}{d \log \alpha_s} 
= \frac{1}{\alpha_s} \; \frac{d \alpha_s}{d \log \mu^2} \; 
  \frac{d}{d \log \alpha_s} 
= - \alpha_s b_0 \; 
  \frac{d}{d \log \alpha_s}  \; .
\end{alignat}
This additional factor of $\alpha_s$ on the left hand side will
cancel the factor $\alpha_s$ on the right hand side of the DGLAP
equation Eq.\eqref{eq:dglap_mellin}
\begin{alignat}{5}
  \frac{d \mat[\qns](m,\mu)}{d \log \alpha_s} 
  &= - \frac{1}{2 \pi b_0} 
     \gamma(m) \; \mat[\qns](m,\mu) 
\notag \\
  \mat[\qns](m,\mu)
  &= \mat[\qns](m,\mu_0) \;
     \exp \left( - \frac{1}{2 \pi b_0} \;
                 \gamma(m) 
                 \log \frac{\alpha_s(\mu^2)}{\alpha_s(\mu_0^2)}
          \right)
\notag \\
  &= \mat[\qns](m,\mu_{F,0} \;
     \left( \frac{\alpha_s(\mu_0^2)}{\alpha_s(\mu^2)}
     \right)^{\frac{\gamma(m)}{2 \pi b_0}} \; .
\label{eq:qcd_pdf_solution}
\end{alignat}
Among other things, in this derivation we neglect that some splitting
functions have singularities and therefore the Mellin transform is not
obviously well defined. Our convolution is not really a convolution
either, because we cut it off at $Q_0^2$ etc; but the final structure
in Eq.\eqref{eq:qcd_pdf_solution} really holds.

Because we will need it in the next section we emphasize that the same
kind of solution appears in pure Yang--Mills theory, \ie in QCD
without quarks. Looking at the different color factors in QCD this
limit can also be derived as the leading terms in $N_c$. In that case
there also exists only one splitting kernel defining an anomalous
dimension $\gamma$. We find in complete analogy to
Eq.\eqref{eq:qcd_pdf_solution}
\begin{alignat}{5}
\boxed{
  \mat[f_g](m,\mu)
  = \mat[f_g](m,\mu_0) \;
     \left( \frac{\alpha_s(\mu_0^2)}{\alpha_s(\mu^2)}
     \right)^{\frac{\gamma(m)}{2 \pi b_0}}
} \; .
\label{eq:qcd_dglap_mellin}
\end{alignat}
To remind ourselves that in this derivation we unify the
renormalization and factorization scales we denote them just as $\mu$.
This solution to the DGLAP equation is not completely determined: as a
solution to a differential equation it also includes an integration
constant which we express in terms of $\mu_0$.  The DGLAP equation
therefore does not determine parton densities, it only describes their
evolution from one scale $\mu_F$ to another, just like a
renormalization group equation in the ultraviolet.\bigskip

The structure of Eq.\eqref{eq:qcd_dglap_mellin} already shows
something we will in more detail discuss in the following
Section~\ref{sec:qcd_resum_collinear}: the splitting probability we
find in the exponent. To make sense of such a structure we remind
ourselves that such ratios of $\alpha_s$ values to some power can
appear as a result of a resummed series.  Such a series would need to
include powers of $(\mat [\hat{P}])^n$ summed over $n$ which
corresponds to a sum over splittings with a varying number of partons
in the final state. Parton densities cannot be formulated in terms of
a fixed final state because they include effects from any number of
collinearly radiated partons summed over the number of such
partons. For the processes we can evaluate using parton densities
fulfilling the DGLAP equation this means that they always have the
form
\begin{alignat}{5}
\boxed{pp \to \mu^+ \mu^- + X}
\qqqquad \text{where $X$ includes any number of collinear jets.}
\label{eq:qcd_crossx}
\end{alignat}
\bigskip

Why is $\gamma$ is referred to as the
\underline{anomalous dimension}\index{anomalous dimension} of the parton density? This is best
illustrated using a running coupling with a finite mass dimension, like
the gravitational coupling $G_\text{Planck} \sim
1/M_\text{Planck}^2$. When we attach a renormalization constant $Z$ to
this coupling we first define a dimensionless running bare coupling
$g$. In $n$ dimensions this gives us 
\begin{alignat}{5}
g^\text{bare} = M^{n-2} \; G_\text{Planck} \to Z g(M^2) \; .
\end{alignat}
For the dimensionless gravitational coupling we can compute the running
\begin{alignat}{5}
\frac{d g(M^2)}{d \log M} 
&= \frac{d}{d \log M} \left( \frac{1}{Z} M^{n-2} \; G_\text{Planck}
                      \right)
\notag \\
&= G_\text{Planck} 
   \left( \frac{1}{Z} M \frac{dM^{n-2}}{d M}
        - \frac{1}{Z^2} \frac{dZ}{d \log M} M^{n-2} 
   \right)
\notag \\
&= g(M) \left( n - 2 + \eta \right)
\qqquad \text{with} \quad \eta = - \frac{1}{Z} \, \frac{dZ}{d \log M} 
\end{alignat}
Hence, there are two sources of running for the renormalized coupling
$g(M^2)$: first, there is the mass dimension of the bare coupling
$n-2$, and secondly there is $\eta$, a quantum effect from the
coupling renormalization. For obvious reasons we call $\eta$ the
anomalous dimension of $G_\text{Planck}$. 

This is similar to the running of the parton densities in Mellin
space, as shown in Eq.\eqref{eq:dglap_mellin}, and with $\gamma(m)$ defined
in Eq.\eqref{eq:def_anodim}, so we refer to $\gamma$ as an anomalous
dimension as well.  The entire running of the transformed parton
density arises from collinear splitting, parameterized by a finite
$\gamma$. There is only a slight stumbling step in this analogy:
usually, an anomalous dimension arises through renormalization
involving a ultraviolet divergence and the renormalization scale. In
our case we are discussing an infrared divergence and the
factorization scale dependence.

\subsubsection{Resumming collinear logarithms}
\label{sec:qcd_resum_collinear}

Remembering how we arrive at the DGLAP equation we notice an analogy
to the case of ultraviolet divergences and the running coupling.  We
start from universal infrared divergences. We describe them in terms
of splitting functions which we regularize using the plus
prescription. The DGLAP equation plays the role of a renormalization
group equation for example for the running coupling. It links parton
densities evaluated at different scales $\mu_F$.

In analogy to the scaling logarithms considered in
Section~\ref{sec:qcd_scale_logs} we now test if we can point to a
\underline{type of logarithm}\index{resummation!collinear logarithms}
the DGLAP equation resums by reorganizing our perturbative series of
parton splitting. To identify these resummed logarithms we build a
physical model based on collinear splitting, but without using the
DGLAP equation. We then solve it to see the resulting structure of the
solutions and compare it to the structure of the DGLAP solutions in
Eq.\eqref{eq:qcd_dglap_mellin}.\bigskip

We start from the basic equation defining the physical picture of
parton splitting in Eq.\eqref{eq:qcd_factorize}\index{splitting kernel}. Only taking into account gluons in pure Yang--Mills theory
it precisely corresponds to the starting point of our discussion
leading to the DGLAP equation, schematically written as
\begin{alignat}{5}
\sigma_{n+1} 
 = \int \sigma_n \; 
   \frac{d t}{t} dz \; 
   \frac{\alpha_s}{2 \pi} \; \hat{P}_{g \leftarrow g}(z) \; .
\label{eq:ir_logs_start}
\end{alignat}
This form of collinear factorization does not include parton
densities and only applies to final state splittings.  To include
initial state splittings we need a definition of the virtuality
variable $t$. If we remember that $t=p_b^2<0$ we can follow
Eq.\eqref{eq:qcd_sudakovdec2} and introduce a positive transverse
momentum variable $\vec{p}_T^2$ in the usual Sudakov decomposition,
such that
\begin{alignat}{5}
-t 
= - \frac{p_T^2}{1-z} 
= \frac{\vec{p}_T^2}{1-z} > 0  
\qquad \qquad \Rightarrow \quad 
\frac{dt}{t} 
= \frac{d p_T^2}{p_T^2} 
= \frac{d \vec{p}_T^2}{\vec{p}_T^2} \; .
\label{eq:ir_logs_start2}
\end{alignat}
From the definition of $p_T$ in Eq.\eqref{eq:qcd_sudakovdec} we see
that $\vec{p}_T^2$ is really the transverse three-momentum of of the
parton pair after splitting.\bigskip

Beyond the single parton radiation discussed in
Section~\ref{sec:qcd_single_jet} we consider a ladder of successive
splittings of one gluon into two. For a moment, we forget about the
actual parton densities and assume that they are part of the hadronic
cross section $\sigma_{n}$. In the collinear limit the appropriate
convolution gives us
\begin{alignat}{5}
\sigma_{n+1}(x,\mu_F)
&= \int_{x_0}^1 \frac{dx_n}{x_n} \;
   \hat{P}_{g \leftarrow g}\left( \frac{x}{x_n} \right) \sigma_n(x_n,\mu_0)
   \int_{\mu_0^2}^{\mu_F^2} \frac{d \vec{p}_{T,n}^2}{\vec{p}_{T,n}^2} 
   \frac{\alpha_s(\mu_R^2)}{2 \pi} \; .
\label{eq:gluon_split2}
\end{alignat}
The $d z$ in Eq.\eqref{eq:ir_logs_start} we replace by the proper
convolution $\hat{P} \otimes \sigma_n$, evaluated at the momentum
fraction $x$.  Because the splitting kernel is infrared divergent we
cut off the convolution integral at $x_0$. Similarly, the transverse
momentum integral is bounded by an infrared cutoff $\mu_0$ and the
physical external scale $\mu_F$. This is the range in which an
additional collinear radiation is included in $\sigma_{n+1}$.

For splitting the two integrals in Eq.\eqref{eq:gluon_split2} it is
crucial that $\mu_0$ is the only scale the matrix element $\sigma_n$
depends on. The other integration variable, the transverse momentum,
does not feature in $\sigma_n$ because collinear factorization is
defined in the limit $\vec{p}_T^2 \to 0$. For $\alpha_s$ we will see
in the next step how $\mu_R$ can depend on the transverse
momentum. All through the argument of this subsection we should keep
in mind that we are looking for assumptions which allow us to solve
Eq.\eqref{eq:gluon_split2} and compare the result to the solution of
the DGLAP equation. In other words, these assumptions we will turn
into a physics picture of the DGLAP equation and its
solutions.\bigskip

Making $\mu_F$ the \underline{global upper boundary of the transverse
  momentum} integration for collinear splitting is our first
assumption. We can then apply the recursion formula in
Eq.\eqref{eq:gluon_split2} iteratively
\begin{alignat}{5}
\sigma_{n+1}(x,\mu_F)
\sim& 
   \int_{x_0}^1 \frac{dx_n}{x_n} \; 
   \hat{P}_{g \leftarrow g}\left( \frac{x}{x_n} \right) \; \cdots \;
   \int_{x_0}^1 \frac{dx_1}{x_1} \; 
   \hat{P}_{g \leftarrow g}\left( \frac{x_2}{x_1} \right) \; 
   \sigma_1(x_1,\mu_0)
\notag \\
\times&  \int_{\mu_0}^{\mu_F} \frac{d \vec{p}_{T,n}^2}{\vec{p}_{T,n}^2} 
   \frac{\alpha_s(\mu_R^2)}{2 \pi} \; \cdots \;
   \int_{\mu_0} \frac{d \vec{p}_{T,1}^2}{\vec{p}_{T,1}^2} 
   \frac{\alpha_s(\mu_R^2)}{2 \pi} \; .
\label{eq:qcd_successive}
\end{alignat}
The two sets of integrals in this equation we will solve one by one,
starting with the $\vec{p}_T$ integrals.\bigskip

To be able to make sense of the $\vec{p}_T^2$ integration in
Eq.\eqref{eq:qcd_successive} and solve it we have to make two more
assumptions in our multiple-splitting model. First, we identify the
\underline{scale of the strong coupling $\alpha_s$} with the
transverse momentum scale of the splitting $\mu_R^2 =
\vec{p}_T^2$. This way we can fully integrate the integrand
$\alpha_s/(2\pi)$ and link the final result to the global boundary
$\mu_F$.

In addition, we assume \underline{strongly ordered
  splittings}\index{transverse momentum ordering} in terms of the
transverse momentum. If the ordering of the splitting is fixed
externally by the chain of momentum fractions $x_j$, the first
splitting, integrated over $\vec{p}_{T,1}^2$, is now bounded from
above by the next external scale $\vec{p}_{T,2}^2$, which is then
bounded by $\vec{p}_{T,3}^2$, etc. For the $n$-fold $\vec{p}_T^2$
integration this means
\begin{alignat}{5}
\mu_0^2 < \vec{p}_{T,1}^2 < \vec{p}_{T,2}^2 < \cdots < \mu_F^2 \; 
\end{alignat}
We will study motivations for this {\sl ad hoc} assumptions in
Section~\ref{sec:qcd_ordered}.\bigskip

Under these three assumptions the transverse momentum integrals in
Eq.\eqref{eq:qcd_successive} become
\begin{alignat}{5}
 & \int_{\mu_0}^{\mu_F} \frac{d \vec{p}_{T,n}^2}{\vec{p}_{T,n}^2} 
   \frac{\alpha_s(\vec{p}_{T,n}^2)}{2 \pi} \; \cdots \;
   \int_{\mu_0}^{p_{T,3}} \frac{d \vec{p}_{T,2}^2}{\vec{p}_{T,2}^2} 
   \frac{\alpha_s(\vec{p}_{T,2}^2)}{2 \pi} \; 
   \int_{\mu_0}^{p_{T,2}} \frac{d \vec{p}_{T,1}^2}{\vec{p}_{T,1}^2} 
   \frac{\alpha_s(\vec{p}_{T,1}^2)}{2 \pi} \; \cdots
\notag \\
=& \int_{\mu_0}^{\mu_F} \frac{d \vec{p}_{T,n}^2}{\vec{p}_{T,n}^2} 
   \frac{1}{2 \pi b_0 \log \dfrac{\vec{p}_{T,n}^2}{\lqcd^2}} \; \cdots \;
   \int_{\mu_0}^{p_{T,3}} \frac{d \vec{p}_{T,2}^2}{\vec{p}_{T,2}^2} 
   \frac{1}{2 \pi b_0 \log \dfrac{\vec{p}_{T,2}^2}{\lqcd^2}} \; 
   \int_{\mu_0}^{p_{T,2}} \frac{d \vec{p}_{T,1}^2}{\vec{p}_{T,1}^2} 
   \frac{1}{2 \pi b_0 \log \dfrac{\vec{p}_{T,1}^2}{\lqcd^2}} \; \cdots \;
\notag \\
=& \frac{1}{(2 \pi b_0)^n} 
   \int_{\mu_0}^{\mu_F} \frac{d \vec{p}_{T,n}^2}{\vec{p}_{T,n}^2} 
   \frac{1}{\log \dfrac{\vec{p}_{T,n}^2}{\lqcd^2}} \; \cdots \;
   \int_{\mu_0}^{p_{T,3}} \frac{d \vec{p}_{T,2}^2}{\vec{p}_{T,2}^2} 
   \frac{1}{\log \dfrac{\vec{p}_{T,2}^2}{\lqcd^2}} \; 
   \int_{\mu_0}^{p_{T,2}} \frac{d \vec{p}_{T,1}^2}{\vec{p}_{T,1}^2} 
   \frac{1}{\log \dfrac{\vec{p}_{T,1}^2}{\lqcd^2}} \; \cdots \; .
\end{alignat}
We can solve the individual integrals by switching variables, for
example in the last integral
\begin{alignat}{5}
   \int_{\mu_0}^{p_{T,2}}  \frac{d \vec{p}_{T,1}^2}{\vec{p}_{T,1}^2} 
   \frac{1}{\log \dfrac{\vec{p}_{T,1}^2}{\lqcd^2}}
&= \int_{\log \log \mu_0^2/\Lambda^2}^{\log \log p_{T,2}^2/\Lambda^2}  d \log \log \dfrac{\vec{p}_{T,1}^2}{\lqcd^2} 
   \qquad \text{with} \qquad \frac{d (ax)}{(ax) \log x} = d \log \log x 
 \notag \\ 
&= \int_0^{\log \log p_{T,2}^2/\Lambda^2 - \log\log \mu_0^2/\Lambda^2}  
   d \left( \log \log \dfrac{\vec{p}_{T,1}^2}{\lqcd^2} 
          - \log \log \dfrac{\mu_0^2}{\lqcd^2} \right) 
 \notag \\ 
&= \log \dfrac{\log \vec{p}_{T,1}^2/\lqcd^2}
              {\log \mu_0^2/\lqcd^2}
   \Bigg|_0^{\vec{p}_{T,1}^2 \equiv \vec{p}_{T,2}^2} 
= \log \frac{\log \vec{p}_{T,2}^2/\lqcd^2}{\log \mu_0^2/\lqcd^2} \; .
\end{alignat}
This gives us for the chain of transverse momentum integrals
\begin{alignat}{5}
   \int_0^{p_{T,n} \equiv \mu_F} &
        d \log \frac{\log \vec{p}_{T,n}^2/\lqcd^2}{\log \mu_0^2/\lqcd^2}
          \; \cdots 
   \int_0^{p_{T,2} \equiv p_{T,3}} 
        d \log \frac{\log \vec{p}_{T,2}^2/\lqcd^2}{\log \mu_0^2/\lqcd^2}
   \int_0^{p_{T,1} \equiv p_{T,2}} 
        d \log \frac{\log \vec{p}_{T,1}^2/\lqcd^2}{\log \mu_0^2/\lqcd^2}
\notag \\
&= 
   \int_0^{p_{T,n} \equiv \mu_F} 
        d \log \frac{\log \vec{p}_{T,n}^2/\lqcd^2}{\log \mu_0^2/\lqcd^2}
          \; \cdots 
   \int_0^{p_{T,2} \equiv p_{T,3}} 
        d \log \frac{\log \vec{p}_{T,2}^2/\lqcd^2}{\log \mu_0^2/\lqcd^2}
        \left( \log \frac{\log \vec{p}_{T,2}^2/\lqcd^2}{\log \mu_0^2/\lqcd^2}
          \right)
\notag \\
&= 
   \int_0^{p_{T,n} \equiv \mu_F} 
        d \log \frac{\log \vec{p}_{T,n}^2/\lqcd^2}{\log \mu_0^2/\lqcd^2}
          \; \cdots 
   \frac{1}{2}
          \left( \log \frac{\log \vec{p}_{T,3}^2/\lqcd^2}{\log \mu_0^2/\lqcd^2}
          \right)^2
\notag \\
&= 
   \int_0^{p_{T,n} \equiv \mu_F} 
        d \log \frac{\log \vec{p}_{T,n}^2/\lqcd^2}{\log \mu_0^2/\lqcd^2}
   \left( \frac{1}{2} \cdots \frac{1}{n-1} \right)
          \left( \log \frac{\log \vec{p}_{T,n}^2/\lqcd^2}{\log \mu_0^2/\lqcd^2}
          \right)^{n-1}
\notag \\
&= \frac{1}{n!} 
   \left( \log \frac{\log \mu_F^2/\lqcd^2}{\log \mu_0^2/\lqcd^2}
   \right)^n 
 = \frac{1}{n!} 
   \left( \log \frac{\alpha_s(\mu_0^2)}{\alpha_s(\mu_F^2)}
   \right)^n \; .
\end{alignat}
This is the final result for the chain of transverse momentum
integrals in Eq.\eqref{eq:qcd_successive}. By assumption, the
strong coupling is evaluated at the factorization scale $\mu_F$, which
means we identify $\mu_R \equiv \mu_F$.\bigskip

To compute the convolution integrals over the momentum fractions in
Eq.\eqref{eq:qcd_successive},
\begin{alignat}{5}
\sigma_{n+1}(x,\mu)
\sim& 
   \frac{1}{n!} 
   \left( \frac{1}{2 \pi b_0} \log \frac{\alpha_s(\mu_0^2)}{\alpha_s(\mu^2)}
   \right)^n \;
   \int_{x_0}^1 \frac{dx_n}{x_n} \; 
   \hat{P}_{g \leftarrow g}\left( \frac{x}{x_n} \right) \; \cdots \;
   \int_{x_0}^1 \frac{dx_1}{x_1} \; 
   \hat{P}_{g \leftarrow g}\left( \frac{x_2}{x_1} \right) \; 
   \sigma_1(x_1,\mu_0) \; ,
\end{alignat}
we again Mellin transform the equation into moment space
\begin{alignat}{5}
\mat[\sigma_{n+1}](m,\mu)
\sim& 
   \frac{1}{n!} 
   \left( \frac{1}{2 \pi b_0} \log \frac{\alpha_s(\mu_0^2)}{\alpha_s(\mu^2)}
   \right)^n \;
   \mat \left[
   \int_{x_0}^1 \frac{dx_n}{x_n} \; 
   \hat{P}_{g \leftarrow g}\left( \frac{x}{x_n} \right) \; \cdots \;
   \int_{x_0}^1 \frac{dx_1}{x_1} \; 
   \hat{P}_{g \leftarrow g}\left( \frac{x_2}{x_1} \right) \; 
   \sigma_1(x_1,\mu_0) \right](m)
\notag \\
=& \frac{1}{n!} 
   \left( \frac{1}{2 \pi b_0} \log \frac{\alpha_s(\mu_0^2)}{\alpha_s(\mu^2)}
   \right)^n \;
   \gamma(m)^n \; 
   \mat[\sigma_1](m,\mu_0)
\qqquad \text{using} \; \gamma(m) \equiv \mat[P](m)
\notag \\
=& \frac{1}{n!} 
   \left( \frac{1}{2 \pi b_0} \log \frac{\alpha_s(\mu_0^2)}{\alpha_s(\mu^2)}
   \; \gamma(m) \right)^n \; 
   \mat[\sigma_1](m,\mu_0) \; .
\end{alignat}
We can now sum the production cross sections for $n$ collinear jets
and obtain
\begin{alignat}{5}
\sum_{n=0}^\infty \mat[\sigma_{n+1}](m,\mu)
=& \mat[\sigma_1](m,\mu_0) \;
   \sum_n \frac{1}{n!} 
   \left( \frac{1}{2 \pi b_0} \log \frac{\alpha_s(\mu_0^2)}{\alpha_s(\mu^2)}
   \; \gamma(m) \right)^n \; 
\notag \\
=& \mat[\sigma_1](m,\mu_0) \;
   \exp \left( 
      \frac{\gamma(m)}{2 \pi b_0} 
      \; \log \frac{\alpha_s(\mu_0^2)}{\alpha_s(\mu^2)} \right) \; .
\end{alignat}
This way we can write the Mellin transform of the $(n+1)$ particle
production rate as the product of the $n$-particle rate times a ratio
of the strong coupling at two scales
\begin{alignat}{5}
\boxed{
\sum_{n=0}^\infty \mat[\sigma_{n+1}](m,\mu)
= \mat[\sigma_1](m,\mu_0) \;
   \left( \frac{\alpha_s(\mu_0^2)}{\alpha_s(\mu^2)} 
   \right)^\frac{\gamma(m)}{2 \pi b_0} 
} \; .
\label{eq:qcd_collinear_summation}
\end{alignat}
This is the same structure as the DGLAP equation's solution in
Eq.\eqref{eq:qcd_dglap_mellin}. It means that we should be able to
understand the physics of the DGLAP equation using our model
calculation of a gluon ladder emission, including the generically
variable number of collinear jets in the form of $pp \to \mu^+ \mu^- +
X$, as shown in Eq.\eqref{eq:qcd_crossx}.

We should remind ourselves of the three assumptions we need to make to arrive
at this form. There are two assumptions which concern the transverse
momenta of the successive radiation: first, the global upper limit on
all transverse momenta should be the factorization scale $\mu_F$, with
a strong ordering in the transverse momenta. This gives us a physical
picture of the successive splittings as well as a physical
interpretation of the factorization scale\index{scales!factorization scale}. Second, the strong coupling should be evaluated at the
transverse momentum or factorization scale, so all scales are unified,
in accordance with the derivation of the DGLAP equation.\bigskip

Bending the rules of pure Yang--Mills QCD we can come back to the hard
process $\sigma_1$ as the Drell--Yan process $q \bar{q} \to Z$. Each
step in $n$ means an additional parton in the final state, so
$\sigma_{n+1}$ is $Z$ production with $n$ collinear partons On the
left hand side of Eq.\eqref{eq:qcd_collinear_summation} we have the
sum over any number of additional collinear partons; on the right hand
side we see fixed order Drell--Yan production without any additional
partons, but with an exponentiated correction factor. Comparing this
to the running parton densities we can draw the analogy that any
process computed with a scale dependent parton density where the scale
dependence is governed by the DGLAP equation includes \underline{any
  number} of collinear partons.  

We can also identify the logarithms which are resummed 
by scale dependent parton densities. Going back
to Eq.\eqref{eq:qcd_collinear} reminds us that we start from the
divergent collinear logarithms $\log p_T^\text{max}/p_T^\text{min}$
arising from the collinear phase space integration. In our model for
successive splitting we replace the upper boundary by $\mu_F$. The
collinear logarithm of successive initial--state parton splitting
diverges for $\mu_0 \to 0$, but it gets absorbed into the parton densities
and determines the structure of the DGLAP equation and its solutions.
The upper boundary $\mu_F$ tells us to what extent we assume incoming
quarks and gluons to be a coupled system of splitting partons and what
the maximum momentum scale of these splittings is. Transverse momenta
$p_T > \mu_F$ generated by hard parton splitting are not covered by
the DGLAP equation and hence not a feature of the incoming partons
anymore. They belong to the hard process and have to be consistently
simulated, as we will see in Sections~\ref{sec:qcd_ckkw}
and~\ref{sec:qcd_merge}.  While this scale can be chosen freely we
have to make sure that it does not become too large, because at some
point the \underline{collinear approximation}\index{collinear limit}
$C \simeq$~constant in Eq.\eqref{eq:qcd_collinear} ceases to hold and
with it our entire argument. Only if we do everything correctly, the DGLAP equation resums
logarithms of the maximal \underline{transverse momentum
  size}\index{transverse momentum size} of the incoming gluon. They
are universal and arise from simple kinematics.\bigskip

The ordering of the splittings we have to assume is not relevant
unless we simulate this splitting, as we will see in the next
section. For the details of this we have to remember that our argument
follows from the leading collinear approximation introduced in
Section~\ref{sec:qcd_single_jet}. Therefore, the strong $p_T$-ordering
can in practice mean angular ordering or rapidity ordering,
just applying a linear transformation.

\subsection{Scales in LHC processes}
\label{sec:qcd_scales}

\begin{table}[t]
\begin{center}
\begin{small}
\begin{tabular}{l||l|l}
   & renormalization scale $\mu_R$
   & factorization scale $\mu_F$ \\ \hline
 source
   & ultraviolet divergence
   & collinear (infrared) divergence \\[2mm]
 poles cancelled
   & counter terms 
   & parton densities \\
   & (renormalization)
   & (mass factorization) \\
 summation
   & resum self energy bubbles
   & resum parton splittings \\
 parameter 
   & running coupling $\alpha_s(\mu_R^2)$
   & running parton density $f_j(x, \mu_F)$ \\
 evolution 
   & RGE for $\alpha_s$
   & DGLAP equation \\[2mm]
 large scales
   & decrease of $\sigma_\text{tot}$
   & increase of $\sigma_\text{tot}$ for gluons/sea quarks \\[2mm]
 theory background
   & renormalizability
   & factorization \\
   & proven for gauge theories
   & proven all orders for DIS \\
   && proven order-by-order DY... \\
\end{tabular}
\end{small}
\end{center}
\caption{Comparison of renormalization and factorization scales 
  appearing in LHC cross sections.}
\label{tab:qcd_scales}
\end{table}

Looking back at Sections~\ref{sec:qcd_uv} and~\ref{sec:qcd_ir} we
introduce the factorization and renormalization scales step by step
completely in parallel: first, computing perturbative higher order
contributions to scattering amplitudes we encounter ultraviolet and
infrared divergences. We regularize both of them using dimensional
regularization with $n=4 - 2 \epsilon<4$ for ultraviolet and $n>4$ for
infrared divergences, linked by analytic continuation. Both kinds
of divergences are universal, which means that they are not process or
observable dependent. This allows us to absorb ultraviolet and
infrared divergences into a re-definition of the strong coupling and
the parton density. This nominally infinite shift of parameters we
refer to as renormalization for example of the strong coupling or as
mass factorization absorbing infrared divergences into the parton
distributions.

After renormalization as well as after mass factorization we are left
with a \underline{scale artifact}\index{scale artifact}. Scales arise as part of a the pole
subtraction: together with the pole $1/\epsilon$ we have a choice of
finite contributions which we subtract with this pole. Logarithms of
the renormalization and factorization scales will always be part of
these finite terms.  Moreover, in both cases the re-definition of
parameters is not based on fixed order perturbation theory. Instead,
it involves summing logarithms which otherwise can become large and
spoil the convergence of our perturbative series in $\alpha_s$. The
only special feature of infrared divergences as compared to
ultraviolet divergences is that to identify the resummed logarithms we
have to unify both scales to one.

The hadronic production cross section for the Drell--Yan process or
other LHC production channels, now including both scales, reads
\begin{alignat}{5}
\sigma_\text{tot}(\mu_F,\mu_R) 
             = \int_0^1 dx_1 \int_0^1 dx_2 \;
               \sum_{ij} \; f_i(x_1,\mu_F) \, f_j(x_2, \mu_F) \;
               \hat{\sigma}_{ij}(x_1 x_2 S, \alpha_s(\mu_R^2), \mu_F, \mu_R) \; .
\label{eq:qcd_hadronic}
\end{alignat}
The Drell--Yan process has the particular feature that at leading
order $\hat{\sigma}_{q\bar{q}}$ only involves weak couplings, it does
not include $\alpha_s$ with its implicit renormalization scale
dependence at leading order. Strictly speaking, in
Eq.\eqref{eq:qcd_hadronic} the parton densities also depend on the
renormalization scale because in their extraction we identify both
scales. Carefully following their extraction we can separate the two
scales if we need to. Lepton pair production and Higgs production in
weak boson fusion are the two prominent electroweak production
processes at the LHC.\bigskip

The evolution of all running parameters from one
renormalization/factorization scale to another is described either by
renormalization group equation in terms of a beta function in the case
of renormalization or by the DGLAP equation in the case of mass
factorization. Our renormalization group equation for $\alpha_s$ is a
single equation, but in general they are sets of coupled differential
equations for all relevant parameters, which again makes them more
similar to the DGLAP equation. 

There is one formal difference between these two otherwise very
similar approaches. The fact that we can absorb ultraviolet
divergences into process--independent, universal counter terms is
called renormalizability and has been proven to all orders for the
kind of gauge theories we are dealing with. The universality of
infrared splitting kernels has not (yet) in general been proven, but
on the other hand we have never seen an example where is fails for
sufficiently inclusive observables like production rates.\index{factorization}  For a while
we thought there might be a problem with factorization in
supersymmetric theories using the $\msbar$ scheme, but this issue has
been resolved. A summary of the properties of the two relevant scales
for LHC physics we show in Table~\ref{tab:qcd_scales}.\bigskip

The way we introduce factorization and renormalization scales clearly
labels them as an artifact of perturbation theories with
divergences. What actually happens if we include \underline{all
  orders} in perturbation theory? For example, the resummation of the
self energy bubbles simply deals with one class of diagrams which have
to be included, either order-by-order or rearranged into a
resummation. Once we include all orders in perturbation theory it
does not matter according to which combination of couplings and
logarithms we order it.  An LHC production rate will then not depend
on arbitrarily chosen renormalization or factorization scales $\mu$.

Practically, in Eq.\eqref{eq:qcd_hadronic} we evaluate the
renormalized parameters and the parton densities at some scale. This
scale dependence will only cancel once we include all implicit and
explicit appearances of the scales at all orders. Whatever scale we
choose for the strong coupling or parton densities will eventually be
compensated by explicit scale logarithms. In the ideal case, these
logarithms are small and do not spoil perturbation theory.  In a
process with one distinct external scale, like the $Z$ mass, we know
that all scale logarithms should have the form $\log (\mu/m_Z)$. This
logarithm vanishes if we evaluate everything at the `correct'
external energy scale, namely $m_Z$. In that sense we can think of the
running coupling as a proper \underline{running
  observable}\index{running coupling} which depends on the external
energy of the process. This dependence on the external energy is not a
perturbative artifact, because a cross section even to all orders does
depend on the energy. The problem in particular for LHC analyses is
that after analysis cuts every process will have more than one
external energy scale.\bigskip

We can turn around the argument of vanishing scale dependence to all
orders in perturbation theory. This gives us an estimate of the
minimum \underline{theoretical error}\index{error!theoretical error} on a rate prediction set by the
scale dependence. The appropriate interval of what we consider
reasonable scale choices depends on the process and the taste of the
people doing this analysis. This error estimate is not at all
conservative; for example the renormalization scale dependence of the
Drell--Yan production rate or Higgs production in weak boson fusion is
zero because $\alpha_s$ only enters are next--to--leading order. At the
same time we know that the next--to--leading order correction to the
Drell--Yan cross section is of the order of 30\%, which far exceeds
the factorization scale dependence. Moreover, the different scaling
behavior of a hadronic cross section shown in
Table~\ref{tab:qcd_scales} implies that for example gluon--induced
processes at typical $x$ values around $10^{-2}$ show a cancellation
of the factorization and renormalization scale variation. Estimating
theoretical uncertainties from scale dependence therefore requires a
good understanding of the individual process and the way it is
affected by the two scales.\bigskip

Guessing the right scale choice for a process is hard, often
impossible. For example in Drell--Yan production at leading order
there exists only one scale, $m_Z$. If we
set $\mu = m_Z$ all scale logarithms vanish. In reality, LHC
observables include several different scales. Some of them 
appear in the hard process, for example in the production of two or
three particles with different masses. Others enter through the
QCD environment where at the LHC we only consider final--state jets
above a certain minimal transverse momentum. Even others appear though
background rejection cuts in a specific analysis, for example when we
only consider the Drell--Yan background for $m_{\mu \mu} > 1$~TeV to
Kaluza--Klein graviton production. Using likelihood methods does not
improve the situation because the phase space regions dominated by the
signal will introduce specific energy scales which affect the perturbative prediction of the 
backgrounds. This is one of the reasons why an automatic comparison of
LHC events with signal or background predictions is bound to fail once
it requires an estimate of the theoretical uncertainty on the
background simulation.

All that means that in practice there is no way to define a `correct'
scale. On the other hand, there are definitely 
\underline{poor scale choices}. For example, using $1000 \times m_Z$
as a typical scale in the Drell--Yan process will if nothing else lead
to logarithms of the size $\log 1000$ whenever a scale logarithm
appears. These logarithms eventually have to be cancelled to all
orders in perturbation theory, inducing unreasonably large
higher order corrections.

When describing jet radiation, we usually introduce a phase space
dependent renormalization scale, evaluating the strong coupling at the
transverse momentum of the radiated jet $\alpha_s(\vec{p}_{T,j}^2)$. This
choice gives the best kinematic distributions for the additional
partons because in Section~\ref{sec:qcd_resum_collinear} we have shown
that it resums large collinear logarithms.  

The transverse momentum of a final--state particle is one of scale
choices allowed by factorization; in addition to poor scale choices
there also exist \underline{wrong scale choices}, \ie scale choices
violating physical properties we need. Factorization or the
Kinoshita--Lee--Nauenberg theorem which ensures that soft divergences
cancel between real and virtual emission diagrams are such properties
we should not violate --- in QED the same property is called the
Bloch--Nordsieck cancellation. Imagine picking a factorization scale
defined by the partonic initial state, for example the partonic
center--of--mass energy $s = x_1 x_2 S$.  
We know that this definition is not unique: for any final
state it corresponds to the well defined sum of all momenta
squared. However, virtual and real gluon emission generate different
multiplicities in the final state, which means that the two sources of
soft divergences only cancel until we multiply each of them with
numerically different parton densities. Only scales which are uniquely
defined in the final state can serve as factorization scales.  For the
Drell--Yan process such a scale could be $m_Z$, or the mass of heavy
new-physics states in their production process. So while there is no such
thing as a correct scale choice, there are more or less smart choices,
and there are definitely very poor choices, which usually lead to an
unstable perturbative behavior.

\subsection{Parton shower}
\label{sec:qcd_shower}

In LHC phenomenology we are usually less interested in fixed-order
perturbation theory than in logarithmically enhanced QCD
effects. Therefore, we will not deepen our discussion of hadronic
rates as shown in Eq.\eqref{eq:qcd_hadronic} based on fixed-order
partonic cross sections convoluted with parton densities obeying the
DGLAP equation.  In Section~\ref{sec:qcd_resum_collinear} we have
already seen that there exist more functions with the same structure
as solutions to the DGLAP equation.  In Section~\ref{sec:qcd_sudakov}
we will derive one such object which we can use to describe jet
radiation of incoming and outgoing partons in hard processes. These
Sudakov factors will immediately lead us to a parton shower. They are
based on universal patterns in jet radiation which we will in detail
study in Section~\ref{sec:qcd_multiple} and \ref{sec:qcd_dipoles}.  In
Section~\ref{sec:qcd_ordered} we will introduce a key property of the
parton shower, the ordered splitting of partons in several
approximation. 

\subsubsection{Sudakov form factor}
\label{sec:qcd_sudakov}

After introducing the kernels $\hat{P}_{i \leftarrow j}(z)$ 
as something like splitting probabilities we never applied a
probabilistic approach to parton splitting\index{splitting kernel}. The basis of such an interpretation are \underline{Sudakov
  form factors}\index{Sudakov factor} describing the splitting of a
parton $i$ into any of the partons $j$ based on the factorized form
Eq.\eqref{eq:qcd_factorize}
\begin{alignat}{5}
\boxed{
 \Delta_i(t) \equiv
 \Delta_i(t,t_0) = 
  \exp \left( - \sum_j \int_{t_0}^t \frac{dt'}{t'} 
                \int_0^1 dy \ \frac{\alpha_s}{2 \pi} 
                \hat{P}_{j \leftarrow i}(y) 
        \right) 
} \; .
\label{eq:sudakov1}
\end{alignat}
Sudakov factors are an excellent example for technical terms hiding 
very simple concepts. If we instead referred to them as simple 
non--splitting probabilities everyone would immediately understand 
what we are talking about, taking away some of the mythical powers of 
theoretical physicists.
The only parton
splitting affecting a hard quark leg is $P_{q \leftarrow q}$ while a
gluon leg can either radiate a gluon via $P_{g \leftarrow g}$ or $P_{q
  \leftarrow g}$. The fourth allowed splitting $P_{g \leftarrow q}$
also splits a quark into a quark--gluon pair, so we can decide to
follow the quark direction instead of switching over to the gluon.
We derive the form of the Sudakov factors for $\Delta_q$,
\begin{alignat}{5}
\Delta_q(t) =& 
  \exp \left( - \int_{t_0}^{t} \frac{dt'}{t'} 
                \int_0^1 dy \ \frac{\alpha_s}{2 \pi} 
                \hat{P}_{q \leftarrow q}(y) 
       \right) \; .
\end{alignat}
%
The unregularized splitting kernel is given by Eq.\eqref{eq:qcd_pqq},
so we can compute
\begin{alignat}{5}
\int_0^1 dz \; \frac{\alpha_s}{2\pi} \;
   \hat{P}_{q \leftarrow q}(y) 
&= \frac{C_F}{2\pi} \int_0^1 dy \; \alpha_s \;
     \frac{1+y^2}{1-y} 
\notag \\
&= \frac{C_F}{2\pi} \int_0^1 dy \; \alpha_s \;
    \frac{-(1-y)(1+y)+2}{1-y}
\notag \\
&= \frac{C_F}{2\pi} \int_0^1 dy \; \alpha_s \;
   \left( \frac{2}{1-y} - 1-y \right) \; .
\label{eq:splitting1}
\end{alignat}
To compute the divergent first term we 
shift the $y$ integration to $t'' = (1-y)^2t$, which gives us 
\begin{alignat}{5}
\frac{dt''}{dy} 
= \frac{d}{dy} (1-y)^2 t
= 2 (1-y) (-1) t
= - 2 \frac{t''}{1-y}
\qquad \Leftrightarrow \qquad 
\frac{dy}{1-y} = - \frac{1}{2} \; \frac{dt''}{t''} \; .
\label{eq:splitting2}
\end{alignat}
Without derivation we quote that the $t''$ integration, which naively
has a lower boundary at zero is cut off at $t'$. In addition, we
approximate $y \to 1$ wherever possible and find for the leading
contributions to the splitting integral
\begin{alignat}{5}
\int_0^1 dz \; \frac{\alpha_s}{2\pi} \;
   \hat{P}_{q \leftarrow q}(y) 
&= \frac{C_F}{2\pi} \int_0^1 dy \; \alpha_s \;
   \left( \frac{2}{1-y} - 1-y \right) \notag \\
&= \frac{C_F}{2\pi} 
   \left( \int_{t'}^t dt'' \; \frac{\alpha_s}{t''}
        - \int_0^1 dy \; \alpha_s \; (1+y) 
   \right) \notag \\
&= \frac{C_F}{2\pi} \alpha_s
   \left( \int_{t'}^t dt'' \; \frac{1}{t''}
        - \int_0^1 dy \; (1+y) 
   \right) 
   \qquad \text{leading power dependence on $y$ and $t''$} \notag \\
&= \frac{C_F}{2\pi} \alpha_s
   \left( \log \frac{t}{t'} 
        - \frac{3}{2} 
   \right) 
\equiv t' \; \Gamma_{q \leftarrow q}(t,t') \; .
\label{eq:splitting3}
\end{alignat}
The argument of the strong coupling was originally $y^2(1-y)^2t'$,
which turns into $t'$ in the limit $y \to 1$. This way we can
express all Sudakov factors in terms of splitting functions
$\Gamma_j$,
\begin{alignat}{5}
\Delta_q(t) =& 
\exp \left( - \int_{t_0}^{t} dt' \;        \Gamma_{q \leftarrow q}(t,t')
       \right)
\notag \\
\Delta_g(t) =& 
  \exp \left( - \int_{t_0}^{t} dt' \;
       \left[ \Gamma_{g \leftarrow g}(t,t')
             +\Gamma_{q \leftarrow g}(t') \right]
       \right) \; ,
\end{alignat}
which to leading logarithm in $\log t/t'$ read
\begin{alignat}{5}
  \Gamma_{q \leftarrow q}(t,t') 
         &= \frac{C_F}{2\pi} \; \frac{\alpha_s(t')}{t'} 
             \left( \log \frac{t}{t'} 
                  - \frac{3}{2} 
             \right) \notag \\
  \Gamma_{g \leftarrow g}(t,t') 
         &= \frac{C_A}{2\pi} \; \frac{\alpha_s(t')}{t'} 
             \left( \log \frac{t}{t'} 
                  - \frac{11}{6} 
             \right) \notag \\
  \Gamma_{q \leftarrow g}(t') 
         &= \frac{n_f}{6\pi} \; \frac{\alpha_s(t')}{t'}  \; . 
\label{eq:qcd_split_const}
\end{alignat}
These formulas have a slight problem: terms arising from
next--to--leading logarithms spoil the limit $t' \to t$, where a
splitting probability should vanish.  Technically, we can deal with
the finite terms in the Sudakov factors by requiring them to be
positive semi--definite, \ie by replacing $\Gamma(t,t') < 0$ with
zero. For the general argument this problem with the analytic
expressions for the splitting functions is irrelevant.\bigskip

Before we check that the Sudakov factors obey the DGLAP equation we
confirm that such exponentials appear in probabilistic arguments,
similar to our discussion of the central jet veto in
Section~\ref{sec:higgs_cjv}.  Using Poisson statistics for something
expected to occur $p$ times, the probability of observing it $n$ times
is given by
\begin{alignat}{5}
 \mathcal{P}(n;p) = \frac{p^n \, e^{-p}}{n!} 
 \qqqquad 
 \mathcal{P}(0;p) = e^{-p} \; .
\label{eq:qcd_poisson}
\end{alignat}
If the exponent in the Sudakov form factor in Eq.\eqref{eq:sudakov1}
describes the integrated splitting probability of a parton $i$ this
means that the Sudakov itself describes a 
\underline{non--splitting probability}\index{splitting!no-splitting probability} of the parton $i$ into any final state $j$.

Based on such probabilistic Sudakov factors we can use a Monte Carlo, which is a
\underline{Markov process}\index{Monte Carlo!Markov process} without a memory of individual past steps,
to compute a chain of parton splittings as depicted in
Figure~\ref{fig:qcd_tx}.  This will describe a quark or a gluon
propagating forward in time. Starting from a point $(x_1,t_1)$ in
momentum--virtuality space we step by step move to the next splitting
point $(x_j,t_j)$. Following the original discussion $t_2$ is the
target virtuality at $x_2$, and for time--like final--state branching
the virtuality is positive $t_j>0$ in all points $j$. The Sudakov
factor is a function of $t$, so it gives us the probability of not
seeing any branching between $t_1$ and $t_2$ as
$\Delta(t_1)/\Delta(t_2)<1$. The appropriate cutoff scale
$t_0$ drops out of this ratio. Using a flat random number $r_t$ the
$t_2$ distribution is implicitly given by the solution to
\begin{alignat}{5}
\frac{\Delta(t_1)}{\Delta(t_2)} = r_t \; \epsilon \; [0,1]
\qqqquad \text{with} \quad
t_1 > t_2 > t_0 > 0  \; .
\label{eq:sudakov_rand1}
\end{alignat}
Beyond the absolute cutoff scale $t_0$ we assume
that no resolvable branching occurs.\bigskip

In a second step we need to compute the matching energy fraction $x_2$
or the ratio $x_2/x_1$ describing the momentum fraction which is kept
in the splitting at $x_2$. The $y$ integral in the Sudakov factor in
Eq.\eqref{eq:sudakov1} gives us this probability distribution which we
can again implicitly solve for $x_2$ using a flat random number $r_x$
\begin{alignat}{5}
\frac{\int_0^{x_2/x_1} dy \dfrac{\alpha_s}{2\pi} \hat{P}(y)}
      {\int_0^1  dy \dfrac{\alpha_s}{2\pi} \hat{P}(y)}
= r_x \; \epsilon \; [0,1]
\qqqquad \text{with} \;
x_1 > x_2 > 0  \; .
\label{eq:sudakov_rand2}
\end{alignat}
For splitting kernels with soft divergences at $y=0$ or $y=1$ we
should include a numerical cutoff in the integration because the
probabilistic Sudakov factor and the parton shower do not involve the regularized
splitting kernels.

Of the four momentum entries of the radiated parton the two equations
Eqs.\eqref{eq:sudakov_rand1} and~\eqref{eq:sudakov_rand2} give us
two. The on--shell mass constraint fixes a third, so all we are left is
the azimuthal angle distribution. We know from symmetry arguments that
QCD splitting is insensitive to this angle, so we can generate it
randomly between zero and $2 \pi$. For final--state radiation this
describes probabilistic branching in a \underline{Monte Carlo
  program}\index{Monte Carlo!Monte Carlo generator}, just based on Sudakov form factors.\bigskip

The same statement for initial--state radiation including parton
densities we will put on a more solid or mathematical footing. The
derivative of the Sudakov form factor Eq.\eqref{eq:sudakov1}
\begin{alignat}{5}
 \frac{1}{\Delta_i(t)} \frac{d \Delta_i(t)}{dt} &=
              - \sum_j \frac{1}{t}
                \int_0^1 dy \ \frac{\alpha_s}{2 \pi} 
                \hat{P}_{j \leftarrow i}(y) 
\label{eq:sudakov_derive}
\end{alignat}
is precisely the second term in $d f(x,t)/dt$ for diagonal splitting,
as shown in Eq.\eqref{eq:qcd_almost_dglap}
\begin{alignat}{5}
 \frac{d f_i(x,t)}{d t}
&= \frac{1}{t} \sum_j \left[
   \int_0^1 \frac{dz}{z}  \; \frac{\alpha_s}{2\pi} \;
   \hat{P}_{i \leftarrow j}(z) \; f_j\left(\frac{x}{z},t\right) 
 - \int_0^1 dy  \; \frac{\alpha_s}{2\pi} \;
   \hat{P}_{j \leftarrow i}(y) \; f_i\left(x,t\right) \right]
 \notag \\
&= \frac{1}{t} \sum_j
   \int_0^1 \frac{dz}{z}  \; \frac{\alpha_s}{2\pi} \;
   \hat{P}_{i \leftarrow j}(z) \; f_j\left(\frac{x}{z},t\right) 
 + \frac{f_i(x,t)}{\Delta_i(t)} \frac{d \Delta_i(t)}{d t} \; .
\end{alignat}
This relation suggests to consider the derivative of the
$f_i/\Delta_i$ instead of the Sudakov factor alone to obtain something
like the DGLAP equation
\begin{alignat}{5}
 \frac{d}{d t} \frac{f_i(x,t)}{\Delta_i(t)}
&= \frac{1}{\Delta_i(t)} \; \frac{d f_i(x,t)}{d t} 
 - \frac{f_i(x,t)}{\Delta_i(t)^2} \; \frac{d \Delta_i(t)}{d t} 
 \notag \\
&= \frac{1}{\Delta_i(t)} \; \left(
   \frac{1}{t} \sum_j
   \int_0^1 \frac{dz}{z}  \; \frac{\alpha_s}{2\pi} \;
   \hat{P}_{i \leftarrow j}(z) \; f_j\left(\frac{x}{z},t\right) 
 + \frac{f_i(x,t)}{\Delta_i(t)} \frac{d \Delta_i(t)}{d t} 
   \right)
 - \frac{f_i(x,t)}{\Delta_i(t)^2} \; \frac{d \Delta_i(t)}{d t} 
 \notag \\
&= \frac{1}{\Delta_i(t)} \; \frac{1}{t} \; \sum_j
   \int_0^{1-\epsilon} \frac{dz}{z}  \; \frac{\alpha_s}{2\pi} \;
   \hat{P}_{i \leftarrow j}(z) \; f_j\left(\frac{x}{z},t\right) \; .
\end{alignat}
In the last step we cancel what corresponds to the plus
prescription for diagonal splitting, which means we remove the
regularization of the splitting kernel at $z \to 1$. Therefore, we
need to modify the upper integration boundary by a small parameter
$\epsilon$ which can in principle depend on $t$.  The resulting
equation is the diagonal DGLAP equation\index{DGLAP equation!parton shower} with unsubtracted splitting
kernels, solved by the ratio of parton densities and Sudakov factors
\begin{alignat}{5}
\boxed{
 t \frac{d}{d t} \frac{f_i(x,t)}{\Delta_i(t)}
= \frac{d}{d \log t} \frac{f_i(x,t)}{\Delta_i(t)}
= \sum_j \int_0^{1-\epsilon} \frac{dz}{z}  \; \frac{\alpha_s}{2\pi} \;
   \hat{P}_{i \leftarrow j}(z) \; \dfrac{f_j\left(\dfrac{x}{z},t\right)}{\Delta_i(t)}
} \; .
\label{eq:qcd_sudakov_dglap}
\end{alignat}
\bigskip

We can study the structure of these solutions of the unsubtracted
DGLAP equation by integrating $f/\Delta$ between appropriate points in
$t$
\begin{alignat}{5}
   \frac{f_i(x,t)}{\Delta_i(t)}
 - \frac{f_i(x,t_0)}{\Delta_i(t_0)}
&= \int_{t_0}^t \frac{dt'}{t'} \sum_j \int_0^{1-\epsilon} \frac{dz}{z}  \; \frac{\alpha_s}{2\pi} \;
   \hat{P}_{i \leftarrow j}(z) \; \dfrac{f_j\left(\dfrac{x}{z},t'\right)}{\Delta_i(t')} 
\notag \\
   f_i(x,t)
&= \frac{\Delta_i(t)}{\Delta_i(t_0)} f_i(x,t_0)
 + \int_{t_0}^t \frac{dt'}{t'} \; \frac{\Delta_i(t)}{\Delta_i(t')} 
   \sum_j \int_0^{1-\epsilon} \frac{dz}{z}  \; \frac{\alpha_s}{2\pi} \;
   \hat{P}_{i \leftarrow j}(z) \; f_j\left(\dfrac{x}{z},t'\right)
\notag \\
&= \Delta_i(t) f_i(x,t_0)
 + \int_{t_0}^t \frac{dt'}{t'} \; \frac{\Delta_i(t)}{\Delta_i(t')} 
   \sum_j \int_0^{1-\epsilon} \frac{dz}{z}  \; \frac{\alpha_s}{2\pi} \;
   \hat{P}_{i \leftarrow j}(z) \; f_j\left(\dfrac{x}{z},t'\right) 
\notag \\
&\equiv \Delta_i(t,t_0) f_i(x,t_0)
 + \int_{t_0}^t \frac{dt'}{t'} \; \Delta_i(t,t') 
   \sum_j \int_0^{1-\epsilon} \frac{dz}{z}  \; \frac{\alpha_s}{2\pi} \;
   \hat{P}_{i \leftarrow j}(z) \; f_j\left(\dfrac{x}{z},t'\right) \; ,
\label{eq:qcd_sudakov_interpret}
\end{alignat}
where we choose $t_0$ such that $\Delta(t_0) = 1$ and introduce the
notation $\Delta(t_1,t_2) = \Delta(t_1,t_0)/\Delta(t_2,t_0)$ for the
ratio of two Sudakov factors in the last line.  This equation is a
\underline{Bethe--Salpeter equation}\index{Bethe--Salpeter equation}
describing the dependence of the parton density $f_i(x,t)$ on $x$ and
$t$.It has a suggestive interpretation: corresponding to
Eq.\eqref{eq:qcd_poisson} the first term can be interpreted as
`nothing happening to $f$ between $t_0$ and $t$' because it is
weighted by the Sudakov no-branching probability
$\Delta_i(t,t_0)$. The second term includes the ratio of Sudakov
factors which just like in Eq.\eqref{eq:sudakov_rand1} means no
branching between $t'$ and $t$. Integrating this factor times the
splitting probability over $t' \epsilon [t_0,t]$ implies at least one
branching between $t_0$ and $t$.\bigskip

The key to using this probabilistic interpretation of the Sudakov form factor in
conjunction with the parton densities is its numerical usability in a
probabilistic approach: starting from a parton density somewhere in
$(x-t)$ space we need to evolve it to a fixed point
$(x_n,t_n)$ given by the hard subprocess, \eg $q \bar q \to Z$ with
$m_Z$ giving the scale and energy fraction of the two
quarks. Numerically it would be much easier to simulate
\underline{backwards evolution}\index{parton shower!backwards evolution} where we start from the known
kinematics of the hard process and the corresponding point in the
$(x-t)$ plane and evolve towards the partons in the proton, ideally to
a point where the probabilistic picture of collinear, stable,
non--radiating quarks and gluons in the proton holds. This means we
need to define a probability that a parton evolved backwards from a
space--like $t_2<0$ to $t_1<0$ with $|t_2| > |t_1|$ does not radiate or
split.

For this final step we define a probability measure for the backwards
evolution of partons $\Pi(t_1,t_2;x)$. Just like the two terms in
Eq.\eqref{eq:qcd_sudakov_interpret} it links the splitting probability
to a probability of an undisturbed evolution. For example, we can
write the probability that a parton is generated by a splitting in the
interval $[t, t+\delta t]$, evaluated at $(t_2,x)$, as
$dF(t;t_2)$. The measure corresponding to a Sudakov survival
probability is then
\begin{alignat}{5}
 \Pi(t_1,t_2;x) = 1 - \int_{t_1}^{t_2} dF(t;t_2) \; .
\end{alignat}
Comparing the definition of $dF$ to the relevant terms in
Eq.\eqref{eq:qcd_sudakov_interpret} and replacing $t \to t_2$ and $t'
\to t$ we know what happens for the combination
\begin{alignat}{5}
  f_i(x,t_2) dF(t;t_2) 
&= \frac{dt}{t} \; \frac{\Delta_i(t_2)}{\Delta_i(t)} \;
   \sum_j \int_0^{1-\epsilon} \frac{dz}{z}  \; \frac{\alpha_s}{2\pi} \;
   \hat{P}_{i \leftarrow j}(z) \; f_j\left(\dfrac{x}{z},t \right)
\notag \\
&= dt \; \Delta_i(t_2) \;
   \frac{1}{t} 
   \sum_j \int_0^{1-\epsilon} \frac{dz}{z}  \; \frac{\alpha_s}{2\pi} \;
   \hat{P}_{i \leftarrow j}(z) \; 
   \dfrac{f_j\left(\dfrac{x}{z},t \right)}{\Delta_i(t)}
\notag \\
&= dt \; \Delta_i(t_2) \;
   \frac{d}{d t} \frac{f_i(x,t)}{\Delta_i(t)}
\qqqquad \text{using Eq.\eqref{eq:qcd_sudakov_dglap}} \; .
\end{alignat}
This means 
\begin{alignat}{5}
 \Pi(t_1,t_2;x) 
= 1 - 
   \frac{f_i(x,t) \Delta_i(t_2)}{f_i(x,t_2) \Delta_i(t)}
   \Bigg|_{t_1}^{t_2}
 = \frac{f_i(x,t_1) \Delta_i(t_2)}{f_i(x,t_2) \Delta_i(t_1)} \; ,
\end{alignat}
and gives us a probability measure for backwards evolution: the
probability of evolving back from $t_2$ to $t_1$ is described by
a Markov process with a flat random number as
\begin{alignat}{5}
\frac{f_i(x,t_1) \Delta_i(t_2)}{f_i(x,t_2) \Delta_i(t_1)}
= r \; \epsilon \; [0,1]
\qqqquad \text{with} \;
|t_2| > |t_1| \; .
\label{eq:sudakov_rand3}
\end{alignat}
While we cannot write down this procedure in a closed form, it shows
how we can algorithmically generate initial state as well as final
state parton radiation patterns based on the unregularized DGLAP
equation and the Sudakov factors solving this equation. One remaining
issue is that in our derivation of the collinear resummation
interpretation of the parton shower we assume some a strong
ordering of the radiated partons which we will discuss in the next
section.

\subsubsection{Multiple gluon radiation}
\label{sec:qcd_multiple}

Following Eqs.\eqref{eq:sudakov_rand1} and~\eqref{eq:sudakov_rand3}
the parton shower is fundamentally a statistical approach. 
Sudakov form factors are nothing by no-emission probabilities.
If we limit ourselves only to abelian splitting, \ie radiating gluons
off hard quark legs, the parton shower generates a statistical
distribution of the number of radiated gluons. This guarantees finite
results even in the presence of different infrared divergences. To
understand the picture of parton splitting in terms of a Poisson
process it is most instructive to consider soft gluon emission of a
quark, ignoring the gluon self coupling. In other words, we study soft
photon emission off an electron leg simply adding color factors
$C_F$.\bigskip

To this point we have built our parton shower on collinear parton
splitting or radiation and its universal properties indicated by
Eq.\eqref{eq:qcd_factorize}. Deriving the diagonal splitting kernels
in Eqs.\eqref{eq:qcd_pgg} and~\eqref{eq:qcd_pqq} we encounter an
additional source of infrared divergences, namely \underline{soft
  gluon emission}\index{soft gluon emission} corresponding to energy
fractions $z \to 0,1$. Its radiation pattern is also universal, just
like the collinear case. One way to study this soft divergence without
an overlapping collinear pole is gluon radiation off a massless or
massive hard quark leg

\begin{center}
\begin{fmfgraph*}(120,40)
 \fmfset{arrow_len}{2mm}
 \fmfleft{in1}
 \fmf{fermion,width=0.5,label=$p+k$}{in1,v1}
 \fmf{fermion,width=0.5,label=$p$,tension=1}{v1,out1}
 \fmf{gluon,width=0.5,label=$k$,tension=0.3}{v1,out2}
 \fmfright{out1,out2}
\end{fmfgraph*}
\end{center}

The original massive quark leg with momentum $p+k$ and mass $m$ could
be attached to some hard process as a splitting final state. It splits
into a hard quark $p$ and a soft gluon $k$. The general matrix element
without any approximation reads
\begin{alignat}{5}
\mat_{n+1}
&= g_s T^a \; \epsilon_\mu^*(k) \; 
  \bar{u}(p) \gamma^\mu 
  \frac{\slashchar{p} + \slashchar{k} + m}{(p+k)^2 - m^2} \; 
  \mat_n
\notag \\
&= g_s T^a \; \epsilon_\mu^*(k) \; 
  \bar{u}(p) 
  \left[ -\slashchar{p} \gamma^\mu 
         +2 p^\mu 
         +m \gamma^\mu
         + \gamma^\mu \slashchar{k}
  \right] \;
  \frac{1}{2(p k) + k^2}  \; \mat_n 
\notag \\
&= g_s T^a \; \epsilon_\mu^*(k) \; 
  \bar{u}(p) 
  \frac{2p^\mu + \gamma^\mu \slashchar{k}}
       {2(p k) + k^2} \;  \mat_n \; ,
\end{alignat}
using the Dirac equation $\bar{u}(p) (\slashchar{p}-m ) = 0$. At this
level, a further simplification requires for example the soft gluon
limit.  In the presence of only hard momenta, except for the gluon, we
can define it for example as $k^\mu = \lambda p^\mu$, where $p^\mu$ is
an arbitrary four-vector combination of the surrounding hard momenta. The
small parameter $\lambda$ then characterizes the soft limit. For the
invariant mass of the gluon we assume $k^2 = \ope(\lambda^2)$,
allowing for a slightly off--shell gluon. We find
\begin{alignat}{5}
\mat_{n+1}
&= g_s T^a \; \epsilon_\mu^*(k) \; 
  \bar{u}(p) 
  \frac{p^\mu + \ope(\lambda)}
       {(p k) + \ope(\lambda^2)} \;  \mat_n
\notag \\
&\sim g_s T^a \; \epsilon_\mu^*(k) \; 
  \frac{p^\mu}{(p k)} \;
  \bar{u}(p) \; \mat_n
\notag \\
&\to g_s \; \epsilon_\mu^*(k) \; 
  \left( \sum_j \hat{T}^a_j \frac{p_j^\mu}{(p_j k)} 
  \right) \; 
  \bar{u}(p) \; \mat_n
\label{eq:softgluon1}
\end{alignat}
The conventions are similar to Eq.\eqref{eq:qcd_pgg}, so $\mat_n$
includes all additional terms except for the spinor of the outgoing
quark with momentum $p+k$. Neglecting the gluon momentum altogether defines
the leading term of the \underline{eikonal approximation}\index{eikonal approximation}. 

In the last step of Eq.\eqref{eq:softgluon1} we simply add all
possible sources $j$ of gluon radiation. This defines a color operator
which we insert into the matrix element and which assumes values of
$+T_{ij}^a$ for radiation off a quark, $-T_{ji}^a$ for radiation off
an antiquark and $-i f_{abc}$ for radiation off a gluon. For a color
neutral process like our favorite Drell--Yan process adding an
additional soft gluon $q \bar{q} \to Z g$ it returns $\sum_j \hat{T}_j
= 0$.  For a full QCD calculation, we would need to add single gluon
radiation with a subsequent gluon splitting via the self
interaction. This diagram does not appear in the QED case, it spoils
our argument below, and it is not suppressed by any good arguments.
In the following, we nevertheless strictly limit ourselves to the
abelian part of QCD, \ie gluon radiation off quarks and $f_{abc} \to
0$. This also means that all color factors are real. For the argument
below we can think of gluon radiation off the process $e^+ e^- \to q
\bar{q}$:

\begin{center}
\begin{fmfgraph*}(150,80)
 \fmfset{arrow_len}{2mm}
 \fmfleft{in1,in2}
 \fmf{fermion,width=0.5,label=$e^-$}{in1,v1}
 \fmf{fermion,width=0.5,label=$e^+$}{v1,in2}
 \fmf{photon,width=0.5}{v1,v2}
 \fmf{fermion,width=0.5,label=$q$}{out1,v2}
 \fmf{fermion,width=0.5}{v2,v3}
 \fmf{fermion,width=0.5,label=$\bar{q}$}{v3,out2}
 \fmf{gluon,width=0.5,tension=0.0}{v3,out3}
 \fmfright{out1,out3,out2}
\end{fmfgraph*}
\end{center}

The sum over the gluon radiation dipoles in Eq.\eqref{eq:softgluon1}
covers the two quarks in the final state.  Next, we need to square
this matrix element. It includes a polarization sum and will therefore
depend on the gauge.  We choose the general \underline{axial
  gauge}\index{axial gauge} for massless gauge bosons
\begin{alignat}{5}
\sum_\text{pol} \epsilon_\mu^*(k) \epsilon_\nu(k) 
= - g_{\mu \nu} 
  +\frac{k_\mu n_\nu + n_\mu k_\nu}{(n k)}
  -n^2 \frac{k_\mu k_\nu}{(n k)^2}
= - g_{\mu \nu} 
  +\frac{k_\mu n_\nu + n_\mu k_\nu}{(n k)} \; ,
\end{alignat}
with a light-like reference vector $n$ obeying $n^2=0$. The matrix
element squared then reads
\begin{alignat}{5}
\overline{|\mat_{n+1}|^2}
&= g_s^2 \; \left( - g_{\mu \nu} +\frac{k_\mu n_\nu + n_\mu k_\nu}{(n k)} 
  \right) \;
  \left( \sum_j \hat{T}^a_j \frac{p_j^\mu}{(p_j k)} 
  \right)^\dag \; 
  \left( \sum_j \hat{T}^a_j \frac{p_j^\nu}{(p_j k)} 
  \right) \; 
\overline{|\mat_n|^2}
\notag \\
&= 
g_s^2 \; \left(
-\left( \sum_j \hat{T}^a_j \frac{p_j^\mu}{(p_j k)} \right)^\dag \; 
  \left( \sum_j \hat{T}^a_j \frac{p_{j \mu}}{(p_j k)} \right) \; 
+ \frac{2}{(n k)}
  \left( \sum_j \hat{T}^a_j \right)^\dag \; 
  \left( \sum_j \hat{T}^a_j \frac{(p_j n)}{(p_j k)} \right) \; 
\right)
\overline{|\mat_n|^2}
\notag \\
&= 
- g_s^2 \; \left( \sum_j \hat{T}^a_j \frac{p_j^\mu}{(p_j k)} \right)^\dag \; 
 \left( \sum_j \hat{T}^a_j \frac{p_{j \mu}}{(p_j k)} \right) \; 
\overline{|\mat_n|^2} \; .
\label{eq:qcd_soft_rad}
\end{alignat}
The insertion operator in the matrix element has the form of an
insertion current multiplied by its hermitian conjugate. This current
describes the universal form of soft gluon radiation off an
$n$-particle process
\begin{alignat}{5}
\boxed{
\overline{|\mat_{n+1}|^2}
\equiv - g_s^2 \; (J^\dag \cdot J) \; \overline{|\mat_n|^2}}
\qqquad \text{with} \quad
J^{a \mu}(k,\{p_j\} = \sum_j \hat{T}^a_j \;
\frac{p_j^\mu}{(p_j k)}  \; .
\label{eq:def_softcurrent}
\end{alignat}
\bigskip

We can further simplify the squared current appearing in
Eq.\eqref{eq:qcd_soft_rad} to
\begin{alignat}{5}
(J^\dag \cdot J) 
&= \sum_j \hat{T}^a_j \hat{T}^a_j \;
          \frac{p_j^2}{(p_j k)^2} 
 + 2 \sum_{i < j} \hat{T}^a_i \hat{T}^a_j \;
          \frac{(p_i p_j)}{(p_i k)(p_j k)} 
\notag \\
&= \sum_j \hat{T}^a_j 
          \left( - \sum_{i \ne j} \hat{T}^a_i
          \right) 
          \frac{p_j^2}{(p_j k)^2} 
 + 2 \sum_{i < j} \hat{T}^a_i \hat{T}^a_j \;
          \frac{(p_i p_j)}{(p_i k)(p_j k)} 
\notag \\
&= - \left( \sum_{i < j} + \sum_{i > j} \right)
          \hat{T}^a_i \hat{T}^a_j 
          \frac{p_j^2}{(p_j k)^2} 
 + 2 \sum_{i < j} \hat{T}^a_i \hat{T}^a_j \;
          \frac{(p_i p_j)}{(p_i k)(p_j k)} 
\notag \\
&= 2 \sum_{i < j} \hat{T}^a_i \hat{T}^a_j \;
     \left( \frac{(p_i p_j)}{(p_i k)(p_j k)} 
           -\frac{p_i^2}{2(p_i k)^2} 
           -\frac{p_j^2}{2(p_j k)^2} 
     \right) \qqquad &&\text{in the general massive case}
\notag \\
&= 2 \sum_{i < j} \hat{T}^a_i \hat{T}^a_j \;
     \frac{(p_i p_j)}{(p_i k)(p_j k)} 
\qqqquad &&\text{for massless partons} \notag \\
&= 2 \sum_{i < j} \hat{T}^a_i \hat{T}^a_j \;
 \frac{(p_i p_j)}{(p_i k) + (p_j k)} \; 
  \left( \frac{1}{(p_i k)} + \frac{1}{(p_j k)} \right)
\; .
\label{eq:qcd_dipole2a}
\end{alignat}
In the last step we only bring the eikonal factor into a different form
which sometimes comes in handy because it separates the two
divergences associated with $p_i$ and with $p_j$\bigskip

At this point we return to massless QCD partons, keeping in
mind that the ansatz Eq.\eqref{eq:softgluon1} ensures that the
insertion currents only model soft, not collinear radiation.  Just as
a side remark at this stage --- our definition of the insertion
current $J^{a \mu}$ in Eq.\eqref{eq:def_softcurrent} can be generalized
to colored processes, where the current becomes dependent on the gauge
vector $n$ to cancel the $n$ dependence of the polarization sum
\begin{alignat}{5}
J^{a \mu}(k,\{p_j\} = \sum_j \hat{T}^a_j \;
\left( \frac{p_j}{(p_j k)} - \frac{n}{(n k)} 
\right)
\label{eq:qcd_dipole2}
\end{alignat}
We will study this \underline{dipole radiation term}\index{dipole radiation} in
Eqs.\eqref{eq:def_softcurrent} and~\eqref{eq:qcd_dipole2} 
in detail. Calling them dipoles is a little bit of a
stretch if we compare it to a multipole series. To see the actual
dipole structure we would need to look at the color structure.\bigskip

Based on Eq.\eqref{eq:qcd_soft_rad} for the eikonal limit we can write
out the infinitesimal cross sections for soft gluon emission off a
massless quark. The difference between the usual QED calculation and
our QCD version of it are the color factors. If we take for example
the Feynman diagram for the Drell--Yan process we see that the color
factor for gluon ration off the outgoing quark diagram squared is
$\hat{T}^a \hat{T}^a = - \tr(T^a T^a) = -(N_c^2-1)/2$. Similarly to
the derivation of the splitting kernels we need to include a factor
$1/(2 N_c)$ to account for the averaging over the states of the
intermediate quark. The phase space factor for any number of final
state gluons we postpone at this stage. In terms of the momenta $p_1$
and $p_2$ of the outgoing quark and antiquark
Eq.\eqref{eq:qcd_dipole2a} gives us for the fully massive case
\begin{alignat}{5}
d\sigma_{n+1} 
&=  g_s^2 \tr(T^a T^a) \; \frac{1}{2 N_c} \; d\sigma_n \; 
\int \frac{d^3k}{(2\pi)^3 2 k_0}
     \left( \frac{(p_1 p_2)}{(p_1 k)(p_2 k)} 
           -\frac{p_1^2}{2(p_1 k)^2} 
           -\frac{p_2^2}{2(p_2 k)^2} 
     \right) \notag \\
&=  g_s^2 d\sigma_n \; \frac{N_c^2-1}{4 N_c} \;
\int \frac{d^3k}{(2\pi)^3 2 k_0}
     \left( \frac{(p_1 p_2)}{(p_1 k)(p_2 k)} 
           -\frac{m^2}{2(p_1 k)^2} 
           -\frac{m^2}{2(p_2 k)^2} 
     \right) \; .
\label{eq:qcd_soft1}
\end{alignat}
Because we are interested in the dependence on the gluon energy it is
not convenient to stick to kinematic invariants. Instead, we compute
the phase space integral over the additional gluon momentum in a
specific reference frame. We choose both, the quark and the antiquark
energy to be the same $E_1 = E_2 \equiv E_q$, with the corresponding
three-momenta $E_q \vec{v}_j$. The three-momentum of the gluon has the
direction $\hat{k}$. In this frame we find
\begin{alignat}{5}
\frac{(p_1 p_2)}{(p_1 k)(p_2 k)} 
           -\frac{m^2}{2(p_1 k)^2} 
           -\frac{m^2}{2(p_2 k)^2} 
&= \frac{E_q^2 (1 - \vec{v}_1 \vec{v}_2)}
       {E_q^2 k_0^2(1 - \vec{v}_1 \hat{k})(1 - \vec{v}_2 \hat{k})}
- \frac{m^2}{2E_q^2 k_0^2 (1 - \vec{v}_1 \hat{k})^2}
- \frac{m^2}{2E_q^2 k_0^2 (1 - \vec{v}_2 \hat{k})^2}
\notag \\
&= \frac{1}{k_0^2} \left( 
\frac{(1 - \vec{v}_1 \vec{v}_2)}
       {(1 - \vec{v}_1 \hat{k})(1 - \vec{v}_2 \hat{k})}
- \frac{m^2}{2E_q^2 (1 - \vec{v}_1 \hat{k})^2}
- \frac{m^2}{2E_q^2 (1 - \vec{v}_2 \hat{k})^2}
\right) \; .
\label{eq:qcd_soft2a}
\end{alignat}
The numerical most relevant axes in the angular integral over the
gluon momentum direction $\hat{k}$ appear when the scalar products in
three dimensions give $\vec{v}_j \hat{k} =1$. We first deal with the
second and third integrals in Eq.\eqref{eq:qcd_soft2a}, using
massless polar coordinates 
$d^3k = k_0^2 dk_0 d \cos \theta_k d \phi_k = 2 \pi k_0^2 dk_0 d \cos \theta_k $,
\begin{alignat}{5}
\int \frac{d^3k}{(2\pi)^3 2 k_0}
\; \frac{1}{k_0^2} \; \frac{m^2}{2E_q^2 (1 - \vec{v} \hat{k})^2}
&= \frac{1}{8 \pi^2} \frac{m^2}{E_q^2} \int \frac{d k_0}{2 k_0} \; \int_{-1}^1 d \cos \theta_k 
\frac{1}{(1- |\vec{v}| \cos \theta_k)^2}
\notag \\
&= \frac{1}{8 \pi^2} \frac{m^2}{E_q^2} \int \frac{d k_0}{2 k_0} \; 
\left[ \frac{(-1)}{|\vec{v}|} \; \frac{1}{|\vec{v}| \cos \theta_k -1}
\right]_{-1}^1 
\notag \\
&= - \frac{1}{8 \pi^2} \frac{m^2}{E_q^2} \int \frac{d k_0}{2 k_0} \; \frac{1}{|\vec{v}|} \; 
\left( 
  \frac{1}{|\vec{v}| -1} + \frac{1}{|\vec{v}| +1}
\right)
\notag \\
&= - \frac{1}{8 \pi^2} \frac{m^2}{E_q^2} \int \frac{d k_0}{2 k_0} \; \frac{1}{|\vec{v}|} \; 
  \frac{2|\vec{v}|}{|\vec{v}|^2 -1} 
\notag \\
&= - \frac{m^2}{8 \pi^2} \int \frac{d k_0}{k_0} \; 
  \frac{1}{|\vec{p}_q|^2 -E_q^2} 
= \frac{1}{4 \pi^2} \int \frac{d k_0}{k_0}
= \frac{1}{4 \pi^2} \log \frac{k_0^\text{max}}{k_0^\text{min}}  \; .
\label{eq:qcd_soft2b}
\end{alignat}
This result is logarithmically divergent in the limit of soft gluon
radiation, but there are not issues with the angular integration over
the gluon phase space and collinear configurations.

The first term in Eq.\eqref{eq:qcd_soft2a} has a more complex
divergence structure. We can separate the two poles in the integrand
and approximate the integrand by the respective residues,
\begin{alignat}{5}
\int \frac{d^3k}{(2\pi)^3 2 k_0} \; \frac{1}{k_0^2} 
\; \frac{(1 - \vec{v}_1 \vec{v}_2)}
        {(1 - \vec{v}_1 \hat{k})(1 - \vec{v}_2 \hat{k})}
&\simeq
\int \frac{d^3k}{(2\pi)^3 2 k_0} \; \frac{1}{k_0^2} 
\left(
  \frac{(1 - \vec{v}_1 \vec{v}_2)}
       {(1 - \vec{v}_1 \hat{k}_\text{pole})(1 - \vec{v}_2 \vec{v}_1)}
+ \frac{(1 - \vec{v}_1 \vec{v}_2)}
       {(1 - \vec{v}_1 \vec{v}_2)(1 - \vec{v}_2 \hat{k}_\text{pole})}
\right)
\notag \\
&=
\int \frac{d^3k}{(2\pi)^3 2 k_0} \; \frac{1}{k_0^2} 
\left(
  \frac{1}{1 - \vec{v}_1 \hat{k}}
+ \frac{1}{1 - \vec{v}_2 \hat{k}}
\right)_\text{pole}
\notag \\
&= \frac{1}{8\pi^2} \int \frac{dk_0}{k_0} \int d \cos \theta_k 
\left(
  \frac{1}{1 - \vec{v}_1 \hat{k}}
+ \frac{1}{1 - \vec{v}_2 \hat{k}}
\right)_\text{pole}
\notag \\
&= \frac{1}{4\pi^2} \int \frac{dk_0}{k_0} \int d \cos \theta_k \;
  \frac{1}{1 - \cos \theta_k}
\qqquad \text{with appropriate reference axes}
\notag \\
&= \frac{1}{4\pi^2} \; \log \frac{k_0^\text{max}}{k_0^\text{min}}
    \; \log \frac{1-\cos \theta_k^\text{max}}{1 - \cos \theta_k^\text{min}} \; .
\label{eq:qcd_soft3}
\end{alignat}
In this calculation we neglect effects of order $m^2/E_q^2$ when
identifying $\vec{v} \hat{k} = \cos \theta_k$.  We see that in the
double integral of Eq.\eqref{eq:qcd_soft3} both parts diverge
logarithmically, usually referred to as the \underline{Sudakov double
  logarithm}. The first integral develops an infrared divergence in
the gluon energy $k_0$ when the gluon becomes soft, $k_0^\text{min}
\to 0$.  The second integral diverges when the gluon is radiated
collinearly with the hard quark or antiquark, $\cos
\theta_k^\text{max} \to 1$. The integrals in Eq.\eqref{eq:qcd_soft2b}
are less divergent, so we can neglect them in the corresponding
differential cross sections of Eq.\eqref{eq:qcd_soft1}
\begin{alignat}{5}
d\sigma_{n+1} 
&= d\sigma_n \; \frac{g_s^2 C_F}{2}
  \frac{1}{4\pi^2} \; \log \frac{k_0^\text{max}}{k_0^\text{min}}
    \; \log \frac{1-\cos \theta_k^\text{max}}{1 - \cos \theta_k^\text{min}} 
\notag \\
&= d\sigma_n \; \frac{\alpha_s C_F}{2 \pi} 
    \; \log \frac{k_0^\text{max}}{k_0^\text{min}}
    \; \log \frac{1-\cos \theta_k^\text{max}}{1 - \cos \theta_k^\text{min}} \; .
\label{eq:qcd_soft4}
\end{alignat}
To be able to continue with our calculation we resort to an obvious
regularization scheme --- the detector. Arbitrarily soft photons leave
no trace in a calorimeter, so they are not observable. The detector
threshold acts as a finite cutoff $k_0^\text{min} > 0$. Similarly, a
tracker cannot separate two tracks which are arbitrarily close to each
other, which means that its resolution limits the $\cos \theta_k$
range.\bigskip

From the general principles of field theory we know that soft
divergences cancel once we combine virtual gluon exchange diagrams and
real gluon emission at the same order in perturbation theory. The soft
cutoff in the $k_0$ integration we assume to be linked between real
and virtual diagrams using a Wick rotation.  For QCD this is called
the Kinoshita--Lee--Nauenberg theorem. The relevant Feynman diagrams
are propagator corrections for the $m^2$ terms in
Eq.\eqref{eq:qcd_soft1} and vertex corrections for the double
divergences. Certainly, the leading overlapping soft and collinear
divergences in Eq.\eqref{eq:qcd_soft3} should vanish after we combine
real and virtual QCD corrections for the Drell--Yan process. This
means that after adding virtual corrections we can assume
$k_0^\text{min}$ to be an experimental constraint without any issues
in the limit $k_0^\text{min} \to 0$. If everything is well defined we
can exponentiate the successive dependence of
Eq.\eqref{eq:qcd_soft4}. The only complication is that now we have to
include the correction for the phase space integration of the many
gluons in the final state. If we declare $n$ the number of gluons
radiated this factor is $1/n!$. The observable we are interested in is
the cross section for any number of radiated gluons, for which we find
\begin{alignat}{5}
d\sigma_{n+1} 
&= d\sigma_0 \; \frac{1}{(n+1)!} \; 
\left( - \frac{\alpha_s C_F}{2 \pi} 
    \; \log \frac{k_0^\text{max}}{k_0^\text{min}}
    \; \log \frac{1-\cos \theta_k^\text{min}}{1 - \cos \theta_k^\text{max}} 
\right)^{n+1} 
\notag \\
\Rightarrow \qquad 
&\sigma_\text{tot} 
= \sigma_0 \; \sum_n 
\; \frac{1}{n!} \; 
\left( - \frac{\alpha_s C_F}{2 \pi} 
    \; \log \frac{k_0^\text{max}}{k_0^\text{min}}
    \; \log \frac{1-\cos \theta_k^\text{min}}{1 - \cos \theta_k^\text{max}} 
\right)^n
\notag \\
& \boxed{ \sigma_\text{tot} 
= \sigma_0 \; \exp
\left( - \frac{\alpha_s C_F}{2 \pi} 
    \; \log \frac{k_0^\text{max}}{k_0^\text{min}}
    \; \log \frac{1-\cos \theta_k^\text{min}}{1 - \cos \theta_k^\text{max}} 
\right)} \; .
\label{eq:qcd_soft5}
\end{alignat}
This pattern implies that the number of radiated gluons in the
Drell--Yan process, neglecting the triple gluon vertex and only taking
into account the leading logarithms, follows a Poisson pattern. The
total cross section as well as the distribution of the radiated gluons
are both well defined even in the limit of $k_0^\text{min} \to
0$. 

\subsubsection{Catani--Seymour dipoles}
\label{sec:qcd_dipoles}

From the previous discussions we know that parton or jet radiation is
dominated by the collinear and soft limits and the double enhancement
shown in Eq.\eqref{eq:qcd_soft5}.  The universal collinear limit of
the different parton splittings is described by the unregularized
splitting kernels $\hat{P}_{i \leftarrow j}(z)$. They are the basis of
the parton shower description of jet radiation. The problem with the
unregularized splittings is that part of them are divergent in the
soft limit $z \to 0$. The question is if we can find an approximate
description of parton splitting including the soft divergence in
addition to the collinear enhancement. Such a description is given by
the Catani--Seymour dipoles and serves as the basis of the shower in
the SHERPA\index{event generators!SHERPA} event generator.\bigskip

Radiating a soft gluon off a hard quark leg is kinematically easy: the
eikonal\index{eikonal approximation} limit shown in Eq.\eqref{eq:softgluon1} leaves the quark
momentum untouched, for example allowing us to define all three
particles involved in the splitting to remain on their mass
shells. For the collinear splitting the situation is less simple.  To
describe the splitting of a quark into a quark and a gluon $(k^\mu)$
we use the Sudakov decomposition of
Eq.\eqref{eq:qcd_sudakovdec}. During our computation of the splitting
kernels it becomes obvious that this parameterization of the momenta
has its shortcomings. The missing on--shell conditions for the partons
involved in the splitting are a serious problem for the implementation
of the splitting processes in a parton shower and its comparison to
data. The question is if we can define the momenta involved in a
parton splitting more appropriately.\bigskip

Clearly, just moving around momentum definitions for example starting
from the Sudakov decomposition will not be helpful. We are missing the
necessary degrees of freedom to allow for all on--shell
conditions. The trick is to include a third parton in the picture: let
us assume an \underline{emitter parton} splitting $\tilde{p}_{1k} \sim
p_1 + k$ together with another, \underline{spectator parton}
$\tilde{p}_s$, where the splitting process and the spectator can
exchange momentum \medskip

\begin{center}
\begin{fmfgraph*}(120,50)
 \fmfset{arrow_len}{2mm}
 \fmfleft{in1,in2}
 \fmf{fermion,width=0.5,lab.side=right,label=$\tilde{p}_{1k}$}{in1,v1}
 \fmf{fermion,width=0.5,lab.side=right,label=$p_1+k$}{v1,v2}
 \fmf{fermion,width=0.5,lab.side=left,label=$p_1$}{v2,out2}
 \fmf{gluon,width=0.5,label=$k$,tension=0.5}{v2,out1}
 \fmf{fermion,width=0.5,lab.side=left,label=$\tilde{p}_s$}{in2,v3}
 \fmf{fermion,width=0.5,tension=0.5,lab.side=left,label=$p_s$}{v3,out3}
 \fmf{photon,width=0.5,lab.side=left,label=re-shuffle,tension=0.3}{v1,v3}
 \fmfright{out1,out2,out3}
\end{fmfgraph*}
\end{center}

The momentum exchange between emitter and spectator respects momentum
conservation,
\begin{alignat}{5}
\boxed{\tilde{p}_{1k}^\mu + \tilde{p}_s^\mu 
= p_1^\mu + k^\mu + p_s^\mu} \; .
\label{eq:cs_momenta}
\end{alignat}
In this picture the splitting process $\tilde{p}_{1k} \to p_1 + k$
does not conserve momentum. Instead, we prefer to require that also
the emitter is on its mass shell, forgetting phase space factors in
the splitting process for a moment. This means for massless partons
that we simultaneously postulate
\begin{alignat}{5}
\tilde{p}_{1k}^2 = \tilde{p}_s^2 = p_1^2 = k^2 = p_s^2 = 0 \; .
\label{eq:cs_onshell}
\end{alignat}
The last three conditions we assume can be fulfilled.  The second
condition we can realize defining an appropriate momentum exchange
with $\tilde{p}_s^2 \propto p_s^2$,
\begin{alignat}{5}
\boxed{p_s^\mu = (1-y) \; \tilde{p}_s^\mu}  
\qquad \Rightarrow \qquad 
\tilde{p}_{1k}^\mu 
&= p_1^\mu + k^\mu + p_s^\mu - \tilde{p}_s^\mu
\notag \\
&= p_1^\mu + k^\mu - y \tilde{p}_s^\mu
\end{alignat}
The exchanged momentum fraction $y$ we can compute using the one
remaining on--shell condition
\begin{alignat}{5}
&& 0 \really \tilde{p}_{1k}^2
&= \left( p_1 + k + p_s - \frac{p_s}{1-y} \right)^2 
 = \left( p_1 + k - \frac{y}{1-y} \; p_s \right)^2 
 = \left( p_1 + k - \frac{1}{y^{-1}-1} \; p_s \right)^2 
\notag \\
\qquad &\Leftrightarrow&\qquad
0
&= \left( (y^{-1}-1) p_1 + (y^{-1}-1)  k - p_s \right)^2 
 = 2 (y^{-1}-1) \left[  
   (y^{-1}-1) (p_1 k) - (p_1 p_s) - (p_s k) \right]
\notag \\
\qquad &\Leftrightarrow& \qquad
\frac{(p_1 k)}{y}
&= (p_1 k) + (p_1 p_s) + (p_s k) 
\notag \\
\qquad &\Leftrightarrow& \qquad
y &= 
\frac{(p_1 k)}{(p_1 k) + (p_1 p_s) + (p_s k)} \; .
\label{eq:cs_ydef}
\end{alignat}
Next, we need to define the momentum fraction $\tilde{z}$ which the
final--state particle $p_1$ carries away from the emitter. We define it
in terms of the projection onto $\tilde{p}_s$ instead of the
four-momentum itself,
\begin{alignat}{5}
(p_1 \tilde{p}_s) = 
\tilde{z}_1 \; (\tilde{p}_{1k} \tilde{p}_s)
\qquad \Leftrightarrow \qquad
\tilde{z}_1 
= \frac{(p_1 \tilde{p}_s)}{(\tilde{p}_{1k} \tilde{p}_s)}
= \frac{(p_1 \tilde{p}_s)}{(p_1 \tilde{p}_s) + (k \tilde{p}_s) -y \tilde{p}_s^2}
= \frac{(p_1 p_s)}{(p_1 p_s) + (p_s k)} \; .
\label{eq:cs_z1a}
\end{alignat}
The second momentum fraction then fulfills 
\begin{alignat}{5}
(k \tilde{p}_s) = 
\tilde{z}_k \; (\tilde{p}_{1k} \tilde{p}_s)
\qquad \Leftrightarrow \qquad
\tilde{z}_1 + \tilde{z}_k
= \frac{(p_1 p_s)}{(p_1 p_s) + (p_s k)} 
+ \frac{(p_s k)}{(p_1 p_s) + (p_s k)} 
= 1 \; .
\label{eq:cs_z1b}
\end{alignat}
In this parameterization we can look at the soft and collinear limits.
According to Eq.\eqref{eq:qcd_dipole2a} the relevant kinematic
variable is $(p_1 k) + (p_s k)$, which we want to express in terms
of $\tilde{z}_{1,k}$ and $y$. We find for the leading pole
\begin{alignat}{5}
 1 - y 
&= \frac{(p_1 p_s)+(p_s k)}{(p_1 k) + (p_1 p_s) + (p_s k)}
\notag \\
\tilde{z}_1 (1-y) 
&= \frac{(p_1 p_s)}{(p_1 k) + (p_1 p_s) + (p_s k)}
\notag \\
\frac{1}{1 - \tilde{z}_1 (1-y)}
&= \frac{(p_1 k) + (p_1 p_s) + (p_s k)}{(p_1k)+(p_sk)} \; ,
\label{eq:cs_divergent}
\end{alignat}
so the divergence of the intermediate propagator is described by the
combination $1/(1 - \tilde{z}_1 (1-y))$. At this point we have
convinced ourselves that the kinematic description including a
spectator quark solves the problems with the on--shell partons
consistently. The only prize we have to pay is the slightly more
complicated form of the divergent kinematic variable in
Eq.\eqref{eq:cs_divergent}. The second question is if anything
unexpected happens in the soft and collinear limits. As in many
instances, we limit ourselves to final state gluon radiation, because
the combination of initial--state and final--state partons leads to many
different cases which are technically more involved.\bigskip

In the \underline{soft limit} the structure of the radiation matrix
element is given by Eq.\eqref{eq:def_softcurrent}. In the presence of
only hard momenta of the kind $p^\mu$, except for the gluon, we can
define the soft limit as $k^\mu = \lambda p^\mu$. The small parameter
$\lambda$ then characterizes the soft limit. From Eq.\eqref{eq:cs_z1b}
we know that in this limit $\tilde{z}_k \to 0$ while $\tilde{z}_1 \to
1$. The parameter $y$ computed in Eq.\eqref{eq:cs_ydef} scales like
\begin{alignat}{5}
&&y 
&= \frac{(p_1 k)}{(p_1 k) + (p_1 p_s) + (p_s k)} 
= \lambda \; \frac{(p p_1)}{(p_s k) + \ope(\lambda)} 
\to 0
\notag \\
&\Leftrightarrow& \qquad 
\tilde{p}_s^\mu &= p_s^\mu + \ope(\lambda) \notag \\
&\Leftrightarrow& \qquad 
\tilde{p}_{1k}^\mu &= p_1^\mu + \ope(\lambda) \; .
\label{eq:cs_soft}
\end{alignat}
This is precisely the leading term in the eikonal approximation\index{eikonal approximation},
assuming that the radiation of a soft gluon does not change the hard
radiating quark leg. We can compute the form of the divergence
following Eq.\eqref{eq:cs_divergent}, namely
\begin{alignat}{5}
\frac{1}{1 - \tilde{z}_1 (1-y)}
= \frac{(p_1 k) + (p_1 p_s) + (p_s k)}{(p_1k)+(p_sk)} 
= \frac{1}{\lambda} \; \frac{(p_1 p_s) + \ope(\lambda)}{(p_1p)+(p_sp)} \; ,
\end{alignat}
with an appropriate choice of a hard reference momentum $p$. This
expression we can use to compute the soft splitting kernel for example
for quark splitting into a hard quark and a soft gluon.\bigskip

In the \underline{collinear limit} we define a transverse momentum
component, similar to Eq.\eqref{eq:qcd_sudakovdec}. The only
difference is that now we can require all participating partons to be
on--shell, as seen in Eq.\eqref{eq:cs_onshell}. We write the two
momenta in the final state as
\begin{alignat}{5}
p_1 = z p + p_T - \frac{p_T^2}{z} \frac{n}{2(pn)}
\qqqquad
k = (1-z) p - p_T - \frac{p_T^2}{1-z} \frac{n}{2(pn)} \; ,
\label{eq:cs_momenta_split}
\end{alignat}
where we postulate $p^2=0$, $n^2=0$, and $(p p_T) = 0 = (n p_T)$. The
momentum $p$ is defined as the sum of $p_1$ and $k$, modulo
contributions of order $p_T$. Aside from the on--shell conditions, this
corresponds to the original Sudakov decomposition. We can confirm that
the condition $p^2 = \ope(p_T^2)$ in addition to the exact relations
$p_1^2 = 0 = k^2$ is allowed,
\begin{alignat}{5}
(p_1+k)^2 = 2 (p_1 k)
&= 2 \left( z p + p_T - \frac{p_T^2}{z} \frac{n}{2(pn)} \right) \;
   \left( (1-z) p - p_T - \frac{p_T^2}{1-z} \frac{n}{2(pn)} \right) 
\notag \\
&= - 2p_T^2 
- \frac{2p_T^2}{z} \frac{(1-z)(pn)}{2(pn)}
- \frac{2p_T^2}{1-z} \frac{z(pn)}{2(pn)} 
\notag \\
&= - p_T^2
\left( 2 + \frac{1-z}{z} + \frac{z}{1-z} \right)
\notag \\
&= - p_T^2 \; \frac{2z -2z^2 + (1 -2 z +z^2) + z^2}{z(1-z)}
 = - \frac{1}{2z(1-z)} \; .
\end{alignat}
This is also the relevant small Mandelstam variable for the collinear
splitting. The simple result also motivates the choice of pre-factors
of $n$ in the ansatz of Eq.\eqref{eq:cs_momenta_split}. The additional
factor $1/(z(1-z))$ is needed once we compute the proper divergence
defined in Eq.\eqref{eq:cs_divergent}.  Again, we now describe the
splitting kinematics in the collinear limit as
\begin{alignat}{5}
y 
&= \frac{(p_1 k)}{(p_1 k) + (p_1 p_s) + (p_s k)} 
 = - \frac{p_T^2}{2z(1-z)} \; \frac{1}{(p_s(p_1+ k)) + \ope(p_T^2)} \to 0
\notag \\
\tilde{p}_s^\mu &= p_s^\mu + \ope(p_T^2) \notag \\
\tilde{p}_{1k}^\mu &= p_1^\mu + k^\mu + \ope(p_T^2) \notag \\
\tilde{z}_1 &= z + \ope(p_T^2) \; .
\label{eq:cs_collinear}
\end{alignat}
As for the soft case, the momentum re-shuffling does not affect the
leading terms in the collinear limit.  Our divergent Mandelstam
variable becomes
\begin{alignat}{5}
\frac{1}{1 - \tilde{z}_1 (1-y)}
&= \dfrac{1}{1 - z} \left( 1 + \ope(p_T^2) \right) \; ,
\end{alignat}
which is exactly the $z$ behavior of the unregularized spitting kernel
$\hat{P}_{q \leftarrow q}$ in Eq.\eqref{eq:qcd_pqq}. We see that the
Catani--Seymour description of parton splitting with its spectator
parton not only allows us to keep all participating particles on their
mass shell, it also correctly describes the soft as well as the
collinear splitting point--by--point in phase space. Without going into
the reasons we should mention that this soft--collinear description of
jet matrix elements turns out to be much more successful than one
would expect. The momentum regime in which the Catani--Seymour dipoles
describe LHC results extends far beyond $p_T \lesssim m_Z$.\bigskip

We will see in Section~\ref{sec:qcd_nlo} that the correct modelling of
parton splittings in the soft and collinear limits is a key ingredient
to higher order calculations of LHC cross sections. These calculations
are the main application of Catani--Seymour dipoles. However, for this
calculation we need to also integrate the expressions for soft and
collinear splitting amplitudes over phase space. The specific
parameterization which allows us to assume that all particles in the
splitting process are on their mass shells. When we include the mother
and the spectator momenta $\tilde{p}_{1k}$ and $\tilde{p}_s$ in the
factorized form of the $n$-particle and $(n+1)$-particle phase space
the re-mapping in Eq.\eqref{eq:cs_momenta} leads to an additional
Jacobian $(1-y)/(1 - \tilde{z}_1)$.

\subsubsection{Ordered emission}
\label{sec:qcd_ordered}

From the derivation of the Catani--Seymour dipoles we know that for
example the emission of a gluon off a hard quark line is governed by
distinctive soft and collinear phase space regimes. In our argument
for the exponentiation of gluon radiation matrix elements in
Eq.\eqref{eq:qcd_soft5} there is one piece missing: multiple gluon
emission has to be ordered by some parameter, such that in squaring
the multiple emission matrix element we can neglect interference
terms.  These interference diagrams contributing to the full amplitude
squared are called non--planar diagrams.  The question is if we can
justify to neglect them from first principles field theory and
QCD. There are three reasons to do this, even though none of them
gives exactly zero for soft and collinear splittings. On the other
hand, in combination they make for a very good reason.\bigskip

First, an arguments for a strongly ordered gluon emission comes from
the \underline{divergence structure} of soft and collinear gluon
  emission.  Two successively radiated gluons look like

\begin{center}
\begin{fmfgraph*}(120,50)
 \fmfset{arrow_len}{2mm}
 \fmfleft{in1}
 \fmf{fermion,width=0.5,label=$p+k_1+k_2$}{in1,v1}
 \fmf{fermion,width=0.5,label=$p+k_2$}{v1,v2}
 \fmf{fermion,width=0.5,label=$p$,tension=1}{v2,out1}
 \fmf{gluon,width=0.5,label=$k_1$,tension=0.3}{v1,out3}
 \fmf{gluon,width=0.5,label=$k_2$,tension=0.3}{v2,out2}
 \fmfright{out1,out2,out3}
\end{fmfgraph*}
\qqquad 
\begin{fmfgraph*}(120,50)
 \fmfset{arrow_len}{2mm}
 \fmfleft{in1}
 \fmf{fermion,width=0.5,label=$p+k_1+k_2$}{in1,v1}
 \fmf{fermion,width=0.5,label=$p+k_1$}{v1,v2}
 \fmf{fermion,width=0.5,label=$p$,tension=1}{v2,out1}
 \fmf{gluon,width=0.5,label=$k_2$,tension=0.3}{v1,out3}
 \fmf{gluon,width=0.5,label=$k_1$,tension=0.3}{v2,out2}
 \fmfright{out1,out2,out3}
\end{fmfgraph*}
\end{center}

According to Eq.\eqref{eq:softgluon1} single gluon radiation with
momentum $k$ off a hard quark with momentum $p$ is described by a
kinematic term $(\epsilon^* p)(pk)$. For successive radiation the two
Feynman diagrams give us the combined kinetic terms
\begin{alignat}{5}
&\frac{(\epsilon_1 p)}{(p+k_1+k_2)^2 - m^2} 
  \frac{(\epsilon_2 p)}{(p+k_2)^2-m^2}
+ \frac{(\epsilon_2 p)}{(p+k_1+k_2)^2 - m^2} 
  \frac{(\epsilon_1 p)}{(p+k_1)^2-m^2} \notag \\
&\qquad = 
  \frac{(\epsilon_1 p)}{2(pk_1) + 2 (pk_2) + (k_1+k_2)^2} 
  \frac{(\epsilon_2 p)}{2(pk_2)}
+ \frac{(\epsilon_2 p)}{2(pk_1) + 2 (pk_2) + (k_1+k_2)^2} 
  \frac{(\epsilon_1 p)}{2(pk_1)} \qqquad k_1^2 = 0 = k_2^2 \notag \\
&\qquad \simeq
  \frac{(\epsilon_1 p)}{2 \max_j (pk_j)} 
  \frac{(\epsilon_2 p)}{2(pk_2)}
+ \frac{(\epsilon_2 p)}{2 \max_j (pk_j)} 
  \frac{(\epsilon_1 p)}{2(pk_1)} \qqqquad (pk_j) \; \text{strongly ordered} \notag \\[5mm]
&\qquad \simeq
\left\{%
\begin{array}{lll}
  \dfrac{(\epsilon_1 p)(\epsilon_2 p)}{2 \max_j (pk_j)} \;
  \dfrac{1}{2(pk_2)} \qquad & (p k_2) \ll (p k_1) \qquad & k_2 \; \text{softer}
\\[5mm]
  \dfrac{(\epsilon_1 p)(\epsilon_2 p)}{2 \max_j (pk_j)} \;
  \dfrac{1}{2(pk_1)} \qquad & (p k_1) \ll (p k_2) \qquad & k_1 \; \text{softer} \; .
\\
\end{array}
\right.
\label{eq:ordered_emission}
\end{alignat}
Going back to the two Feynman diagrams this means that once one of the
gluons is significantly softer than the other the Feynman diagrams
with the later soft emission dominates. After squaring the amplitude
there will be no phase space regime where interference
terms between the two diagrams are numerically relevant. The coherent
sum over gluon radiation channels reduces to a incoherent sum,
ordered by the softness of the gluon.

This argument can be generalized to multiple gluon emission by
recognizing that the kinematics will always be dominated by the more
divergent propagators towards the final state quark with momentum
$p$. Note, however, that it is based on an ordering of the scalar
products $(p k_j)$ interpreted as the softness of the gluons. We
already know that a small value of $(p k_j)$ can as well point to a
collinear divergence; every step in the argument of
Eq.\eqref{eq:ordered_emission} still applies.\bigskip

Second, we can derive ordered multiple gluon emission from the phase
space integration in the \underline{soft or eikonal approximation}.
There, gluon radiation is governed by the so-called radiation dipoles
given in Eq.\eqref{eq:qcd_dipole2a}. Because each dipole includes a
sum over all radiating legs in the amplitude, the square includes a
double sum over the hard legs. Diagonal terms vanish at least for
over--all color--neutral processes. Because the following argument is
purely based on kinematics we will ignore all color charges and other
factors.

For successive gluon radiation off a quark leg the question we are
interested in is where the soft gluon $k$ is radiated, for example in
relation to the hard quark $p_1$ and the harder gluon $p_2$. The
kinematics of this process is the same as soft gluon radiation of a
quark--antiquark pair produced in an electroweak process.  For the
dipoles we let the indices $i,j$ run over the harder quark, antiquark,
and possibly gluon legs.
A well--defined process with all momenta defined as outgoing is
\begin{center}
\begin{fmfgraph*}(120,50)
 \fmfset{arrow_len}{2mm}
 \fmfleft{in1}
 \fmf{photon,width=0.5}{in1,v1}
 \fmf{fermion,width=0.5,label=$p+k_1$}{v1,v2}
 \fmf{fermion,width=0.5,label=$p_1$,tension=1}{v2,out1}
 \fmf{fermion,width=0.5,lab.side=right,label=$p_2$,tension=0.3}{out3,v1}
 \fmf{gluon,width=0.5,label=$k$,tension=0}{v2,out2}
 \fmfright{out1,out2,out3}
\end{fmfgraph*}
\qqquad 
\begin{fmfgraph*}(120,50)
 \fmfset{arrow_len}{2mm}
 \fmfleft{in1}
 \fmf{photon,width=0.5}{in1,v1}
 \fmf{fermion,width=0.5,lab.side=right,label=$p_1$}{v1,out1}
 \fmf{fermion,width=0.5,label=$p_2+k$,tension=0.3}{v2,v1}
 \fmf{fermion,width=0.5,lab.side=right,label=$p_2$,tension=0.3}{out3,v2}
 \fmf{gluon,width=0.5,label=$k$,tension=0}{v2,out2}
 \fmfright{out1,out2,out3}
\end{fmfgraph*}
\end{center}

in the approximation of abelian QCD, \ie no triple gluon vertices.  We
start by symmetrizing the leading soft radiation dipole with respect to the
two hard momenta in a particular way,
\begin{alignat}{5}
(J^\dag \cdot J)_{12} 
&= \frac{(p_1 p_2)}{(p_1 k)(p_2 k)}
  \notag \\
&= \frac{1}{k_0^2} \frac{1- \cos \theta_{12}}{(1-\cos \theta_{1k})(1-\cos \theta_{2k})}
  \qqqquad \text{in terms of opening angles} \; \theta
  \notag \\
&= \frac{1}{2k_0^2} 
   \left( \frac{1- \cos \theta_{12}}{(1-\cos \theta_{1k})(1-\cos \theta_{2k})}
        + \frac{1}{1-\cos \theta_{1k}}
        - \frac{1}{1-\cos \theta_{2k}}
   \right) + (1 \leftrightarrow 2)
  \notag \\
&\equiv \frac{W^{[1]}_{12} + W^{[2]}_{12}}{k_0^2} \; .
\label{eq:def_wi}
\end{alignat}
The last term is an implicit definition of the two terms
$W^{[1]}_{12}$. The pre-factor $1/k_0^2$ is given by the leading soft
divergence.  The original form of $(J^\dagger J)$ is symmetric in the
two indices, which means that both hard partons can take the role of
the hard parton and the interference partner.  In the new form the
symmetry in each of the two terms is broken.  Each of the two terms we
need to integrate over the gluon's phase space, including the
azimuthal angle\index{azimuthal angle} $\phi_{1k}$. Note, however,
that this splitting into two contributions is not the standard
separation into the two diagrams. It is a specific ansatz to show the
ordering patterns we will see below.\bigskip

To compute the actual integral we express the three parton vectors in
polar coordinates where the initial parton $p_1$ propagates into the
$x$ direction, the interference partner $p_2$ in the $(x-y)$ plane,
and the soft gluon in the full three-dimensional space described by
polar coordinates,
\begin{alignat}{5}
\hat{p}_1 &= (1,0,0) 
  &&\text{hard parton}
  \notag \\
\hat{p}_2 &= (\cos \theta_{12}, \sin \theta_{12}, 0)
  &&\text{interference partner}
  \notag \\
\hat{k} &= (\cos \theta_{1k}, \sin \theta_{1k} \cos\phi_{1k}, \sin \theta_{1k} \sin \phi_{1k} 
  \qquad &&\text{soft gluon}
  \notag \\
\Rightarrow \qquad 
\cos \theta_{2k} \equiv (\hat{p}_2 \hat{k}
&= \cos \theta_{12} \cos \theta_{1k}
  +\sin \theta_{12} \sin \theta_{1k} \cos\phi_{1k} \; .
\end{alignat}
From the scalar product between these four-vectors we see that of the
terms appearing in Eq.\eqref{eq:def_wi} only the opening angle
$\theta_{2k}$ includes $\phi_{1k}$, which for the azimuthal angle
integration means
\begin{alignat}{5}
  \int_0^{2 \pi} d \phi_{1k} \;  W^{[1]}_{12}
&= \frac{1}{2} \int_0^{2 \pi} d \phi_{1k} \;
   \left( \frac{1- \cos \theta_{12}}{(1-\cos \theta_{1k})(1-\cos \theta_{2k})}
        + \frac{1}{1-\cos \theta_{1k}}
        - \frac{1}{1-\cos \theta_{2k}}
   \right) \; .
\notag \\
&= \frac{1}{2} \; \frac{1}{1-\cos \theta_{1k}}
   \int_0^{2 \pi} d \phi_{1k} \;
   \left( \frac{1- \cos \theta_{12}}{1-\cos \theta_{2k}}
          + 1 
          - \frac{1- \cos \theta_{1k}}{1-\cos \theta_{2k}}
   \right) \notag \\
&= \frac{1}{2} \; \frac{1}{1-\cos \theta_{1k}}
 \left( 2\pi 
  + \left( \cos \theta_{1k} - \cos \theta_{12} \right) \;
   \int_0^{2 \pi} d \phi_{1k} \; \frac{1}{1-\cos \theta_{2k}}
 \right) \; .
\label{eq:integ_wi2}
\end{alignat}
The azimuthal angle integral in this expression for $W^{[i]}_{12}$ we
can solve
\begin{alignat}{5}
\int_0^{2 \pi} d \phi_{1k} \frac{1}{1-\cos \theta_{2k}}
&= \int_0^{2 \pi} d \phi_{1k}
   \frac{1}{1- \cos \theta_{12} \cos \theta_{1k}
             +\sin \theta_{12} \sin \theta_{1k} \cos\phi_{1k}}
  \notag \\
&= \int_0^{2 \pi} d \phi_{1k}
   \frac{1}{a - b \cos\phi_{1k}}
  \notag \\
&= \oint_\text{unit circle} d z \; \frac{1}{iz}  
   \dfrac{1}{a - b \dfrac{z+1/z}{2}}
  &&\text{with} \quad z = e^{i \phi_{1k}}, \; \cos \phi_{1k} = \frac{z+1/z}{2}
  \notag \\
&= \frac{2}{i} \oint d z \; 
   \dfrac{1}{2az - b - b z^2}
  \notag \\
&= \frac{2 i}{b} \oint \; 
   \dfrac{dz}{(z-z_-)(z-z_+)}
  &&\text{with} \quad z_\pm = \frac{a}{b} \pm \sqrt{\frac{a^2}{b^2} - 1} \; .
\end{alignat}
This integral is related to the sum of all residues of poles
inside the closed integration contour. Of the two poles $z_-$ is the one which typically lies within the unit circle, so we find
\begin{alignat}{5}
\int_0^{2 \pi} d \phi_{1k} \frac{1}{1-\cos \theta_{2k}}
&= \frac{2 i}{b} \; 2 \pi i
   \frac{1}{z_--z_+}
 = \frac{2 \pi}{\sqrt{a^2-b^2}} \notag \\
&= \frac{2 \pi}{\sqrt{ (\cos \theta_{1k} - \cos \theta_{12} )^2 }}
 = \frac{2 \pi}{|\cos \theta_{1k} - \cos \theta_{12}|} \; .
\end{alignat}
The entire integral in Eq.\eqref{eq:integ_wi2} then becomes
\begin{alignat}{5}
  \int_0^{2 \pi} d \phi_{1k} \;  W^{[1]}_{12}
&= \frac{1}{2} \; \frac{1}{1-\cos \theta_{1k}}
 \left( 2 \pi 
  + \left( \cos \theta_{1k} - \cos \theta_{12} \right) \;
    \frac{2 \pi}{|\cos \theta_{1k} - \cos \theta_{12}|}
 \right)
 \notag \\
&= \frac{\pi}{1-\cos \theta_{1k}}
 \left( 1
  + \sign(\cos \theta_{1k} - \cos \theta_{12}
 \right)
 \notag \\
&= \begin{cases}
   \dfrac{2\pi}{1-\cos \theta_{1k}} & \text{if} \quad \theta_{1k} < \theta_{12} \\[2mm]
   0 & \text{else .}
   \end{cases}
\end{alignat}
The soft gluon is only radiated at angles between zero and the opening
angle of the initial parton $p_1$ and its hard interference partner or
spectator $p_2$. The same integral over $W_{12}^{[2]}$ gives the same
result, with switched roles of $p_1$ and $p_2$. Combining the two
permutations this means that the soft gluon is always radiated within
a cone centered around one of the hard partons and with a radius given
by the distance between the two hard partons. Again, the coherent sum
of diagrams reduces to an incoherent sum. This derivation
\underline{angular ordering}\index{angular ordering} is exact in the
soft limit. 

There is a simple physical argument for this suppressed radiation
outside a cone defined by the radiating legs. Part of the deviation is
that the over--all process is color--neutral. This means that once the
gluon is far enough from the two quark legs it will not resolve their
individual charges but only feel the combined charge. This screening
leads to an additional suppression factor of the kind
$\theta_{12}^2/\theta_{1k}^2$. This effect is called
coherence.\bigskip

The third argument for ordered emission comes from \underline{color
  factors}.  Crossed successive splittings or interference terms
between different orderings are color suppressed. For example in the
squared diagram for three jet production in $e^+ e^-$ collisions the
additional gluon contributes a color factor
\begin{alignat}{5}
\tr (T^a T^a) = \frac{N_c^2 -1}{2} = N_c C_F
\end{alignat}
When we consider the successive radiation of two gluons the ordering
matters. As long as the gluon legs do not cross each other we find the
color factor
\begin{alignat}{5}
\tr (T^a T^a T^b T^b) 
&= (T^a T^a)_{il} (T^b T^b)_{li} \notag \\
&= \frac{1}{4} 
\left( \delta_{il} \delta_{jj} - \frac{\delta_{ij} \delta_{jl}}{N_c}
\right) 
\left( \delta_{il} \delta_{jj} - \frac{\delta_{ij} \delta_{jl}}{N_c}
\right) 
\qquad \text{using} \quad 
T^a_{ij} T^a_{kl} = 
\frac{1}{2}
\left( \delta_{il} \delta_{jk} - \frac{\delta_{ij} \delta_{kl}}{N_c}
\right)
\notag \\
&= \frac{1}{4} 
\left( \delta_{il} N_c - \frac{\delta_{il}}{N_c}
\right) 
\left( \delta_{il} N_c - \frac{\delta_{il}}{N_c}
\right) 
\notag \\
&= 
N_c \; \left( \frac{N_c^2-1}{2 N_c} \right)^2 
= N_c C_F^2
= \frac{16}{3}
\end{alignat}
Similarly, we can compute the color factor when the two gluon lines
cross. We find
\begin{alignat}{5}
\tr (T^a T^b T^a T^b) 
= - \frac{N_c^2-1}{4N_c}
= - \frac{C_F}{2}
= - \frac{2}{3} \; .
\end{alignat}
Numerically, this color factor is suppressed compared to $16/3$. This
kind of behavior is usually quoted in powers of $N_c$ where we assume
$N_c$ to be large. In those terms non--planar diagrams are suppressed
by a factor $1/N_c^2$ compared to the planar diagrams.

Once we also include the triple gluon vertex we can radiate two gluons
off a quark leg with the color factor
\begin{alignat}{5}
\tr (T^a T^b) \; f^{acd} f^{bcd}
&= \frac{\delta^{ab}}{2}  \; N_c \delta^{ab} 
 = \frac{N_c (N_c^2-1)}{2} = N_c^2 C_F = \frac{36}{3} \; .
\end{alignat}
This is not suppressed compared to successive planar gluon emission,
neither in actual numbers not in the large-$N_c$ limit.\bigskip

We can try the same argument for a purely gluonic theory, \ie
radiating gluons off two hard gluons in the final state. The color
factor for single gluon emission after squaring is
\begin{alignat}{5}
f^{abc} f^{abc} = N_c \delta^{aa} = N_c (N_c^2-1) \sim N_c^3 \; ,
\end{alignat}
using the large-$N_c$ limit in the last step.  For planar double gluon
emission with the exchanged gluon indices $b$ and $f$ we find
\begin{alignat}{5}
f^{abd} f^{abe} f^{dfg} f^{efg}
= N_c \delta^{de} \; N_c \delta^{de}
= N_c^3 \; .
\end{alignat}
Splitting one radiated gluon into two gives 
\begin{alignat}{5}
f^{abc} \; f^{cef} f^{def} \; f^{abd}
= N_c \delta^{cd} \; N_c \delta^{cd}
= N_c^3 \; .
\end{alignat}
This means that planar emission and successive splittings cannot be
separated based on the color factor for either hard radiating quarks
or gluons. We can use the color factor argument only for abelian
splittings to justify ordered gluon emission.

\subsection{Multi--jet events}
\label{sec:qcd_jets}

Up to now we have derived and established the parton shower as a
probabilistic tool to simulate the successive emission of jets in hard
processes. This includes a careful look at the collinear and soft
structure of parton splitting as well as the crucial assumption of
ordered emission. The starting point of this whole argument was that
the Sudakov factors obey the DGLAP equation, as shown in
Eq.\eqref{eq:qcd_sudakov_dglap}.

In the following we will introduce an alternative object which allows
us to the compute rates and patterns of jet radiation.  In
Section~\ref{sec:qcd_radiation} we will introduce generating
functionals and their evolution equations. Their underlying
approximations are related to the parton shower, but their main
results hold more generally. We have
already used some of the key features which will derive here
in Higgs physics applications
in Section~\ref{sec:higgs_cjv}   {\sl Per se} it
is not clear how jet radiation described by the parton shower and jet
radiation described by fixed-order QCD processes are linked. In
Section~\ref{sec:qcd_ckkw} we will discuss ways to combine the two
approaches in realistic LHC simulations, bringing us very close to
contemporary research topics.

\subsubsection{Jet radiation patterns}
\label{sec:qcd_radiation}

From Section~\ref{sec:qcd_multiple} we know that unlike other
observables related to multi--jet events the number of radiated jets is
well defined after a simple resummation.
Generating functionals for the jet multiplicity allow us to calculate
resummed jet quantities from first principles in QCD. We construct a
generating functional in an arbitrary parameter $u$ by demanding that
repeated differentiation at $u=0$ gives \underline{exclusive multiplicity}
distributions $P_n \equiv \sigma_n/\sigma_\text{tot}$,
\begin{alignat}{5}
\boxed{  \Phi = \sum_{n=1}^\infty u^n P_{n-1} }
\qqquad 
\text{with}
\quad 
  P_{n-1} 
  = \frac{\sigma_{n-1}}{\sigma_\text{tot}}
  = \left. \frac{1}{n!} \frac{d^n}{du^n} \Phi \right|_{u=0} \; .
\label{eq:def_gf}
\end{alignat}
For the generating functional $\Phi$ we will suppress the argument $u$.  In
the application to gluon emission the explicit factor $1/n!$
corresponds to the phase space factor for identical bosons.  Because in $P_n$
we only count radiated jets, our definition uses $P_{n-1}$
where other conventions use $P_n$. A second observable we can
extract from $\Phi$ is the \underline{average jet multiplicity},
\begin{alignat}{5}
\left. \frac{d \Phi}{d u} \right|_{u=1}
= \left. \sum_{n=1}^\infty n \; u^{n-1} \; \frac{\sigma_{n-1}}{\sigma_\text{tot}} \right|_{u=1}
= 1 + \frac{1}{\sigma_\text{tot}} \sum_{n=1}^\infty (n-1) \; \sigma_{n-1} \; .
\end{alignat}
Note again that $P_{n-1}$ describes $n-1$ radiated jets,
in the simplest case corresponding to $n$ observed jets in the final
state.\bigskip

The question is what we can say about such generating functionals. In
analogy to the DGLAP equation we can derive an evolution equation for
$\Phi$. We start by reminding ourselves that for the parton densities
and the Sudakov factors\index{Sudakov factor} the integrated version of the evolution
equation given in Eq.\eqref{eq:qcd_sudakov_interpret} reads
\begin{alignat}{5}
   f_i(x,t)
&= \Delta_i(t,t_0) f_i(x,t_0)
 + \int_{t_0}^t \frac{dt'}{t'} \; \Delta_i(t,t') 
   \sum_j \int_0^{1-\epsilon} \frac{dz}{z}  \; \frac{\alpha_s}{2\pi} \;
   \hat{P}_{i \leftarrow j}(z) \; f_j\left(\dfrac{x}{z},t'\right) \; .
\end{alignat}
The sum over the splittings is organized by initial states $j$ which
turn into the relevant parton $i$ in the collinear approximation. The
third particle involved in the splitting $j \rightarrow i$ follows
automatically.  

Instead of deriving the corresponding equation for the generating
functional $\Phi$ we motivate it by analogy.  In the Sudakov picture
we can apply our probabilistic picture to parton splittings $i \to jk$
in the final state. This should correspond to an evolution equation for the
generating functionals for the number of jets. All three external
particles are then described by generating functionals $\Phi$ instead
of parton densities, giving us
\begin{alignat}{5}
\boxed{
\Phi_i(t) 
= \Delta_i(t,t_0) \Phi_i(t_0) \;
   + \int_{t_0}^t
    \frac{dt'}{t'} \Delta_i(t,t')
    \sum_{i \rightarrow j,k} 
    \int_0^1 dz \; \frac{\alpha_s}{2\pi} \;
    \hat{P}_{i \rightarrow jk}(z) \; \Phi_j(z^2t') \Phi_k((1-z)^2t') 
} \; .
\label{eq:evolution_eq}
\end{alignat}
This evolution equation for general functionals is the same DGLAP
equation we use for parton densities in the initial state. The
difference is that the generating functionals count jets in the final
state. The precise link between the generating functionals $\Phi$ and
a parton--density--inspired partition function we skip at this stage.
Similarly, we introduce the argument of the strong coupling without
any further motivation as $\alpha_s(z^2(1-z)^2t')$. It will become
clear during our computation that this scale choice is
appropriate.\bigskip

The argument in this section will go two ways: first, we write down a
proper \underline{differential evolution equation} for $\Phi_q(t)$. Then, we solve
this equation for quarks, only including the abelian splitting $q
\rightarrow qg$. This solution will give us the known \underline{jet scaling
patterns}.  To start with, we insert the unregularized splitting kernel
from Eq.\eqref{eq:qcd_pqq} into the evolution equation,
\begin{alignat}{5}
\Phi_q(t) 
&= \Delta_q(t,t_0) \Phi_q(t_0) \;
   + \int_{t_0}^t
    \frac{dt'}{t'} \Delta_q(t,t')
    \int_0^1 dz \; \frac{\alpha_s}{2\pi} \;
    C_F \frac{1+z^2}{1-z} \; \Phi_q(z^2t') \Phi_g((1-z)^2t') 
\notag \\
&= \Delta_q(t,t_0) \Phi_q(t_0) \;
   + \int_{t_0}^t
    \frac{dt'}{t'} \Delta_q(t,t')
    \int_0^1 dz \; \frac{\alpha_s C_F}{2\pi} \;
    \frac{-(1-z)(1+z)+2}{1-z} \; \Phi_q(z^2t') \Phi_g((1-z)^2t') 
\notag \\
&= \Delta_q(t,t_0) \Phi_q(t_0) \;
   + \int_{t_0}^t
    \frac{dt'}{t'} \Delta_q(t,t')
    \int_0^1 dz \; \frac{\alpha_s C_F}{2\pi} \;
    \left( \frac{2}{1-z} - 1 - z \right) \; 
    \Phi_q(z^2t') \Phi_g((1-z)^2t') \; .
\label{eq:quarkeq}
\end{alignat}
First, we simplify the divergent part of
Eq.\eqref{eq:quarkeq}, using the new integration parameter $t'' =
(1-z)^2t'$. This gives us the same Jacobian as in
Eq.\eqref{eq:splitting2},
\begin{alignat}{5}
\frac{dt''}{dz} 
= \frac{d}{dz} (1-z)^2t'
= 2 (1-z) (-1) t'
= - 2 \frac{t''}{1-z}
\qquad \Leftrightarrow \qquad 
\frac{dz}{1-z} = - \frac{1}{2} \; \frac{dt''}{t''} \; .
\end{alignat}
In addition, we approximate $z \to 1$ wherever possible and cut off
all $t$ integrations at the infrared resolution scale $t_0$,
\begin{alignat}{5}
\int_0^1 dz \; \frac{\alpha_s(z^2(1-z)^2t') C_F}{2\pi} \; 
\frac{2}{1-z} \; \Phi_q(z^2t') \Phi_g((1-z)^2t') 
= 
\Phi_q(t') \int_{t_0}^{t'} dt'' \;
\frac{\alpha_s(t'') C_F}{2\pi} \frac{1}{t''} \Phi_g(t'') \; .
\label{eq:quarkeq2}
\end{alignat}
For the finite part in Eq.\eqref{eq:quarkeq} we neglect the
logarithmic $z$ dependence of all functions and integrate the leading
power dependence $1+z$ to $3/2$,
\begin{alignat}{5}
- \int_0^1 dz \; \frac{\alpha_s (z^2(1-z)^2t') C_F}{2\pi} \;
\left( 1 +z \right) \; \Phi_q(z^2t') \Phi_g((1-z)^2t') 
\simeq
- \frac{\alpha_s(t') C_F}{2\pi} \; \frac{3}{2} \;
  \Phi_q(t') \Phi_g(t') \; .
\label{eq:quarkeq3}
\end{alignat}
After these two simplifying steps Eq.\eqref{eq:quarkeq} reads
\begin{alignat}{5}
\Phi_q(t) 
&= 
\Delta_q(t,t_0) \Phi_q(t_0) 
+ \frac{C_F}{2\pi} \int_{t_0}^t \frac{dt'}{t'} \;
  \Delta_q(t,t') 
  \left( \int_{t_0}^{t'} dt'' \; \frac{\alpha_s(t'')}{t''} \Phi_q(t') \Phi_g(t'') 
       - \frac{3}{2} \alpha_s(t') \Phi_q(t') \Phi_g(t') \right)
\label{eq:quarkeq4}
\\
&= 
\Delta_q(t,t_0) \Phi_q(t_0) 
+ \frac{C_F}{2\pi} \Delta_q(t,t_0) \int_{t_0}^t \frac{dt'}{t'} \;
  \frac{1}{\Delta_q(t',t_0)} \; \Phi_q(t') 
  \left( \int_{t_0}^{t'} dt'' \; \frac{\alpha_s(t'')}{t''} \Phi_g(t'') 
       - \frac{3}{2} \alpha_s(t') \Phi_g(t') \right) \; .
\notag 
\end{alignat}
The original Sudakov factor $\Delta_q(t,t')$ is split into a ratio of
two Sudakov factors. This allows us to differentiate both sides with
respect to $t$, 
\begin{alignat}{5}
\frac{d}{dt} \Phi_q(t) 
&= \frac{d \Delta_q(t,t_0)}{dt} \Phi_q(t_0) 
+ \frac{C_F}{2\pi} \frac{d \Delta_q(t,t_0)}{dt} \int_{t_0}^t \frac{dt'}{t'} 
  \frac{1}{\Delta_q(t',t_0)} \Phi_q(t') 
  \left( \int_{t_0}^{t'} dt'' \frac{\alpha_s(t'')}{t''} \Phi_g(t'') 
       - \frac{3}{2} \alpha_s(t') \Phi_g(t') \right)
\notag \\ 
&\qqqquad \qquad \quad
+ \frac{C_F}{2\pi} \; \Delta_q(t,t_0) 
  \frac{1}{t} \frac{1}{\Delta_q(t,t_0)} \Phi_q(t) 
  \left( \int_{t_0}^{t} dt'' \; \frac{\alpha_s(t'')}{t''} \Phi_g(t'') 
       - \frac{3}{2} \alpha_s(t) \Phi_g(t) \right)
\notag\\ 
&= \frac{d \Delta_q(t,t_0)}{dt} 
   \left[ \Phi_q(t_0) 
        + \frac{C_F}{2\pi} \int_{t_0}^t \frac{dt'}{t'} 
           \frac{1}{\Delta_q(t',t_0)} \Phi_q(t') 
           \left( \int_{t_0}^{t'} dt'' \frac{\alpha_s(t'')}{t''} \Phi_g(t'') 
         - \frac{3}{2} \alpha_s(t') \Phi_g(t') \right) \right]
\notag \\ 
&
\qqqquad \qquad \qquad
+ \frac{C_F}{2\pi} \; \frac{1}{t} \; \Phi_q(t) \;
  \left( \int_{t_0}^{t} dt'' \; \frac{\alpha_s(t'')}{t''} \Phi_g(t'') 
       - \frac{3}{2} \alpha_s(t) \Phi_g(t) \right) 
\notag\\ 
&= \frac{1}{\Delta_q(t,t_0)} \; \frac{d \Delta_q(t,t_0)}{dt} \; \Phi_q(t) 
  + \Phi_q(t) \frac{C_F}{2\pi} \; \frac{1}{t} 
     \left( 
     \int_{t_0}^{t} dt'' \;
     \frac{\alpha_s(t'')}{t''} \Phi_g(t') - \frac{3}{2} \alpha_s(t) \Phi_g(t) 
     \right) \; .
\label{eq:quarkeq5}
\end{alignat}
In the last step we use the definition in Eq.\eqref{eq:quarkeq4}. This
simplified equation has a \underline{solution} which we can write in a closed
form, namely
\begin{alignat}{5}
\Phi_q(t) 
&= \Phi_q(t_0) \; \Delta_q(t,t_0) \;
\exp \left[ \frac{C_F}{2\pi} \;
\int^t_{t_0} dt' \; \frac{\alpha_s(t')}{t'} 
           \left( \log \frac{t}{t'} -\frac{3}{2} \right) \Phi_g(t') 
      \right] \notag \\
&= \Phi_q(t_0) \; 
\exp \left[ - \int_{t_0}^{t} dt' \; 
       \Gamma_{q \leftarrow q}(t,t')
       \right] \; 
\exp \left[ \int^t_{t_0} dt' \;  \Gamma_{q \leftarrow q}(t,t') \Phi_g(t') 
      \right] \notag \\
&= \Phi_q(t_0) \; 
\exp \left[ \int^t_{t_0} dt' \;  \Gamma_{q \leftarrow q}(t,t') \left( \Phi_g(t') -1 \right) 
      \right] \; .
\label{eq:quarkeq6}
\end{alignat}
We can prove this by straightforward differentiation of the first line
in Eq.\eqref{eq:quarkeq6},
\begin{alignat}{5}
\frac{d \Phi_q(t)}{dt} 
&= \Phi_q(t_0) \; \frac{d \Delta_q(t,t_0)}{dt} \; 
\exp \left[ \frac{C_F}{2\pi} \;
\int^t_{t_0} dt' \; \frac{\alpha_s(t')}{t'} 
           \left( \log \frac{t}{t'} -\frac{3}{2} \right) \Phi_g(t') 
      \right] 
\notag \\
&\qqquad + \Phi_q(t) 
\frac{d}{dt} \left[ \frac{C_F}{2\pi} \;
\int^t_{t_0} dt' \; \frac{\alpha_s(t')}{t'} 
           \left( \log t - \log t' -\frac{3}{2} \right) \Phi_g(t') 
      \right] 
\notag \\
&= \frac{1}{\Delta_q(t,t_0)} \; \frac{d \Delta_q(t,t_0)}{dt} \; 
   \Phi_q(t) 
 + \Phi_q(t) \; \frac{C_F}{2\pi} \;
\frac{\alpha_s(t)}{t} 
           \left( - \log t -\frac{3}{2} \right) \Phi_g(t) 
\notag \\
&\qqquad + \Phi_q(t) \; \frac{C_F}{2\pi} \; \frac{1}{t} 
\int^t_{t_0} dt' \; \frac{\alpha_s(t')}{t'} \; \Phi_g(t') 
 + \Phi_q(t) \; \frac{C_F}{2\pi} \; \log t \; 
\frac{\alpha_s(t)}{t} \; \Phi_g(t) 
\notag \\
&= \frac{1}{\Delta_q(t,t_0)} \; \frac{d \Delta_q(t,t_0)}{dt} \; 
   \Phi_q(t) 
 + \Phi_q(t) \; \frac{C_F}{2\pi} \; \frac{1}{t}
   \left( \int^t_{t_0} dt' \; \frac{\alpha_s(t')}{t'} \; \Phi_g(t') 
        - \alpha_s(t) \frac{3}{2} \Phi_g(t) \right) \; .
\label{eq:quarkeq7}
\end{alignat}
The expression given in Eq.\eqref{eq:quarkeq6} indeed solves the
evolution equation in Eq.\eqref{eq:quarkeq5}.  The corresponding
computation for $\Phi_g(t)$ follows the same path.\bigskip

By definition, the generating functional evaluated at the resolution
scale $t_0$ describes an ensemble of jets which have had no
opportunity to split. This means $\Phi_{q,g}(t_0) = u$.  The quark and
gluon generating functionals to next--to--leading logarithmic accuracy
are
\begin{alignat}{5}
\Phi_q(t) &= u \; \exp 
\left[ \int_{t_0}^{t} dt' \; \Gamma_{q \leftarrow q}(t,t') 
       \left( \Phi_g(t') -1 \right) \right] 
\notag \\ 
\Phi_g(t) &= u \; \exp
\left[ \int_{t_0}^{t} dt' \left( \Gamma_{g \leftarrow g} (t,t')
    \left( \Phi_g(t') -1 \right) + \Gamma_{q \leftarrow g}(t') \left( \frac
        {\Phi_q^2(t')} {\Phi_g(t')} -1 \right) \right) \right] \; .
\label{eq:gf_evolution}
\end{alignat}
The splitting kernels are defined in Eq.\eqref{eq:qcd_split_const};
gluon splitting to quarks described by $\Gamma_{q \leftarrow g}$ is
suppressed by a power of the logarithm $\log t/t'$.\bigskip

The logarithm $\log t/t'$ combined with the coupling constant
$\alpha_s$ included in the splitting kernels is the small parameter
which we will use for the following argument. If this logarithmically 
enhanced term dominates the physics, the evolution
equations for quark and gluons are structurally identical. In both
cases, the $\Phi$ dependence of the exponent spoils an effective
solution of Eq.\eqref{eq:gf_evolution}. However, the general form of
$\Gamma(t,t')$ ensures that the main contribution to the $t'$ integral
comes from the region where $t' \sim t_0$. Unless something drastic
happens with the integrands in Eq.\eqref{eq:gf_evolution} this means
that under the integral we can approximate $\Phi_{q,g}(t_0) = u$ and,
if necessary, iteratively insert the solution for $\Phi(t)$ into the
differential equation. The leading terms for both, quark and gluon
evolution equations turn into the closed form
\begin{alignat}{5}
\Phi_{q,g}(t)
= u \; \exp 
\left[ \int_{t_0}^{t} dt' \; \Gamma_{q,g}(t,t') 
       \left( u -1 \right) \right] 
= u \; \exp 
\left[ - (1-u) \int_{t_0}^{t} dt' \; \Gamma_{q,g}(t,t') \right] \; .
\end{alignat}
Using the Sudakov factor defined in Eq.\eqref{eq:sudakov1}
the generating functional in the approximation of large
logarithmically enhanced parton splitting is
\begin{alignat}{5}
\boxed{ 
\Phi_{q,g}(t) = u \; \Delta_{q,g}(t)^{1-u}} \; .
\label{eq:gf_poisson_approx}
\end{alignat}
For the jet rates this corresponds to a \underline{Poisson
  distribution}
\begin{alignat}{5}
P_{n-1} 
&= \Delta_{q,g}(t) \; \frac{|\log \Delta_{q,g}(t)|^{n-1}}{(n-1)!}
\qqquad \text{or} \qqquad 
\boxed{R_{(n+1)/n} = \frac{|\log \Delta_{q,g}(t)|}{n+1}} \; .
\label{eq:gf_poisson}
\end{alignat}
We can prove this result by induction. For general values of $u$ we
will show that the $n$-th derivative of the generating functional
$\Phi_{q,g}$ reads
\begin{alignat}{5}
\frac{1}{n!} \; \frac{d^n}{du^n} \; \Phi_{q,g}(t)
&= \frac{( - \log \Delta_{q,g} )^{n-1}}{n!} \; \Delta_{q,g} \;
  \left( n - u \log \Delta_{q,g} \right) \; e^{-u \log \Delta_{q,g}} \; \notag \\
&\stackrel{u = 0}{=}
 \frac{( - \log \Delta_{q,g} )^{n-1}}{(n-1)!} \; \Delta_{q,g} \;
= P_{n-1} \; .
\label{eq:gf_poisson_gen}
\end{alignat}
By construction, the Sudakov factor is smaller than unity, so $|\log
\Delta_{q,g}| = - \log \Delta_{q,g}$. For the case $n=1$ and general
values of $u$ Eq.\eqref{eq:gf_poisson_approx} indeed gives us
\begin{alignat}{5}
\frac{d}{du} \; \Phi_{q,g}(t)
&= \Delta_{q,g} \frac{d}{du} u  e^{-u \log \Delta_{q,g}} \notag \\
&= \Delta_{q,g} \left[ e^{-u \log \Delta_{q,g}}  
               + u \left( - \log \Delta_{q,g} \right) e^{-u \log \Delta_{q,g}}  
               \right] \notag \\
&= \Delta_{q,g} \left( 1
              - u \log \Delta_{q,g} \right) e^{-u \log \Delta_{q,g}} \; ,
\end{alignat}
in agreement with our aim in Eq.\eqref{eq:gf_poisson_gen}. Next, we
compute the step
\begin{alignat}{5}
\frac{1}{n!} \; \frac{d^n}{du^n} \Phi_{q,g}(t) 
&= \frac{1}{n} \; \frac{d}{du} \; 
   \left( \frac{1}{(n-1)!} \; \frac{d^{n-1}}{du^{n-1}} \Phi_{q,g}(t) 
   \right)
\notag \\
&= \frac{1}{n} \; \frac{d}{du} 
\left[ \frac{( - \log \Delta_{q,g} )^{n-2}}{(n-1)!} \;  \Delta_{q,g} \;
  \left( n -1 - u \log \Delta_{q,g} \right) \; e^{-u \log \Delta_{q,g}} 
\right] 
\qqquad \text{using Eq.\eqref{eq:gf_poisson_gen}} \notag \\
&= \frac{ ( - \log \Delta_{q,g} )^{n-2}}{n!} \; \Delta_{q,g} 
\left[ \left( - \log \Delta_{q,g} \right) e^{-u \log \Delta_{q,g}} 
  +   \left( n -1 - u \log \Delta_{q,g} \right) (- \log \Delta_{q,g} e^{-u \log \Delta_{q,g}} 
\right] \notag \\
&= \frac{( - \log \Delta_{q,g} )^{n-1}}{n!} \;  \Delta_{q,g} \;
   \left[ 1 + n- 1 - u \log \Delta_{q,g} \right] \; e^{-u \log \Delta_{q,g}} \; .
\end{alignat}
This is indeed Eq.\eqref{eq:gf_poisson_gen}, completing our proof of
this general solution and the special case $u=0$ in
Eq.\eqref{eq:gf_poisson}.\index{Poisson scaling}\bigskip 

In addition to this Poisson case we can find a second, recursive
solution for the generating functionals. It holds in the limit of
small emission probabilities. The emission probability is governed by
$\Gamma_{i \leftarrow j}(t,t')$, as defined in
Eq.\eqref{eq:qcd_split_const}. We can make it small by avoiding a
logarithmic enhancement, corresponding to no large scale ratios $t/t_0$.
In addition, we would like to get rid of $\Gamma_{q \leftarrow
  g}$ while keeping $\Gamma_{g \leftarrow g}$.  Theoretically, this
means removing the gluon splitting into two quarks and limiting
ourselves to pure Yang-Mills theory.  In that case the scale
derivative of Eq.\eqref{eq:gf_evolution} reads
\begin{alignat}{5}
\frac{d\Phi_g(t)}{dt} 
&= u \; \frac{d}{dt} 
\; \exp
\left[ \int_{t_0}^{t} dt' \; \Gamma_{g \leftarrow g} (t,t')
    \left( \Phi_g(t') -1 \right) \right] \notag \\
& = \Phi_g(t) \; 
    \frac{C_A}{2\pi} \frac{d}{dt} \; \int_{t_0}^{t} dt' \; \frac{\alpha_s(t')}{t'} 
    \left( \log t - \log t' - \frac{11}{6} \right) \; 
    \left( \Phi_g(t') -1 \right) \qqquad \text{inserting Eq.\eqref{eq:qcd_split_const}} \notag \\
& = \Phi_g(t) \frac{C_A}{2\pi} 
\left[  \frac{\alpha_s(t)}{t} 
    \left( - \log t - \frac{11}{6} \right) \; 
    \left( \Phi_g(t) -1 \right)
    + \frac{1}{t}  \int_{t_0}^{t} dt' \; \frac{\alpha_s(t')}{t'} 
    \left( \Phi_g(t') -1 \right)
    + \log t \; \frac{\alpha_s(t)}{t} 
    \left( \Phi_g(t) -1 \right) \right] \notag \\
& = \Phi_g(t) \frac{C_A}{2\pi t} \left[ 
    - \frac{11}{6} \; \alpha_s(t) 
    \left( \Phi_g(t) -1 \right) 
    + \int_{t_0}^{t} dt' \; \frac{\alpha_s(t')}{t'} 
    \left( \Phi_g(t') -1 \right) \right] \; .
\label{eq:staircase1}
\end{alignat}
This form is already greatly simplified, but in the combination of the
integral and the running strong coupling it is not clear what the
limit of small but finite $\log t/t_0$ would be. Integrating by parts
we find a form which we can estimate systematically,
\begin{alignat}{5}
\frac{d\Phi_g(t)}{dt} 
& = \Phi_g(t) \frac{C_A}{2\pi t} \left[ 
    - \frac{11}{6} \alpha_s(t) 
        \left( \Phi_g(t) -1 \right) 
    - \int_{t_0}^{t} dt'  \log \frac{t'}{t_0} 
    \frac{d}{dt'} \left( \alpha_s(t') 
    \left( \Phi_g(t') -1 \right) \right) 
    + \log \frac{t'}{t_0} 
    \alpha_s(t') 
    \left( \Phi_g(t') -1 \right) \Big|_{t_0}^t
    \right] \notag \\
& = \Phi_g(t) \frac{C_A}{2\pi t} \left[ 
    \alpha_s(t) 
    \left( \log \frac{t}{t_0} - \frac{11}{6} \right)
    \left( \Phi_g(t) -1 \right) 
    - \int_{t_0}^{t} dt' \; \log \frac{t'}{t_0} \; 
    \frac{d}{dt'} \left( \alpha_s(t') 
    \left( \Phi_g(t') -1 \right) \right) 
    \right] \; .
\label{eq:staircase2}
\end{alignat}
We can evaluate this expression in the limit of $t = t_0 + \delta$
or $t_0/t = 1 - \delta/t$. The two leading terms read
\begin{alignat}{5}
\frac{d\Phi_g(t)}{dt} 
& = \Phi_g(t) \frac{C_A}{2\pi t} \left[ 
    \alpha_s(t) 
    \left( \frac{\delta}{t} - \frac{11}{6} \right)
    \left( \Phi_g(t) -1 \right) 
    - (t - t_0) \; \frac{\delta}{t} \; 
    \frac{d}{dt} \left( \alpha_s(t) 
    \left( \Phi_g(t) -1 \right) \right) 
    \right] \notag \\
& = \Phi_g(t) \; \frac{C_A}{2\pi} \;
    \frac{\alpha_s(t)}{t}
    \left( \frac{\delta}{t} - \frac{11}{6} \right)
    \left( \Phi_g(t) -1 \right) 
    + \ope \left( \frac{\delta^2}{t^2} \right) \; .
\label{eq:staircase3}
\end{alignat}
To next--to--leading order in $\delta/t$ the equation for the generating
functional becomes
\begin{alignat}{5}
\frac{d\Phi_g(t)}{dt} 
&= \Phi_g(t) \,
    \tilde{\Gamma}_{g \leftarrow g}(t,t_0) \,\left( \Phi_g(t) -1 \right) 
\qqquad \text{with} \qquad 
\tilde{\Gamma}_{g \leftarrow g} (t,t_0)
= \frac{C_A}{2\pi} \,\frac{\alpha_s(t)}{t}
\left( \log \frac{t}{t_0} - \frac{11}{6} \right) \; .
\label{eq:staircase4}
\end{alignat}
With $\tilde{\Gamma}$ we define a slightly modified splitting kernel,
where the prefactor $\alpha_s/t$ is evaluated at the first argument
$t$ instead of the second argument $t_0$.  Including the boundary
condition $\Phi_g(t_0) = u$ we can solve this equation for the
generating functional, again using the method of the known solution,
\begin{alignat}{5}
\boxed{ 
\Phi_g(t) 
= \dfrac{1}{1 + \dfrac{1-u}{u \tilde{\Delta}_g(t)}} }
\qqquad \text{with} \qquad 
\tilde{\Delta}_g(t) = 
\exp \left( - \int_{t_0}^{t} dt' \tilde{\Gamma}_{g \leftarrow g}(t',t_0) \right) \; .
\label{eq:gf_solution}
\end{alignat}
The derivative of this solution is
\begin{alignat}{5}
\frac{d \Phi_g(t)}{t} 
&=  \frac{d}{dt} 
    \left( 1 + \dfrac{1-u}{u \tilde{\Delta}_g(t)} \right)^{-1} 
\notag \\
&= - \Phi_g(t)^2 \; 
   \dfrac{1-u}{u} \; \frac{d}{dt} 
   \exp \left( + \int_{t_0}^{t} dt' \; \tilde{\Gamma}_{g \leftarrow g}(t',t_0) \right)
\notag \\
&= - \Phi_g(t)^2 \; 
   \dfrac{1-u}{u \tilde{\Delta}_g(t)} \;
   \frac{d}{dt} 
   \int_{t_0}^{t} dt' \; \tilde{\Gamma}_{g \leftarrow g}(t',t_0) 
\notag \\
&= - \Phi_g(t)^2 \; 
   \left( \dfrac{1}{\Phi_g(t)} -1 \right) \;
   \frac{d}{dt} 
   \int_{t_0}^{t} dt' \; \tilde{\Gamma}_{g \leftarrow g}(t',t_0)  
 = \Phi_g(t) \; \left( \Phi_g(t)-1 \right) \;
   \tilde{\Gamma}_{g \leftarrow g}(t,t_0)  \; ,
\end{alignat}
which is precisely the evolution equation in
Eq.\eqref{eq:staircase4}.\bigskip

While we have suggestively defined a modified splitting kernel
$\tilde{\Gamma}$ in Eq.\eqref{eq:staircase4} and even extended this
analogy to a Sudakov-like factor in Eq.\eqref{eq:gf_solution} it is
not entirely clear what this object represents. In the limit of large
$\log t/t_0 \gg 1$ or $t \gg t_0$, which is not the limit we rely on 
for the pure Yang--Mills case, we find
\begin{alignat}{5}
 \int_{t_0}^{t} dt' \; \tilde{\Gamma}_{g \leftarrow g}(t',t_0)
 -  \int_{t_0}^{t} dt' \; \Gamma_{g \leftarrow g}(t',t_0)
&= - \frac{C_A}{2\pi} \; \alpha_s(t_0)\int_{t_0}^{t} d \frac{t'}{t_0} \; 
   \log \frac{t'}{t_0} \notag \\
&= - \frac{C_A}{2\pi} \; \alpha_s(t_0) \; 
   \left[ \frac{t'}{t_0} \log \frac{t'}{t_0} - \frac{t'}{t_0} \right]_1^{t/t_0} \notag \\
&= - \frac{C_A}{2\pi} \; \alpha_s(t_0) \; 
     \frac{t}{t_0} \log \frac{t}{t_0} \; . 
\end{alignat}
In the staircase limit $t \sim t_0$ and consistently neglecting $\log
t/t_0$ the two kernels $\Gamma_{g \leftarrow g}$ and
$\tilde{\Gamma}_{g \leftarrow g}$ become identical. In the same limit
we find $\Delta_g \sim \tilde{\Delta}_g \sim 1$. Again using $t' = t_0
+ \delta$ and only keeping the leading terms in $\delta$ we can
compute the leading difference
\begin{alignat}{5}
\tilde{\Gamma}_{g \leftarrow g}(t',t_0) - \Gamma_{g \leftarrow g}(t',t_0)
&= \frac{C_A}{2 \pi} \;
   \left( \frac{\alpha_s(t')}{t'} - \frac{\alpha_s(t_0)}{t_0} \right) \;
   \left( \frac{\delta}{t'} - \frac{11}{6} \right) \notag \\
&= - \frac{C_A}{2 \pi} \; \frac{11}{6} (t'-t_0) \;
   \frac{d}{dt} \frac{\alpha_s(t)}{t} \Bigg|_{t_0} 
 = - \frac{C_A}{2 \pi} \;  \frac{11}{6} \delta \;
   \left[ \frac{1}{t} \; \frac{d\alpha_s(t)}{dt} 
         -\frac{\alpha_s(t)}{t^2} \right]_{t_0} \notag \\
&= - \frac{C_A}{2 \pi} \; \frac{11}{6} \delta \;
   \left[ - \frac{1}{t} \; \frac{\alpha_s^2(t) b_0}{t} 
         -\frac{\alpha_s(t)}{t^2} \right]_{t_0} 
   \qquad \text{using Eq.\eqref{eq:run_alphas4}} \notag \\
&= \frac{C_A \alpha_s(t_0)}{2 \pi} \; \frac{11}{6} \;
   \frac{\delta}{t_0^2} \;
   \left( 1 + b_0 \alpha_s(t_0) \right) \notag \\
 \int_{t_0}^{t} dt' \; \tilde{\Gamma}_{g \leftarrow g}(t',t_0)
 -  \int_{t_0}^{t} dt' \; \Gamma_{g \leftarrow g}(t',t_0)
&= \frac{C_A \alpha_s(t_0)}{2 \pi} \; \frac{11}{6} \;
   \frac{\delta^2}{t_0^2} \;
   \left( 1 + b_0 \alpha_s(t_0) \right) \; .
\end{alignat}
In the pure Yang--Mills theory the running of the strong
coupling is described by $b_0 = 1/(4\pi) 11 N_c/3$. 
In both limits the true and the modified splitting kernels differ by
the respective small parameter.\bigskip

The closed form for the generating functional in
Eq.\eqref{eq:gf_solution} allows us to compute the number of jets in
purely gluonic events. The first derivative is 
\begin{alignat}{5}
\frac{d}{du} \Phi_g(t) 
&= \frac{d}{du} \; 
    u \left( u + \dfrac{1-u}{\tilde{\Delta}_g(t)} \right)^{-1} 
\notag \\
&= \left( u + \frac{1-u}{\tilde{\Delta}_g} \right)^{-1} 
   + u (-1) \left(u + \frac{1-u}{\tilde{\Delta}_g^2} \right)^{-2} 
     \left( 1 - \frac{1}{\tilde{\Delta}_g} \right) \; .
\end{alignat}
The form of the $n$-th derivative we can again prove by
induction. Clearly, for $n=1$ the above result is identical with the
general solution
\begin{alignat}{5}
\frac{d^n}{du^n} \Phi_g(t) 
&= n!  \left( \frac{1}{\tilde{\Delta}_g} -1 \right)^{n-1} 
   \left(u + \frac{1-u}{\tilde{\Delta}_g} \right)^{-n} 
   \left[ 1 + u \left(u + \frac{1-u}{\tilde{\Delta}_g} \right)^{-1} 
                 \left( \frac{1}{\tilde{\Delta}_g} -1 \right) \right] \; .
\label{eq:gf_staircase_derivative}
\end{alignat}
The induction step from $n$ to $n+1$ is
\begin{alignat}{5}
\frac{d^{n+1}}{du^{n+1}} \Phi_g(t) 
&= \frac{d}{du} \; n!  
   \left( \frac{1}{\tilde{\Delta}_g} -1 \right)^{n-1} 
   \left(u + \frac{1-u}{\tilde{\Delta}_g} \right)^{-n} 
   \left[ 1 + u ~ \left(u + \frac{1-u}{\tilde{\Delta}_g} \right)^{-1} 
         \left( \frac{1}{\tilde{\Delta}_g} -1 \right) \right] \notag\\ 
&= n! \left( \frac{1}{\tilde{\Delta}_g} -1 \right)^{n-1} \;
\left\{ 
  - n \left( u + \frac{1-u}{\tilde{\Delta}_g} \right)^{-n-1}
      \left( 1 - \frac{1}{\tilde{\Delta}_g} \right) 
      \left[ 1 + u ~ \left(u + \frac{1-u}{\tilde{\Delta}_g} \right)^{-1} 
              \left( \frac{1}{\tilde{\Delta}_g} -1 \right) \right]
    \right. \notag \\ 
&\qqquad + \left. \left(u + \frac{1-u}{\tilde{\Delta}_g} \right)^{-n} 
         \left[ \left(u + \frac{1-u}{\tilde{\Delta}_g} \right)^{-1} 
                \left( \frac{1}{\tilde{\Delta}_g} -1 \right) 
              + u \left(u + \frac{1-u}{\tilde{\Delta}_g} \right)^{-2} 
                \left( \frac{1}{\tilde{\Delta}_g} -1 \right)^2 \right] \right\} \notag\\ 
&= n! \left( \frac{1}{\tilde{\Delta}_g} -1 \right)^{n-1} 
   \left\{ 
    n \left( u + \frac{1-u)}{\tilde{\Delta}_g} \right)^{-n-1}
      \left( \frac{1}{\tilde{\Delta}_g} - 1 \right) 
      \left[ 1 + u ~ \left(u + \frac{1-u}{\tilde{\Delta}_g} \right)^{-1} 
              \left( \frac{1}{\tilde{\Delta}_g} -1 \right) \right]
    \right. \notag \\ 
&\qqqquad + \left. \left(u + \frac{1-u}{\tilde{\Delta}_g} \right)^{-n-1} 
                   \left( \frac{1}{\tilde{\Delta}_g} -1 \right) 
         \left[ 1
              +  u \left(u + \frac{1-u}{\tilde{\Delta}_g} \right)^{-1} 
                \left( \frac{1}{\tilde{\Delta}_g} -1 \right) \right] \right\} \notag\\ 
&= (n+1)! \left( \frac{1}{\tilde{\Delta}_g} -1 \right)^n 
          \left(u + \frac{1-u}{\tilde{\Delta}_g}\right)^{-n-1} 
          \left[ 1 + u \left(u + \frac{1-u}{\tilde{\Delta}_g} \right)^{-1} 
                    \left( \frac{1}{\tilde{\Delta}_g} -1 \right) \right] \; .
\end{alignat}
Evaluating the solution given by Eq.\eqref{eq:gf_staircase_derivative}
for $u=0$ gives us the jet rates
\begin{alignat}{5}
P_{n-1} 
= \frac{1}{n!} \frac{d^n}{du^n} \Phi_g(t) \Bigg|_{u=0}
= \left( \frac{1}{\tilde{\Delta}_g} -1 \right)^{n-1} 
  \tilde{\Delta}_g^n
= \tilde{\Delta}_g
  \left( 1 - \tilde{\Delta}_g \right)^{n-1} \; ,
\end{alignat}
which predicts constant ratios
\begin{alignat}{5}
\boxed{ R_{(n+1)/n} = 1 - \tilde{\Delta}_g(t) } \; .
\label{eq:gf_staircase}
\end{alignat}
Such constant ratios define a \underline{staircase pattern}\index{staircase scaling}. 
The name describes the form of the $n$-distribution on a logarithmic
scale. This pattern was first seen in $W$+jets production at UA1 in
1985. It has for a long time been considered an accidental sweet spot
where many QCD effects cancel each other to produce constant ratios of
successive exclusive $n$-jet rates. Our derivation from the generating
functionals suggest that staircase scaling is one of two
\underline{pure jet scaling patterns}:
\begin{enumerate}
\item in the presence of large scale differences abelian splittings
  generate a Poisson pattern with $R_{(n+1)/n} \propto 1/(n+1)$, as
  seen in Eq.\eqref{eq:gf_poisson}.\index{Poisson scaling}
\item for democratic scales non--abelian splittings generate a
  staircase pattern with constant $R_{(n+1)/n}$ shown in
  Eq.\eqref{eq:gf_staircase}.
\end{enumerate}
We have shown them for final state radiation only, so they should be
observable in $e^+ e^- \to$ jets events. Our derivation of the scaling
patterns is exclusively based on the parton shower. However, it turns out that
corrections from hard matrix element corrections, described in the
next section, do not change the staircase scaling patterns.\bigskip

To generalize the final--state jet scaling patterns to initial state
radiation we need to include parton densities. For simplicity, we
again resort to the Drell--Yan process. We can approximate the parton
densities based on two assumptions: threshold kinematics and $x_1
\approx x_2$.
In the absence of additional jets the hadronic and partonic energy
scales in on--shell $Z$ production are linked by $x^{(0)} \approx m_Z /
\sqrt{s}$. 
For other jet configurations we denote the threshold value as $x^{(n)}$. 
Putting everything together we find for the generating function
\begin{alignat}{5}
\Phi_\text{Drell--Yan} &= 
\sum_{i,j} f_i\left( \frac{x^{(n)}}{2}\right) \; \tilde{\Phi}_i 
    \times f_j\left(\frac{x^{(n)}}{2}\right) \; \tilde{\Phi}_j \; .
\end{alignat}
This means we can use the jet rates from the $e^+e^-$ case multiplied
by a parton density factor.  The structural new feature in
$\Phi_\text{Drell--Yan}$ is the $n$-dependence from the parton
densities. This is different from the resummed and $n$-independent
definition in Eq.\eqref{eq:def_gf}.  However, the generating function
is not a physical object.  Using our usual formalism and notation we
find
\begin{alignat}{5}
P_{n-1} 
&=  \left.  \frac{1}{n!} \frac{d^n}{du^n} \Phi_\text{Drell--Yan}
    \right|_{u=0} \notag \\ 
&= \left.  \frac{1}{n!} \sum_{i,j} 
   f_i\left( \frac{x^{(n)}}{2} \right) 
   f_j\left( \frac{x^{(n)}}{2} \right) 
   \times \frac{d^n}{du^n} \tilde{\Phi}_i \tilde{\Phi}_j
  \right|_{u=0} .
\end{alignat}
%

\subsubsection{CKKW and MLM schemes}
\label{sec:qcd_ckkw}

The main problem with QCD at the LHC is the range of energy scales of
the jets we encounter.  \underline{Collinear jets}\index{collinear radiation} with their small
transverse momenta are well described by a parton shower. From
Section~\ref{sec:qcd_resum_collinear} we know that strictly speaking
the parton shower only fills the phase space region up to a maximum
transverse momentum $p_T < \mu_F$.  In contrast,
\underline{hard jets} with large transverse momentum are described by
matrix elements which we compute using the QCD Feynman rules.  They
fill the non--collinear part of phase space which is not covered by the
parton shower.  Because of the collinear logarithmic enhancement we
discussed in Section~\ref{sec:qcd_resum_collinear} we expect many
more collinear and soft jets than hard jets at the LHC.

The natural question then becomes: what is the range of `soft' or
`collinear' and what is `hard'? Applying a consistency condition we
can define collinear jet radiation by the validity of the collinear
approximation in Eq.\eqref{eq:qcd_collinear}. The maximum $p_T$ of a
collinear jet is the upper end of the region for which the jet
radiation cross section behaves like $1/p_T$, or the point where the
distribution $p_T d\sigma/dp_T$ leaves its plateau. For harder and
harder jets we will at some point become limited by the partonic
energy available at the LHC, which means the $p_T$ distribution of
additional jets will start dropping faster than $1/p_T$. Collinear
logarithms will become numerically irrelevant and jets will be
described by the regular matrix element squared without any
resummation.\bigskip

Quarks and gluons produced in association with gauge bosons at the
Tevatron behave like collinear jets for $p_T \lesssim 20$~GeV, because
quarks at the Tevatron are limited in energy. At the LHC, jets
produced in association with tops behave like collinear jets to $p_T
\sim 150$~GeV, jets produced with new particles of mass 500~GeV behave like
collinear jets to $p_T$ scales larger than 300~GeV. This is not good
news, because collinear jets means many jets, and many jets produce
\underline{combinatorial backgrounds}\index{combinatorial background} and ruin the missing momentum
resolution of the detector: if we are looking for example for two jets
to reconstruct an invariant mass you can simply plot all events as a
function of this invariant mass and remove the backgrounds by requiring
all event to sit around a peak in $m_{jj}$. If we have for example
three jets in the event we have to decide which of the three jet--jet
combinations should go into this distribution. If this is not
possible we have to  consider two of the three combinations as
uncorrelated `background' events. In other words, we make three
histogram entries out of each signal or background event and consider
all three background events plus two of the three signal combinations as
background. This way the signal--to--background ratio decreases from
$N_S/N_B$ to $N_S/(3N_B+2N_S)$. A famous victim of such combinatorics 
is the (former) Higgs discovery channel $pp \to
t\bar{t}H$ with $H \to b\bar{b}$.\bigskip

For theorists this means that at the LHC we have to reliably model
collinear and hard jets.  For simplicity, in this section we will
first limit our discussion to final--state radiation, for example off
the $R$-ratio process $e^+ e^- \to q \bar{q}$ from
Section~\ref{sec:qcd_dy_r}. Combining collinear and hard jets in the
final state has to proceed in two steps. The first of them has nothing
to do with the actual jet simulation. If we categorize the generated
events by counting the number of jets in the final state we can refer
to an \underline{exclusive rate}\index{cross section!exclusive rate},
which requires a process to have exactly a given number of jets, or an
\underline{inclusive rate}\index{cross section!inclusive rate}, where
we for example identify $n$ jets and ignore everything else appearing
in the event. Additional collinear jets which we usually
denote as `$+X$' will be included.  We already know that a total rate
for any hard process we compute as $e^+ e^- \to q \bar{q} +X$, with
any additional number of collinear jets in the final
state. Predictions involving parton densities and the DGLAP equation
are jet--inclusive. Any scheme combining the parton shower and hard
matrix elements for events with arbitrary jet multiplicity has to
follow the path
\begin{enumerate}
\item define jet--exclusive events from the hard matrix elements
  and the parton shower
\item combine final states with \underline{different numbers of
  final--state particles}
\item reproduce matrix element results in high-$p_T$ and well
  separated phase space region
\item reproduce parton shower results for collinear and soft
  radiation
\item interpolate smoothly and avoid double counting of events
\end{enumerate}
For specific processes at the Tevatron the third and fourth point on
this list have actually been tackled by so-called matrix element
corrections in the parton shower Monte Carlos 
PYTHIA\index{event generators!PYTHIA} and 
HERWIG\index{event generators!HERWIG}. At the LHC this structure of 
event generation has become standard.\bigskip

The final state of the process $e^+ e^- \to q\bar{q} +X$
often involves more than two jets due to final state
splitting. Even for the first step of defining jet--exclusive
predictions from the matrix element we have to briefly consider the
geometry of different jets. To separate jet--inclusive event samples
into jet--exclusive event samples we have to define some kind of jet
separation parameter. If we radiate a gluon off one of the
quark legs, it gives us a $q \bar{q} g$ final state. This
additional gluon can be collinear with and hence geometrically
close to one of the quarks or not.  Jet algorithms which decide if we
count such a splitting as one or two jets we describe in detail in
Section~\ref{sec:sim_jetalgo}. They are based on a choice of 
collinearity measure $y_{ij}$ which we can for example construct as a
function of the distance in $R$ space, introduced in 
Eq.\eqref{eq:eta_phi}, and the transverse momenta. We define two jets
as collinear and hence as one jet if $y_{ij} < y_\text{resol}$ where
$y_\text{resol}$ is a free parameter in the algorithm.  As a result, the number of
jets in an event will depend on this 
\underline{resolution parameter $y_\text{resol}$}.\bigskip

For the second step of combining hard and collinear jet simulation the
same resolution parameter appears in a form where it becomes a
collinear vs \underline{hard matching parameter $y_\text{match}$}. It allows
us to clearly assign each hadron collider event a number of collinear
jets and a number of hard jets. Such an event with its given number of
more or less hard jets we can then describe either using matrix
elements or using a parton shower, where `describe' means computing
the relative probability of different phase space configurations. The
parton shower will do well for jets with $y_{ij} < y_\text{match}$. In
contrast, if for our closest jets we find $y_{ij} > y_\text{match}$,
we know that collinear logarithms did not play a major role, so we
should use the hard matrix element. If we assign the hard process a
typical energy or virtuality scale $t_\text{hard}$ we can translate
the matching parameter $y_\text{match}$ into a virtuality scale
$t_\text{match} = y_\text{match}^2 t_\text{hard}$, below which we do
not trust the hard matrix element. For example for the Drell--Yan
process the hard scale would be something like the $Z$ mass.\bigskip

The CKKW jet combination scheme first tackles the problem of defining
and combining jet--exclusive final states with different numbers of
jets.  The main ingredient to translating one into the other are
non--splitting probabilities called Sudakov factors\index{Sudakov factor}.  They can transform inclusive $n$-particle rates into exact
$n$-particle rates, with no additional final--state jet outside a given
resolution scale. We can compute integrated splitting
probabilities $\Gamma_j(t_\text{hard},t)$ which for quarks and gluons
are implicitly defined through the Sudakov factors introduced
in Eq.\eqref{eq:sudakov1}
\begin{alignat}{5}
\Delta_q(t_\text{hard},t_\text{match}) =& 
 \exp \left( - \int_{t_\text{match}}^{t_\text{hard}} d t \; 
       \Gamma_{q \leftarrow q}(t_\text{hard},t)
       \right)
\notag \\
\Delta_g(t_\text{hard},t_\text{match}) =& 
  \exp \left( - \int_{t_\text{match}}^{t_\text{hard}} d t \;
       \left[ \Gamma_{g \leftarrow g}(t_\text{hard},t)
             +\Gamma_{q \leftarrow g}(t) \right]
       \right) \; .
\end{alignat}
For final--state radiation $t$ corresponds to the original
$\sqrt{p_a^2}$. Moving forward in time it is ordered according to
$t_\text{hard} > t > t_\text{match}$. The resolution of individual
jets we identify with the matrix element--shower matching scale
$t_\text{match}$.  To leading logarithm the explicit form of the
splitting kernels is given in Eq.\eqref{eq:qcd_split_const}.
The virtualities $t_\text{hard} > t$ correspond to the incoming
(mother) and outgoing (daughter) parton. Note that the wrong limit for
$\Gamma_j(t_\text{hard},t_\text{hard}) \neq 0$ can be circumvented
technically.  To avoid unnecessary approximations in the $y$
integration more recent CKKW implementations integrate the splitting
kernels numerically.\bigskip

\begin{figure}[t]
\begin{center}
\includegraphics[width=0.20\hsize]{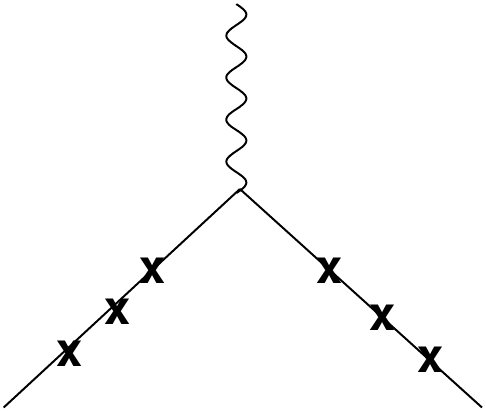} 
\hspace*{0.15\hsize}
\includegraphics[width=0.20\hsize]{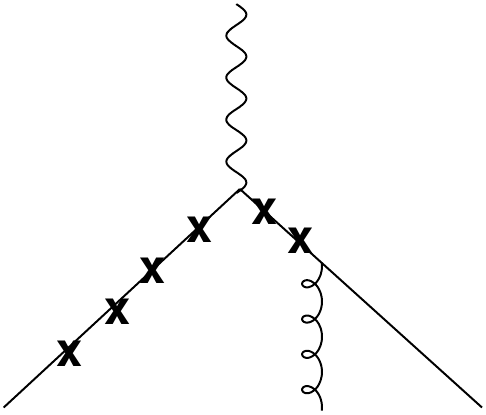} 
\hspace*{0.15\hsize}
\includegraphics[width=0.20\hsize]{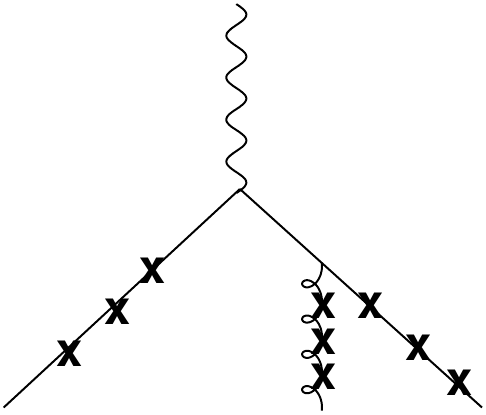} 
\end{center}
\caption{Vetoed showers on two-jet and three-jet contributions. The
  scale at the gauge boson vertex is $t_\text{hard}$. The two-jet
  (three-jet) diagram implies exactly two (three) jets at the resolution
  scale $t_\text{match}$, below which we rely on the parton
  shower. Figure from Ref.~\cite{Catani:2001cc}.}
\label{fig:qcd_ckkw_feyn}
\end{figure}

To get a first idea how to transform inclusive into exact $n$-jet
rates we compute the probability of seeing exactly \underline{two
  jets} in the process $e^+ e^- \to q\bar{q}$.  Looking at
Figure~\ref{fig:qcd_ckkw_feyn} this means that none of the two quarks
in the final state radiate a resolved gluon between the virtualities
$t_\text{hard}$ (given by the $qqZ$ vertex) and $t_\text{match} <
t_\text{hard}$. As will become important later, we specify that this
no-radiation statement assumes a jet resolution as given by end point
of the external quark and gluon legs. The probability we have to
multiply the inclusive two-jet rate with is then 
\begin{equation}
\left[ \Delta_q(t_\text{hard},t_\text{match}) \right]^2 \; , 
\end{equation}
once for each
quark. Whatever happens at virtualities below $t_\text{match}$ will be
governed by the parton shower and does not matter
anymore. Technically, this requires us to define a so-called vetoed
parton shower\index{parton shower!vetoed shower} which we will
describe in Section~\ref{sec:qcd_powheg}.\bigskip

What is the probability that the initially two-jet final state
evolves exactly into \underline{three jets}, again following
Figure~\ref{fig:qcd_ckkw_feyn}? We know that it contains a factor
$\Delta_q(t_\text{hard},t_\text{match})$ for one untouched quark. 

After splitting at $t_q$ with the probability
$\Gamma_{q \leftarrow q}(t_q,t_\text{hard}$ the second quark survives to
$t_\text{match}$, giving us a factor $\Delta_q(t_q,t_\text{match})$.
If we assign the virtuality $t_g$ to the radiated gluon at the
splitting point we find the gluon's survival probability as
$\Delta_g(t_g,t_\text{match})$. Together with the quark Sudakovs this gives us
\begin{alignat}{5}
 \Delta_q(t_\text{hard},t_\text{match}) \;
 \Gamma_{q \leftarrow q}(t_\text{hard},t_q) \;
 \Delta_q(t_q,t_\text{match}) \;
 \Delta_g(t_g,t_\text{match}) \cdots
\label{eq:qcd_veto1}
\end{alignat}
That's all there is, with the exception of the intermediate quark.  
There has to appear another factor describing that the quark, starting
from $t_\text{hard}$, gets to the splitting point $t_q$
untouched. Naively we would guess that this probability is given by
$\Delta_q(t_\text{hard},t_q)$. However, this Sudakov factor describes
no splittings resolved at the lower scale $t_q$. What we really mean
is no splitting between $t_\text{hard}$ and $t_q$ resolved at a third
scale $t_\text{match} < t_q$ given by the quark leg hitting the parton
shower regime. We get this information by computing the probability of
no splitting between $t_\text{hard}$ and $t_q$, namely
$\Delta_q(t_\text{hard},t_\text{match})$, but under the condition that
splittings from $t_q$ down to $t_\text{match}$ are explicitly allowed.

If zero splittings give us a probability factor 
$\Delta_q(t_\text{hard},t_\text{match})$, to describe exactly 
one splitting from $t_q$ on, we add a factor $\Gamma(t_q,t)$ with
an unknown splitting point $t$. This point $t$ we integrate over
between the resolution point $t_\text{match}$ and the endpoint of the 
no-splitting range, $t_q$. This is the same argument as in our physical 
interpretation of the Sudakov factors solving the DGLAP equation 
Eq.\eqref{eq:qcd_sudakov_interpret}. For an arbitrary number of
possible splittings between $t_q$ and $t_\text{match}$ we find the sum
\begin{alignat}{5}
 &\Delta_q(t_\text{hard},t_\text{match}) \left[ 1 
                         + \int_{t_\text{match}}^{t_q} d t \; \Gamma_{q \leftarrow q}(t_q,t)
                         + \text{more splittings} 
                   \right] =
  \notag \\
 & \qquad \qquad \quad = \Delta_q(t_\text{hard},t_\text{match}) \;
   \exp \left[  \int_{t_\text{match}}^{t_q} d t \; \Gamma_{q \leftarrow q}(t_q,t) \right]
 = \frac{\Delta_q(t_\text{hard},t_\text{match})}{\Delta_q(t_q,t_\text{match})} \; .
\label{eq:qcd_veto2}
\end{alignat}
The factors $1/n!$ in the Taylor series appear because for example
radiating two ordered jets in the same $t$ interval can proceed two
ways, both of which lead to the same final state. Once again: we compute the
probability of nothing happening between $t_\text{hard}$ and $t_q$
from the probability of nothing happening between
$t_\text{hard}$ and $t_\text{match}$ times any number of possible
splittings between $t_q$ and $t_\text{match}$. \bigskip

Collecting all factors from Eq.\eqref{eq:qcd_veto1} and
Eq.\eqref{eq:qcd_veto2} gives us the probability to find
exactly three partons resolved at $t_\text{match}$ as part of the
inclusive sample
\begin{alignat}{5}
  \Delta_q(t_\text{hard},t_\text{match}) \;
  \Gamma_{q \leftarrow q}(t_\text{hard},t_q) \; 
  \Delta_q(t_q,t_\text{match}) \; 
  \Delta_g(t_g,t_\text{match}) \; 
  \frac{\Delta_q(t_\text{hard},t_\text{match})}{\Delta_q(t_q,t_\text{match})} \notag \\
 = \Gamma_{q \leftarrow q}(t_\text{hard},t_q)  \; [\Delta_q(t_\text{hard},t_\text{match})]^2 \; \Delta_g(t_g,t_\text{match}) \; .
\end{alignat}
This result is what we expect: both quarks go
through untouched, just like in the two-parton case. In addition, we
need exactly one splitting producing a gluon, and this gluon cannot
split further. This example illustrates how we can compute these
probabilities using Sudakov factors: adding a gluon corresponds to
adding a splitting probability times the survival probability for this
gluon, everything else magically drops out. At the end, we 
integrate over the splitting point $t_q$.\bigskip

This discussion allows us to write down the first step of the
\underline{CKKW algorithm}, combining different hard $n$-jet channels
into one consistent set of events. One by one we turn inclusive
$n$-jet events into exact $n$-jet events.  We can write down the
slightly simplified algorithm for final--state radiation. As a starting
point, we generate events and compute leading order cross sections for
all $n$-jet processes. A universal lower jet radiation cutoff
$t_\text{match}$ ensures that all jets are hard and that all
corresponding cross sections $\sigma_{n,i}$ are finite. The second
index $i$ describes different non--interfering parton configurations
for a given number of final--state jets, like $q\bar{q} gg$ and
$q\bar{q} q\bar{q}$ for $n=4$.  The purpose of the algorithm is to
assign a weight (probability, matrix element squared,...) to a given
phase space point, statistically picking the correct process and
combining them properly. It proceeds event by event:
\begin{enumerate}
\item for each jet final state $(n,i)$ compute the relative
  probability $P_{n,i} = \sigma_{n,i}/\sum_{k,j} \sigma_{k,j}$ \\
  and select a final state $(n,i)$ with this probability $P_{n,i}$
\item assign the momenta from the phase space generator to, assumed,
  hard external particles \\
  and compute the transition matrix element $\matx$ including parton
  shower below $t_\text{match}$
\item use a jet algorithm to compute the shower history, \ie 
  all splitting points $t_j$ in each event \\
  and check that this history corresponds to possible Feynman diagrams
  and does not violate any symmetries
\item for each internal and external line compute
  the Sudakov non--splitting probability down to $t_\text{match}$
\item re-weight the $\alpha_s$ values of each splitting using the
  $k_T$ scale from the shower history
\item combine matrix element, Sudakovs, and $\alpha_s$ into a final weight
\end{enumerate}
We can use this final event weight to compute distributions from
weighted events or to decide if to keep or discard an event when
producing unweighted events. The construction ensures that the
relative weight of the different $n$--jet rates is identical to the
probabilities we initially computed. In step~2 the CKKW event
generation first chooses the appropriate hard scale in the event;
in step~3 we compute the individual starting scale for
the parton shower applied to each of the legs. Following our example,
this might be $t_\text{hard}$ for partons leaving the hard process 
itself or $t_g$ for a parton appearing via later splitting.\bigskip

In a second step of the CKKW scheme we match this combined hard matrix
element with the parton shower, given the matching point
$t_\text{match}$. From the final experimental resolution scale 
$t_\text{resol}$ up to a
matching scale $t_\text{match}$ we rely on the parton shower to
describe jet radiation while above the matching scale jet radiation is
explicitly forbidden by the Sudakov non--splitting
probabilities. Individually, both regimes consistently combine
different $n$--jet processes. All we need to make sure is that there
is no double counting.

From the discussion of Eq.\eqref{eq:qcd_veto2} we know that Sudakovs
describing the evolution between two scales and using a third scale as
the resolution are going to be the problem.  Carefully distinguishing
the scale of the actual splitting from the scale of jet resolution is
the key.  The CKKW scheme starts each parton shower at the point where
the parton first appears, and it turns out that we can use this
argument to keep the regimes $y > y_\text{match}$ and $y <
y_\text{match}$ separate. There is a simple way to check this, namely
if the \underline{$y_\text{match}$ dependence} drops out of the final
combined probabilities. The answer for final--state radiation is yes,
as proven in the original paper, including a hypothetical
next--to--leading logarithm parton shower. A modified CKKW scheme is
implemented in the publicly available SHERPA\index{event generators!SHERPA} 
event generator.\bigskip

An alternative to the CKKW scheme which has been developed
independently but incorporates essentially the same physics is the
\underline{MLM scheme}, for example implemented in 
ALPGEN\index{event generators!ALPGEN} or 
Madgraph\index{event generators!Madgraph}. Its
main difference to the CKKW scheme is that it avoids computing the
survival properties using Sudakov form factors. Instead, it vetos
those events which CKKW removes by applying the Sudakov non--splitting
probabilities. This way MLM avoids problems with splitting
probabilities beyond the leading logarithms, for example the finite
terms appearing in Eq.\eqref{eq:qcd_split_const}, which can otherwise
lead to a mismatch between the actual shower evolution and the
analytic expressions of the Sudakov factors. In addition, the veto
approach allows the MLM scheme to combine a set of independently
generated $n$--parton events, which can be convenient.

In the MLM scheme we again start by independently simulating $n$-jet
events including hard jet radiation as well as the parton shower.  In
this set of complete events we then veto events which are simulated
the wrong way.  This avoids double counting of events which on the one
hand are generated with $n$ hard jets from the matrix element and on
the other hand appear for example as $(n-1)$ hard jets with an
additional jet from the parton shower.  

After applying a jet algorithm, which in the case of
ALPGEN\index{event generators!ALPGEN} is a cone algorithm and in case
of Madgraph\index{event generators!Madgraph} is a $k_T$ algorithm, we
compare the showered event with the un-showered hard event by
identifying each reconstructed showered jet with the partons we
started from. If all jet--parton combinations match and there exist no
additional resolved jets we know that the showering has not altered
the hard structure of the event. If there is a significant change
between the original hard parton event and the showered event this
event has to go.  This choice corresponds to an event weight including
the Sudakov non--splitting probabilities in the CKKW scheme.  The only
exception to this rule is the set of events with the highest jet
multiplicity for which additional jets can only come from the parton
shower.  After defining the proper exclusive $n$-jet event sets we can
again use the parton shower to describe more collinear jet radiation
between $t_\text{match}$ and $t_\text{resol}$.

After combining the samples we still need a backwards evolution of a
generated event to know the virtuality scales which fix
$\alpha_s(Q^2)$.  As a side effect, if we also know the Feynman
diagrams describing an event we can check that a certain splitting
with its color structure is actually possible. For the parton shower
or splitting simulation we need to know the interval of virtualities
over which for example the additional gluon in the previous two-jet
example can split. The lower end of this interval is universally given
by $t_\text{match}$, but the upper end we cannot extract from the
record event by event.  Therefore, to compute the $\alpha_s$ values at
each splitting point we start the parton shower at an universal hard
scale, chosen as the hard(est) scale of the process.\bigskip

Aside from such technical details all merging schemes are conceptually
similar enough that we should expect them to reproduce each others'
results, and they largely do. But the devil is in the details, so 
experiment tells us which scheme as part of which event generator
produces the most usable results for a given LHC measurement.\bigskip

\begin{figure}[t]
\begin{center}
\includegraphics[width=0.29\hsize]{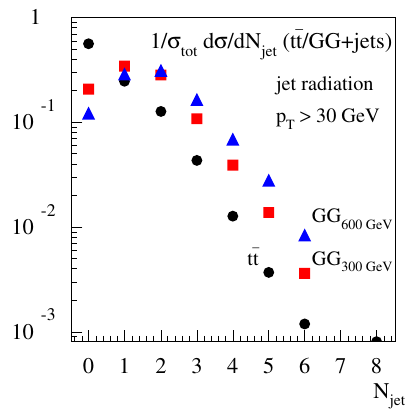} 
 \hspace*{0.05\hsize}
\includegraphics[width=0.29\hsize]{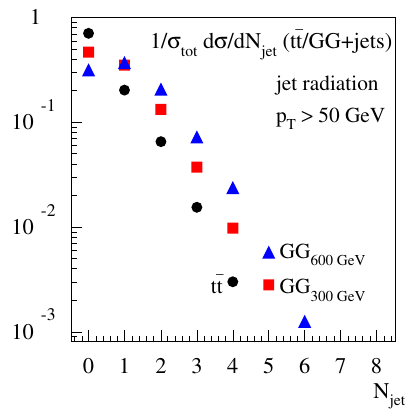}
 \hspace*{0.05\hsize}
\includegraphics[width=0.29\hsize]{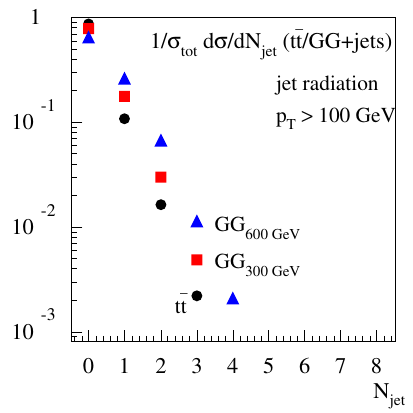}
\end{center}
\caption{Number of additional jets with a transverse momentum of at
  least 30, 50 or 100~GeV radiated off top pair production and the
  production of heavy states at the LHC. An example for such heavy
  states are scalar gluons with a mass of 300 or 600~GeV,
  pair-produced in gluon fusion. Figures from
  Ref.~\cite{Plehn:2008ae}.}
\label{fig:qcd_madgraph}
\end{figure}

To summarize, we can use the CKKW and MLM schemes to first combine
$n$-jet events with variable $n$ and then consistently add the parton
shower. In other words, we can for example simulate $Z+n$~jets
production at the LHC to arbitrarily large numbers of jets, limited
only by computational resources and the physical argument that at some
point any additional jet radiation will be described by the
parton shower.  This combination will describe all jets correctly over
the entire collinear and hard phase space. In
Figure~\ref{fig:qcd_madgraph} we show the number of jets expected to
be produced in association with a pair of top quarks and a pair of
heavy new states at the LHC. The details of these heavy scalar gluons
are secondary for the basic features of these distributions. The only
parameter which matters is their mass serving as the hard scale of the
process, setting the factorization scale, and defining the upper limit
of collinearly enhanced initial--state radiation. We see that heavy
states come with many jets radiated at $p_T \lesssim 30$~GeV, where
most of these jets vanish once we require transverse momenta of at
least 100~GeV. This figure tells us that an analysis which asks for a
reconstruction of two $W$-decay jets may well be
swamped by combinatorics\index{combinatorial background}.

Looking at the individual columns in Figure~\ref{fig:qcd_madgraph}
there is one thing we have to keep in mind: each of the merged matrix
elements combined into this sample is computed at leading order. The
emission of real particles is included, virtual corrections are not.
In other words, the CKKW and MLM schemes give us all \underline{jet
  distributions}, but only to leading order in the strong coupling.
When we combine the different jet multiplicities to evaluate total
rates, jet merging improves the rate prediction because it includes
contributions from all orders in $\alpha_s$, provided they come with a
potentially large logarithm from jet emission. From all we know, these
leading logarithms dominate the higher order QCD corrections for most
LHC processes, but it is not obvious how general this feature is and
how we can quantify it. This is certainly true for all cases where
higher order effects appear unexpectedly large and can be traced back
to new partonic processes or phase space configurations opening up at
higher jet multiplicities. Systematically curing some of this
shortcoming (but at a prize) will be the topic of the next
section. \bigskip

Before moving on to an alternative scheme we will illustrate why Higgs
or exotics searches at the LHC really care about progress in QCD
simulations: one way to look for heavy particles decaying into jets,
leptons and missing energy is the variable
\begin{alignat}{5}
 m_\text{eff} &= \met + \sum_j E_{T,j} + \sum_\ell E_{T,\ell} 
 \notag \\
     &= \slashchar{p}_T + \sum_j p_{T,j} + \sum_\ell p_{T,\ell}
 \qquad \text{(for massless quarks, leptons)}
\end{alignat}
This variable and its relatives we will discuss in detail in
Section~\ref{sec:sim_met_sm}. For gluon--induced QCD processes the
effective mass should be small while the new physics signal's
effective mass scale will be determined by the heavy masses.

For QCD jets as well as for $W$ and $Z$ plus jets backgrounds we can
study the production of many jets using the CKKW scheme.
Figure~\ref{fig:qcd_ckkw} shows the two critical distributions. First,
in the number of hard jets we see the so-called staircase scaling\index{staircase scaling}
behavior, namely constant ratios of exclusive ($n+1$)-jet and $n$-jet
rates $\sigma_{n+1}/\sigma_n$. Such a scaling is closely related to
the pattern we discuss in Eq.\eqref{eq:staircase_incl}, in the context
of the central jet veto of Section~\ref{sec:higgs_cjv}.  The
particularly interesting aspect of staircase scaling is that the
constant ratio is the same for jet--inclusive and jet--exclusive cross
sections $R^\text{incl}_{(n+1)/n} = R_{(n+1)/n}$, as shown in
Eq.\eqref{eq:staircase_excl}.

The consistent variation of $\alpha_s$
gives a small parametric uncertainty on these rates. A common scaling
factor $\mu/\mu_0$ for all factorization, renormalization and shower
scales in the process following our argument of
Section~\ref{sec:qcd_scales} is strictly speaking not fixed by 
our physical interpretation in terms of resummation; such a factor 
as part of the leading logarithm can be factored out as a subleading 
finite term, so it should really be considered a tuning
parameters for each simulation tool. Using the same simulation we also
show the effective mass and observe a drop towards large values of
$m_\text{eff}$. However, this drop is nowhere as pronounced as in
some parton shower predictions.  This analysis shows that the naive parton
shower is not a good description of QCD background processes to the
production of heavy particles. Using a very pragmatic approach and
tune the parton shower to correctly describe LHC data even in this
parameter region will most likely violate basic concepts like
factorization, so we would be well advised to use merging schemes like
CKKW or MLM for such predictions.

\begin{figure}[t]
\begin{center}
 \includegraphics[width=0.4\hsize]{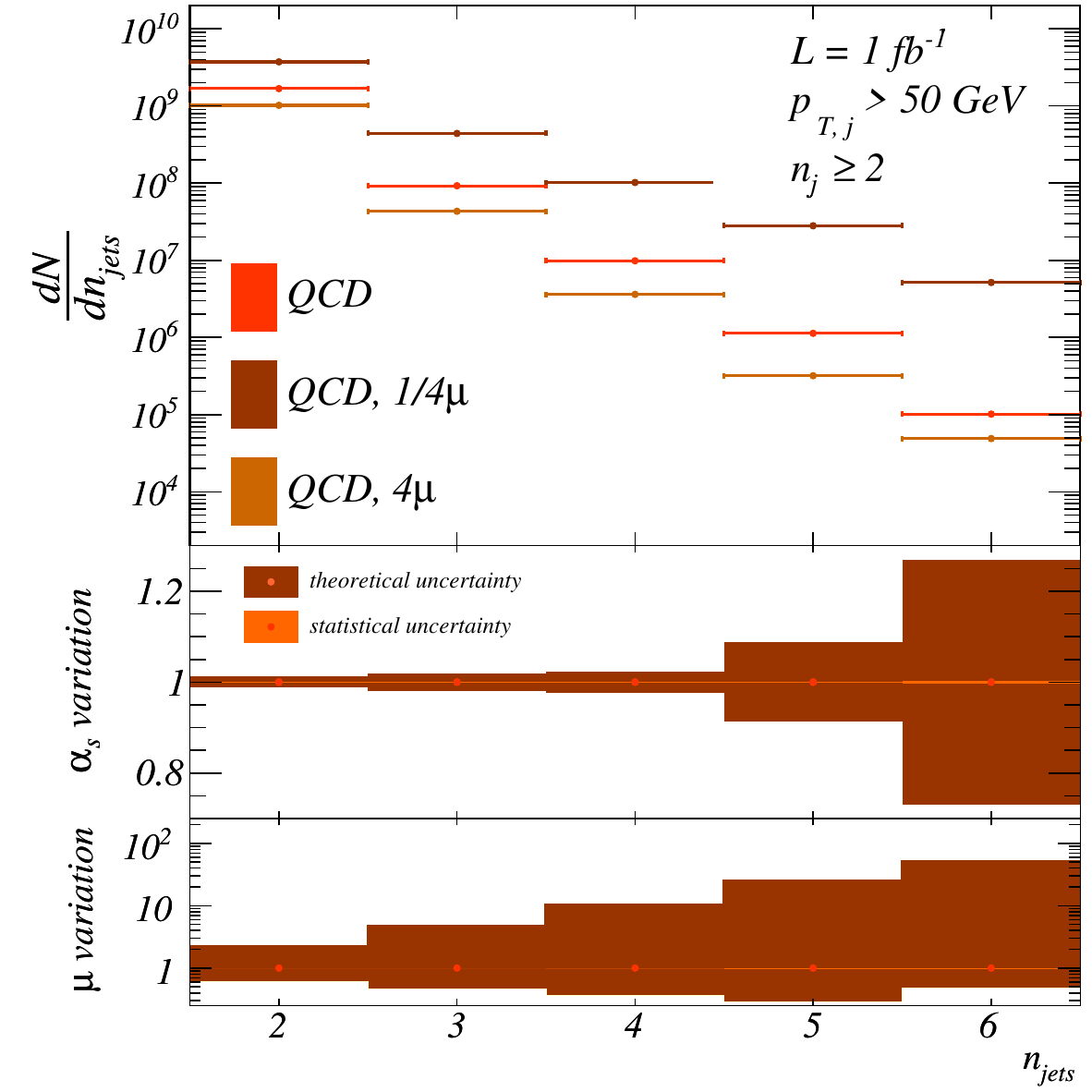} 
 \hspace*{0.1\hsize}
 \includegraphics[width=0.4\hsize]{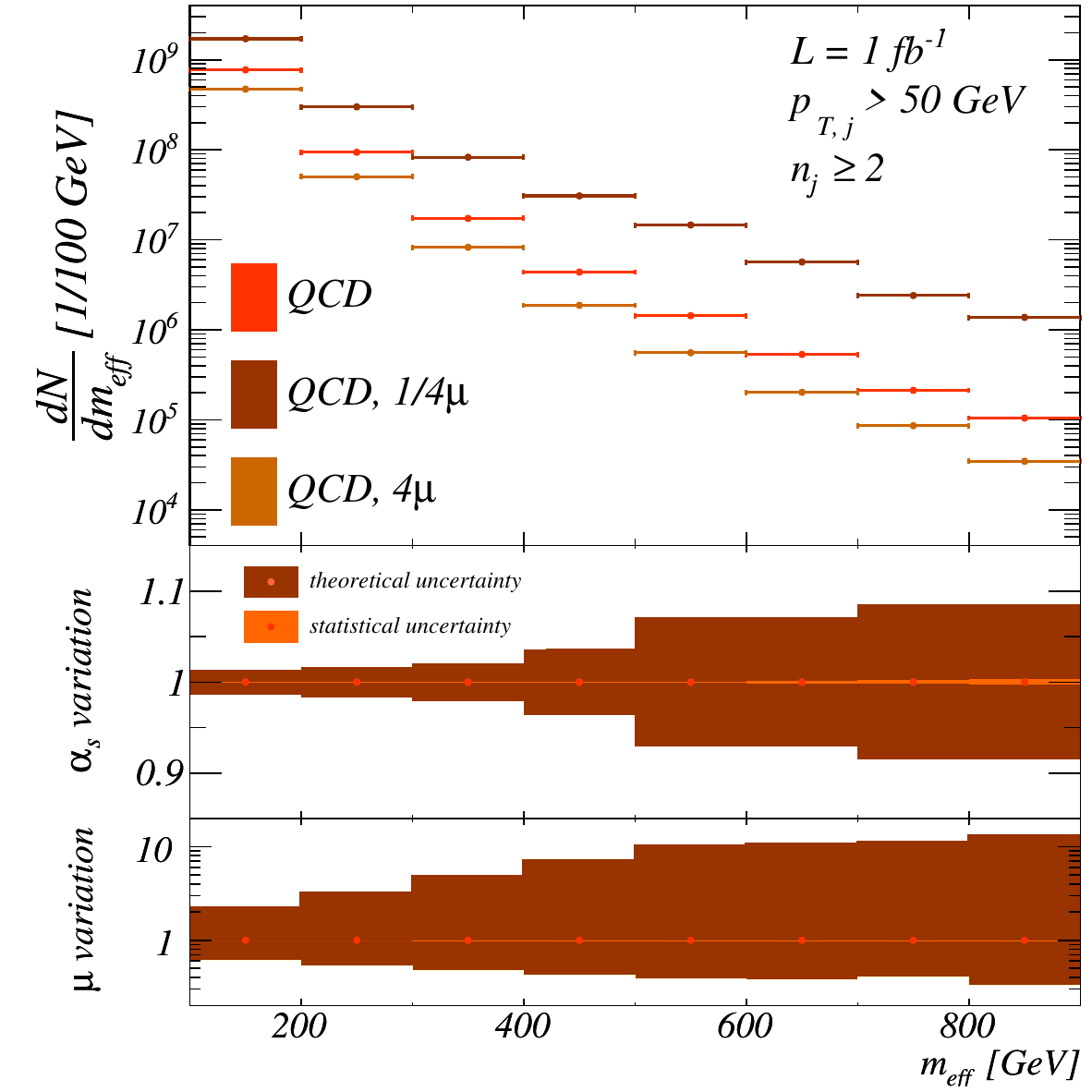}
\end{center}
\caption{Exclusive number of jets and effective mass distributions for pure QCD
  jet events at the LHC with a center--of--mass energy of 7~TeV and
  $p_{T,j} > 50$~GeV. The curves including the $\alpha_s$ uncertainty
  and a scale variation (tuning parameter) are computed with
  SHERPA\index{event generators!SHERPA} and a fully merged sample
  including up to six hard jets. These distributions describe typical
  backgrounds for searches for jets plus missing energy with fake
  missing energy, which could originate from supersymmetric squark and
  gluino production\index{supersymmetry}. Figures from
  Ref.~\cite{Englert:2011cg}.}
\label{fig:qcd_ckkw}
\end{figure}

\subsection{Next--to--leading orders and parton shower}
\label{sec:qcd_merge}

As we know for example for the $R$ ratio 
from Section~\ref{sec:qcd_dy_r} the precision of a leading
order QCD calculation in terms of the strong coupling constant
$\alpha_s$ is not always sufficient to match the experimental
accuracy. In such a case we need to compute observables to higher
order in QCD. On the other hand, in
Section~\ref{sec:qcd_resum_collinear} we have seen that the parton
shower does not fit into fixed order perturbation theory. With its
collinear logarithm it sums particular terms to all orders in
$\alpha_s$. So how can we on the one hand compute higher order
corrections to for example the Drell--Yan cross section and
distributions and in addition consistently combine them with the
parton shower?\bigskip

Such a combination would remove one of the historic shortcomings of
parton shower Monte Carlos.  Apart from the collinear approximation
for jet radiation they were always limited by the fact that 
in the words of one of the authors they `only do
shapes'. In other words, the normalization of the simulated event
sample will always be leading order in perturbative QCD and hence
subject to large theoretical uncertainties.  The reason for this
shortcoming is that collinear jet radiation relies on a hard process
and the corresponding production cross section and works with
splitting probabilities, but never touches the total cross section it
started from.

As a solution we compute higher order cross sections to normalize the
total cross section entering the respective Monte Carlo
simulation. This is what we call a \underline{$K$ factor}\index{K factor}: $K =
\sigma^\text{improved}/\sigma^\text{MC} =
\sigma^\text{improved}/\sigma^\text{LO}$. It is crucial to remember
that higher order cross sections integrate over unobserved additional
jets in the final state.  So when we normalize the Monte Carlo we
assume that we can first integrate over additional jets and obtain
$\sigma^\text{improved}$ and then just normalize the Monte Carlo which
puts back these jets in the collinear approximation. Obviously, we
should try to do better than that, and there are two ways to improve
this traditional Monte Carlo approach, the MC\@@NLO scheme and the POWHEG scheme.

\subsubsection{Next--to--leading order in QCD}
\label{sec:qcd_nlo}

When we compute the next--to--leading order correction to a cross
section, for example to Drell--Yan production, we consider all
contributions of the order $G_F^2 \alpha_s$. There are three obvious
sets of Feynman diagrams we have to add and then square: the
Born contribution $q \bar{q} \to Z$, the virtual gluon exchange for
example between the incoming quarks, and the real gluon emission $q
\bar{q} \to Zg$. An additional set of diagrams we should not forget are
the crossed channels $q g \to Zq$ and $\bar{q} g \to Z \bar{q}$.  Only
amplitudes with the same external particles can be squared, so we find
the matrix-element-squared contributions
\begin{alignat}{5}
  |\mathcal{M}_B|^2 &\propto G_F^2 \notag \\
  2 \text{Re} \; \mathcal{M}_V^* \mathcal{M}_B &\propto G_F^2 \alpha_s
   \quad \quad
  |\mathcal{M}_{Zg}|^2  &\propto G_F^2 \alpha_s
   \quad \quad
  |\mathcal{M}_{Zq}|^2, |\mathcal{M}_{Z\bar{q}}|^2 \propto G_F^2 \alpha_s \; .
\end{alignat}
Strictly speaking, we have to include counter terms, which following
Eq.\eqref{eq:renorm} are a modification of $|\mathcal{M}_B|^2$. We add these
counter terms to the interference of Born and virtual gluon
diagrams to remove the ultraviolet divergences. However, this is not
the issue we want to discuss.\bigskip

Infrared poles arise from two sources, soft and collinear divergences.
To avoid the complication of overlapping collinear and soft
divergences we will follow a toy model by Bryan Webber. It describes
simplified particle radiation off a hard process: the energy of the
system before radiation is $x_s$ and the energy of the outgoing
particle, call it photon or gluon, is $x$, so $x<x_s<1$. When we
compute \underline{next--to--leading order corrections}\index{next--to--leading order corrections} to a hard
process, the different contributions, neglecting crossed channels,
are
\begin{alignat}{5}
            \frac{d \sigma}{dx} \Big|_B = B \, \delta(x) 
            \quad \quad
            \frac{d \sigma}{dx} \Big|_V = \alpha_s \left( -\frac{B}{2\epsilon} + V
                                             \right) \delta(x) 
            \quad \quad
            \frac{d \sigma}{dx} \Big|_R = \alpha_s \frac{R(x)}{x} \; .
\label{eq:qcd_bvr_contrib}
\end{alignat}
The constant $B$ describes the Born process and the factorizing poles
in the virtual contribution. The coupling constant $\alpha_s$ ignores
factors 2 and $\pi$ or color factors.  We immediately see that the
integral over $x$ in the real emission rate is logarithmically
divergent in the soft limit, similar to the collinear divergences we
know. Because we are interested in infrared divergences we choose
$n=4+2 \epsilon$ dimensions with $\epsilon >0$, just like in
Section~\ref{sec:qcd_single_jet}, which will regularize the real emission
and compensate the resulting pole $1/\epsilon$ with the virtual
corrections. This means that the Kinoshita--Lee--Nauenberg theorem is
built into our toy model.

From factorization which we have illustrated based on the
universality of the leading splitting kernels we know that in the
collinear and soft limits the real emission has to follow the Born
matrix element
\begin{alignat}{5}
\boxed{\lim_{x\to 0} R(x) = B} \; .
\label{eq:qcd_born_real}
\end{alignat}
An \underline{observable} computed beyond leading order includes
contributions from real gluon emission and virtual gluon exchange. If
the observable is infrared safe it will have a smooth limit towards
vanishing gluon energy $O(x) \to O(0)$. The virtual corrections alone
diverge, but the expectation value including virtual and real gluon
contributions after dimensional regularization is finite.  In
dimensional regularization this cancellation schematically reads
\begin{alignat}{5}
\langle O \rangle
\sim \int_0^1 dx \; \frac{O(x)}{x^{1-2\epsilon}}
- \frac{O(0)}{2 \epsilon} \; .
\end{alignat}
This kind of combination has a finite limit for $\epsilon \to
0$. However, for numerical applications and event simulation we need
to implement this cancellation differently.\bigskip

The expectation value of any \underline{infrared safe observable}\index{infrared safety} over
  the entire phase space, including Born terms, virtual corrections
  and real emission, is given by
\begin{alignat}{5}
           \langle O \rangle
            \equiv 
           \langle O \rangle_B + \langle O \rangle_V + \langle O \rangle_R
            = \int_0^1 \; dx \; O(x) 
                       \left[ \frac{d \sigma}{dx} \Big|_B
                             +\frac{d \sigma}{dx} \Big|_V
                             +\frac{1}{x^{-2\epsilon}}
                              \frac{d \sigma}{dx} \Big|_R
                       \right] \; .
\label{eq:qcd_exp1}
\end{alignat}
The same way in which the renormalization and factorization scales appear,
dimensional regularization now yields an additional factor
$1/x^{-2\epsilon}$. Because we know its structure, we will omit 
the factorization scale factor in the following.\bigskip

When we compute for example a distribution of the energy of one of the
heavy particles in the process, we can extract a histogram from of the
integral for $\langle O \rangle$ in Eq.\eqref{eq:qcd_exp1} and obtain
a normalized distribution. The problem is that we have to numerically
integrate over $x$, and the individual parts of the integrand in
Eq.\eqref{eq:qcd_exp1} are not integrable.

There exist two methods to combine the virtual and real contributions
to an observable and produce a finite physics result. The first way
historically introduced by the Dutch loop school for example to
compute QCD corrections to top pair production is 
\underline{phase space slicing}\index{phase space!phase space slicing}: we divide the
divergent phase space integral into a finite part and a pole, by
introducing a small parameter $\Delta$, which acts like
\begin{alignat}{5}
   \langle O \rangle_R + \langle O \rangle_V
   =&  \int_0^1 \; dx \; \frac{O(x)}{x^{-2\epsilon}} \;
                         \frac{d \sigma}{dx} \Big|_R 
    + \langle O \rangle_V \notag \\
   =&  \left( \int_0^\Delta + \int_\Delta^1 \right) 
       \; dx \; \alpha_s \frac{R(x) O(x)}{x^{1-2\epsilon}}
    + \langle O \rangle_V \notag \\
=& \alpha_s R(0) \; O(0) \int_0^\Delta dx \; \frac{1}{x^{1-2\epsilon}}
+ \alpha_s \int_\Delta^1 dx \; \frac{R(x) O(x)}{x}
    + \langle O \rangle_V 
    &&\text{with $\Delta \ll 1$} \notag \\
=& \alpha_s B \; O(0) \; \frac{\Delta^{2\epsilon}}{2\epsilon}
+ \alpha_s \int_\Delta^1 dx \; \frac{R(x) O(x)}{x}
    + \langle O \rangle_V 
    &&\text{using Eq.\eqref{eq:qcd_born_real}} \notag \\
=& \alpha_s \frac{B\; O(0)}{2} \; 
   \frac{2 \epsilon \log \Delta + \ope(\epsilon^2)}{\epsilon}
+ \alpha_s \int_\Delta^1 dx \; \frac{R(x) O(x)}{x}
+ \alpha_s V O(0)
      \qquad &&\text{using Eq.\eqref{eq:qcd_bvr_contrib}}
\notag \\
=& \alpha_s B O(0) \; \log \Delta
+ \alpha_s \int_\Delta^1 dx \; \frac{R(x) O(x)}{x}
+ \alpha_s V O(0)   + \ope(\epsilon) \; .
\label{eq:qcd_slicing}
\end{alignat}
The two sources of $\log \Delta$ dependence have to cancel in the
final expression, so we can evaluate the integral at finite but small
values of $\Delta$. An amusing numerical trick is to re-write the
explicit $\log \Delta$ contribution into a real--emission--type phase space
integral. If the eikonal approximation\index{eikonal approximation} is
given in terms of a Mandelstam variable $\delta(s_4)$ and the cutoff
has mass dimension two we can write
\begin{alignat}{5}
\log \frac{\Delta}{\mu^2}
= \int_0^{s_4^\text{max}} d s_4 \;
  \log \frac{\Delta}{\mu^2} \; \delta(s_4) 
= \int_\Delta^{s_4^\text{max}} d s_4 \;
  \left[ \dfrac{\log \dfrac{s_4^\text{max}}{\mu^2}}{s_4^\text{max} - \Delta}
        -\dfrac{1}{s_4}
  \right]
\end{alignat}
and similarly for $\log^2 \Delta$. We can conveniently integrate this
representation along with the real emission phase space. The result
will be a finite value for the next--to--leading order rate in the
limit $\Delta \to 0$ and exactly $\epsilon = 0$. This means that using
phase space slicing we have exchanged dimensional regularization for
an energy cutoff. The advantage is that we can compute cross section
more easily, the disadvantage is that numerically the large
cancellation between the real and virtual correction appears in the
single $x=0$ bin.\bigskip

To avoid such cancellations between integrals and replace
them by cancellations among integrands we use a \underline{subtraction
  method}\index{phase space!phase space subtraction} to define
integrable functions under the $x$ integral in
Eq.\eqref{eq:qcd_exp1}. While our toy model appears more similar to
the {Frixione--Kunszt--Signer} subtraction scheme than to the
\underline{Catani--Seymour} scheme, both of them really are equivalent
at the level of the soft--collinear toy model. The special features of
the Catani--Seymour dipoles only feature when we include the full
modelling of the soft and collinear divergences described in
Section~\ref{sec:qcd_dipoles}.

Starting from the individually divergent virtual and real
contributions we first subtract and then add again a smartly chosen
term, in this toy model identical to a plus--subtraction following
Eq.\eqref{eq:plus_subtraction}
\begin{alignat}{5}
   \langle O \rangle_R + \langle O \rangle_V
   =&  \int_0^1 \; dx \; \alpha_s \frac{R(x) O(x)}{x^{1-2\epsilon}}
    + \langle O \rangle_V \notag \\
   =&  \int_0^1 dx \; \left( \frac{\alpha_s R(x)O(x)}{x^{1-2\epsilon}} 
                            -\frac{\alpha_s R(0) O(0)}{x^{1-2\epsilon}}
                      \right) 
     +  \alpha_s \; B \; O(0) \int_0^1 \; dx \frac{1}{x^{1-2\epsilon}}
     + \langle O \rangle_V \notag \\
   =&  \alpha_s \int_0^1 dx \; \frac{R(x)O(x)-BO(0)}{x}
     + \alpha_s \frac{B \; O(0)}{2\epsilon}
     + \langle O \rangle_V \notag \\
   =&  \alpha_s \int_0^1 dx \; \frac{R(x)O(x)-BO(0)}{x}
     + \alpha_s V O(0)
      \qqquad \text{using Eq.\eqref{eq:qcd_bvr_contrib}} \; .
\label{eq:qcd_subtract}
\end{alignat}
In the subtracted real emission integral we take the limit $\epsilon
\to 0$ because the asymptotic behavior of $R(x \to 0)$ regularizes
this integral without any dimensional regularization required. 
In our toy model we omit finite contributions from the
integrated subtraction term which will have to be added to the finite virtual
corrections.  In proper QCD exactly the same happens with the
Catani--Seymour dipoles and their integrated form.  We end up with a
perfectly finite $x$ integral for the sum of all three contributions,
so even in the limit $\epsilon =0$ there is no numerically small
parameter in the expression
\begin{alignat}{5}
\langle O \rangle
= \langle O \rangle_B + \langle O \rangle_V +  \langle O \rangle_R 
&= B \; O(0)
 + \alpha_s V \; O(0)
 + \alpha_s \int_0^1 \; dx \; \frac{R(x)\; O(x) -B \; O(0)}{x}
 \notag \\
&= \int_0^1 \; dx \; 
 \left[ 
        O(0) \; \left( B + \alpha_s V - \alpha_s \frac{B}{x} \right) 
      + O(x) \; \alpha_s \; \frac{R(x)}{x} 
 \right] \; .
\label{eq:qcd_exp2}
\end{alignat}
This subtraction procedure is a standard method to compute
next--to--leading order corrections involving one-loop virtual
contributions and the emission of one additional parton.\bigskip

As a side remark, we can numerically improve this expression using a
distribution relation
\begin{alignat}{5}
\int_0^1 dx \; \frac{f(x)}{x^{1-2 \epsilon}}
&= \int_0^1 dx \; \frac{f(x)- \theta(x_c-x) \, f(0)}{x^{1-2 \epsilon}}
 + f(0) \int_0^{x_c} dx \; x^{-1+2 \epsilon} \notag \\
&= \int_0^1 dx \; \frac{f(x)-\theta(x_c-x) \, f(0)}{x}
   \left( 1 + 2 \epsilon \log x + \ope(\epsilon^2) \right)
 + f(0) \; \frac{x_c^{2 \epsilon}}{2 \epsilon} \notag \\
&= \int_0^1 dx \; \left( 
   \frac{f(x)-\theta(x_c-x) \, f(0)}{x}
   + 2 \epsilon \; \frac{f(x)-\theta(x_c-x) \, f(0)}{x}  \log x
   + \frac{x_c^{2 \epsilon}}{2 \epsilon} \; f(x) \delta(x) 
   \right) \notag \\
\Leftrightarrow \qquad 
\frac{1}{x^{1-2 \epsilon}}
&= \frac{x_c^{2 \epsilon}}{2 \epsilon} \; \delta(x) 
   + \left( \frac{1}{x} \right)_c
   + 2 \epsilon \left( \frac{\log x}{x} \right)_c \; , 
\end{alignat}
where the last line is a relation between appropriately defined
distributions. This $c$-subtraction first introduced as part of the
Frixione--Kunszt--Signer subtraction scheme is defined as
\begin{alignat}{5}
\int_0^1 dx \; f(x) \; g(x)_c
= \int_0^1 dx \; 
  \left[ f(x) g(x) - f(0) g(x) \theta(x_c -x) \right] \; .
\end{alignat}
It is a generalization of the plus subtraction defined in
Eq.\eqref{eq:plus_subtraction} which we reproduce for $x_c =
1$.\index{plus subtraction} Linking the delta distribution to the
divergent integral over $1/x$ it is also reminiscent of the principal
value integration, but for an endpoint singularity and a dimensionally
regularized phase space.  Effectively combining phase space
subtraction Eq.\eqref{eq:qcd_subtract} and phase space slicing
Eq.\eqref{eq:qcd_slicing}, we include a cutoff in the integrals
holding the subtraction terms
\begin{alignat}{5}
   \langle O \rangle_R 
   =&  \alpha_s \int_0^1 \; dx \; \frac{R(x) O(x)}{x^{1-2\epsilon}}
   \notag \\
   =&  \alpha_s \int_0^1 dx \; 
       \frac{R(x)O(x)-\theta(x_c-x) B \, O(0)}{x}
       \left( 1 + 2 \epsilon \log x \right) 
     + \alpha_s B O(0) \; \frac{x_c^{2\epsilon}}{2\epsilon}
     + \ope(\epsilon^2) \; .
\end{alignat}
The dependence on the cutoff parameter $x_c$ drops out of the
final result. The numerical behavior, however, should be improved if
we subtract the infrared divergence only close to the actual pole where
following Eq.\eqref{eq:qcd_born_real} we understand the behavior of
the real emission amplitude.\bigskip

The formula Eq.\eqref{eq:qcd_exp2} is, in fact, a little tricky:
usually, the Born--type kinematics would come with an explicit factor
$\delta(x)$, which in this special case we can omit because of the
integration boundaries. We can re-write the same formula in a more
appropriate way to compute distributions, possibly including
experimental cuts
\begin{alignat}{5}
\boxed{
 \frac{d \sigma}{d O} 
= \int_0^1 \; dx \; 
 \left[ I(O)_\text{LO} \;
        \left( B + \alpha_s V - \alpha_s \frac{B}{x} 
        \right) 
      + I(O)_\text{NLO} \;
        \alpha_s \; \frac{R(x)}{x} 
 \right] 
} \; .
\label{eq:qcd_naive}
\end{alignat}
The \underline{transfer function}\index{transfer function} $I(O)$ is defined in a way that
formally does precisely what we require: at leading
order we evaluate $I(O)$ using the Born kinematics $x=0$ while for the
real emission kinematics it allows for general $x=0 \cdots 1$.

\subsubsection{MC\@@NLO method}
\label{sec:qcd_mcnlo}

For example in Eq.\eqref{eq:qcd_exp1} we integrate over the entire
phase space of the additional parton. For a hard additional parton or
jet the cross section looks well defined and finite, provided we
fully combine real and virtual corrections.
An infrared divergence appears after integrating over small
but finite $x \to 0$ from real emission, and we
cancel it with an infrared divergence in the virtual corrections
proportional to a Born--type momentum configuration $\delta(x)$.  In
terms of a histogram in $x$ we encounter the real emission divergence
at small $x$, and this divergence is cancelled by a negative delta
distribution at $x=0$.  Obviously, this will only give a
well behaved distribution after integrating over at least 
a range of $x$ values just above zero.

This soft and collinear subtraction scheme for next--to--leading order
calculations leads us to the first method of combining or matching
next--to--leading order calculations with a parton shower.  Instead of
the contribution from the virtual corrections contributing at
$\delta(x)$ what we would rather want is a smeared virtual corrections
pole which coincides with the justified collinear approximation and
cancels the real emission over the entire low-$x$ range. We can view this
contribution as events with a negative weight or
counter--events. Negative events\index{event generation!negative weights} trigger
negative reactions with
experimentalists, because they cause problems 
in a chain of probabilistic statements like a detector 
simulation. Fundamentally, there is no problem with them as long as
any physical prediction we make after adding all leading order and
next--to--leading order contributions gives a positive cross section.

Because we know they describe collinear jet radiation correctly such a
modification will make use of Sudakov
factors\index{Sudakov factor}. We can write them as a function of the energy fraction $z$
and define the associated probability as 
$d \mathcal{P} = \alpha_s P(z)/z \, dz$. Note that we avoid
the complicated proper two-dimensional description of
Eq.\eqref{eq:sudakov1} in favor of the simpler picture just in terms
of particle energy fractions as introduced in the last section.\bigskip

Once we integrate over the entire phase space this modified
subtraction scheme has to give the same result as the
\underline{next--to--leading order rate}\index{next--to--leading order corrections}.  Smearing the integrated
soft--collinear subtraction term using the splitting probabilities
entering the parton shower means that the MC\@@NLO subtraction scheme
has to be adjusted to the parton shower we use.\bigskip

Let us consider the perturbatively critical but otherwise perfectly
fine observable, the radiated \underline{photon spectrum} as a
function of the external energy scale $z$. We know what this
spectrum looks like for the collinear and hard kinematic
configurations
\begin{alignat}{5}
 \frac{d \sigma}{d z} \Big|_\text{LO} = \alpha_s \; \frac{B P(z)}{z}
 \qqqquad 
 \frac{d \sigma}{d z} \Big|_\text{NLO} = \alpha_s \; \frac{R(z)}{z} \; .
\label{eq:split_mcnlo}
\end{alignat}
The first term describes parton shower radiation from the Born diagram
at order $\alpha_s$, while the second term is the hard real emission
defined in Eq.\eqref{eq:qcd_bvr_contrib}. According to
Eq.\eqref{eq:qcd_naive} the transfer functions read
\begin{alignat}{5}
 I(z,1) \Big|_\text{LO} &= \alpha_s \; \frac{P(z)}{z}
 \notag \\
 I(z,x_M) \Big|_\text{NLO} &= \delta(z-x) 
                      + \alpha_s \; \frac{P(z)}{z} \; \theta(x_M(x)-z) \; .
\end{alignat}
The second term in the real radiation transfer function arises because
at the next order in perturbation theory the parton shower also acts
on the real emission process. It requires that enough energy to
radiate a photon with an energy $z$ be available, where $x_M$
is the energy available at the respective stage of showering, \ie $z <
x_M$.

We can include these transfer functions in Eq.\eqref{eq:qcd_naive} and obtain
\begin{alignat}{5}
 \frac{d \sigma}{d z} 
&= \int_0^1 dx \; 
 \left[ I(z,1) \;
        \left( B + \alpha_s V - \alpha_s \frac{B}{x} 
        \right) 
      + I(z,x_M) \;
        \alpha_s \; \frac{R(x)}{x} 
 \right] \notag \\
&= \int_0^1 dx \; 
 \left[ \alpha_s \frac{P(z)}{z}
        \left( B + \alpha_s V - \alpha_s \frac{B}{x}
        \right)
      + \left( \delta(x-z) + \ope(\alpha_s)
        \right)
        \; \alpha_s \frac{R(x)}{x} 
   \right] \notag \\
&= \int_0^1 dx \; 
 \left[ \alpha_s \; \frac{B P(z)}{z}
      + \alpha_s \; \frac{R(z)}{z} 
   \right] + \ope(\alpha_s^2) \notag \\
&= \alpha_s \; \frac{B P(z) + R(z)}{z}
 + \ope(\alpha_s^2) \; .
\label{eq:mcnlo1}
\end{alignat}
All Born terms proportional to $\delta(z)$ vanish because their
contributions would be unphysical. This already fulfills the first
requirement for our scheme, without having done anything except for
including a transfer function. Now, we can integrate over $z$ and
calculate the total cross section $\sigma_\text{tot}$ with a cutoff
$z_\text{min}$ for consistency. However, Eq.\eqref{eq:mcnlo1} includes
an additional term which spoils the result: the same kind of jet
radiation is included twice, once through the matrix element and once
through the shower. This is precisely the double counting which we
avoid in the CKKW scheme. So we are still missing something.\bigskip

We also knew we would fall short, because our
strategy includes a smeared virtual subtraction term which for finite
$x$ should cancel the real emission. This subtraction is not yet
included. Factorization tells us how to write such a subtraction term
using the splitting function $P$ as defined in
Eq.\eqref{eq:split_mcnlo}, to turn the real emission term into a finite
contribution
\begin{alignat}{5}
\frac{R(x)}{x} \quad \longrightarrow \quad
\frac{R(x) - B P(x)}{x} \; .
\end{alignat}
This {\sl ad hoc} subtraction term we have to add again
to the Born--type contribution. This
leads us to a modified version of Eq.\eqref{eq:qcd_naive}, now written
for general observables
\begin{alignat}{5}
\boxed{
  \frac{d \sigma}{d O} = \int_0^1 \; dx \; 
    \left[ I(O,1) \; \left( B 
                        + \alpha_s V 
                        - \frac{\alpha_s B}{x} 
                        + \frac{\alpha_s B P(x)}{x} 
                   \right) 
       + I(O,x_M) \; \alpha_s \frac{R(x)-B P(x)}{x}
  \right]
} \; .
\end{alignat}
Looking back at different methods of removing ultraviolet divergences
this modification from the minimal soft and collinear subtraction in
Eq.\eqref{eq:qcd_naive} to a physical subtraction term corresponding
to the known radiation pattern reminds us of different renormalization
schemes. The minimal $\msbar$ scheme will always guarantee finite results, but
for example the on--shell scheme with its additional finite terms has
at least to a certain degree beneficial properties when it comes to
understanding its physical meaning. This is the same for the MC\@@NLO
method: we replace the minimal subtraction terms by physically
motivated non--minimal subtraction terms such that the radiation
pattern of the additional parton is described correctly.

When we use this form to compute the $z$ spectrum to order
$\alpha_s$ it will in addition to Eq.\eqref{eq:mcnlo1}
include an integrated subtraction term contributing
to the Born--type kinematics
\begin{alignat}{5}
 \frac{d \sigma}{d z} 
&\longrightarrow 
   \int_0^1 \; dx \; \left[ \alpha_s \; \frac{B P(z)}{z}
                          + \alpha_s \; \delta(x-z) \; 
                            \left( \frac{R(x)}{x} - \frac{B P(x)}{x} \right)
                     \right] 
    + \ope(\alpha_s^2) \notag \\
&= \int_0^1 \; dx \; \alpha_s \; \frac{B P(z) + R(z) - B P(z)}{z} 
    + \ope(\alpha_s^2) \notag \\
&= \alpha_s \; \frac{R(z)}{z}  + \ope(\alpha_s^2) \; .
\label{eq:mcnlo2}
\end{alignat}
This is exactly the distribution we expect.\bigskip

Following the above argument the subtraction scheme implemented in the
MC\@@NLO Monte Carlo describes hard emission just like a
next--to--leading order calculation. This includes the next--to--leading
order normalization of the rate as well as the
\underline{next--to--leading order distributions}\index{next--to--leading order corrections} for those particles produced in the original hard
process. For example for $W$+jets production such corrections to the
$W$ and leading jet distributions matter, while for the production of
heavy new particles their distributions hardly change at
next--to--leading order. The distribution of the first radiated parton
is included at leading order, as we see in
Eq.\eqref{eq:mcnlo2}. Finally, additional collinear particle emissions
is simulated using Sudakov factors, precisely like a parton shower.

Most importantly, this scheme avoids double counting between the first hard
emission and the collinear jets, which means it describes the entire
$p_T$ range of jet emission for the \underline{first and hardest}
radiated jet consistently. Those additional jets, which do not feature
in the next--to--leading order calculation, are added through the parton
shower, \ie in the collinear approximation. As usually, what looked fairly
easy in our toy example is much harder in QCD reality, but the setup
is the same.\bigskip

\subsubsection{POWHEG method}
\label{sec:qcd_powheg}

As described in Section~\ref{sec:qcd_mcnlo} the MC\@@NLO matching scheme for a
next--to--leading order correction and the parton shower is based on an
extended subtraction scheme. It starts from a given parton shower and
avoids double counting by modifying the next--to--leading
corrections. An interesting question is: can we also combine a
next--to--leading order calculation by keeping the next--to--leading order
structure and apply a modified parton shower? The main ingredient to
this structure are Sudakov factors introduced in
Section~\ref{sec:qcd_sudakov} and used for the CKKW merging scheme in
Section~\ref{sec:qcd_ckkw}. 

In contrast to the MC\@@NLO scheme the POWHEG (Positive Weight
Hardest Emission Generator) scheme does not introduce counter--events
or subtraction terms. It considers the next--to--leading order
calculation of a cross section a combination of an $n$-particle and
an $(n+1)$-particle process and attempts to adjust the parton shower
attached to each of these two contributions such that there is no
double counting.\bigskip

Our starting point is the next--to--leading order computation of a cross
section following Eq.\eqref{eq:qcd_bvr_contrib}.  We can combine it
with appropriate soft and collinear subtraction terms $C$ in the
factorized $(n+1)$-particle phase space where for simplicity we assume
that the integrated subtraction terms exactly cancel the divergences
from the virtual corrections. In our simplified model where the extra
radiation is only described by an integral over the energy fraction
$x$ we find
\begin{alignat}{5}
\frac{d \sigma}{dx} &= B \; \delta(x)
          + \alpha_s \left( - \frac{B}{2\epsilon} +V \right) \; \delta(x) 
          + \alpha_s R 
\notag \\  
         &= B \; \delta(x)
          + \alpha_s V \; \delta(x)
          + \alpha_s \left( R - C \pro \right)  
\qquad \qquad \text{after soft--collinear subtraction}
\notag \\  
         &=  B \;
            \left[ \delta(x) + \frac{\alpha_s R}{B} \, \left( 1 - \pro \right) \
            \right]
          + \alpha_s \left[ V \delta(x) + (R - C) \, \pro \right]\; .
\label{eq:powheg_cxn1}
\end{alignat}
The projector $\pro$ maps the nominal $(n+1)$-particle phase space of
the real emission onto the $n$-particle phase space of the leading
order process. We keep it separate from the factor $\delta(x)$ and
define $\pro \delta(x) = \delta(x)$.\bigskip

The first term in Eq.\eqref{eq:powheg_cxn1} consists of the Born
contribution and the hard emission of one parton, so we have to avoid
double counting when defining the appropriate Sudakov factors.  The
second term is suppressed by one power of $\alpha_s$, so we can add a
parton shower to it without any worry. A serious problem
appears in Eq.\eqref{eq:powheg_cxn1} when we interpret it
probabilistically: nothing forces the combination of virtual and
subtracted real emission in the second bracket to be positive.  To
cure this shortcoming we can instead combine all $n$-particle
contributions into one term
\begin{alignat}{5}
\frac{d \sigma}{dx}  &= 
\left[ \delta(x) + \frac{\alpha_s R}{B} \, \left( 1 - \pro \right) \right]
\left[ B + \alpha_s V + \alpha_s (R - C) \pro \right] 
            + \ope(\alpha_s^2)
\notag \\
&\equiv \overline{B} \;
 \left[ \delta(x) + \frac{\alpha_s R}{B} \, \left( 1 - \pro \right) \right]
 + \ope(\alpha_s^2)
\notag \\
&= \overline{B} \;
 \left[ \delta(x) + \frac{\alpha_s R}{B} \, \theta \left( p_T(x) - p_T^\text{min} \right) \right]
            + \ope(\alpha_s^2) \; .
\label{eq:powheg_cxn2}
\end{alignat}
Defined like this the combination $\overline{B}$ can only become
negative if the regularized next--to--leading contribution
over--compensates the Born term which would indicate a breakdown of
perturbation theory. 
If we replace the symbolic projection $( 1 - \pro )$ by a step
function in terms of the transverse momentum of the radiated parton
$p_T(x)$ we can ensure that it really only appears for hard radiation
above $p_T^\text{min}$ and at the same time keep the integral over the
radiation phase space finite.\bigskip

From CKKW jet merging we know what we have to do to combine an
$n$-particle process with an $(n+1)$-particle process, even in the
presence of the parton shower: the $n$-particle process has to be
exclusive, which means we need to attach a Sudakov factor $\Delta$ to
veto additional jet radiation to the first term in the brackets of
Eq.\eqref{eq:powheg_cxn2}.  In the CKKW scheme the factor in the front
of the brackets would be $B$ and not $\overline{B}$.  The introduction
of $\overline{B}$ is nothing but a re-weighting factor\index{event generation!re-weighting} for the events contributing to the
$n$-particle configuration which we need to maintain the
next--to--leading order normalization of the combined $n$-particle and
$(n+1)$-particle rates. The second factor $\alpha_s R/B$ is
essentially the multiplicative PYTHIA\index{event generators!PYTHIA}
or HERWIG\index{event generators!HERWIG} matrix element correction
used for an improved simulation for example of $W+$jet events. The
only technical issue with such a re-weighted shower is that the
generating shower has to cover the entire radiation phase space. From
Section~\ref{sec:qcd_resum_collinear} we know that for a proper
resummation the collinear logarithms should only be integrated up to
the combined renormalization and factorization scales, $p_T < \mu_R
\equiv \mu_F$. This additional constraint needs to be addressed in the
POWHEG approach.\bigskip

The appropriate Sudakov factor for the real emission has to veto only
hard jet radiation from an additional parton shower. This way we
ensure that for the $(n+1)$-particle contribution the hardest jet
radiation is given by the matrix element $R$, which means no splitting occurs
in the hard regime $p_T > p_T^\text{min}$. Such a \underline{vetoed
  shower}\index{parton shower!vetoed shower} we can define in analogy
to the (diagonal) Sudakov survival probability Eq.\eqref{eq:sudakov1}
by adding a step function which limits the unwanted splittings to $p_T
> p_T^\text{min}$ 
\begin{alignat}{5}
 \Delta(t,p_T^\text{min}) 
 &= \exp \left( - \int_{t_0}^t \frac{dt'}{t'} 
                \int_0^1 dz \ \frac{\alpha_s}{2 \pi} 
                \hat{P}(z) \; 
                \theta \left( p_T(t',z) - p_T^\text{min} \right)
        \right) \; ,
\label{eq:veto_sudakov}
\end{alignat}
omitting the resolution $t_0$ in the argument and switching back to
the proper real emission phase space in terms of $z$ and $t'$.  This
modified Sudakov factor indicates that in contrast to the MC\@@NLO
method we now modify the structure of the parton shower which we
combine with the higher order matrix elements.

For the vetoed Sudakov factors to make sense we need to show that they
obey a DGLAP equation\index{DGLAP equation} like
Eq.\eqref{eq:qcd_sudakov_interpret}, including the veto condition in
the splitting kernel\index{splitting kernel}
\begin{alignat}{5}
   f(x,t)
&= \Delta(t,p_T^\text{min}) f(x,t_0)
 + \int_{t_0}^t \frac{dt'}{t'} \; \Delta(t,t',p_T^\text{min}) 
   \int_0^1 \frac{dz}{z}  \; \frac{\alpha_s}{2\pi} \;
   \hat{P}(z) \; \theta \left( p_T(t',z) - p_T^\text{min} \right)
   f\left(\dfrac{x}{z},t'\right) \; .
\label{eq:veto_dglap}
\end{alignat}
Again, we show the diagonal case to simplify the notation. The
proof of this formula starts from Eq.\eqref{eq:qcd_sudakov_interpret}
with the modification of an explicit veto. Using $1 = \theta(g) + (1 -
\theta(g))$ we find Eq.\eqref{eq:veto_dglap} more or less straight
away. The bottom line is that we can consistently write down vetoed
Sudakov probabilities and build a parton shower\index{parton shower!vetoed shower} out of them.\bigskip

Inserting both Sudakov factors into Eq.\eqref{eq:powheg_cxn2} gives us
for the combined next--to--leading order exclusive contributions
\begin{alignat}{5}
\boxed{
\frac{d \sigma}{d\Phi_n}  = \overline{B} \; 
            \left[ \Delta(t,0) + 
                   \Delta(t',p_T^\text{min}) \frac{\alpha_s R}{B} \,  \theta \left( p_T(t',z) - p_T^\text{min} \right) d t' dz
            \right]
            + \ope(\alpha_s^2) 
} \; .
\label{eq:powheg_cxn4}
\end{alignat}
The first Sudakov factor is not vetoed
which means it is evaluated at $p_T^\text{min} = 0$.\bigskip

Based on the next--to--leading order normalization of the integrated
form of Eq.\eqref{eq:powheg_cxn4} we can determine the form of the
splitting probability entering the Sudakov factor from the
perturbative series: the term in brackets integrated over the entire
phase space has to give unity.  Starting from Eq.\eqref{eq:veto_dglap}
we first compute the derivative of the Sudakov factor with respect to
one of its integration boundaries, just like in
Eq.\eqref{eq:sudakov_derive}
\begin{alignat}{5}
  \frac{d \Delta(t,p_T^\text{min})}{d t}
&= \frac{d}{d t}
   \exp \left( - \int_{t_0}^t \frac{dt'}{t'} 
                \int_0^1 dz \ \frac{\alpha_s}{2 \pi} 
                \hat{P}(z) \; 
                \theta \left( p_T(t',z) - p_T^\text{min} \right)
        \right) \notag \\
&= \Delta(t,p_T^\text{min}) \; 
   \frac{(-1)}{t} \int_0^1 dz \ \frac{\alpha_s}{2 \pi} \hat{P}(z) \; 
   \theta \left( p_T(t,z) - p_T^\text{min} \right) \; .
\end{alignat}
Using this relation we indeed find for the integral over the second
term in the brackets of Eq.\eqref{eq:powheg_cxn4}
\begin{alignat}{5}
 \int_{t_0}^t d t' dz \, \Delta(t',p_T^\text{min}) \frac{\alpha_s R}{B} \, 
 \theta \left( p_T(t',z) - p_T^\text{min} \right) 
&= - \int_{t_0}^t dt' \;
  \frac{d \Delta(t',p_T^\text{min})}{d t'} \; 
    \dfrac{\int dz \dfrac{\alpha_s R}{B}
           \theta \left( p_T(t',z) - p_T^\text{min} \right)}
          {\int dz \dfrac{\alpha_s}{2 \pi t'} \hat{P}(z) \; 
           \theta \left( p_T(t',z) - p_T^\text{min} \right)}
\notag \\
&= - \int_{t_0}^t dt' \;
  \frac{d \Delta(t',p_T^\text{min})}{d t'} 
\notag \\
&= - \Delta(t,p_T^\text{min}) 
\qqquad \Leftrightarrow \qqquad 
\boxed{\dfrac{\alpha_s R}{B} = 
\dfrac{\alpha_s}{2 \pi t'} \hat{P}(z)} \; .
\end{alignat}
Looking back at Eq.\eqref{eq:qcd_factorize} this corresponds to
identifying $B = \sigma_n$ and $\alpha_s R = \sigma_{n+1}$.  In the
POWHEG scheme the Sudakov factors are based on the simulated splitting
probability $\alpha_s R/B$ instead of the splitting kernels. This
replacement is nothing new, though. We can already read it off
Eq.\eqref{eq:qcd_factorize}.\bigskip

A technical detail which we have not mentioned yet is that the POWHEG
scheme assumes that our Sudakov factors can be ordered in such a
way that the hardest emission always occurs first.  Following the
discussion in Section~\ref{sec:qcd_ordered} we expect any collinear
transverse momentum ordering to be disrupted by soft radiation,
ordered by the angle\index{angular ordering}. The first emission of the
parton shower might well appear at large angles but with small energy,
which means it will not be particularly hard. 

For the POWHEG shower this soft radiation has to be removed or moved
to a lower place in the ordering of the splittings. The condition to
treat soft emission separately we know from CKKW merging, namely
Eq.\eqref{eq:qcd_veto2}: the scale at which we resolve a parton
splitting does not have to identical with the lower boundary of
simulated splittings. We can construct a parton shower taking into
account such splitting kernels, defining a \underline{truncated
  shower}\index{parton shower!truncated shower}. This modified shower
is the big difference between the MC\@@NLO scheme and the POWHEG
scheme in combining next--to--leading order corrections with a parton
shower. In the MC\@@NLO scheme we modify the next--to--leading order
correction for a given shower, but the shower stays the same. In the
POWHEG scheme the events get re-weighted according to standard
building blocks of a next--to--leading order calculation, but the shower
has to be adapted.\bigskip

\begin{table}[t]
\begin{center} \begin{tabular}{l||l|l}
   & MC\@@NLO/POWHEG matching
   & CKKW/MLM merging \\ \hline 
hard jets
   & first jet correct
   & all jets correct \\
collinear jets
   & all jets correct, tuned
   & all jets correct, tuned \\
normalization
   & correct to NLO
   & correct to LO plus real emission \\
implementations
   & MC\@@NLO, POWHEG, SHERPA, HERWIG
   & SHERPA, Alpgen, Madgraph,... \\
\end{tabular} \end{center}
\caption{Comparison of the MC\@@NLO and CKKW schemes combining
  collinear and hard jets.}
\label{tab:qcd_merging}
\end{table}

In Sections~\ref{sec:qcd_ckkw} and
Sections~\ref{sec:qcd_mcnlo}-\ref{sec:qcd_powheg} we have introduced
different ways to simulate jet radiation at the LHC.  The main
features and shortcomings of the matching and merging approaches we
summarize in Table~\ref{tab:qcd_merging}.

At this stage it is up to the competent user to pick the scheme which
describes their analysis best. First of all, if there is a well
defined and sufficiently hard scale in the process, the old-fashioned
Monte Carlo with a tuned parton shower will be fine, and it is by far
the fastest method. When for some reason we are mainly interested in
one hard jet we can use MC\@@NLO or POWHEG and benefit from the
next--to--leading order normalization. This is the case for example when
a gluon splits into two bottoms in the initial state and we are
interested in the radiated bottom jet and its kinematics. In cases
where we really need a large number of jets correctly described we
will end up with CKKW or MLM simulations. However, just like the
old-fashioned parton shower Monte Carlo we need to include the
normalization of the rate by hand. Or we are lucky and combined
versions of CKKW and POWHEG, as currently developed by both groups, 
will be available.\bigskip

I am not getting tired of emphasizing that the conceptual progress in
QCD describing jet radiation for all transverse momenta is
absolutely crucial for LHC analyses. If I were a string theorist I
would definitely call this achievement a revolution or even two, like
1917 but with the trombones and cannons of Tchaikovsky's 1812. In
contrast to a lot of other progress in theoretical physics jet merging
solves a problem which would otherwise have limited our ability to
understand LHC data, no matter what kind of Higgs or new physics we
are looking for. 

\subsection{Further reading}

Just like the Higgs part, the QCD part of these lecture notes is 
something in between a
text book chapter and a review of QCD and mostly focused on LHC
searches. I cut some corners, in particular when calculations
do not affect the main topic, namely the resummation of logarithms in
QCD and the physical meaning of these logarithms.  There is no point
in giving a list of original references, but I will list a few
books and review articles which should come in handy if you would like
to know more:

\begin{itemize}
\item[--] I started learning high energy theory including QCD from
  Otto Nachtmann's book. I still use his appendices for Feynman rules
  because I have not seen another book with as few (if not zero)
  typos~\cite{Nachtmann:1990ta}. 

\item[--] similar, but maybe a little more modern is the Standard
  Model primer by Cliff Burgess and Guy
  Moore~\cite{Burgess:2007zi}. At the end of it you will find more
  literature.
  
\item[--] the best source to learn QCD at colliders is the pink book
  by Keith Ellis, James Stirling, and Bryan Webber~\cite{Ellis:1991qj}. It includes
  everything you ever wanted to know about QCD and
  more. This QCD section essentially follows its Chapter~5.

\item[--] a little more phenomenology you can find in G\"unther
  Dissertori, Ian Knowles and Michael Schmelling's book~\cite{Dissertori:2003pj}. 
  Again, I borrowed some of
  the discussions in the QCD section from there. In the same direction but more
  theory oriented is the QCD book by Ioffe, Fadin, and Lipatov~\cite{Ioffe:2010zz}.
  
\item[--] if you would like to learn how to for example compute
  higher order cross sections to Drell--Yan production, Rick Field
  works it all out~\cite{Field:1989uq}.

\item[--] for those of you who are now hooked on QCD and jet physics
  at hadron colliders there are two comprehensive reviews by Steve
  Ellis etal.~\cite{Ellis:2007ib} and by Gavin Salam~\cite{Salam:2009jx}.

\item[--] aimed more at perturbative QCD at the LHC is the QCD primer
  by John Campbell, Joey Huston, and James
  Stirling~\cite{Campbell:2006wx}.

\item[--] coming to the usual brilliant TASI lectures, there are
  Dave Soper's~\cite{Soper:2000kt} and George
  Sterman's~\cite{Sterman:2004pd} notes. Both of them do not exactly
  use my kind of notations and are comparably formal, but they are a
  great read if you know something about QCD already. More on the
  phenomenological side there are Mike Seymour's lecture
  notes~\cite{Seymour:2005hs}.
  
\item[--] for a more complete discussion of the Catani--Seymour
  dipoles the very brief discussion in this writeup should allow you
  to follow the original long paper~\cite{Catani:1996vz}.

\item[--] the only review on leading order jet merging is by
  Michelangelo Mangano and Tim Stelzer~\cite{Mangano:2005dj}.  The
  original CKKW paper beautifully explains the general idea for final
  state radiation, and I follow their analysis~\cite{Catani:2001cc}.
  For other approaches there is a very concise discussion included
  with the comparison of the different models~\cite{Alwall:2007fs}.
  
\item[--] to understand MC\@@NLO there is nothing like the original
  papers by Bryan Webber and Stefano Frixione~\cite{Frixione:2002ik}.

\item[--] the POWHEG method is really nicely described in the original
  paper by Paolo Nason~\cite{Nason:2004rx}. Different processes you
  can find discussed in detail in a later paper by Stefano Frixione,
  Paolo Nason, and Carlo Oleari~\cite{Frixione:2007vw}.

\item[--] even though they are just hand written and do not include a
  lot of text it might be useful to have a look at Michael Spira's QCD
  lecture notes~\cite{spirix_notes} to view some of the topics from a
  different angle.

\end{itemize}

\newpage

\section{LHC phenomenology}
\label{sec:sim}

While the first two parts of these lecture notes focus on Higgs
physics and on QCD, biased towards aspects relevant to the LHC, they
hardly justify the title of the lecture notes. In addition, both
introductions really are theoretical physics.  The third section will
introduce other aspects which theorists working on LHC topics need to know. It
goes beyond what you find in theoretical physics text books and is
usually referred to as `phenomenology'.
\footnote{The term `phenomenology' is borrowed from
  philosophy where it means exactly the opposite from what it means in physics.
  Originally, phenomenology is a
  school based on Edmund Husserl, who were interested 
  not in observations
  but the actual nature of things. 
  Doing exactly the opposite, physicist phenomenologists are theorists who 
  really care about measurements.}

This terms indicates that these topics are not really theoretical
physics in the sense that they rely on for example field theory.
They are not experimental physics either, because they go beyond
understanding the direct results of the LHC detectors. Instead, they
lie in between the two fields and need to be well understood to allow
theorists and experimentalists to interact with each other.

Sometimes, phenomenology has the reputation of not being proper
theoretical physics. From these lecture notes it is clear that LHC
physics is field theory, either electroweak symmetry breaking, QCD, or
--- not covered in these notes --- physics models describing
extensions of our Standard Model at the TeV scale. This chapter
supplements the pure theory aspects and links them to experimental
issues of the ATLAS and CMS experiments. In Section~\ref{sec:sim_jets} we fill in some blanks
from Section~\ref{sec:higgs_gf_lhc}, \ref{sec:higgs_wh},
and~\ref{sec:qcd_ckkw}. We first discuss jets and how to link the
asymptotic final states of QCD amplitudes, partons, to experimentally
observed QCD objects, jets. Then, we turn to a current field of
research, so-called fat jets. In Section~\ref{sec:sim_hel} we
introduce a particularly efficient way of computing transition
amplitudes from Feynman rules, the helicity method. Most professional
tools for the computation of LHC cross sections or for simulating LHC events use this
method instead of squaring amplitudes analytically.
Section~\ref{sec:sim_met} discusses how to reconstruct particles which
interact too weakly to be observed in LHC detectors. In the Standard
Model those would be neutrinos, but as part of the LHC program we hope
to find dark matter candidates that way. Finally, in
Section~\ref{sec:sim_errors} we very briefly discuss LHC uncertainties
from a theory point of view. In the public arXiv version more sections
will follow, hopefully triggered by LHC measurements challenging
theorists and their simulations.

\subsection{Jets and fat jets}
\label{sec:sim_jets}

Throughout Section~\ref{sec:qcd} we pretend that quarks and gluons
produced at the LHC are what we observe in the LHC
detectors. In perturbative QCD they are assumed to form the initial and final states,
even though they cannot exist individually as long as QCD is
asymptotically free. In Eq.\eqref{eq:qcd_wf_renorm} we even
apply wave function renormalization factors to their quantum fields. On the other
hand, in Section~\ref{sec:qcd_run_coup} we see that the strong
coupling explodes at small energy scales around $\lqcd$ which means that
something has to happen with quarks and gluons on their way through
the detectors. Indeed, the gluon and all quarks except for the top
quark \underline{hadronize} before they decay and form bunches of
baryons and mesons which in turn decay in many stages. At the LHC these particles
carry a lot of energy, typically around the electroweak scale. 
Relativistic kinematics then tells us
that these baryons and mesons are strongly boosted together to form
\underline{jets}\index{jet}. Those jets we measure at hadron colliders
and link to the partons produced in the hard interaction.\bigskip

Consequently, in Section~\ref{sec:qcd} we use the terms parton and jet
synonymously, essentially assuming that each parton at the LHC turns
into a jet and that the measured jet four-momentum can be linked to the
parton four-momentum.  The way we usually define jets is based on so-called
\underline{recombination algorithms}, including for example the Cambridge--Aachen
or (anti-) $k_T$ algorithms. Imagine we observe a large number of
energy depositions in the ATLAS or CMS calorimeter which we would like
to combine into jets.  We know that they come from a small number of
partons which originate in the hard QCD process and which since have
undergone a sizeable number of splittings, hadronized and decayed to
stable particles. Can we try to reconstruct the original partons?

The answer is yes, in the sense that we can combine a large number of
subjets into smaller numbers, where unfortunately nothing tells us
what the final number of jets should be. We know from
Section~\ref{sec:qcd} that in QCD we can produce an arbitrary number
of hard jets in a hard matrix element and another arbitrary number of jets via
soft or collinear radiation. Therefore, we need to tell the jet
algorithm either how many jets it should arrive at or what 
the resolution of the smallest subjets we consider partons should be,
whatever the measure for this resolution might be. Below
we will therefore discuss what criteria exist for a subjet
recombination to correspond to an assumed physical jet.

\subsubsection{Jet algorithms}
\label{sec:sim_jetalgo}

The basic idea of recombination algorithms is to ask if a given
subjet has a soft or collinear partner. This follows from
Section~\ref{sec:qcd}: we know that
partons produced in a hard process preferably turn into collinear
pairs of partons as approximately described by the parton shower.  To
decide if two subjets have arisen from one parton leaving the hard
process we have to define a collinearity measure. This measure will on
the one hand include the distance in $R$ space as introduced in
Eq.\eqref{eq:eta_phi} and on the other hand the transverse momentum of
one subjet with respect to another or to the beam axis. Explicit
measures weighted by the relative power of the two ingredients are
\begin{alignat}{8}
&k_T \qquad \qquad &
y_{ij} &= \frac{\Delta R_{ij}}{R} \;
          \min \left( p_{T,i}, p_{T,j} \right) \qquad  \qquad  \qquad &
y_{iB} &= p_{T,i} \notag \\
&\text{C/A} &
y_{ij} &=\frac{\Delta R_{ij}}{R} &
y_{iB} &= 1 \notag \\
&\text{anti-}k_T &
y_{ij} &=\frac{\Delta R_{ij}}{R} 
          \min \left( p_{T,i}^{-1}, p_{T,j}^{-1} \right) &
y_{iB} &= p_{T,i}^{-1} \; .
\label{eq:qcd_jet_measure}
\end{alignat}
The parameter $R$ 
balances the jet--jet and jet--beam criteria.  In an exclusive jet
algorithm we define two subjets as coming from one jet if $y_{ij} <
y_\text{cut}$, where $y_\text{cut}$ is a reference scale we give to the
algorithm. Such an exclusive jet algorithm then proceeds as
\begin{itemize}
\item[(1)] for all combinations of two subjets in the event find
  the minimum $y^\text{min} = \text{min}_{ij} (y_{ij},y_{iB})$
\item[(2a)] if $y^\text{min} = y_{ij} < y_\text{cut}$ merge subjets
  $i$ and $j$ and their momenta, keep only the new subjet $i$, go
  back to (1)
\item[(2b)] if $y^\text{min} = y_{iB} < y_\text{cut}$ remove subjet
  $i$, call it beam radiation, go back to (1)
\item[(2c)] if $y^\text{min} > y_\text{cut}$ keep all subjets, call
  them jets, done
\end{itemize}
The result of the algorithm will of course depend on the resolution
$y_\text{cut}$. Alternatively, we can give the algorithm the minimum
number of physical jets and stop there.\bigskip

In an inclusive jet algorithm we do not introduce $y_\text{cut}$. We
can postpone the decision if want to include a jet in our analysis to
the point where all jets are defined. Instead, $y_{iB}$ acts as
the cutoff:
\begin{itemize}
\item[(1)] for all combinations of two subjets in the event find
  the minimum $y^\text{min} = \text{min}_{ij} (y_{ij},y_{iB})$
\item[(2a)] if $y^\text{min} = y_{ij}$ merge subjets
  $i$ and $j$ and their momenta, keep only the new subjet $i$, go
  back to (1)
\item[(2b)] if $y^\text{min} = y_{iB}$ remove subjet
  $i$ and call it a final state jet, go back to (1)
\end{itemize}
The algorithm ends when condition (2a) has left no particles or subjets in the event. Now, the
smallest jet--beam distance $y_{iB}$ sets the scale for all jet--jet
separations. In the C/A example we immediately see that this translates
into a geometric jet size given by $R$. For regular QCD jets we choose
values of $R=0.4...0.7$. For the C/A and $k_T$ cases we see that an
inclusive jet algorithm produces jets arbitrarily close to the beam
axis. Those are hard to observe and often not theoretically well
defined, as we know from our discussion of collinear
divergences. Therefore, inclusive jet algorithms have to include a
final minimum cut on $p_{T, \text{jet}}$ which at the LHC can be
anything from 20~GeV to more than 100~GeV, depending on the
analysis.\bigskip

A technical question is what `combine jets' means in terms of the
four-momentum of the new jet. The three-momentum vectors we simply add
$\vec{k}_i + \vec{k}_j \to \vec{k}_i$. For the zero component we can
assume that the new physical jet have zero invariant mass, which is
inspired by the massless parton we are usually looking for. If instead
we add the four-momenta we can compute the invariant mass of the jet
constituents, the \underline{jet mass}\index{jet!jet mass}. As we will
see in the next section this allows us to extend the concept of jet
algorithms to massive particles like a $W$ or $Z$ boson, the Higgs
boson, or the top quark.\bigskip

All jet algorithms them have in common that they link
physical objects, namely calorimeter towers, to other more or less
physical objects, namely partons from the hard process. As we can see
from the different choices in Eq.\eqref{eq:qcd_jet_measure} we have
all the freedom in the world to weight the angular and transverse
momentum distances relative to each other.
As determined by their power dependence on the
transverse momenta, the three algorithms start with soft constituents
($k_T$), purely geometric (\underline{Cambridge--Aachen
  C/A})\index{jet!Cambridge-Aachen algorithm} or hard constituents
(anti-$k_T$) to form a jet. 
 While for the $k_T$ and the
$C/A$ algorithms it is fairly clear that the intermediate steps have a
physical interpretation, this is not clear at all for the anti-$k_T$
algorithm.

From Section~\ref{sec:qcd} and the derivation of the collinear
splitting kernels it is obvious why theorists working on perturbative
QCD often prefer the $k_T$ algorithm: we know that the showering
probability or the collinear splitting probability is best described
in terms of virtuality or transverse momentum. A transverse momentum
distance between jets is therefore best suited to combine the correct
subjets into the original parton from the hard interaction, following a
series of actual physical intermediate splittings. Moreover, this
transverse momentum measure is intrinsically infrared safe, which means the
radiation of an additional soft parton cannot affect the global
structure of the reconstructed jets. For other algorithms we have to
ensure this property explicitly, and you can find examples in QCD
lectures by Mike Seymour.

The problem of the $k_T$ algorithm arises with pile--up or underlying
event, \ie very soft QCD activity entering the detectors
undirectionally or from secondary partonic vertices.  Such noise is easiest understood geometrically in a
probabilistic picture. Basically, the low energy jet activity is
constant all over the detector, so we \underline{subtract it from each
  event}. How much energy deposition we have to subtract from a
reconstructed jet depends on the area the jet covers in the detector.
Therefore, it is a major step that even for the $k_T$ algorithm we can
compute an IR--safe geometric jet size. The C/A and anti-$k_T$
algorithms are more compact and easier to interpret experimentally.

\subsubsection{Fat jets}
\label{sec:sim_fatjet}

Starting from the way the experiments at the Tevatron and the LHC
search for bottom jets, including several detailed requirements on the
content of such jets, the question arises if we can look for other
\underline{heavy objects} inside a jet. Such jets involving heavy
particles and (usually) a large geometrical size are referred to as
fat jets. For example, looking for boosted top quarks a fat jet
algorithm will try to distinguish between two splitting histories,
where we mark the massive splittings from boosted top decays:

\begin{center} 

\begin{fmfgraph*}(170,80)
 \fmfset{arrow_len}{2mm}
 \fmfset{curly_len}{2mm}
 \fmfleft{in1}
 \fmf{gluon,width=0.2,tension=8}{in1,v1}
 \fmf{gluon,width=0.5,tension=3}{v1,v2}
 \fmf{gluon,width=0.5}{out2,v2}
 \fmf{gluon,width=0.5}{v2,out1}
 \fmf{gluon,width=0.5,tension=6}{v3,v1}
 \fmf{fermion,width=0.5,tension=3}{v3,v8}
 \fmf{fermion,width=0.5}{v8,out3a}
 \fmf{gluon,width=0.5}{v8,out3}
 \fmf{fermion,width=0.5,lab.side=left,label=$u$,tension=5}{v3,v4}
 \fmf{fermion,width=0.5,tension=4}{v4,v6}
 \fmf{fermion,width=0.5}{v6,out4}
 \fmf{gluon,width=0.5,tension=2}{v7,v6}
 \fmf{fermion,width=0.5}{v7,out5}
 \fmf{fermion,width=0.5}{out6,v7} 
 \fmf{gluon,width=0.5,tension=2}{v5,v4}
 \fmf{gluon,width=0.5}{v5,out7}
 \fmf{gluon,width=0.5}{out8,v5}
 \fmfright{out1,out2,out3,out3a,out4,out5,out6,out7,out8}
\end{fmfgraph*} 
\hspace*{2cm}
\begin{fmfgraph*}(170,80)
 \fmfset{arrow_len}{2mm}
 \fmfset{curly_len}{2mm}
 \fmfleft{in1}
 \fmf{fermion,width=0.5,lab.side=right,label=$t$,tension=8}{in1,v1}
 \fmf{fermion,width=0.5,lab.side=right,label=$b$,tension=3}{v1,v2}
 \fmf{fermion,width=0.5,lab.side=right}{v2,out2}
 \fmf{gluon,width=0.5}{v2,out1}
 \fmf{photon,width=0.5,lab.side=left,label=$W$,tension=6}{v1,v3}
 \fmf{fermion,width=0.5,tension=3}{v3,v8}
 \fmf{fermion,width=0.5}{v8,out3a}
 \fmf{gluon,width=0.5}{v8,out3}
 \fmf{fermion,width=0.5,tension=5}{v3,v4}
 \fmf{fermion,width=0.5,tension=4}{v4,v6} 
 \fmf{fermion,width=0.5}{v6,out4}
 \fmf{gluon,width=0.5,tension=2}{v7,v6}
 \fmf{fermion,width=0.5}{v7,out5}
 \fmf{fermion,width=0.5}{out6,v7}
 \fmf{gluon,width=0.5,tension=2}{v5,v4}
 \fmf{gluon,width=0.5}{v5,out7}
 \fmf{gluon,width=0.5}{out8,v5}
 \fmfdot{v1}
 \fmfdot{v3}
 \fmfright{out1,out2,out3,out3a,out4,out5,out6,out7,out8}
\end{fmfgraph*} 

\end{center}

The splittings inside the light--flavor jet are predicted by the soft
and collinear QCD structure. The splittings in the top decays differ
because some of the particles involved have masses. This is the jet
substructure pattern a fat jet algorithm looks for.\bigskip

Three main motivations lead us into the direction of fat
jets: first, dependent on our physics model heavy objects like $W$
bosons or top quarks will be boosted enough to fit into a regular jet
of angular size $R \lesssim 0.7$. Secondly, a jet algorithm might
include hadronic decay products which we would not trust to include in
a regular mass reconstruction based on reconstructed detector
objects. And finally, even if only a fraction of the heavy particles
we are searching for are sufficiently boosted such an algorithm
automatically resolves signal combinatorics known to limit some LHC
analyses\index{combinatorial background}.

At the LHC, we are guaranteed to encounter the experimental situation
$p_T/m \gtrsim 1$ for electroweak gauge bosons, Higgs bosons, and top
quarks.  The more extreme case of $p_T \gg m$, for example searching
for top quarks with a transverse momentum in excess of 1~TeV, is
unlikely to appear in the Standard Model and will only become
interesting if we encounter very heavy resonances decaying to a pair
of top quarks. This is why we focus on the moderate
scenario. Amusingly, the identification of $W$ and top jets was part
of the original paper studying the pattern of splittings $y_{ij}$
defining the $k_T$ algorithm. At the time this was mostly a
gedankenexperiment to test the consistency of the general $k_T$
algorithm approach. Only later reality caught up with it.\bigskip

Historically, fat jet searches were first designed to look for
strongly interacting $W$ bosons. Based on the $k_T$ algorithm they
look for structures in the chain of $y$ values introduced in
Eq.\eqref{eq:qcd_jet_measure}, which define the kinematics of each
jet. For such an analysis of $y$ values it is helpful but not crucial
that the intermediate steps of the jet algorithm have a physics
interpretation. More recent fat jet algorithms looking for not too
highly boosted heavy particles are based on the C/A algorithm which
appears to be best suited to extract massive splittings inside the jet
clustering history. A comparison of different jet algorithms can be
found in the original paper on associated Higgs and gauge boson
production.  Using a C/A algorithm we can search for hadronically
decaying boosted $W$ and $Z$ bosons. The problem is that for those we
only have one hard criterion based on which we can reject QCD
backgrounds: the mass of the $W/Z$ resonance. Adding a second $W/Z$
boson and possibly the mass of a resonance decaying to these two, like
a heavy Higgs boson, adds to the necessary QCD rejection. For Higgs
and top decays the situation is significantly more promising.\bigskip

Starting with the \underline{Higgs tagger}\index{jet!Higgs tagger} we search for jets which
include two bottom quarks coming from a massive Higgs boson with $m_H
\gtrsim 120$~GeV. First, we run the C/A algorithm over the event,
choosing a large geometric size $R=1.2$ estimated to cover
\begin{alignat}{5}
  R_{b\bar b} \sim \frac{1}{\sqrt{z(1-z)}}\;  \frac{m_H}{p_{T,H}}
             > \frac{2 m_H}{p_{T,H}} \; ,
\end{alignat}
in terms of the transverse momentum of the boosted Higgs and the
momentum fractions $z$ and $1-z$ of the two bottom jets.

We then uncluster again this \underline{fat jet}\index{jet!fat jet},
searching for a drop in jet mass indicating the decay of the massive
Higgs to two essentially massless quarks. The iterative unclustering
we start by undoing the last clustering of the jet $j$, giving us two
subjets $j_1,j_2$ ordered such that $m_{j_1} > m_{j_2}$. If the mass
drop between the original jet and its more massive splitting product 
is small, \ie $m_{j_1}> 0.8~m_j$, we conclude that $j_2$ is soft
enough to come from the underlying event or soft--collinear QCD
emission and discard $j_2$ while keeping $j_1$; otherwise we keep both
$j_1$ and $j_2$; each surviving subjet $j_i$ we further decompose
recursively until it reaches some minimum value, $m_{j_i} <
30$~GeV, ensuring it does not involve heavy states. This way we obtain
a splitting pattern which should only include massive splittings and
which for the Higgs jet uniquely identifies the $H \to b\bar{b}$
decay. Making use of the scalar nature of the Higgs boson we can add
an additional requirement on the balance based on $\min(p_{T j_1}^2, p_{T
  j_2}^2) \Delta R_{j_1 j_2}^2$. Of course, all actual numbers in this
selection are subject to experimental scrutiny and can only be
determined after testing the algorithm on LHC data.\bigskip

Experimentally, the goal of such a Higgs search is a distribution of
the invariant mass of the bottom quarks which gives us a signal peak
and side bins to estimate the background.  However, applying jet
algorithms with very large $R$ size makes us increasingly vulnerable
to underlying event, pile--up, or even regular initial--state radiation
as described in Section~\ref{sec:qcd_splitting}. Therefore, we cannot
simply use the mass of a set of fat jet constituents. Instead,
we apply a \underline{filtering} procedure looking at the same
constituent with a higher resolution which can for example be
$R_\text{filt} = \min(0.3, R_{b\bar b}/2)$. This filtering
significantly reduces the $y$-$\phi$ surface area of the relevant
constituents and thereby the destructive power of the underlying event
and pile--up.  The invariant mass we include in the histogram is the
mass of the three hardest filtered constituents, the two bottom quarks and
possibly a radiated gluon.\bigskip

\begin{figure}[t]
\begin{center}
  \includegraphics[width=0.27\hsize]{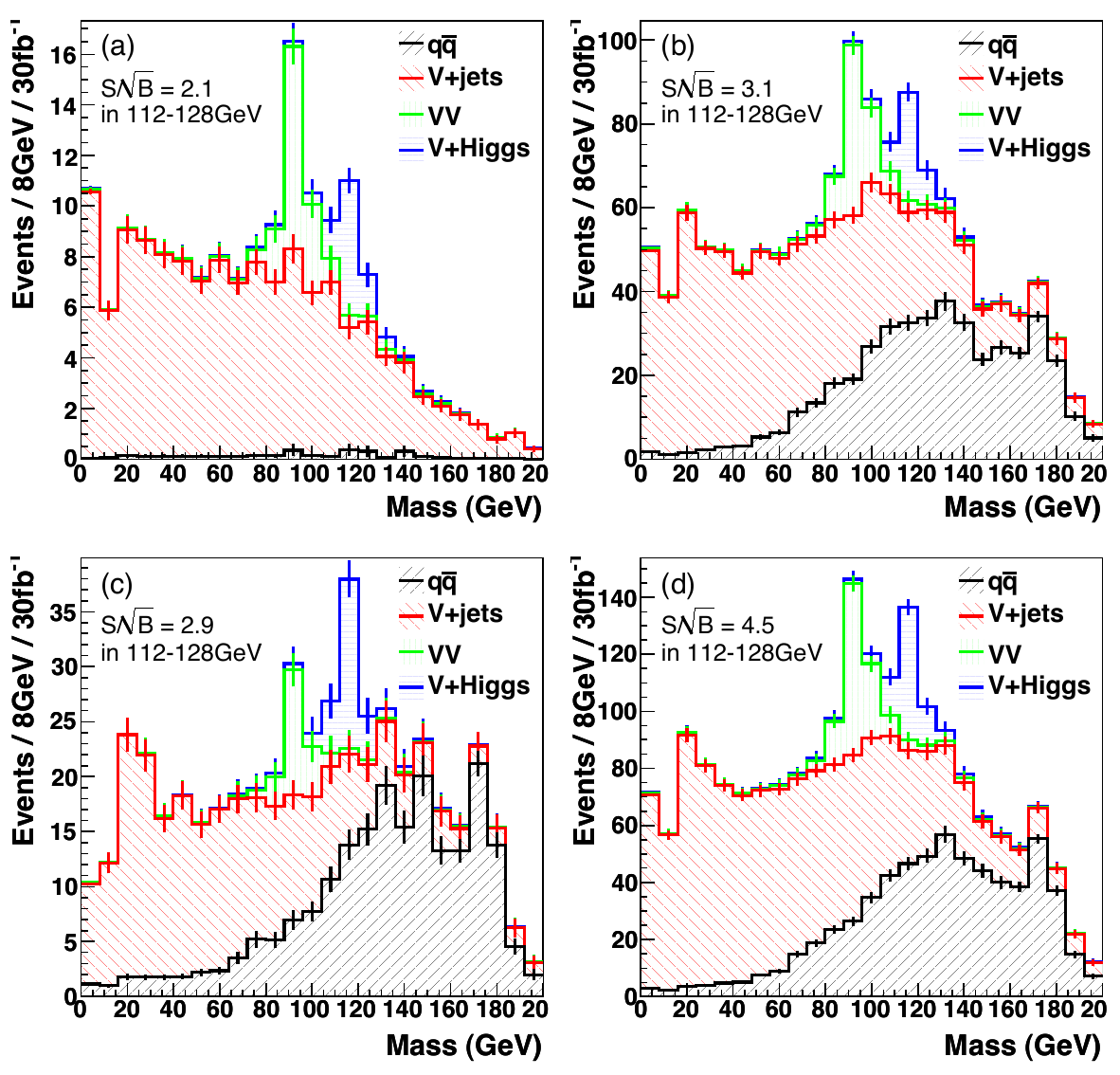}
  \hspace*{0.05\hsize}
  \raisebox{-0.8mm}{\includegraphics[width=0.28\hsize]{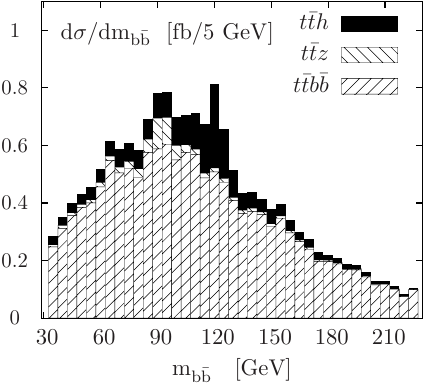}}
  \hspace*{0.05\hsize}
  \raisebox{-0mm}{\includegraphics[width=0.32\hsize]{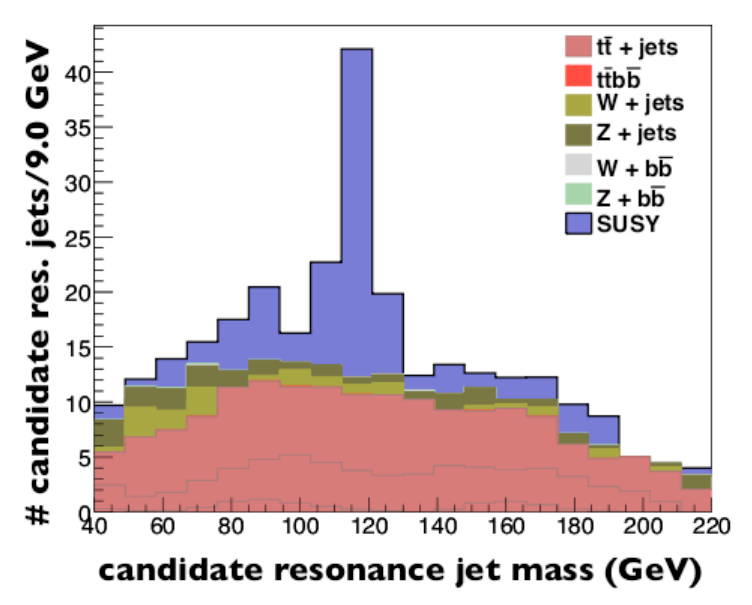}}
\end{center}
\caption{Invariant mass distributions for Higgs searches using fat
  jets from $H \to b\bar{b}$ decays. For a Standard Model Higgs boson
  the production mechanisms are $pp \to WH/ZH$ (left) and $pp \to
  t\bar{t}H$ (center). In cascade decays of supersymmetric squarks
  and gluinos we can apply the same search for the light Higgs boson
  (right). Figures from Refs.~\cite{Butterworth:2008iy},
  \cite{fatjet_tth} and \cite{fatjet_susy} (left to right).}
\label{fig:sim_fat}
\end{figure}

In a \underline{busy QCD environment} another problem arises: errand
jets from initial--state radiation or other particles in the final
state enter the fat jet algorithm and give us several mass drops in
the fat jet history. To avoid shaping the background side bins we can
include several (filtered) subjet combinations, ordered in the
modified Jade distance $p_{T,1} p_{T,2} ( \Delta R_{12} )^4$ --- the
original Jade distance is given by $p_{T,1} p_{T,2} (\Delta
R_{12} )^2$. The invariant mass distributions for
different Higgs search channels in
Figure~\ref{fig:sim_fat} include Standard Model Higgs searches
in $WH/ZH$ production, in $t \bar{t}H$ production, and 
in decays of squarks and gluinos.

From the above discussion we see that Higgs taggers rely only on
\underline{one kinematic criterion}, the mass of the $b\bar{b}$
pair. In terms of background rejection this is not much, so we usually
add two bottom tags on the constituents which according to detector
simulations can be very efficient. The two combined add to a QCD
rejection of at least $10^{-4}$, which might even allows us to run a
Higgs tagger over any kind of event sample and see if we find any
Higgs bosons for example in new physics decays.\bigskip

While fat jet Higgs searches are targeted mostly at the Standard
Model, looking for other boosted heavy particles is usually motivated
by new physics scenarios. Looking for massive particles decaying to
heavy quarks \underline{top taggers}\index{jet!top tagger} should be the next step. Starting from a C/A
jet of size $R=1.5-1.8$ we again search for mass drops, this time
corresponding to the top and $W$ masses. After appropriate filtering
we apply two mass window conditions: first, the entire fat jet has to
reproduce the top mass. Second, we require a mass drop corresponding to
the $W$ decay and effectively constrain a second combination of two
decay jets evaluating the helicity angle of the left handed $W$ decay.
Instead of these two distinct steps we can also apply a
two-dimensional condition on the kinematics of the three top decay
products which avoids assigning the two $W$ decay jets in cases where
two combinations of decay jets give similar invariant masses. On the
simulation level both methods give similar results.

Applying these three kinematic conditions for example in the
HEPTopTagger implementation gives a QCD rejection of a few
per-cent. If this should not be sufficient for a given analysis we can
increase the rejection rate by requiring a bottom tag which as a bonus
also tells us which of the three top decay jets should reconstruct the
$W$ mass. When we use top taggers to look for new particles decaying
to top quarks we are not only interested in finding boosted top
quarks, but we would like to know their invariant mass. This means we
would like to reconstruct their direction and their energy. Such a
reconstruction is possible in a reasonably clean sample, 
provided the top quarks have large enough energy to
boost all three decay jets into a small enough cone.\bigskip

While it seems like the C/A jet algorithm with its purely geometric
selection has the best potential to search for massive particles in
its jet history there exists a multitude of algorithms searching for
boosted top quarks. Once the top quarks have very large transverse
momenta the two-step mass drop criterion becomes less critical because
the three decay jets are too close to be cleanly resolved. In this
situation analyses based on the $k_T$ or anti-$k_T$ algorithms can be
very promising, as could be event shapes which do not involve any jet
algorithm.

\subsection{Helicity amplitudes}
\label{sec:sim_hel}

When we simulate LHC events we do not actually rely on the approach
usually described in text books. This is most obvious when it comes to
the computation of a transition matrix elements in modern LHC Monte
Carlo tools, which you will not even recognize when looking at the
codes. In Section~\ref{sec:qcd_dy} we compute the cross section for
$Z$ production by writing down all external spinors, external
polarization vectors, interaction vertices and propagators and
squaring the amplitude analytically. The amplitude itself inherits
external indices for example from the polarization vectors, while
$\matx$ is a real positive number with a fixed mass dimension
depending on the number of external particles.

For the LHC nobody calculates gamma matrix traces by hand
anymore. Instead, we use powerful tools like FORM\index{FORM}
to compute traces of
Dirac matrices in the calculation of $\matx$.  Nevertheless, a major
problem with squaring Feynman diagrams and computing polarization sums
and gamma matrix traces is that once we include more than three
particles in the final state, the number of terms appearing in 
$\matx$ soon becomes very large. Moreover, this approach requires
symbolic computer manipulation instead of pure numerics.  In this
section we illustrate how we can transform the computation of $\matx$
at the tree level into a purely numerical problem.\bigskip

As an example, we consider our usual toy process
\begin{alignat}{5}
u \bar{u} \to \gamma^* \to \mu^+ \mu^- \; .
\label{eq:proc_hel}
\end{alignat}
The structure of the amplitude $\mat$ with two internal Dirac indices
$\mu$ and $\nu$ involves one vector current on each side $(\bar{u}_f
\gamma_\mu u_f)$ where $f=u,\mu$ are to good approximation massless, so we do not
have to be careful with the different spinors $u$ and $v$. The entries
in the external spinors are given by the spin of the massless fermions
obeying the Dirac equation.  For each value of $\mu = 0 \cdots 3$ each
current is a complex number, computed from the four component of each
spinor and the respective $4 \times 4$ gamma matrix $\gamma^\mu$ shown
in Eq.\eqref{eq:dirac_matrices}.  The intermediate photon propagator
has the form $g_{\mu \nu}/s$, which is a real number for each value of
$\mu=\nu$. Summing over $\mu$ and $\nu$ in both currents forms the
matrix element. To square this matrix element we need to sum $\mat^*
\times \mat$ over all possible spin directions of the external
fermions.\bigskip

Instead of squaring this amplitude symbolically we can follow exactly
the steps described above and compute an array of numbers for
different spin and helicity combinations numerically.  Summing over
the internal Dirac indices we compute the matrix element; however, to
compute the matrix element squared we need to sum over external
fermion spin directions or gauge boson polarizations. The helicity
basis we have to specify externally. This is why this method is called
helicity amplitude approach. To explain the way this method works, we
illustrate it for muon pair production based on the implementation in
the \underline{Madgraph/Helas}\index{helicity amplitudes, HELAS}
package.

Madgraph\index{event generators!Madgraph} is a tool to compute matrix elements this
way. Other event generators have corresponding codes serving the same
purposes.  In our case, Madgraph5 automatically produces a Fortran
routine which then calls functions to compute spinors, polarization
vectors, currents of all kinds, etc.  These functions are available as
the so-called Helas library. For our toy process
Eq.\eqref{eq:proc_hel} the slightly shortened Madgraph5 output reads

{\scriptsize
\begin{verbatim} 
      REAL*8 FUNCTION MATRIX1(P,NHEL,IC)
C     
C     Generated by Madgraph 5 
C     
C     Returns amplitude squared summed/avg over colors
C     for the point with external lines W(0:6,NEXTERNAL)
C     
C     Process: u u~ > mu+ mu- / z WEIGHTED=4 @1
C     
      INTEGER    NGRAPHS, NWAVEFUNCS, NCOLOR
      PARAMETER (NGRAPHS=1, NWAVEFUNCS=5, NCOLOR=1)

      REAL*8 P(0:3,NEXTERNAL)
      INTEGER NHEL(NEXTERNAL), IC(NEXTERNAL)

      INCLUDE 'coupl.inc'

      DATA DENOM(1)/1/
      DATA (CF(I,  1),I=  1,  1) /    3/

      CALL IXXXXX(P(0,1),ZERO,NHEL(1),+1*IC(1),W(1,1))
      CALL OXXXXX(P(0,2),ZERO,NHEL(2),-1*IC(2),W(1,2))
      CALL IXXXXX(P(0,3),ZERO,NHEL(3),-1*IC(3),W(1,3))
      CALL OXXXXX(P(0,4),ZERO,NHEL(4),+1*IC(4),W(1,4))
      CALL FFV1_3(W(1,1),W(1,2),GC_2,ZERO, ZERO, W(1,5))
      CALL FFV1_0(W(1,3),W(1,4),W(1,5),GC_3,AMP(1))
      JAMP(1)=+AMP(1)

      DO I = 1, NCOLOR
        DO J = 1, NCOLOR
          ZTEMP = ZTEMP + CF(J,I)*JAMP(J)
        ENDDO
        MATRIX1 = MATRIX1 + ZTEMP*DCONJG(JAMP(I))/DENOM(I)
      ENDDO

      END
\end{verbatim}
}

The input to this function are the external four-momenta $p(0:3,1:4)$
and the helicities of all fermions $n_\text{hel}(1:4)$ in the
process. Remember that helicity and chirality are identical only for
massless fermions because chirality is defined as the eigenvalue of
the projectors $(\one \pm \gamma_5)/2$, while helicity is defined as
the projection of the spin onto the momentum direction, \ie as the
left or right handedness. We give the exact definition of these two properties
in Section~\ref{sec:field_theory}.
The entries of $n_\text{hel}$ will be either
$+1$ or $-1$.  For each point in phase space and each helicity
combination the Madgraph subroutine {\tt MATRIX1} computes the
matrix element using standard \underline{Helas routines}\index{helicity amplitudes, HELAS}.
\begin{itemize}
\item[$\cdot$] {\tt IXXXXX}($p,m,n_\text{hel},n_\text{sf},F$) computes
  the wave function of a fermion with incoming fermion number, so
  either an incoming fermion or an outgoing anti--fermion. As input it
  requires the four-momentum, the mass and the helicity of this
  fermion. Moreover, $n_\text{sf} = +1$ marks the incoming fermion $u$
  and $n_\text{sf} = -1$ the outgoing anti--fermion $\mu^+$,
  because by convention Madgraph defines its particles as $u$ and
  $\mu^-$. 

  The fermion wave function output is a complex array $F(1:6)$.  Its
  first two entries are the left--chiral part of the fermionic spinor,
  \ie $F(1:2) = (\one -\gamma_5)/2 \; u$ or $F(1:2) = (\one
  -\gamma_5)/2 \; v$ for $n_\text{sf} = \pm 1$. The entries $F(3:4)$
  are the right--chiral spinor. These four numbers can directly be
  computed from the four-momentum if we know the helicity.  The four
  entries correspond to the size of one $\gamma$ matrix, so we can
  compute the trace of the chain of gamma matrices.  Because for
  massless particles helicity and chirality are identical, our quarks
  and leptons will only have finite entries $F(1:2)$ for
  $n_\text{hel}=-1$ and $F(3:4)$ for $n_\text{hel}=+1$.

  The last two entries of $F$ contain the four-momentum in the
  direction of the fermion flow, namely $F(5) = n_\text{sf}
  (p(0)+ip(3))$ and $F(6) = n_\text{sf} (p(1)+ip(2))$.

\item[$\cdot$] {\tt OXXXXX}($p,m,n_\text{hel},n_\text{sf},F$) does the 
  same for a fermion with outgoing fermion flow, \ie our incoming
  $\bar{u}$ and our outgoing $\mu^-$. The left--chiral and
  right--chiral components now read $F(1:2) = \bar{u} (\one -
  \gamma_5)/2$ and $F(3:4) = \bar{u} (\one + \gamma_5)/2$, and
  similarly for the spinor $\bar{v}$.  The last two entries are $F(5)
  = n_\text{sf} (p(0)+ip(3))$ and $F(6) = n_\text{sf} (p(1)+ip(2))$.

\item[$\cdot$] {\tt FFV1\_3}($F_i,F_o,g,m,\Gamma,J_{io}$) computes the
  (off--shell) current for the vector boson attached to the two
  external fermions $F_i$ and $F_o$. The coupling $g(1:2)$ is a
  complex array with the interaction of the left--chiral and
  right--chiral fermion in the upper and lower index. For a general
  Breit--Wigner propagator\index{particle width!Breit--Wigner propagator} we need to know the mass $m$ and the width
  $\Gamma$ of the intermediate vector boson. The output array $J_{io}$
  again has six components which for the photon with momentum $q$ are
\begin{alignat}{5}
J_{io}(\mu+1)&= - \frac{i}{q^2} \; F_o^T \; \gamma^\mu
                \left(  g(1) \; \frac{\one - \gamma_5}{2} \;
                      + g(2) \; \frac{\one + \gamma_5}{2} \;
                \right) \;
                F_i 
\qqqquad \mu = 0,1,2,3 \notag \\
J_{io}(5) &= -F_i(5) + F_o(5) 
           \sim - p_i(0) + p_o(0) +i \left( -p_i(3) - p_o(3) 
                                     \right) \notag \\
J_{io}(6) &= -F_i(6) + F_o(6)
           \sim - p_i(1) + p_o(1) +i \left( -p_i(2) + p_o(2) 
                                     \right) \; .
\end{alignat}
  The first four entries in $J_{io}$ correspond to the index $\mu$ or
  the dimensionality of the Dirac matrices in this vector current.
  The spinor index is contracted between $F_o^T$ and $F_i$.

  As two more arguments $J_{io}$ includes the four-momentum flowing
  through the gauge boson propagator. They allow us to reconstruct
  $q^\mu$ from the last two entries
\begin{alignat}{5}
q^\mu = \left( \text{Re} J_{io}(5), \text{Re} J_{io}(6),
                \text{Im} J_{io}(6), \text{Im} J_{io}(5)
        \right) \; .
\end{alignat}

\item[$\cdot$] {\tt FFV1\_0}($F_i,F_o,J,g,V$) computes the amplitude of
  a fermion--fermion--vector coupling using the two external fermionic
  spinors $F_i$ and $F_o$ and an incoming vector current $J$ which in
  our case comes from {\tt FFV1\_3}. Again, the coupling $g(1:2)$ is a
  complex array, so we numerically compute
\begin{alignat}{5}
   F_o^T \; \slashchar{J} \left( g(1) \; \frac{\one - \gamma_5}{2} \;
                         + g(2) \; \frac{\one + \gamma_5}{2} \;
                    \right) \; F_i \; .
\end{alignat}
  All spinor and Dirac indices of the three input
  arguments are contracted in the final result. Momentum conservation
  is not enforced by {\tt FFV1\_0}, so we have to take care of it by
  hand.
\end{itemize}

Given the list above it is easy to follow how Madgraph computes the
amplitude for $u \bar{u} \to \gamma^* \to \mu^+ \mu^-$. First, it
calls wave functions for all external particles and puts
them into the array $W(1:6,1:4)$. The vectors $W(*,1)$ and $W(*,3)$
correspond to $F_i(u)$ and $F_i(\mu^+)$, while $W(*,2)$ and $W(*,4)$
mean $F_o(\bar{u}$ and $F_o(\mu^-)$. 

The first vertex we evaluate is the incoming quark--photon vertex. Given
the wave functions $F_i = W(*,1)$ and $F_o = W(*,2)$ {\tt FFV1\_3}
computes the vector current for the massless photon in the 
$s$-channel. Not much changes if we instead choose a massive $Z$ boson,
except for the arguments $m$ and $\Gamma$ in the {\tt FFV1\_3}
call. Its output is the photon current $J_{io} \equiv W(*,5)$.

The last step combines this current with the two outgoing muons
coupling to the photon. Since this number gives the final amplitude,
it should return a complex number, not an array. Madgraph calls {\tt
  FFV1\_0} with $F_i = W(*,3)$ and $F_o = W(*,4)$, combined with the
photon current $J = W(*,5)$. The result {\tt AMP} is copied into {\tt
  JAMP} without an additional sign which could have come from the
relative ordering of external fermions in different Feynman diagrams
contributing to the same process. 

The only remaining sum left to compute before we square {\tt JAMP} is
the color structure, which in our simple case means one color
structure with a color factor $N_c=3$.

As an added bonus Madgraph produces a file with all Feynman diagrams
in which the numbering of the external particles corresponds to the
second argument of $W$ and the numbering of the Feynman diagrams
corresponds to the argument of {\tt AMP}. This helps us identify
intermediate results $W$, each of which is only computed once, even if
is appears several times in the different Feynman diagrams.

As mentioned above, to calculate the transition amplitude Madgraph
requires all masses and couplings. They are transferred through common
blocks in the file coupl.inc and computed elsewhere.  In general,
Madgraph uses unitary gauge for all vector bosons, because in the
helicity amplitude approach it is easy to accommodate complicated
tensors, in exchange for a large number of Feynman diagrams\index{unitary gauge}.\bigskip

The function {\tt MATRIX1} described above is not yet the full
story. When we square $\mat$ symbolically we need to sum over the
spins of the outgoing states to transform a spinor product of the kind
$u \bar{u}$ into the residue or numerator of a fermion propagator. To
obtain the full transition amplitude numerically we correspondingly
sum over all \underline{helicity combinations}\index{helicity amplitudes, HELAS} of the external
fermions, in our case $2^4 = 16$ combinations\index{event generators!Madgraph}.

{\scriptsize
\begin{verbatim} 
      SUBROUTINE SMATRIX1(P,ANS)
C     
C     Generated by Madgraph 5
C     
C     Returns amplitude squared summed/avg over colors
C     and helicities for the point in phase space P(0:3,NEXTERNAL)
C     
C     Process: u u~ > mu+ mu- / z 
C     
      INTEGER     NCOMB, NGRAPHS, NDIAGS, THEL
      PARAMETER (NCOMB=16, NGRAPHS=1, NDIAGS=1, THEL=2*NCOMB)

      REAL*8 P(0:3,NEXTERNAL)

      INTEGER I,J,IDEN
      INTEGER NHEL(NEXTERNAL,NCOMB),NTRY(2),ISHEL(2),JHEL(2)
      INTEGER JC(NEXTERNAL),NGOOD(2), IGOOD(NCOMB,2)
      REAL*8 T,MATRIX1
      LOGICAL GOODHEL(NCOMB,2)

      DATA NGOOD /0,0/
      DATA ISHEL/0,0/
      DATA GOODHEL/THEL*.FALSE./

      DATA (NHEL(I,   1),I=1,4) /-1,-1,-1,-1/
      DATA (NHEL(I,   2),I=1,4) /-1,-1,-1, 1/
      DATA (NHEL(I,   3),I=1,4) /-1,-1, 1,-1/
      DATA (NHEL(I,   4),I=1,4) /-1,-1, 1, 1/
      DATA (NHEL(I,   5),I=1,4) /-1, 1,-1,-1/
      DATA (NHEL(I,   6),I=1,4) /-1, 1,-1, 1/
      DATA (NHEL(I,   7),I=1,4) /-1, 1, 1,-1/
      DATA (NHEL(I,   8),I=1,4) /-1, 1, 1, 1/
      DATA (NHEL(I,   9),I=1,4) / 1,-1,-1,-1/
      DATA (NHEL(I,  10),I=1,4) / 1,-1,-1, 1/
      DATA (NHEL(I,  11),I=1,4) / 1,-1, 1,-1/
      DATA (NHEL(I,  12),I=1,4) / 1,-1, 1, 1/
      DATA (NHEL(I,  13),I=1,4) / 1, 1,-1,-1/
      DATA (NHEL(I,  14),I=1,4) / 1, 1,-1, 1/
      DATA (NHEL(I,  15),I=1,4) / 1, 1, 1,-1/
      DATA (NHEL(I,  16),I=1,4) / 1, 1, 1, 1/
      DATA IDEN/36/

      DO I=1,NEXTERNAL
        JC(I) = +1
      ENDDO

      DO I=1,NCOMB
        IF (GOODHEL(I,IMIRROR) .OR. NTRY(IMIRROR).LE.MAXTRIES) THEN
          T = MATRIX1(P ,NHEL(1,I),JC(1))
          ANS = ANS+T
        ENDIF
      ENDDO

      ANS = ANS/DBLE(IDEN)
      END
\end{verbatim}
}

The important part of this subroutine is the list of possible helicity
combinations stored in the array $n_\text{hel}(1:4,1:16)$. Adding all
different helicity combinations means a loop over the second argument
and a call of {\tt MATRIX1} with the respective helicity
combination.  Because of the naive helicity combinations many are
not allowed the array {\tt GOODHEL} keeps track of
valid combinations. After an initialization to all `false' this array
is only switched to `true' if {\tt MATRIX1} returns a finite
value, otherwise Madgraph does not waste time to compute the matrix
element.  At the very end, a complete spin--color averaging factor is
included as {\tt IDEN} and in our case given by $2 \times 2 \times
N_c^2 = 36$.\bigskip

Altogether, Madgraph provides us with the subroutine {\tt
  SMATRIX1} and the function {\tt MATRIX1} which together
compute $\overline{\matx}$ for each phase space point given as an
external momentum configuration.  This helicity method might not seem
particularly appealing for a simple $(2 \to 2)$ process, but it makes it
possible to compute processes with many particles in the
final state and typically up to 10000 Feynman diagrams which we could
never square symbolically, no matter how many graduate students' live
times we throw in.

\subsection{Missing transverse energy} 
\label{sec:sim_met}

Some of the most interesting signatures at the LHC involve dark matter
particles.  From cosmological constraints we know that dark matter
definitely interacts gravitationally and that it cannot carry
electromagnetic or color charges. Weak 
interactions\index{weak interaction} are allowed because of their limited reach. It turns
out that a weakly interacting particle with a mass around the
electroweak scale typically gives the observed relic density in the
universe. This is called the \underline{WIMP miracle}\index{dark matter, WIMP miracle}. It it the reason why in modern TeV-scale
model building every model (and its dog) predict a stable WIMP. From
supersymmetry\index{supersymmetry} we know that this is not hard to
achieve: all we need is a $\mathbb{Z}_2$ symmetry to induce a
\underline{multiplicative quantum number} for a sector of newly
introduced particles. In supersymmetry this is called $R$ parity, in
little-Higgs models $T$ parity, and in extra-dimensional models
Kaluza--Klein parity.\bigskip

At the LHC we typically produce strongly interacting new particles,
provided they exist. In the presence of a conserved dark matter quantum number exists
we will always produce them in pairs.  Each of them decays to the
weakly interacting sector which includes a stable dark matter
agent. On the way, the originally produced particles have to radiate
quarks or gluons to shed their color charge. If in some kind of
cascade decays\index{cascade decay} they also radiate leptons or
photons those can be very useful to trigger\index{trigger} on and to
reduce QCD backgrounds, but this depends on the details of the weakly
interacting new physics sector. The decay steps ideally
are two body decays from on--shell particle to on--shell particle, but
they do not have to be.  What we therefore look for is jets in
association with pairs of only weakly interacting, hence invisible
particles in the ATLAS and CMS detectors.\bigskip

From Eq.\eqref{eq:qcd_sigtot} and the discussion of parton densities
we remember that at hadron colliders we do not know the 
kinematics of the initial state. While in the transverse plane by definition the
incoming partons have zero momentum, in beam direction we only know its
boost statistically. The way to look for invisible
particles therefore is a mis-balance of three-momentum in the
transverse plane.  The actual observable is the \underline{missing
  transverse momentum} defined as the vector sum of the transverse momenta
of all invisible particles. We can convert it into a missing
transverse energy which in the absence of any information on particle
masses is defined as the absolute value of the two-dimensional missing
momentum vector.  LHC events including dark matter are characterized
by a high jet multiplicity and large missing transverse energy.

At the end of Section~\ref{sec:qcd_ckkw} we focus on the proper
simulation of $W$+jets and $Z$+jets samples, which are the Standard
Model backgrounds to such signals. It will turn out that jet merging
is needed to reliably predict the missing transverse momentum
distributions in Standard Model processes. After all our
studies in Section~\ref{sec:qcd} we are at least theoretically on safe
ground. However, this is not the whole story of missing transverse
momentum.

\subsubsection{Measuring missing energy}
\label{sec:sim_met_ex}

\begin{figure}[t]
\begin{center}
  \includegraphics[width=0.4\hsize]{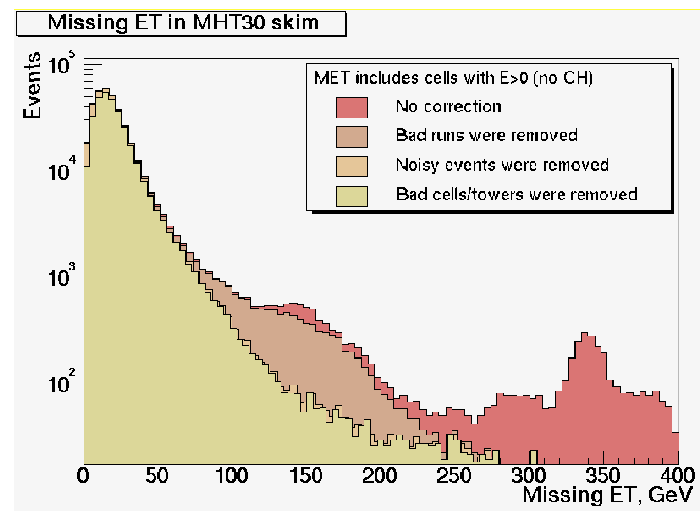}
  \hspace*{0.1\hsize}
  \raisebox{-2.5mm}{\includegraphics[width=0.41\hsize]{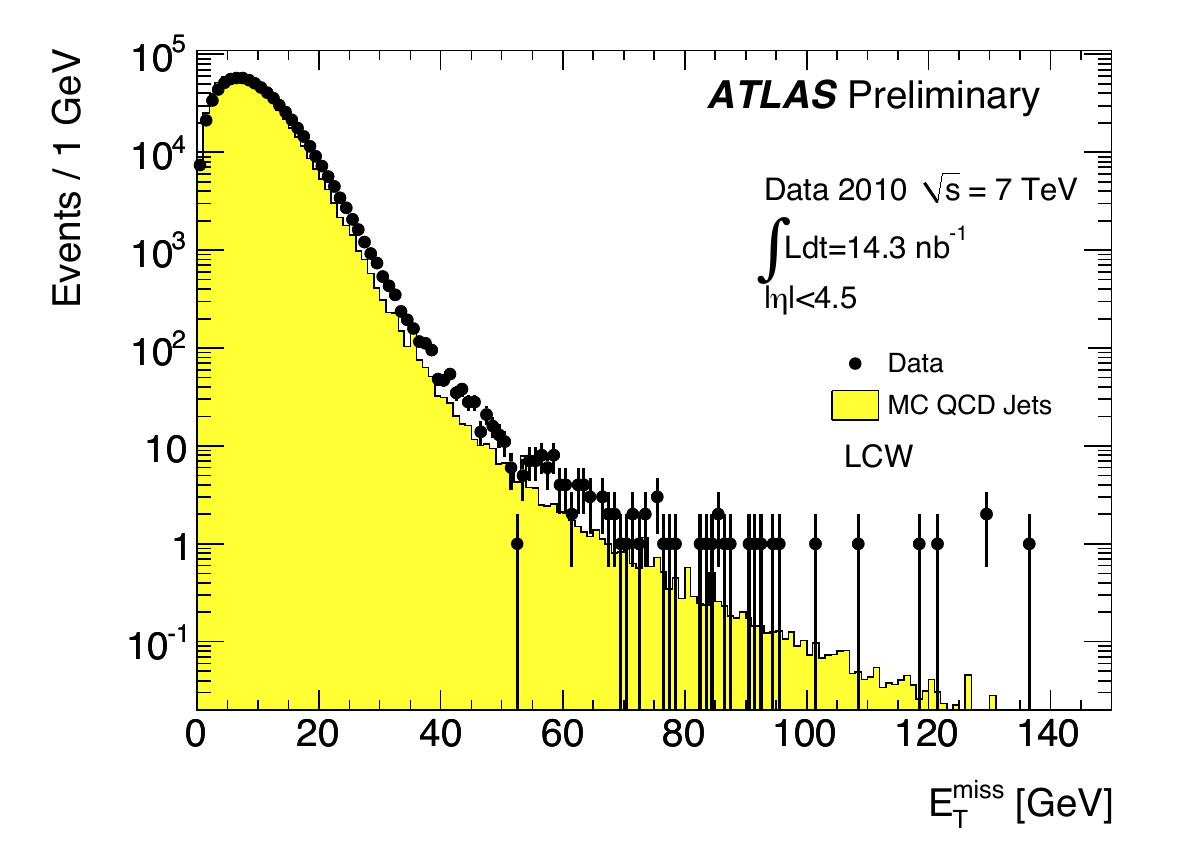}}
\end{center}
\caption{Left: missing energy distribution from the early running
  phase of the DZero experiment at the Tevatron. Figure from Beate
  Heinemann. Right: corrected missing energy distribution in QCD
  events at ATLAS using only data from April/May 2010 at 7~TeV
  collider energy. Figure from Ref.\cite{atlas_met}.} 
\label{fig:sim_met}
\end{figure}

The left panel of Figure~\ref{fig:sim_met} is a historic distribution of missing
transverse energy from DZero. It nicely illustrates that
by just measuring missing transverse energy, Tevatron would have
discovered supersymmetry based on two beautiful peaks 
around 150~GeV and around 350~GeV. However, this
preliminary experimental result has nothing to do with physics, it
is purely a detector effect.

We can illustrate the problem of missing energy using a simple number:
to identify and measure a lepton we need around 500 out of 200000
calorimeter cells in an experiment like ATLAS, while for missing
transverse energy we need all of them. To cut on a variable like
missing transverse momentum we need to understand our detectors really
well, and this level of understanding needs a lot of time and effort.\bigskip

There are several sources of missing energy which we have to understand
before we get to search for new physics:
\begin{itemize}
\item[--] First, we have to subtract bad runs. They happen if for a
  few hours parts of the detector do not work properly. We can
  identify them by looking at the detector response and its
  correlation. One example is a so-called ring of fire where we see
  coherent effects in detector modules of circular shape around the
  beam axis. 

\item[--] Next, there will be coherent noise in the calorimeter. With
  200000 cells we know that some of them will individually fail or
  produce noise. Some sources of noise, like leaking voltage or other
  electronic noise can be correlated geometrically and lead to
  beautiful missing momentum signals. The way to get rid of such noise
  event by event is to again look for usual detector response.
  Combined with bad runs such events can constitute
  $\ope(0.1\%)$ of all events and get removed from the data
  sample.

\item[--] In addition, there might be particles crossing our
  detector, but not coming from the interaction point. They can be
  cosmic rays and lead to unbalanced energy deposition as well. Such
  events will have reconstructed particle tracks which are not
  compatible with the measured primary vertex.

\item[--] Another source of fake missing energy is failing calorimeter
  cells, like continuously hot cells or dead cells. ATLAS for example
  has developed such a hole by 2010. Events where missing energy
  points into such a region can often be removed once we understand
  the detector.

\item[--] While not really a detector fake the main source of missing
  energy at hadron colliders are mis-measured QCD jets. If parts of
  jets point into regions with poor calorimetry, like support
  structures, the jet energy will be wrongly measured, and the
  corresponding QCD event will show missing transverse energy. One way
  to tackle this problem is to require a geometric separation of the
  missing momentum vector and hard jets in the event. ATLAS detector
  studies indicate that up to $\ope(0.1\% - 1\%)$ of all hard QCD
  events at the LHC lead to more than 100~GeV of well separated fake
  missing transverse energy. Figure~\ref{fig:higgs_lhcall} in
  Section~\ref{sec:higgs_gf_lhc} shows that this is not at all a
  negligible number of events.

\end{itemize}
\bigskip

Once we understand all sources of fake missing momentum we can
focus on real missing momentum. This missing transverse momentum we
compute from the momenta of all tracks seen in the detector. This
means that any uncertainty on these measurements, like the jet or
lepton energy scale will smear the missing momentum. Moreover, we know
that there is for example dead matter in the detector, so we have to
compensate for this. This compensation is a global correction to
individual events, which means it will generally smear the missing
energy distribution. The right panel of Figure~\ref{fig:sim_met} shows
a very early missing transverse energy distribution of ATLAS after
some of the corrections described above.

To simulate a realistic missing transverse momentum distribution
at the LHC we have to smear all jet and lepton momenta, and in
addition apply a Gaussian \underline{smearing}\index{detector smearing} of the order
\begin{alignat}{5}
\frac{\Delta \slashchar{E}_T}{\text{GeV}}
\sim 
\frac{1}{2} \; \sqrt{ \frac{\sum E_T}{\text{GeV}}} \;
\gtrsim 20 \; .
\end{alignat}
While this sounds like a trivial piece of information it is impossible to count
the number of papers where people forget this smearing and discover
great channels to look for Higgs bosons or new physics. They fall
apart when experimentalists take a careful look. The simple rule is:
phenomenological studies are right or wrong based on if they can be
reproduced by real experimentalists and real detectors or not.

\subsubsection{Missing energy in the Standard Model}
\label{sec:sim_met_sm}

In the Standard Model there exists a particle which only interacts
weakly: the neutrino. At colliders we produce them in reasonably large
numbers in $W$ decays.  This means that in $W+$~jets production we can
learn how to reconstruct the mass of a leptonically decaying $W$ from
one observed and one missing particle. We construct a
\underline{transverse mass}\index{mass!transverse mass} in analogy to an invariant mass, but
neglecting the longitudinal momenta of the decay products
\begin{alignat}{5}
m_{T}^2 &= \left( E_{T,\text{miss}} + E_{T, \ell} \right)^2
       - \left( \vec{p}_{T,\text{miss}} + \vec{p}_{T,\ell} \right)^2 
\notag \\
      &= m_{\ell}^2 + m_\text{miss}^2
       + 2 \left( E_{T,\ell} E_{T,\text{miss}} 
       - \vec{p}_{T,\ell} \cdot \vec{p}_{T,\text{miss}} \right) \; ,
\label{eq:sig_mt}
\end{alignat}
in terms of a transverse energy $E_T^2 = \vec{p}_T^2 + m^2$.  Since
the transverse mass is always smaller than the actual $W$ mass and
reaches this limit for realistic phase space regions we can extract
$m_W$ from the upper edge in the $m_{T,W}$ distribution. Obviously, we
can define the transverse mass in many different reference
frames. However, its value is invariant under --- or better
independent of --- longitudinal boosts. Moreover, given that we
construct it as the transverse projection of an invariant mass it is
also invariant under transverse boosts. By construction we cannot
analyze the transverse mass event by event, so this $W$ mass
measurement only uses the fraction of events which populate the upper
end of the transverse mass distribution.\bigskip

Alternatively, from single top production and the production of mixed
leptonically and hadronically decaying top pairs we know another
method to conditionally reconstruct masses and momenta involving one
invisible particle: from a leptonically decaying top quark we only
miss the longitudinal momentum of the neutrino. On the other hand, we
know at least for the signal events that the neutrino and the lepton
come from an on--shell $W$ boson, so we can use this
\underline{on--shell condition} to reconstruct the longitudinal
neutrino momentum under the assumption that the neutrino has zero
mass. Recently, we have seen that sufficiently boosted top quarks 
with leptonic decays can be fully reconstructed even without using the 
measured missing energy vector. Instead, we rely on the $W$ and $t$ 
on--shell conditions and on an assumption about the neutrino momentum 
in relation to the bottom-lepton decay plane.\bigskip

From Higgs searches we know how to extend the transverse mass to two
leptonic $W$ decays with two neutrinos in the final state. The
definition of this transverse mass 
\begin{alignat}{5}
 m^2_{T,WW} &= \left( E_{T,\text{miss}} + E_{T, \ell \ell} \right)^2
            - \left( \vec{p}_{T,\text{miss}} + \vec{p}_{T, \ell \ell} \right)^2 
\notag \\
      &= m_{\ell \ell}^2 + m_\text{miss}^2
       + 2 \left( E_{T,\ell \ell} E_{T,\text{miss}} 
       - \vec{p}_{T,\ell \ell} \cdot \vec{p}_{T,\text{miss}} \right) 
\end{alignat}
is not unique because it is not clear how to define $m_\text{miss}$,
which also enters the definition of $E_{T,\text{miss}}$. From Monte
Carlo studies it seems that identifying $m_\text{miss} \equiv m_{\ell
  \ell}$, which is correct at threshold, is most strongly peaked. On
the other hand, setting $m_\text{miss}=0$ to define a proper
bounded--from--above transverse mass variable seems to improve the Higgs
mass extraction.\bigskip

For an unspecified number of visible and invisible particles in the
final state there also exist global observables we can rely on.  The
\underline{visible mass}\index{mass!visible mass} is based on the
assumption that we are looking for the decay of two heavy new states
where the parton densities will ensure that these two particles are
non--relativistic. We can then approximate the partonic energy
$\sqrt{\hat{s}} \sim m_1 + m_2$ by some kind of visible energy. If
the heavy states are produced with little energy, boost
invariance is not required for this construction.  Without taking into
account missing energy and adding leptons $\ell$ and jets $j$ the
visible mass looks like
\begin{alignat}{5}
m^2_\text{visible} = \left[ \sum_{\ell,j} \; E \right]^2 
                   - \left[ \sum_{\ell,j} \; \vec{p} \right]^2 \; .
\end{alignat}
Similarly, the Tevatron experiments have for a long time used an
effective transverse mass scale which is usually evaluated for jets
only, but can trivially be extended to leptons:
\begin{alignat}{5}
 H_T = \sum_{\ell,j} \; E_T 
     = \sum_{\ell,j} \; p_T  \; ,
\end{alignat}
where the last step assumes that all final--state particles are
massless. In an alternative definition of $H_T$ we sum over a number
of jets plus the missing energy and skip the hardest jet in this sum.

When combining such a measure with missing transverse momentum the
question arises: do we want to pair up the missing transverse momentum
with the visible transverse momenta or with the complete visible
momenta? For example, we can use the scalar sum of all transverse
momenta in the event, now including the missing transverse momentum
\begin{alignat}{5}
 m_\text{eff} = \sum_{\ell,j,\text{miss}} \; E_T 
             = \sum_{\ell,j,\text{miss}} \; p_T  \; .
\end{alignat}
This effective mass is known to trace the mass of the heavy new
particles decaying for example to jets and missing energy. This
interpretation relies on the non--relativistic nature of the production
process and our confidence that all jets included are really decay
jets.

\subsubsection{Missing energy and new physics}
\label{sec:sim_met_bsm}

The methods described in the last section are well studied for
different Standard Model processes and can be applied in new physics
searches for various lengths of decay chains. However, there is need
for one significant modification, namely to account for a finite
\underline{unknown mass} of the missing energy particle. This is
a problem of relativistic kinematics and at leading order does not
require any knowledge of QCD or new physics models. 

\begin{figure}[t]
\begin{center}
  \includegraphics[width=0.30\textwidth]{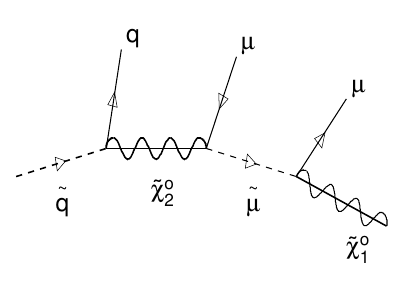}
\end{center}
\caption{Feynman diagram for the long decay chain shown in
  Eq.\eqref{eq:sim_squarkchain}.}
\label{fig:feyn_sq}
\end{figure}

The chain of three successive three-body decays shown in
Figure~\ref{fig:feyn_sq} is the typical left handed \underline{squark
  cascade decay}\index{cascade decay} in supersymmetry.  The same topology we can interpret
in extra-dimensional models (universal extra dimensions or UED) as the decay of a Kaluza--Klein quark
excitation
\begin{alignat}{5}
  \tilde{q}_L \to \tilde{\chi}_2^0 q \to \tilde{\ell}^\pm
  \ell^\mp q \to \tilde{\chi}_1^0 \ell^+\ell^-q
 \qqqquad
  Q^{(1)}_L \to Z^{(1)} q \to {\ell^{(1)}}^\pm
  \ell^\mp q \to \gamma^{(1)} \ell^+\ell^-q \; .
\label{eq:sim_squarkchain}
\end{alignat}
In both cases the last particle in the chain, the lightest neutralino
or the Kaluza--Klein photon excitation pass the detectors unobserved.
The branching ratio for such decays might not be particularly large;
for example in the supersymmetric\index{supersymmetry} parameter point
SPS1a with a mass spectrum we will discuss later in
Figure~\ref{fig:sim_sig_gluinochain} the long squark decay ranges
around $4\%$. On the other hand, strongly interacting new particles
should in principle be generously produced at the LHC, so we usually assume that
there will be enough events to study.  The question is how we can then
extract the four masses of the new particles appearing in this decay
from the three observed external momenta.\bigskip

The proposals to solve this problem can be broadly classified into
four classes.  While all of them should in principle work and would
then differ mostly by statistics, we only know
how QCD and detector smearing affect the first strategy.
\begin{enumerate}
\item \underline{Endpoint methods} extract masses from lower
  (threshold) and upper (edge) kinematic endpoints of invariant mass
  distributions of visible decay products.  This method is best suited
  for long decay chains, where the number of independent endpoint
  measurements in one leg at least matches the number of unknown
  masses in the cascade\index{cascade decay}. An implicit assumption of these endpoint
  techniques is that the form of the matrix element populates the
  phase space close to the endpoint well. Otherwise, the endpoint will
  be soft and difficult to identify on top of the continuum background.

  The squark decay Eq.\eqref{eq:sim_squarkchain} has a particular
  kinematic feature: the invariant mass distributions of the two
  leptons $m_{\ell \ell}$. Looked at in the rest frame of the
  intermediate slepton it is a current--current interaction, similar to
  the Drell--Yan process computed in Eq.\eqref{eq:dy_sigma1}. Because
  in the $s$-channel there now appears a scalar particle there cannot
  be any angular correlations between the two currents, which means
  the $m_{\ell \ell}$ distribution will have a triangular shape. We can compute its
  upper limit, called the \underline{dilepton edge}\index{kinematic endpoint}:
  in the rest frame of the scalar lepton the three-momenta of the
  incoming and outgoing pair of particles have the absolute values
  $|\vec{p}| = |m_{\tilde{\chi}^0_{1,2}}^2 - m_{\tilde{\ell}}^2|/(2
  m_{\tilde{\ell}})$. The lepton mass we set to zero. The invariant
  mass of the two lepton reaches its maximum if the two leptons are
  back--to--back and the scattering angle is $\cos \theta = -1$
\begin{alignat}{5}
m_{\ell \ell}^2 
&= (p_{\ell^+} + p_{\ell^-})^2 \notag \\
&= 2 \, \left( E_{\ell^+} E_{\ell^-}
             - |\vec{p}_{\ell^+}| |\vec{p}_{\ell^-}| \cos \theta \right) \notag \\
&< 2 \, \left( E_{\ell^+} E_{\ell^-}
             + |\vec{p}_{\ell^+}| |\vec{p}_{\ell^-}| \right) \notag \\
&= 4 \, \frac{m_{\tilde{\chi}^0_2}^2 - m_{\tilde{\ell}}^2}{2 m_{\tilde{\ell}}}
     \; \frac{m_{\tilde{\ell}}^2 - m_{\tilde{\chi}^0_1}^2}{2 m_{\tilde{\ell}}} 
\qqqquad \text{using} \quad E_{\ell^\pm}^2 = \vec{p}_{\ell^\pm}^2 \; .
\end{alignat}
  This kinematic statement is independent of the shape of the $m_{\ell
    \ell}$ distribution. For the particle assignments shown in
  Eq.\eqref{eq:sim_squarkchain} the kinematic endpoints are given by
\begin{alignat}{5}
0 < m^2_{\ell\ell} 
  < \frac{(m_{\tilde{\chi}^0_2}^2-m_{\tilde{\ell}}^2)
          (m_{\tilde{\ell}}^2-m_{\tilde{\chi}^0_1}^2)}{m_{\tilde{\ell}}^2}
\qqqquad 
0 < m^2_{\ell\ell} 
  < \frac{(m_{Z^{(1)}}^2-m_{\ell^{(1)}}^2)
          (m_{\ell^{(1)}}^2-m_{\gamma^{(1)}}^2)}{m_{\ell^{(1)}}^2} \; .
\label{eq:sig_mll}
\end{alignat}
  A problem in realistic applications of endpoint methods is 
  combinatorics. We need to either clearly separate the decays of
  two heavy new states, or we need to combine a short decay chain on
  one side with a long chain on the other side. In supersymmetry this
  is naturally the case for associated squark and gluino production. A
  right handed squark often decays directly to the lightest neutralino
  which is the dark matter candidate in the model. The gluino has to
  radiate two quark jets to reach the weakly interacting sector of the
  model and can then further decay in many different ways. In other
  models this feature is less generic. The impressive potential of
  endpoint methods in the case of supersymmetry we will illustrate
  later in this section.

  When looking at long cascade decays for example with two leptons we
  usually cannot tell which of the two leptons is radiated first.
  Therefore, endpoint techniques will
  always be plagued with \underline{combinatorial background} from the mapping of the
  particle momenta on the decay topology. The same applies to QCD jet
  radiation vs decay jets. In this situation it is useful to consider
  the correlation of different invariant masses and their endpoints.
  The endpoint method can be extended to use invariant mass
  distributions from both sides of the event (hidden threshold
  techniques), and correlations between the distributions from each
  leg (wedgebox techniques).

\item \underline{Mass relation methods} generalize the single top
  example in Section~\ref{sec:sim_met_sm} and completely reconstruct
  the kinematics event by event. For each event this construction
  provides a set of kinematic constraints.  While for one event the
  number of parameters can be larger than the number of measurements,
  adding signal events increases the number of measurements while
  keeping the number of unknowns constant. Eventually, the system of
  equations will solve, provided all events are really signal
  events. Implicitly, we always assume that all decaying particles are
  on--shell.

\begin{figure}[t]
\begin{center}
  \includegraphics[width=0.30\textwidth]{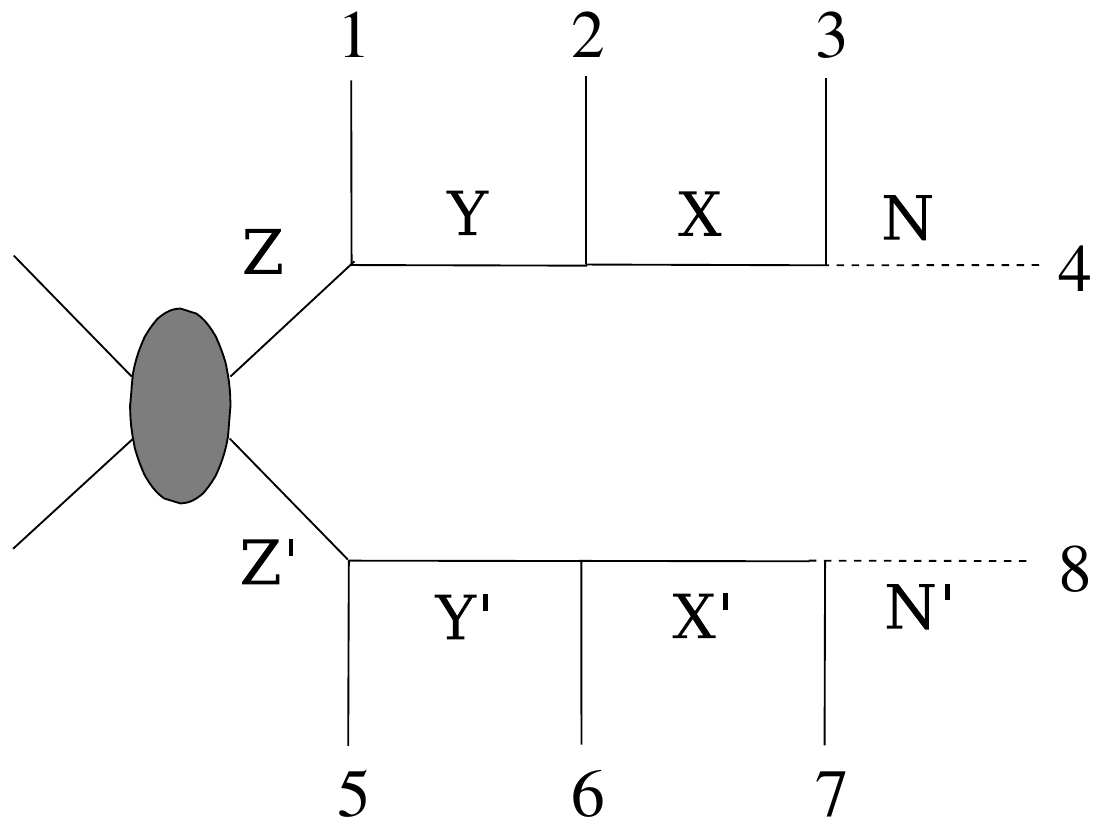}
  \hspace*{0.1\textwidth}
  \includegraphics[width=0.45\textwidth]{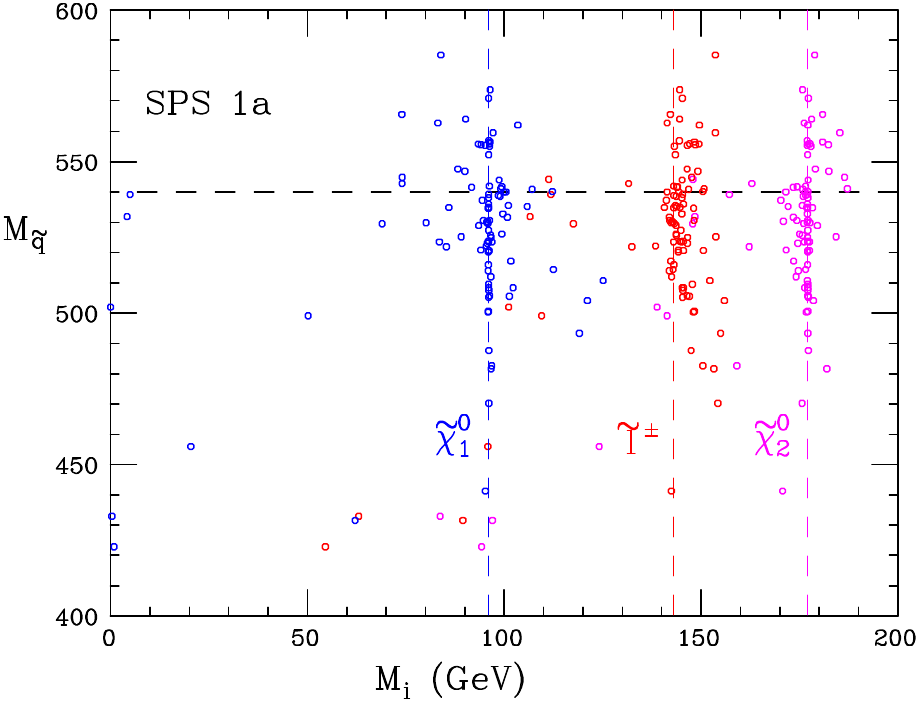}
\end{center}
\caption{Example for the mass relation method using three successive
  two-body decays on both sides of the events (left). After detector
  smearing we can reconstruct the masses for the supersymmetric
  parameter point SPS1a with the squark decay chain shown in
  Eq.\eqref{eq:sim_squarkchain} (right). Figure from
  Ref.~\cite{bryan}.}
\label{fig:sim_massrel}
\end{figure}

  In Figure~\ref{fig:sim_massrel} we show the general topology of a
  three-step cascade decay on each side of the event, like we expect
  it for a pair of left handed squarks following
  Eq.\eqref{eq:sim_squarkchain}. To extract the masses of the new
  particles we need to solve the system of equations
\begin{alignat}{5}
(p_1+p_2+p_3+p_4)^2 &&= m_Z^2 \notag \\
(p_2+p_3+p_4)^2 &&= m_Y^2 \notag \\
(p_3+p_4)^2 &&= m_X^2 \notag \\
p_4^2 &&= m_N^2 \; ,
\label{eq:massrel1}
\end{alignat}
  for each side of the event. For each event there are eight unknown
  masses and six unknown missing momentum components of which we
  measure two combinations as the missing transverse momentum. All of
  these 12 unknowns we can determine if we add a sufficiently large
  number of events.

  One strategy to solve this problem is to assume eight test masses $m
  = (m_Z^2, m_Y^2, m_X^2, m_N^2,...)$, use the three first equations
  in Eq.\eqref{eq:massrel1} for each event plus the two missing
  transverse momentum components to determine both missing
  four-momenta, and test the consistency of this solution using the
  last line of Eq.\eqref{eq:massrel1} for each of the two legs. In
  this consistency test we combine the information from several
  events.

  We can conveniently
  solve the first three lines in Eq.\eqref{eq:massrel1} for the missing momentum $p_4$
\begin{alignat}{5}
-2 (p_1 p_4) &\equiv s_1 = m_Y^2-m_Z^2+2(p_1 p_2) +2(p_1 p_3)  \notag \\
-2 (p_2 p_4) &\equiv s_2 = m_X^2-m_Y^2+2(p_2 p_3)  \notag \\
-2 (p_3 p_4) &\equiv s_3 = m_N^2-m_X^2  \; ,
\label{eq:massrel2}
\end{alignat}
  for simplicity assuming massless Standard Model decay products.
  Similarly, we define the measured combinations $s_{5,6,7}$ from the
  opposite chain.  In addition, we measure the two-dimensional
  missing transverse
  momentum, so 
  we can collect the two missing four-momenta into $p_\text{miss} =
  (\vec{p}_4, E_4, \vec{p}_8, E_8)$ and define two additional entries
  of the vector $s$ in terms of measured quantities and masses like
\begin{alignat}{5}
(\hat{x} p_4) + (\hat{x} p_8) &= s_4 \notag \\
(\hat{y} p_4) + (\hat{y} p_8) &= s_8 \; .
\label{eq:massrel3}
\end{alignat}
  Combining the first equal signs of 
  Eqs.\eqref{eq:massrel2} and~\eqref{eq:massrel3} for both halves of the events reads 
$A \cdot p_\text{miss} = s$, where 
  the matrix $A$ includes only components of measured momenta and is
  almost block diagonal, so it can be inverted.  Following the second
  equal sign in Eq.\eqref{eq:massrel2} we can then write $s = B \cdot
  m + c$, where the matrix $B$ only contains non-zero entries $\pm 1$
  and the vector $c$ consists of measured quantities. Together, this
  gives us
\begin{alignat}{5}
  p_\text{miss} = A^{-1} s = A^{-1} B \, m + A^{-1} \, c \; .
\end{alignat}
  We show the result for all masses in the decay chain using 25 events per set
  and including all combinatorics in
  Figure~\ref{fig:sim_massrel}. The mass relation method has also
  been successfully applied to single legs as well as in combination
  with kinematic endpoints.

\begin{figure}[t]
\begin{center}
  \includegraphics[width=0.25\textwidth]{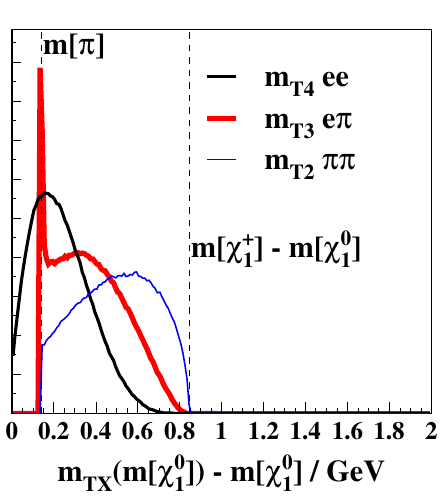}
\end{center}
\caption{Simulations for different $m_{TX}$, for the decay
  $\tilde{\chi}^+_1 \to \tilde{\chi}^0_1 \pi$ or $\tilde{\chi}^+_1 \to
  \tilde{\chi}^0_1 e^+ \nu$. The blue $m_{T2}$ line applies to the
  two-body decay. Figure from Ref.~\cite{Barr:2003rg}.}
\label{fig:sim_mt2}
\end{figure}

\item \underline{MT2 methods}\index{mass!MT2 construction} are based on a global variable
  $m_{T2}$. It generalizes the transverse mass known from $W$ decays
  to the case of two massive invisible particles, one from each leg of
  the event. The observed missing energy in the event we can divide
  into two scalar fractions $p_{T,\text{miss}} = q_1 + q_2$. Given the
  two fractions $q_j$ we can construct a transverse mass for each side
  of the event, assuming we know the invisible particle's mass
  $m_{T,j}(q_j;\hat{m}_\text{miss})$; the second argument is an
  external assumption, so $\hat{m}_\text{miss}$ is an assumed value
  for $m_\text{miss}$. 

  Inspired by the usual transverse mass we are interested in a mass
  variable with a well--defined upper edge, so we need to construct some kind
  of minimum of $m_{T,j}$ as a function of the splitting of ${p}_{T,
    \text{miss}}$. Naively, this minimum will simply be the zero
  transverse momentum limit of $m_T$ on one leg, which is not very
  interesting. On the other hand, in this case the transverse mass
  from the other leg reaches a maximum, so we can instead define
\begin{alignat}{5}
m_{T2}(\hat{m}_\text{miss})
     = \min_{p_{T,\text{miss}} = q_1 + q_2}
       \left[ \max_j \; m_{T,j}(q_j;\hat{m}_\text{miss}) \right]
\end{alignat}
  as a function of the unknown missing particle mass. There are two
  properties we know by construction
\begin{alignat}{5}
m_\text{light} + \hat{m}_\text{miss} &< m_{T2}(\hat{m}_\text{miss})  
\notag \\
m_\text{light} + m_\text{miss} &< m_{T2}(m_\text{miss}) 
                                  < m_\text{heavy} \; .
\end{alignat}
  The first line means that each of the $m_{T,j}$ lie between the sum
  of the two decay products' masses and the mass of the decaying
  particle, so for massless Standard Model decay products there will
  be a global $m_{T2}$ threshold at the missing particle's
  mass. 

  Moreover, for the correct value of $m_\text{miss}$ the $m_{T2}$
  distribution has a sharp edge at the mass of the parent particle. In
  favorable cases $m_{T2}$ may allow the measurement of both the
  parent particle and the LSP based on a single--step decay
  chain. These two aspects for the correct value $\hat{m}_\text{miss}
  = m_\text{miss}$ we can see in Figure~\ref{fig:sim_mt2}: the lower
  threshold is indeed given by $m_{T2} - m_{\tilde{\chi}^0_1} =
  m_\pi$.  while the upper edge of $m_{T2} - m_{\tilde{\chi}^0_1}$
  coincides with the dashed line for $m_{\tilde{\chi}^+_1} -
  m_{\tilde{\chi}^0_1}$.

  An interesting aspect of $m_{T2}$ is that it is boost invariant if
  and only if $\hat{m}_\text{miss} = m_\text{miss}$. For a wrong
  assignment of $m_\text{miss}$ it has nothing to do with the actual
  kinematics and hence with any kind of invariant, and house numbers
  are not boost invariant. We can exploit this aspect by scanning over
  $m_\text{miss}$ and looking for so-called kinks, defined as points where
  different events kinematics all return the same value for $m_{T2}$.

  Similar to the more global $m_\text{eff}$ variable we can generalize
  $m_{T2}$ to the case where we do not have a clear assignment of the
  two decay chains involved. This modification $M_{T \text{Gen}}$ again has an
  upper edge, which unfortunately is not as sharp as the one in
  $m_{T2}$.
  Similarly, the procedure can be generalized to any one-step decay,
  for example a three-body decay with either one or two missing
  particles on each side of the event. Such $M_{TX}$ distributions are
  useful as long as they have a sharp enough edge, as illustrated in
  Figure~\ref{fig:sim_mt2}.

\item \underline{Extreme kinematics} can also give us a handle on 
  mass reconstruction from an incomplete set of observables.
  One such phase space region are points close
  kinematic endpoints where particles are produced at rest. Other
  examples are the approximate collinear Higgs mass reconstruction in a
  decay to boosted tau pairs described in
  Section~\ref{sec:higgs_approx_mass} or the boosted leptonic 
  top decays mentioned before.
\end{enumerate}
\bigskip

\begin{figure}[t]
\begin{center}
  \includegraphics[width=0.40\textwidth]{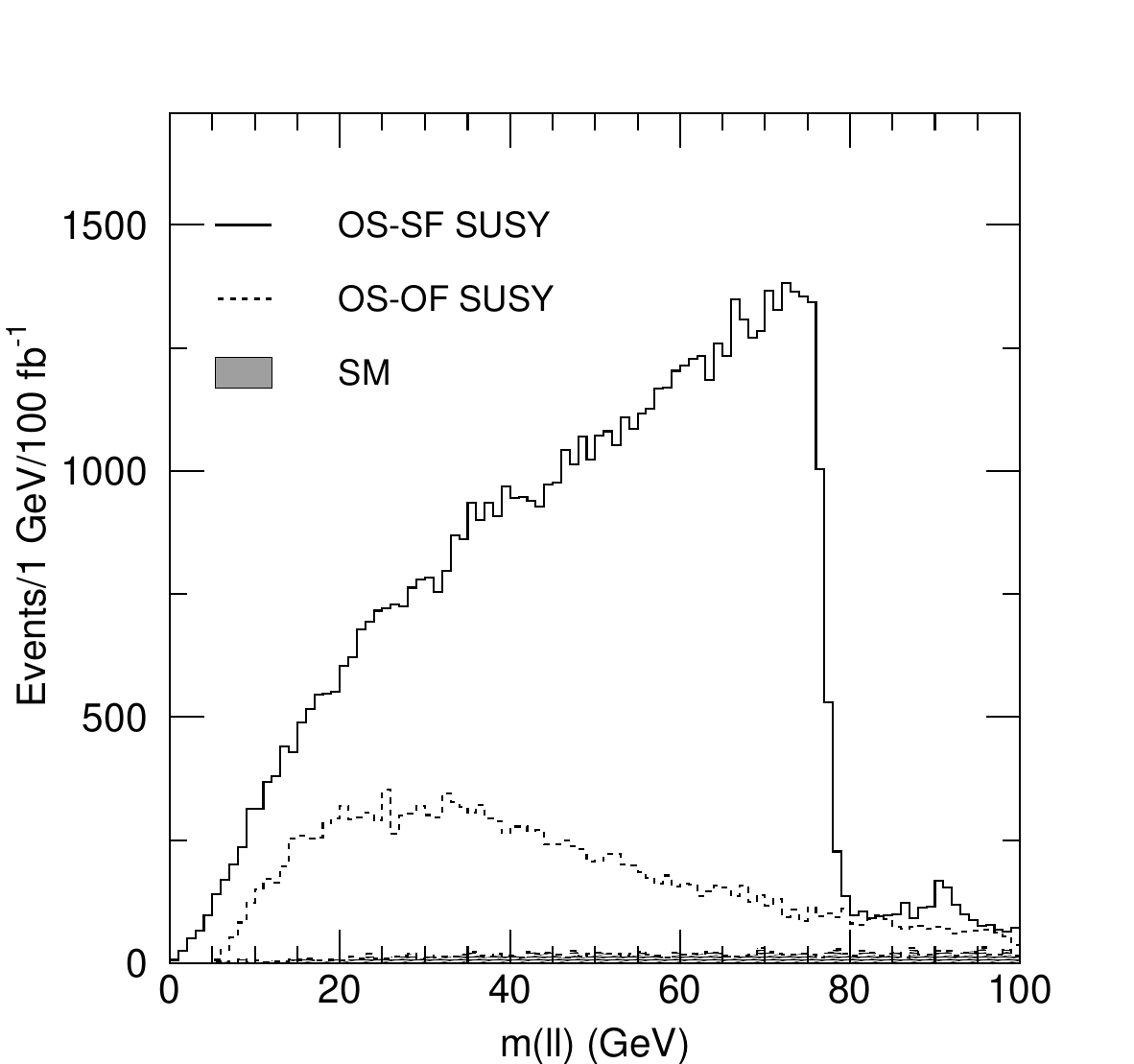}
\end{center}
\caption{Invariant mass of two leptons after selection cuts for the
  SPS1a parameter point: SUSY signal Opposite Sign Same Flavor
  (OS-SF): full line; SUSY signal Opposite Sign Opposite Flavor
  (OS-OF): dotted line; Standard Model background: grey. Figure from
  Giacomo Polesello (ATLAS).}
\label{fig:sim_sig_mll}
\end{figure}

The way mass measurements can lead to proper \underline{model
  reconstruction} we sketch for one scenario. The classic example for
the endpoint method is the long supersymmetric left handed squark
decay chain shown in Eq.\eqref{eq:sim_squarkchain} and in
Figure~\ref{fig:feyn_sq}. The quoted supersymmetric partner masses
are by now ruled out, but in the absence of more recent studies we stick to
their historic values. When we use such kinematic endpoints or
other methods to extract mass parameters it is crucial to start from a
signal--rich sample to avoid combinatorics and washed out endpoints
vanishing in a fluctuating or even sculptured background.  For jets,
leptons and missing energy a major background will be top pairs. The
key observation is that in long cascade decays\index{cascade decay} the leptons are
flavor--locked, which means the combination $e^+e^- + \mu^+\mu^- -
e^-\mu^+- e^+\mu^-$ becomes roughly twice $\mu^+\mu^-$ for the signal,
while it cancels for top pairs. Using such a combination for the
endpoint analysis means the top background is subtracted purely from
data, as illustrated in Figure~\ref{fig:sim_sig_mll}.

The long squark decay in by now ruled out SPS1a-like parameter points with squark
masses in the 500 to 600~GeV range has an important advantage: for
a large \underline{mass hierarchy} we should be able to isolate
the one decay jet just based on its energy.  In complete analogy to
the dilepton edge shown in Eq.\eqref{eq:sig_mll}, but with somewhat
reduced elegance we can measure the threshold and edge of the
$\ell^+\ell^- q$ distribution and the edges of the two $\ell^\pm q$
combinations. Then, we solve the system for the intermediate masses
without any model assumption, which allows us to even measure the dark
matter mass to $\ope(10\%)$. The limiting factors will likely be our
chances to observe enough endpoints in addition to $m_{\ell
  \ell}^\text{max}$ and the jet energy scale uncertainty. An
interesting question is how well we will do with tau leptons, where
the edge is softened by neutrinos from tau decays.\bigskip

\begin{figure}[t]
\begin{center}
  \includegraphics[width=0.9\textwidth]{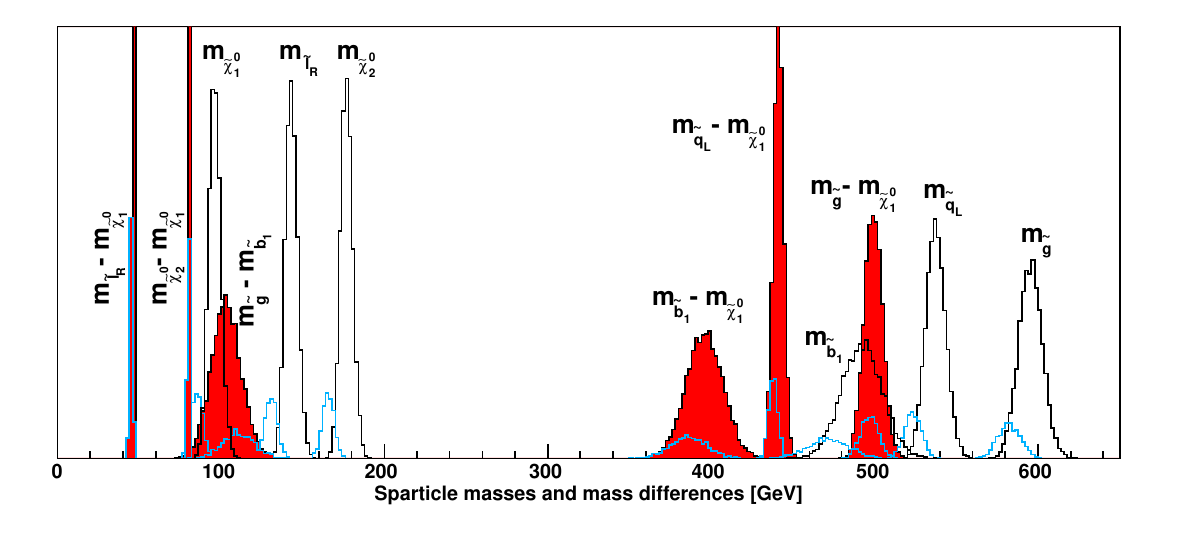}
\end{center}
\caption{Masses extracted from the gluino-sbottom decay chain,
  including estimated errors. The faint blue lines indicate wrong
  solutions when inverting the endpoint--mass relations. The
  supersymmetric mass spectrum is given by the SPS1a parameter
  point. Figure from Ref.~\cite{Gjelsten:2005aw}.}
\label{fig:sim_sig_gluinochain}
\end{figure}

Provided the gluino or heavy gluon is heavier than the squarks or
heavy quarks we can measure its mass by extending the squark chain by
one step: $\tilde{g}\to q\tilde{q}$.  This measurement is hard if one
of the two jets from the gluino decay is not very hard, because its
information will be buried by the combinatorial error due to QCD jet
radiation. The way around is to ask for two bottom jets from the
strongly interacting decay: $\tilde{g} \to b \tilde{b}^*$ or $G^{(1)}
\to b \bar{B}^{(1)}$. The summary of all measurements in
Figure~\ref{fig:sim_sig_gluinochain} shows that we can extract for example the gluino
mass at the per-cent level, a point at which we might
have to start thinking about off--shell propagators and at some point
even define what exactly we mean by `masses as appearing in cascade
decays'.

A generic feature or all methods relying on decay kinematics is that
it is easier to constrain the differences of squared masses than the
absolute mass scale. This is also visible in
Figure~\ref{fig:sim_sig_gluinochain}. It is due to the form of the
endpoint formulas which involve the difference of mass squares $m_1^2
- m_2^2 = (m_1+m_2)(m_1-m_2)$. This combination is much more sensitive
to $(m_1-m_2)$ than it is to $(m_1+m_2)$. Experimentally, correlated
jet and lepton energy scale uncertainties do not make life easier
either. Nevertheless, the common lore that kinematics only constrain
mass differences is obviously not true for two body decays.\bigskip

Alternatively, we can use the same gluino decay to first reconstruct
the intermediate neutralino or Kaluza--Klein $Z$ momentum for lepton pairs
residing near the $m_{\ell \ell}$ edge. In that case the invisible
heavy state is produced approximately at rest, and the momenta are
correlated as
\begin{alignat}{5}
  \vec{p}_{\tilde{\chi}^0_2} = \left( 1 - \frac{m_{\tilde{\chi}_1^0}}{m_{\ell\ell}}
                      \right) \; \vec{p}_{\ell\ell}
\qqqquad
  \vec{p}_{Z^{(1)}} = \left( 1 - \frac{m_{\gamma^{(1)}}}{m_{\ell\ell}}
                      \right) \; \vec{p}_{\ell\ell}
\end{alignat}
If both neutralino masses (or the Kaluza--Klein photon and $Z$ masses)
are known, we can extract the sbottom (Kaluza--Klein bottom) and gluino
(Kaluza--Klein gluon) masses by adding the measured bottom momenta to
this neutralino (Kaluza--Klein photon) momentum.  Again, for the mass
spectrum shown in Figure~\ref{fig:sim_sig_gluinochain} we can measure
the gluino mass to few per-cent, depending on the systematic
errors. 

For a complete analysis, kinematic endpoints can be supplemented by
any other method to measure new physics masses. For short decay chains
$m_{T2}$ is best suited to measure the masses of particles decaying
directly to the dark matter agent. In supersymmetry, this happens for
right handed sleptons or right handed squarks. The issue with short
decay chains is that they often require on some kind of jet veto,
which following Sections~\ref{sec:qcd_ckkw} and~\ref{sec:higgs_cjv} is
problematic for low-$p_T$ jets.\bigskip

\begin{figure}[t]
\begin{center}
  \includegraphics[width=0.35\textwidth]{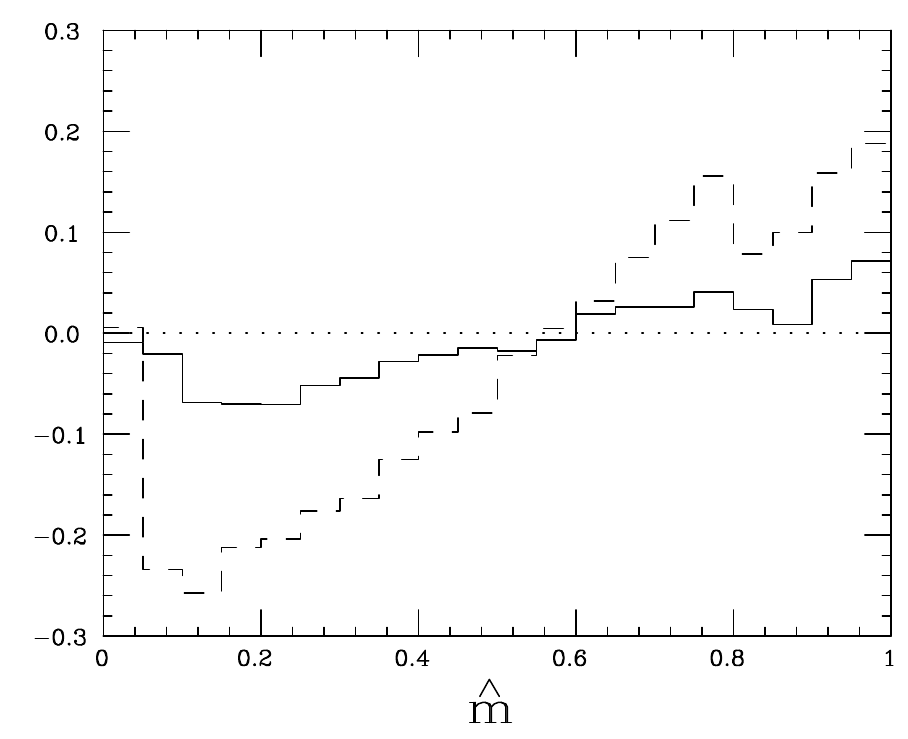}
\end{center}
\caption{Asymmetry in $m_{j \ell}/m_{j \ell}^\text{max}$ for
  supersymmetry (dashed) and universal extra dimensions (solid). The
  spectrum is assumed to be hierarchical, typical for supersymmetric
  theories. Figure taken from Ref.~\cite{Smillie:2005ar}.}
\label{fig:sim_sig_bryan}
\end{figure}

Keeping in mind that endpoint analyses only use a small fraction of
the events, namely those with extreme kinematics, an obvious way to
improve their precision is to include the complete shape of the
invariant mass distributions. However, this strategy bears a serious
danger. Invariant masses are just an invariant way of writing angular
correlations between outgoing particles. Those depend on the
\underline{spin and quantum numbers}\index{cascade decay} of all
particles involved. For example, in the case of the $m_{\ell \ell}$
endpoint the triangular shape implies the absence of angular
correlations, because the intermediate particle is a scalar.  This
means that we should be careful when extracting information for
example from kinematic endpoints we do not observe. Depending on the
quantum numbers and mixing angles in the new physics scenario,
kinematic endpoints can for example be softened, so they vanish in the
background noise.\bigskip

This argument we can turn around. Measuring discrete quantum numbers,
like the spin of new particles, is hard in the absence of fully
reconstructed events. The usual threshold behavior is not observable at
hadron colliders, in particular when the final state includes missing
transverse energy. Instead, we rely on angular correlation in
decays. For the squark decay chain given in
Eq.\eqref{eq:sim_squarkchain} there exists a promising method to
simultaneously determine the spin of all new particles in the chain:
\begin{enumerate}
\item Instead of trying to measure spins in a general parameterization
  we start from the observation that cascade decays radiate particles
  with known spins. This is most obvious for long gluino decays where
  we know that the radiated bottom quarks as well as muons are
  fermions. The spins inside the decay chain can only alternate
  between fermions and bosons. Supersymmetry\index{supersymmetry} switches this
  fermion/boson nature compared to the corresponding Standard Model
  particle, so we can contrast it with another hypothesis where the
  spins in the decay chain follow the Standard Model assignments.
  Such a model are Universal Extra Dimensions, where
  each Standard Model particle acquires a Kaluza--Klein partner from
  the propagation in the bulk of the additional dimensions.

\item Thresholds and edges of all invariant masses of the radiated
  fermions are completely determined by the masses inside the decays
  chain. Kinematic endpoints cannot distinguish between supersymmetry
  and universal extra dimensions. In contrast, the shape of the
  distribution between the endpoints is nothing but an angular
  correlation in some reference frame. For example, the $m_{j \ell}$
  distribution in principle allows us to analyze spin correlations in
  squark decays in a Lorentz invariant manner. The only problem is the
  link between $\ell^\pm$ and their ordering in decay chain.

\item a proton--proton collider like the LHC produces considerably
  more squarks than antisquarks in the squark--gluino associated
  channel. For the SPS1a spectrum at 14~TeV collider energy their ratio ranges around 2:1. A
  decaying squark radiates a quark while an antisquark radiates an
  antiquark, which means that we can define a non-zero production-side
  asymmetry between $m_{j \ell^+}$ and $m_{j \ell^-}$. Such an
  asymmetry we show in Figure~\ref{fig:sim_sig_bryan}, for the SUSY
  and for the UED hypotheses. Provided the masses in the decay chain
  are not too degenerate we can indeed distinguish the two hypotheses.
\end{enumerate}
This basic idea has since been applied to many similar
situations, like decays including gauge bosons, three-body decays,
gluino decays with decay--side asymmetries, cascades including
charginos, weak boson fusion signatures, etc. They show that the LHC
can do much more than just discover some kind of particle beyond the
Standard Model. It actually allows us to study underlying models 
and symmetries.

\subsection{Uncertainties}
\label{sec:sim_errors}

As we argue in the very beginning of the lecture, LHC physics always
means extracting signals from often large backgrounds. This means, a
correct error estimate is crucial. For LHC calculations we are usually
confronted with three types of errors.
\begin{enumerate}
\item The first and easiest one are the \underline{statistical
  errors}\index{error!statistical error}. For small numbers of events
  these experimental errors are described by Poisson statistics, 
\begin{alignat}{5}
f(N;N_\text{theo}) = \frac{e^{-N_\text{theo}} N_\text{theo}^{N}}{\Gamma(N+1)} \; ,
\end{alignat}
  where $N_\text{theo}$ is the expected number of events and $N$ the
  observed numbers. The mean value as well as the variance of this
  distribution is $N_\text{theo}$. For large event numbers $N'$ they
  converge to the Gaussian limit, Eq.\eqref{eq:higgs_gaussian}. This
  is an example of the central limit theorem which says that a
  sufficiently large number of independent random variables will
  eventually follow a Gaussian shape.  In that limit the number of
  standard deviations in terms of the number of signal and background
  events is $S/\sqrt{B}$. The two event numbers are proportional to the
  integrated luminosity $\lumi$\index{luminosity} which means that the
  statistical significance in the Gaussian limit increases with
  $\sqrt{\lumi}$.  In high energy physics five standard deviations
  above a known background we call a discovery, three sigma is often
  referred to as an evidence. The Poisson region is the only
  complication we encounter for statistical errors. It means that for
  small number of signal and background events we need more luminosity
  than the Gaussian limit suggests.\bigskip
\item The second set of errors are \underline{systematic
  errors}\index{error!systematic error}, like the calibration of the
  jet and lepton energy scales, the measurements of the luminosity, or
  the efficiencies to for instance identify a muon as a muon. Some
  readers might remember a bunch of theorists mistaking a forward pion
  for an electron --- that happened right around my TASI in 1997 and
  people not only discovered supersymmetry but also identified its
  breaking mechanism. Of course, our experimentalist CDF lecturer told
  us that the whole thing was a problem of identifying a particle in
  the detector with an efficiency which does not have to be zero or
  one.

  Naively, we would not assume that systematic errors follow a
  Gaussian distribution, but experimentally we determine efficiencies
  and scaling factors largely from well understood background
  processes. Such counting experiments in background channels like $Z
  \to$~leptons and their extracted parameters also follow a Gaussian
  distribution. The only caveat is the shape of far-away tails, which
  often turn out to be higher than the exponentially suppressed
  Gaussian tails.

  Systematic errors which do not follow a Gaussian distribution can scale 
  like $S/B$, which means they do not improve with
  increasing luminosity\index{luminosity}. Again, five standard
  deviations are required to claim a discovery, and once we are
  systematics dominated waiting for more data does not help.\bigskip
\item The third source of errors are \underline{theoretical
  errors}\index{error!theoretical error}. They are hardest to model
  because they are often dominated by higher--order QCD effects in
  fixed order perturbation theory. From Section~\ref{sec:qcd} we know
  that higher order corrections for example to LHC production rates do
  not follow a naive power counting in $\alpha_s$ but are enhanced by
  large logarithms. If we can compute or reliably estimate some higher
  order terms in the perturbative QCD series we call this a
  prediction. In other words, once we consider a statement about
  perturbative QCD a statement about its uncertainty we are probably
  only giving a wild guess.

  To model theoretical uncertainties it is crucial to realize that
  higher order effects are not any more likely to give a $K$ factor of
  1.0 than 0.9 or 1.1. In other words, likelihood distributions
  accounting for theoretical errors do not have a peak and are
  definitely not Gaussian. Strictly speaking, all we know 
  is a range of theoretically acceptable values. There is a good reason to choose the
  Gaussian short cut, which is that folding three Gaussian shapes for
  statistical, systematic and theoretical errors gives us a Gaussian
  distribution altogether, which makes things numerically much
  easier. But this approach assumes that we know much more about QCD
  than we actually do which means it is not conservative at all.

  On the other hand, we also know that theoretical errors cannot be
  arbitrarily large. Unless there is a very good reason, a $K$ factor
  for a total LHC cross section should not be larger than something
  like two or three. If that happens we need to conclude that
  perturbative QCD breaks down, and the proper description of error
  bars is our smallest problem. In other words, the centrally flat
  theory probability distribution for an LHC observable has to go to
  zero for very large deviations from the currently best
  value. Strictly speaking, even this minimalist distribution is not
  well defined, because there is not frequentist interpretation of 
  a range of theory uncertainties which could be used to define or
  test such a distribution.
\end{enumerate}
\bigskip

For example in the case if Higgs coupling measurements these different
sources of errors pose two problems: first, we need to construct a
probability measure combining all these three sources. Second, this
exclusive probability or likelihood has to be reduced in
dimensionality such that we can show an error bar on one of the Higgs
couplings in a well defined manner. Both of these problems lead us to
the same objects, likelihoods vs probabilities:

A \underline{likelihood}\index{likelihood} is defined as a probability
to obtain an experimental outcome $\vec{x}_\text{meas}$ given model
predictions $\vec{x}_\text{mod}$, varied over the model space. We can
link it to the corresponding probability as
\begin{alignat}{5}
P(\vec{x}_\text{meas}|\vec{x}_\text{mod}) = 
\lumi(\vec{x}_\text{mod}|\vec{x}_\text{meas}) \; ,
\label{eq:sim_deflikelihood}
\end{alignat}
where in both expressions are meant to be evaluated over model
parameter space. In Section~\ref{sec:higgs_couplings} we introduce the
logarithm of a likelihood Eq.\eqref{eq:higgs_chi2simple} as a
generalization of the usual $\chi^2$ distribution
\begin{alignat}{5}
\chi^2 (\vec{m}) &= 
\sum_{j=1}^{n_\text{meas}} 
\frac{\left| \vec{x}_\text{meas} - \vec{x}_\text{mod}(\vec{m}) \right|_j^2}{\sigma_j^2} \; ,
\label{eq:sim_chi2simple}
\end{alignat}
which is expressed in terms of the measurements $\vec{x}_\text{meas}$,
the model predictions $\vec{x}_\text{mod}$, and the variance
$\sigma^2$. This definition is really only useful in the Gaussian
limit where we know what the variance is. 

According to the definition Eq.\eqref{eq:sim_deflikelihood} we can
replace the Gaussian form of $\chi^2$ by any other estimated shape for
the statistical distribution of $\vec{x}_\text{meas}$. This includes
the Poisson, Gaussian, and box shapes discussed above, as well as any
combination of the three. Before we discuss in detail how to construct
a likelihood for example for a Higgs couplings measurement we should
link this likelihood to a mathematically properly defined
probability.\bigskip

\underline{Bayes' theorem} tells us how to convert the likelihood
Eq.\eqref{eq:sim_deflikelihood} into the \underline{probability} that
a choice of model parameters $\vec{x}_\text{mod}$ is true given the
experimental data, $\vec{x}_\text{meas}$. This is what we are
actually interested in when we measure for example Higgs couplings
\begin{alignat}{5}
P(\vec{x}_\text{mod}|\vec{x}_\text{meas})
 = P(\vec{x}_\text{meas}|\vec{x}_\text{mod}) \; 
  \frac{P(\vec{x}_\text{mod})}{P(\vec{x}_\text{meas})}
\equiv \lumi(\vec{x}_\text{mod}|\vec{x}_\text{meas}) \; 
  \frac{P(\vec{x}_\text{mod})}{P(\vec{x}_\text{meas})} \; .
\label{eq:sim_bayes}
\end{alignat}
In this relation $P(\vec{x}_\text{meas})$ is a normalization constant
which might be hard to evaluate but which ensures that the probability
$P(\vec{x}_\text{mod}|\vec{x}_\text{meas})$ summed over all possible
experimental outcomes is normalized to unity. The problem is the prior
$P(\vec{x}_\text{mod})$ which is a statement about the model or the
model parameter choice and which obviously cannot be determined from
experiment. If we bring it to the other side of
Eq.\eqref{eq:sim_bayes} it ensures that the conditional probability
$P(\vec{x}_\text{meas}|\vec{x}_\text{mod})$ integrated over model
space is unity. This implies some kind of measure in model space or
model parameter space. As an example, if we want to measure the mass
of a particle we can integrate $m$ over the entire allowed or
interesting range, but we can also integrate $\log m$ instead.
In an ideal world of perfect measurement the
difference between these two measures will not affect the final answer
for $P(\vec{x}_\text{mod}|\vec{x}_\text{meas})$. The problem is that
Higgs coupling measurements are far from ideal, so we have to decide
on a measure in Higgs couplings space to compute a probability for a
set of couplings to be true.

One aspect we can immediately learn from this Bayesian argument is how
to combine different uncertainties, \ie statistical, systematic, and
theoretical uncertainties for the same observable: we introduce one
so-called nuisance parameter for each of these errors, describing the
deviation of the measured value from the expected value for the given
observable. All three nuisance parameters combined correspond to an
actual observable, individually they are not interesting. This means
that we want to remove them as dimensions or degrees of freedom from
our big exclusive likelihood map. If the integral over model parameter
space, including nuisance parameters, is well defined we simply
integrate them out, leaving for example the normalization of our
probability intact. If we write out this integration it turns out to
be a \underline{convolution}.

As is well known, the convolution of several Gaussians is again a
Gaussian.  The convolution of a Gaussian experimental error with a
flat theory error returns a two-sided error distribution which has a
peak again. While we started with the assumption that theory errors
should not give a preferred value within the error band, the measure
in model space after convolution again returns such a maximum.\bigskip

The frequentist construction to reduce the number of model parameters
avoids any measure in model parameter space, but leads to
mathematical problem: to keep the mathematical properties of a
likelihood as a probability measure, including the normalization, we
would indeed prefer to integrate over unwanted directions. In the
frequentist approach such a measure is not justified. An alternative
solution which is defined to keep track of the best--fitting points in
model space is the \underline{profile likelihood}.  It projects the
best fitting point of the unwanted direction onto the new parameter
space; for each binned parameter point in the $(n-1)$-dimensional
space we explore the $n$th direction which is to be removed
$\lumi(x_{1,...,n-1},x_n)$. Along this direction we pick the best
value and identify it with the lower--dimensional parameter point
$\lumi(x_{1,...,n-1}) \equiv \lumi^{\text{max}(n)}
(x_{1,...,n-1},x_n)$.  Such a projection avoids defining a measure but
it does not maintain for example the normalization of the likelihood
distribution.

We first compute the profile likelihood for two one-dimensional
Gaussians affecting the same measurement $x$, removing the nuisance
parameter $y$, and ignoring the normalization. The form of the
combined likelihood is the same as for a convolution, except that the
integral over $y$ is replaced by the maximization,
\begin{alignat}{5}
\lumi(x) 
&\sim \max_y \; e^{-y^2/(2 \sigma_1^2)} \; e^{-(x-y)^2/(2 \sigma_2^2)}
\notag \\
&= \max_y \; \exp
\left[ - \frac{\sigma_1^2 + \sigma_2^2}{2\sigma_1^2 \sigma_2^2}
       \left( y^2 - 2xy \dfrac{\sigma_1^2}{\sigma_1^2+\sigma_2^2}
                  + x^2 \dfrac{\sigma_1^2}{\sigma_1^2 + \sigma_2^2}
      \right)
\right]
\notag \\
&= \max_y \; \exp
\left[ - \frac{\sigma^2}{2\sigma_1^2 \sigma_2^2}
       \left( \left(y - x \dfrac{\sigma_1^2}{\sigma^2} \right)^2 
              - x^2 \dfrac{\sigma_1^4}{\sigma^4} + x^2 \dfrac{\sigma_1^2}{\sigma^2}
      \right)
\right]
\qqquad \text{with} \quad \sigma^2 = \sigma_1^2 + \sigma_2^2 
\notag \\
&= \max_{y'} \; \exp
\left[ - \frac{\sigma^2}{2\sigma_1^2 \sigma_2^2}
       \left( y'^2 
              - x^2 \dfrac{\sigma_1^4}{\sigma^4} + x^2 \dfrac{\sigma_1^2}{\sigma^2}
      \right)
\right]
\qqqquad \qquad \text{with} \quad y' = y - x \dfrac{\sigma_1^2}{\sigma^2}
\notag \\
&= \max_{y'} \; \exp
\left[ - \frac{\sigma^2}{2\sigma_1^2 \sigma_2^2}
       \left( y'^2 
              + x^2 \dfrac{\sigma_1^2 \sigma_2^2}{\sigma^4} 
      \right)
\right]
\notag \\
&= e^{-x^2/(2\sigma^2)}
\max_{y'} \; e^{- y'^2 \sigma^2/(2\sigma_1^2 \sigma_2^2)}
\; = \; e^{-x^2/(2\sigma^2)} \; .
\label{eq:sim_profilegg}
\end{alignat}
We use that the profile likelihood over $y$ and $y'$ is the same
after the linear transformation. Just like in the case of the
convolution, the profile likelihood of two Gaussian is again a
Gaussian with $\sigma^2 = \sigma_1^2 + \sigma_2^2$.  Next, we use the
same reasoning to see what happens if we combine two sources of flat
errors with identical widths,
\begin{alignat}{5}
\lumi(x) 
&= \max_y \; 
  \Theta(x_\text{max}-y) \; \Theta(y-x_\text{min}) \; 
  \Theta(x_\text{max}-x+y) \; \Theta(x-y-x_\text{min}) 
\notag \\
&= \max_{y\in[x_\text{min},x_\text{max}]}
  \Theta((x_\text{max}+y)-x) \; \Theta(x-(x_\text{min}+y)) 
\notag \\
&= 
  \Theta(2x_\text{max}-x) \; \Theta(x- 2x_\text{min}) \; .
\label{eq:sim_twobox}
\end{alignat}
Each of the original boxes starts with a of $x_\text{max} -
x_\text{min}$.  The width of the box covering the allowed values for
$x$ after computing the profile likelihood is $2 (x_\text{max} -
x_\text{min})$, so unlike for the Gaussian case the two flat errors
get \underline{added linearly}, even though they are assumed to be uncorrelated.
We can follow  the same kind of calculation for the combination of a
Gaussian and a flat box--shaped distribution,
\begin{alignat}{5}
\lumi(x) 
&= \max_y \; 
  \Theta(x_\text{max}-y) \; \Theta(y-x_\text{min}) \; 
  e^{-(x-y)^2/(2\sigma^2)}
\notag \\
&= \max_{y\in[x_\text{min},x_\text{max}]}
  e^{-(x-y)^2/(2\sigma^2)}
\notag \\ 
&= 
\begin{cases}
e^{-(x-x_\text{min}^2/(2\sigma^2)}
\qquad & x < x_\text{min} \\ 
\qquad 1 
\qquad & x \in [x_\text{min},x_\text{max}] \\
e^{-(x-x_\text{max}^2/(2\sigma^2)}
\qquad & x > x_\text{max} \; . 
\end{cases}
\label{eq:sim_profilegb}
\end{alignat}
This profile likelihood construction is called \underline{Rfit
  scheme}\index{error!RFit scheme} and is used for example by
CKMfitter or SFitter. 
We obtain the combined distribution  by cutting
open the experimental Gaussian distribution and inserting a flat
theory piece. Exactly the same happens for the profile likelihood
combination of a Poisson distribution and a flat box. The last
combination we need to compute is a Gaussian with widths $\sigma$ with
a Poisson with expectation value $N$.  This projection is not trivial
to compute,
\begin{alignat}{5}
\lumi(x) 
&= \max_y \; 
  \frac{e^{-N} N^y}{y!} \; 
  e^{-(x-y)^2/(2\sigma^2)}
\notag \\
&= \max_y \; 
  \exp \left[ 
  -N + y \log N - \log y! - \frac{(x-y)^2}{2\sigma^2} \right]
\notag \\
&= \max_y \; 
  \exp \left[ 
  -N + y \log N 
  - \frac{1}{2} \log \frac{2\pi}{y+1} 
  - (y+1) \log \frac{y+1}{e}
  - \frac{(x-y)^2}{2\sigma^2} \right] \; .
\end{alignat}
However, we can approximate the result numerically as
\begin{alignat}{5}
\dfrac{1}{\log \lumi(x)} = 
\dfrac{1}{\log \lumi_\text{Poisson}} +
\dfrac{1}{\log \lumi_\text{Gauss}} =
\dfrac{1}{\log \dfrac{e^{-N} N^x}{x!}} + 
\dfrac{1}{- \dfrac{x^2}{2\sigma^2}} \; .
\end{alignat}
We can check this formula for the case of two Gaussians
\begin{alignat}{5}
\frac{1}{\log \lumi(x)} 
= 
- \dfrac{2\sigma_1^2}{x^2}
- \dfrac{2\sigma_2^2}{x^2}
=
- \dfrac{2 \sigma^2}{x^2} 
\qqquad \Leftrightarrow \qquad
\lumi = e^{-x^2/(2\sigma^2)} \; ,
\label{eq:sim_profile_pg}
\end{alignat}
with $\sigma^2 = \sigma_1^2 + \sigma_2^2$. This is precisely the
result of Eq.\eqref{eq:sim_profilegg}. Another sanity check is that if
one of the likelihoods becomes very large it decouples from the final
results and the combined likelihood is dominated by the bigger
deviation. We can test that Eq.\eqref{eq:sim_profile_pg} reproduces
the full result to a few per-cent.\bigskip

Numerically, we usually compute the logarithm of the likelihood
instead of the likelihood itself. The reason is that for many channels
we need to multiply all individual likelihoods, leading to a vast
numerical range of our likelihood map. It is numerically much more
stable to use the logarithm instead and add the
\underline{log-likelihoods} instead. In the Gaussian limit this is
related to the $\chi^2$ value via $\chi^2 = -2 \log \lumi$. If we
allow for a general correlation matrix $C$ between the entries in the
measurements vector $\vec{x}_\text{meas}$ and a symmetric theory error
$x \pm \sigma^\text{(theo)}$ we find the RFit expression
\begin{alignat}{7}
-2\log \lumi     &= {\vec{x}}^T \; C^{-1} \; \vec{x}  \notag \\ 
x_i &=
  \begin{cases}
  \dfrac{x_{\text{meas,i}} -x_{\text{mod},i}+ \sigma^{\text{(theo)}}_i}{\sigma^{\text{(exp)}}_i}
          & \; x_{\text{meas,i}}  < x_{\text{mod},i} - \sigma^{\text{(theo)}}_i \\
  \qquad 0  
          &|x_{\text{meas,i}}-x_{\text{mod},i} | <   \sigma^{\text{(theo)}}_i \\
  \dfrac{x_{\text{meas,i}} - x_{\text{mod},i}- \sigma^{\text{(theo)}}_i}{\sigma^{\text{(exp)}}_i}
  \qquad  & \; x_{\text{meas,i}} >  x_{\text{mod},i} + \sigma^{\text{(theo)}}_i \; .
  \end{cases}
\label{eq:sim_flat_errors}
\end{alignat}
This distribution implies that for very large deviations there will
always be tails from the experimental errors, so we can neglect the
impact of the theoretical errors on this range. In the center the
distribution is flat, reflecting our ignorance of the theory
prediction. The impact of the size of the flat box we need to
test. 

This concludes our construction of the multi--dimensional
correlated likelihood map with different types of errors, which we can
apply for example in the Higgs couplings analysis introduced in
Section~\ref{sec:higgs_couplings}. In principle, it is possible to
compute an exclusive likelihood map even more generally by keeping all
the nuisance parameters, avoiding any of the profile constructions
described below, and then removing the nuisance parameter alongside the
unwanted couplings at the end. However, this hugely increases the
number of dimensions we initially encounter, so it is numerically more
economical to first apply analytical profiling as done in SFitter.

\subsection{Further reading}

Again, there exist several good review articles with more in-depth discussions of different aspects touched in this Section:

\begin{itemize}
\item[--] as mentioned in Section~\ref{sec:qcd}, two very useful
  reviews of jet physics are available by Steve Ellis and
  collaborators~\cite{Ellis:2007ib} and by Gavin
  Salam~\cite{Salam:2009jx}. 

\item[--] if you are interested in top identification using fat jet
  techniques we wrote a short pedagogical review article illustrating
  the different techniques and tools available~\cite{Plehn:XXXX}.

\item[--] for the general phenomenology of the heaviest Standard Model
  particles, the top quark, have a look at Sally Dawson's
  TASI lectures~\cite{Dawson:2003uc}.

\item[--] if you use Madgraph/HELAS to compute helicity amplitudes
  there is the original documentation which describes every
  routine~\cite{Murayama:1992gi}.

\item[--] a lot of experimental knowledge on new physics searches well
  described and theoretically sound you can find in the CMS technical
  design report. Some key analyses are described in detail while most
  of the documentation focuses on the physics expectations~\cite{cms_tdr}.

\item[--] more on the magical variable $m_{T2}$ you can find in an
  article by Alan Barr, Chris Lester and Phil
  Stephens~\cite{Barr:2003rg}. Chris Lester's
  thesis~\cite{chris_thesis} is a good point to start with. Recently,
  Alan Barr and Chris Lester published a broad review on techniques to
  measure masses in models with dark matter
  particles~\cite{alan_chris}.

\item[--] as mentioned in the introduction, there is our more advanced
  review on new physics at the LHC which includes an extensive chapter
  on LHC signatures~\cite{bsm_review}.

\item[--] a lot of insight into new physics searches at the LHC and at a
  linear collider you can find in a huge review
  article collected by Georg Weiglein~\cite{Weiglein:2004hn}.

\item[--] the pivotal work on determining spins in cascade decays is
  Jennie Smillie's PhD thesis~\cite{Smillie:2005ar}.  On the same
  topic there exists a nicely written review by Liantao Wang and Itay
  Yavin~\cite{Wang:2008sw}.

\item[--] many useful pieces of information on mass extraction,
  errors, and the statistical treatment of new-physics parameter
  spaces you can find in the big SFitter
  publication~\cite{Lafaye:2007vs}. The SFitter analysis of the Higgs
  sector~\cite{Lafaye:2009vr} is very similar in structure, but
  different in the physics application.

\item[--] if you are interested in a recent discussion of experimental
  and theoretical errors and how to factorize them, you can try a
  recent paper we wrote with Kyle Cranmer, Sven Kreiss, and David
  Lopez--Val~\cite{Cranmer:2013hia}.
\end{itemize}

\newpage

\section{Not using Feynman rules} 
\label{sec:field_theory}

In these LHC lecture notes we always assume some field theory
background which allows us to compute transition amplitudes on the
basis of Feynman rules and phase space integrals. In other words, the
corresponding lectures are meant to be heard after Quantum Field
Theory~I and~II. Nevertheless, it can be useful to briefly repeat the
steps which we have to go through to compute a transition amplitude
from an action or from first principles. In this short section we will
sketch this calculation and indicate where in the calculation Feynman
ruled come in and significantly simplify our lives. In that sense this
section is not actually needed to understand the other parts of these
notes, but it might come in handy at times.\bigskip

When we compute transition amplitudes for collider like LEP or LHC, we
usually combine building blocks defined by Feynman rules in a way
which does not make it obvious that we are dealing with a quantum
field theory. For example, in Section~\ref{sec:qcd_dy} we compute the
transition amplitude for the process $e^+ e^- \to \gamma^* \to q
\bar{q}$ through a photon, all starting from these Feynman rules. In
this section, we will give a brief sketch of what we have to do to
describe the quantum fields involved and to compute this transition
amplitude without using Feynman rules.\bigskip

From theoretical mechanics we remember that there are several ways to
describe a physical system and calculate the time evolution of the
states. One object to start from is the action as a
function of one degree of freedom or field $\phi$
\begin{alignat}{5}
\boxed{ S = \int d^4x \, \lag(\phi,\p_\mu \phi) } \; .
\end{alignat}
With $x$ we denote the four-vector including the time component
$(x_0,\vx)$.  The action has to be invariant under a variation $\delta
S = 0$. We can translate this condition into the Euler-Lagrange
equations
\begin{alignat}{5}
\p_\mu \left( \frac{\p \lag}{\p (\p_\mu \phi)}
             \right)
= \frac{\p \lag}{\p \phi} \; ,
\label{eq:qft_euler}
\end{alignat}
Using the notation $\p_\mu = \p/\p x^\mu$.  A
convenient second parameter in addition to $x$ is the conjugate
momentum $\pi(x) = \p \lag/\p \dot{\phi}$. With this
coordinate we define the third object which we can use to describe
the dynamics of a system, the Hamiltonian density
\begin{alignat}{5}
\int d^3x \, \mathcal{H}(x) =
\int d^3x \, \left( \pi \dot{\phi} - \lag \right) \; .
\end{alignat}
While for example in quantum mechanics this Hamiltonian or energy
functional is the basis of most calculations, in field theory we
usually start from the Lagrangian.\bigskip

We already know that for our scattering process we need to compute a
transition amplitude between two kinds of matter particles, namely
incoming electrons and outgoing quarks, interacting via their electric
charges.  The interaction is classically described by the
electromagnetic Lagrangian based on the abelian $U(1)$ field theory,
\begin{alignat}{5}
\lag \supset -\frac{1}{4} F_{\mu \nu} F^{\mu \nu}
\qquad \qquad 
\text{with}
\qquad 
F_{\mu \nu} = \p_\mu A_\nu - \p_\nu A_\mu \; ,
\end{alignat}
in terms of a photon vector field $A_\mu$. The Maxwell equations 
\begin{alignat}{5}
0 
= \p^\mu F_{\mu \nu} 
= \p^\mu \p_\mu A_\nu 
 -\p^\mu \p_\nu A_\mu 
= \Box A_\nu
\label{eq:qft_dalambert}
\end{alignat}
are the equations of motion for this photon field. In the
last step we assume the Lorentz gauge condition $\p_\mu A^\mu =
0$ and find the d'Alembert equation for the vector potential
$A_\mu$.\bigskip

To omit the vector index of the photon field we instead use a
real scalar field $\phi$ to illustrate bosonic fields, their
quantization, their equation of motions, and how they enter a
calculation. Including a mass for this real scalar field we can write
down its equation of motion which is the same for a spin-zero scalar
boson as for the spin-one vector boson of Eq.\eqref{eq:qft_dalambert}
\begin{alignat}{5}
\left( \Box + m^2 \right) \, \phi(x) = 0 \; .
\label{eq:qft_kge}
\end{alignat}
This \underline{Klein--Gordon equation}\index{Klein--Gordon equation} corresponds
to the d'Alembert equation for the electromagnetic vector potential in
Lorentz gauge. This equation of motion corresponds to a contribution to
the Lagrangian of the form
\begin{alignat}{5}
\lag \supset \frac{1}{2} (\p_\mu \phi) (\p^\mu \phi)
            -\frac{m^2}{2} \phi^2 \; ,
\end{alignat}
which we can easily confirm using the Euler-Lagrange equation
Eq.\eqref{eq:qft_euler}.\bigskip

Under a Lorentz transformation of the d'Alembert operator and of the scalar field,
\begin{alignat}{5}
x^\mu &\to {x'}^\mu = {\Lambda^\mu}_\nu \, x^\nu 
\notag \\
\phi(x) &\to \phi'(x) = \phi(\Lambda^{-1}x) \; ,
\label{eq:qft_lorenz}
\end{alignat}
the Klein--Gordon equation keeps its form in the new
coordinates. It is a standard wave equation which we can solve using
plane waves, and which modulo prefactors gives us
\begin{alignat}{5}
\phi(x) = \int \, \frac{d^3k}{(2 \pi)^3 2 k_0} \; 
\left(
e^{i k \cdot x} a^*(\vk) + e^{-i k \cdot x} a(\vk) 
\right) \; .
\label{eq:qft_scalar1}
\end{alignat}
Complex conjugates $a$ and $a^*$ are required for a real field
$\phi$. The value of $k_0$ is given by the dispersion relation, which
means that if $\phi$ fulfills the Klein--Gordon equation it is $k_0^2 =
\vk^2 + m^2$. The Fourier transform $a$ therefore explicitly `only
depends on $\vk$. Up to this point the field $\phi$ is a real function
of the space coordinates, \ie it is not quantized.\bigskip

Because in the remaining lecture we will only use quantized operator
valued fields we will not change notation at this point.  Switching
from a field function to a field operator $\phi$ leaves
the Klein--Gordon equation Eq.\eqref{eq:qft_kge} the same, except that
now it constrains an operator $\phi(x)$ which for example we cannot
simply commute with other operators. $\phi$ used to be a real field,
so now it will be a hermitian operator $\phi^\dag = \phi$. In the
plane wave solution Eq.\eqref{eq:qft_scalar1} the coefficients $a$ and
(now) $a^\dag$ are promoted to operators as well.

The Hilbert space in which our system resides includes a vacuum state
$\ket{0}$ which has to be normalized, $\braket{0}{0} = 1$, and which has zero
energy and momentum. We can show that the on--shell state $\ket{k}
\equiv a^\dag(\vk) \ket{0}$ is an eigenvector of the
energy--momentum operator with eigenvalues $k^\mu$. We can
interpret $a^\dag$ as a creation operator for a state with
four-momentum $k^\mu$. Successive application of $a^\dag(\vk_j)$ on the
vacuum gives a basis of states with a varying number of states,
defining the Fock space. Acting with the energy--momentum operator on
$a(\vk) \ket{0}$ gives us a negative energy eigenvalue, which means
that we have to set $a(\vk) \ket{0} = 0$; the operator $a(\vk)$ is an
annihilation operator.

For operators it is crucial that we define their commutation
relations. The basic property which we use to fix the commutators of
$a$ and $a^\dag$ is \underline{causality}, namely that field
configurations outside each other's light cone cannot affect each
other and therefore their field operators commute
\begin{alignat}{5}
\left[ \phi(x), \phi(x') \right] &= 0
\qquad \qquad &&\text{for} \quad
(x-x')^2 < 0 
\notag \\
\text{or}\qquad 
\left[ \phi(\vx), \phi(\vx') \right] &= 0
&&\text{for} \quad
t=t', \; \left| \vx-\vx' \right| > 0 \; .
\label{eq:qft_causality}
\end{alignat}
We can insert the operator--valued form of Eq.\eqref{eq:qft_scalar1} 
into the equal-time commutators $(t=t', \vx \ne \vx')$ of $\phi$ and
$\dot{\phi}$, where the latter allows us to vary the relative factor
between $a$ and $a^\dag$ and derive several independent
relations. One of them reads
\begin{alignat}{5}
0 &= \left[ \phi(\vx), \dot{\phi}(\vx') \right] \\
&= \int \, \frac{d^3k}{(2 \pi)^3 2 k_0} \, \frac{d^3k'}{(2 \pi)^3 2 k'_0} \; 
   e^{-i (\vk \vx + \vk' \vx')} (i k'_0) 
   \left[ \left( e^{i k_0 t} a^\dag(\vk) + e^{-i k_0 t} a(\vk) \right),
          \left( e^{i k'_0 t} a^\dag(\vk') - e^{-i k'_0 t} a(-\vk') \right) \right] \; . \notag 
\end{alignat}
In this Fourier transform for free $k_0$ and $k_0'$ the integral
vanishes only if the integrand, and therefore all commutators under
the integral vanish, at least as long as $\vx \ne \vx'$. For
$\vx = \vx'$ the condition $[\phi(\vx,t),
  \dot{\phi}(\vx',t)] = i \delta^3(\vx-\vx')$ fixes the
third commutation relations for bosonic creation and annihilation
operators
\begin{alignat}{5}
\left[ a^\dag(\vk), a^\dag(\vk') \right] &= 0 
\notag \\
\left[ a(\vk), a(\vk') \right] &= 0 
\notag \\
\left[ a(\vk), a^\dag(\vk') \right] &= (2\pi)^3 2 k_0 \; \delta^3(\vk - \vk')  \; .
\label{eq:qft_commute_boson}
\end{alignat}
Using these commutators we can analyze specific configurations values
of the kind $\bra{0} a(\vk) a^\dag(\vk) \ket{0}$, which do not
vanish. It turns out that the integral defining the expectation value
for the energy $k^0$ in such a basis diverges. This problem we can
solve {\sl ad hoc} by postulating that sandwiched between vacuum
states we only evaluate combinations where all annihilation operators
$a$ are moved to the right and all creation operators $a^\dag$ to
the left
\begin{alignat}{5}
: a^\dag a : &= a^\dag a    \notag \\
: a^\dag a^\dag : &= a^\dag a^\dag \notag \\
: a a : &= a a     \notag \\
: a a^\dag : &= a^\dag a \; .
\label{eq:qft_normal_boson}
\end{alignat}
We can interpret this \underline{normal--ordering} as ordering the
operators in a sensible way and neglecting the corresponding commutators
sandwiched between vacuum states, for example
$\bra{0} [a, a^\dag] \ket{0} \equiv 0$. If we want a Hamilton operator to
only give positive but finite energy states we need to include a 
time--ordering into its definition. The Wick theorem links normal--ordering
and time--ordering in a way which makes them identical as long as we
only compute tree level leading order processes.\bigskip

Looking at the scattering process we want to evaluate we 
need to describe is four external fermions, their
coupling to a photon, and the propagation of this boson from the
$e^+e^-$ annihilation to the point where is splits into a quark and
antiquark pair. Let us start with this propagator. Such a propagator
is defined as a time--ordered product of two field operators sandwiched
between vacuum states. For scalar fields it reads
\begin{alignat}{5}
\Delta(x-x') \equiv 
i \; \bra{0} T \left( \phi(x) \phi(x') \right) \ket{0}  \; .
\end{alignat}
The time--ordered product of two operators is defined as
\begin{alignat}{5}
T \left( A(x) B(x') \right) = 
\theta(x_0-x'_0) A(x) B(x') +  \theta(x'_0-x_0) B(x') A(x) \; .
\end{alignat}
We again use the operator version of
Eq.\eqref{eq:qft_scalar1} to evaluate this combination for free fields
\begin{alignat}{5}
\bra{0} \phi(x) \phi(x') \ket{0} \Bigg|_{x_0>x'_0}
&=
\int \, \frac{d^3k}{(2 \pi)^3 2 k_0} \, \frac{d^3k'}{(2 \pi)^3 2 k'_0} \; 
\bra{0} \, 
\left(
e^{i k x} a^\dag(\vk) + e^{-i k x} a(\vk) 
\right) \; 
\left(
e^{i k' x'} a^\dag(\vk') + e^{-i k' x'} a(\vk') 
\right) 
\, \ket{0} 
\notag \\
&=
\int \, \frac{d^3k}{(2 \pi)^3 2 k_0} \, \frac{d^3k'}{(2 \pi)^3 2 k'_0} \; 
e^{-i (k x-k'x')}
\bra{0} a(\vk) a^\dag(\vk') \ket{0} 
\qquad \text{with vacuum} \; \bra{0} a^\dag = 0 = a \ket{0} 
\notag \\
&=
\int \, \frac{d^3k}{(2 \pi)^3 2 k_0} \, \frac{d^3k'}{(2 \pi)^3 2 k'_0} \; 
e^{-i (k x-k'x')}
\bra{0} [a(\vk) a^\dag(\vk')] \ket{0} 
\notag \\
&=
\int \, \frac{d^3k}{(2 \pi)^3 2 k_0} \; e^{-i k (x-x')} \braket{0}{0}
\qquad \qquad \qquad \text{with} \left[ a, a^\dag \right] = (2\pi)^3 2 k_0 \; \delta^3(\vk - \vk')
\notag \\
&=
\int \, \frac{d^3k}{(2 \pi)^3}\; e^{i \vk (\vx-\vx')} \; 
        \frac{e^{-i k_0 (x_0-x_0')}}{2 k_0}  \; .
\label{eq:qft_prop1}
\end{alignat}
In the integral $k_0$ is given by its the on--shell value, so for $\vk
= \vk'$ we also have $k_0 = k_0'$.  Under the assumption $x_0 > x'_0$
the last ratio under the integral can be viewed as the result of a
contour integration over the $k_0$ component, to allow us to write the
propagator as a four-dimensional integral over $d^4k$. We
discuss this integral in detail in Section~\ref{sec:qcd_bw}, where we find in
Eq.\eqref{eq:prop_final}
\begin{alignat}{5}
\frac{i}{2 \pi} \int dk_0 \frac{e^{-i k_0 (x_0 - x'_0)}}{k^2 - m^2 + i \varepsilon}
= \theta(x_0 - x'_0) \frac{e^{-i k_0^\text{(os)} (x_0 - x'_0)}}{2 k_0^\text{(os)}}
+ \theta(x'_0 - x_0) \frac{e^{i k_0^\text{(os)} (x_0 - x'_0)}}{2 k_0^\text{(os)}} \; .
\end{alignat}
As in Eq.\eqref{eq:qft_prop1} but now explicitly denoted
$k_0^\text{(os)} = \sqrt{\vk^2 + m^2}$ is the on--shell value. Using
this result and slightly abusing our notation by now writing $k_0$ for
the zero component of the four-dimensional $k$ integration we can write
\begin{alignat}{5}
\bra{0} \phi(x) \phi(x') \ket{0} \Bigg|_{x_0>x'_0}
=& 
\int \, \frac{d^3k}{(2 \pi)^3}\; e^{i \vk (\vx-\vx')} \; 
        \frac{e^{-i k_0 (x_0-x_0')}}{2 k_0}  \; 
        \theta(x_0 - x'_0)
\notag \\
=& 
i \int \, \frac{d^4k}{(2 \pi)^4}\; e^{i \vk (\vx-\vx')} \; 
        \frac{e^{-i k_0 (x_0 - x'_0)}}{k^2 - m^2 + i \varepsilon}
\notag \\
=& 
i \int \, \frac{d^4k}{(2 \pi)^4}\; e^{-i k (x-x')} \; 
        \frac{1}{k^2 - m^2 + i \varepsilon} \; .
\end{alignat}
Similarly, we can show the same relation for $x_0 < x'_0$, picking up
the other theta function and returning for the combination of the two
\begin{alignat}{5}
\Delta(x-x') = 
- \int \, \frac{d^4k}{(2 \pi)^4}\; e^{-i k (x-x')} \; 
          \frac{1}{k^2 - m^2 + i \varepsilon} \; .
\end{alignat}
This Feynman propagator is a Green function for the Klein--Gordon
equation Eq.\eqref{eq:qft_kge}, which we can explicitly confirm
to read
\begin{alignat}{5}
\left( \Box + m^2 \right) \Delta(x-x') &= - \int \frac{d^4 k}{(2 \pi)^4} \; 
  \left( \Box + m^2 \right) e^{-i k \cdot (x-x')} \; \frac{1}{k^2-m^2}
 \notag \\
 &= \int \frac{d^4 k}{(2 \pi)^4} \; 
  \left( (i k)^2 + m^2 \right) e^{-i k \cdot (x-x')} \; \frac{(-1)}{k^2-m^2}
 \notag \\
 &= \int \frac{d^4 k}{(2 \pi)^4} \; 
  e^{-i k \cdot (x-x')} 
 = \delta^4 (x-x') \; .
\end{alignat}
This concludes our discussion of the bosonic propagator. Using a
scalar field instead of a vector field we have shown how the field can
be expressed in terms of creation and annihilation operators and what
the commutation rules for the scalar fields as well as for the
creation and annihilation operators are. A major problem arises when
we sandwich these operators between vacuum states, which means that
such insertions have to be normal--ordered. In addition, the time--ordered 
correlation function of two scalar fields is the Feynman
propagator, defining an inverse of the Klein--Gordon equation over
the entire position space.

All these properties we will later use for the photon field. The
photon $A^\mu$ is a vector field, where each of the components obey
the Klein--Gordon equation. The commutation relations and the photon
propagator will not change, they will simply be dressed up by
factors $g_{\mu \nu}$ where appropriate. For the propagator this
generalization is strictly speaking gauge dependent, $g^{\mu \nu}$
corresponds to Feynman gauge. Nevertheless, from
Section~\ref{sec:qcd_bw} we know that additional terms from other
gauge choices do not contribute to our scattering amplitude.\bigskip

The second object we need to describe for our scattering process are
the external fermion fields. Most generally, scalars and gauge bosons
are not the only particle we find in Nature. Matter particles or
fermions, like electrons or quarks have a different equation of motion
and a different contribution to the Lagrangian. No matter how it
looks, the equation of motion for fermion fields $\psi$ has to be
invariant under a Lorentz transformation
\begin{alignat}{5}
\psi(x)  \to \psi'(x) = \Lambda_{1/2} \, \psi(\Lambda^{-1}x) \; ,
\end{alignat}
where $\Lambda_{1/2}$ is a different representation of the Lorentz
transformations. It is called \underline{spinor representation} and we
can define it using the four Dirac matrices $\gamma^\mu$ with their
anti--commutator Clifford algebra
\begin{alignat}{5}
\{ \gamma^\mu, \gamma^\nu \} = 2 g^{\mu \nu} \, \one \; .
\end{alignat}
The unit matrix has the same size as the $\gamma$ matrices. That we
usually write them as $4\times4$ matrices has nothing to do with the
number of --- also four --- $\gamma$ matrices. The explicit form of
the $\gamma_\mu$ matrices is not relevant because it never appears in
actual calculations. All we need is a few trace relations arising from
their commutators. A representation of the Lorentz algebra in terms of
the Dirac matrices is
\begin{alignat}{5}
S^{\mu \nu} = \frac{i}{4} \, \left[ \gamma^\mu, \gamma^\nu \right] 
\qquad \qquad \text{implying} \quad 
\Lambda_{1/2} = \exp \left( -\frac{i}{2} \omega_{\mu \nu} S^{\mu \nu} \right) 
\; .
\end{alignat}
Using the transformation property $\Lambda_{1/2}^{-1} \gamma^\mu
\Lambda_{1/2} = {\Lambda^\mu}_\nu \gamma^\nu$ we can postulate an
equation of motion for the fermions, the \underline{Dirac equation}
\begin{alignat}{5}
\left( i \gamma^\mu \p_\mu - m \one
\right) \, \psi(x) \equiv
\left( i \slashchar{\p} - m \one
\right) \, \psi(x) = 0 \; .
\label{eq:qft_dirac_eq}
\end{alignat}
The unit matrix in the mass term is a four-by-four matrix, just like
the Dirac matrices.  Of course, we need to check that this equation is
invariant under Lorentz transformations, keeping in mind that
$\Lambda_{1/2}$ commutes with everything except for the Dirac matrices
\begin{alignat}{5}
\left( i \gamma^\mu \p_\mu - m \one \right) \, \psi(x) 
&\to 
\left( i \gamma^\mu {(\Lambda^{-1})^\nu}_\mu \p_\nu 
       - m \one \right) \, \Lambda_{1/2} \psi(\Lambda^{-1}x)  
\notag \\
&= 
\Lambda_{1/2} \Lambda_{1/2}^{-1}
\left( i \gamma^\mu {(\Lambda^{-1})^\nu}_\mu \p_\nu 
       - m \one \right) \, \Lambda_{1/2} \psi(\Lambda^{-1}x)  
\notag \\
&= 
\Lambda_{1/2} 
\left( i \Lambda_{1/2}^{-1} \gamma^\mu  \Lambda_{1/2} {(\Lambda^{-1})^\nu}_\mu \p_\nu 
       - m \one \right) \, \psi(\Lambda^{-1}x)  
\notag \\
&= 
\Lambda_{1/2} 
\left( i {\Lambda^\mu}_\rho \gamma^\rho {(\Lambda^{-1})^\nu}_\mu \p_\nu 
       - m \one \right) \, \psi(\Lambda^{-1}x)  
\notag \\
&= 
\Lambda_{1/2} 
\left( i {g_\nu}^\rho \gamma^\rho \p_\nu 
       - m \one \right) \, \psi(\Lambda^{-1}x)  
\notag \\
&= 
\Lambda_{1/2} 
\left( i \gamma^\nu \p_\nu 
       - m \one \right) \, \psi(\Lambda^{-1}x)  
= 0 \; .
\end{alignat}
An interesting side remark is that we can multiply the Dirac equation
with $(-i \gamma^\mu \p_\mu -m \one)$ and obtain the Klein--Gordon
equation $(\p^2 +m^2) \psi = 0$, which will be useful when we
construct a fermion propagator.

An additional problem is that for example to define a mass term in the
Lagrangian we need to form Lorentz scalars or invariants out of the
fermion fields $\psi$. Naively, $(\psi^\dag \psi)$ would work if
the Lorentz transformations in $(\psi^\dag \Lambda_{1/2}^\dag
\Lambda_{1/2} \psi)$ cancelled. Unfortunately $\Lambda_{1/2}$ is not a
unitary transformation, which means we have to go beyond $(\psi^\dag
\psi)$. One can show that the \underline{Dirac adjoint}
\begin{alignat}{5}
\psib = \psi^\dag \gamma^0 
\qquad \qquad \text{with} \quad 
\psib \to \psib \Lambda_{1/2}^{-1} 
\qquad \text{and} \quad 
\psib \psi \to \psib \psi
\end{alignat}
has the correct transformation property. This allows us to write down
the Lagrangian for a fermion field
\begin{alignat}{5}
\lag \supset \psib ( i \slashchar{\p} - m \one ) \psi \; .
\end{alignat}
Just like for bosons we can show that this term produces the Dirac
equation of motion. Because we will later need the fermion--photon
interaction in the form of a Hamiltonian or Lagrangian we introduce
the convenient form of the covariant derivative
\begin{alignat}{5}
\lag \supset 
\psib \left( i \slashchar{D} - m \one \right) \psi
\equiv 
\psib \left( i ( \slashchar{\p}  + i e Q \slashchar{A} ) - m \one \right) \psi
=
\psib \left( i ( \slashchar{\p} - m \one \right) \psi
+ e Q A_\mu \, \psib \gamma^\mu \psi
\label{eq:qft_covariant}
\end{alignat}
The last term describes the coupling of a vector photon field $A_\mu$
to a vector-like expression $\psib \gamma^\mu \psi$ which we call a
vector current.\bigskip

Everything written above we could as well apply to classical
fields. Just like in the bosonic case we need to define the Dirac
field operator in terms of plane wave coefficients
\begin{alignat}{5}
\psi(x) &= 
\int \, \frac{d^3k}{(2 \pi)^3 2 k_0} \; 
\sum_\text{spin} \left(
  e^{i k x} v_s(k) b^\dag(\vk) 
+ e^{-i k x} u_s(k) a(\vk) 
\right) 
\notag \\
\psib(x) &= 
\int \, \frac{d^3k}{(2 \pi)^2 2 k_0} \; 
\sum_\text{spin} \left( e^{ikx} \bar{u}(k) a^\dag(\vk) 
        + e^{-ikx} \bar{v}(k) b(\vk)  \right) \; ,
\label{eq:qft_fermion1}
\end{alignat}
where the fermion spin can be $\pm 1/2$. In the absence of any other
constraints we have four generating operators, $a,a^\dag, b,
b^\dag$. Acting on the vacuum $a$ and $b$ are annihilation
operators and $a^\dag, b^\dag$ are creation operators, $a$ for
particles and $b$ for anti--particles. These operators only depend on
the momentum three-vector because the fourth component follows from the
dispersion relation of the on--shell particles. The way we introduce
the spinors $u$ and $v$ the same would hold for them, but there are
instances where we use them also for off--shell states and we have to
take into account their dependence on the complete momentum
four-vector. Again following causality we postulate the anti--commutation
relations, for example at equal time $t = t'$
\begin{alignat}{5} 
\left\{ \psi(x), \psi(x') \right\}
&= 0 = 
\left\{ \psib(x), \psib(x') \right\}
\notag \\
\left\{ \psi(x), \psib(x') \right\}
&= \gamma^0  \delta^3(\vx - \vx') \; .
\end{alignat}
Trying the same thing with commutators simply does not work, as
Michael Peskin nicely shows in his book.  These anti--commutators we
can link to anti--commutators for the creation and annihilation
operators in momentum space
\begin{alignat}{5}
\left\{ a_r(\vk), a^\dag_s(\vk') \right\} 
&= \delta_{rs} (2\pi)^3 \, 2 k^0 \; \delta^3(\vk - \vk')
\notag \\
\left\{ b_r(\vk), b^\dag_s(\vk') \right\} 
&= \delta_{rs} (2\pi)^3 \, 2 k^0 \; \delta^3(\vk - \vk')
\notag \\
\left\{ a_r(\vk), b_s(\vk') \right\} 
&= 0 \qquad \qquad \text{for all other $a^{(\dag)}$ and $b^{(\dag)}$,}
\end{alignat}
provided we know the spin sums for the spinors $u$ and $v$
and their Dirac adjoints
\begin{alignat}{5}
\sum_\text{spin} u_s(k) \bar{u}_s(k) &= \slashchar{k} + m \one 
\notag \\
\sum_\text{spin} v_s(k) \bar{v}_s(k) &= \slashchar{k} - m \one  \; .
\label{eq:qft_spin_sum}
\end{alignat}
Strictly speaking, $\slashchar{k}$ is a $(4 \times 4)$ matrix, so in
the mass term we need to include a unit matrix which is often
omitted. Most of the time this is not a problem, unless we for example
compute traces of chains of Dirac matrices and need to remember that
$\tr \one \ne 1$. To produce such a matrix $u$ and $v$ are four-dimensional
objects.

These anti--commutator relation have the fundamental consequence that
for two fermion states generated from the vacuum we have to keep track
of the ordering
\begin{alignat}{5}
\ket{e^-(k,r) e^-(k',r')}
= a^\dag_r(k) a^\dag_{r'}(k') \ket{0} 
= - a^\dag_{r'}(k') a^\dag_r(k) \ket{0} 
= - \ket{e^-(k',r') e^-(k,r)} \; .
\end{alignat}
This factor $(-1)$ needs to be taken into account when we apply 
normal--ordering to fermions. For $k=k'$ and $r=r'$ this leads to Pauli's
exclusion principle: two identical fermion states cannot co-exist
exactly in the same point.

Again, this is all we need to say about fermions to compute our
electron--positron scattering process. We know the Dirac equation and
the corresponding contribution to the Lagrangian, including the
definition of the Dirac adjoint to construct Lorentz scalars. The
quantized fermion field obeys anti--commutation relations, as do its
creation and annihilation operators. To link them we need to know the
form of the spin sums over the spinors $u$ and $v$.\bigskip

To illustrate how we can compute a \underline{transition
  amplitude} without using Feynman rules we use our usual scattering
process
\begin{alignat}{5}
\boxed{
e^-(k_1,s_1) + e^+(k_2,s_2) \to q(p_1,s_3) + \bar{q}(p_2,s_4) 
} \; ,
\end{alignat}
where $k_j$, $p_j$ and $s_j$ are the four-momenta and spin orientations
of the external fermions. In the future, or more specifically 
asymptotically for $t \to + \infty$, the initial state $\lim_{t \to -
  \infty} \ket{t} \equiv \ket{i}$ will have evolved into the final state
$\lim_{t \to \infty} \ket{t} = \sop \ket{i}$ via a yet unknown linear
operator $\sop$. To describe this scattering into a
final state $\bra{f}$ we need to compute the transition
amplitude
\begin{alignat}{5}
S \equiv
\braket{f}{t\to \infty} = 
\bra{f} \sop \ket{i} = 
\bra{q_3 \bar{q}_4} \sop \ket{e_1^+ e_2^-}  =
\bra{0} a_3 b_4 \, \sop \, a^\dag_1 b^\dag_2 \ket{0} 
\; .
\end{alignat}
We use one index to indicate the momenta and spins of the
external particles.

The transition matrix element $\sop$ we compute from the time evolution of the initial state $i
\p_t \ket{t} = \hop(t) \ket{t}$ in the interaction picture with
a time dependent Hamilton operator. The evolution equation then reads
\begin{alignat}{5}
\ket{t} 
&= \ket{i}
    - i \int_{-\infty}^t dt' \, \hop(t') \ket{t'} 
\notag \\
&= \ket{i}
    - i \int_{-\infty}^t dt' \, \hop(t') 
   \left[  \ket{i}
    - i \int_{-\infty}^{t'} dt'' \, \hop(t'') \ket{t''} 
   \right]
\notag \\
&= \ket{i}
    - i \int_{-\infty}^t dt' \, \hop(t') 
   \left[  \ket{i}
    - i \int_{-\infty}^{t'} dt'' \, \hop(t'') 
   \left[ \ket{i}
    - i \int_{-\infty}^{t''} dt''' \, \hop(t''') \ket{t'''} 
   \right]
   \right] \; ,
\end{alignat}
just inserting the same evolution twice. The problem with
this form is that it still involves $\ket{t'''}$ at the very end. What
we instead want is something that is only proportional to
$\ket{i}$. We can achieve this by looking at the integration
boundaries: the integration range becomes smaller in each step of
primed variables. In the limit of infinitely many insertions the
remaining integrals should be over less and less time, starting at
$t \to -\infty$. Neglecting higher powers of the Hamilton
operator $\hop$ or, as we will see later, neglecting powers of a
coupling mediating the interaction between the states involves we can
rewrite this form as
\begin{alignat}{5}
\ket{t \to \infty} 
&= \ket{i}
    + (-i) \; \int_{-\infty}^\infty dt' \, \hop(t') \; \ket{i} 
    + (-i)^2 \; \int_{-\infty}^\infty dt' \hop(t')
             \; \int_{-\infty}^{t'} dt'' \, \hop(t'') \; \ket{i}
    + \ope(\hop^3)
\notag \\
&= \ket{i}
    + (-i) \; \int_{-\infty}^\infty dt' \, \hop(t') \; \ket{i} 
    + (-i)^2 \; \int_{-\infty}^\infty dt' dt'' \, \theta(t'-t'') 
             \;  \hop(t') \hop(t'') \; \ket{i}
    + \ope(\hop^3)
\notag \\
&= \ket{i}
    + (-i) \; \int_{-\infty}^\infty dt' \, \hop(t') \; \ket{i} 
    + \frac{(-i)^2}{2} \; \int_{-\infty}^\infty dt' dt'' \, 
      T ( \hop(t') \hop(t'') ) \; \ket{i}
    + \ope(\hop^3)  
\notag \\
&\really \sop \, \ket{i} \; ,
\label{eq:qft_perturbe}
\end{alignat}
where the time--ordered product only contributes a factor two for two
identical and hence commuting operators.  The last line of
Eq.\eqref{eq:qft_perturbe} with the time--ordered Hamilton operators 
and the corresponding factor $1/(2!)$ is
important because it means that we can sum $S$ to an exponential series 
\begin{alignat}{5}
\sop = T \left( e^{-i \int  dt \, \hop(t)} \right) \; ,
\end{alignat}
and ensure that it generates a unitary transformation.  For our
computation we will be fine with the quadratic term which we
explicitly list.\bigskip

The form of the interaction Hamiltonian for two fermionic currents
each involving a different particle species $j$ with charge $Q_j$
follows from the covariant derivative Eq.\eqref{eq:qft_covariant}
\begin{alignat}{5}
\hop_\text{int}(t) 
= - \int d^3x \, \lag_\text{int}(x)
\supset 
\sum_{j} -e Q_j \int d^3x \, A_\mu \, : \psib_j \gamma^\mu \psi_j : \; ,
\end{alignat}
in terms of the four-vector $x$ including its first entry $t = x_0$, the
fermion current $(\psib \gamma_\mu \psi)$, and the photon vector field
$A_\mu$. The current is normal--ordered, which means that annihilation operators
$a,b$ are moved to the right and creation operators $a^\dag,
b^\dag$ are moved to the left. For fermions an exchange of fields includes a
minus sign, while for bosons the two operators are simply exchanged.

To connect four creation and annihilation operators arising from the external
states we need four such operators from $\sop$, which means the first
term which will contribute to the scattering process is the quadratic
term in $\hop$. The two Hamiltonians contribute one electron and one
quark current each. It is not hard to check that the two possible assignments give
the same result, so we only follow one of them and include an
additional factor two in the formula for
\begin{alignat}{5}
S 
&= 2 \times \frac{(-i)^2}{2} \; \int dt' dt''
\bra{0} a_3 b_4 \,       
T ( \hop(t') \hop(t'') ) 
\, a^\dag_1 b^\dag_2 \ket{0} 
\notag \\
&= - e^2 Q_e Q_q \; \int d^4x' d^4x''
\bra{0} a_3 b_4 \,       
T \left( :  \psib_q(x') \gamma_\mu \psi_q(x') : \, A^\mu(x')
         :  \psib_e(x'') \gamma_\nu \psi_e(x'') : \, A^\nu(x'')
  \right)
\, a^\dag_1 b^\dag_2 \ket{0} 
\notag \\
&= - e^2 Q_e Q_q \; \int d^4x' d^4x''
\bra{0} T \left( A^\mu(x') A^\nu(x'') \right) \ket{0} 
\notag \\
& \phantom{haaallllllllooooooooooooo}
\bra{0} a_3 b_4 
T \left( : \psib_q(x') \gamma_\mu \psi_q(x') : \,
         : \psib_e(x'') \gamma_\nu \psi_e(x'') :
  \right)
\, b^\dag_1 a_2^\dag \ket{0} \; .
\label{eq:nofeyn_trans1}
\end{alignat}
The first of the time--ordered products is a gauge boson propagator in
Feynman gauge
\begin{alignat}{5}
\bra{0} T \left( A^\mu(x') A^\nu(x'') \right) \ket{0} 
= -i \int \frac{d^4q}{(2\pi)^4} \; e^{-iq(x'-x'')} \frac{g^{\mu \nu}}{q^2}
\equiv g^{\mu \nu} \Delta(x'-x'') \; .
\end{alignat}
We still have to evaluate the second time--ordered product by properly
combining the creation and annihilation operators with the fermion
fields. For example, we can write
\begin{alignat}{5}
\bra{0} \psi(x) a^\dag(\vp) \ket{0} 
&= \int \frac{d^3k}{(2 \pi)^2 2 E} \sum_\text{spins}
\bra{0} \left( e^{ikx} b^\dag(\vk) v(k)
             + e^{-ikx} a(\vk) u(k) \right) \; 
a^\dag(\vp) \ket{0}
\notag \\
&= \int \frac{d^3k}{(2 \pi)^2 2 E} \sum_\text{spins}
e^{-ikx} u(k) \; \bra{0} a(\vk) a^\dag(\vp) \ket{0}
\notag \\
&= \int \frac{d^3k}{(2 \pi)^2 2 E} \sum_\text{spins}
e^{-ikx} u(k) \; 
\bra{0} [ a(\vk) a^\dag(\vp) ] \ket{0}
\notag \\
&= \int \frac{d^3k}{(2 \pi)^2 2 E} \sum_\text{spins}
e^{-ikx} u(k) \; 
(2 \pi)^3  2 E \, \delta^3(\vk - \vp) \, \braket{0}{0}
\; \sim \; e^{-ipx} \, u(p) \; .
\end{alignat}
In the last step we remove the spin sum to later add it
to the transition amplitude.  The normal--ordering of the fermion
currents in this case is never really needed after properly defining
the interaction Hamiltonian.  Correspondingly, we find the other
non-zero normal--ordered combinations
\begin{alignat}{5}
\bra{0}  b(\vp) \psi(x) \ket{0} &= e^{ipx} \, v(p) 
\notag \\
\bra{0} a(\vp) \psib(x)  \ket{0} &= e^{ipx} \, \bar{u}(p) 
\notag \\
\bra{0} \psib(x) b^\dag(\vp)  \ket{0} &= e^{-ipx} \, \bar{v}(p) \; .
\end{alignat}
All other combinations of $a,a^\dag,b,b^\dag$ with $\psi$ and
$\psib$ vanish when we sandwich them between vacua.  Before we
contract the four creation and annihilation operators we need to keep
in mind that electromagnetic currents only link one set of particle,
they do not convert quarks into electrons. This limits the number of
permutations we need to take into account. We find one unique
non--vanishing combination of external states and current creators and
annihilators, namely matching $a_3 \psib$ and $b_4 \psib$ for the
quarks and $\psib b_1^\dag$ as well as $\psi a_1^\dag$ for the
electrons.
\begin{alignat}{5}
\bra{0} a_3 b_4 : \psib_q(x') \gamma_\mu \psi_q(x') : \,
                : \psib_e(x'') \gamma_\nu \psi_e(x'') :
\, a^\dag_1 b^\dag_2 \ket{0} 
&= e^{i (p_1+p_2) x'} \bar{u}_3 \gamma_\mu v_4 \; 
   e^{-i (k_1+k_2) x''} \bar{v}_2 \gamma_\nu  u_1 \; .
\label{eq:qft_nofeyn1}
\end{alignat}
Inserting the different contributions into Eq.\eqref{eq:nofeyn_trans1}
we find
\begin{alignat}{5}
S &=
- e^2 Q_e Q_q \; \int d^4x' d^4x''
\; \int \frac{d^4q}{(2\pi)^4} \; e^{iq(x'-x'')} \frac{-ig^{\mu \nu}}{q^2}
\; e^{i (p_1+p_2) x'} \bar{u}_3 \gamma_\mu v_4 \; 
   e^{-i (k_1+k_2) x''} \bar{v}_2 \gamma_\nu  u_1 
\notag \\
&= i e^2 Q_e Q_q \; 
\; \int \frac{d^4q}{(2\pi)^4} \; 
\bar{u}_3 \gamma_\mu v_4 
\frac{1}{q^2}
\bar{v}_2 \gamma^\mu  u_1 \;
\int d^4x' \; e^{i (q + p_1+p_2) x'} 
\int d^4x'' \; e^{-i (q + k_1+k_2) x''} 
\notag \\
&= i e^2 Q_e Q_q \; (2 \pi)^8
\; \int \frac{d^4q}{(2\pi)^4} \; 
\bar{u}_3 \gamma_\mu v_4 
\frac{1}{q^2}
\bar{v}_2 \gamma^\mu  u_1 \;
\delta^4(q+k_1+k_2)
\delta^4(-q-p_1-p_2)
\notag \\
&= i (2 \pi)^4
\delta^4(k_1+k_2-p_1-p_2) \;
e^2 Q_e Q_q \; 
\bar{u}_3 \gamma_\mu v_4 
\frac{1}{(k_1+k_2)^2}
\bar{v}_2 \gamma^\mu  u_1 \; .
\label{eq:qft_nofeyn1a}
\end{alignat}
Stripping off unwanted prefactors we can define the transition
\underline{matrix element} for quark--antiquark production in QED as
\begin{alignat}{5}
\mat = 
e^2 Q_e Q_q \; 
( \bar{u}_3 \gamma_\mu v_4 )
\frac{1}{(k_1+k_2)^2}
( \bar{v}_2 \gamma^\mu  u_1 ) \; ,
\end{alignat}
with $(k_1+k_2)^2 = (p_1+p_2)^2$. This matrix element or transition
amplitude we have to square to compute the transition
probability. Part of the squaring is the sum over all spins which uses
the spin sums Eq.\eqref{eq:qft_spin_sum} to get rid of the spinors and
then some trace rules to get rid of all Dirac matrices. Neither for
the spinors nor for the Dirac matrices we need to know their explicit
form
\begin{alignat}{5}
|\mat|^2 
=& \sum_\text{spin, color} 
e^4 Q_e^2 Q_q^2 \; 
\frac{1}{(k_1+k_2)^4} \;
( \bar{v}_4 \gamma_\nu u_3 ) 
( \bar{u}_1 \gamma^\nu  v_2 ) \;
( \bar{u}_3 \gamma_\mu v_4 ) \; 
( \bar{v}_2 \gamma^\mu  u_1 ) 
\notag \\
=& e^4 Q_e^2 Q_q^2 N_c \; 
\frac{1}{(k_1+k_2)^4} \;
\sum_\text{spin} 
( \bar{v}_4 \gamma_\nu u_3 ) 
( \bar{u}_1 \gamma^\nu  v_2 ) \;
( \bar{u}_3 \gamma_\mu v_4 ) \; 
( \bar{v}_2 \gamma^\mu  u_1 )  \; .
\end{alignat}
The color factor $N_c$ is the number of color singlet states we can
form out of a quark and an antiquark with 
opposite color charges. Because color only appears in the final state
we sum over all possible color states or multiply by $N_c$. In the
next step we can observe how the crucial structure of transition
amplitudes with external fermions, namely traces of chains of Dirac
matrices, magically form:
\begin{alignat}{5}
|\mat|^2 
=& e^4 Q_e^2 Q_q^2 N_c \; 
\frac{1}{(k_1+k_2)^4} \;
\sum_\text{spin} 
(\bar{v}_4)_i (\gamma_\nu)_{ij} (u_3)_j
(\bar{u}_3)_k (\gamma_\mu)_{kl} (v_4)_l \cdots 
\qquad &&\text{for one trace}
\notag \\
=& e^4 Q_e^2 Q_q^2 N_c \; 
\frac{1}{(k_1+k_2)^4} \;
\left( \sum_\text{spin} (v_4)_l (\bar{v}_4)_i \right) \; 
\left( \sum_\text{spin} (u_3)_j (\bar{u}_3)_k \right) \;
 (\gamma_\nu)_{ij} 
 (\gamma_\mu)_{kl} \cdots 
\notag \\
=& e^4 Q_e^2 Q_q^2 N_c \; 
\frac{1}{(k_1+k_2)^4} \;
(\slashchar{p}_4)_{li}
(\slashchar{p}_3)_{jk}
(\gamma_\nu)_{ij} 
(\gamma_\mu)_{kl} \cdots 
\qquad &&\text{using Eq.\eqref{eq:qft_spin_sum}} \notag \\
=& e^4 Q_e^2 Q_q^2 N_c \; 
\frac{1}{(k_1+k_2)^4} \;
\tr \left( \slashchar{p}_4 \gamma_\nu \slashchar{p}_3 \gamma_\mu \right) \; 
\tr \left( \slashchar{p}_1 \gamma^\nu \slashchar{p}_2 \gamma^\mu \right) 
\qquad &&\text{both traces again.}
\end{alignat}
In the final step we need to use a know expression for the Dirac
trace. More complicated and longer traces become very complicated very
fast and we use FORM\index{FORM}
to evaluate them on the computer. We find
\begin{alignat}{5}
|\mat|^2 
=& e^4 Q_e^2 Q_q^2 N_c \; 
\frac{1}{(k_1+k_2)^4} \;
4 \left( p_{3 \nu} p_{4 \mu} + p_{3 \mu} p_{4 \nu} - g_{\mu \nu} (p_1 p_2) \right) \;
4 \left( k_1^\nu k_2^\mu + k_1^\mu k_2^\nu - g_{\mu \nu} (k_1 k_2) \right)
\notag \\
=& 16 e^4 Q_e^2 Q_q^2 N_c \; 
\frac{1}{(k_1+k_2)^4} \;
\left[ 
2 (k_1 p_1) (k_2 p_2) + 2 (k_1 p_2) (k_2 p_1) + 0 \times (p_1 p_2) (k_1 k_2)
\right]
\quad &&\text{with} \; g_{\mu \nu} g^{\mu \nu} =4 
\notag \\
=& 32 e^4 Q_e^2 Q_q^2 N_c \; 
\frac{1}{(k_1+k_2)^4} \;
\left[ (k_1 p_1) (k_2 p_2) + (k_1 p_2) (k_2 p_1) \right] \; ,
\label{eq:qft_nofeyn2}
\end{alignat}
This result for the matrix element and the matrix element squared is
the same expression as we obtain from Feynman rules in
Eq.\eqref{eq:dy_me4}.\bigskip

\underline{Feynman rules} are calculational rules which we can extract
from the Lagrangian. These building blocks representing external and
internal particles we combine to construct $\mat$. To compute the
matrix element in Eq.\eqref{eq:qft_nofeyn1} while skipping everything we did
to get this formula, we start by drawing Feynman diagrams 
representing all ways we can link the given initial and final
states through interaction vertices and internal propagators. For
$q\bar{q}$ production in $e^+ e^-$ scattering described by QED there
exists only one such diagram:
\begin{center} \begin{fmfgraph*}(80,60)
 \fmfset{arrow_len}{2mm}
 \fmfleft{in1,in2}
 \fmf{fermion,width=0.5}{in1,v1}
 \fmf{fermion,width=0.5}{in2,v1}
 \fmf{photon,width=0.5}{v1,v2}
 \fmf{fermion,width=0.5}{out1,v2}
 \fmf{fermion,width=0.5}{out2,v2}
 \fmfright{out1,out2}
\end{fmfgraph*} \end{center}
It consist of four external fermions, one internal photon, and two
interaction vertices. From Eq.\eqref{eq:qft_nofeyn1} we
know how to describe external fermions in terms of spinors:

\begin{center} \begin{tabular}{l|l}
symbol         & diagram 
\\ \hline
$u(p,s)$       & incoming fermion $(e^-,q)$ with momentum $p$ and spin $s$
\\
$\bar{v}(p,s)$ & incoming anti--fermion $(e^+,\bar{q})$
\\
$\bar{u}(p,s)$ & outgoing fermion $(e^-,q)$
\\
$v(p,s)$       & outgoing anti--fermion $(e^+,\bar{q})$
\end{tabular} \end{center}

Spin sums are the only way to get rid of spinors in the
computation. Equation~\eqref{eq:qft_spin_sum} shows that as long as we
neglect fermion masses the two spinors $u$ and $v$ for particles and
antiparticles are identical. To link external particles to each other
and to internal propagators we need vertices. In
Eq.\eqref{eq:qft_nofeyn1} we see that two fermions and a gauge boson
interact via a vector current proportional to $\gamma^\mu$. As a
convention, we add one factor $i$, so the vertex rule in QED  becomes
\begin{alignat}{5}
 i e Q_f \, \gamma^\mu
\qquad \qquad 
(f-\bar{f}-\gamma).
\end{alignat}
This factor $i$ we can consistently change for
all three-point and four-point vertices in our theory. Finally, there is the
intermediate photon which propagates between the $\gamma^\mu$ and the
$\gamma^\nu$ vertices. The wave line in the Feynman diagram
corresponds to
\begin{alignat}{5}
\frac{- i g^{\mu \nu}}{p^2 + i \epsilon} \; .
\end{alignat}
Again, the factor $-i$ is conventional. For a bosonic propagator
it does not matter in which direction the momentum flows. Blindly
combining these Feynman rules gives us directly
Eq.\eqref{eq:qft_nofeyn1}, so all we need to do is square the matrix
element, insert the spin sums and compute the Dirac trace.

Whenever we compute such a matrix element starting from a Feynman
diagram nothing tells us that the lines in the Feynman diagrams are
not actual physical states propagating from the left to the
right. Even including loop diagrams will still look completely
reasonably from a semi--classical point of view. Feynman rules define
an algorithm which hides all field theory input in the calculation of
scattering amplitudes and are therefore perfectly suited to compute
the differential and total cross sections on the computer.\bigskip

The vector structure of the QED couplings, for example mediated by a
covariant derivative Eq.\eqref{eq:qft_covariant} we did not actually
motivate. It happens to work on the Lagrangian level and agrees with
data, so it is correct. We can write a completely general interaction
of two fermions with a boson in terms of basis elements
\begin{alignat}{5} 
g \, \psib M \psi
= \sum_{\text{basis} \; j} g_j \, \psib M_j \psi \; .
\label{eq:qft_interaction}
\end{alignat}
For a real $(4 \times 4)$ matrix $M$ the necessary 16 basis
elements can be organized such that they are easy to keep
track of using Lorentz transformation properties. This eventually 
leads to the Fierz transformation used in 
Section~\ref{sec:higgs_technicolor}.\index{Fierz transformation}
The vector
$\gamma^\mu$ from the QED interaction gives us four such basis
elements, the unit matrix a fifth. Another six we already know as
well, they are the generators of the spinor representation
$[\gamma^\mu, \gamma^\nu]$. We can check that all of them are linearly
independent. Five basis elements in a handy form are still missing.

To define them, we need to know that there exists another $(4\times
4)$ matrix which is invariant under proper Lorentz
transformations. We can write it in terms of the four Dirac matrices
in two equivalent forms
\begin{alignat}{5}
\gamma_5 
= i \gamma^0 \gamma^1 \gamma^2 \gamma^3 \gamma^4 
= \frac{i}{4!} \, \epsilon_{\mu \nu \rho \sigma}
  \gamma^\mu \gamma^\nu \gamma^\rho \gamma^\sigma \; ,
\end{alignat}
using the totally anti--symmetric Levi--Civita tensor $\epsilon_{\mu \nu
  \rho \sigma}$.\index{Levi--Civita tensor} This form already shows a major
technical complication in dealing with $\gamma_5$: in other than four
space--time dimensions we do not know how to define the Levi--Civita
tensor, which means that for example for regularization purposes we
cannot analytically continue our calculation to $n = 4-2\epsilon$
dimensions. The main property of $\gamma_5$ is equivalent to that fact
that it is another basis element of our $(4 \times 4)$ matrices, it
commutes with the other four Dirac matrices $[ \gamma_\mu, \gamma_5] =
0$. Combining this new object as $(\gamma^\mu \gamma_5)$ and $i
\gamma_5$ gives us all 16 basis element for the interaction of two
spinors with a third scalar, vector, or tensor field:

\begin{center} \begin{tabular}{l|c|c}
& degrees of freedom  & basis elements $M_j$
\\ \hline
scalar & 1       & $\one$ 
\\[1mm]
vector & 4       & $\gamma^\mu$ 
\\[1mm]
pseudoscalar & 1 & $i \gamma_5$
\\[1mm]
axialvector & 4 & $\gamma^\mu \gamma_5$
\\[1mm]
tensor & 6       & $\dfrac{i}{2} [\gamma^\mu, \gamma^\nu]$ 
\end{tabular} \end{center}

The field indices need to contract with the indices of the object
$\psib M \psi$.  Again, the factors $i$ are conventional. In the
Standard Model as a fundamental theory, tensor interactions do not play
a major role. The reason is the dimensionality of the Lagrangian. The
mass dimension of a fermion field $\psi$ or $\psib$ is $m^{3/2}$ while
the mass dimension of a scalar field, a photon field, or a derivative
is $m$. For example from Eq.\eqref{eq:qft_covariant} we see that every
term in the QED Lagrangian is of mass dimension four. This is required
for a renormalizable fundamental field theory. Introducing a tensor
coupling we have to contract two indices, $\mu$ and $\nu$, and not
with the metric tensor. The only other objects coming to mind have
mass dimension $m^2$, which means that together with the fermion fields
the term in the Lagrangian has mass dimension of at least $m^5$ and is
therefore not allowed.\bigskip

The second obvious question is: what does it mean to include a factor
$\gamma_5$ in the interaction, \ie what distinguishes a scalar from a
pseudoscalar and a vector from an axialvector? We can
give an easy answer by defining three transformations of our field in space and
time. The first one is the \underline{parity transformation} $P$ which
mirrors the three spatial coordinates $(t, \vx) \to (t, -\vx)$. The
second is \underline{charge conjugation} $C$ which converts particles
into their anti--particles. Both of them leave the Dirac equation
intact and can be represented by a unitary transformation. The third
transformation is \underline{time reversal} $T$ which converts $(t,
\vx) \to (-t, \vx)$, also leaves the Dirac equation intact, but only
has an anti--unitary representation. Every single one of them is
violated in our Standard Model.

Instead of writing out the representation of these transformations in
terms of Dirac matrices we characterize them using the basic
interactions from Eq.\eqref{eq:qft_interaction}. Parity symmetry does
not allow any interaction including $\gamma_5$, which means it forbids 
pseudoscalars and axialvectors. Time reversal symmetry does not
allow any complex couplings $g_j$. Because any field theory described
by a Lagrangian not including some kind of external field is invariant
under $CPT$, and we have never observed $CPT$ violation, a combined
$CP$ invariance is essentially the same as $T$ invariance.\bigskip

To again look at the same question we rotate the $\{ \one, \gamma_5
\}$ plane and define the two $(4 \times 4)$ matrix valued objects
which we already use in Eq.\eqref{eq:def_prolr},\index{chiral projectors}
\begin{alignat}{5}
\prorl = \frac{1}{2} \, \left( \one \pm \gamma_5 \right) \; .
\end{alignat}
It is easy to show that the two are orthogonal 
\begin{alignat}{5}
\prol \pror
= \frac{1}{4} \, \left( \one-\gamma_5 \right) \, \left( \one+\gamma_5 \right)
= \frac{1}{4} \, \left( \one - \gamma_5^2  \right)
= 0 
\qquad \qquad \text{using} \quad
\gamma_5^2 = \one \; ,
\end{alignat}
and projectors 
\begin{alignat}{5}
\prorl^2 
= \frac{1}{4} \, \left( \one \pm 2 \gamma_5 + \gamma_5^2 \right) 
= \frac{1}{4} \, \left( 2 \one \pm 2 \gamma_5  \right) 
= \frac{1}{2} \, \left( \one \pm \gamma_5  \right) 
= \prorl \; .
\end{alignat}
In QED these combinations do not play any role. Their effect on
kinetic and mass terms we compute in Eqs.\eqref{eq:higgs_chiral_mass}
and~\eqref{eq:higgs_chiral_kin}.  Looking at interactions, we can for
example define a combined vector--axialvector coupling as $\gamma^\mu
\pm \gamma^\mu \gamma_5 = 2 \gamma_\mu \prorl$. Sandwiching this
coupling between fermion fields gives for example
\begin{alignat}{5}
\psib \, \gamma_\mu \prol \, \psi
&= \psib \, \gamma_\mu \prol^2 \, \psi
\notag \\
&= \psi^\dag \prol \, \gamma_0 \, \gamma_\mu \, \prol \psi
      \qqquad
      &\text{with}& \qquad \{ \gamma_5, \gamma_\mu \} = 0 
\notag \\
&= \left( \prol \psi \right)^\dag \gamma_0 \, \gamma_\mu \, \prol \psi
      \qqquad
      &\text{with}& \qquad \gamma_5^\dag = \gamma_5
\notag \\
&= \psib_L \, \gamma_\mu \, \psi_L
      \qqquad
      &\text{with}& \qquad \psi_{L,R} \equiv \prolr \, \psi \; .
\label{eq:qft_chiral}
\end{alignat}
If we call the eigenstates of $\prorl$ right handed and left
handed fermions $\psi_{L,R}$ this 
\underline{chirality} allows us to define a vector coupling
between only left handed fermions by combining the vector and the
axialvector couplings with a relative minus sign. The same is of
course true for right handed couplings. In
Section~\ref{sec:higgs_doublets} we show that kinetic terms can also
be defined independently for left and right handed fermions, while
mass terms or scalar interactions mix the two chiralities
\begin{alignat}{5}
\psib \, \slashchar{\p} \, \psi  
&= \psib_R \, \slashchar{\p} \, \psi_R +  \psib_L \, \slashchar{\p} \,  \psi_L
\notag \\
\psib \, \one \, \psi  
&= \psib_R \, \one \, \psi_L +  \psib_L \, \one \, \psi_R  \; .
\end{alignat}
In other words, we can write for example QED in terms of independent
left and right handed fields as long as we neglect all fermion
masses. This defines the chiral limit where the Lagrangian is
symmetric under $\psi_L \leftrightarrow \psi_R$. Introducing fermion
masses breaks this chiral symmetry, or turning the argument around, to
introduce fermion masses we need to combine a left handed and a right
handed fermion fields and give them one common \underline{Dirac
  mass}.\bigskip

At this stage it is not obvious at all what chirality means in physics
terms. However, we will see that in the Standard Model the left handed
fermions play a special role: the massive $W$ bosons only couple to
them and not to their right handed counter parts. So chirality is a
property of fermions known to one gauge interaction of the Standard
Model as part of the corresponding charge. The Higgs mechanism breaks
it and only leaves the QCD--like gauge symmetry intact.\bigskip

There exists a property which is identical to chirality for massless
fermions and has an easy physical interpretation: the
\underline{helicity}. It is defined as the projection of the particle
spin onto its three-momentum direction
\begin{alignat}{5}
h 
= \vec{s} \cdot \frac{\vp}{|\vp|}
= \left( \vec{s} + \vec{L} \right) \cdot \frac{\vp}{|\vp|}
= \vec{J} \cdot \frac{\vp}{|\vp|} 
\qquad \qquad 
\text{with} \quad \vp \perp \vec{L} \; ,
\end{alignat}
or equivalently the projection of the combined orbital angular
momentum and the spin on the momentum direction. From quantum
mechanics we know that there exist discrete eigenvalues for the $z$
component of the angular momentum operator, symmetric around
zero. Applied to fermions this gives us two spin states with the
eigenvalues of $h$ being $\pm 1/2$. Unfortunately, there is no really
nice way to show this identity. What we need to know is that the spin
operator is in general given by
\begin{alignat}{5}
\vec{s} = \gamma_5 \gamma^0 \, \vec{\gamma} \, .
\end{alignat}
We can show this by writing it out in terms of Pauli matrices, but we
will skip this here and instead just accept this general form.  We
then write the solution $\psi$ to the massless Dirac equation after
transforming it into momentum space $\psi(\vx) = u(\vp) \exp(-i p \cdot
x)$
\begin{alignat}{5}
\left( \gamma^0 p_0 - \vec{\gamma} \vp \, \right) \; u(\vp) 
&= 0
\notag \\
\gamma_5 \gamma^0 \; \gamma^0 p_0 \; u(\vp) 
&= \gamma_5 \gamma^0 \;  \vec{\gamma} \vp \; u(\vp) 
\notag \\
\gamma_5 p_0 \; u(\vp) 
&= \vec{s} \cdot \vp \; u(\vp)
\qquad \qquad \text{with} \; \left( \gamma^0 \right)^2 = \one
\notag \\
\gamma_5 \; u(\vp) 
&= \frac{\vec{s} \cdot \vp}{p_0} \; u(\vp)
\notag \\
\gamma_5 \; u(\vp) 
&= \pm \frac{\vec{s} \cdot \vp}{|\vp|} \; u(\vp)
 = \pm h \; u(\vp) \; .
\end{alignat}
In other words, the chirality operator $\gamma_5$ indeed gives us the
helicity of the particle state, modulo a sign depending on the sign of
the energy. For the helicity it is easy to argue why for massive
particles this property is not Lorentz invariant and hence not a well
defined property: massless particles propagate with the speed of
light, which means we can never boost into their rest frame or pass
them. For massive particles we can do that and this way switch the
sign of $\vp$ and the sign of $h$. Luckily, for almost all Standard
Model fermions we can at the LHC neglect their masses.